%% file: o3a_catalog.tex
\RequirePackage[pagewise]{lineno}

\documentclass[floatfix,aps,prd,twocolumn,letterpaper,lengthcheck,superscriptaddress,showpacs,amssymb,amsmath,amsfonts,nofootinbib,altaffilletter,nopreprintnumbers,showpacs,longbibliography]{revtex4-2}

\pdfoutput=1
\usepackage[utf8]{inputenc}
\usepackage{xspace}
\usepackage{acronym}
\usepackage{booktabs}
\usepackage{amsmath,amssymb,graphicx}
\usepackage{color}
\usepackage[colorlinks, linkcolor=blue, pdfborder={0 0 0}, breaklinks=true]{hyperref}
\usepackage[table]{xcolor}
\definecolor{lightgray}{gray}{0.9} 
\usepackage{tabularx}
\usepackage{multirow}
\usepackage{bigstrut}

\newcolumntype{Y}{>{\centering\arraybackslash}X}
\newcolumntype{Z}{>{\raggedright\arraybackslash}X}
\newcolumntype{K}{>{\raggedleft\arraybackslash}X}
\newcolumntype{U}{>{\hsize=1.01\hsize}Y}
\newcolumntype{V}{>{\hsize=1.2\hsize}Y}
\newcolumntype{W}{>{\hsize=0.71\hsize}Y}

\def\mystrut{\vrule height 9.3pt depth 3.1pt width 0pt}
\newenvironment{PE_table}{\setlength{\tabcolsep}{0pt}}{}
\newenvironment{event_table}{\setlength{\tabcolsep}{0pt}\setlength{\aboverulesep}{1pt}\setlength{\belowrulesep}{1pt}}{}

\usepackage{scrextend}
\usepackage{textgreek}
\usepackage{graphicx}
\usepackage[caption=false]{subfig}

\usepackage{makerobust}

\renewcommand{\today}{\number\day\space\ifcase\month\or
  January\or February\or March\or April\or May\or June\or
  July\or August\or September\or October\or November\or December\fi
  \space\number\year}

\definecolor{NOTECOLOR}{rgb}{0.4, 0.2, 0.1}
\definecolor{notecolor}{rgb}{0.4, 0.2, 0.1}
\newcommand{\note}[1]{}
\newcommand{\fixme}[1]{\textcolor{black}{#1}}

\begin{document}

\input{event_macros}
\input{user_macros}

\input{PE_macros}

\input{superlative_macros}

\input{detchar_macros}
\input{name_macros}
\input{JS_macros}
\input{localization_macros}
\input{BF_macros}
\input{IMRPv2_macros}
\input{GW190425A_macros}
\input{GW190425_lambda_macros_RW}
\input{chip_js}
\input{extremal_spin_macros}

\title{GWTC-2: Compact Binary Coalescences Observed by LIGO and Virgo
During the First Half of the Third Observing Run}

\input{o3a_catalog_authors.tex}

\date[\relax]{Compiled: \today}

\begin{abstract}
We report on gravitational wave discoveries from compact binary coalescences
detected by Advanced LIGO and Advanced Virgo in the first half of the third observing run (O3a) between \RUNSTART{} and
\fixme{\RUNEND{}}.  By imposing a false-alarm-rate threshold of two per year in
each of the four search pipelines that constitute our search, we present
\NUMEVENTS{} candidate gravitational wave events.  At this threshold, we expect a contamination fraction
of less than \fixme{\CONTAMINATION{}}.  Of these, \PREVIOUSLYREPORTED{}
candidate events were reported previously in near real-time through GCN Notices and
Circulars; \NEWEVENTS{} are reported here for the first time.  The catalog
contains events whose sources are black hole binary mergers up to a redshift of $\sim 0.8$,
as well as events
whose components could not be unambiguously identified as black holes or neutron stars.
    For the latter group, we are unable to determine the nature
based on estimates of the component masses and spins from gravitational wave data alone.
The range of candidate event masses which are
unambiguously identified as binary black holes (both objects $\geq 3~\Msun$)
is increased compared to GWTC-1, with total masses from
$\sim 14\Msun$ for \NAME{GW190924A}\ to $\sim 150\Msun$ for \NAME{GW190521A}.
For the first time, this
catalog includes binary systems with significantly asymmetric mass ratios, which had not been
observed in data taken before April 2019.
We also find that 11 of the \NUMEVENTS{} events detected since April 2019
have positive effective inspiral spins under our default prior (at 90\% credibility), while
none exhibit negative effective inspiral spin. 
Given the increased
sensitivity of Advanced LIGO and Advanced Virgo, the detection of \NUMEVENTS{}
candidate events in $\sim$26 weeks of data (\fixme{\DETECTIONRATE{}}) is consistent
with GWTC-1.
\end{abstract}


\maketitle

\acrodef{LSC}[LSC]{LIGO Scientific Collaboration}
\acrodef{LVC}[LVC]{LIGO Scientific Collaboration and Virgo Collaboration}
\acrodef{aLIGO}{Advanced Laser Interferometer Gravitational wave Observatory}
\acrodef{aVirgo}{Advanced Virgo}
\acrodef{LIGO}[LIGO]{Laser Interferometer Gravitational-Wave Observatory}
\acrodef{IFO}[IFO]{interferometer}
\acrodef{LHO}[LHO]{LIGO-Hanford}
\acrodef{LLO}[LLO]{LIGO-Livingston}
\acrodef{O2}[O2]{second observing run}
\acrodef{O1}[O1]{first observing run}

\acrodef{BH}[BH]{black hole}
\acrodef{BBH}[BBH]{binary black hole}
\acrodef{BNS}[BNS]{binary neutron star}
\acrodef{NS}[NS]{neutron star}
\acrodef{BHNS}[BHNS]{black hole--neutron star binaries}
\acrodef{NSBH}[NSBH]{neutron star--black hole binary}
\acrodefplural{NSBH}[NSBHs]{neutron star--black hole binaries}
\acrodef{PBH}[PBH]{primordial black hole binaries}
\acrodef{CBC}[CBC]{compact binary coalescence}
\acrodef{GW}[GW]{gravitational wave}

\acrodef{CWB}[cWB]{coherent WaveBurst}

\acrodef{SNR}[SNR]{signal-to-noise ratio}
\acrodef{FAR}[FAR]{false alarm rate}
\acrodef{IFAR}[IFAR]{inverse false alarm rate}
\acrodef{FAP}[FAP]{false alarm probability}
\acrodef{PSD}[PSD]{power spectral density}
\acrodefplural{PSD}[PSDs]{power spectral densities}

\acrodef{GR}[GR]{general relativity}
\acrodef{NR}[NR]{numerical relativity}
\acrodef{PN}[PN]{post-Newtonian}
\acrodef{EOB}[EOB]{effective-one-body}
\acrodef{ROM}[ROM]{reduced-order-model}
\acrodef{IMR}[IMR]{inspiral-merger-ringdown}

\acrodef{PDF}[PDF]{probability density function}
\acrodef{PE}[PE]{parameter estimation}
\acrodef{CL}[CL]{credible level}

\acrodef{EOS}[EOS]{equation of state}

\acrodef{LAL}[LAL]{LIGO Algorithm Library}

\acrodef{KLD}[KLD]{Kullback-Leibler divergence}
\acrodef{JSD}[JSD]{Jensen-Shannon divergence}

\newcommand{\PN}[0]{\ac{PN}\xspace}
\newcommand{\BBH}[0]{\ac{BBH}\xspace}
\newcommand{\BNS}[0]{\ac{BNS}\xspace}
\newcommand{\BH}[0]{\ac{BH}\xspace}
\newcommand{\NR}[0]{\ac{NR}\xspace}
\newcommand{\GW}[0]{\ac{GW}\xspace}
\newcommand{\SNR}[0]{\ac{SNR}\xspace}
\newcommand{\aLIGO}[0]{\ac{aLIGO}\xspace}
\newcommand{\PE}[0]{\ac{PE}\xspace}
\newcommand{\IMR}[0]{\ac{IMR}\xspace}
\newcommand{\PDF}[0]{\ac{PDF}\xspace}
\newcommand{\GR}[0]{\ac{GR}\xspace}
\newcommand{\PSD}[0]{\ac{PSD}\xspace}
\newcommand{\EOS}[0]{\ac{EOS}\xspace}
\newcommand{\LVC}[0]{\ac{LVC}\xspace}

\input{introduction.tex}

\input{instruments.tex}

\input{data.tex}

\input{methods.tex}

\input{candidate_list.tex}

\input{source_properties.tex}

\input{conclusion.tex}

\acknowledgments
\input{LVCack.tex}

The detection of the signals and subsequent significance evaluations in this catalog 
were performed with the \GSTLAL{}-based inspiral software 
pipeline~\cite{Sachdev:2019vvd, Hanna:2019ezx, Messick:2016aqy}
built on the \textsc{LALSuite} software library~\cite{LALSuite}, and with the 
\PYCBC{}~\cite{Allen:2005fk, Allen:2004gu, Canton:2014ena, Usman:2015kfa,Nitz:2017svb}
and \CWB{}~\cite{Klimenko:2004qh,Klimenko:2005xv,Klimenko:2006rh,
Klimenko:2011hz,Klimenko:2015ypf} packages. 
Parameter estimation was performed with
the \LALINFERENCE{} \cite{Veitch:2014wba} and LALSimulation libraries within
LALSuite \cite{LALSuite}, the 
Bilby and Parallel Bilby libraries~\cite{Ashton:2018jfp, Smith:2019ucc,Romero-Shaw:2020owr} 
using the dynesty nested sampling package~\cite{Speagle:2020},
and the RIFT library~\cite{Pankow:2015cra,Lange:2017wki,Wysocki:2019grj}.
PESummary was used to post-process and collate parameter estimation
results~\cite{Hoy:2020vys}.
Estimates of the noise spectra and glitch models were obtained using 
BayesWave~\cite{Cornish:2014kda,Littenberg:2015kpb,Chatziioannou:2019zvs}. 
The \CWB{} and BayesWave packages were also used 
to generate waveform reconstructions. 
Plots were prepared with Matplotlib~\cite{Hunter:2007ouj}, Seaborn~\cite{seaborn}
and gwpy~\cite{gwpy-software}. NumPy~\cite{numpy}
and SciPy~\cite{scipy} were used in the preparation of the
manuscript.

\appendix
\input{systematics_appendix.tex}

\input{waveform_recon_appendix.tex}
\input{source_parameters_appendix.tex}

\clearpage
\bibliography{o3catalog.bib}
\end{document}

%% file: event_macros.tex
\newcommand{\NUMEVENTS}{39}

\newcommand{\TIME}[1]{\IfEqCase{#1}{{GW190413A}{05:29:54}{GW190719A}{21:55:14}{GW190620A}{03:04:21}{GW190514A}{06:54:16}{GW190731A}{14:09:36}{GW190503A}{18:54:04}{GW190602A}{17:59:27}{GW190929A}{01:21:49}{GW190517A}{05:51:01}{GW190915A}{23:57:02}{GW190425A}{08:18:05}{GW190512A}{18:07:14}{GW190630A}{18:52:05}{GW190521A}{03:02:29}{GW190413B}{13:43:08}{GW190924A}{02:18:46}{GW190930A}{13:35:41}{GW190706A}{22:26:41}{GW190408A}{18:18:02}{GW190909A}{11:41:49}{GW190728A}{06:45:10}{GW190426A}{15:21:55}{GW190412A}{05:30:44}{GW190720A}{00:08:36}{GW190521B}{07:43:59}{GW190910A}{11:28:07}{GW190803A}{02:27:01}{GW190519A}{15:35:44}{GW190708A}{23:24:57}{GW190527A}{09:20:55}{GW190513A}{20:54:28}{GW190424A}{18:06:48}{GW190727A}{06:03:33}{GW190814A}{21:10:39}{GW190707A}{09:33:26}{GW190828A}{06:34:05}{GW190828B}{06:55:09}{GW190701A}{20:33:06}{GW190421A}{21:38:56}}}

\newcommand{\DATE}[1]{\IfEqCase{#1}{{GW190413A}{2019-04-13 05:29:54}{GW190719A}{2019-07-19 21:55:14}{GW190620A}{2019-06-20 03:04:21}{GW190514A}{2019-05-14 06:54:16}{GW190731A}{2019-07-31 14:09:36}{GW190503A}{2019-05-03 18:54:04}{GW190602A}{2019-06-02 17:59:27}{GW190929A}{2019-09-29 01:21:49}{GW190517A}{2019-05-17 05:51:01}{GW190915A}{2019-09-15 23:57:02}{GW190425A}{2019-04-25 08:18:05}{GW190512A}{2019-05-12 18:07:14}{GW190630A}{2019-06-30 18:52:05}{GW190521A}{2019-05-21 03:02:29}{GW190413B}{2019-04-13 13:43:08}{GW190924A}{2019-09-24 02:18:46}{GW190930A}{2019-09-30 13:35:41}{GW190706A}{2019-07-06 22:26:41}{GW190408A}{2019-04-08 18:18:02}{GW190909A}{2019-09-09 11:41:49}{GW190728A}{2019-07-28 06:45:10}{GW190426A}{2019-04-26 15:21:55}{GW190412A}{2019-04-12 05:30:44}{GW190720A}{2019-07-20 00:08:36}{GW190521B}{2019-05-21 07:43:59}{GW190910A}{2019-09-10 11:28:07}{GW190803A}{2019-08-03 02:27:01}{GW190519A}{2019-05-19 15:35:44}{GW190708A}{2019-07-08 23:24:57}{GW190527A}{2019-05-27 09:20:55}{GW190513A}{2019-05-13 20:54:28}{GW190424A}{2019-04-24 18:06:48}{GW190727A}{2019-07-27 06:03:33}{GW190814A}{2019-08-14 21:10:39}{GW190707A}{2019-07-07 09:33:26}{GW190828A}{2019-08-28 06:34:05}{GW190828B}{2019-08-28 06:55:09}{GW190701A}{2019-07-01 20:33:06}{GW190421A}{2019-04-21 21:38:56}}}

\newcommand{\NAME}[1]{\IfEqCase{#1}{{GW190413A}{GW190413\_052954}{GW190719A}{GW190719\_215514}{GW190620A}{GW190620\_030421}{GW190514A}{GW190514\_065416}{GW190731A}{GW190731\_140936}{GW190503A}{GW190503\_185404}{GW190602A}{GW190602\_175927}{GW190929A}{GW190929\_012149}{GW190517A}{GW190517\_055101}{GW190915A}{GW190915\_235702}{GW190425A}{GW190425}{GW190512A}{GW190512\_180714}{GW190630A}{GW190630\_185205}{GW190521A}{GW190521}{GW190413B}{GW190413\_134308}{GW190924A}{GW190924\_021846}{GW190930A}{GW190930\_133541}{GW190706A}{GW190706\_222641}{GW190408A}{GW190408\_181802}{GW190909A}{GW190909\_114149}{GW190728A}{GW190728\_064510}{GW190426A}{GW190426\_152155}{GW190412A}{GW190412}{GW190720A}{GW190720\_000836}{GW190521B}{GW190521\_074359}{GW190910A}{GW190910\_112807}{GW190803A}{GW190803\_022701}{GW190519A}{GW190519\_153544}{GW190708A}{GW190708\_232457}{GW190527A}{GW190527\_092055}{GW190513A}{GW190513\_205428}{GW190424A}{GW190424\_180648}{GW190727A}{GW190727\_060333}{GW190814A}{GW190814}{GW190707A}{GW190707\_093326}{GW190828A}{GW190828\_063405}{GW190828B}{GW190828\_065509}{GW190701A}{GW190701\_203306}{GW190421A}{GW190421\_213856}}}

\newcommand{\SID}[1]{\IfEqCase{#1}{{GW190413A}{S190413i}{GW190719A}{S190719an}{GW190620A}{S190620e}{GW190514A}{S190514n}{GW190731A}{S190731aa}{GW190503A}{S190503bf}{GW190602A}{S190602aq}{GW190929A}{S190929d}{GW190517A}{S190517h}{GW190915A}{S190915ak}{GW190425A}{S190425z}{GW190512A}{S190512at}{GW190630A}{S190630ag}{GW190521A}{S190521g}{GW190413B}{S190413ac}{GW190924A}{S190924h}{GW190930A}{S190930s}{GW190706A}{S190706ai}{GW190408A}{S190408an}{GW190909A}{S190909w}{GW190728A}{S190728q}{GW190426A}{S190426c}{GW190412A}{S190412m}{GW190720A}{S190720a}{GW190521B}{S190521r}{GW190910A}{S190910s}{GW190803A}{S190803e}{GW190519A}{S190519bj}{GW190708A}{S190708ap}{GW190527A}{S190527w}{GW190513A}{S190513bm}{GW190424A}{S190424ao}{GW190727A}{S190727h}{GW190814A}{S190814bv}{GW190707A}{S190707q}{GW190828A}{S190828j}{GW190828B}{S190828l}{GW190701A}{S190701ah}{GW190421A}{S190421ar}}}

\newcommand{\PUBLIC}[1]{\IfEqCase{#1}{{GW190413A}{\bf}{GW190719A}{\bf}{GW190620A}{\bf}{GW190514A}{\bf}{GW190731A}{\bf}{GW190503A}{}{GW190602A}{}{GW190929A}{\bf}{GW190517A}{}{GW190915A}{}{GW190425A}{}{GW190512A}{}{GW190630A}{}{GW190521A}{}{GW190413B}{\bf}{GW190924A}{}{GW190930A}{}{GW190706A}{}{GW190408A}{}{GW190909A}{\bf}{GW190728A}{}{GW190426A}{}{GW190412A}{}{GW190720A}{}{GW190521B}{}{GW190910A}{\bf}{GW190803A}{\bf}{GW190519A}{}{GW190708A}{\bf}{GW190527A}{\bf}{GW190513A}{}{GW190424A}{\bf}{GW190727A}{}{GW190814A}{}{GW190707A}{}{GW190828A}{}{GW190828B}{}{GW190701A}{}{GW190421A}{}}}

\newcommand{\INSTRUMENTS}[1]{\IfEqCase{#1}{{GW190413A}{HL}{GW190719A}{HL}{GW190620A}{LV}{GW190514A}{HL}{GW190731A}{HL}{GW190503A}{HLV}{GW190602A}{HLV}{GW190929A}{HLV}{GW190517A}{HLV}{GW190915A}{HLV}{GW190425A}{LV}{GW190512A}{HLV}{GW190630A}{LV}{GW190521A}{HLV}{GW190413B}{HLV}{GW190924A}{HLV}{GW190930A}{HL}{GW190706A}{HLV}{GW190408A}{HLV}{GW190909A}{HL}{GW190728A}{HLV}{GW190426A}{HLV}{GW190412A}{HLV}{GW190720A}{HLV}{GW190521B}{HL}{GW190910A}{LV}{GW190803A}{HLV}{GW190519A}{HLV}{GW190708A}{LV}{GW190527A}{HL}{GW190513A}{HLV}{GW190424A}{L}{GW190727A}{HLV}{GW190814A}{LV}{GW190707A}{HL}{GW190828A}{HLV}{GW190828B}{HLV}{GW190701A}{HLV}{GW190421A}{HL}}}

\newcommand{\PARTINSTRUMENTS}[1]{\IfEqCase{#1}{{GW190413A}{HL}{GW190719A}{HL}{GW190620A}{L}{GW190514A}{HL}{GW190731A}{HL}{GW190503A}{HL}{GW190602A}{HL}{GW190929A}{HL}{GW190517A}{HL}{GW190915A}{HL}{GW190425A}{L}{GW190512A}{HL}{GW190630A}{LV}{GW190521A}{HL}{GW190413B}{HL}{GW190924A}{HL}{GW190930A}{HL}{GW190706A}{HL}{GW190408A}{HL}{GW190909A}{HL}{GW190728A}{HL}{GW190426A}{HL}{GW190412A}{HL}{GW190720A}{HLV}{GW190521B}{HL}{GW190910A}{L}{GW190803A}{HL}{GW190519A}{HL}{GW190708A}{L}{GW190527A}{HL}{GW190513A}{HL}{GW190424A}{L}{GW190727A}{HL}{GW190814A}{LV}{GW190707A}{HL}{GW190828A}{HL}{GW190828B}{HL}{GW190701A}{HLV}{GW190421A}{HL}}}

\newcommand{\FARTHRESHYR}{2.0}

\newcommand{\NUMSNGLS}{5}

\newcommand{\PREVIOUSLYREPORTED}{26}

\newcommand{\NEWEVENTS}{13}

\newcommand{\RETRACTIONS}{8}

\newcommand{\CWBALLSKYPTERRES}[1]{\IfEqCase{#1}{{GW190413A}{--}{GW190719A}{--}{GW190620A}{--}{GW190514A}{--}{GW190731A}{--}{GW190503A}{--}{GW190602A}{--}{GW190929A}{--}{GW190517A}{--}{GW190915A}{--}{GW190425A}{--}{GW190512A}{--}{GW190630A}{--}{GW190521A}{--}{GW190413B}{--}{GW190924A}{--}{GW190930A}{--}{GW190706A}{--}{GW190408A}{--}{GW190909A}{--}{GW190728A}{--}{GW190426A}{--}{GW190412A}{--}{GW190720A}{--}{GW190521B}{--}{GW190910A}{--}{GW190803A}{--}{GW190519A}{--}{GW190708A}{--}{GW190527A}{--}{GW190513A}{--}{GW190424A}{--}{GW190727A}{--}{GW190814A}{--}{GW190707A}{--}{GW190828A}{--}{GW190828B}{--}{GW190701A}{--}{GW190421A}{--}}}

\newcommand{\CWBALLSKYPASTRO}[1]{\IfEqCase{#1}{{GW190413A}{--}{GW190719A}{--}{GW190620A}{~}{GW190514A}{--}{GW190731A}{--}{GW190503A}{--}{GW190602A}{--}{GW190929A}{--}{GW190517A}{--}{GW190915A}{--}{GW190425A}{~}{GW190512A}{--}{GW190630A}{~}{GW190521A}{--}{GW190413B}{--}{GW190924A}{--}{GW190930A}{--}{GW190706A}{--}{GW190408A}{--}{GW190909A}{--}{GW190728A}{--}{GW190426A}{--}{GW190412A}{--}{GW190720A}{--}{GW190521B}{--}{GW190910A}{~}{GW190803A}{--}{GW190519A}{--}{GW190708A}{~}{GW190527A}{--}{GW190513A}{--}{GW190424A}{~}{GW190727A}{--}{GW190814A}{~}{GW190707A}{--}{GW190828A}{--}{GW190828B}{--}{GW190701A}{--}{GW190421A}{--}}}

\newcommand{\PYCBCHIGHMASSPTERRES}[1]{\IfEqCase{#1}{{GW190413A}{$0.02$}{GW190719A}{$0.18$}{GW190620A}{--}{GW190514A}{$0.04$}{GW190731A}{$0.04$}{GW190503A}{$0.00$}{GW190602A}{$0.00$}{GW190929A}{--}{GW190517A}{$0.00$}{GW190915A}{$0.00$}{GW190425A}{--}{GW190512A}{$0.00$}{GW190630A}{--}{GW190521A}{--}{GW190413B}{$0.02$}{GW190924A}{$0.00$}{GW190930A}{$0.01$}{GW190706A}{$0.00$}{GW190408A}{$0.00$}{GW190909A}{--}{GW190728A}{$0.00$}{GW190426A}{--}{GW190412A}{$0.00$}{GW190720A}{$0.00$}{GW190521B}{$0.00$}{GW190910A}{--}{GW190803A}{$0.01$}{GW190519A}{$0.00$}{GW190708A}{--}{GW190527A}{--}{GW190513A}{$0.00$}{GW190424A}{--}{GW190727A}{$0.00$}{GW190814A}{--}{GW190707A}{$0.00$}{GW190828A}{$0.00$}{GW190828B}{$0.00$}{GW190701A}{--}{GW190421A}{$0.00$}}}

\newcommand{\PYCBCHIGHMASSPASTRO}[1]{\IfEqCase{#1}{{GW190413A}{$0.98$}{GW190719A}{$0.82$}{GW190620A}{~}{GW190514A}{$0.96$}{GW190731A}{$0.96$}{GW190503A}{$1.00$}{GW190602A}{$1.00$}{GW190929A}{--}{GW190517A}{$1.00$}{GW190915A}{$1.00$}{GW190425A}{~}{GW190512A}{$1.00$}{GW190630A}{~}{GW190521A}{--}{GW190413B}{$0.98$}{GW190924A}{$1.00$}{GW190930A}{$0.99$}{GW190706A}{$1.00$}{GW190408A}{$1.00$}{GW190909A}{--}{GW190728A}{$1.00$}{GW190426A}{--}{GW190412A}{$1.00$}{GW190720A}{$1.00$}{GW190521B}{$1.00$}{GW190910A}{~}{GW190803A}{$0.99$}{GW190519A}{$1.00$}{GW190708A}{~}{GW190527A}{--}{GW190513A}{$1.00$}{GW190424A}{~}{GW190727A}{$1.00$}{GW190814A}{~}{GW190707A}{$1.00$}{GW190828A}{$1.00$}{GW190828B}{$1.00$}{GW190701A}{--}{GW190421A}{$1.00$}}}

\newcommand{\PYCBCALLSKYPTERRES}[1]{\IfEqCase{#1}{{GW190413A}{--}{GW190719A}{--}{GW190620A}{--}{GW190514A}{--}{GW190731A}{--}{GW190503A}{$0.00$}{GW190602A}{--}{GW190929A}{--}{GW190517A}{$0.00$}{GW190915A}{$0.00$}{GW190425A}{--}{GW190512A}{$0.00$}{GW190630A}{--}{GW190521A}{$0.07$}{GW190413B}{--}{GW190924A}{$0.00$}{GW190930A}{$0.00$}{GW190706A}{$0.00$}{GW190408A}{$0.00$}{GW190909A}{--}{GW190728A}{$0.00$}{GW190426A}{--}{GW190412A}{$0.00$}{GW190720A}{$0.00$}{GW190521B}{$0.00$}{GW190910A}{--}{GW190803A}{--}{GW190519A}{$0.00$}{GW190708A}{--}{GW190527A}{--}{GW190513A}{$0.00$}{GW190424A}{--}{GW190727A}{$0.00$}{GW190814A}{--}{GW190707A}{$0.00$}{GW190828A}{$0.00$}{GW190828B}{$0.00$}{GW190701A}{--}{GW190421A}{$0.11$}}}

\newcommand{\PYCBCALLSKYPASTRO}[1]{\IfEqCase{#1}{{GW190413A}{--}{GW190719A}{--}{GW190620A}{~}{GW190514A}{--}{GW190731A}{--}{GW190503A}{$1.00$}{GW190602A}{--}{GW190929A}{--}{GW190517A}{$1.00$}{GW190915A}{$1.00$}{GW190425A}{~}{GW190512A}{$1.00$}{GW190630A}{~}{GW190521A}{$0.93$}{GW190413B}{--}{GW190924A}{$1.00$}{GW190930A}{$1.00$}{GW190706A}{$1.00$}{GW190408A}{$1.00$}{GW190909A}{--}{GW190728A}{$1.00$}{GW190426A}{--}{GW190412A}{$1.00$}{GW190720A}{$1.00$}{GW190521B}{$1.00$}{GW190910A}{~}{GW190803A}{--}{GW190519A}{$1.00$}{GW190708A}{~}{GW190527A}{--}{GW190513A}{$1.00$}{GW190424A}{~}{GW190727A}{$1.00$}{GW190814A}{~}{GW190707A}{$1.00$}{GW190828A}{$1.00$}{GW190828B}{$1.00$}{GW190701A}{--}{GW190421A}{$0.89$}}}

\newcommand{\GSTLALALLSKYPTERRES}[1]{\IfEqCase{#1}{{GW190413A}{--}{GW190719A}{--}{GW190620A}{$0.00$}{GW190514A}{--}{GW190731A}{$0.03$}{GW190503A}{$0.00$}{GW190602A}{$0.00$}{GW190929A}{$0.00$}{GW190517A}{$0.00$}{GW190915A}{$0.00$}{GW190425A}{--}{GW190512A}{$0.00$}{GW190630A}{$0.00$}{GW190521A}{$0.00$}{GW190413B}{$0.05$}{GW190924A}{$0.00$}{GW190930A}{$0.08$}{GW190706A}{$0.00$}{GW190408A}{$0.00$}{GW190909A}{$0.11$}{GW190728A}{$0.00$}{GW190426A}{--}{GW190412A}{$0.00$}{GW190720A}{$0.00$}{GW190521B}{$0.00$}{GW190910A}{$0.00$}{GW190803A}{$0.01$}{GW190519A}{$0.00$}{GW190708A}{$0.00$}{GW190527A}{$0.01$}{GW190513A}{$0.00$}{GW190424A}{$0.09$}{GW190727A}{$0.00$}{GW190814A}{$0.00$}{GW190707A}{$0.00$}{GW190828A}{$0.00$}{GW190828B}{$0.00$}{GW190701A}{$0.00$}{GW190421A}{$0.00$}}}

\newcommand{\GSTLALALLSKYPASTRO}[1]{\IfEqCase{#1}{{GW190413A}{--}{GW190719A}{--}{GW190620A}{$1.00$}{GW190514A}{--}{GW190731A}{$0.97$}{GW190503A}{$1.00$}{GW190602A}{$1.00$}{GW190929A}{$1.00$}{GW190517A}{$1.00$}{GW190915A}{$1.00$}{GW190425A}{--}{GW190512A}{$1.00$}{GW190630A}{$1.00$}{GW190521A}{$1.00$}{GW190413B}{$0.95$}{GW190924A}{$1.00$}{GW190930A}{$0.92$}{GW190706A}{$1.00$}{GW190408A}{$1.00$}{GW190909A}{$0.89$}{GW190728A}{$1.00$}{GW190426A}{--}{GW190412A}{$1.00$}{GW190720A}{$1.00$}{GW190521B}{$1.00$}{GW190910A}{$1.00$}{GW190803A}{$0.99$}{GW190519A}{$1.00$}{GW190708A}{$1.00$}{GW190527A}{$0.99$}{GW190513A}{$1.00$}{GW190424A}{$0.91$}{GW190727A}{$1.00$}{GW190814A}{$1.00$}{GW190707A}{$1.00$}{GW190828A}{$1.00$}{GW190828B}{$1.00$}{GW190701A}{$1.00$}{GW190421A}{$1.00$}}}

\newcommand{\GSTLALSNRTHRESH}{4.0}

\newcommand{\NUMCWB}{15}

\newcommand{\NUMGSTLAL}{36}

\newcommand{\NUMPYCBC}{27}

\newcommand{\TWOPIPES}{25}

\newcommand{\MINFAR}[1]{\IfEqCase{#1}{{GW190413A}{$7.2 \times 10^{-2}$}{GW190719A}{$1.6 \times 10^{0}$}{GW190620A}{$2.9 \times 10^{-3}$}{GW190514A}{$5.3 \times 10^{-1}$}{GW190731A}{$2.1 \times 10^{-1}$}{GW190503A}{$1.0 \times 10^{-5}$}{GW190602A}{$1.1 \times 10^{-5}$}{GW190929A}{$2.0 \times 10^{-2}$}{GW190517A}{$5.7 \times 10^{-5}$}{GW190915A}{$1.0 \times 10^{-5}$}{GW190425A}{$7.5 \times 10^{-4}$}{GW190512A}{$1.0 \times 10^{-5}$}{GW190630A}{$1.0 \times 10^{-5}$}{GW190521A}{$2.0 \times 10^{-4}$}{GW190413B}{$4.4 \times 10^{-2}$}{GW190924A}{$1.0 \times 10^{-5}$}{GW190930A}{$3.3 \times 10^{-2}$}{GW190706A}{$1.0 \times 10^{-5}$}{GW190408A}{$1.0 \times 10^{-5}$}{GW190909A}{$1.1 \times 10^{0}$}{GW190728A}{$1.0 \times 10^{-5}$}{GW190426A}{$1.4 \times 10^{0}$}{GW190412A}{$1.0 \times 10^{-5}$}{GW190720A}{$1.0 \times 10^{-5}$}{GW190521B}{$1.0 \times 10^{-5}$}{GW190910A}{$1.9 \times 10^{-5}$}{GW190803A}{$2.7 \times 10^{-2}$}{GW190519A}{$1.0 \times 10^{-5}$}{GW190708A}{$2.8 \times 10^{-5}$}{GW190527A}{$6.2 \times 10^{-2}$}{GW190513A}{$1.0 \times 10^{-5}$}{GW190424A}{$7.8 \times 10^{-1}$}{GW190727A}{$1.0 \times 10^{-5}$}{GW190814A}{$1.0 \times 10^{-5}$}{GW190707A}{$1.0 \times 10^{-5}$}{GW190828A}{$1.0 \times 10^{-5}$}{GW190828B}{$1.0 \times 10^{-5}$}{GW190701A}{$1.1 \times 10^{-2}$}{GW190421A}{$7.7 \times 10^{-4}$}}}

\newcommand{\CWBALLSKYFAR}[1]{\IfEqCase{#1}{{GW190413A}{--}{GW190719A}{--}{GW190620A}{~}{GW190514A}{--}{GW190731A}{--}{GW190503A}{$1.8 \times 10^{-3}$}{GW190602A}{$1.5 \times 10^{-2}$}{GW190929A}{--}{GW190517A}{$6.5 \times 10^{-3}$}{GW190915A}{$<$ $1.0 \times 10^{-3}$}{GW190425A}{~}{GW190512A}{$8.8 \times 10^{-1}$}{GW190630A}{~}{GW190521A}{$2.0 \times 10^{-4}$}{GW190413B}{--}{GW190924A}{--}{GW190930A}{--}{GW190706A}{$<$ $1.0 \times 10^{-3}$}{GW190408A}{$<$ $9.5 \times 10^{-4}$}{GW190909A}{--}{GW190728A}{--}{GW190426A}{--}{GW190412A}{$<$ $9.5 \times 10^{-4}$}{GW190720A}{--}{GW190521B}{$<$ $1.0 \times 10^{-4}$}{GW190910A}{~}{GW190803A}{--}{GW190519A}{$3.1 \times 10^{-4}$}{GW190708A}{~}{GW190527A}{--}{GW190513A}{--}{GW190424A}{~}{GW190727A}{$8.8 \times 10^{-2}$}{GW190814A}{~}{GW190707A}{--}{GW190828A}{$<$ $9.6 \times 10^{-4}$}{GW190828B}{--}{GW190701A}{$5.5 \times 10^{-1}$}{GW190421A}{$3.0 \times 10^{-1}$}}}

\newcommand{\CWBALLSKYIFAR}[1]{\IfEqCase{#1}{{GW190413A}{--}{GW190719A}{--}{GW190620A}{~}{GW190514A}{--}{GW190731A}{--}{GW190503A}{2.7}{GW190602A}{1.8}{GW190929A}{--}{GW190517A}{2.2}{GW190915A}{3.0}{GW190425A}{~}{GW190512A}{0.1}{GW190630A}{~}{GW190521A}{3.7}{GW190413B}{--}{GW190924A}{--}{GW190930A}{--}{GW190706A}{3.0}{GW190408A}{3.0}{GW190909A}{--}{GW190728A}{--}{GW190426A}{--}{GW190412A}{3.0}{GW190720A}{--}{GW190521B}{4.0}{GW190910A}{~}{GW190803A}{--}{GW190519A}{3.5}{GW190708A}{~}{GW190527A}{--}{GW190513A}{--}{GW190424A}{~}{GW190727A}{1.1}{GW190814A}{~}{GW190707A}{--}{GW190828A}{3.0}{GW190828B}{--}{GW190701A}{0.3}{GW190421A}{0.5}}}

\newcommand{\CWBALLSKYSNR}[1]{\IfEqCase{#1}{{GW190413A}{--}{GW190719A}{--}{GW190620A}{~}{GW190514A}{--}{GW190731A}{--}{GW190503A}{11.5}{GW190602A}{11.1}{GW190929A}{--}{GW190517A}{10.7}{GW190915A}{12.3}{GW190425A}{~}{GW190512A}{10.7}{GW190630A}{~}{GW190521A}{14.4}{GW190413B}{--}{GW190924A}{--}{GW190930A}{--}{GW190706A}{12.7}{GW190408A}{14.8}{GW190909A}{--}{GW190728A}{--}{GW190426A}{--}{GW190412A}{19.7}{GW190720A}{--}{GW190521B}{24.7}{GW190910A}{~}{GW190803A}{--}{GW190519A}{14.0}{GW190708A}{~}{GW190527A}{--}{GW190513A}{--}{GW190424A}{~}{GW190727A}{11.4}{GW190814A}{~}{GW190707A}{--}{GW190828A}{16.6}{GW190828B}{--}{GW190701A}{10.2}{GW190421A}{9.3}}}

\newcommand{\PYCBCHIGHMASSFAR}[1]{\IfEqCase{#1}{{GW190413A}{$7.2 \times 10^{-2}$}{GW190719A}{$1.6 \times 10^{0}$}{GW190620A}{~}{GW190514A}{$5.3 \times 10^{-1}$}{GW190731A}{$2.8 \times 10^{-1}$}{GW190503A}{$<$ $7.9 \times 10^{-5}$}{GW190602A}{$1.5 \times 10^{-2}$}{GW190929A}{--}{GW190517A}{$<$ $5.7 \times 10^{-5}$}{GW190915A}{$<$ $3.3 \times 10^{-5}$}{GW190425A}{~}{GW190512A}{$<$ $5.7 \times 10^{-5}$}{GW190630A}{~}{GW190521A}{--}{GW190413B}{$4.4 \times 10^{-2}$}{GW190924A}{$<$ $3.3 \times 10^{-5}$}{GW190930A}{$3.3 \times 10^{-2}$}{GW190706A}{$<$ $4.6 \times 10^{-5}$}{GW190408A}{$<$ $7.9 \times 10^{-5}$}{GW190909A}{--}{GW190728A}{$<$ $3.7 \times 10^{-5}$}{GW190426A}{--}{GW190412A}{$<$ $7.9 \times 10^{-5}$}{GW190720A}{$<$ $3.7 \times 10^{-5}$}{GW190521B}{$<$ $5.7 \times 10^{-5}$}{GW190910A}{~}{GW190803A}{$2.7 \times 10^{-2}$}{GW190519A}{$<$ $5.7 \times 10^{-5}$}{GW190708A}{~}{GW190527A}{--}{GW190513A}{$<$ $5.7 \times 10^{-5}$}{GW190424A}{~}{GW190727A}{$<$ $3.7 \times 10^{-5}$}{GW190814A}{~}{GW190707A}{$<$ $4.6 \times 10^{-5}$}{GW190828A}{$<$ $3.3 \times 10^{-5}$}{GW190828B}{$<$ $3.3 \times 10^{-5}$}{GW190701A}{--}{GW190421A}{$6.6 \times 10^{-3}$}}}

\newcommand{\PYCBCHIGHMASSIFAR}[1]{\IfEqCase{#1}{{GW190413A}{1.1}{GW190719A}{-0.2}{GW190620A}{~}{GW190514A}{0.3}{GW190731A}{0.6}{GW190503A}{4.1}{GW190602A}{1.8}{GW190929A}{--}{GW190517A}{4.2}{GW190915A}{4.5}{GW190425A}{~}{GW190512A}{4.2}{GW190630A}{~}{GW190521A}{--}{GW190413B}{1.4}{GW190924A}{4.5}{GW190930A}{1.5}{GW190706A}{4.3}{GW190408A}{4.1}{GW190909A}{--}{GW190728A}{4.4}{GW190426A}{--}{GW190412A}{4.1}{GW190720A}{4.4}{GW190521B}{4.2}{GW190910A}{~}{GW190803A}{1.6}{GW190519A}{4.2}{GW190708A}{~}{GW190527A}{--}{GW190513A}{4.2}{GW190424A}{~}{GW190727A}{4.4}{GW190814A}{~}{GW190707A}{4.3}{GW190828A}{4.5}{GW190828B}{4.5}{GW190701A}{--}{GW190421A}{2.2}}}

\newcommand{\PYCBCHIGHMASSSNR}[1]{\IfEqCase{#1}{{GW190413A}{8.6}{GW190719A}{8.0}{GW190620A}{~}{GW190514A}{8.3}{GW190731A}{8.2}{GW190503A}{12.2}{GW190602A}{11.4}{GW190929A}{--}{GW190517A}{10.2}{GW190915A}{12.7}{GW190425A}{~}{GW190512A}{12.2}{GW190630A}{~}{GW190521A}{--}{GW190413B}{9.0}{GW190924A}{12.4}{GW190930A}{9.8}{GW190706A}{12.3}{GW190408A}{13.6}{GW190909A}{--}{GW190728A}{13.4}{GW190426A}{--}{GW190412A}{17.8}{GW190720A}{10.5}{GW190521B}{24.0}{GW190910A}{~}{GW190803A}{8.6}{GW190519A}{13.0}{GW190708A}{~}{GW190527A}{--}{GW190513A}{11.9}{GW190424A}{~}{GW190727A}{11.8}{GW190814A}{~}{GW190707A}{12.8}{GW190828A}{15.3}{GW190828B}{10.8}{GW190701A}{--}{GW190421A}{10.2}}}

\newcommand{\PYCBCALLSKYFAR}[1]{\IfEqCase{#1}{{GW190413A}{--}{GW190719A}{--}{GW190620A}{~}{GW190514A}{--}{GW190731A}{--}{GW190503A}{$3.7 \times 10^{-2}$}{GW190602A}{--}{GW190929A}{--}{GW190517A}{$1.8 \times 10^{-2}$}{GW190915A}{$8.6 \times 10^{-4}$}{GW190425A}{~}{GW190512A}{$3.8 \times 10^{-5}$}{GW190630A}{~}{GW190521A}{$1.1 \times 10^{0}$}{GW190413B}{--}{GW190924A}{$<$ $6.3 \times 10^{-5}$}{GW190930A}{$3.4 \times 10^{-2}$}{GW190706A}{$6.7 \times 10^{-5}$}{GW190408A}{$<$ $2.5 \times 10^{-5}$}{GW190909A}{--}{GW190728A}{$<$ $1.6 \times 10^{-5}$}{GW190426A}{--}{GW190412A}{$<$ $3.1 \times 10^{-5}$}{GW190720A}{$<$ $2.0 \times 10^{-5}$}{GW190521B}{$<$ $1.8 \times 10^{-5}$}{GW190910A}{~}{GW190803A}{--}{GW190519A}{$<$ $1.8 \times 10^{-5}$}{GW190708A}{~}{GW190527A}{--}{GW190513A}{$3.7 \times 10^{-4}$}{GW190424A}{~}{GW190727A}{$3.5 \times 10^{-3}$}{GW190814A}{~}{GW190707A}{$<$ $1.0 \times 10^{-5}$}{GW190828A}{$<$ $1.5 \times 10^{-5}$}{GW190828B}{$5.8 \times 10^{-5}$}{GW190701A}{--}{GW190421A}{$1.9 \times 10^{0}$}}}

\newcommand{\PYCBCALLSKYIFAR}[1]{\IfEqCase{#1}{{GW190413A}{--}{GW190719A}{--}{GW190620A}{~}{GW190514A}{--}{GW190731A}{--}{GW190503A}{1.4}{GW190602A}{--}{GW190929A}{--}{GW190517A}{1.8}{GW190915A}{3.1}{GW190425A}{~}{GW190512A}{4.4}{GW190630A}{~}{GW190521A}{-0.0}{GW190413B}{--}{GW190924A}{4.2}{GW190930A}{1.5}{GW190706A}{4.2}{GW190408A}{4.6}{GW190909A}{--}{GW190728A}{4.8}{GW190426A}{--}{GW190412A}{4.5}{GW190720A}{4.7}{GW190521B}{4.8}{GW190910A}{~}{GW190803A}{--}{GW190519A}{4.8}{GW190708A}{~}{GW190527A}{--}{GW190513A}{3.4}{GW190424A}{~}{GW190727A}{2.5}{GW190814A}{~}{GW190707A}{5.0}{GW190828A}{4.8}{GW190828B}{4.2}{GW190701A}{--}{GW190421A}{-0.3}}}

\newcommand{\PYCBCALLSKYSNR}[1]{\IfEqCase{#1}{{GW190413A}{--}{GW190719A}{--}{GW190620A}{~}{GW190514A}{--}{GW190731A}{--}{GW190503A}{12.2}{GW190602A}{--}{GW190929A}{--}{GW190517A}{10.4}{GW190915A}{13.0}{GW190425A}{~}{GW190512A}{12.2}{GW190630A}{~}{GW190521A}{12.6}{GW190413B}{--}{GW190924A}{12.5}{GW190930A}{9.7}{GW190706A}{11.7}{GW190408A}{13.5}{GW190909A}{--}{GW190728A}{13.4}{GW190426A}{--}{GW190412A}{17.9}{GW190720A}{10.6}{GW190521B}{24.0}{GW190910A}{~}{GW190803A}{--}{GW190519A}{13.0}{GW190708A}{~}{GW190527A}{--}{GW190513A}{11.8}{GW190424A}{~}{GW190727A}{11.5}{GW190814A}{~}{GW190707A}{12.8}{GW190828A}{15.3}{GW190828B}{10.8}{GW190701A}{--}{GW190421A}{10.2}}}

\newcommand{\GSTLALALLSKYFAR}[1]{\IfEqCase{#1}{{GW190413A}{--}{GW190719A}{--}{GW190620A}{$2.9 \times 10^{-3}$$^\dagger$}{GW190514A}{--}{GW190731A}{$2.1 \times 10^{-1}$}{GW190503A}{$<$ $1.0 \times 10^{-5}$}{GW190602A}{$1.1 \times 10^{-5}$}{GW190929A}{$2.0 \times 10^{-2}$}{GW190517A}{$9.6 \times 10^{-4}$}{GW190915A}{$<$ $1.0 \times 10^{-5}$}{GW190425A}{$7.5 \times 10^{-4}$$^\dagger$}{GW190512A}{$<$ $1.0 \times 10^{-5}$}{GW190630A}{$<$ $1.0 \times 10^{-5}$}{GW190521A}{$1.2 \times 10^{-3}$}{GW190413B}{$3.8 \times 10^{-1}$}{GW190924A}{$<$ $1.0 \times 10^{-5}$}{GW190930A}{$5.8 \times 10^{-1}$}{GW190706A}{$<$ $1.0 \times 10^{-5}$}{GW190408A}{$<$ $1.0 \times 10^{-5}$}{GW190909A}{$1.1 \times 10^{0}$}{GW190728A}{$<$ $1.0 \times 10^{-5}$}{GW190426A}{$1.4 \times 10^{0}$}{GW190412A}{$<$ $1.0 \times 10^{-5}$}{GW190720A}{$<$ $1.0 \times 10^{-5}$}{GW190521B}{$<$ $1.0 \times 10^{-5}$}{GW190910A}{$1.9 \times 10^{-5}$$^\dagger$}{GW190803A}{$3.2 \times 10^{-2}$}{GW190519A}{$<$ $1.0 \times 10^{-5}$}{GW190708A}{$2.8 \times 10^{-5}$$^\dagger$}{GW190527A}{$6.2 \times 10^{-2}$}{GW190513A}{$<$ $1.0 \times 10^{-5}$}{GW190424A}{$7.8 \times 10^{-1}$$^\dagger$}{GW190727A}{$<$ $1.0 \times 10^{-5}$}{GW190814A}{$<$ $1.0 \times 10^{-5}$}{GW190707A}{$<$ $1.0 \times 10^{-5}$}{GW190828A}{$<$ $1.0 \times 10^{-5}$}{GW190828B}{$<$ $1.0 \times 10^{-5}$}{GW190701A}{$1.1 \times 10^{-2}$}{GW190421A}{$7.7 \times 10^{-4}$}}}

\newcommand{\GSTLALALLSKYIFAR}[1]{\IfEqCase{#1}{{GW190413A}{--}{GW190719A}{--}{GW190620A}{2.5$^\dagger$}{GW190514A}{--}{GW190731A}{0.7}{GW190503A}{5.0}{GW190602A}{4.9}{GW190929A}{1.7}{GW190517A}{3.0}{GW190915A}{5.0}{GW190425A}{3.1$^\dagger$}{GW190512A}{5.0}{GW190630A}{5.0}{GW190521A}{2.9}{GW190413B}{0.4}{GW190924A}{5.0}{GW190930A}{0.2}{GW190706A}{5.0}{GW190408A}{5.0}{GW190909A}{-0.0}{GW190728A}{5.0}{GW190426A}{-0.2}{GW190412A}{5.0}{GW190720A}{5.0}{GW190521B}{5.0}{GW190910A}{4.7$^\dagger$}{GW190803A}{1.5}{GW190519A}{5.0}{GW190708A}{4.5$^\dagger$}{GW190527A}{1.2}{GW190513A}{5.0}{GW190424A}{0.1$^\dagger$}{GW190727A}{5.0}{GW190814A}{5.0}{GW190707A}{5.0}{GW190828A}{5.0}{GW190828B}{5.0}{GW190701A}{2.0}{GW190421A}{3.1}}}

\newcommand{\GSTLALALLSKYSNR}[1]{\IfEqCase{#1}{{GW190413A}{--}{GW190719A}{--}{GW190620A}{10.9}{GW190514A}{--}{GW190731A}{8.5}{GW190503A}{12.1}{GW190602A}{12.1}{GW190929A}{9.9}{GW190517A}{10.6}{GW190915A}{13.1}{GW190425A}{13.0}{GW190512A}{12.3}{GW190630A}{15.6}{GW190521A}{15.0}{GW190413B}{10.0}{GW190924A}{13.2}{GW190930A}{10.0}{GW190706A}{12.3}{GW190408A}{14.7}{GW190909A}{8.5}{GW190728A}{13.6}{GW190426A}{10.1}{GW190412A}{18.9}{GW190720A}{11.7}{GW190521B}{24.4}{GW190910A}{13.4}{GW190803A}{9.0}{GW190519A}{12.0}{GW190708A}{13.1}{GW190527A}{8.9}{GW190513A}{12.3}{GW190424A}{10.0}{GW190727A}{12.3}{GW190814A}{22.2}{GW190707A}{13.0}{GW190828A}{16.0}{GW190828B}{11.1}{GW190701A}{11.6}{GW190421A}{10.6}}}

%% file: user_macros.tex
\newcommand{\RUNSTART}{1 April 2019 15:00 UTC}
\newcommand{\RUNEND}{1 October 2019 15:00 UTC}

\newcommand{\CONTAMINATION}{10\%}

\newcommand{\DETECTIONRATE}{$\sim$1.5 per week}
\newcommand{\NUMOPA}{33}
\newcommand{\NUMOPANOTFOUND}{7}

\newcommand{\GSTLAL}{GstLAL\xspace{}}
\newcommand{\CWB}{cWB\xspace{}}
\newcommand{\PYCBC}{PyCBC\xspace{}}
\newcommand{\LALINFERENCE}{LALInference\xspace{}}

\newcommand{\VIRGORANGE}{\fixme{45 Mpc}}
\newcommand{\HANFORDRANGE}{\fixme{108 Mpc}}
\newcommand{\LIVINGSTONRANGE}{\fixme{135 Mpc}}

\newcommand{\VIRGORANGEOTWO}{\fixme{26 Mpc}}
\newcommand{\HANFORDRANGEOTWO}{\fixme{66 Mpc}}
\newcommand{\LIVINGSTONRANGEOTWO}{\fixme{88 Mpc}}

\newcommand{\VIRGORANGEINCREASE}{\fixme{1.73}}
\newcommand{\HANFORDRANGEINCREASE}{\fixme{1.64}}
\newcommand{\LIVINGSTONRANGEINCREASE}{\fixme{1.53}}

\newcommand{\VIRGODUTYCYCLE}{\fixme{76}}
\newcommand{\HANFORDDUTYCYCLE}{\fixme{71}}
\newcommand{\LIVINGSTONDUTYCYCLE}{\fixme{76}}
\newcommand{\ONEDETECTORDUTYCYCLE}{\fixme{96.9}}
\newcommand{\TWODETECTORSDUTYCYCLE}{\fixme{81.9}}
\newcommand{\THREEDETECTORSDUTYCYCLE}{\fixme{44.5}}

\newcommand{\HANFORDDUTYCYCLEOTWO}{\fixme{62}}
\newcommand{\LIVINGSTONDUTYCYCLEOTWO}{\fixme{61}}
\newcommand{\ONEDETECTORDUTYCYCLEOTWO}{\fixme{75.6}}
\newcommand{\TWODETECTORSDUTYCYCLEOTWO}{\fixme{46.4}}

\newcommand{\VIRGODAYS}{\fixme{139.5}}
\newcommand{\HANFORDDAYS}{\fixme{130.3}}
\newcommand{\LIVINGSTONDAYS}{\fixme{138.5}}
\newcommand{\ONEDETECTORDAYS}{\fixme{177.3}}
\newcommand{\TWODETECTORSDAYS}{\fixme{149.9}}
\newcommand{\THREEDETECTORSDAYS}{\fixme{81.4}}

\newcommand{\HANFORDPOWER}{\fixme{37}}
\newcommand{\HANFORDPOWEROTWO}{\fixme{30}}
\newcommand{\LIVINGSTONPOWER}{\fixme{40}}
\newcommand{\LIVINGSTONPOWEROTWO}{\fixme{25}}
\newcommand{\VIRGOPOWER}{\fixme{19}}
\newcommand{\VIRGOPOWEROTWO}{\fixme{10}}
\newcommand{\VIRGOPOWERMAX}{\fixme{65}}

\newcommand{\VIRGORANGEPLOT}{\fixme{50 Mpc}}
\newcommand{\HANFORDRANGEPLOT}{\fixme{109 Mpc}}
\newcommand{\LIVINGSTONRANGEPLOT}{\fixme{136 Mpc}}

\newcommand{\HANFORDIMPROVEMENTHF}{\fixme{1.68}}
\newcommand{\LIVINGSTONIMPROVEMENTHF}{\fixme{1.96}}

\newcommand{\LLOOTWOGLITCHRATE}{\fixme{0.2 per minute}}
\newcommand{\LLOOTHREEGLITCHRATE}{\fixme{0.8 per minute}}

\newcommand{\BLIPRATE}{\fixme{1.4 per hour}}

\newcommand{\SCATTERINGEVENTS}{\fixme{7}}

\newcommand{\LHOCATONE}{\fixme{0.27}}
\newcommand{\LLOCATONE}{\fixme{0.08}}
\newcommand{\VIRGOCATONE}{\fixme{0.15}}

\newcommand{\LHOCATTWOCBC}{\fixme{0.37}}
\newcommand{\LLOCATTWOCBC}{\fixme{0.10}}
\newcommand{\VIRGOCATTWOCBC}{\fixme{--}}

\newcommand{\LHOCATTWOBURST}{\fixme{0.83}}
\newcommand{\LLOCATTWOBURST}{\fixme{0.64}}
\newcommand{\VIRGOCATTWOBURST}{\fixme{--}}

\newcommand{\LHOCATTHREEBURST}{\fixme{0.19}}
\newcommand{\LLOCATTHREEBURST}{\fixme{0.15}}
\newcommand{\VIRGOCATTHREEBURST}{\fixme{--}}

\newcommand{\VIRGONOISESUBRANGEINCREASE}{\fixme{3~Mpc}}

\newcommand{\Mc}{\ensuremath{\mathcal{M}}}
\newcommand{\Mtot}{\ensuremath{M}}
\newcommand{\Msun}{\ensuremath{\text{M}_\odot}}

\newcommand\perMpcyr{\ensuremath{\mathrm{Mpc}^{-3}\,\mathrm{yr}^{-1}}}
\newcommand\perGpcyr{\ensuremath{\mathrm{Gpc}^{-3}\,\mathrm{yr}^{-1}}}
\newcommand{\chieff}{\ensuremath{\chi_\mathrm{eff}}}
\newcommand{\chip}{\ensuremath{\chi_\mathrm{p}}}
\newcommand{\DKLchip}{D_\mathrm{KL}^{\chi_\mathrm{p}}}
\newcommand{\DKLchieff}{D_\mathrm{KL}^{\chi_\mathrm{eff}}}
\newcommand{\VT}{\ensuremath{\langle VT \rangle}}
\newcommand{\DL}{\ensuremath{D_{\mathrm{L}}}}
\newcommand{\spintilt}{\ensuremath{\theta_{{LS}}}}

\newcommand{\NUMINJORIG}{\fixme{76982499}}
\newcommand{\NUMINJORIGAPPX}{\fixme{\ensuremath{7.70\times10^7}}}
\newcommand{\NUMINJ}{\fixme{156,878}}
\newcommand{\NUMINJAPPX}{\fixme{$\sim$157,000}}
\newcommand{\GSTLALVT}{\fixme{$0.456$~Gpc$^3$\,yr}}
\newcommand{\PYCBCVT}{\fixme{$0.296$~Gpc$^3$\,yr}}
\newcommand{\PYCBCBBHVT}{\fixme{$0.386$~Gpc$^3$\,yr}}
\newcommand{\ANYCBCVT}{\fixme{$0.567$~Gpc$^3$\,yr}}

\newcommand{\RBBH}{$23.9_{-8.6}^{+14.9}$~Gpc$^{-3}$\,yr$^{-1}$}
\newcommand{\RBNS}{$320_{-240}^{+490}$~Gpc$^{-3}$\,yr$^{-1}$}

%% file: PE_macros.tex
\newcommand{\loglikelihoodminus}[1]{\IfEqCase{#1}{{GW190930A}{8.4}{GW190929A}{10.9}{GW190924A}{8.6}{GW190915A}{8.1}{GW190910A}{4.8}{GW190909A}{4.5}{GW190828B}{5.3}{GW190828A}{5.0}{GW190814A}{4.9}{GW190803A}{4.4}{GW190731A}{4.0}{GW190728A}{48007.6}{GW190727A}{5.1}{GW190720A}{9.4}{GW190719A}{4.2}{GW190708A}{4.8}{GW190707A}{7.1}{GW190706A}{5.2}{GW190701A}{3.9}{GW190630A}{5.3}{GW190620A}{5.0}{GW190602A}{4.3}{GW190527A}{6.7}{GW190521B}{5.9}{GW190521A}{11.1}{GW190519A}{17.8}{GW190517A}{5.9}{GW190514A}{4.6}{GW190513A}{4.7}{GW190512A}{5.5}{GW190503A}{4.4}{GW190426A}{5.6}{GW190425A}{5.7}{GW190424A}{3.9}{GW190421A}{4.0}{GW190413B}{5.7}{GW190413A}{4.8}{GW190412A}{10.1}{GW190408A}{5.0}}}
\newcommand{\loglikelihoodmed}[1]{\IfEqCase{#1}{{GW190930A}{-15934.9}{GW190929A}{-11962.2}{GW190924A}{-97031.8}{GW190915A}{-2809.6}{GW190910A}{93.5}{GW190909A}{25.8}{GW190828B}{43.6}{GW190828A}{121.4}{GW190814A}{298.6}{GW190803A}{28.7}{GW190731A}{29.8}{GW190728A}{64.1}{GW190727A}{58.4}{GW190720A}{-23904.6}{GW190719A}{27.0}{GW190708A}{76.7}{GW190707A}{-15883.0}{GW190706A}{72.6}{GW190701A}{55.9}{GW190630A}{114.8}{GW190620A}{65.2}{GW190602A}{73.6}{GW190527A}{22.9}{GW190521B}{322.2}{GW190521A}{-11913.6}{GW190519A}{111.0}{GW190517A}{48.8}{GW190514A}{25.9}{GW190513A}{74.3}{GW190512A}{67.5}{GW190503A}{68.5}{GW190426A}{-389547.0}{GW190425A}{-500483.9}{GW190424A}{45.7}{GW190421A}{48.8}{GW190413B}{42.7}{GW190413A}{28.4}{GW190412A}{-22827.3}{GW190408A}{108.8}}}
\newcommand{\loglikelihoodplus}[1]{\IfEqCase{#1}{{GW190930A}{15973.8}{GW190929A}{12007.8}{GW190924A}{97091.0}{GW190915A}{2897.5}{GW190910A}{4.0}{GW190909A}{3.6}{GW190828B}{4.0}{GW190828A}{4.2}{GW190814A}{3.0}{GW190803A}{2.7}{GW190731A}{2.5}{GW190728A}{13.4}{GW190727A}{5.2}{GW190720A}{23953.2}{GW190719A}{2.8}{GW190708A}{3.3}{GW190707A}{15964.2}{GW190706A}{4.0}{GW190701A}{2.6}{GW190630A}{3.8}{GW190620A}{4.1}{GW190602A}{3.3}{GW190527A}{3.0}{GW190521B}{4.7}{GW190521A}{12013.1}{GW190519A}{9.5}{GW190517A}{4.4}{GW190514A}{2.7}{GW190513A}{4.2}{GW190512A}{3.8}{GW190503A}{3.4}{GW190426A}{4.6}{GW190425A}{4.5}{GW190424A}{2.8}{GW190421A}{2.6}{GW190413B}{3.6}{GW190413A}{4.0}{GW190412A}{23002.9}{GW190408A}{3.7}}}
\newcommand{\chieffminus}[1]{\IfEqCase{#1}{{GW190930A}{0.15}{GW190929A}{0.33}{GW190924A}{0.09}{GW190915A}{0.25}{GW190910A}{0.18}{GW190909A}{0.36}{GW190828B}{0.16}{GW190828A}{0.16}{GW190814A}{0.06}{GW190803A}{0.27}{GW190731A}{0.24}{GW190728A}{0.07}{GW190727A}{0.25}{GW190720A}{0.12}{GW190719A}{0.31}{GW190708A}{0.08}{GW190707A}{0.08}{GW190706A}{0.29}{GW190701A}{0.29}{GW190630A}{0.13}{GW190620A}{0.25}{GW190602A}{0.24}{GW190527A}{0.28}{GW190521B}{0.13}{GW190521A}{0.39}{GW190519A}{0.22}{GW190517A}{0.19}{GW190514A}{0.32}{GW190513A}{0.17}{GW190512A}{0.13}{GW190503A}{0.26}{GW190426A}{0.30}{GW190425A}{0.05}{GW190424A}{0.22}{GW190421A}{0.27}{GW190413B}{0.29}{GW190413A}{0.34}{GW190412A}{0.11}{GW190408A}{0.19}}}
\newcommand{\chieffmed}[1]{\IfEqCase{#1}{{GW190930A}{0.14}{GW190929A}{0.01}{GW190924A}{0.03}{GW190915A}{0.02}{GW190910A}{0.02}{GW190909A}{-0.06}{GW190828B}{0.08}{GW190828A}{0.19}{GW190814A}{0.00}{GW190803A}{-0.03}{GW190731A}{0.06}{GW190728A}{0.12}{GW190727A}{0.11}{GW190720A}{0.18}{GW190719A}{0.32}{GW190708A}{0.02}{GW190707A}{-0.05}{GW190706A}{0.28}{GW190701A}{-0.07}{GW190630A}{0.10}{GW190620A}{0.33}{GW190602A}{0.07}{GW190527A}{0.11}{GW190521B}{0.09}{GW190521A}{0.03}{GW190519A}{0.31}{GW190517A}{0.52}{GW190514A}{-0.19}{GW190513A}{0.11}{GW190512A}{0.03}{GW190503A}{-0.03}{GW190426A}{-0.03}{GW190425A}{0.06}{GW190424A}{0.13}{GW190421A}{-0.06}{GW190413B}{-0.03}{GW190413A}{-0.01}{GW190412A}{0.25}{GW190408A}{-0.03}}}
\newcommand{\chieffplus}[1]{\IfEqCase{#1}{{GW190930A}{0.31}{GW190929A}{0.34}{GW190924A}{0.30}{GW190915A}{0.20}{GW190910A}{0.18}{GW190909A}{0.37}{GW190828B}{0.16}{GW190828A}{0.15}{GW190814A}{0.06}{GW190803A}{0.24}{GW190731A}{0.24}{GW190728A}{0.20}{GW190727A}{0.26}{GW190720A}{0.14}{GW190719A}{0.29}{GW190708A}{0.10}{GW190707A}{0.10}{GW190706A}{0.26}{GW190701A}{0.23}{GW190630A}{0.12}{GW190620A}{0.22}{GW190602A}{0.25}{GW190527A}{0.28}{GW190521B}{0.10}{GW190521A}{0.32}{GW190519A}{0.20}{GW190517A}{0.19}{GW190514A}{0.29}{GW190513A}{0.28}{GW190512A}{0.12}{GW190503A}{0.20}{GW190426A}{0.32}{GW190425A}{0.11}{GW190424A}{0.22}{GW190421A}{0.22}{GW190413B}{0.25}{GW190413A}{0.29}{GW190412A}{0.08}{GW190408A}{0.14}}}
\newcommand{\totalmasssourceminus}[1]{\IfEqCase{#1}{{GW190930A}{1.5}{GW190929A}{25.2}{GW190924A}{1.0}{GW190915A}{6.4}{GW190910A}{9.1}{GW190909A}{17.6}{GW190828B}{4.4}{GW190828A}{4.8}{GW190814A}{0.9}{GW190803A}{9.0}{GW190731A}{11.3}{GW190728A}{1.3}{GW190727A}{8.0}{GW190720A}{2.3}{GW190719A}{10.7}{GW190708A}{1.8}{GW190707A}{1.3}{GW190706A}{13.9}{GW190701A}{9.5}{GW190630A}{4.8}{GW190620A}{13.1}{GW190602A}{15.6}{GW190527A}{9.8}{GW190521B}{4.8}{GW190521A}{23.5}{GW190519A}{14.8}{GW190517A}{9.6}{GW190514A}{10.8}{GW190513A}{5.9}{GW190512A}{3.5}{GW190503A}{8.3}{GW190426A}{1.5}{GW190425A}{0.1}{GW190424A}{10.7}{GW190421A}{9.2}{GW190413B}{11.9}{GW190413A}{9.7}{GW190412A}{3.7}{GW190408A}{3.0}}}
\newcommand{\totalmasssourcemed}[1]{\IfEqCase{#1}{{GW190930A}{20.3}{GW190929A}{104.3}{GW190924A}{13.9}{GW190915A}{59.9}{GW190910A}{79.6}{GW190909A}{75.0}{GW190828B}{34.4}{GW190828A}{58.0}{GW190814A}{25.8}{GW190803A}{64.5}{GW190731A}{70.1}{GW190728A}{20.6}{GW190727A}{67.1}{GW190720A}{21.5}{GW190719A}{57.8}{GW190708A}{30.9}{GW190707A}{20.1}{GW190706A}{104.1}{GW190701A}{94.3}{GW190630A}{59.1}{GW190620A}{92.1}{GW190602A}{116.3}{GW190527A}{59.1}{GW190521B}{74.7}{GW190521A}{163.9}{GW190519A}{106.6}{GW190517A}{63.5}{GW190514A}{67.2}{GW190513A}{53.9}{GW190512A}{35.9}{GW190503A}{71.7}{GW190426A}{7.2}{GW190425A}{3.4}{GW190424A}{72.6}{GW190421A}{72.9}{GW190413B}{78.8}{GW190413A}{58.6}{GW190412A}{38.4}{GW190408A}{43.0}}}
\newcommand{\totalmasssourceplus}[1]{\IfEqCase{#1}{{GW190930A}{8.9}{GW190929A}{34.9}{GW190924A}{5.1}{GW190915A}{7.5}{GW190910A}{9.3}{GW190909A}{55.9}{GW190828B}{5.4}{GW190828A}{7.7}{GW190814A}{1.0}{GW190803A}{12.6}{GW190731A}{15.8}{GW190728A}{4.5}{GW190727A}{11.7}{GW190720A}{4.3}{GW190719A}{18.3}{GW190708A}{2.5}{GW190707A}{1.9}{GW190706A}{20.2}{GW190701A}{12.1}{GW190630A}{4.6}{GW190620A}{18.5}{GW190602A}{19.0}{GW190527A}{21.3}{GW190521B}{7.0}{GW190521A}{39.2}{GW190519A}{13.5}{GW190517A}{9.6}{GW190514A}{18.7}{GW190513A}{8.6}{GW190512A}{3.8}{GW190503A}{9.4}{GW190426A}{3.5}{GW190425A}{0.3}{GW190424A}{13.3}{GW190421A}{13.4}{GW190413B}{17.4}{GW190413A}{13.3}{GW190412A}{3.8}{GW190408A}{4.2}}}
\newcommand{\chipminus}[1]{\IfEqCase{#1}{{GW190930A}{0.24}{GW190929A}{0.45}{GW190924A}{0.18}{GW190915A}{0.39}{GW190910A}{0.32}{GW190909A}{0.38}{GW190828B}{0.23}{GW190828A}{0.31}{GW190814A}{0.03}{GW190803A}{0.33}{GW190731A}{0.30}{GW190728A}{0.20}{GW190727A}{0.36}{GW190720A}{0.22}{GW190719A}{0.30}{GW190708A}{0.24}{GW190707A}{0.23}{GW190706A}{0.28}{GW190701A}{0.31}{GW190630A}{0.23}{GW190620A}{0.28}{GW190602A}{0.31}{GW190527A}{0.34}{GW190521B}{0.29}{GW190521A}{0.44}{GW190519A}{0.29}{GW190517A}{0.29}{GW190514A}{0.34}{GW190513A}{0.22}{GW190512A}{0.17}{GW190503A}{0.29}{GW190426A}{0.00}{GW190425A}{0.27}{GW190424A}{0.38}{GW190421A}{0.36}{GW190413B}{0.41}{GW190413A}{0.31}{GW190412A}{0.16}{GW190408A}{0.31}}}
\newcommand{\chipmed}[1]{\IfEqCase{#1}{{GW190930A}{0.34}{GW190929A}{0.59}{GW190924A}{0.24}{GW190915A}{0.55}{GW190910A}{0.40}{GW190909A}{0.52}{GW190828B}{0.31}{GW190828A}{0.43}{GW190814A}{0.04}{GW190803A}{0.43}{GW190731A}{0.39}{GW190728A}{0.29}{GW190727A}{0.47}{GW190720A}{0.33}{GW190719A}{0.43}{GW190708A}{0.29}{GW190707A}{0.29}{GW190706A}{0.39}{GW190701A}{0.42}{GW190630A}{0.32}{GW190620A}{0.43}{GW190602A}{0.41}{GW190527A}{0.44}{GW190521B}{0.40}{GW190521A}{0.68}{GW190519A}{0.44}{GW190517A}{0.49}{GW190514A}{0.47}{GW190513A}{0.30}{GW190512A}{0.22}{GW190503A}{0.38}{GW190426A}{0.00}{GW190425A}{0.34}{GW190424A}{0.52}{GW190421A}{0.48}{GW190413B}{0.56}{GW190413A}{0.41}{GW190412A}{0.30}{GW190408A}{0.39}}}
\newcommand{\chipplus}[1]{\IfEqCase{#1}{{GW190930A}{0.40}{GW190929A}{0.32}{GW190924A}{0.40}{GW190915A}{0.36}{GW190910A}{0.39}{GW190909A}{0.39}{GW190828B}{0.38}{GW190828A}{0.36}{GW190814A}{0.04}{GW190803A}{0.42}{GW190731A}{0.46}{GW190728A}{0.37}{GW190727A}{0.40}{GW190720A}{0.43}{GW190719A}{0.37}{GW190708A}{0.43}{GW190707A}{0.39}{GW190706A}{0.39}{GW190701A}{0.41}{GW190630A}{0.32}{GW190620A}{0.37}{GW190602A}{0.42}{GW190527A}{0.43}{GW190521B}{0.32}{GW190521A}{0.26}{GW190519A}{0.34}{GW190517A}{0.30}{GW190514A}{0.39}{GW190513A}{0.39}{GW190512A}{0.36}{GW190503A}{0.41}{GW190426A}{0.00}{GW190425A}{0.43}{GW190424A}{0.38}{GW190421A}{0.39}{GW190413B}{0.37}{GW190413A}{0.41}{GW190412A}{0.19}{GW190408A}{0.38}}}
\newcommand{\spinoneyminus}[1]{\IfEqCase{#1}{{GW190930A}{0.47}{GW190929A}{0.71}{GW190924A}{0.35}{GW190915A}{0.68}{GW190910A}{0.48}{GW190909A}{0.66}{GW190828B}{0.41}{GW190828A}{0.51}{GW190814A}{0.04}{GW190803A}{0.58}{GW190731A}{0.52}{GW190728A}{0.37}{GW190727A}{0.61}{GW190720A}{0.49}{GW190719A}{0.55}{GW190708A}{0.43}{GW190707A}{0.39}{GW190706A}{0.50}{GW190701A}{0.52}{GW190630A}{0.36}{GW190620A}{0.53}{GW190602A}{0.52}{GW190527A}{0.61}{GW190521B}{0.44}{GW190521A}{0.76}{GW190519A}{0.55}{GW190517A}{0.58}{GW190514A}{0.56}{GW190513A}{0.41}{GW190512A}{0.29}{GW190503A}{0.48}{GW190426A}{0.00}{GW190425A}{0.48}{GW190424A}{0.64}{GW190421A}{0.58}{GW190413B}{0.70}{GW190413A}{0.54}{GW190412A}{0.39}{GW190408A}{0.48}}}
\newcommand{\spinoneymed}[1]{\IfEqCase{#1}{{GW190930A}{0.002}{GW190929A}{0.005}{GW190924A}{0.0009}{GW190915A}{0.00}{GW190910A}{0.0008}{GW190909A}{-0.01}{GW190828B}{0.0010}{GW190828A}{0.00}{GW190814A}{0.0008}{GW190803A}{0.0007}{GW190731A}{0.0007}{GW190728A}{0.004}{GW190727A}{0.0002}{GW190720A}{0.0009}{GW190719A}{0.004}{GW190708A}{0.00}{GW190707A}{0.00}{GW190706A}{0.00}{GW190701A}{0.003}{GW190630A}{0.0002}{GW190620A}{-0.01}{GW190602A}{0.00}{GW190527A}{0.00}{GW190521B}{0.00}{GW190521A}{0.0003}{GW190519A}{0.00}{GW190517A}{-0.01}{GW190514A}{0.00}{GW190513A}{0.0005}{GW190512A}{0.00}{GW190503A}{0.00}{GW190426A}{0.00}{GW190425A}{0.003}{GW190424A}{0.00}{GW190421A}{0.0008}{GW190413B}{0.00}{GW190413A}{0.003}{GW190412A}{0.06}{GW190408A}{0.001}}}
\newcommand{\spinoneyplus}[1]{\IfEqCase{#1}{{GW190930A}{0.48}{GW190929A}{0.70}{GW190924A}{0.36}{GW190915A}{0.68}{GW190910A}{0.50}{GW190909A}{0.64}{GW190828B}{0.43}{GW190828A}{0.53}{GW190814A}{0.04}{GW190803A}{0.55}{GW190731A}{0.54}{GW190728A}{0.39}{GW190727A}{0.59}{GW190720A}{0.44}{GW190719A}{0.55}{GW190708A}{0.41}{GW190707A}{0.38}{GW190706A}{0.51}{GW190701A}{0.53}{GW190630A}{0.37}{GW190620A}{0.55}{GW190602A}{0.54}{GW190527A}{0.59}{GW190521B}{0.45}{GW190521A}{0.75}{GW190519A}{0.54}{GW190517A}{0.57}{GW190514A}{0.59}{GW190513A}{0.41}{GW190512A}{0.29}{GW190503A}{0.49}{GW190426A}{0.00}{GW190425A}{0.48}{GW190424A}{0.63}{GW190421A}{0.60}{GW190413B}{0.69}{GW190413A}{0.55}{GW190412A}{0.33}{GW190408A}{0.49}}}
\newcommand{\finalmassdetminus}[1]{\IfEqCase{#1}{{GW190930A}{0.9}{GW190929A}{24.5}{GW190924A}{0.8}{GW190915A}{7.4}{GW190910A}{7.1}{GW190909A}{21.3}{GW190828B}{4.2}{GW190828A}{5.2}{GW190814A}{1.0}{GW190803A}{10.9}{GW190731A}{12.8}{GW190728A}{0.7}{GW190727A}{9.8}{GW190720A}{1.2}{GW190719A}{14.1}{GW190708A}{0.7}{GW190707A}{0.5}{GW190706A}{23.7}{GW190701A}{13.4}{GW190630A}{3.3}{GW190620A}{16.2}{GW190602A}{18.3}{GW190527A}{9.5}{GW190521B}{4.8}{GW190521A}{30.4}{GW190519A}{15.4}{GW190517A}{6.4}{GW190514A}{13.9}{GW190513A}{6.7}{GW190512A}{2.8}{GW190503A}{10.8}{GW190424A}{10.0}{GW190421A}{11.3}{GW190413B}{16.7}{GW190413A}{14.0}{GW190412A}{4.7}{GW190408A}{3.4}}}
\newcommand{\finalmassdetmed}[1]{\IfEqCase{#1}{{GW190930A}{22.1}{GW190929A}{144.3}{GW190924A}{14.8}{GW190915A}{74.8}{GW190910A}{97.0}{GW190909A}{114.5}{GW190828B}{42.7}{GW190828A}{75.7}{GW190814A}{26.9}{GW190803A}{95.8}{GW190731A}{104.6}{GW190728A}{22.7}{GW190727A}{99.2}{GW190720A}{23.7}{GW190719A}{90.0}{GW190708A}{34.4}{GW190707A}{22.1}{GW190706A}{171.1}{GW190701A}{124.0}{GW190630A}{66.3}{GW190620A}{130.3}{GW190602A}{163.8}{GW190527A}{80.3}{GW190521B}{88.0}{GW190521A}{256.6}{GW190519A}{146.8}{GW190517A}{79.8}{GW190514A}{108.3}{GW190513A}{70.6}{GW190512A}{43.5}{GW190503A}{87.6}{GW190424A}{96.0}{GW190421A}{103.9}{GW190413B}{129.8}{GW190413A}{89.6}{GW190412A}{42.9}{GW190408A}{53.0}}}
\newcommand{\finalmassdetplus}[1]{\IfEqCase{#1}{{GW190930A}{10.8}{GW190929A}{36.4}{GW190924A}{5.9}{GW190915A}{7.9}{GW190910A}{9.3}{GW190909A}{92.0}{GW190828B}{6.6}{GW190828A}{6.0}{GW190814A}{1.1}{GW190803A}{13.1}{GW190731A}{12.8}{GW190728A}{5.5}{GW190727A}{10.7}{GW190720A}{5.2}{GW190719A}{22.5}{GW190708A}{2.7}{GW190707A}{1.9}{GW190706A}{20.0}{GW190701A}{15.1}{GW190630A}{4.2}{GW190620A}{17.7}{GW190602A}{20.7}{GW190527A}{51.0}{GW190521B}{4.3}{GW190521A}{36.6}{GW190519A}{14.7}{GW190517A}{8.8}{GW190514A}{16.6}{GW190513A}{11.5}{GW190512A}{4.0}{GW190503A}{10.2}{GW190424A}{13.0}{GW190421A}{14.1}{GW190413B}{16.4}{GW190413A}{16.3}{GW190412A}{4.6}{GW190408A}{3.2}}}
\newcommand{\phioneminus}[1]{\IfEqCase{#1}{{GW190930A}{2.79}{GW190929A}{2.78}{GW190924A}{2.82}{GW190915A}{2.86}{GW190910A}{2.79}{GW190909A}{2.98}{GW190828B}{2.78}{GW190828A}{2.84}{GW190814A}{2.65}{GW190803A}{2.80}{GW190731A}{2.83}{GW190728A}{2.76}{GW190727A}{2.83}{GW190720A}{2.81}{GW190719A}{2.77}{GW190708A}{2.90}{GW190707A}{2.85}{GW190706A}{2.84}{GW190701A}{2.77}{GW190630A}{2.80}{GW190620A}{2.89}{GW190602A}{2.85}{GW190527A}{2.88}{GW190521B}{2.87}{GW190521A}{2.79}{GW190519A}{2.84}{GW190517A}{2.90}{GW190514A}{2.82}{GW190513A}{2.82}{GW190512A}{2.84}{GW190503A}{2.83}{GW190426A}{0.00}{GW190425A}{2.73}{GW190424A}{2.84}{GW190421A}{2.82}{GW190413B}{2.83}{GW190413A}{2.76}{GW190412A}{2.36}{GW190408A}{2.77}}}
\newcommand{\phionemed}[1]{\IfEqCase{#1}{{GW190930A}{3.10}{GW190929A}{3.09}{GW190924A}{3.09}{GW190915A}{3.17}{GW190910A}{3.12}{GW190909A}{3.26}{GW190828B}{3.09}{GW190828A}{3.16}{GW190814A}{2.97}{GW190803A}{3.12}{GW190731A}{3.12}{GW190728A}{3.05}{GW190727A}{3.13}{GW190720A}{3.12}{GW190719A}{3.09}{GW190708A}{3.21}{GW190707A}{3.16}{GW190706A}{3.15}{GW190701A}{3.07}{GW190630A}{3.13}{GW190620A}{3.23}{GW190602A}{3.18}{GW190527A}{3.16}{GW190521B}{3.17}{GW190521A}{3.14}{GW190519A}{3.15}{GW190517A}{3.23}{GW190514A}{3.18}{GW190513A}{3.11}{GW190512A}{3.15}{GW190503A}{3.15}{GW190426A}{0.00}{GW190425A}{3.05}{GW190424A}{3.16}{GW190421A}{3.13}{GW190413B}{3.16}{GW190413A}{3.06}{GW190412A}{2.69}{GW190408A}{3.08}}}
\newcommand{\phioneplus}[1]{\IfEqCase{#1}{{GW190930A}{2.86}{GW190929A}{2.90}{GW190924A}{2.86}{GW190915A}{2.79}{GW190910A}{2.85}{GW190909A}{2.71}{GW190828B}{2.87}{GW190828A}{2.82}{GW190814A}{2.95}{GW190803A}{2.85}{GW190731A}{2.88}{GW190728A}{2.92}{GW190727A}{2.83}{GW190720A}{2.84}{GW190719A}{2.86}{GW190708A}{2.79}{GW190707A}{2.83}{GW190706A}{2.83}{GW190701A}{2.88}{GW190630A}{2.83}{GW190620A}{2.75}{GW190602A}{2.77}{GW190527A}{2.83}{GW190521B}{2.80}{GW190521A}{2.80}{GW190519A}{2.83}{GW190517A}{2.76}{GW190514A}{2.77}{GW190513A}{2.86}{GW190512A}{2.84}{GW190503A}{2.81}{GW190426A}{0.00}{GW190425A}{2.90}{GW190424A}{2.78}{GW190421A}{2.82}{GW190413B}{2.79}{GW190413A}{2.90}{GW190412A}{3.20}{GW190408A}{2.87}}}
\newcommand{\phitwominus}[1]{\IfEqCase{#1}{{GW190930A}{2.89}{GW190929A}{2.80}{GW190924A}{2.83}{GW190915A}{2.82}{GW190910A}{2.83}{GW190909A}{2.79}{GW190828B}{2.84}{GW190828A}{2.83}{GW190814A}{2.74}{GW190803A}{2.82}{GW190731A}{2.88}{GW190728A}{2.85}{GW190727A}{2.83}{GW190720A}{2.82}{GW190719A}{2.74}{GW190708A}{2.76}{GW190707A}{2.79}{GW190706A}{2.81}{GW190701A}{2.97}{GW190630A}{2.80}{GW190620A}{2.84}{GW190602A}{2.84}{GW190527A}{2.79}{GW190521B}{2.82}{GW190521A}{2.86}{GW190519A}{2.79}{GW190517A}{2.82}{GW190514A}{2.89}{GW190513A}{2.80}{GW190512A}{2.75}{GW190503A}{2.80}{GW190426A}{0.00}{GW190425A}{2.84}{GW190424A}{2.84}{GW190421A}{2.81}{GW190413B}{2.97}{GW190413A}{2.83}{GW190412A}{2.75}{GW190408A}{2.80}}}
\newcommand{\phitwomed}[1]{\IfEqCase{#1}{{GW190930A}{3.21}{GW190929A}{3.12}{GW190924A}{3.15}{GW190915A}{3.16}{GW190910A}{3.14}{GW190909A}{3.09}{GW190828B}{3.16}{GW190828A}{3.13}{GW190814A}{3.07}{GW190803A}{3.14}{GW190731A}{3.21}{GW190728A}{3.15}{GW190727A}{3.12}{GW190720A}{3.15}{GW190719A}{3.08}{GW190708A}{3.05}{GW190707A}{3.09}{GW190706A}{3.12}{GW190701A}{3.27}{GW190630A}{3.12}{GW190620A}{3.14}{GW190602A}{3.15}{GW190527A}{3.06}{GW190521B}{3.15}{GW190521A}{3.17}{GW190519A}{3.12}{GW190517A}{3.14}{GW190514A}{3.19}{GW190513A}{3.14}{GW190512A}{3.07}{GW190503A}{3.14}{GW190426A}{0.00}{GW190425A}{3.15}{GW190424A}{3.15}{GW190421A}{3.14}{GW190413B}{3.28}{GW190413A}{3.16}{GW190412A}{3.09}{GW190408A}{3.11}}}
\newcommand{\phitwoplus}[1]{\IfEqCase{#1}{{GW190930A}{2.74}{GW190929A}{2.83}{GW190924A}{2.82}{GW190915A}{2.80}{GW190910A}{2.82}{GW190909A}{2.86}{GW190828B}{2.84}{GW190828A}{2.84}{GW190814A}{2.86}{GW190803A}{2.87}{GW190731A}{2.77}{GW190728A}{2.78}{GW190727A}{2.88}{GW190720A}{2.84}{GW190719A}{2.90}{GW190708A}{2.90}{GW190707A}{2.88}{GW190706A}{2.88}{GW190701A}{2.70}{GW190630A}{2.84}{GW190620A}{2.81}{GW190602A}{2.79}{GW190527A}{2.91}{GW190521B}{2.82}{GW190521A}{2.82}{GW190519A}{2.84}{GW190517A}{2.83}{GW190514A}{2.75}{GW190513A}{2.85}{GW190512A}{2.89}{GW190503A}{2.81}{GW190426A}{0.00}{GW190425A}{2.83}{GW190424A}{2.83}{GW190421A}{2.81}{GW190413B}{2.72}{GW190413A}{2.82}{GW190412A}{2.85}{GW190408A}{2.88}}}
\newcommand{\phionetwominus}[1]{\IfEqCase{#1}{{GW190930A}{2.84}{GW190929A}{2.79}{GW190924A}{2.83}{GW190915A}{2.89}{GW190910A}{2.78}{GW190909A}{2.85}{GW190828B}{2.83}{GW190828A}{2.63}{GW190814A}{2.71}{GW190803A}{2.82}{GW190731A}{2.73}{GW190728A}{2.79}{GW190727A}{2.83}{GW190720A}{2.74}{GW190719A}{2.75}{GW190708A}{2.66}{GW190707A}{2.79}{GW190706A}{2.80}{GW190701A}{2.82}{GW190630A}{2.74}{GW190620A}{2.78}{GW190602A}{2.78}{GW190527A}{2.77}{GW190521B}{2.78}{GW190521A}{3.09}{GW190519A}{2.92}{GW190517A}{2.91}{GW190514A}{2.83}{GW190513A}{2.84}{GW190512A}{2.80}{GW190503A}{2.70}{GW190426A}{0.00}{GW190425A}{2.87}{GW190424A}{2.85}{GW190421A}{2.83}{GW190413B}{2.67}{GW190413A}{2.84}{GW190412A}{3.10}{GW190408A}{2.77}}}
\newcommand{\phionetwomed}[1]{\IfEqCase{#1}{{GW190930A}{3.17}{GW190929A}{3.11}{GW190924A}{3.17}{GW190915A}{3.18}{GW190910A}{3.15}{GW190909A}{3.17}{GW190828B}{3.16}{GW190828A}{2.94}{GW190814A}{3.01}{GW190803A}{3.12}{GW190731A}{3.03}{GW190728A}{3.13}{GW190727A}{3.13}{GW190720A}{3.09}{GW190719A}{3.13}{GW190708A}{3.08}{GW190707A}{3.13}{GW190706A}{3.11}{GW190701A}{3.15}{GW190630A}{3.06}{GW190620A}{3.16}{GW190602A}{3.09}{GW190527A}{3.08}{GW190521B}{3.16}{GW190521A}{3.35}{GW190519A}{3.23}{GW190517A}{3.22}{GW190514A}{3.15}{GW190513A}{3.15}{GW190512A}{3.14}{GW190503A}{3.02}{GW190426A}{0.00}{GW190425A}{3.18}{GW190424A}{3.15}{GW190421A}{3.17}{GW190413B}{3.00}{GW190413A}{3.15}{GW190412A}{3.44}{GW190408A}{3.11}}}
\newcommand{\phionetwoplus}[1]{\IfEqCase{#1}{{GW190930A}{2.77}{GW190929A}{2.87}{GW190924A}{2.80}{GW190915A}{2.78}{GW190910A}{2.80}{GW190909A}{2.82}{GW190828B}{2.78}{GW190828A}{3.01}{GW190814A}{2.95}{GW190803A}{2.83}{GW190731A}{2.88}{GW190728A}{2.82}{GW190727A}{2.86}{GW190720A}{2.84}{GW190719A}{2.80}{GW190708A}{2.83}{GW190707A}{2.81}{GW190706A}{2.86}{GW190701A}{2.81}{GW190630A}{2.89}{GW190620A}{2.77}{GW190602A}{2.84}{GW190527A}{2.89}{GW190521B}{2.73}{GW190521A}{2.68}{GW190519A}{2.71}{GW190517A}{2.77}{GW190514A}{2.80}{GW190513A}{2.79}{GW190512A}{2.82}{GW190503A}{2.94}{GW190426A}{0.00}{GW190425A}{2.76}{GW190424A}{2.84}{GW190421A}{2.81}{GW190413B}{2.95}{GW190413A}{2.82}{GW190412A}{2.54}{GW190408A}{2.84}}}
\newcommand{\raminus}[1]{\IfEqCase{#1}{{GW190930A}{5.23224}{GW190929A}{3.05568}{GW190924A}{0.14434}{GW190915A}{0.08865}{GW190910A}{2.79654}{GW190909A}{1.21034}{GW190828B}{0.25804}{GW190828A}{0.19728}{GW190814A}{0.02832}{GW190803A}{0.32018}{GW190731A}{2.05429}{GW190728A}{3.93966}{GW190727A}{0.28275}{GW190720A}{5.03310}{GW190719A}{1.97394}{GW190708A}{2.51716}{GW190707A}{1.47850}{GW190706A}{0.11777}{GW190701A}{0.02938}{GW190630A}{3.00747}{GW190620A}{3.79912}{GW190602A}{0.16873}{GW190527A}{4.81181}{GW190521B}{0.49181}{GW190521A}{3.83925}{GW190519A}{3.53823}{GW190517A}{0.11026}{GW190514A}{2.80859}{GW190513A}{0.16674}{GW190512A}{0.34154}{GW190503A}{0.07308}{GW190426A}{5.11703}{GW190425A}{1.14713}{GW190424A}{2.87584}{GW190421A}{1.90797}{GW190413B}{0.19894}{GW190413A}{2.38305}{GW190412A}{0.06390}{GW190408A}{0.23336}}}
\newcommand{\ramed}[1]{\IfEqCase{#1}{{GW190930A}{5.56800}{GW190929A}{4.57328}{GW190924A}{2.28446}{GW190915A}{3.41338}{GW190910A}{3.71091}{GW190909A}{1.53780}{GW190828B}{2.46668}{GW190828A}{2.55563}{GW190814A}{0.22230}{GW190803A}{1.64028}{GW190731A}{3.03870}{GW190728A}{5.47833}{GW190727A}{1.82934}{GW190720A}{5.19166}{GW190719A}{2.73410}{GW190708A}{2.97596}{GW190707A}{3.54979}{GW190706A}{2.59111}{GW190701A}{0.66145}{GW190630A}{5.86181}{GW190620A}{4.24971}{GW190602A}{1.30570}{GW190527A}{5.13405}{GW190521B}{4.87951}{GW190521A}{3.88408}{GW190519A}{3.58629}{GW190517A}{4.07092}{GW190514A}{3.58258}{GW190513A}{0.89431}{GW190512A}{4.37140}{GW190503A}{1.65900}{GW190426A}{5.27391}{GW190425A}{1.62833}{GW190424A}{3.13318}{GW190421A}{3.50423}{GW190413B}{2.70054}{GW190413A}{2.53690}{GW190412A}{3.81244}{GW190408A}{6.08868}}}
\newcommand{\raplus}[1]{\IfEqCase{#1}{{GW190930A}{0.45017}{GW190929A}{0.94968}{GW190924A}{0.18201}{GW190915A}{0.07354}{GW190910A}{1.06677}{GW190909A}{4.24495}{GW190828B}{3.42033}{GW190828A}{3.27852}{GW190814A}{0.16914}{GW190803A}{1.69538}{GW190731A}{0.37410}{GW190728A}{0.52075}{GW190727A}{4.30388}{GW190720A}{0.96175}{GW190719A}{3.20570}{GW190708A}{2.84449}{GW190707A}{2.17972}{GW190706A}{3.23517}{GW190701A}{0.02994}{GW190630A}{0.12425}{GW190620A}{0.37483}{GW190602A}{0.27462}{GW190527A}{0.80861}{GW190521B}{0.59768}{GW190521A}{2.35925}{GW190519A}{2.66341}{GW190517A}{1.75963}{GW190514A}{1.66161}{GW190513A}{4.13656}{GW190512A}{0.17389}{GW190503A}{0.07931}{GW190426A}{0.85181}{GW190425A}{3.12911}{GW190424A}{2.81897}{GW190421A}{0.15768}{GW190413B}{1.98722}{GW190413A}{0.89215}{GW190412A}{0.03095}{GW190408A}{0.08337}}}
\newcommand{\phijlminus}[1]{\IfEqCase{#1}{{GW190930A}{2.81}{GW190929A}{3.08}{GW190924A}{2.69}{GW190915A}{2.76}{GW190910A}{2.83}{GW190909A}{2.84}{GW190828B}{2.87}{GW190828A}{3.02}{GW190814A}{1.87}{GW190803A}{2.70}{GW190731A}{2.69}{GW190728A}{2.87}{GW190727A}{2.85}{GW190720A}{2.86}{GW190719A}{2.89}{GW190708A}{2.84}{GW190707A}{2.89}{GW190706A}{2.70}{GW190701A}{2.57}{GW190630A}{2.93}{GW190620A}{3.11}{GW190602A}{2.78}{GW190527A}{2.80}{GW190521B}{2.82}{GW190521A}{2.81}{GW190519A}{2.86}{GW190517A}{2.16}{GW190514A}{2.83}{GW190513A}{2.69}{GW190512A}{2.82}{GW190503A}{3.64}{GW190426A}{1.43}{GW190425A}{2.88}{GW190424A}{2.83}{GW190421A}{2.83}{GW190413B}{3.08}{GW190413A}{2.65}{GW190412A}{3.64}{GW190408A}{2.54}}}
\newcommand{\phijlmed}[1]{\IfEqCase{#1}{{GW190930A}{3.14}{GW190929A}{3.37}{GW190924A}{3.02}{GW190915A}{3.39}{GW190910A}{3.15}{GW190909A}{3.16}{GW190828B}{3.19}{GW190828A}{3.35}{GW190814A}{2.28}{GW190803A}{3.05}{GW190731A}{3.03}{GW190728A}{3.19}{GW190727A}{3.21}{GW190720A}{3.15}{GW190719A}{3.21}{GW190708A}{3.17}{GW190707A}{3.21}{GW190706A}{3.03}{GW190701A}{3.05}{GW190630A}{3.23}{GW190620A}{3.51}{GW190602A}{3.12}{GW190527A}{3.13}{GW190521B}{3.17}{GW190521A}{3.14}{GW190519A}{3.17}{GW190517A}{2.38}{GW190514A}{3.15}{GW190513A}{3.00}{GW190512A}{3.16}{GW190503A}{3.90}{GW190426A}{1.70}{GW190425A}{3.23}{GW190424A}{3.16}{GW190421A}{3.14}{GW190413B}{3.36}{GW190413A}{2.98}{GW190412A}{3.82}{GW190408A}{2.90}}}
\newcommand{\phijlplus}[1]{\IfEqCase{#1}{{GW190930A}{2.86}{GW190929A}{2.65}{GW190924A}{2.94}{GW190915A}{2.29}{GW190910A}{2.84}{GW190909A}{2.81}{GW190828B}{2.77}{GW190828A}{2.59}{GW190814A}{3.50}{GW190803A}{2.90}{GW190731A}{2.89}{GW190728A}{2.76}{GW190727A}{2.79}{GW190720A}{2.84}{GW190719A}{2.78}{GW190708A}{2.80}{GW190707A}{2.77}{GW190706A}{2.92}{GW190701A}{2.72}{GW190630A}{2.76}{GW190620A}{2.40}{GW190602A}{2.83}{GW190527A}{2.85}{GW190521B}{2.86}{GW190521A}{2.80}{GW190519A}{2.79}{GW190517A}{3.65}{GW190514A}{2.83}{GW190513A}{2.95}{GW190512A}{2.80}{GW190503A}{2.15}{GW190426A}{1.19}{GW190425A}{2.76}{GW190424A}{2.82}{GW190421A}{2.84}{GW190413B}{2.68}{GW190413A}{2.97}{GW190412A}{2.23}{GW190408A}{3.00}}}
\newcommand{\tilttwominus}[1]{\IfEqCase{#1}{{GW190930A}{0.90}{GW190929A}{1.09}{GW190924A}{1.01}{GW190915A}{1.03}{GW190910A}{1.01}{GW190909A}{1.23}{GW190828B}{0.96}{GW190828A}{0.84}{GW190814A}{1.02}{GW190803A}{1.12}{GW190731A}{1.02}{GW190728A}{0.85}{GW190727A}{0.97}{GW190720A}{0.90}{GW190719A}{0.81}{GW190708A}{1.00}{GW190707A}{1.18}{GW190706A}{0.88}{GW190701A}{1.14}{GW190630A}{0.87}{GW190620A}{0.78}{GW190602A}{0.97}{GW190527A}{1.01}{GW190521B}{0.85}{GW190521A}{1.02}{GW190519A}{0.77}{GW190517A}{0.67}{GW190514A}{1.21}{GW190513A}{0.94}{GW190512A}{0.99}{GW190503A}{1.06}{GW190426A}{0.00}{GW190425A}{0.87}{GW190424A}{0.96}{GW190421A}{1.10}{GW190413B}{1.18}{GW190413A}{1.13}{GW190412A}{0.90}{GW190408A}{1.06}}}
\newcommand{\tilttwomed}[1]{\IfEqCase{#1}{{GW190930A}{1.26}{GW190929A}{1.53}{GW190924A}{1.42}{GW190915A}{1.56}{GW190910A}{1.53}{GW190909A}{1.74}{GW190828B}{1.32}{GW190828A}{1.19}{GW190814A}{1.60}{GW190803A}{1.63}{GW190731A}{1.42}{GW190728A}{1.17}{GW190727A}{1.37}{GW190720A}{1.25}{GW190719A}{1.11}{GW190708A}{1.43}{GW190707A}{1.76}{GW190706A}{1.19}{GW190701A}{1.74}{GW190630A}{1.23}{GW190620A}{1.05}{GW190602A}{1.38}{GW190527A}{1.41}{GW190521B}{1.27}{GW190521A}{1.59}{GW190519A}{1.05}{GW190517A}{0.91}{GW190514A}{1.94}{GW190513A}{1.32}{GW190512A}{1.41}{GW190503A}{1.59}{GW190426A}{0.00}{GW190425A}{1.41}{GW190424A}{1.37}{GW190421A}{1.73}{GW190413B}{1.70}{GW190413A}{1.65}{GW190412A}{1.32}{GW190408A}{1.59}}}
\newcommand{\tilttwoplus}[1]{\IfEqCase{#1}{{GW190930A}{1.21}{GW190929A}{1.13}{GW190924A}{1.16}{GW190915A}{1.09}{GW190910A}{1.04}{GW190909A}{1.01}{GW190828B}{1.20}{GW190828A}{1.25}{GW190814A}{1.00}{GW190803A}{1.05}{GW190731A}{1.16}{GW190728A}{1.29}{GW190727A}{1.20}{GW190720A}{1.25}{GW190719A}{1.31}{GW190708A}{1.13}{GW190707A}{0.92}{GW190706A}{1.29}{GW190701A}{0.97}{GW190630A}{1.18}{GW190620A}{1.32}{GW190602A}{1.15}{GW190527A}{1.20}{GW190521B}{1.13}{GW190521A}{1.05}{GW190519A}{1.26}{GW190517A}{1.30}{GW190514A}{0.88}{GW190513A}{1.19}{GW190512A}{1.11}{GW190503A}{1.07}{GW190426A}{3.14}{GW190425A}{0.94}{GW190424A}{1.16}{GW190421A}{0.99}{GW190413B}{1.00}{GW190413A}{1.04}{GW190412A}{1.12}{GW190408A}{1.03}}}
\newcommand{\costhetajnminus}[1]{\IfEqCase{#1}{{GW190930A}{1.54}{GW190929A}{0.76}{GW190924A}{1.67}{GW190915A}{0.50}{GW190910A}{0.88}{GW190909A}{1.06}{GW190828B}{0.71}{GW190828A}{0.36}{GW190814A}{1.35}{GW190803A}{1.39}{GW190731A}{1.33}{GW190728A}{1.30}{GW190727A}{0.98}{GW190720A}{0.20}{GW190719A}{0.92}{GW190708A}{1.17}{GW190707A}{0.43}{GW190706A}{1.14}{GW190701A}{0.41}{GW190630A}{1.29}{GW190620A}{0.60}{GW190602A}{0.82}{GW190527A}{1.25}{GW190521B}{1.05}{GW190521A}{1.35}{GW190519A}{0.81}{GW190517A}{0.34}{GW190514A}{1.02}{GW190513A}{1.65}{GW190512A}{0.91}{GW190503A}{0.23}{GW190426A}{0.84}{GW190425A}{1.43}{GW190424A}{1.02}{GW190421A}{0.80}{GW190413B}{0.63}{GW190413A}{1.35}{GW190412A}{0.35}{GW190408A}{1.65}}}
\newcommand{\costhetajnmed}[1]{\IfEqCase{#1}{{GW190930A}{0.59}{GW190929A}{-0.13}{GW190924A}{0.74}{GW190915A}{-0.45}{GW190910A}{-0.05}{GW190909A}{0.12}{GW190828B}{-0.25}{GW190828A}{-0.62}{GW190814A}{0.65}{GW190803A}{0.44}{GW190731A}{0.37}{GW190728A}{0.33}{GW190727A}{0.02}{GW190720A}{-0.79}{GW190719A}{-0.04}{GW190708A}{0.20}{GW190707A}{-0.55}{GW190706A}{0.20}{GW190701A}{0.84}{GW190630A}{0.34}{GW190620A}{-0.36}{GW190602A}{-0.14}{GW190527A}{0.30}{GW190521B}{0.09}{GW190521A}{0.41}{GW190519A}{-0.01}{GW190517A}{-0.64}{GW190514A}{0.07}{GW190513A}{0.70}{GW190512A}{-0.04}{GW190503A}{-0.75}{GW190426A}{-0.13}{GW190425A}{0.47}{GW190424A}{0.05}{GW190421A}{-0.17}{GW190413B}{-0.33}{GW190413A}{0.41}{GW190412A}{0.75}{GW190408A}{0.70}}}
\newcommand{\costhetajnplus}[1]{\IfEqCase{#1}{{GW190930A}{0.39}{GW190929A}{1.01}{GW190924A}{0.25}{GW190915A}{1.32}{GW190910A}{0.96}{GW190909A}{0.82}{GW190828B}{1.20}{GW190828A}{1.57}{GW190814A}{0.16}{GW190803A}{0.53}{GW190731A}{0.59}{GW190728A}{0.65}{GW190727A}{0.94}{GW190720A}{1.68}{GW190719A}{1.01}{GW190708A}{0.78}{GW190707A}{1.51}{GW190706A}{0.75}{GW190701A}{0.15}{GW190630A}{0.62}{GW190620A}{1.27}{GW190602A}{1.10}{GW190527A}{0.66}{GW190521B}{0.87}{GW190521A}{0.55}{GW190519A}{0.81}{GW190517A}{1.38}{GW190514A}{0.89}{GW190513A}{0.27}{GW190512A}{0.99}{GW190503A}{0.48}{GW190426A}{1.09}{GW190425A}{0.50}{GW190424A}{0.91}{GW190421A}{1.12}{GW190413B}{1.27}{GW190413A}{0.55}{GW190412A}{0.14}{GW190408A}{0.28}}}
\newcommand{\spintwominus}[1]{\IfEqCase{#1}{{GW190930A}{0.37}{GW190929A}{0.44}{GW190924A}{0.32}{GW190915A}{0.43}{GW190910A}{0.33}{GW190909A}{0.43}{GW190828B}{0.38}{GW190828A}{0.37}{GW190814A}{0.46}{GW190803A}{0.40}{GW190731A}{0.40}{GW190728A}{0.35}{GW190727A}{0.41}{GW190720A}{0.45}{GW190719A}{0.49}{GW190708A}{0.28}{GW190707A}{0.28}{GW190706A}{0.44}{GW190701A}{0.40}{GW190630A}{0.34}{GW190620A}{0.50}{GW190602A}{0.45}{GW190527A}{0.45}{GW190521B}{0.37}{GW190521A}{0.52}{GW190519A}{0.48}{GW190517A}{0.52}{GW190514A}{0.48}{GW190513A}{0.39}{GW190512A}{0.32}{GW190503A}{0.40}{GW190426A}{0.009}{GW190425A}{0.25}{GW190424A}{0.42}{GW190421A}{0.42}{GW190413B}{0.45}{GW190413A}{0.41}{GW190412A}{0.43}{GW190408A}{0.32}}}
\newcommand{\spintwomed}[1]{\IfEqCase{#1}{{GW190930A}{0.42}{GW190929A}{0.49}{GW190924A}{0.35}{GW190915A}{0.48}{GW190910A}{0.37}{GW190909A}{0.49}{GW190828B}{0.42}{GW190828A}{0.41}{GW190814A}{0.52}{GW190803A}{0.45}{GW190731A}{0.45}{GW190728A}{0.39}{GW190727A}{0.45}{GW190720A}{0.51}{GW190719A}{0.55}{GW190708A}{0.30}{GW190707A}{0.31}{GW190706A}{0.49}{GW190701A}{0.44}{GW190630A}{0.38}{GW190620A}{0.56}{GW190602A}{0.50}{GW190527A}{0.50}{GW190521B}{0.42}{GW190521A}{0.58}{GW190519A}{0.54}{GW190517A}{0.58}{GW190514A}{0.54}{GW190513A}{0.43}{GW190512A}{0.36}{GW190503A}{0.44}{GW190426A}{0.009}{GW190425A}{0.28}{GW190424A}{0.47}{GW190421A}{0.46}{GW190413B}{0.50}{GW190413A}{0.45}{GW190412A}{0.49}{GW190408A}{0.36}}}
\newcommand{\spintwoplus}[1]{\IfEqCase{#1}{{GW190930A}{0.49}{GW190929A}{0.45}{GW190924A}{0.51}{GW190915A}{0.46}{GW190910A}{0.51}{GW190909A}{0.45}{GW190828B}{0.49}{GW190828A}{0.46}{GW190814A}{0.41}{GW190803A}{0.49}{GW190731A}{0.48}{GW190728A}{0.50}{GW190727A}{0.46}{GW190720A}{0.43}{GW190719A}{0.40}{GW190708A}{0.50}{GW190707A}{0.52}{GW190706A}{0.45}{GW190701A}{0.48}{GW190630A}{0.46}{GW190620A}{0.40}{GW190602A}{0.44}{GW190527A}{0.45}{GW190521B}{0.39}{GW190521A}{0.38}{GW190519A}{0.41}{GW190517A}{0.38}{GW190514A}{0.42}{GW190513A}{0.48}{GW190512A}{0.51}{GW190503A}{0.48}{GW190426A}{0.03}{GW190425A}{0.51}{GW190424A}{0.47}{GW190421A}{0.47}{GW190413B}{0.44}{GW190413A}{0.48}{GW190412A}{0.44}{GW190408A}{0.53}}}
\newcommand{\massonedetminus}[1]{\IfEqCase{#1}{{GW190930A}{2.6}{GW190929A}{28.9}{GW190924A}{2.2}{GW190915A}{7.7}{GW190910A}{6.2}{GW190909A}{17.5}{GW190828B}{8.7}{GW190828A}{4.7}{GW190814A}{1.0}{GW190803A}{9.0}{GW190731A}{10.5}{GW190728A}{2.5}{GW190727A}{8.2}{GW190720A}{3.5}{GW190719A}{16.4}{GW190708A}{2.4}{GW190707A}{1.8}{GW190706A}{17.6}{GW190701A}{10.9}{GW190630A}{6.4}{GW190620A}{15.4}{GW190602A}{15.7}{GW190527A}{11.0}{GW190521B}{5.4}{GW190521A}{20.8}{GW190519A}{12.9}{GW190517A}{8.4}{GW190514A}{10.8}{GW190513A}{12.3}{GW190512A}{6.8}{GW190503A}{9.9}{GW190426A}{2.5}{GW190425A}{0.4}{GW190424A}{8.0}{GW190421A}{8.8}{GW190413B}{14.6}{GW190413A}{12.0}{GW190412A}{6.2}{GW190408A}{4.0}}}
\newcommand{\massonedetmed}[1]{\IfEqCase{#1}{{GW190930A}{14.2}{GW190929A}{111.3}{GW190924A}{9.9}{GW190915A}{46.0}{GW190910A}{56.3}{GW190909A}{73.0}{GW190828B}{31.1}{GW190828A}{43.9}{GW190814A}{24.4}{GW190803A}{57.6}{GW190731A}{64.4}{GW190728A}{14.4}{GW190727A}{58.8}{GW190720A}{15.7}{GW190719A}{60.3}{GW190708A}{20.6}{GW190707A}{13.4}{GW190706A}{112.9}{GW190701A}{74.1}{GW190630A}{41.4}{GW190620A}{84.6}{GW190602A}{101.7}{GW190527A}{52.0}{GW190521B}{51.9}{GW190521A}{153.4}{GW190519A}{95.4}{GW190517A}{50.5}{GW190514A}{64.9}{GW190513A}{49.0}{GW190512A}{29.4}{GW190503A}{55.2}{GW190426A}{6.2}{GW190425A}{2.1}{GW190424A}{56.1}{GW190421A}{61.1}{GW190413B}{80.7}{GW190413A}{55.3}{GW190412A}{34.6}{GW190408A}{31.5}}}
\newcommand{\massonedetplus}[1]{\IfEqCase{#1}{{GW190930A}{14.3}{GW190929A}{39.3}{GW190924A}{7.8}{GW190915A}{11.6}{GW190910A}{8.8}{GW190909A}{84.2}{GW190828B}{8.8}{GW190828A}{7.5}{GW190814A}{1.2}{GW190803A}{14.2}{GW190731A}{13.4}{GW190728A}{8.4}{GW190727A}{12.8}{GW190720A}{7.7}{GW190719A}{26.5}{GW190708A}{5.7}{GW190707A}{3.7}{GW190706A}{20.5}{GW190701A}{15.1}{GW190630A}{8.3}{GW190620A}{20.0}{GW190602A}{19.9}{GW190527A}{32.9}{GW190521B}{7.1}{GW190521A}{45.9}{GW190519A}{14.9}{GW190517A}{14.7}{GW190514A}{17.6}{GW190513A}{12.5}{GW190512A}{6.5}{GW190503A}{10.6}{GW190426A}{4.2}{GW190425A}{0.6}{GW190424A}{14.9}{GW190421A}{14.7}{GW190413B}{19.0}{GW190413A}{17.1}{GW190412A}{5.5}{GW190408A}{6.3}}}
\newcommand{\massratiominus}[1]{\IfEqCase{#1}{{GW190930A}{0.46}{GW190929A}{0.16}{GW190924A}{0.37}{GW190915A}{0.26}{GW190910A}{0.23}{GW190909A}{0.39}{GW190828B}{0.16}{GW190828A}{0.23}{GW190814A}{0.009}{GW190803A}{0.31}{GW190731A}{0.31}{GW190728A}{0.38}{GW190727A}{0.32}{GW190720A}{0.30}{GW190719A}{0.29}{GW190708A}{0.28}{GW190707A}{0.27}{GW190706A}{0.25}{GW190701A}{0.30}{GW190630A}{0.22}{GW190620A}{0.27}{GW190602A}{0.33}{GW190527A}{0.32}{GW190521B}{0.21}{GW190521A}{0.34}{GW190519A}{0.19}{GW190517A}{0.29}{GW190514A}{0.33}{GW190513A}{0.18}{GW190512A}{0.18}{GW190503A}{0.23}{GW190426A}{0.15}{GW190425A}{0.25}{GW190424A}{0.29}{GW190421A}{0.30}{GW190413B}{0.31}{GW190413A}{0.28}{GW190412A}{0.06}{GW190408A}{0.25}}}
\newcommand{\massratiomed}[1]{\IfEqCase{#1}{{GW190930A}{0.64}{GW190929A}{0.30}{GW190924A}{0.57}{GW190915A}{0.69}{GW190910A}{0.82}{GW190909A}{0.62}{GW190828B}{0.42}{GW190828A}{0.82}{GW190814A}{0.112}{GW190803A}{0.75}{GW190731A}{0.72}{GW190728A}{0.66}{GW190727A}{0.79}{GW190720A}{0.59}{GW190719A}{0.58}{GW190708A}{0.75}{GW190707A}{0.72}{GW190706A}{0.58}{GW190701A}{0.76}{GW190630A}{0.68}{GW190620A}{0.62}{GW190602A}{0.71}{GW190527A}{0.64}{GW190521B}{0.78}{GW190521A}{0.75}{GW190519A}{0.61}{GW190517A}{0.68}{GW190514A}{0.75}{GW190513A}{0.50}{GW190512A}{0.54}{GW190503A}{0.65}{GW190426A}{0.25}{GW190425A}{0.67}{GW190424A}{0.81}{GW190421A}{0.79}{GW190413B}{0.69}{GW190413A}{0.69}{GW190412A}{0.28}{GW190408A}{0.76}}}
\newcommand{\massratioplus}[1]{\IfEqCase{#1}{{GW190930A}{0.30}{GW190929A}{0.52}{GW190924A}{0.36}{GW190915A}{0.27}{GW190910A}{0.15}{GW190909A}{0.33}{GW190828B}{0.38}{GW190828A}{0.15}{GW190814A}{0.008}{GW190803A}{0.22}{GW190731A}{0.25}{GW190728A}{0.29}{GW190727A}{0.18}{GW190720A}{0.36}{GW190719A}{0.37}{GW190708A}{0.21}{GW190707A}{0.24}{GW190706A}{0.34}{GW190701A}{0.21}{GW190630A}{0.27}{GW190620A}{0.33}{GW190602A}{0.25}{GW190527A}{0.32}{GW190521B}{0.19}{GW190521A}{0.23}{GW190519A}{0.28}{GW190517A}{0.27}{GW190514A}{0.21}{GW190513A}{0.42}{GW190512A}{0.37}{GW190503A}{0.29}{GW190426A}{0.41}{GW190425A}{0.29}{GW190424A}{0.17}{GW190421A}{0.18}{GW190413B}{0.28}{GW190413A}{0.28}{GW190412A}{0.12}{GW190408A}{0.21}}}
\newcommand{\spinoneminus}[1]{\IfEqCase{#1}{{GW190930A}{0.35}{GW190929A}{0.54}{GW190924A}{0.21}{GW190915A}{0.49}{GW190910A}{0.30}{GW190909A}{0.52}{GW190828B}{0.26}{GW190828A}{0.40}{GW190814A}{0.03}{GW190803A}{0.37}{GW190731A}{0.34}{GW190728A}{0.28}{GW190727A}{0.42}{GW190720A}{0.35}{GW190719A}{0.53}{GW190708A}{0.20}{GW190707A}{0.21}{GW190706A}{0.48}{GW190701A}{0.36}{GW190630A}{0.23}{GW190620A}{0.50}{GW190602A}{0.34}{GW190527A}{0.43}{GW190521B}{0.28}{GW190521A}{0.63}{GW190519A}{0.50}{GW190517A}{0.35}{GW190514A}{0.46}{GW190513A}{0.28}{GW190512A}{0.16}{GW190503A}{0.31}{GW190426A}{0.14}{GW190425A}{0.25}{GW190424A}{0.47}{GW190421A}{0.41}{GW190413B}{0.52}{GW190413A}{0.36}{GW190412A}{0.22}{GW190408A}{0.31}}}
\newcommand{\spinonemed}[1]{\IfEqCase{#1}{{GW190930A}{0.39}{GW190929A}{0.64}{GW190924A}{0.24}{GW190915A}{0.55}{GW190910A}{0.34}{GW190909A}{0.58}{GW190828B}{0.28}{GW190828A}{0.44}{GW190814A}{0.03}{GW190803A}{0.41}{GW190731A}{0.37}{GW190728A}{0.32}{GW190727A}{0.46}{GW190720A}{0.40}{GW190719A}{0.62}{GW190708A}{0.22}{GW190707A}{0.24}{GW190706A}{0.55}{GW190701A}{0.40}{GW190630A}{0.26}{GW190620A}{0.61}{GW190602A}{0.38}{GW190527A}{0.47}{GW190521B}{0.31}{GW190521A}{0.73}{GW190519A}{0.60}{GW190517A}{0.86}{GW190514A}{0.52}{GW190513A}{0.30}{GW190512A}{0.17}{GW190503A}{0.34}{GW190426A}{0.14}{GW190425A}{0.27}{GW190424A}{0.53}{GW190421A}{0.46}{GW190413B}{0.58}{GW190413A}{0.40}{GW190412A}{0.44}{GW190408A}{0.34}}}
\newcommand{\spinoneplus}[1]{\IfEqCase{#1}{{GW190930A}{0.40}{GW190929A}{0.32}{GW190924A}{0.43}{GW190915A}{0.39}{GW190910A}{0.50}{GW190909A}{0.37}{GW190828B}{0.43}{GW190828A}{0.45}{GW190814A}{0.05}{GW190803A}{0.51}{GW190731A}{0.54}{GW190728A}{0.37}{GW190727A}{0.47}{GW190720A}{0.40}{GW190719A}{0.34}{GW190708A}{0.52}{GW190707A}{0.47}{GW190706A}{0.39}{GW190701A}{0.50}{GW190630A}{0.37}{GW190620A}{0.34}{GW190602A}{0.51}{GW190527A}{0.47}{GW190521B}{0.42}{GW190521A}{0.25}{GW190519A}{0.33}{GW190517A}{0.13}{GW190514A}{0.43}{GW190513A}{0.51}{GW190512A}{0.44}{GW190503A}{0.51}{GW190426A}{0.40}{GW190425A}{0.51}{GW190424A}{0.42}{GW190421A}{0.47}{GW190413B}{0.38}{GW190413A}{0.51}{GW190412A}{0.16}{GW190408A}{0.47}}}
\newcommand{\costiltoneminus}[1]{\IfEqCase{#1}{{GW190930A}{1.08}{GW190929A}{0.83}{GW190924A}{0.97}{GW190915A}{0.82}{GW190910A}{0.93}{GW190909A}{0.78}{GW190828B}{0.99}{GW190828A}{1.09}{GW190814A}{0.90}{GW190803A}{0.80}{GW190731A}{0.99}{GW190728A}{1.13}{GW190727A}{1.02}{GW190720A}{0.97}{GW190719A}{1.02}{GW190708A}{0.88}{GW190707A}{0.68}{GW190706A}{0.99}{GW190701A}{0.70}{GW190630A}{1.02}{GW190620A}{0.84}{GW190602A}{0.99}{GW190527A}{1.02}{GW190521B}{0.97}{GW190521A}{0.89}{GW190519A}{0.74}{GW190517A}{0.37}{GW190514A}{0.50}{GW190513A}{1.10}{GW190512A}{0.93}{GW190503A}{0.75}{GW190426A}{0.00}{GW190425A}{0.65}{GW190424A}{1.00}{GW190421A}{0.75}{GW190413B}{0.76}{GW190413A}{0.90}{GW190412A}{0.47}{GW190408A}{0.73}}}
\newcommand{\costiltonemed}[1]{\IfEqCase{#1}{{GW190930A}{0.47}{GW190929A}{0.02}{GW190924A}{0.19}{GW190915A}{0.06}{GW190910A}{0.08}{GW190909A}{-0.10}{GW190828B}{0.26}{GW190828A}{0.51}{GW190814A}{0.01}{GW190803A}{-0.07}{GW190731A}{0.16}{GW190728A}{0.49}{GW190727A}{0.30}{GW190720A}{0.54}{GW190719A}{0.66}{GW190708A}{0.07}{GW190707A}{-0.19}{GW190706A}{0.66}{GW190701A}{-0.22}{GW190630A}{0.29}{GW190620A}{0.68}{GW190602A}{0.18}{GW190527A}{0.28}{GW190521B}{0.17}{GW190521A}{0.05}{GW190519A}{0.65}{GW190517A}{0.83}{GW190514A}{-0.45}{GW190513A}{0.41}{GW190512A}{0.07}{GW190503A}{-0.16}{GW190426A}{-1.00}{GW190425A}{0.26}{GW190424A}{0.36}{GW190421A}{-0.15}{GW190413B}{-0.06}{GW190413A}{0.01}{GW190412A}{0.70}{GW190408A}{-0.17}}}
\newcommand{\costiltoneplus}[1]{\IfEqCase{#1}{{GW190930A}{0.49}{GW190929A}{0.65}{GW190924A}{0.76}{GW190915A}{0.73}{GW190910A}{0.79}{GW190909A}{0.96}{GW190828B}{0.63}{GW190828A}{0.44}{GW190814A}{0.87}{GW190803A}{0.91}{GW190731A}{0.74}{GW190728A}{0.47}{GW190727A}{0.61}{GW190720A}{0.42}{GW190719A}{0.31}{GW190708A}{0.80}{GW190707A}{0.97}{GW190706A}{0.31}{GW190701A}{1.01}{GW190630A}{0.63}{GW190620A}{0.29}{GW190602A}{0.72}{GW190527A}{0.65}{GW190521B}{0.71}{GW190521A}{0.78}{GW190519A}{0.32}{GW190517A}{0.16}{GW190514A}{1.08}{GW190513A}{0.53}{GW190512A}{0.82}{GW190503A}{0.96}{GW190426A}{2.00}{GW190425A}{0.61}{GW190424A}{0.56}{GW190421A}{0.94}{GW190413B}{0.82}{GW190413A}{0.87}{GW190412A}{0.20}{GW190408A}{0.94}}}
\newcommand{\finalmasssourceminus}[1]{\IfEqCase{#1}{{GW190930A}{1.5}{GW190929A}{25.3}{GW190924A}{1.0}{GW190915A}{6.0}{GW190910A}{8.6}{GW190909A}{16.8}{GW190828B}{4.5}{GW190828A}{4.3}{GW190814A}{0.9}{GW190803A}{8.5}{GW190731A}{10.8}{GW190728A}{1.3}{GW190727A}{7.5}{GW190720A}{2.2}{GW190719A}{10.2}{GW190708A}{1.8}{GW190707A}{1.3}{GW190706A}{13.5}{GW190701A}{8.9}{GW190630A}{4.6}{GW190620A}{12.1}{GW190602A}{14.9}{GW190527A}{9.3}{GW190521B}{4.4}{GW190521A}{22.4}{GW190519A}{13.8}{GW190517A}{8.9}{GW190514A}{10.4}{GW190513A}{5.8}{GW190512A}{3.5}{GW190503A}{7.7}{GW190424A}{10.1}{GW190421A}{8.7}{GW190413B}{11.4}{GW190413A}{9.2}{GW190412A}{3.8}{GW190408A}{2.8}}}
\newcommand{\finalmasssourcemed}[1]{\IfEqCase{#1}{{GW190930A}{19.4}{GW190929A}{101.5}{GW190924A}{13.3}{GW190915A}{57.2}{GW190910A}{75.8}{GW190909A}{72.0}{GW190828B}{33.1}{GW190828A}{54.9}{GW190814A}{25.6}{GW190803A}{61.7}{GW190731A}{67.0}{GW190728A}{19.6}{GW190727A}{63.8}{GW190720A}{20.4}{GW190719A}{54.9}{GW190708A}{29.5}{GW190707A}{19.2}{GW190706A}{99.0}{GW190701A}{90.2}{GW190630A}{56.4}{GW190620A}{87.2}{GW190602A}{110.9}{GW190527A}{56.4}{GW190521B}{71.0}{GW190521A}{156.3}{GW190519A}{101.0}{GW190517A}{59.3}{GW190514A}{64.5}{GW190513A}{51.6}{GW190512A}{34.5}{GW190503A}{68.6}{GW190424A}{68.9}{GW190421A}{69.7}{GW190413B}{75.5}{GW190413A}{56.0}{GW190412A}{37.3}{GW190408A}{41.1}}}
\newcommand{\finalmasssourceplus}[1]{\IfEqCase{#1}{{GW190930A}{9.2}{GW190929A}{33.6}{GW190924A}{5.2}{GW190915A}{7.1}{GW190910A}{8.5}{GW190909A}{54.9}{GW190828B}{5.5}{GW190828A}{7.2}{GW190814A}{1.1}{GW190803A}{11.8}{GW190731A}{14.6}{GW190728A}{4.7}{GW190727A}{10.9}{GW190720A}{4.5}{GW190719A}{17.3}{GW190708A}{2.5}{GW190707A}{1.9}{GW190706A}{18.3}{GW190701A}{11.3}{GW190630A}{4.4}{GW190620A}{16.8}{GW190602A}{17.7}{GW190527A}{20.2}{GW190521B}{6.5}{GW190521A}{36.8}{GW190519A}{12.4}{GW190517A}{9.1}{GW190514A}{17.9}{GW190513A}{8.2}{GW190512A}{3.8}{GW190503A}{8.8}{GW190424A}{12.4}{GW190421A}{12.5}{GW190413B}{16.4}{GW190413A}{12.5}{GW190412A}{3.9}{GW190408A}{3.9}}}
\newcommand{\phaseminus}[1]{\IfEqCase{#1}{{GW190930A}{2.91}{GW190929A}{2.79}{GW190924A}{2.77}{GW190915A}{2.96}{GW190910A}{2.81}{GW190909A}{2.80}{GW190828B}{2.90}{GW190828A}{2.89}{GW190814A}{2.76}{GW190803A}{2.85}{GW190731A}{2.83}{GW190728A}{2.80}{GW190727A}{2.88}{GW190720A}{2.81}{GW190719A}{2.84}{GW190708A}{2.83}{GW190707A}{2.91}{GW190706A}{2.45}{GW190701A}{2.60}{GW190630A}{3.46}{GW190620A}{2.68}{GW190602A}{2.80}{GW190527A}{2.85}{GW190521B}{1.76}{GW190521A}{2.87}{GW190519A}{2.80}{GW190517A}{2.79}{GW190514A}{2.81}{GW190513A}{2.78}{GW190512A}{2.87}{GW190503A}{2.88}{GW190426A}{2.79}{GW190425A}{2.82}{GW190424A}{2.86}{GW190421A}{2.89}{GW190413B}{2.82}{GW190413A}{2.81}{GW190412A}{1.89}{GW190408A}{2.79}}}
\newcommand{\phasemed}[1]{\IfEqCase{#1}{{GW190930A}{3.22}{GW190929A}{3.08}{GW190924A}{3.08}{GW190915A}{3.30}{GW190910A}{3.15}{GW190909A}{3.15}{GW190828B}{3.23}{GW190828A}{3.17}{GW190814A}{3.13}{GW190803A}{3.14}{GW190731A}{3.15}{GW190728A}{3.11}{GW190727A}{3.21}{GW190720A}{3.12}{GW190719A}{3.15}{GW190708A}{3.14}{GW190707A}{3.24}{GW190706A}{3.01}{GW190701A}{2.89}{GW190630A}{3.82}{GW190620A}{2.97}{GW190602A}{3.13}{GW190527A}{3.13}{GW190521B}{2.03}{GW190521A}{3.12}{GW190519A}{3.13}{GW190517A}{3.10}{GW190514A}{3.13}{GW190513A}{3.08}{GW190512A}{3.18}{GW190503A}{3.16}{GW190426A}{3.09}{GW190425A}{3.12}{GW190424A}{3.16}{GW190421A}{3.20}{GW190413B}{3.11}{GW190413A}{3.12}{GW190412A}{2.15}{GW190408A}{3.11}}}
\newcommand{\phaseplus}[1]{\IfEqCase{#1}{{GW190930A}{2.77}{GW190929A}{2.89}{GW190924A}{2.88}{GW190915A}{2.70}{GW190910A}{2.84}{GW190909A}{2.82}{GW190828B}{2.73}{GW190828A}{2.85}{GW190814A}{2.81}{GW190803A}{2.81}{GW190731A}{2.80}{GW190728A}{2.86}{GW190727A}{2.76}{GW190720A}{2.85}{GW190719A}{2.83}{GW190708A}{2.83}{GW190707A}{2.76}{GW190706A}{2.65}{GW190701A}{3.10}{GW190630A}{2.15}{GW190620A}{2.97}{GW190602A}{2.84}{GW190527A}{2.81}{GW190521B}{3.94}{GW190521A}{2.87}{GW190519A}{2.83}{GW190517A}{2.88}{GW190514A}{2.82}{GW190513A}{2.90}{GW190512A}{2.82}{GW190503A}{2.84}{GW190426A}{2.85}{GW190425A}{2.87}{GW190424A}{2.81}{GW190421A}{2.77}{GW190413B}{2.84}{GW190413A}{2.84}{GW190412A}{3.84}{GW190408A}{2.86}}}
\newcommand{\radiatedenergyminus}[1]{\IfEqCase{#1}{{GW190930A}{0.2}{GW190929A}{1.4}{GW190924A}{0.1}{GW190915A}{0.7}{GW190910A}{0.7}{GW190909A}{1.4}{GW190828B}{0.2}{GW190828A}{0.5}{GW190814A}{0.007}{GW190803A}{0.8}{GW190731A}{1.1}{GW190728A}{0.2}{GW190727A}{0.9}{GW190720A}{0.2}{GW190719A}{1.1}{GW190708A}{0.2}{GW190707A}{0.09}{GW190706A}{2.1}{GW190701A}{1.1}{GW190630A}{0.5}{GW190620A}{1.9}{GW190602A}{1.9}{GW190527A}{1.0}{GW190521B}{0.7}{GW190521A}{2.4}{GW190519A}{1.7}{GW190517A}{1.2}{GW190514A}{0.8}{GW190513A}{0.6}{GW190512A}{0.3}{GW190503A}{1.0}{GW190424A}{0.9}{GW190421A}{0.9}{GW190413B}{1.1}{GW190413A}{0.8}{GW190412A}{0.1}{GW190408A}{0.3}}}
\newcommand{\radiatedenergymed}[1]{\IfEqCase{#1}{{GW190930A}{0.9}{GW190929A}{2.7}{GW190924A}{0.6}{GW190915A}{2.7}{GW190910A}{3.8}{GW190909A}{3.0}{GW190828B}{1.2}{GW190828A}{3.1}{GW190814A}{0.2}{GW190803A}{2.9}{GW190731A}{3.2}{GW190728A}{1.0}{GW190727A}{3.3}{GW190720A}{1.0}{GW190719A}{2.9}{GW190708A}{1.4}{GW190707A}{0.9}{GW190706A}{5.3}{GW190701A}{4.1}{GW190630A}{2.8}{GW190620A}{4.9}{GW190602A}{5.4}{GW190527A}{2.7}{GW190521B}{3.7}{GW190521A}{7.6}{GW190519A}{5.6}{GW190517A}{4.1}{GW190514A}{2.7}{GW190513A}{2.2}{GW190512A}{1.5}{GW190503A}{3.1}{GW190424A}{3.6}{GW190421A}{3.3}{GW190413B}{3.4}{GW190413A}{2.6}{GW190412A}{1.1}{GW190408A}{1.9}}}
\newcommand{\radiatedenergyplus}[1]{\IfEqCase{#1}{{GW190930A}{0.1}{GW190929A}{2.8}{GW190924A}{0.06}{GW190915A}{0.7}{GW190910A}{0.9}{GW190909A}{2.2}{GW190828B}{0.3}{GW190828A}{0.7}{GW190814A}{0.006}{GW190803A}{0.9}{GW190731A}{1.2}{GW190728A}{0.09}{GW190727A}{1.1}{GW190720A}{0.1}{GW190719A}{1.7}{GW190708A}{0.1}{GW190707A}{0.08}{GW190706A}{2.3}{GW190701A}{1.1}{GW190630A}{0.5}{GW190620A}{2.0}{GW190602A}{1.8}{GW190527A}{1.5}{GW190521B}{0.6}{GW190521A}{2.9}{GW190519A}{1.7}{GW190517A}{1.3}{GW190514A}{1.1}{GW190513A}{1.1}{GW190512A}{0.3}{GW190503A}{0.9}{GW190424A}{1.2}{GW190421A}{1.0}{GW190413B}{1.1}{GW190413A}{1.0}{GW190412A}{0.2}{GW190408A}{0.3}}}
\newcommand{\masstwodetminus}[1]{\IfEqCase{#1}{{GW190930A}{3.9}{GW190929A}{16.4}{GW190924A}{2.1}{GW190915A}{8.5}{GW190910A}{9.1}{GW190909A}{23.3}{GW190828B}{2.7}{GW190828A}{6.5}{GW190814A}{0.09}{GW190803A}{13.5}{GW190731A}{16.4}{GW190728A}{3.0}{GW190727A}{13.8}{GW190720A}{2.6}{GW190719A}{12.7}{GW190708A}{3.1}{GW190707A}{1.9}{GW190706A}{27.0}{GW190701A}{17.5}{GW190630A}{5.5}{GW190620A}{19.7}{GW190602A}{27.8}{GW190527A}{12.3}{GW190521B}{7.6}{GW190521A}{39.7}{GW190519A}{16.9}{GW190517A}{9.5}{GW190514A}{15.9}{GW190513A}{6.0}{GW190512A}{2.9}{GW190503A}{10.6}{GW190426A}{0.5}{GW190425A}{0.3}{GW190424A}{10.3}{GW190421A}{13.2}{GW190413B}{20.0}{GW190413A}{11.5}{GW190412A}{1.0}{GW190408A}{4.6}}}
\newcommand{\masstwodetmed}[1]{\IfEqCase{#1}{{GW190930A}{9.1}{GW190929A}{33.9}{GW190924A}{5.6}{GW190915A}{31.9}{GW190910A}{45.9}{GW190909A}{47.1}{GW190828B}{13.3}{GW190828A}{36.1}{GW190814A}{2.72}{GW190803A}{42.7}{GW190731A}{45.6}{GW190728A}{9.5}{GW190727A}{46.0}{GW190720A}{9.2}{GW190719A}{34.5}{GW190708A}{15.5}{GW190707A}{9.7}{GW190706A}{66.6}{GW190701A}{56.2}{GW190630A}{28.0}{GW190620A}{53.1}{GW190602A}{71.5}{GW190527A}{32.8}{GW190521B}{40.5}{GW190521A}{114.8}{GW190519A}{59.3}{GW190517A}{34.4}{GW190514A}{48.0}{GW190513A}{24.7}{GW190512A}{15.8}{GW190503A}{36.2}{GW190426A}{1.6}{GW190425A}{1.4}{GW190424A}{44.8}{GW190421A}{47.8}{GW190413B}{55.2}{GW190413A}{38.0}{GW190412A}{9.6}{GW190408A}{23.7}}}
\newcommand{\masstwodetplus}[1]{\IfEqCase{#1}{{GW190930A}{1.8}{GW190929A}{37.2}{GW190924A}{1.5}{GW190915A}{7.0}{GW190910A}{7.0}{GW190909A}{20.9}{GW190828B}{4.9}{GW190828A}{4.6}{GW190814A}{0.08}{GW190803A}{9.8}{GW190731A}{11.8}{GW190728A}{1.8}{GW190727A}{8.4}{GW190720A}{2.4}{GW190719A}{13.3}{GW190708A}{2.0}{GW190707A}{1.4}{GW190706A}{23.9}{GW190701A}{11.7}{GW190630A}{5.5}{GW190620A}{17.1}{GW190602A}{18.6}{GW190527A}{19.4}{GW190521B}{5.7}{GW190521A}{26.0}{GW190519A}{16.1}{GW190517A}{8.2}{GW190514A}{11.1}{GW190513A}{10.6}{GW190512A}{4.7}{GW190503A}{10.4}{GW190426A}{0.9}{GW190425A}{0.3}{GW190424A}{8.3}{GW190421A}{9.4}{GW190413B}{14.9}{GW190413A}{10.7}{GW190412A}{1.7}{GW190408A}{3.6}}}
\newcommand{\masstwosourceminus}[1]{\IfEqCase{#1}{{GW190930A}{3.3}{GW190929A}{10.6}{GW190924A}{1.9}{GW190915A}{6.1}{GW190910A}{7.2}{GW190909A}{12.7}{GW190828B}{2.1}{GW190828A}{4.8}{GW190814A}{0.09}{GW190803A}{8.2}{GW190731A}{9.5}{GW190728A}{2.6}{GW190727A}{8.4}{GW190720A}{2.2}{GW190719A}{7.2}{GW190708A}{2.7}{GW190707A}{1.7}{GW190706A}{13.3}{GW190701A}{12.0}{GW190630A}{5.1}{GW190620A}{12.3}{GW190602A}{17.4}{GW190527A}{8.1}{GW190521B}{6.4}{GW190521A}{23.1}{GW190519A}{11.1}{GW190517A}{7.3}{GW190514A}{8.8}{GW190513A}{4.1}{GW190512A}{2.5}{GW190503A}{8.0}{GW190426A}{0.5}{GW190425A}{0.3}{GW190424A}{7.7}{GW190421A}{8.8}{GW190413B}{10.8}{GW190413A}{6.7}{GW190412A}{0.9}{GW190408A}{3.6}}}
\newcommand{\masstwosourcemed}[1]{\IfEqCase{#1}{{GW190930A}{7.8}{GW190929A}{24.1}{GW190924A}{5.0}{GW190915A}{24.4}{GW190910A}{35.6}{GW190909A}{28.3}{GW190828B}{10.2}{GW190828A}{26.2}{GW190814A}{2.59}{GW190803A}{27.3}{GW190731A}{28.8}{GW190728A}{8.1}{GW190727A}{29.4}{GW190720A}{7.8}{GW190719A}{20.8}{GW190708A}{13.2}{GW190707A}{8.4}{GW190706A}{38.2}{GW190701A}{40.8}{GW190630A}{23.7}{GW190620A}{35.5}{GW190602A}{47.8}{GW190527A}{22.6}{GW190521B}{32.8}{GW190521A}{69.0}{GW190519A}{40.5}{GW190517A}{25.3}{GW190514A}{28.4}{GW190513A}{18.0}{GW190512A}{12.6}{GW190503A}{28.4}{GW190426A}{1.5}{GW190425A}{1.4}{GW190424A}{31.8}{GW190421A}{31.9}{GW190413B}{31.8}{GW190413A}{23.7}{GW190412A}{8.3}{GW190408A}{18.4}}}
\newcommand{\masstwosourceplus}[1]{\IfEqCase{#1}{{GW190930A}{1.7}{GW190929A}{19.3}{GW190924A}{1.4}{GW190915A}{5.6}{GW190910A}{6.3}{GW190909A}{13.4}{GW190828B}{3.6}{GW190828A}{4.6}{GW190814A}{0.08}{GW190803A}{7.8}{GW190731A}{9.7}{GW190728A}{1.7}{GW190727A}{7.1}{GW190720A}{2.3}{GW190719A}{9.0}{GW190708A}{2.0}{GW190707A}{1.4}{GW190706A}{14.6}{GW190701A}{8.7}{GW190630A}{5.2}{GW190620A}{12.2}{GW190602A}{14.3}{GW190527A}{10.5}{GW190521B}{5.4}{GW190521A}{22.7}{GW190519A}{11.0}{GW190517A}{7.0}{GW190514A}{9.3}{GW190513A}{7.7}{GW190512A}{3.6}{GW190503A}{7.7}{GW190426A}{0.8}{GW190425A}{0.3}{GW190424A}{7.6}{GW190421A}{8.0}{GW190413B}{11.7}{GW190413A}{7.3}{GW190412A}{1.6}{GW190408A}{3.3}}}
\newcommand{\decminus}[1]{\IfEqCase{#1}{{GW190930A}{0.66804}{GW190929A}{1.09453}{GW190924A}{0.31095}{GW190915A}{0.43700}{GW190910A}{0.78759}{GW190909A}{1.37456}{GW190828B}{0.42413}{GW190828A}{0.45822}{GW190814A}{0.12765}{GW190803A}{0.76320}{GW190731A}{0.54299}{GW190728A}{1.45903}{GW190727A}{0.53050}{GW190720A}{1.79814}{GW190719A}{1.52271}{GW190708A}{1.12280}{GW190707A}{0.66130}{GW190706A}{1.13851}{GW190701A}{0.08561}{GW190630A}{0.88066}{GW190620A}{1.15741}{GW190602A}{0.22445}{GW190527A}{0.63740}{GW190521B}{0.61963}{GW190521A}{0.40205}{GW190519A}{1.22807}{GW190517A}{0.23138}{GW190514A}{1.30610}{GW190513A}{1.20492}{GW190512A}{0.07267}{GW190503A}{0.08741}{GW190426A}{1.53579}{GW190425A}{0.89977}{GW190424A}{1.08941}{GW190421A}{0.52647}{GW190413B}{0.10088}{GW190413A}{1.21171}{GW190412A}{0.03938}{GW190408A}{0.33259}}}
\newcommand{\decmed}[1]{\IfEqCase{#1}{{GW190930A}{0.64489}{GW190929A}{0.12643}{GW190924A}{0.16308}{GW190915A}{0.64820}{GW190910A}{-0.19299}{GW190909A}{0.45798}{GW190828B}{-0.70225}{GW190828A}{-0.38234}{GW190814A}{-0.43746}{GW190803A}{0.56728}{GW190731A}{-0.84452}{GW190728A}{0.14876}{GW190727A}{-0.69575}{GW190720A}{0.61861}{GW190719A}{0.60387}{GW190708A}{0.31278}{GW190707A}{-0.26771}{GW190706A}{0.49342}{GW190701A}{-0.11450}{GW190630A}{-0.17794}{GW190620A}{0.40157}{GW190602A}{-0.60585}{GW190527A}{-0.67031}{GW190521B}{0.31697}{GW190521A}{-0.79351}{GW190519A}{0.62138}{GW190517A}{-0.77834}{GW190514A}{0.75342}{GW190513A}{0.66794}{GW190512A}{-0.46498}{GW190503A}{-0.88291}{GW190426A}{0.90282}{GW190425A}{-0.13006}{GW190424A}{-0.00051}{GW190421A}{-0.82328}{GW190413B}{-0.53598}{GW190413A}{0.45042}{GW190412A}{0.63309}{GW190408A}{0.92018}}}
\newcommand{\decplus}[1]{\IfEqCase{#1}{{GW190930A}{0.44937}{GW190929A}{0.92641}{GW190924A}{0.26435}{GW190915A}{0.49993}{GW190910A}{0.99066}{GW190909A}{0.80671}{GW190828B}{1.28711}{GW190828A}{1.25309}{GW190814A}{0.03175}{GW190803A}{0.63137}{GW190731A}{1.11036}{GW190728A}{0.43956}{GW190727A}{1.58271}{GW190720A}{0.05969}{GW190719A}{0.55220}{GW190708A}{0.86640}{GW190707A}{1.42731}{GW190706A}{0.43610}{GW190701A}{0.08803}{GW190630A}{0.78734}{GW190620A}{0.77559}{GW190602A}{0.57550}{GW190527A}{0.97030}{GW190521B}{0.25837}{GW190521A}{1.54032}{GW190519A}{0.40124}{GW190517A}{0.96906}{GW190514A}{0.60827}{GW190513A}{0.37574}{GW190512A}{0.32502}{GW190503A}{0.10637}{GW190426A}{0.61722}{GW190425A}{0.96811}{GW190424A}{1.08246}{GW190421A}{0.53382}{GW190413B}{1.13504}{GW190413A}{0.90354}{GW190412A}{0.02643}{GW190408A}{0.08290}}}
\newcommand{\psiminus}[1]{\IfEqCase{#1}{{GW190930A}{1.82}{GW190929A}{1.43}{GW190924A}{1.79}{GW190915A}{1.85}{GW190910A}{2.23}{GW190909A}{1.42}{GW190828B}{1.30}{GW190828A}{2.05}{GW190814A}{0.32}{GW190803A}{2.82}{GW190731A}{2.78}{GW190728A}{1.92}{GW190727A}{2.89}{GW190720A}{1.90}{GW190719A}{1.84}{GW190708A}{2.83}{GW190707A}{1.83}{GW190706A}{2.03}{GW190701A}{1.87}{GW190630A}{1.25}{GW190620A}{1.54}{GW190602A}{2.89}{GW190527A}{2.77}{GW190521B}{1.24}{GW190521A}{1.42}{GW190519A}{2.65}{GW190517A}{2.20}{GW190514A}{2.82}{GW190513A}{2.03}{GW190512A}{2.81}{GW190503A}{2.51}{GW190426A}{1.40}{GW190425A}{1.46}{GW190424A}{2.78}{GW190421A}{2.77}{GW190413B}{2.84}{GW190413A}{2.79}{GW190412A}{2.36}{GW190408A}{2.73}}}
\newcommand{\psimed}[1]{\IfEqCase{#1}{{GW190930A}{2.02}{GW190929A}{1.62}{GW190924A}{2.00}{GW190915A}{2.06}{GW190910A}{3.19}{GW190909A}{1.60}{GW190828B}{1.45}{GW190828A}{2.21}{GW190814A}{0.39}{GW190803A}{3.18}{GW190731A}{3.13}{GW190728A}{2.16}{GW190727A}{3.16}{GW190720A}{2.09}{GW190719A}{2.05}{GW190708A}{3.14}{GW190707A}{2.05}{GW190706A}{2.19}{GW190701A}{2.03}{GW190630A}{1.81}{GW190620A}{1.82}{GW190602A}{3.13}{GW190527A}{3.10}{GW190521B}{1.73}{GW190521A}{1.59}{GW190519A}{3.30}{GW190517A}{2.37}{GW190514A}{3.18}{GW190513A}{2.26}{GW190512A}{3.14}{GW190503A}{3.12}{GW190426A}{1.58}{GW190425A}{1.62}{GW190424A}{3.15}{GW190421A}{3.11}{GW190413B}{3.10}{GW190413A}{3.15}{GW190412A}{2.56}{GW190408A}{3.14}}}
\newcommand{\psiplus}[1]{\IfEqCase{#1}{{GW190930A}{3.53}{GW190929A}{1.36}{GW190924A}{3.51}{GW190915A}{3.52}{GW190910A}{2.39}{GW190909A}{1.40}{GW190828B}{1.52}{GW190828A}{3.56}{GW190814A}{2.62}{GW190803A}{2.75}{GW190731A}{2.84}{GW190728A}{3.57}{GW190727A}{2.83}{GW190720A}{3.47}{GW190719A}{3.54}{GW190708A}{2.87}{GW190707A}{3.56}{GW190706A}{3.34}{GW190701A}{3.64}{GW190630A}{3.38}{GW190620A}{3.56}{GW190602A}{2.94}{GW190527A}{2.81}{GW190521B}{3.41}{GW190521A}{1.38}{GW190519A}{2.15}{GW190517A}{3.46}{GW190514A}{2.79}{GW190513A}{3.41}{GW190512A}{2.71}{GW190503A}{2.59}{GW190426A}{1.36}{GW190425A}{1.38}{GW190424A}{2.77}{GW190421A}{2.85}{GW190413B}{2.87}{GW190413A}{2.74}{GW190412A}{0.44}{GW190408A}{2.70}}}
\newcommand{\totalmassdetminus}[1]{\IfEqCase{#1}{{GW190930A}{1.0}{GW190929A}{26.3}{GW190924A}{0.7}{GW190915A}{8.1}{GW190910A}{7.8}{GW190909A}{22.9}{GW190828B}{4.0}{GW190828A}{5.9}{GW190814A}{1.0}{GW190803A}{11.9}{GW190731A}{14.3}{GW190728A}{0.7}{GW190727A}{10.9}{GW190720A}{1.2}{GW190719A}{15.5}{GW190708A}{0.8}{GW190707A}{0.5}{GW190706A}{27.7}{GW190701A}{14.8}{GW190630A}{3.5}{GW190620A}{18.4}{GW190602A}{20.6}{GW190527A}{10.3}{GW190521B}{5.4}{GW190521A}{34.6}{GW190519A}{17.9}{GW190517A}{7.3}{GW190514A}{15.1}{GW190513A}{6.7}{GW190512A}{2.8}{GW190503A}{11.8}{GW190426A}{1.6}{GW190425A}{0.08}{GW190424A}{10.9}{GW190421A}{12.4}{GW190413B}{18.0}{GW190413A}{15.3}{GW190412A}{4.6}{GW190408A}{3.8}}}
\newcommand{\totalmassdetmed}[1]{\IfEqCase{#1}{{GW190930A}{23.2}{GW190929A}{148.8}{GW190924A}{15.5}{GW190915A}{78.3}{GW190910A}{101.9}{GW190909A}{119.7}{GW190828B}{44.4}{GW190828A}{79.9}{GW190814A}{27.1}{GW190803A}{100.3}{GW190731A}{109.7}{GW190728A}{23.9}{GW190727A}{104.4}{GW190720A}{24.9}{GW190719A}{94.9}{GW190708A}{36.1}{GW190707A}{23.1}{GW190706A}{180.3}{GW190701A}{129.7}{GW190630A}{69.6}{GW190620A}{137.6}{GW190602A}{171.8}{GW190527A}{84.1}{GW190521B}{92.6}{GW190521A}{269.4}{GW190519A}{155.1}{GW190517A}{85.4}{GW190514A}{112.9}{GW190513A}{73.6}{GW190512A}{45.3}{GW190503A}{91.6}{GW190426A}{7.8}{GW190425A}{3.50}{GW190424A}{101.1}{GW190421A}{108.7}{GW190413B}{135.4}{GW190413A}{93.7}{GW190412A}{44.2}{GW190408A}{55.5}}}
\newcommand{\totalmassdetplus}[1]{\IfEqCase{#1}{{GW190930A}{10.5}{GW190929A}{38.6}{GW190924A}{5.7}{GW190915A}{8.4}{GW190910A}{10.4}{GW190909A}{95.3}{GW190828B}{6.4}{GW190828A}{6.9}{GW190814A}{1.1}{GW190803A}{14.1}{GW190731A}{14.3}{GW190728A}{5.3}{GW190727A}{11.9}{GW190720A}{5.0}{GW190719A}{24.4}{GW190708A}{2.5}{GW190707A}{1.8}{GW190706A}{23.3}{GW190701A}{16.4}{GW190630A}{4.2}{GW190620A}{20.1}{GW190602A}{23.2}{GW190527A}{53.7}{GW190521B}{4.8}{GW190521A}{39.8}{GW190519A}{16.7}{GW190517A}{9.6}{GW190514A}{17.8}{GW190513A}{12.7}{GW190512A}{3.9}{GW190503A}{11.2}{GW190426A}{3.7}{GW190425A}{0.3}{GW190424A}{14.4}{GW190421A}{15.3}{GW190413B}{17.9}{GW190413A}{17.8}{GW190412A}{4.5}{GW190408A}{3.5}}}
\newcommand{\thetajnminus}[1]{\IfEqCase{#1}{{GW190930A}{0.74}{GW190929A}{1.20}{GW190924A}{0.57}{GW190915A}{1.52}{GW190910A}{1.19}{GW190909A}{1.12}{GW190828B}{1.51}{GW190828A}{1.92}{GW190814A}{0.24}{GW190803A}{0.87}{GW190731A}{0.93}{GW190728A}{1.02}{GW190727A}{1.28}{GW190720A}{2.01}{GW190719A}{1.34}{GW190708A}{1.15}{GW190707A}{1.87}{GW190706A}{1.06}{GW190701A}{0.42}{GW190630A}{0.97}{GW190620A}{1.52}{GW190602A}{1.43}{GW190527A}{0.99}{GW190521B}{1.20}{GW190521A}{0.87}{GW190519A}{0.94}{GW190517A}{1.52}{GW190514A}{1.23}{GW190513A}{0.58}{GW190512A}{1.30}{GW190503A}{0.57}{GW190426A}{1.43}{GW190425A}{0.85}{GW190424A}{1.26}{GW190421A}{1.44}{GW190413B}{1.55}{GW190413A}{0.87}{GW190412A}{0.25}{GW190408A}{0.59}}}
\newcommand{\thetajnmed}[1]{\IfEqCase{#1}{{GW190930A}{0.94}{GW190929A}{1.70}{GW190924A}{0.74}{GW190915A}{2.03}{GW190910A}{1.62}{GW190909A}{1.45}{GW190828B}{1.83}{GW190828A}{2.24}{GW190814A}{0.86}{GW190803A}{1.12}{GW190731A}{1.19}{GW190728A}{1.23}{GW190727A}{1.55}{GW190720A}{2.48}{GW190719A}{1.61}{GW190708A}{1.37}{GW190707A}{2.15}{GW190706A}{1.37}{GW190701A}{0.58}{GW190630A}{1.22}{GW190620A}{1.94}{GW190602A}{1.71}{GW190527A}{1.26}{GW190521B}{1.48}{GW190521A}{1.15}{GW190519A}{1.58}{GW190517A}{2.26}{GW190514A}{1.50}{GW190513A}{0.79}{GW190512A}{1.61}{GW190503A}{2.41}{GW190426A}{1.70}{GW190425A}{1.08}{GW190424A}{1.52}{GW190421A}{1.74}{GW190413B}{1.90}{GW190413A}{1.15}{GW190412A}{0.72}{GW190408A}{0.79}}}
\newcommand{\thetajnplus}[1]{\IfEqCase{#1}{{GW190930A}{1.90}{GW190929A}{0.97}{GW190924A}{2.04}{GW190915A}{0.78}{GW190910A}{1.13}{GW190909A}{1.34}{GW190828B}{1.02}{GW190828A}{0.70}{GW190814A}{1.48}{GW190803A}{1.73}{GW190731A}{1.66}{GW190728A}{1.66}{GW190727A}{1.31}{GW190720A}{0.50}{GW190719A}{1.25}{GW190708A}{1.54}{GW190707A}{0.77}{GW190706A}{1.43}{GW190701A}{0.55}{GW190630A}{1.60}{GW190620A}{0.90}{GW190602A}{1.17}{GW190527A}{1.55}{GW190521B}{1.37}{GW190521A}{1.65}{GW190519A}{0.95}{GW190517A}{0.64}{GW190514A}{1.33}{GW190513A}{2.02}{GW190512A}{1.22}{GW190503A}{0.52}{GW190426A}{1.19}{GW190425A}{1.77}{GW190424A}{1.36}{GW190421A}{1.13}{GW190413B}{0.95}{GW190413A}{1.64}{GW190412A}{0.44}{GW190408A}{2.03}}}
\newcommand{\redshiftminus}[1]{\IfEqCase{#1}{{GW190930A}{0.06}{GW190929A}{0.17}{GW190924A}{0.04}{GW190915A}{0.10}{GW190910A}{0.10}{GW190909A}{0.33}{GW190828B}{0.10}{GW190828A}{0.15}{GW190814A}{0.010}{GW190803A}{0.24}{GW190731A}{0.26}{GW190728A}{0.07}{GW190727A}{0.22}{GW190720A}{0.06}{GW190719A}{0.29}{GW190708A}{0.07}{GW190707A}{0.07}{GW190706A}{0.27}{GW190701A}{0.12}{GW190630A}{0.07}{GW190620A}{0.20}{GW190602A}{0.17}{GW190527A}{0.20}{GW190521B}{0.10}{GW190521A}{0.28}{GW190519A}{0.14}{GW190517A}{0.14}{GW190514A}{0.31}{GW190513A}{0.13}{GW190512A}{0.10}{GW190503A}{0.11}{GW190426A}{0.03}{GW190425A}{0.02}{GW190424A}{0.19}{GW190421A}{0.21}{GW190413B}{0.30}{GW190413A}{0.24}{GW190412A}{0.03}{GW190408A}{0.10}}}
\newcommand{\redshiftmed}[1]{\IfEqCase{#1}{{GW190930A}{0.15}{GW190929A}{0.38}{GW190924A}{0.12}{GW190915A}{0.30}{GW190910A}{0.28}{GW190909A}{0.62}{GW190828B}{0.30}{GW190828A}{0.38}{GW190814A}{0.05}{GW190803A}{0.55}{GW190731A}{0.55}{GW190728A}{0.18}{GW190727A}{0.55}{GW190720A}{0.16}{GW190719A}{0.64}{GW190708A}{0.18}{GW190707A}{0.16}{GW190706A}{0.71}{GW190701A}{0.37}{GW190630A}{0.18}{GW190620A}{0.49}{GW190602A}{0.47}{GW190527A}{0.44}{GW190521B}{0.24}{GW190521A}{0.64}{GW190519A}{0.44}{GW190517A}{0.34}{GW190514A}{0.67}{GW190513A}{0.37}{GW190512A}{0.27}{GW190503A}{0.27}{GW190426A}{0.08}{GW190425A}{0.03}{GW190424A}{0.39}{GW190421A}{0.49}{GW190413B}{0.71}{GW190413A}{0.59}{GW190412A}{0.15}{GW190408A}{0.29}}}
\newcommand{\redshiftplus}[1]{\IfEqCase{#1}{{GW190930A}{0.06}{GW190929A}{0.49}{GW190924A}{0.04}{GW190915A}{0.11}{GW190910A}{0.16}{GW190909A}{0.41}{GW190828B}{0.10}{GW190828A}{0.10}{GW190814A}{0.009}{GW190803A}{0.26}{GW190731A}{0.31}{GW190728A}{0.05}{GW190727A}{0.21}{GW190720A}{0.12}{GW190719A}{0.33}{GW190708A}{0.06}{GW190707A}{0.07}{GW190706A}{0.32}{GW190701A}{0.11}{GW190630A}{0.10}{GW190620A}{0.23}{GW190602A}{0.25}{GW190527A}{0.34}{GW190521B}{0.07}{GW190521A}{0.28}{GW190519A}{0.25}{GW190517A}{0.24}{GW190514A}{0.33}{GW190513A}{0.13}{GW190512A}{0.09}{GW190503A}{0.11}{GW190426A}{0.04}{GW190425A}{0.01}{GW190424A}{0.23}{GW190421A}{0.19}{GW190413B}{0.31}{GW190413A}{0.29}{GW190412A}{0.03}{GW190408A}{0.06}}}
\newcommand{\iotaminus}[1]{\IfEqCase{#1}{{GW190930A}{0.73}{GW190929A}{1.24}{GW190924A}{0.56}{GW190915A}{1.73}{GW190910A}{1.19}{GW190909A}{1.11}{GW190828B}{1.55}{GW190828A}{1.97}{GW190814A}{0.27}{GW190803A}{0.85}{GW190731A}{0.90}{GW190728A}{1.01}{GW190727A}{1.26}{GW190720A}{2.01}{GW190719A}{1.34}{GW190708A}{1.14}{GW190707A}{1.88}{GW190706A}{1.04}{GW190701A}{0.43}{GW190630A}{0.98}{GW190620A}{1.69}{GW190602A}{1.53}{GW190527A}{0.97}{GW190521B}{1.19}{GW190521A}{0.87}{GW190519A}{0.93}{GW190517A}{1.40}{GW190514A}{1.22}{GW190513A}{0.58}{GW190512A}{1.29}{GW190503A}{0.57}{GW190426A}{1.43}{GW190425A}{0.85}{GW190424A}{1.26}{GW190421A}{1.48}{GW190413B}{1.61}{GW190413A}{0.86}{GW190412A}{0.35}{GW190408A}{0.59}}}
\newcommand{\iotamed}[1]{\IfEqCase{#1}{{GW190930A}{0.94}{GW190929A}{1.70}{GW190924A}{0.74}{GW190915A}{2.13}{GW190910A}{1.62}{GW190909A}{1.47}{GW190828B}{1.84}{GW190828A}{2.27}{GW190814A}{0.85}{GW190803A}{1.09}{GW190731A}{1.15}{GW190728A}{1.23}{GW190727A}{1.53}{GW190720A}{2.47}{GW190719A}{1.61}{GW190708A}{1.37}{GW190707A}{2.15}{GW190706A}{1.33}{GW190701A}{0.59}{GW190630A}{1.24}{GW190620A}{2.03}{GW190602A}{1.79}{GW190527A}{1.24}{GW190521B}{1.46}{GW190521A}{1.15}{GW190519A}{1.59}{GW190517A}{2.21}{GW190514A}{1.48}{GW190513A}{0.79}{GW190512A}{1.61}{GW190503A}{2.43}{GW190426A}{1.70}{GW190425A}{1.09}{GW190424A}{1.52}{GW190421A}{1.76}{GW190413B}{1.97}{GW190413A}{1.13}{GW190412A}{0.84}{GW190408A}{0.78}}}
\newcommand{\iotaplus}[1]{\IfEqCase{#1}{{GW190930A}{1.89}{GW190929A}{1.03}{GW190924A}{2.04}{GW190915A}{0.74}{GW190910A}{1.14}{GW190909A}{1.30}{GW190828B}{1.01}{GW190828A}{0.67}{GW190814A}{1.51}{GW190803A}{1.76}{GW190731A}{1.71}{GW190728A}{1.66}{GW190727A}{1.34}{GW190720A}{0.49}{GW190719A}{1.27}{GW190708A}{1.55}{GW190707A}{0.78}{GW190706A}{1.49}{GW190701A}{0.57}{GW190630A}{1.55}{GW190620A}{0.85}{GW190602A}{1.11}{GW190527A}{1.57}{GW190521B}{1.40}{GW190521A}{1.65}{GW190519A}{0.92}{GW190517A}{0.66}{GW190514A}{1.37}{GW190513A}{2.03}{GW190512A}{1.22}{GW190503A}{0.51}{GW190426A}{1.19}{GW190425A}{1.77}{GW190424A}{1.38}{GW190421A}{1.11}{GW190413B}{0.88}{GW190413A}{1.66}{GW190412A}{0.38}{GW190408A}{2.06}}}
\newcommand{\spinonexminus}[1]{\IfEqCase{#1}{{GW190930A}{0.44}{GW190929A}{0.71}{GW190924A}{0.35}{GW190915A}{0.67}{GW190910A}{0.51}{GW190909A}{0.63}{GW190828B}{0.42}{GW190828A}{0.52}{GW190814A}{0.04}{GW190803A}{0.57}{GW190731A}{0.56}{GW190728A}{0.37}{GW190727A}{0.58}{GW190720A}{0.43}{GW190719A}{0.55}{GW190708A}{0.42}{GW190707A}{0.39}{GW190706A}{0.53}{GW190701A}{0.52}{GW190630A}{0.36}{GW190620A}{0.55}{GW190602A}{0.52}{GW190527A}{0.59}{GW190521B}{0.44}{GW190521A}{0.74}{GW190519A}{0.54}{GW190517A}{0.57}{GW190514A}{0.57}{GW190513A}{0.42}{GW190512A}{0.30}{GW190503A}{0.49}{GW190426A}{0.00}{GW190425A}{0.50}{GW190424A}{0.64}{GW190421A}{0.59}{GW190413B}{0.68}{GW190413A}{0.53}{GW190412A}{0.33}{GW190408A}{0.47}}}
\newcommand{\spinonexmed}[1]{\IfEqCase{#1}{{GW190930A}{0.002}{GW190929A}{0.007}{GW190924A}{0.0001}{GW190915A}{0.00}{GW190910A}{0.00}{GW190909A}{0.002}{GW190828B}{0.00}{GW190828A}{0.00}{GW190814A}{0.00}{GW190803A}{0.00}{GW190731A}{0.0007}{GW190728A}{0.0008}{GW190727A}{0.002}{GW190720A}{0.003}{GW190719A}{0.004}{GW190708A}{0.004}{GW190707A}{0.003}{GW190706A}{0.00}{GW190701A}{0.00}{GW190630A}{0.00}{GW190620A}{0.00}{GW190602A}{0.00}{GW190527A}{0.002}{GW190521B}{0.001}{GW190521A}{-0.02}{GW190519A}{0.004}{GW190517A}{0.0009}{GW190514A}{0.00007}{GW190513A}{0.0006}{GW190512A}{0.0010}{GW190503A}{0.00}{GW190426A}{0.00}{GW190425A}{0.00}{GW190424A}{0.00}{GW190421A}{0.00005}{GW190413B}{0.00}{GW190413A}{0.0005}{GW190412A}{-0.02}{GW190408A}{0.003}}}
\newcommand{\spinonexplus}[1]{\IfEqCase{#1}{{GW190930A}{0.47}{GW190929A}{0.69}{GW190924A}{0.36}{GW190915A}{0.66}{GW190910A}{0.51}{GW190909A}{0.68}{GW190828B}{0.43}{GW190828A}{0.51}{GW190814A}{0.04}{GW190803A}{0.55}{GW190731A}{0.51}{GW190728A}{0.40}{GW190727A}{0.58}{GW190720A}{0.45}{GW190719A}{0.57}{GW190708A}{0.47}{GW190707A}{0.42}{GW190706A}{0.53}{GW190701A}{0.52}{GW190630A}{0.36}{GW190620A}{0.53}{GW190602A}{0.53}{GW190527A}{0.59}{GW190521B}{0.45}{GW190521A}{0.75}{GW190519A}{0.55}{GW190517A}{0.57}{GW190514A}{0.60}{GW190513A}{0.43}{GW190512A}{0.28}{GW190503A}{0.50}{GW190426A}{0.00}{GW190425A}{0.47}{GW190424A}{0.63}{GW190421A}{0.60}{GW190413B}{0.71}{GW190413A}{0.53}{GW190412A}{0.39}{GW190408A}{0.49}}}
\newcommand{\chirpmassdetminus}[1]{\IfEqCase{#1}{{GW190930A}{0.2}{GW190929A}{15.4}{GW190924A}{0.03}{GW190915A}{3.9}{GW190910A}{3.6}{GW190909A}{12.4}{GW190828B}{0.7}{GW190828A}{2.8}{GW190814A}{0.02}{GW190803A}{6.1}{GW190731A}{8.2}{GW190728A}{0.08}{GW190727A}{5.7}{GW190720A}{0.1}{GW190719A}{6.6}{GW190708A}{0.2}{GW190707A}{0.09}{GW190706A}{17.5}{GW190701A}{8.1}{GW190630A}{1.5}{GW190620A}{11.2}{GW190602A}{13.7}{GW190527A}{5.5}{GW190521B}{3.0}{GW190521A}{17.6}{GW190519A}{10.3}{GW190517A}{3.4}{GW190514A}{7.7}{GW190513A}{2.5}{GW190512A}{0.8}{GW190503A}{6.0}{GW190426A}{0.01}{GW190425A}{0.0006}{GW190424A}{4.8}{GW190421A}{6.0}{GW190413B}{9.8}{GW190413A}{6.6}{GW190412A}{0.2}{GW190408A}{1.7}}}
\newcommand{\chirpmassdetmed}[1]{\IfEqCase{#1}{{GW190930A}{9.8}{GW190929A}{52.2}{GW190924A}{6.44}{GW190915A}{33.1}{GW190910A}{43.9}{GW190909A}{49.8}{GW190828B}{17.4}{GW190828A}{34.5}{GW190814A}{6.41}{GW190803A}{42.7}{GW190731A}{46.6}{GW190728A}{10.1}{GW190727A}{44.7}{GW190720A}{10.4}{GW190719A}{38.7}{GW190708A}{15.5}{GW190707A}{9.89}{GW190706A}{75.1}{GW190701A}{55.5}{GW190630A}{29.4}{GW190620A}{57.5}{GW190602A}{72.9}{GW190527A}{34.9}{GW190521B}{39.8}{GW190521A}{114.8}{GW190519A}{65.1}{GW190517A}{35.9}{GW190514A}{48.1}{GW190513A}{29.5}{GW190512A}{18.6}{GW190503A}{38.6}{GW190426A}{2.60}{GW190425A}{1.49}{GW190424A}{43.4}{GW190421A}{46.6}{GW190413B}{57.0}{GW190413A}{39.4}{GW190412A}{15.2}{GW190408A}{23.7}}}
\newcommand{\chirpmassdetplus}[1]{\IfEqCase{#1}{{GW190930A}{0.2}{GW190929A}{19.9}{GW190924A}{0.04}{GW190915A}{3.3}{GW190910A}{4.6}{GW190909A}{32.2}{GW190828B}{0.6}{GW190828A}{2.9}{GW190814A}{0.02}{GW190803A}{6.3}{GW190731A}{6.8}{GW190728A}{0.09}{GW190727A}{5.3}{GW190720A}{0.2}{GW190719A}{9.2}{GW190708A}{0.3}{GW190707A}{0.1}{GW190706A}{11.0}{GW190701A}{7.3}{GW190630A}{1.6}{GW190620A}{9.0}{GW190602A}{10.8}{GW190527A}{21.7}{GW190521B}{2.2}{GW190521A}{15.2}{GW190519A}{7.7}{GW190517A}{4.0}{GW190514A}{7.5}{GW190513A}{5.6}{GW190512A}{0.9}{GW190503A}{5.3}{GW190426A}{0.01}{GW190425A}{0.0008}{GW190424A}{6.0}{GW190421A}{6.6}{GW190413B}{8.6}{GW190413A}{7.7}{GW190412A}{0.2}{GW190408A}{1.4}}}
\newcommand{\cosiotaminus}[1]{\IfEqCase{#1}{{GW190930A}{1.54}{GW190929A}{0.79}{GW190924A}{1.67}{GW190915A}{0.43}{GW190910A}{0.88}{GW190909A}{1.03}{GW190828B}{0.69}{GW190828A}{0.33}{GW190814A}{1.37}{GW190803A}{1.42}{GW190731A}{1.37}{GW190728A}{1.30}{GW190727A}{1.00}{GW190720A}{0.20}{GW190719A}{0.93}{GW190708A}{1.18}{GW190707A}{0.43}{GW190706A}{1.19}{GW190701A}{0.44}{GW190630A}{1.26}{GW190620A}{0.52}{GW190602A}{0.75}{GW190527A}{1.27}{GW190521B}{1.07}{GW190521A}{1.35}{GW190519A}{0.79}{GW190517A}{0.37}{GW190514A}{1.05}{GW190513A}{1.65}{GW190512A}{0.91}{GW190503A}{0.22}{GW190426A}{0.84}{GW190425A}{1.42}{GW190424A}{1.02}{GW190421A}{0.78}{GW190413B}{0.57}{GW190413A}{1.37}{GW190412A}{0.32}{GW190408A}{1.67}}}
\newcommand{\cosiotamed}[1]{\IfEqCase{#1}{{GW190930A}{0.59}{GW190929A}{-0.13}{GW190924A}{0.74}{GW190915A}{-0.53}{GW190910A}{-0.05}{GW190909A}{0.10}{GW190828B}{-0.27}{GW190828A}{-0.65}{GW190814A}{0.66}{GW190803A}{0.47}{GW190731A}{0.41}{GW190728A}{0.34}{GW190727A}{0.04}{GW190720A}{-0.78}{GW190719A}{-0.04}{GW190708A}{0.20}{GW190707A}{-0.55}{GW190706A}{0.24}{GW190701A}{0.83}{GW190630A}{0.32}{GW190620A}{-0.45}{GW190602A}{-0.22}{GW190527A}{0.32}{GW190521B}{0.11}{GW190521A}{0.41}{GW190519A}{-0.02}{GW190517A}{-0.59}{GW190514A}{0.09}{GW190513A}{0.71}{GW190512A}{-0.04}{GW190503A}{-0.76}{GW190426A}{-0.13}{GW190425A}{0.46}{GW190424A}{0.05}{GW190421A}{-0.19}{GW190413B}{-0.39}{GW190413A}{0.42}{GW190412A}{0.67}{GW190408A}{0.71}}}
\newcommand{\cosiotaplus}[1]{\IfEqCase{#1}{{GW190930A}{0.39}{GW190929A}{1.03}{GW190924A}{0.24}{GW190915A}{1.45}{GW190910A}{0.96}{GW190909A}{0.84}{GW190828B}{1.23}{GW190828A}{1.60}{GW190814A}{0.17}{GW190803A}{0.51}{GW190731A}{0.56}{GW190728A}{0.64}{GW190727A}{0.93}{GW190720A}{1.68}{GW190719A}{1.00}{GW190708A}{0.77}{GW190707A}{1.51}{GW190706A}{0.72}{GW190701A}{0.16}{GW190630A}{0.65}{GW190620A}{1.39}{GW190602A}{1.18}{GW190527A}{0.64}{GW190521B}{0.85}{GW190521A}{0.55}{GW190519A}{0.81}{GW190517A}{1.29}{GW190514A}{0.88}{GW190513A}{0.27}{GW190512A}{0.99}{GW190503A}{0.47}{GW190426A}{1.09}{GW190425A}{0.51}{GW190424A}{0.91}{GW190421A}{1.15}{GW190413B}{1.32}{GW190413A}{0.54}{GW190412A}{0.22}{GW190408A}{0.27}}}
\newcommand{\comovingdistminus}[1]{\IfEqCase{#1}{{GW190930A}{257}{GW190929A}{650}{GW190924A}{184}{GW190915A}{401}{GW190910A}{396}{GW190909A}{1128}{GW190828B}{395}{GW190828A}{567}{GW190814A}{41}{GW190803A}{824}{GW190731A}{906}{GW190728A}{285}{GW190727A}{774}{GW190720A}{254}{GW190719A}{965}{GW190708A}{307}{GW190707A}{295}{GW190706A}{859}{GW190701A}{438}{GW190630A}{289}{GW190620A}{723}{GW190602A}{621}{GW190527A}{725}{GW190521B}{409}{GW190521A}{943}{GW190519A}{518}{GW190517A}{538}{GW190514A}{1034}{GW190513A}{484}{GW190512A}{378}{GW190503A}{431}{GW190426A}{143}{GW190425A}{67}{GW190424A}{717}{GW190421A}{759}{GW190413B}{957}{GW190413A}{829}{GW190412A}{134}{GW190408A}{399}}}
\newcommand{\comovingdistmed}[1]{\IfEqCase{#1}{{GW190930A}{658}{GW190929A}{1541}{GW190924A}{507}{GW190915A}{1244}{GW190910A}{1139}{GW190909A}{2329}{GW190828B}{1228}{GW190828A}{1539}{GW190814A}{229}{GW190803A}{2108}{GW190731A}{2120}{GW190728A}{742}{GW190727A}{2119}{GW190720A}{678}{GW190719A}{2399}{GW190708A}{748}{GW190707A}{667}{GW190706A}{2594}{GW190701A}{1498}{GW190630A}{752}{GW190620A}{1893}{GW190602A}{1832}{GW190527A}{1729}{GW190521B}{1003}{GW190521A}{2390}{GW190519A}{1754}{GW190517A}{1389}{GW190514A}{2475}{GW190513A}{1501}{GW190512A}{1120}{GW190503A}{1133}{GW190426A}{344}{GW190425A}{151}{GW190424A}{1578}{GW190421A}{1923}{GW190413B}{2603}{GW190413A}{2232}{GW190412A}{640}{GW190408A}{1198}}}
\newcommand{\comovingdistplus}[1]{\IfEqCase{#1}{{GW190930A}{258}{GW190929A}{1537}{GW190924A}{174}{GW190915A}{403}{GW190910A}{588}{GW190909A}{1139}{GW190828B}{359}{GW190828A}{342}{GW190814A}{37}{GW190803A}{778}{GW190731A}{925}{GW190728A}{182}{GW190727A}{627}{GW190720A}{472}{GW190719A}{915}{GW190708A}{235}{GW190707A}{272}{GW190706A}{867}{GW190701A}{400}{GW190630A}{379}{GW190620A}{725}{GW190602A}{784}{GW190527A}{1068}{GW190521B}{253}{GW190521A}{795}{GW190519A}{816}{GW190517A}{816}{GW190514A}{915}{GW190513A}{453}{GW190512A}{330}{GW190503A}{406}{GW190426A}{154}{GW190425A}{64}{GW190424A}{755}{GW190421A}{598}{GW190413B}{831}{GW190413A}{856}{GW190412A}{105}{GW190408A}{240}}}
\newcommand{\spintwoyminus}[1]{\IfEqCase{#1}{{GW190930A}{0.54}{GW190929A}{0.57}{GW190924A}{0.48}{GW190915A}{0.61}{GW190910A}{0.52}{GW190909A}{0.61}{GW190828B}{0.54}{GW190828A}{0.50}{GW190814A}{0.61}{GW190803A}{0.57}{GW190731A}{0.58}{GW190728A}{0.51}{GW190727A}{0.59}{GW190720A}{0.55}{GW190719A}{0.55}{GW190708A}{0.44}{GW190707A}{0.47}{GW190706A}{0.51}{GW190701A}{0.58}{GW190630A}{0.48}{GW190620A}{0.56}{GW190602A}{0.60}{GW190527A}{0.61}{GW190521B}{0.53}{GW190521A}{0.68}{GW190519A}{0.55}{GW190517A}{0.54}{GW190514A}{0.60}{GW190513A}{0.54}{GW190512A}{0.51}{GW190503A}{0.57}{GW190426A}{0.00}{GW190425A}{0.48}{GW190424A}{0.60}{GW190421A}{0.59}{GW190413B}{0.59}{GW190413A}{0.57}{GW190412A}{0.57}{GW190408A}{0.52}}}
\newcommand{\spintwoymed}[1]{\IfEqCase{#1}{{GW190930A}{0.00}{GW190929A}{0.0008}{GW190924A}{0.00}{GW190915A}{0.00}{GW190910A}{0.00}{GW190909A}{0.003}{GW190828B}{0.00}{GW190828A}{0.0004}{GW190814A}{0.005}{GW190803A}{0.0002}{GW190731A}{0.00}{GW190728A}{0.00}{GW190727A}{0.0009}{GW190720A}{0.00}{GW190719A}{0.003}{GW190708A}{0.003}{GW190707A}{0.002}{GW190706A}{0.001}{GW190701A}{-0.01}{GW190630A}{0.0006}{GW190620A}{0.00010}{GW190602A}{0.00}{GW190527A}{0.004}{GW190521B}{0.00}{GW190521A}{0.00}{GW190519A}{0.001}{GW190517A}{0.00}{GW190514A}{0.00}{GW190513A}{0.00}{GW190512A}{0.002}{GW190503A}{0.00}{GW190426A}{0.00}{GW190425A}{0.00}{GW190424A}{0.00}{GW190421A}{0.00}{GW190413B}{-0.01}{GW190413A}{0.00}{GW190412A}{0.004}{GW190408A}{0.0008}}}
\newcommand{\spintwoyplus}[1]{\IfEqCase{#1}{{GW190930A}{0.51}{GW190929A}{0.59}{GW190924A}{0.49}{GW190915A}{0.60}{GW190910A}{0.52}{GW190909A}{0.57}{GW190828B}{0.54}{GW190828A}{0.50}{GW190814A}{0.61}{GW190803A}{0.58}{GW190731A}{0.55}{GW190728A}{0.50}{GW190727A}{0.56}{GW190720A}{0.56}{GW190719A}{0.55}{GW190708A}{0.46}{GW190707A}{0.48}{GW190706A}{0.53}{GW190701A}{0.55}{GW190630A}{0.49}{GW190620A}{0.54}{GW190602A}{0.58}{GW190527A}{0.59}{GW190521B}{0.52}{GW190521A}{0.69}{GW190519A}{0.55}{GW190517A}{0.55}{GW190514A}{0.60}{GW190513A}{0.55}{GW190512A}{0.55}{GW190503A}{0.58}{GW190426A}{0.00}{GW190425A}{0.48}{GW190424A}{0.60}{GW190421A}{0.59}{GW190413B}{0.62}{GW190413A}{0.56}{GW190412A}{0.58}{GW190408A}{0.53}}}
\newcommand{\tiltoneminus}[1]{\IfEqCase{#1}{{GW190930A}{0.79}{GW190929A}{0.72}{GW190924A}{1.05}{GW190915A}{0.85}{GW190910A}{0.97}{GW190909A}{1.14}{GW190828B}{0.83}{GW190828A}{0.72}{GW190814A}{1.08}{GW190803A}{1.06}{GW190731A}{0.96}{GW190728A}{0.77}{GW190727A}{0.85}{GW190720A}{0.72}{GW190719A}{0.59}{GW190708A}{0.98}{GW190707A}{1.08}{GW190706A}{0.61}{GW190701A}{1.12}{GW190630A}{0.87}{GW190620A}{0.59}{GW190602A}{0.94}{GW190527A}{0.89}{GW190521B}{0.90}{GW190521A}{0.93}{GW190519A}{0.60}{GW190517A}{0.42}{GW190514A}{1.15}{GW190513A}{0.81}{GW190512A}{1.03}{GW190503A}{1.10}{GW190426A}{0.00}{GW190425A}{0.80}{GW190424A}{0.80}{GW190421A}{1.06}{GW190413B}{0.93}{GW190413A}{1.06}{GW190412A}{0.35}{GW190408A}{1.06}}}
\newcommand{\tiltonemed}[1]{\IfEqCase{#1}{{GW190930A}{1.08}{GW190929A}{1.55}{GW190924A}{1.38}{GW190915A}{1.51}{GW190910A}{1.49}{GW190909A}{1.67}{GW190828B}{1.31}{GW190828A}{1.04}{GW190814A}{1.56}{GW190803A}{1.64}{GW190731A}{1.41}{GW190728A}{1.06}{GW190727A}{1.26}{GW190720A}{1.00}{GW190719A}{0.85}{GW190708A}{1.50}{GW190707A}{1.77}{GW190706A}{0.85}{GW190701A}{1.79}{GW190630A}{1.28}{GW190620A}{0.82}{GW190602A}{1.39}{GW190527A}{1.29}{GW190521B}{1.40}{GW190521A}{1.52}{GW190519A}{0.87}{GW190517A}{0.59}{GW190514A}{2.03}{GW190513A}{1.14}{GW190512A}{1.50}{GW190503A}{1.73}{GW190426A}{0.00}{GW190425A}{1.31}{GW190424A}{1.20}{GW190421A}{1.72}{GW190413B}{1.63}{GW190413A}{1.56}{GW190412A}{0.80}{GW190408A}{1.74}}}
\newcommand{\tiltoneplus}[1]{\IfEqCase{#1}{{GW190930A}{1.14}{GW190929A}{0.97}{GW190924A}{1.09}{GW190915A}{0.92}{GW190910A}{1.09}{GW190909A}{0.99}{GW190828B}{1.08}{GW190828A}{1.15}{GW190814A}{1.11}{GW190803A}{0.99}{GW190731A}{1.13}{GW190728A}{1.20}{GW190727A}{1.11}{GW190720A}{1.01}{GW190719A}{1.09}{GW190708A}{1.00}{GW190707A}{0.86}{GW190706A}{1.06}{GW190701A}{0.95}{GW190630A}{1.11}{GW190620A}{0.91}{GW190602A}{1.12}{GW190527A}{1.11}{GW190521B}{1.09}{GW190521A}{1.04}{GW190519A}{0.79}{GW190517A}{0.50}{GW190514A}{0.78}{GW190513A}{1.18}{GW190512A}{1.11}{GW190503A}{0.97}{GW190426A}{3.14}{GW190425A}{0.66}{GW190424A}{1.06}{GW190421A}{0.97}{GW190413B}{0.91}{GW190413A}{1.11}{GW190412A}{0.54}{GW190408A}{0.94}}}
\newcommand{\spintwozminus}[1]{\IfEqCase{#1}{{GW190930A}{0.41}{GW190929A}{0.55}{GW190924A}{0.36}{GW190915A}{0.51}{GW190910A}{0.36}{GW190909A}{0.62}{GW190828B}{0.42}{GW190828A}{0.35}{GW190814A}{0.53}{GW190803A}{0.55}{GW190731A}{0.47}{GW190728A}{0.38}{GW190727A}{0.45}{GW190720A}{0.53}{GW190719A}{0.46}{GW190708A}{0.33}{GW190707A}{0.36}{GW190706A}{0.45}{GW190701A}{0.54}{GW190630A}{0.31}{GW190620A}{0.47}{GW190602A}{0.46}{GW190527A}{0.49}{GW190521B}{0.32}{GW190521A}{0.54}{GW190519A}{0.43}{GW190517A}{0.46}{GW190514A}{0.59}{GW190513A}{0.41}{GW190512A}{0.33}{GW190503A}{0.51}{GW190426A}{0.03}{GW190425A}{0.18}{GW190424A}{0.44}{GW190421A}{0.53}{GW190413B}{0.55}{GW190413A}{0.54}{GW190412A}{0.44}{GW190408A}{0.37}}}
\newcommand{\spintwozmed}[1]{\IfEqCase{#1}{{GW190930A}{0.08}{GW190929A}{0.008}{GW190924A}{0.02}{GW190915A}{0.003}{GW190910A}{0.006}{GW190909A}{-0.04}{GW190828B}{0.06}{GW190828A}{0.11}{GW190814A}{-0.01}{GW190803A}{-0.01}{GW190731A}{0.03}{GW190728A}{0.11}{GW190727A}{0.04}{GW190720A}{0.11}{GW190719A}{0.16}{GW190708A}{0.03}{GW190707A}{-0.03}{GW190706A}{0.11}{GW190701A}{-0.04}{GW190630A}{0.09}{GW190620A}{0.21}{GW190602A}{0.05}{GW190527A}{0.04}{GW190521B}{0.09}{GW190521A}{-0.01}{GW190519A}{0.21}{GW190517A}{0.29}{GW190514A}{-0.13}{GW190513A}{0.06}{GW190512A}{0.03}{GW190503A}{0.00}{GW190426A}{0.00}{GW190425A}{0.03}{GW190424A}{0.06}{GW190421A}{-0.04}{GW190413B}{-0.03}{GW190413A}{-0.02}{GW190412A}{0.07}{GW190408A}{0.00}}}
\newcommand{\spintwozplus}[1]{\IfEqCase{#1}{{GW190930A}{0.50}{GW190929A}{0.57}{GW190924A}{0.48}{GW190915A}{0.47}{GW190910A}{0.39}{GW190909A}{0.51}{GW190828B}{0.54}{GW190828A}{0.48}{GW190814A}{0.52}{GW190803A}{0.47}{GW190731A}{0.54}{GW190728A}{0.48}{GW190727A}{0.53}{GW190720A}{0.54}{GW190719A}{0.61}{GW190708A}{0.39}{GW190707A}{0.34}{GW190706A}{0.62}{GW190701A}{0.42}{GW190630A}{0.44}{GW190620A}{0.57}{GW190602A}{0.56}{GW190527A}{0.60}{GW190521B}{0.36}{GW190521A}{0.53}{GW190519A}{0.56}{GW190517A}{0.53}{GW190514A}{0.44}{GW190513A}{0.54}{GW190512A}{0.45}{GW190503A}{0.46}{GW190426A}{0.03}{GW190425A}{0.30}{GW190424A}{0.52}{GW190421A}{0.41}{GW190413B}{0.49}{GW190413A}{0.48}{GW190412A}{0.57}{GW190408A}{0.38}}}
\newcommand{\massonesourceminus}[1]{\IfEqCase{#1}{{GW190930A}{2.3}{GW190929A}{33.2}{GW190924A}{2.0}{GW190915A}{6.4}{GW190910A}{6.1}{GW190909A}{13.3}{GW190828B}{7.2}{GW190828A}{4.0}{GW190814A}{1.0}{GW190803A}{7.0}{GW190731A}{9.0}{GW190728A}{2.2}{GW190727A}{6.2}{GW190720A}{3.0}{GW190719A}{10.3}{GW190708A}{2.3}{GW190707A}{1.7}{GW190706A}{16.2}{GW190701A}{8.0}{GW190630A}{5.6}{GW190620A}{12.7}{GW190602A}{13.0}{GW190527A}{9.0}{GW190521B}{4.8}{GW190521A}{18.9}{GW190519A}{12.0}{GW190517A}{7.6}{GW190514A}{8.2}{GW190513A}{9.2}{GW190512A}{5.8}{GW190503A}{8.1}{GW190426A}{2.3}{GW190425A}{0.3}{GW190424A}{7.3}{GW190421A}{6.9}{GW190413B}{10.7}{GW190413A}{8.1}{GW190412A}{5.1}{GW190408A}{3.4}}}
\newcommand{\massonesourcemed}[1]{\IfEqCase{#1}{{GW190930A}{12.3}{GW190929A}{80.8}{GW190924A}{8.9}{GW190915A}{35.3}{GW190910A}{43.9}{GW190909A}{45.8}{GW190828B}{24.1}{GW190828A}{32.1}{GW190814A}{23.2}{GW190803A}{37.3}{GW190731A}{41.5}{GW190728A}{12.3}{GW190727A}{38.0}{GW190720A}{13.4}{GW190719A}{36.5}{GW190708A}{17.6}{GW190707A}{11.6}{GW190706A}{67.0}{GW190701A}{53.9}{GW190630A}{35.1}{GW190620A}{57.1}{GW190602A}{69.1}{GW190527A}{36.5}{GW190521B}{42.2}{GW190521A}{95.3}{GW190519A}{66.0}{GW190517A}{37.4}{GW190514A}{39.0}{GW190513A}{35.7}{GW190512A}{23.3}{GW190503A}{43.3}{GW190426A}{5.7}{GW190425A}{2.0}{GW190424A}{40.5}{GW190421A}{41.3}{GW190413B}{47.5}{GW190413A}{34.7}{GW190412A}{30.1}{GW190408A}{24.6}}}
\newcommand{\massonesourceplus}[1]{\IfEqCase{#1}{{GW190930A}{12.4}{GW190929A}{33.0}{GW190924A}{7.0}{GW190915A}{9.5}{GW190910A}{7.6}{GW190909A}{52.7}{GW190828B}{7.0}{GW190828A}{5.8}{GW190814A}{1.1}{GW190803A}{10.6}{GW190731A}{12.2}{GW190728A}{7.2}{GW190727A}{9.5}{GW190720A}{6.7}{GW190719A}{18.0}{GW190708A}{4.7}{GW190707A}{3.3}{GW190706A}{14.6}{GW190701A}{11.8}{GW190630A}{6.9}{GW190620A}{16.0}{GW190602A}{15.7}{GW190527A}{16.4}{GW190521B}{5.9}{GW190521A}{28.7}{GW190519A}{10.7}{GW190517A}{11.7}{GW190514A}{14.7}{GW190513A}{9.5}{GW190512A}{5.3}{GW190503A}{9.2}{GW190426A}{3.9}{GW190425A}{0.6}{GW190424A}{11.1}{GW190421A}{10.4}{GW190413B}{13.5}{GW190413A}{12.6}{GW190412A}{4.7}{GW190408A}{5.1}}}
\newcommand{\geocenttimeminus}[1]{\IfEqCase{#1}{{GW190930A}{0.02}{GW190929A}{0.02}{GW190924A}{0.008}{GW190915A}{0.003}{GW190910A}{0.0}{GW190909A}{0.0}{GW190828B}{0.0}{GW190828A}{0.0}{GW190814A}{0.0009}{GW190803A}{0.0}{GW190731A}{0.0}{GW190728A}{0.03}{GW190727A}{0.0}{GW190720A}{0.01}{GW190719A}{0.0}{GW190708A}{0.0}{GW190707A}{0.03}{GW190706A}{0.0000002}{GW190701A}{0.0}{GW190630A}{0.0000002}{GW190620A}{0.0000002}{GW190602A}{0.0}{GW190527A}{0.0}{GW190521B}{0.0}{GW190521A}{0.04}{GW190519A}{0.0}{GW190517A}{0.0}{GW190514A}{0.0}{GW190513A}{0.010}{GW190512A}{0.0}{GW190503A}{0.0}{GW190426A}{0.03}{GW190425A}{0.009}{GW190424A}{0.0}{GW190421A}{0.0}{GW190413B}{0.0}{GW190413A}{0.0}{GW190412A}{0.001}{GW190408A}{0.0}}}
\newcommand{\geocenttimemed}[1]{\IfEqCase{#1}{{GW190930A}{1253885759.2}{GW190929A}{1253755327.5}{GW190924A}{1253326744.8}{GW190915A}{1252627040.7}{GW190910A}{1252150105.3}{GW190909A}{1252064527.7}{GW190828B}{1251010527.9}{GW190828A}{1251009263.8}{GW190814A}{1249852257.0}{GW190803A}{1248834439.9}{GW190731A}{1248617394.6}{GW190728A}{1248331528.6}{GW190727A}{1248242632.0}{GW190720A}{1247616534.7}{GW190719A}{1247608532.9}{GW190708A}{1246663515.4}{GW190707A}{1246527224.2}{GW190706A}{1246487219.3}{GW190701A}{1246048404.6}{GW190630A}{1245955943.2}{GW190620A}{1245035079.3}{GW190602A}{1243533585.1}{GW190527A}{1242984073.8}{GW190521B}{1242459857.5}{GW190521A}{1242442967.4}{GW190519A}{1242315362.4}{GW190517A}{1242107479.8}{GW190514A}{1241852074.8}{GW190513A}{1241816086.8}{GW190512A}{1241719652.4}{GW190503A}{1240944862.3}{GW190426A}{1240327333.4}{GW190425A}{1240215503.0}{GW190424A}{1240164426.1}{GW190421A}{1239917954.2}{GW190413B}{1239198206.7}{GW190413A}{1239168612.5}{GW190412A}{1239082262.2}{GW190408A}{1238782700.3}}}
\newcommand{\geocenttimeplus}[1]{\IfEqCase{#1}{{GW190930A}{0.002}{GW190929A}{0.03}{GW190924A}{0.008}{GW190915A}{0.002}{GW190910A}{0.0}{GW190909A}{0.0}{GW190828B}{0.0}{GW190828A}{0.0000005}{GW190814A}{0.004}{GW190803A}{0.0}{GW190731A}{0.0}{GW190728A}{0.0010}{GW190727A}{0.0}{GW190720A}{0.01}{GW190719A}{0.05}{GW190708A}{0.0}{GW190707A}{0.009}{GW190706A}{0.0}{GW190701A}{0.0000005}{GW190630A}{0.0}{GW190620A}{0.0}{GW190602A}{0.0}{GW190527A}{0.0}{GW190521B}{0.0}{GW190521A}{0.01}{GW190519A}{0.0}{GW190517A}{0.0}{GW190514A}{0.0}{GW190513A}{0.0}{GW190512A}{0.0}{GW190503A}{0.0}{GW190426A}{0.02}{GW190425A}{0.03}{GW190424A}{0.0}{GW190421A}{0.0}{GW190413B}{0.0}{GW190413A}{0.0}{GW190412A}{0.007}{GW190408A}{0.0}}}
\newcommand{\costilttwominus}[1]{\IfEqCase{#1}{{GW190930A}{1.09}{GW190929A}{0.93}{GW190924A}{1.00}{GW190915A}{0.90}{GW190910A}{0.88}{GW190909A}{0.75}{GW190828B}{1.06}{GW190828A}{1.14}{GW190814A}{0.83}{GW190803A}{0.83}{GW190731A}{0.99}{GW190728A}{1.16}{GW190727A}{1.04}{GW190720A}{1.11}{GW190719A}{1.19}{GW190708A}{0.97}{GW190707A}{0.71}{GW190706A}{1.16}{GW190701A}{0.74}{GW190630A}{1.08}{GW190620A}{1.22}{GW190602A}{1.01}{GW190527A}{1.02}{GW190521B}{1.03}{GW190521A}{0.86}{GW190519A}{1.17}{GW190517A}{1.21}{GW190514A}{0.59}{GW190513A}{1.05}{GW190512A}{0.98}{GW190503A}{0.87}{GW190426A}{0.00}{GW190425A}{0.87}{GW190424A}{1.02}{GW190421A}{0.76}{GW190413B}{0.77}{GW190413A}{0.82}{GW190412A}{1.01}{GW190408A}{0.85}}}
\newcommand{\costilttwomed}[1]{\IfEqCase{#1}{{GW190930A}{0.31}{GW190929A}{0.04}{GW190924A}{0.15}{GW190915A}{0.02}{GW190910A}{0.04}{GW190909A}{-0.17}{GW190828B}{0.24}{GW190828A}{0.37}{GW190814A}{-0.03}{GW190803A}{-0.06}{GW190731A}{0.15}{GW190728A}{0.39}{GW190727A}{0.20}{GW190720A}{0.31}{GW190719A}{0.44}{GW190708A}{0.14}{GW190707A}{-0.19}{GW190706A}{0.37}{GW190701A}{-0.17}{GW190630A}{0.33}{GW190620A}{0.50}{GW190602A}{0.19}{GW190527A}{0.16}{GW190521B}{0.29}{GW190521A}{-0.02}{GW190519A}{0.50}{GW190517A}{0.61}{GW190514A}{-0.36}{GW190513A}{0.25}{GW190512A}{0.16}{GW190503A}{-0.02}{GW190426A}{-1.00}{GW190425A}{0.16}{GW190424A}{0.20}{GW190421A}{-0.16}{GW190413B}{-0.13}{GW190413A}{-0.08}{GW190412A}{0.25}{GW190408A}{-0.02}}}
\newcommand{\costilttwoplus}[1]{\IfEqCase{#1}{{GW190930A}{0.63}{GW190929A}{0.86}{GW190924A}{0.77}{GW190915A}{0.85}{GW190910A}{0.83}{GW190909A}{1.04}{GW190828B}{0.69}{GW190828A}{0.57}{GW190814A}{0.87}{GW190803A}{0.93}{GW190731A}{0.77}{GW190728A}{0.56}{GW190727A}{0.72}{GW190720A}{0.62}{GW190719A}{0.51}{GW190708A}{0.77}{GW190707A}{1.02}{GW190706A}{0.58}{GW190701A}{1.00}{GW190630A}{0.60}{GW190620A}{0.47}{GW190602A}{0.73}{GW190527A}{0.76}{GW190521B}{0.62}{GW190521A}{0.86}{GW190519A}{0.46}{GW190517A}{0.36}{GW190514A}{1.11}{GW190513A}{0.68}{GW190512A}{0.75}{GW190503A}{0.88}{GW190426A}{2.00}{GW190425A}{0.70}{GW190424A}{0.72}{GW190421A}{0.96}{GW190413B}{1.00}{GW190413A}{0.95}{GW190412A}{0.67}{GW190408A}{0.88}}}
\newcommand{\finalspinminus}[1]{\IfEqCase{#1}{{GW190930A}{0.06}{GW190929A}{0.31}{GW190924A}{0.05}{GW190915A}{0.11}{GW190910A}{0.07}{GW190909A}{0.20}{GW190828B}{0.08}{GW190828A}{0.07}{GW190814A}{0.02}{GW190803A}{0.11}{GW190731A}{0.13}{GW190728A}{0.04}{GW190727A}{0.10}{GW190720A}{0.05}{GW190719A}{0.17}{GW190708A}{0.04}{GW190707A}{0.04}{GW190706A}{0.18}{GW190701A}{0.13}{GW190630A}{0.07}{GW190620A}{0.15}{GW190602A}{0.14}{GW190527A}{0.16}{GW190521B}{0.07}{GW190521A}{0.16}{GW190519A}{0.13}{GW190517A}{0.07}{GW190514A}{0.15}{GW190513A}{0.12}{GW190512A}{0.07}{GW190503A}{0.12}{GW190424A}{0.09}{GW190421A}{0.11}{GW190413B}{0.12}{GW190413A}{0.13}{GW190412A}{0.06}{GW190408A}{0.07}}}
\newcommand{\finalspinmed}[1]{\IfEqCase{#1}{{GW190930A}{0.72}{GW190929A}{0.66}{GW190924A}{0.67}{GW190915A}{0.70}{GW190910A}{0.70}{GW190909A}{0.66}{GW190828B}{0.65}{GW190828A}{0.75}{GW190814A}{0.28}{GW190803A}{0.68}{GW190731A}{0.70}{GW190728A}{0.71}{GW190727A}{0.73}{GW190720A}{0.72}{GW190719A}{0.78}{GW190708A}{0.69}{GW190707A}{0.66}{GW190706A}{0.78}{GW190701A}{0.66}{GW190630A}{0.70}{GW190620A}{0.79}{GW190602A}{0.70}{GW190527A}{0.71}{GW190521B}{0.72}{GW190521A}{0.71}{GW190519A}{0.79}{GW190517A}{0.87}{GW190514A}{0.63}{GW190513A}{0.68}{GW190512A}{0.65}{GW190503A}{0.66}{GW190424A}{0.74}{GW190421A}{0.67}{GW190413B}{0.68}{GW190413A}{0.68}{GW190412A}{0.67}{GW190408A}{0.67}}}
\newcommand{\finalspinplus}[1]{\IfEqCase{#1}{{GW190930A}{0.07}{GW190929A}{0.20}{GW190924A}{0.05}{GW190915A}{0.09}{GW190910A}{0.08}{GW190909A}{0.15}{GW190828B}{0.08}{GW190828A}{0.06}{GW190814A}{0.02}{GW190803A}{0.10}{GW190731A}{0.10}{GW190728A}{0.04}{GW190727A}{0.10}{GW190720A}{0.06}{GW190719A}{0.11}{GW190708A}{0.04}{GW190707A}{0.03}{GW190706A}{0.09}{GW190701A}{0.09}{GW190630A}{0.05}{GW190620A}{0.08}{GW190602A}{0.10}{GW190527A}{0.12}{GW190521B}{0.05}{GW190521A}{0.12}{GW190519A}{0.07}{GW190517A}{0.05}{GW190514A}{0.11}{GW190513A}{0.14}{GW190512A}{0.07}{GW190503A}{0.09}{GW190424A}{0.09}{GW190421A}{0.10}{GW190413B}{0.10}{GW190413A}{0.12}{GW190412A}{0.05}{GW190408A}{0.06}}}
\newcommand{\luminositydistanceminus}[1]{\IfEqCase{#1}{{GW190930A}{0.32}{GW190929A}{1.05}{GW190924A}{0.22}{GW190915A}{0.61}{GW190910A}{0.58}{GW190909A}{2.22}{GW190828B}{0.60}{GW190828A}{0.93}{GW190814A}{0.05}{GW190803A}{1.58}{GW190731A}{1.72}{GW190728A}{0.37}{GW190727A}{1.50}{GW190720A}{0.32}{GW190719A}{2.00}{GW190708A}{0.39}{GW190707A}{0.37}{GW190706A}{1.93}{GW190701A}{0.73}{GW190630A}{0.37}{GW190620A}{1.31}{GW190602A}{1.12}{GW190527A}{1.24}{GW190521B}{0.57}{GW190521A}{1.95}{GW190519A}{0.92}{GW190517A}{0.84}{GW190514A}{2.17}{GW190513A}{0.80}{GW190512A}{0.55}{GW190503A}{0.63}{GW190426A}{0.16}{GW190425A}{0.07}{GW190424A}{1.16}{GW190421A}{1.38}{GW190413B}{2.12}{GW190413A}{1.66}{GW190412A}{0.17}{GW190408A}{0.60}}}
\newcommand{\luminositydistancemed}[1]{\IfEqCase{#1}{{GW190930A}{0.76}{GW190929A}{2.13}{GW190924A}{0.57}{GW190915A}{1.62}{GW190910A}{1.46}{GW190909A}{3.77}{GW190828B}{1.60}{GW190828A}{2.13}{GW190814A}{0.24}{GW190803A}{3.27}{GW190731A}{3.30}{GW190728A}{0.87}{GW190727A}{3.30}{GW190720A}{0.79}{GW190719A}{3.94}{GW190708A}{0.88}{GW190707A}{0.77}{GW190706A}{4.42}{GW190701A}{2.06}{GW190630A}{0.89}{GW190620A}{2.81}{GW190602A}{2.69}{GW190527A}{2.49}{GW190521B}{1.24}{GW190521A}{3.92}{GW190519A}{2.53}{GW190517A}{1.86}{GW190514A}{4.13}{GW190513A}{2.06}{GW190512A}{1.43}{GW190503A}{1.45}{GW190426A}{0.37}{GW190425A}{0.16}{GW190424A}{2.20}{GW190421A}{2.88}{GW190413B}{4.45}{GW190413A}{3.55}{GW190412A}{0.74}{GW190408A}{1.55}}}
\newcommand{\luminositydistanceplus}[1]{\IfEqCase{#1}{{GW190930A}{0.36}{GW190929A}{3.65}{GW190924A}{0.22}{GW190915A}{0.71}{GW190910A}{1.03}{GW190909A}{3.27}{GW190828B}{0.62}{GW190828A}{0.66}{GW190814A}{0.04}{GW190803A}{1.95}{GW190731A}{2.39}{GW190728A}{0.26}{GW190727A}{1.54}{GW190720A}{0.69}{GW190719A}{2.59}{GW190708A}{0.33}{GW190707A}{0.38}{GW190706A}{2.59}{GW190701A}{0.76}{GW190630A}{0.56}{GW190620A}{1.68}{GW190602A}{1.79}{GW190527A}{2.48}{GW190521B}{0.40}{GW190521A}{2.19}{GW190519A}{1.83}{GW190517A}{1.62}{GW190514A}{2.65}{GW190513A}{0.88}{GW190512A}{0.55}{GW190503A}{0.69}{GW190426A}{0.18}{GW190425A}{0.07}{GW190424A}{1.58}{GW190421A}{1.37}{GW190413B}{2.48}{GW190413A}{2.27}{GW190412A}{0.14}{GW190408A}{0.40}}}
\newcommand{\spinonezminus}[1]{\IfEqCase{#1}{{GW190930A}{0.29}{GW190929A}{0.43}{GW190924A}{0.24}{GW190915A}{0.41}{GW190910A}{0.34}{GW190909A}{0.56}{GW190828B}{0.23}{GW190828A}{0.31}{GW190814A}{0.05}{GW190803A}{0.46}{GW190731A}{0.33}{GW190728A}{0.27}{GW190727A}{0.33}{GW190720A}{0.29}{GW190719A}{0.44}{GW190708A}{0.25}{GW190707A}{0.29}{GW190706A}{0.38}{GW190701A}{0.51}{GW190630A}{0.21}{GW190620A}{0.39}{GW190602A}{0.34}{GW190527A}{0.36}{GW190521B}{0.23}{GW190521A}{0.58}{GW190519A}{0.37}{GW190517A}{0.36}{GW190514A}{0.54}{GW190513A}{0.24}{GW190512A}{0.25}{GW190503A}{0.44}{GW190426A}{0.51}{GW190425A}{0.12}{GW190424A}{0.35}{GW190421A}{0.49}{GW190413B}{0.48}{GW190413A}{0.52}{GW190412A}{0.23}{GW190408A}{0.42}}}
\newcommand{\spinonezmed}[1]{\IfEqCase{#1}{{GW190930A}{0.15}{GW190929A}{0.008}{GW190924A}{0.02}{GW190915A}{0.02}{GW190910A}{0.01}{GW190909A}{-0.03}{GW190828B}{0.06}{GW190828A}{0.20}{GW190814A}{0.0001}{GW190803A}{-0.01}{GW190731A}{0.03}{GW190728A}{0.13}{GW190727A}{0.10}{GW190720A}{0.20}{GW190719A}{0.36}{GW190708A}{0.009}{GW190707A}{-0.03}{GW190706A}{0.33}{GW190701A}{-0.05}{GW190630A}{0.05}{GW190620A}{0.37}{GW190602A}{0.04}{GW190527A}{0.09}{GW190521B}{0.04}{GW190521A}{0.02}{GW190519A}{0.35}{GW190517A}{0.67}{GW190514A}{-0.18}{GW190513A}{0.09}{GW190512A}{0.005}{GW190503A}{-0.03}{GW190426A}{-0.03}{GW190425A}{0.06}{GW190424A}{0.15}{GW190421A}{-0.04}{GW190413B}{-0.02}{GW190413A}{0.001}{GW190412A}{0.30}{GW190408A}{-0.03}}}
\newcommand{\spinonezplus}[1]{\IfEqCase{#1}{{GW190930A}{0.41}{GW190929A}{0.45}{GW190924A}{0.39}{GW190915A}{0.40}{GW190910A}{0.39}{GW190909A}{0.53}{GW190828B}{0.24}{GW190828A}{0.41}{GW190814A}{0.04}{GW190803A}{0.39}{GW190731A}{0.45}{GW190728A}{0.30}{GW190727A}{0.48}{GW190720A}{0.29}{GW190719A}{0.43}{GW190708A}{0.24}{GW190707A}{0.20}{GW190706A}{0.43}{GW190701A}{0.34}{GW190630A}{0.28}{GW190620A}{0.42}{GW190602A}{0.46}{GW190527A}{0.50}{GW190521B}{0.32}{GW190521A}{0.53}{GW190519A}{0.37}{GW190517A}{0.25}{GW190514A}{0.39}{GW190513A}{0.46}{GW190512A}{0.21}{GW190503A}{0.31}{GW190426A}{0.36}{GW190425A}{0.18}{GW190424A}{0.46}{GW190421A}{0.40}{GW190413B}{0.40}{GW190413A}{0.44}{GW190412A}{0.12}{GW190408A}{0.26}}}
\newcommand{\chirpmasssourceminus}[1]{\IfEqCase{#1}{{GW190930A}{0.5}{GW190929A}{8.2}{GW190924A}{0.2}{GW190915A}{2.7}{GW190910A}{4.1}{GW190909A}{7.5}{GW190828B}{1.0}{GW190828A}{2.1}{GW190814A}{0.06}{GW190803A}{4.1}{GW190731A}{5.2}{GW190728A}{0.3}{GW190727A}{3.7}{GW190720A}{0.8}{GW190719A}{4.0}{GW190708A}{0.6}{GW190707A}{0.5}{GW190706A}{7.0}{GW190701A}{4.9}{GW190630A}{2.1}{GW190620A}{6.5}{GW190602A}{8.5}{GW190527A}{4.2}{GW190521B}{2.5}{GW190521A}{10.6}{GW190519A}{7.1}{GW190517A}{4.0}{GW190514A}{4.8}{GW190513A}{1.9}{GW190512A}{1.0}{GW190503A}{4.2}{GW190426A}{0.08}{GW190425A}{0.02}{GW190424A}{4.6}{GW190421A}{4.2}{GW190413B}{5.4}{GW190413A}{4.1}{GW190412A}{0.3}{GW190408A}{1.2}}}
\newcommand{\chirpmasssourcemed}[1]{\IfEqCase{#1}{{GW190930A}{8.5}{GW190929A}{35.8}{GW190924A}{5.8}{GW190915A}{25.3}{GW190910A}{34.3}{GW190909A}{30.9}{GW190828B}{13.3}{GW190828A}{25.0}{GW190814A}{6.09}{GW190803A}{27.3}{GW190731A}{29.5}{GW190728A}{8.6}{GW190727A}{28.6}{GW190720A}{8.9}{GW190719A}{23.5}{GW190708A}{13.2}{GW190707A}{8.5}{GW190706A}{42.7}{GW190701A}{40.3}{GW190630A}{24.9}{GW190620A}{38.3}{GW190602A}{49.1}{GW190527A}{24.3}{GW190521B}{32.1}{GW190521A}{69.2}{GW190519A}{44.5}{GW190517A}{26.6}{GW190514A}{28.5}{GW190513A}{21.6}{GW190512A}{14.6}{GW190503A}{30.2}{GW190426A}{2.41}{GW190425A}{1.44}{GW190424A}{31.0}{GW190421A}{31.2}{GW190413B}{33.0}{GW190413A}{24.6}{GW190412A}{13.3}{GW190408A}{18.3}}}
\newcommand{\chirpmasssourceplus}[1]{\IfEqCase{#1}{{GW190930A}{0.5}{GW190929A}{14.9}{GW190924A}{0.2}{GW190915A}{3.2}{GW190910A}{4.1}{GW190909A}{17.2}{GW190828B}{1.2}{GW190828A}{3.4}{GW190814A}{0.06}{GW190803A}{5.7}{GW190731A}{7.1}{GW190728A}{0.5}{GW190727A}{5.3}{GW190720A}{0.5}{GW190719A}{6.5}{GW190708A}{0.9}{GW190707A}{0.6}{GW190706A}{10.0}{GW190701A}{5.4}{GW190630A}{2.1}{GW190620A}{8.3}{GW190602A}{9.1}{GW190527A}{9.1}{GW190521B}{3.2}{GW190521A}{17.0}{GW190519A}{6.4}{GW190517A}{4.0}{GW190514A}{7.9}{GW190513A}{3.8}{GW190512A}{1.3}{GW190503A}{4.2}{GW190426A}{0.08}{GW190425A}{0.02}{GW190424A}{5.8}{GW190421A}{5.9}{GW190413B}{8.2}{GW190413A}{5.5}{GW190412A}{0.4}{GW190408A}{1.9}}}
\newcommand{\symmetricmassratiominus}[1]{\IfEqCase{#1}{{GW190930A}{0.11}{GW190929A}{0.07}{GW190924A}{0.09}{GW190915A}{0.03}{GW190910A}{0.01}{GW190909A}{0.09}{GW190828B}{0.04}{GW190828A}{0.01}{GW190814A}{0.006}{GW190803A}{0.03}{GW190731A}{0.04}{GW190728A}{0.07}{GW190727A}{0.03}{GW190720A}{0.06}{GW190719A}{0.06}{GW190708A}{0.03}{GW190707A}{0.03}{GW190706A}{0.05}{GW190701A}{0.03}{GW190630A}{0.03}{GW190620A}{0.04}{GW190602A}{0.04}{GW190527A}{0.06}{GW190521B}{0.01}{GW190521A}{0.04}{GW190519A}{0.03}{GW190517A}{0.04}{GW190514A}{0.04}{GW190513A}{0.04}{GW190512A}{0.03}{GW190503A}{0.03}{GW190426A}{0.08}{GW190425A}{0.03}{GW190424A}{0.02}{GW190421A}{0.03}{GW190413B}{0.04}{GW190413A}{0.04}{GW190412A}{0.02}{GW190408A}{0.02}}}
\newcommand{\symmetricmassratiomed}[1]{\IfEqCase{#1}{{GW190930A}{0.24}{GW190929A}{0.18}{GW190924A}{0.23}{GW190915A}{0.242}{GW190910A}{0.248}{GW190909A}{0.24}{GW190828B}{0.21}{GW190828A}{0.248}{GW190814A}{0.090}{GW190803A}{0.245}{GW190731A}{0.243}{GW190728A}{0.24}{GW190727A}{0.247}{GW190720A}{0.23}{GW190719A}{0.23}{GW190708A}{0.245}{GW190707A}{0.244}{GW190706A}{0.23}{GW190701A}{0.246}{GW190630A}{0.241}{GW190620A}{0.24}{GW190602A}{0.243}{GW190527A}{0.24}{GW190521B}{0.246}{GW190521A}{0.245}{GW190519A}{0.24}{GW190517A}{0.241}{GW190514A}{0.245}{GW190513A}{0.22}{GW190512A}{0.23}{GW190503A}{0.24}{GW190426A}{0.16}{GW190425A}{0.240}{GW190424A}{0.247}{GW190421A}{0.247}{GW190413B}{0.241}{GW190413A}{0.241}{GW190412A}{0.17}{GW190408A}{0.245}}}
\newcommand{\symmetricmassratioplus}[1]{\IfEqCase{#1}{{GW190930A}{0.01}{GW190929A}{0.07}{GW190924A}{0.02}{GW190915A}{0.008}{GW190910A}{0.002}{GW190909A}{0.01}{GW190828B}{0.04}{GW190828A}{0.002}{GW190814A}{0.005}{GW190803A}{0.005}{GW190731A}{0.007}{GW190728A}{0.01}{GW190727A}{0.003}{GW190720A}{0.02}{GW190719A}{0.02}{GW190708A}{0.005}{GW190707A}{0.006}{GW190706A}{0.02}{GW190701A}{0.004}{GW190630A}{0.009}{GW190620A}{0.01}{GW190602A}{0.007}{GW190527A}{0.01}{GW190521B}{0.004}{GW190521A}{0.005}{GW190519A}{0.01}{GW190517A}{0.009}{GW190514A}{0.005}{GW190513A}{0.03}{GW190512A}{0.02}{GW190503A}{0.01}{GW190426A}{0.08}{GW190425A}{0.010}{GW190424A}{0.003}{GW190421A}{0.003}{GW190413B}{0.008}{GW190413A}{0.008}{GW190412A}{0.03}{GW190408A}{0.005}}}
\newcommand{\spintwoxminus}[1]{\IfEqCase{#1}{{GW190930A}{0.52}{GW190929A}{0.60}{GW190924A}{0.49}{GW190915A}{0.59}{GW190910A}{0.52}{GW190909A}{0.58}{GW190828B}{0.55}{GW190828A}{0.50}{GW190814A}{0.62}{GW190803A}{0.58}{GW190731A}{0.57}{GW190728A}{0.46}{GW190727A}{0.55}{GW190720A}{0.58}{GW190719A}{0.56}{GW190708A}{0.46}{GW190707A}{0.46}{GW190706A}{0.53}{GW190701A}{0.55}{GW190630A}{0.49}{GW190620A}{0.54}{GW190602A}{0.60}{GW190527A}{0.61}{GW190521B}{0.51}{GW190521A}{0.66}{GW190519A}{0.57}{GW190517A}{0.53}{GW190514A}{0.60}{GW190513A}{0.53}{GW190512A}{0.50}{GW190503A}{0.55}{GW190426A}{0.00}{GW190425A}{0.47}{GW190424A}{0.59}{GW190421A}{0.59}{GW190413B}{0.59}{GW190413A}{0.58}{GW190412A}{0.57}{GW190408A}{0.53}}}
\newcommand{\spintwoxmed}[1]{\IfEqCase{#1}{{GW190930A}{0.00}{GW190929A}{0.0009}{GW190924A}{0.00}{GW190915A}{0.00}{GW190910A}{0.0003}{GW190909A}{0.0003}{GW190828B}{0.00}{GW190828A}{0.002}{GW190814A}{-0.01}{GW190803A}{0.006}{GW190731A}{0.003}{GW190728A}{0.001}{GW190727A}{0.002}{GW190720A}{0.002}{GW190719A}{0.00}{GW190708A}{0.002}{GW190707A}{0.0004}{GW190706A}{0.003}{GW190701A}{0.003}{GW190630A}{0.00}{GW190620A}{0.00006}{GW190602A}{0.00}{GW190527A}{0.005}{GW190521B}{0.00}{GW190521A}{0.002}{GW190519A}{0.0006}{GW190517A}{0.001}{GW190514A}{0.001}{GW190513A}{0.00}{GW190512A}{0.0009}{GW190503A}{0.001}{GW190426A}{0.00}{GW190425A}{0.0006}{GW190424A}{0.00}{GW190421A}{0.00}{GW190413B}{0.0007}{GW190413A}{0.00}{GW190412A}{-0.01}{GW190408A}{0.002}}}
\newcommand{\spintwoxplus}[1]{\IfEqCase{#1}{{GW190930A}{0.52}{GW190929A}{0.59}{GW190924A}{0.48}{GW190915A}{0.60}{GW190910A}{0.54}{GW190909A}{0.57}{GW190828B}{0.54}{GW190828A}{0.51}{GW190814A}{0.59}{GW190803A}{0.57}{GW190731A}{0.58}{GW190728A}{0.48}{GW190727A}{0.55}{GW190720A}{0.57}{GW190719A}{0.56}{GW190708A}{0.43}{GW190707A}{0.46}{GW190706A}{0.54}{GW190701A}{0.56}{GW190630A}{0.48}{GW190620A}{0.56}{GW190602A}{0.60}{GW190527A}{0.59}{GW190521B}{0.51}{GW190521A}{0.69}{GW190519A}{0.55}{GW190517A}{0.53}{GW190514A}{0.59}{GW190513A}{0.54}{GW190512A}{0.49}{GW190503A}{0.58}{GW190426A}{0.00}{GW190425A}{0.47}{GW190424A}{0.59}{GW190421A}{0.58}{GW190413B}{0.60}{GW190413A}{0.57}{GW190412A}{0.56}{GW190408A}{0.53}}}
\newcommand{\networkoptimalsnrminus}[1]{\IfEqCase{#1}{{GW190814A}{1.7}{GW190426A}{1.8}{GW190425A}{1.7}}}
\newcommand{\networkoptimalsnrmed}[1]{\IfEqCase{#1}{{GW190814A}{24.7}{GW190426A}{8.3}{GW190425A}{12.0}}}
\newcommand{\networkoptimalsnrplus}[1]{\IfEqCase{#1}{{GW190814A}{1.7}{GW190426A}{1.8}{GW190425A}{1.7}}}
\newcommand{\networkmatchedfiltersnrminus}[1]{\IfEqCase{#1}{{GW190814A}{0.2}{GW190426A}{0.6}{GW190425A}{0.4}{GW190412A}{0.4}}}
\newcommand{\networkmatchedfiltersnrmed}[1]{\IfEqCase{#1}{{GW190814A}{24.9}{GW190426A}{8.7}{GW190425A}{12.4}{GW190412A}{19.0}}}
\newcommand{\networkmatchedfiltersnrplus}[1]{\IfEqCase{#1}{{GW190814A}{0.1}{GW190426A}{0.5}{GW190425A}{0.3}{GW190412A}{0.2}}}
\newcommand{\logpriorminus}[1]{\IfEqCase{#1}{{GW190426A}{10.5}{GW190425A}{8.6}}}
\newcommand{\logpriormed}[1]{\IfEqCase{#1}{{GW190426A}{161.3}{GW190425A}{98.4}}}
\newcommand{\logpriorplus}[1]{\IfEqCase{#1}{{GW190426A}{8.6}{GW190425A}{6.7}}}
\newcommand{\PEpercentBNS}[1]{\IfEqCase{#1}{{GW190930A}{0}{GW190929A}{0}{GW190924A}{0}{GW190915A}{0}{GW190910A}{0}{GW190909A}{0}{GW190828B}{0}{GW190828A}{0}{GW190814A}{0}{GW190803A}{0}{GW190731A}{0}{GW190728A}{0}{GW190727A}{0}{GW190720A}{0}{GW190719A}{0}{GW190708A}{0}{GW190707A}{0}{GW190706A}{0}{GW190701A}{0}{GW190630A}{0}{GW190620A}{0}{GW190602A}{0}{GW190527A}{0}{GW190521B}{0}{GW190521A}{0}{GW190519A}{0}{GW190517A}{0}{GW190514A}{0}{GW190513A}{0}{GW190512A}{0}{GW190503A}{0}{GW190426A}{1}{GW190425A}{100}{GW190424A}{0}{GW190421A}{0}{GW190413B}{0}{GW190413A}{0}{GW190412A}{0}{GW190408A}{0}}}
\newcommand{\PEpercentNSBH}[1]{\IfEqCase{#1}{{GW190930A}{0}{GW190929A}{0}{GW190924A}{4}{GW190915A}{0}{GW190910A}{0}{GW190909A}{0}{GW190828B}{0}{GW190828A}{0}{GW190814A}{100}{GW190803A}{0}{GW190731A}{0}{GW190728A}{0}{GW190727A}{0}{GW190720A}{0}{GW190719A}{0}{GW190708A}{0}{GW190707A}{0}{GW190706A}{0}{GW190701A}{0}{GW190630A}{0}{GW190620A}{0}{GW190602A}{0}{GW190527A}{0}{GW190521B}{0}{GW190521A}{0}{GW190519A}{0}{GW190517A}{0}{GW190514A}{0}{GW190513A}{0}{GW190512A}{0}{GW190503A}{0}{GW190426A}{99}{GW190425A}{0}{GW190424A}{0}{GW190421A}{0}{GW190413B}{0}{GW190413A}{0}{GW190412A}{0}{GW190408A}{0}}}
\newcommand{\PEpercentBBH}[1]{\IfEqCase{#1}{{GW190930A}{100}{GW190929A}{100}{GW190924A}{96}{GW190915A}{100}{GW190910A}{100}{GW190909A}{100}{GW190828B}{100}{GW190828A}{100}{GW190814A}{0}{GW190803A}{100}{GW190731A}{100}{GW190728A}{100}{GW190727A}{100}{GW190720A}{100}{GW190719A}{100}{GW190708A}{100}{GW190707A}{100}{GW190706A}{100}{GW190701A}{100}{GW190630A}{100}{GW190620A}{100}{GW190602A}{100}{GW190527A}{100}{GW190521B}{100}{GW190521A}{100}{GW190519A}{100}{GW190517A}{100}{GW190514A}{100}{GW190513A}{100}{GW190512A}{100}{GW190503A}{100}{GW190426A}{0}{GW190425A}{0}{GW190424A}{100}{GW190421A}{100}{GW190413B}{100}{GW190413A}{100}{GW190412A}{100}{GW190408A}{100}}}
\newcommand{\PEpercentMassGap}[1]{\IfEqCase{#1}{{GW190930A}{0}{GW190929A}{0}{GW190924A}{0}{GW190915A}{0}{GW190910A}{0}{GW190909A}{0}{GW190828B}{0}{GW190828A}{0}{GW190814A}{0}{GW190803A}{0}{GW190731A}{0}{GW190728A}{0}{GW190727A}{0}{GW190720A}{0}{GW190719A}{0}{GW190708A}{0}{GW190707A}{0}{GW190706A}{0}{GW190701A}{0}{GW190630A}{0}{GW190620A}{0}{GW190602A}{0}{GW190527A}{0}{GW190521B}{0}{GW190521A}{0}{GW190519A}{0}{GW190517A}{0}{GW190514A}{0}{GW190513A}{0}{GW190512A}{0}{GW190503A}{0}{GW190426A}{0}{GW190425A}{0}{GW190424A}{0}{GW190421A}{0}{GW190413B}{0}{GW190413A}{0}{GW190412A}{0}{GW190408A}{0}}}

%% file: superlative_macros.tex
    \newcommand{\luminositydistanceleast}{GW190924A}

    \newcommand{\luminositydistanceleastpercent}{42}

    \newcommand{\luminositydistancemost}{GW190413B}

    \newcommand{\luminositydistancemostpercent}{17}

    \newcommand{\luminositydistanceleastsecond}{GW190707A}

    \newcommand{\luminositydistanceleastpercentsecond}{14}

    \newcommand{\luminositydistancemostsecond}{GW190706A}

    \newcommand{\luminositydistancemostpercentsecond}{17}

    \newcommand{\massonesourceleast}{GW190924A}

    \newcommand{\massonesourceleastpercent}{77}

    \newcommand{\massonesourcemost}{GW190521A}

    \newcommand{\massonesourcemostpercent}{66}

    \newcommand{\massonesourceleastsecond}{GW190707A}

    \newcommand{\massonesourceleastpercentsecond}{ 8}

    \newcommand{\massonesourcemostsecond}{GW190929A}

    \newcommand{\massonesourcemostpercentsecond}{25}

    \newcommand{\masstwosourceleast}{GW190924A}

    \newcommand{\masstwosourceleastpercent}{85}

    \newcommand{\masstwosourcemost}{GW190521A}

    \newcommand{\masstwosourcemostpercent}{88}

    \newcommand{\masstwosourceleastsecond}{GW190930A}

    \newcommand{\masstwosourceleastpercentsecond}{ 9}

    \newcommand{\masstwosourcemostsecond}{GW190602A}

    \newcommand{\masstwosourcemostpercentsecond}{ 8}

    \newcommand{\totalmasssourceleast}{GW190924A}

    \newcommand{\totalmasssourceleastpercent}{96}

    \newcommand{\totalmasssourcemost}{GW190521A}

    \newcommand{\totalmasssourcemostpercent}{98}

    \newcommand{\totalmasssourceleastsecond}{GW190707A}

    \newcommand{\totalmasssourceleastpercentsecond}{ 2}

    \newcommand{\totalmasssourcemostsecond}{GW190909A}

    \newcommand{\totalmasssourcemostpercentsecond}{ 1}

    \newcommand{\massratioleast}{GW190929A}

    \newcommand{\massratioleastpercent}{34}

    \newcommand{\massratiomost}{GW190828A}

    \newcommand{\massratiomostpercent}{ 6}

    \newcommand{\massratioleastsecond}{GW190412A}

    \newcommand{\massratioleastpercentsecond}{30}

    \newcommand{\massratiomostsecond}{GW190424A}

    \newcommand{\massratiomostpercentsecond}{ 5}

    \newcommand{\chieffleast}{GW190514A}

    \newcommand{\chieffleastpercent}{28}

    \newcommand{\chieffmost}{GW190517A}

    \newcommand{\chieffmostpercent}{60}

    \newcommand{\chieffleastsecond}{GW190909A}

    \newcommand{\chieffleastpercentsecond}{12}

    \newcommand{\chieffmostsecond}{GW190719A}

    \newcommand{\chieffmostpercentsecond}{12}

%% file: detchar_macros.tex
\newcommand{\OBSERVINGINSTRUMENTS}[1]{\IfEqCase{#1}{{GW190413A}{HLV}{GW190719A}{HL}{GW190620A}{LV}{GW190514A}{HL}{GW190731A}{HL}{GW190503A}{HLV}{GW190602A}{HLV}{GW190929A}{HLV}{GW190517A}{HLV}{GW190915A}{HLV}{GW190425A}{LV}{GW190512A}{HLV}{GW190630A}{LV}{GW190521A}{HLV}{GW190413B}{HLV}{GW190924A}{HLV}{GW190930A}{HL}{GW190706A}{HLV}{GW190408A}{HLV}{GW190909A}{HL}{GW190728A}{HLV}{GW190426A}{HLV}{GW190412A}{HLV}{GW190720A}{HLV}{GW190521B}{HL}{GW190910A}{LV}{GW190803A}{HLV}{GW190519A}{HLV}{GW190708A}{LV}{GW190527A}{HL}{GW190513A}{HLV}{GW190424A}{L}{GW190727A}{HLV}{GW190814A}{LV}{GW190707A}{HL}{GW190828A}{HLV}{GW190828B}{HLV}{GW190701A}{HLV}{GW190421A}{HL}}}

\newcommand{\PEINSTRUMENTS}[1]{\IfEqCase{#1}{{GW190413A}{HLV}{GW190719A}{HL}{GW190620A}{LV}{GW190514A}{HL}{GW190731A}{HL}{GW190503A}{HLV}{GW190602A}{HLV}{GW190929A}{HLV}{GW190517A}{HLV}{GW190915A}{HLV}{GW190425A}{LV}{GW190512A}{HLV}{GW190630A}{LV}{GW190521A}{HLV}{GW190413B}{HLV}{GW190924A}{HLV}{GW190930A}{HL}{GW190706A}{HLV}{GW190408A}{HLV}{GW190909A}{HL}{GW190728A}{HLV}{GW190426A}{HLV}{GW190412A}{HLV}{GW190720A}{HLV}{GW190521B}{HL}{GW190910A}{(H)LV}{GW190803A}{HLV}{GW190519A}{HLV}{GW190708A}{LV}{GW190527A}{HL}{GW190513A}{HLV}{GW190424A}{L}{GW190727A}{HLV}{GW190814A}{(H)LV}{GW190707A}{HL}{GW190828A}{HLV}{GW190828B}{HLV}{GW190701A}{HLV}{GW190421A}{HL}}}

\newcommand{\MITIGATIONMETHOD}[1]{\IfEqCase{#1}{{GW190413A}{None}{GW190719A}{None}{GW190620A}{None}{GW190514A}{L1 glitch subtraction, glitch-only model}{GW190731A}{None}{GW190503A}{L1 glitch subtraction, glitch-only model}{GW190602A}{None}{GW190929A}{None}{GW190517A}{None}{GW190915A}{None}{GW190425A}{L1 glitch subtraction, glitch-only model}{GW190512A}{None}{GW190630A}{None}{GW190521A}{None}{GW190413B}{L1 glitch subtraction, glitch-only model}{GW190924A}{L1 glitch subtraction, glitch-only model}{GW190930A}{None}{GW190706A}{None}{GW190408A}{None}{GW190909A}{None}{GW190728A}{None}{GW190426A}{None}{GW190412A}{None}{GW190720A}{None}{GW190521B}{None}{GW190910A}{None}{GW190803A}{None}{GW190519A}{None}{GW190708A}{None}{GW190527A}{None}{GW190513A}{L1 glitch subtraction, glitch-only model}{GW190424A}{L1 glitch subtraction, glitch-only model}{GW190727A}{\fixme{L1 $f_\text{min}$: 50 Hz}}{GW190814A}{L1 $f_\text{min}$: 30 Hz; H1 non-observing data used}{GW190707A}{None}{GW190828A}{None}{GW190828B}{None}{GW190701A}{L1 glitch subtraction, glitch+signal model}{GW190421A}{None}}}

\newcommand{\OTHERDQ}[1]{\IfEqCase{#1}{{GW190413A}{--}{GW190719A}{--}{GW190620A}{--}{GW190514A}{--}{GW190731A}{--}{GW190503A}{--}{GW190602A}{--}{GW190929A}{--}{GW190517A}{--}{GW190915A}{--}{GW190425A}{--}{GW190512A}{--}{GW190630A}{--}{GW190521A}{--}{GW190413B}{--}{GW190924A}{--}{GW190930A}{--}{GW190706A}{--}{GW190408A}{--}{GW190909A}{--}{GW190728A}{--}{GW190426A}{--}{GW190412A}{--}{GW190720A}{--}{GW190521B}{--}{GW190910A}{--}{GW190803A}{--}{GW190519A}{--}{GW190708A}{--}{GW190527A}{--}{GW190513A}{--}{GW190424A}{--}{GW190727A}{--}{GW190814A}{--}{GW190707A}{--}{GW190828A}{--}{GW190828B}{--}{GW190701A}{--}{GW190421A}{--}}}

%% file: name_macros.tex
\newcommand{\SNAME}[1]{\IfEqCase{#1}{{GW190930A}{S190930s}{GW190929A}{S190929d}{GW190924A}{S190924h}{GW190915A}{S190915ak}{GW190910A}{S190910s}{GW190909A}{S190909w}{GW190828B}{S190828l}{GW190828A}{S190828j}{GW190814A}{S190814bv}{GW190803A}{S190803e}{GW190731A}{S190731aa}{GW190728A}{S190728q}{GW190727A}{S190727h}{GW190720A}{S190720a}{GW190719A}{S190719an}{GW190708A}{S190708ap}{GW190707A}{S190707q}{GW190706A}{S190706ai}{GW190701A}{S190701ah}{GW190630A}{S190630ag}{GW190620A}{S190620e}{GW190602A}{S190602aq}{GW190527A}{S190527w}{GW190521B}{S190521r}{GW190521A}{S190521g}{GW190519A}{S190519bj}{GW190517A}{S190517h}{GW190514A}{S190514n}{GW190513A}{S190513bm}{GW190512A}{S190512at}{GW190503A}{S190503bf}{GW190426A}{S190426c}{GW190425A}{S190425z}{GW190424A}{S190424ao}{GW190421A}{S190421ar}{GW190413B}{S190413ac}{GW190413A}{S190413i}{GW190412A}{S190412m}{GW190408A}{S190408an}}}
\newcommand{\FULLNAME}[1]{\IfEqCase{#1}{{GW190930A}{GW190930\_133541}{GW190929A}{GW190929\_012149}{GW190924A}{GW190924\_021846}{GW190915A}{GW190915\_235702}{GW190910A}{GW190910\_112807}{GW190909A}{GW190909\_114149}{GW190828B}{GW190828\_065509}{GW190828A}{GW190828\_063405}{GW190814A}{GW190814\_211039}{GW190803A}{GW190803\_022701}{GW190731A}{GW190731\_140936}{GW190728A}{GW190728\_064510}{GW190727A}{GW190727\_060333}{GW190720A}{GW190720\_000836}{GW190719A}{GW190719\_215514}{GW190708A}{GW190708\_232457}{GW190707A}{GW190707\_093326}{GW190706A}{GW190706\_222641}{GW190701A}{GW190701\_203306}{GW190630A}{GW190630\_185205}{GW190620A}{GW190620\_030421}{GW190602A}{GW190602\_175927}{GW190527A}{GW190527\_092055}{GW190521B}{GW190521\_074359}{GW190521A}{GW190521\_030229}{GW190519A}{GW190519\_153544}{GW190517A}{GW190517\_055101}{GW190514A}{GW190514\_065416}{GW190513A}{GW190513\_205428}{GW190512A}{GW190512\_180714}{GW190503A}{GW190503\_185404}{GW190426A}{GW190426\_152155}{GW190425A}{GW190425\_081805}{GW190424A}{GW190424\_180648}{GW190421A}{GW190421\_213856}{GW190413B}{GW190413\_134308}{GW190413A}{GW190413\_052954}{GW190412A}{GW190412\_053044}{GW190408A}{GW190408\_181802}}}
\newcommand{\NNAME}[1]{\IfEqCase{#1}{{GW190930A}{GW190930\_133541}{GW190929A}{GW190929\_012149}{GW190924A}{GW190924\_021846}{GW190915A}{GW190915\_235702}{GW190910A}{GW190910\_112807}{GW190909A}{GW190909\_114149}{GW190828B}{GW190828\_065509}{GW190828A}{GW190828\_063405}{GW190814A}{GW190814}{GW190803A}{GW190803\_022701}{GW190731A}{GW190731\_140936}{GW190728A}{GW190728\_064510}{GW190727A}{GW190727\_060333}{GW190720A}{GW190720\_000836}{GW190719A}{GW190719\_215514}{GW190708A}{GW190708\_232457}{GW190707A}{GW190707\_093326}{GW190706A}{GW190706\_222641}{GW190701A}{GW190701\_203306}{GW190630A}{GW190630\_185205}{GW190620A}{GW190620\_030421}{GW190602A}{GW190602\_175927}{GW190527A}{GW190527\_092055}{GW190521B}{GW190521\_074359}{GW190521A}{GW190521}{GW190519A}{GW190519\_153544}{GW190517A}{GW190517\_055101}{GW190514A}{GW190514\_065416}{GW190513A}{GW190513\_205428}{GW190512A}{GW190512\_180714}{GW190503A}{GW190503\_185404}{GW190426A}{GW190426\_152155}{GW190425A}{GW190425}{GW190424A}{GW190424\_180648}{GW190421A}{GW190421\_213856}{GW190413B}{GW190413\_134308}{GW190413A}{GW190413\_052954}{GW190412A}{GW190412}{GW190408A}{GW190408\_181802}}}
\newcommand{\MINIMALNAME}[1]{\IfEqCase{#1}{{GW190930A}{GW190930}{GW190929A}{GW190929}{GW190924A}{GW190924}{GW190915A}{GW190915}{GW190910A}{GW190910}{GW190909A}{GW190909}{GW190828B}{GW190828\_0655}{GW190828A}{GW190828\_0634}{GW190814A}{GW190814}{GW190803A}{GW190803}{GW190731A}{GW190731}{GW190728A}{GW190728}{GW190727A}{GW190727}{GW190720A}{GW190720}{GW190719A}{GW190719}{GW190708A}{GW190708}{GW190707A}{GW190707}{GW190706A}{GW190706}{GW190701A}{GW190701}{GW190630A}{GW190630}{GW190620A}{GW190620}{GW190602A}{GW190602}{GW190527A}{GW190527}{GW190521B}{GW190521\_07}{GW190521A}{GW190521}{GW190519A}{GW190519}{GW190517A}{GW190517}{GW190514A}{GW190514}{GW190513A}{GW190513}{GW190512A}{GW190512}{GW190503A}{GW190503}{GW190426A}{GW190426}{GW190425A}{GW190425}{GW190424A}{GW190424}{GW190421A}{GW190421}{GW190413B}{GW190413\_13}{GW190413A}{GW190413\_05}{GW190412A}{GW190412}{GW190408A}{GW190408}}}

%% file: JS_macros.tex
\newcommand{\JSIMRPSEOBchirpmassdet}[1]{\IfEqCase{#1}{{GW190930A}{0.0129}{GW190929A}{0.08551}{GW190924A}{0.0574}{GW190915A}{0.00807}{GW190910A}{0.00582}{GW190909A}{0.00495}{GW190828B}{0.00632}{GW190828A}{0.03371}{GW190803A}{0.00562}{GW190731A}{0.03183}{GW190728A}{0.01965}{GW190727A}{0.008}{GW190720A}{0.05532}{GW190719A}{0.05882}{GW190708A}{0.03324}{GW190707A}{0.07465}{GW190706A}{0.01733}{GW190701A}{0.00585}{GW190630A}{0.01743}{GW190620A}{0.01413}{GW190602A}{0.00676}{GW190527A}{0.0495}{GW190521B}{0.0489}{GW190519A}{0.00878}{GW190517A}{0.02901}{GW190514A}{0.008}{GW190513A}{0.03334}{GW190512A}{0.02687}{GW190503A}{0.00962}{GW190424A}{0.00493}{GW190421A}{0.02647}{GW190413B}{0.03779}{GW190413A}{0.00158}{GW190412A}{0.06654}{GW190408A}{0.02324}}}
\newcommand{\JSIMRPSEOBmassratio}[1]{\IfEqCase{#1}{{GW190930A}{0.03033}{GW190929A}{0.04464}{GW190924A}{0.1029}{GW190915A}{0.03143}{GW190910A}{0.00189}{GW190909A}{0.00528}{GW190828B}{0.00251}{GW190828A}{0.0024}{GW190803A}{0.00202}{GW190731A}{0.00281}{GW190728A}{0.03167}{GW190727A}{0.00176}{GW190720A}{0.01805}{GW190719A}{0.00225}{GW190708A}{0.00111}{GW190707A}{0.02141}{GW190706A}{0.00473}{GW190701A}{0.00186}{GW190630A}{0.00202}{GW190620A}{0.00331}{GW190602A}{0.00366}{GW190527A}{0.00399}{GW190521B}{0.00418}{GW190519A}{0.00261}{GW190517A}{0.01631}{GW190514A}{0.00068}{GW190513A}{0.0046}{GW190512A}{0.02218}{GW190503A}{0.00295}{GW190424A}{0.00543}{GW190421A}{0.00136}{GW190413B}{0.00072}{GW190413A}{0.00444}{GW190412A}{0.1448}{GW190408A}{0.0107}}}
\newcommand{\JSIMRPSEOBchieff}[1]{\IfEqCase{#1}{{GW190930A}{0.02636}{GW190929A}{0.04569}{GW190924A}{0.11014}{GW190915A}{0.00588}{GW190910A}{0.01135}{GW190909A}{0.00857}{GW190828B}{0.01915}{GW190828A}{0.04411}{GW190803A}{0.00684}{GW190731A}{0.03543}{GW190728A}{0.04915}{GW190727A}{0.0034}{GW190720A}{0.02282}{GW190719A}{0.0453}{GW190708A}{0.02849}{GW190707A}{0.05627}{GW190706A}{0.00858}{GW190701A}{0.00834}{GW190630A}{0.04798}{GW190620A}{0.00733}{GW190602A}{0.0048}{GW190527A}{0.03188}{GW190521B}{0.10217}{GW190519A}{0.00593}{GW190517A}{0.00817}{GW190514A}{0.00338}{GW190513A}{0.05779}{GW190512A}{0.03075}{GW190503A}{0.00894}{GW190424A}{0.00468}{GW190421A}{0.0334}{GW190413B}{0.01084}{GW190413A}{0.00517}{GW190412A}{0.15241}{GW190408A}{0.00446}}}
\newcommand{\JSIMRPSEOBluminositydistance}[1]{\IfEqCase{#1}{{GW190930A}{0.0009}{GW190929A}{0.08025}{GW190924A}{0.01185}{GW190915A}{0.00156}{GW190910A}{0.00196}{GW190909A}{0.00171}{GW190828B}{0.00176}{GW190828A}{0.01038}{GW190803A}{0.00216}{GW190731A}{0.00301}{GW190728A}{0.00292}{GW190727A}{0.00381}{GW190720A}{0.10464}{GW190719A}{0.0092}{GW190708A}{0.00184}{GW190707A}{0.02769}{GW190706A}{0.00184}{GW190701A}{0.00228}{GW190630A}{0.00218}{GW190620A}{0.00085}{GW190602A}{0.00353}{GW190527A}{0.01098}{GW190521B}{0.04095}{GW190519A}{0.01235}{GW190517A}{0.0056}{GW190514A}{0.00142}{GW190513A}{0.01302}{GW190512A}{0.01483}{GW190503A}{0.00118}{GW190424A}{0.00614}{GW190421A}{0.00902}{GW190413B}{0.00773}{GW190413A}{0.00159}{GW190412A}{0.00767}{GW190408A}{0.00175}}}
\newcommand{\JSIMRPSEOBHMchirpmassdet}[1]{\IfEqCase{#1}{{GW190930A}{0.01375}{GW190929A}{0.03394}{GW190924A}{0.05939}{GW190915A}{0.00337}{GW190910A}{0.00719}{GW190909A}{0.0407}{GW190828B}{0.01874}{GW190828A}{0.05431}{GW190728A}{0.01417}{GW190720A}{0.03855}{GW190719A}{0.05064}{GW190708A}{0.01559}{GW190707A}{0.10849}{GW190706A}{0.0254}{GW190701A}{0.00732}{GW190630A}{0.03541}{GW190620A}{0.03114}{GW190602A}{0.04865}{GW190521B}{0.07125}{GW190519A}{0.0118}{GW190517A}{0.03536}{GW190513A}{0.03819}{GW190512A}{0.01212}{GW190426A}{0.02053}{GW190424A}{0.0139}{GW190421A}{0.03766}{GW190413A}{0.01031}{GW190412A}{0.05007}{GW190408A}{0.00426}}}
\newcommand{\JSIMRPSEOBHMmassratio}[1]{\IfEqCase{#1}{{GW190930A}{0.04113}{GW190929A}{0.07616}{GW190924A}{0.09482}{GW190915A}{0.07458}{GW190910A}{0.01064}{GW190909A}{0.01548}{GW190828B}{0.01352}{GW190828A}{0.01144}{GW190728A}{0.0412}{GW190720A}{0.03253}{GW190719A}{0.01313}{GW190708A}{0.02287}{GW190707A}{0.02161}{GW190706A}{0.04509}{GW190701A}{0.01118}{GW190630A}{0.03165}{GW190620A}{0.02437}{GW190602A}{0.08028}{GW190521B}{0.02216}{GW190519A}{0.07568}{GW190517A}{0.01147}{GW190513A}{0.00147}{GW190512A}{0.01656}{GW190426A}{0.05666}{GW190424A}{0.00369}{GW190421A}{0.02467}{GW190413A}{0.01108}{GW190412A}{0.31528}{GW190408A}{0.01123}}}
\newcommand{\JSIMRPSEOBHMchieff}[1]{\IfEqCase{#1}{{GW190930A}{0.04042}{GW190929A}{0.02615}{GW190924A}{0.10517}{GW190915A}{0.00378}{GW190910A}{0.01641}{GW190909A}{0.03476}{GW190828B}{0.06671}{GW190828A}{0.05445}{GW190728A}{0.05322}{GW190720A}{0.01354}{GW190719A}{0.01203}{GW190708A}{0.02217}{GW190707A}{0.08612}{GW190706A}{0.06447}{GW190701A}{0.00544}{GW190630A}{0.04394}{GW190620A}{0.02724}{GW190602A}{0.00354}{GW190521B}{0.14778}{GW190519A}{0.03108}{GW190517A}{0.01771}{GW190513A}{0.05517}{GW190512A}{0.02032}{GW190426A}{0.0526}{GW190424A}{0.00619}{GW190421A}{0.02456}{GW190413A}{0.01512}{GW190412A}{0.27352}{GW190408A}{0.00996}}}
\newcommand{\JSIMRPSEOBHMluminositydistance}[1]{\IfEqCase{#1}{{GW190930A}{0.00073}{GW190929A}{0.2227}{GW190924A}{0.01181}{GW190915A}{0.00631}{GW190910A}{0.04268}{GW190909A}{0.00209}{GW190828B}{0.00923}{GW190828A}{0.02225}{GW190728A}{0.00207}{GW190720A}{0.09198}{GW190719A}{0.01641}{GW190708A}{0.00492}{GW190707A}{0.03824}{GW190706A}{0.01716}{GW190701A}{0.02973}{GW190630A}{0.00324}{GW190620A}{0.00736}{GW190602A}{0.03954}{GW190521B}{0.05043}{GW190519A}{0.14486}{GW190517A}{0.00103}{GW190513A}{0.01651}{GW190512A}{0.00513}{GW190426A}{0.00879}{GW190424A}{0.01351}{GW190421A}{0.05611}{GW190413A}{0.01694}{GW190412A}{0.20105}{GW190408A}{0.03676}}}
\newcommand{\JSSEOBSEOBHMchirpmassdet}[1]{\IfEqCase{#1}{{GW190930A}{0.00353}{GW190929A}{0.04189}{GW190924A}{0.00347}{GW190915A}{0.01274}{GW190910A}{0.00086}{GW190909A}{0.03961}{GW190828B}{0.00447}{GW190828A}{0.0037}{GW190728A}{0.00777}{GW190720A}{0.00573}{GW190719A}{0.01636}{GW190708A}{0.0101}{GW190707A}{0.00674}{GW190706A}{0.02897}{GW190701A}{0.00594}{GW190630A}{0.02757}{GW190620A}{0.01522}{GW190602A}{0.03991}{GW190521B}{0.00262}{GW190519A}{0.02297}{GW190517A}{0.01177}{GW190513A}{0.0045}{GW190512A}{0.03673}{GW190424A}{0.00839}{GW190421A}{0.00565}{GW190413A}{0.01369}{GW190412A}{0.09829}{GW190408A}{0.03074}}}
\newcommand{\JSSEOBSEOBHMmassratio}[1]{\IfEqCase{#1}{{GW190930A}{0.00602}{GW190929A}{0.03561}{GW190924A}{0.00293}{GW190915A}{0.01204}{GW190910A}{0.01106}{GW190909A}{0.00367}{GW190828B}{0.01549}{GW190828A}{0.01326}{GW190728A}{0.00347}{GW190720A}{0.00311}{GW190719A}{0.02346}{GW190708A}{0.01738}{GW190707A}{0.00306}{GW190706A}{0.03107}{GW190701A}{0.01525}{GW190630A}{0.03218}{GW190620A}{0.02478}{GW190602A}{0.05763}{GW190521B}{0.01437}{GW190519A}{0.09308}{GW190517A}{0.03705}{GW190513A}{0.00196}{GW190512A}{0.04189}{GW190424A}{0.01399}{GW190421A}{0.02166}{GW190413A}{0.00382}{GW190412A}{0.0983}{GW190408A}{0.03784}}}
\newcommand{\JSSEOBSEOBHMchieff}[1]{\IfEqCase{#1}{{GW190930A}{0.00594}{GW190929A}{0.03266}{GW190924A}{0.0045}{GW190915A}{0.00685}{GW190910A}{0.00134}{GW190909A}{0.01236}{GW190828B}{0.01735}{GW190828A}{0.00211}{GW190728A}{0.00548}{GW190720A}{0.00479}{GW190719A}{0.01315}{GW190708A}{0.02179}{GW190707A}{0.00805}{GW190706A}{0.08713}{GW190701A}{0.00121}{GW190630A}{0.00574}{GW190620A}{0.00903}{GW190602A}{0.00576}{GW190521B}{0.00633}{GW190519A}{0.03814}{GW190517A}{0.01492}{GW190513A}{0.00183}{GW190512A}{0.01108}{GW190424A}{0.00309}{GW190421A}{0.00241}{GW190413A}{0.00924}{GW190412A}{0.04268}{GW190408A}{0.00989}}}
\newcommand{\JSSEOBSEOBHMluminositydistance}[1]{\IfEqCase{#1}{{GW190930A}{0.00073}{GW190929A}{0.06644}{GW190924A}{0.00105}{GW190915A}{0.00271}{GW190910A}{0.02829}{GW190909A}{0.00344}{GW190828B}{0.00764}{GW190828A}{0.00419}{GW190728A}{0.00033}{GW190720A}{0.00135}{GW190719A}{0.00454}{GW190708A}{0.00117}{GW190707A}{0.00121}{GW190706A}{0.01739}{GW190701A}{0.01945}{GW190630A}{0.00335}{GW190620A}{0.00983}{GW190602A}{0.02696}{GW190521B}{0.00219}{GW190519A}{0.07861}{GW190517A}{0.00557}{GW190513A}{0.02401}{GW190512A}{0.02541}{GW190424A}{0.00293}{GW190421A}{0.02144}{GW190413A}{0.0196}{GW190412A}{0.15309}{GW190408A}{0.02663}}}
\newcommand{\JSIMRPNRsurchirpmassdet}[1]{\IfEqCase{#1}{{GW190828A}{0.0113}{GW190803A}{0.00162}{GW190731A}{0.00349}{GW190727A}{0.00481}{GW190719A}{0.04791}{GW190706A}{0.00838}{GW190701A}{0.00566}{GW190630A}{0.05289}{GW190620A}{0.01063}{GW190602A}{0.0638}{GW190527A}{0.02057}{GW190521B}{0.01957}{GW190519A}{0.05361}{GW190517A}{0.03336}{GW190514A}{0.00543}{GW190513A}{0.07092}{GW190503A}{0.00319}{GW190424A}{0.00835}{GW190421A}{0.00799}{GW190413B}{0.01209}{GW190413A}{0.00411}}}
\newcommand{\JSIMRPNRsurmassratio}[1]{\IfEqCase{#1}{{GW190828A}{0.01495}{GW190803A}{0.00311}{GW190731A}{0.00421}{GW190727A}{0.01642}{GW190719A}{0.03918}{GW190706A}{0.01516}{GW190701A}{0.01069}{GW190630A}{0.044}{GW190620A}{0.00892}{GW190602A}{0.10078}{GW190527A}{0.04715}{GW190521B}{0.02625}{GW190519A}{0.03914}{GW190517A}{0.00974}{GW190514A}{0.00283}{GW190513A}{0.026}{GW190503A}{0.01}{GW190424A}{0.01735}{GW190421A}{0.02714}{GW190413B}{0.01569}{GW190413A}{0.0069}}}
\newcommand{\JSIMRPNRsurchieff}[1]{\IfEqCase{#1}{{GW190828A}{0.00709}{GW190803A}{0.00203}{GW190731A}{0.00194}{GW190727A}{0.00193}{GW190719A}{0.00278}{GW190706A}{0.04912}{GW190701A}{0.00148}{GW190630A}{0.02515}{GW190620A}{0.00336}{GW190602A}{0.00513}{GW190527A}{0.00551}{GW190521B}{0.03471}{GW190519A}{0.06735}{GW190517A}{0.04309}{GW190514A}{0.00897}{GW190513A}{0.04479}{GW190503A}{0.00289}{GW190424A}{0.00893}{GW190421A}{0.00201}{GW190413B}{0.00893}{GW190413A}{0.00407}}}
\newcommand{\JSIMRPNRsurluminositydistance}[1]{\IfEqCase{#1}{{GW190828A}{0.00724}{GW190803A}{0.01529}{GW190731A}{0.00322}{GW190727A}{0.07302}{GW190719A}{0.04451}{GW190706A}{0.00794}{GW190701A}{0.0242}{GW190630A}{0.00968}{GW190620A}{0.01123}{GW190602A}{0.05142}{GW190527A}{0.01143}{GW190521B}{0.01797}{GW190519A}{0.12155}{GW190517A}{0.00605}{GW190514A}{0.0157}{GW190513A}{0.03916}{GW190503A}{0.00343}{GW190424A}{0.02257}{GW190421A}{0.02792}{GW190413B}{0.04378}{GW190413A}{0.04138}}}
\newcommand{\JSNRsurSEOBchirpmassdet}[1]{\IfEqCase{#1}{{GW190828A}{0.03378}{GW190803A}{0.00828}{GW190731A}{0.0485}{GW190727A}{0.00687}{GW190719A}{0.04061}{GW190706A}{0.03191}{GW190701A}{0.01759}{GW190630A}{0.05794}{GW190620A}{0.03977}{GW190602A}{0.05433}{GW190527A}{0.02034}{GW190521B}{0.01938}{GW190519A}{0.07669}{GW190517A}{0.06446}{GW190514A}{0.01326}{GW190513A}{0.0119}{GW190503A}{0.01753}{GW190424A}{0.01126}{GW190421A}{0.01985}{GW190413B}{0.04629}{GW190413A}{0.0072}}}
\newcommand{\JSNRsurSEOBmassratio}[1]{\IfEqCase{#1}{{GW190828A}{0.0151}{GW190803A}{0.00881}{GW190731A}{0.00817}{GW190727A}{0.02433}{GW190719A}{0.05027}{GW190706A}{0.01889}{GW190701A}{0.01496}{GW190630A}{0.0441}{GW190620A}{0.01553}{GW190602A}{0.07441}{GW190527A}{0.02791}{GW190521B}{0.02064}{GW190519A}{0.04944}{GW190517A}{0.01783}{GW190514A}{0.00453}{GW190513A}{0.01265}{GW190503A}{0.01526}{GW190424A}{0.03543}{GW190421A}{0.024}{GW190413B}{0.01704}{GW190413A}{0.0036}}}
\newcommand{\JSNRsurSEOBchieff}[1]{\IfEqCase{#1}{{GW190828A}{0.02831}{GW190803A}{0.01269}{GW190731A}{0.04422}{GW190727A}{0.00449}{GW190719A}{0.04807}{GW190706A}{0.06873}{GW190701A}{0.01363}{GW190630A}{0.02142}{GW190620A}{0.00583}{GW190602A}{0.01466}{GW190527A}{0.01428}{GW190521B}{0.03045}{GW190519A}{0.07866}{GW190517A}{0.0292}{GW190514A}{0.01648}{GW190513A}{0.00422}{GW190503A}{0.0056}{GW190424A}{0.01005}{GW190421A}{0.02734}{GW190413B}{0.0135}{GW190413A}{0.00376}}}
\newcommand{\JSNRsurSEOBluminositydistance}[1]{\IfEqCase{#1}{{GW190828A}{0.00345}{GW190803A}{0.00946}{GW190731A}{0.00101}{GW190727A}{0.06445}{GW190719A}{0.02073}{GW190706A}{0.00559}{GW190701A}{0.01436}{GW190630A}{0.01128}{GW190620A}{0.01493}{GW190602A}{0.04553}{GW190527A}{0.00915}{GW190521B}{0.00895}{GW190519A}{0.07267}{GW190517A}{0.01171}{GW190514A}{0.00998}{GW190513A}{0.03686}{GW190503A}{0.00299}{GW190424A}{0.00803}{GW190421A}{0.00571}{GW190413B}{0.07556}{GW190413A}{0.0442}}}
\newcommand{\JSNRsurSEOBHMchirpmassdet}[1]{\IfEqCase{#1}{{GW190828A}{0.05119}{GW190719A}{0.01294}{GW190706A}{0.01869}{GW190701A}{0.0157}{GW190630A}{0.01151}{GW190620A}{0.05162}{GW190602A}{0.00276}{GW190521B}{0.03145}{GW190521A}{0.04675}{GW190519A}{0.02084}{GW190517A}{0.07526}{GW190513A}{0.01605}{GW190424A}{0.0148}{GW190421A}{0.02093}{GW190413A}{0.00365}}}
\newcommand{\JSNRsurSEOBHMmassratio}[1]{\IfEqCase{#1}{{GW190828A}{0.00107}{GW190719A}{0.01808}{GW190706A}{0.02115}{GW190701A}{0.00041}{GW190630A}{0.00216}{GW190620A}{0.01003}{GW190602A}{0.00245}{GW190521B}{0.00157}{GW190521A}{0.02361}{GW190519A}{0.00928}{GW190517A}{0.00538}{GW190513A}{0.01802}{GW190424A}{0.00685}{GW190421A}{0.00095}{GW190413A}{0.00263}}}
\newcommand{\JSNRsurSEOBHMchieff}[1]{\IfEqCase{#1}{{GW190828A}{0.0376}{GW190719A}{0.01439}{GW190706A}{0.00253}{GW190701A}{0.01005}{GW190630A}{0.01263}{GW190620A}{0.0242}{GW190602A}{0.00412}{GW190521B}{0.06188}{GW190521A}{0.00819}{GW190519A}{0.01026}{GW190517A}{0.04777}{GW190513A}{0.00276}{GW190424A}{0.01461}{GW190421A}{0.01964}{GW190413A}{0.00446}}}
\newcommand{\JSNRsurSEOBHMluminositydistance}[1]{\IfEqCase{#1}{{GW190828A}{0.01161}{GW190719A}{0.0099}{GW190706A}{0.03756}{GW190701A}{0.00608}{GW190630A}{0.00359}{GW190620A}{0.00696}{GW190602A}{0.01089}{GW190521B}{0.01458}{GW190521A}{0.10292}{GW190519A}{0.03121}{GW190517A}{0.00524}{GW190513A}{0.00985}{GW190424A}{0.00272}{GW190421A}{0.00671}{GW190413A}{0.00728}}}

%% file: localization_macros.tex
\newcommand{\skyarea}[1]{\IfEqCase{#1}{{GW190930A}{1700}{GW190929A}{2200}{GW190924A}{360}{GW190915A}{400}{GW190910A}{11000}{GW190909A}{4700}{GW190828B}{660}{GW190828A}{520}{GW190814A}{19}{GW190803A}{1500}{GW190731A}{3400}{GW190728A}{400}{GW190727A}{830}{GW190720A}{460}{GW190719A}{2900}{GW190708A}{14000}{GW190707A}{1300}{GW190706A}{650}{GW190701A}{46}{GW190630A}{1200}{GW190620A}{7200}{GW190602A}{690}{GW190527A}{3700}{GW190521B}{550}{GW190521A}{1000}{GW190519A}{860}{GW190517A}{470}{GW190514A}{3000}{GW190513A}{520}{GW190512A}{220}{GW190503A}{94}{GW190426A}{1300}{GW190425A}{10000}{GW190424A}{28000}{GW190421A}{1200}{GW190413B}{730}{GW190413A}{1500}{GW190412A}{21}{GW190408A}{150}}}
\newcommand{\skyvol}[1]{\IfEqCase{#1}{{GW190930A}{0.15}{GW190929A}{8.7}{GW190924A}{0.012}{GW190915A}{0.19}{GW190910A}{5.1}{GW190909A}{17}{GW190828B}{0.32}{GW190828A}{0.42}{GW190814A}{3.2\times 10^{-5}}{GW190803A}{3.7}{GW190731A}{8.9}{GW190728A}{0.036}{GW190727A}{1.8}{GW190720A}{0.083}{GW190719A}{9.3}{GW190708A}{1.2}{GW190707A}{0.096}{GW190706A}{3.1}{GW190701A}{0.035}{GW190630A}{0.17}{GW190620A}{13}{GW190602A}{1.2}{GW190527A}{8.9}{GW190521B}{0.11}{GW190521A}{3.7}{GW190519A}{1.4}{GW190517A}{0.51}{GW190514A}{11}{GW190513A}{0.44}{GW190512A}{0.082}{GW190503A}{0.041}{GW190426A}{0.019}{GW190425A}{0.009}{GW190424A}{28}{GW190421A}{1.9}{GW190413B}{3.2}{GW190413A}{4.9}{GW190412A}{0.00084}{GW190408A}{0.054}}}
\newcommand{\minareaevent}{GW190814A}
\newcommand{\maxareaevent}{GW190424A}
\newcommand{\minvolevent}{GW190814A}
\newcommand{\maxvolevent}{GW190424A}

%% file: BF_macros.tex
\newcommand{\lnBFprecession}[1]{\IfEqCase{#1}{{GW190929A}{0.29}{GW190909A}{0.33}{GW190828B}{0.82}{GW190828A}{-0.2}{GW190719A}{-0.17}{GW190706A}{-0.47}{GW190701A}{-0.06}{GW190630A}{0.11}{GW190620A}{-0.28}{GW190602A}{-0.05}{GW190521B}{-0.12}{GW190519A}{0.12}{GW190517A}{-0.05}{GW190513A}{-0.57}}}

%% file: IMRPv2_macros.tex
\newcommand{\costhetajnIMRminus}[1]{\IfEqCase{#1}{{GW190930A}{1.54}{GW190929A}{0.71}{GW190924A}{1.74}{GW190915A}{0.69}{GW190910A}{0.88}{GW190909A}{1.03}{GW190828B}{0.84}{GW190828A}{0.31}{GW190814A}{1.42}{GW190803A}{1.18}{GW190731A}{1.10}{GW190728A}{1.32}{GW190727A}{1.42}{GW190720A}{0.14}{GW190719A}{0.94}{GW190708A}{1.15}{GW190707A}{0.39}{GW190706A}{1.08}{GW190701A}{0.44}{GW190630A}{1.47}{GW190620A}{0.43}{GW190602A}{0.69}{GW190527A}{1.30}{GW190521B}{1.47}{GW190521A}{1.57}{GW190519A}{0.82}{GW190517A}{0.34}{GW190514A}{0.96}{GW190513A}{1.67}{GW190512A}{0.96}{GW190503A}{0.19}{GW190426A}{0.84}{GW190425A}{1.43}{GW190424A}{1.00}{GW190421A}{0.84}{GW190413B}{0.81}{GW190413A}{1.61}{GW190412A}{1.25}{GW190408A}{0.96}}}
\newcommand{\costhetajnIMRmed}[1]{\IfEqCase{#1}{{GW190930A}{0.58}{GW190929A}{-0.16}{GW190924A}{0.82}{GW190915A}{-0.22}{GW190910A}{-0.10}{GW190909A}{0.09}{GW190828B}{-0.13}{GW190828A}{-0.68}{GW190814A}{0.68}{GW190803A}{0.23}{GW190731A}{0.14}{GW190728A}{0.35}{GW190727A}{0.48}{GW190720A}{-0.85}{GW190719A}{-0.02}{GW190708A}{0.18}{GW190707A}{-0.59}{GW190706A}{0.13}{GW190701A}{0.77}{GW190630A}{0.51}{GW190620A}{-0.54}{GW190602A}{-0.27}{GW190527A}{0.37}{GW190521B}{0.53}{GW190521A}{0.62}{GW190519A}{-0.13}{GW190517A}{-0.62}{GW190514A}{0.01}{GW190513A}{0.75}{GW190512A}{-0.01}{GW190503A}{-0.80}{GW190426A}{-0.13}{GW190425A}{0.47}{GW190424A}{0.04}{GW190421A}{-0.11}{GW190413B}{-0.13}{GW190413A}{0.67}{GW190412A}{0.74}{GW190408A}{0.00}}}
\newcommand{\costhetajnIMRplus}[1]{\IfEqCase{#1}{{GW190930A}{0.40}{GW190929A}{0.97}{GW190924A}{0.17}{GW190915A}{1.09}{GW190910A}{1.06}{GW190909A}{0.86}{GW190828B}{1.10}{GW190828A}{1.63}{GW190814A}{0.15}{GW190803A}{0.72}{GW190731A}{0.82}{GW190728A}{0.63}{GW190727A}{0.49}{GW190720A}{1.63}{GW190719A}{0.97}{GW190708A}{0.80}{GW190707A}{1.56}{GW190706A}{0.82}{GW190701A}{0.21}{GW190630A}{0.47}{GW190620A}{1.47}{GW190602A}{1.22}{GW190527A}{0.58}{GW190521B}{0.43}{GW190521A}{0.35}{GW190519A}{1.07}{GW190517A}{1.36}{GW190514A}{0.94}{GW190513A}{0.23}{GW190512A}{0.98}{GW190503A}{0.52}{GW190426A}{1.09}{GW190425A}{0.50}{GW190424A}{0.91}{GW190421A}{1.06}{GW190413B}{1.05}{GW190413A}{0.30}{GW190412A}{0.16}{GW190408A}{0.97}}}
\newcommand{\loglikelihoodIMRminus}[1]{\IfEqCase{#1}{{GW190930A}{4.7}{GW190929A}{7.5}{GW190924A}{5.3}{GW190915A}{4.7}{GW190910A}{4.1}{GW190909A}{5.2}{GW190828B}{4.7}{GW190828A}{4.6}{GW190814A}{4.9}{GW190803A}{4.8}{GW190731A}{4.2}{GW190728A}{5.2}{GW190727A}{5.7}{GW190720A}{7.8}{GW190719A}{6.3}{GW190708A}{4.6}{GW190707A}{5.0}{GW190706A}{5.0}{GW190701A}{4.1}{GW190630A}{5.1}{GW190620A}{4.9}{GW190602A}{4.4}{GW190527A}{7.2}{GW190521B}{4.4}{GW190521A}{5.0}{GW190519A}{4.5}{GW190517A}{6.2}{GW190514A}{4.7}{GW190513A}{5.3}{GW190512A}{4.8}{GW190503A}{4.4}{GW190426A}{5.6}{GW190425A}{5.7}{GW190424A}{4.2}{GW190421A}{4.4}{GW190413B}{5.1}{GW190413A}{6.5}{GW190412A}{6.7}{GW190408A}{4.5}}}
\newcommand{\loglikelihoodIMRmed}[1]{\IfEqCase{#1}{{GW190930A}{-15939.8}{GW190929A}{-11966.5}{GW190924A}{-97036.5}{GW190915A}{-2814.1}{GW190910A}{-7980.5}{GW190909A}{-7987.2}{GW190828B}{-23976.5}{GW190828A}{-11929.3}{GW190814A}{298.0}{GW190803A}{-11706.7}{GW190731A}{-7957.8}{GW190728A}{-47939.6}{GW190727A}{-2700.5}{GW190720A}{-23907.6}{GW190719A}{-3904.4}{GW190708A}{-15917.4}{GW190707A}{-15886.3}{GW190706A}{-2802.0}{GW190701A}{-2790.2}{GW190630A}{-7866.2}{GW190620A}{-3893.6}{GW190602A}{-2793.2}{GW190527A}{-15670.8}{GW190521B}{-1871.5}{GW190521A}{-11921.7}{GW190519A}{-2814.6}{GW190517A}{-5836.2}{GW190514A}{-7948.9}{GW190513A}{-5849.7}{GW190512A}{-11692.6}{GW190503A}{-2783.0}{GW190426A}{-389547.0}{GW190425A}{-500483.9}{GW190424A}{-1961.6}{GW190421A}{-1851.9}{GW190413B}{-11957.9}{GW190413A}{-2782.0}{GW190412A}{-22832.7}{GW190408A}{-5836.3}}}
\newcommand{\loglikelihoodIMRplus}[1]{\IfEqCase{#1}{{GW190930A}{2.9}{GW190929A}{6.2}{GW190924A}{3.2}{GW190915A}{3.6}{GW190910A}{2.7}{GW190909A}{3.2}{GW190828B}{3.1}{GW190828A}{3.4}{GW190814A}{3.1}{GW190803A}{2.8}{GW190731A}{2.3}{GW190728A}{3.1}{GW190727A}{3.6}{GW190720A}{3.4}{GW190719A}{2.9}{GW190708A}{3.2}{GW190707A}{3.0}{GW190706A}{3.2}{GW190701A}{2.4}{GW190630A}{2.9}{GW190620A}{3.2}{GW190602A}{2.9}{GW190527A}{3.0}{GW190521B}{3.1}{GW190521A}{4.5}{GW190519A}{2.9}{GW190517A}{4.1}{GW190514A}{2.7}{GW190513A}{3.9}{GW190512A}{3.2}{GW190503A}{2.5}{GW190426A}{4.6}{GW190425A}{4.5}{GW190424A}{2.7}{GW190421A}{2.6}{GW190413B}{3.8}{GW190413A}{3.9}{GW190412A}{3.8}{GW190408A}{3.4}}}
\newcommand{\logpriorIMRminus}[1]{\IfEqCase{#1}{{GW190930A}{8.4}{GW190929A}{10.1}{GW190924A}{10.1}{GW190915A}{10.1}{GW190910A}{8.4}{GW190909A}{8.3}{GW190828B}{10.1}{GW190828A}{10.1}{GW190803A}{10.2}{GW190731A}{8.3}{GW190728A}{10.0}{GW190727A}{9.9}{GW190720A}{10.3}{GW190719A}{8.3}{GW190708A}{8.5}{GW190707A}{8.6}{GW190706A}{9.7}{GW190701A}{9.9}{GW190630A}{8.6}{GW190620A}{8.4}{GW190602A}{10.0}{GW190527A}{8.7}{GW190521B}{8.7}{GW190521A}{10.2}{GW190519A}{10.1}{GW190517A}{10.1}{GW190514A}{8.5}{GW190513A}{10.1}{GW190512A}{9.9}{GW190503A}{10.0}{GW190426A}{10.5}{GW190425A}{8.6}{GW190424A}{6.7}{GW190421A}{8.5}{GW190413B}{10.0}{GW190413A}{10.0}{GW190412A}{10.2}{GW190408A}{10.2}}}
\newcommand{\logpriorIMRmed}[1]{\IfEqCase{#1}{{GW190930A}{139.6}{GW190929A}{178.4}{GW190924A}{170.0}{GW190915A}{169.3}{GW190910A}{117.3}{GW190909A}{145.8}{GW190828B}{179.1}{GW190828A}{179.6}{GW190803A}{179.4}{GW190731A}{147.8}{GW190728A}{175.8}{GW190727A}{173.7}{GW190720A}{175.8}{GW190719A}{145.3}{GW190708A}{114.0}{GW190707A}{141.9}{GW190706A}{174.5}{GW190701A}{171.7}{GW190630A}{113.8}{GW190620A}{116.8}{GW190602A}{166.2}{GW190527A}{137.5}{GW190521B}{128.4}{GW190521A}{170.4}{GW190519A}{167.0}{GW190517A}{170.4}{GW190514A}{139.5}{GW190513A}{169.5}{GW190512A}{169.9}{GW190503A}{164.6}{GW190426A}{161.3}{GW190425A}{98.4}{GW190424A}{76.9}{GW190421A}{131.7}{GW190413B}{174.1}{GW190413A}{167.1}{GW190412A}{168.7}{GW190408A}{170.1}}}
\newcommand{\logpriorIMRplus}[1]{\IfEqCase{#1}{{GW190930A}{6.6}{GW190929A}{8.4}{GW190924A}{8.2}{GW190915A}{8.1}{GW190910A}{6.7}{GW190909A}{6.9}{GW190828B}{8.3}{GW190828A}{8.3}{GW190803A}{8.3}{GW190731A}{6.7}{GW190728A}{8.3}{GW190727A}{8.1}{GW190720A}{8.4}{GW190719A}{6.9}{GW190708A}{6.6}{GW190707A}{6.7}{GW190706A}{8.2}{GW190701A}{8.2}{GW190630A}{6.8}{GW190620A}{6.8}{GW190602A}{8.3}{GW190527A}{6.8}{GW190521B}{6.7}{GW190521A}{8.3}{GW190519A}{8.2}{GW190517A}{8.3}{GW190514A}{6.8}{GW190513A}{8.2}{GW190512A}{8.2}{GW190503A}{8.3}{GW190426A}{8.6}{GW190425A}{6.7}{GW190424A}{4.9}{GW190421A}{6.5}{GW190413B}{8.5}{GW190413A}{8.4}{GW190412A}{8.2}{GW190408A}{8.3}}}
\newcommand{\networkmatchedfiltersnrIMRminus}[1]{\IfEqCase{#1}{{GW190930A}{0.5}{GW190929A}{0.8}{GW190924A}{0.4}{GW190915A}{0.3}{GW190910A}{0.3}{GW190909A}{0.6}{GW190828B}{0.5}{GW190828A}{0.3}{GW190814A}{0.2}{GW190803A}{0.5}{GW190731A}{0.5}{GW190728A}{0.4}{GW190727A}{0.5}{GW190720A}{0.7}{GW190719A}{0.8}{GW190708A}{0.3}{GW190707A}{0.4}{GW190706A}{0.4}{GW190701A}{0.3}{GW190630A}{0.3}{GW190620A}{0.4}{GW190602A}{0.3}{GW190527A}{0.9}{GW190521B}{0.2}{GW190521A}{0.3}{GW190519A}{0.3}{GW190517A}{0.6}{GW190514A}{0.6}{GW190513A}{0.4}{GW190512A}{0.4}{GW190503A}{0.3}{GW190426A}{0.6}{GW190425A}{0.4}{GW190424A}{0.4}{GW190421A}{0.4}{GW190413B}{0.5}{GW190413A}{0.7}{GW190412A}{0.3}{GW190408A}{0.3}}}
\newcommand{\networkmatchedfiltersnrIMRmed}[1]{\IfEqCase{#1}{{GW190930A}{9.5}{GW190929A}{10.1}{GW190924A}{11.5}{GW190915A}{13.6}{GW190910A}{14.1}{GW190909A}{8.1}{GW190828B}{10.0}{GW190828A}{16.2}{GW190814A}{24.9}{GW190803A}{8.6}{GW190731A}{8.7}{GW190728A}{13.0}{GW190727A}{11.9}{GW190720A}{11.0}{GW190719A}{8.3}{GW190708A}{13.1}{GW190707A}{13.3}{GW190706A}{12.6}{GW190701A}{11.3}{GW190630A}{15.6}{GW190620A}{12.1}{GW190602A}{12.8}{GW190527A}{8.1}{GW190521B}{25.8}{GW190521A}{14.2}{GW190519A}{15.6}{GW190517A}{10.7}{GW190514A}{8.2}{GW190513A}{12.9}{GW190512A}{12.2}{GW190503A}{12.4}{GW190426A}{8.7}{GW190425A}{12.4}{GW190424A}{10.4}{GW190421A}{10.7}{GW190413B}{10.0}{GW190413A}{8.9}{GW190412A}{18.9}{GW190408A}{15.3}}}
\newcommand{\networkmatchedfiltersnrIMRplus}[1]{\IfEqCase{#1}{{GW190930A}{0.3}{GW190929A}{0.6}{GW190924A}{0.3}{GW190915A}{0.2}{GW190910A}{0.2}{GW190909A}{0.4}{GW190828B}{0.3}{GW190828A}{0.2}{GW190814A}{0.1}{GW190803A}{0.3}{GW190731A}{0.2}{GW190728A}{0.2}{GW190727A}{0.3}{GW190720A}{0.3}{GW190719A}{0.3}{GW190708A}{0.2}{GW190707A}{0.2}{GW190706A}{0.2}{GW190701A}{0.2}{GW190630A}{0.2}{GW190620A}{0.3}{GW190602A}{0.2}{GW190527A}{0.3}{GW190521B}{0.1}{GW190521A}{0.3}{GW190519A}{0.2}{GW190517A}{0.4}{GW190514A}{0.3}{GW190513A}{0.3}{GW190512A}{0.2}{GW190503A}{0.2}{GW190426A}{0.5}{GW190425A}{0.3}{GW190424A}{0.2}{GW190421A}{0.2}{GW190413B}{0.4}{GW190413A}{0.4}{GW190412A}{0.2}{GW190408A}{0.2}}}
\newcommand{\networkoptimalsnrIMRminus}[1]{\IfEqCase{#1}{{GW190930A}{1.7}{GW190929A}{1.8}{GW190924A}{1.7}{GW190915A}{1.7}{GW190910A}{1.7}{GW190909A}{1.8}{GW190828B}{1.7}{GW190828A}{1.7}{GW190814A}{1.7}{GW190803A}{1.8}{GW190731A}{1.8}{GW190728A}{1.7}{GW190727A}{1.7}{GW190720A}{1.8}{GW190719A}{1.8}{GW190708A}{1.7}{GW190707A}{1.7}{GW190706A}{1.7}{GW190701A}{1.7}{GW190630A}{1.7}{GW190620A}{1.7}{GW190602A}{1.7}{GW190527A}{1.8}{GW190521B}{1.6}{GW190521A}{1.7}{GW190519A}{1.6}{GW190517A}{1.7}{GW190514A}{1.8}{GW190513A}{1.7}{GW190512A}{1.7}{GW190503A}{1.7}{GW190426A}{1.8}{GW190425A}{1.7}{GW190424A}{1.7}{GW190421A}{1.7}{GW190413B}{1.7}{GW190413A}{1.8}{GW190412A}{1.7}{GW190408A}{1.7}}}
\newcommand{\networkoptimalsnrIMRmed}[1]{\IfEqCase{#1}{{GW190930A}{9.1}{GW190929A}{9.7}{GW190924A}{11.1}{GW190915A}{13.3}{GW190910A}{13.9}{GW190909A}{7.8}{GW190828B}{9.6}{GW190828A}{16.0}{GW190814A}{24.7}{GW190803A}{8.3}{GW190731A}{8.3}{GW190728A}{12.7}{GW190727A}{11.6}{GW190720A}{10.6}{GW190719A}{7.9}{GW190708A}{12.8}{GW190707A}{13.0}{GW190706A}{12.4}{GW190701A}{11.0}{GW190630A}{15.4}{GW190620A}{11.9}{GW190602A}{12.6}{GW190527A}{7.7}{GW190521B}{25.6}{GW190521A}{14.0}{GW190519A}{15.4}{GW190517A}{10.4}{GW190514A}{7.8}{GW190513A}{12.7}{GW190512A}{12.0}{GW190503A}{12.1}{GW190426A}{8.3}{GW190425A}{12.0}{GW190424A}{10.1}{GW190421A}{10.4}{GW190413B}{9.7}{GW190413A}{8.5}{GW190412A}{18.7}{GW190408A}{15.1}}}
\newcommand{\networkoptimalsnrIMRplus}[1]{\IfEqCase{#1}{{GW190930A}{1.7}{GW190929A}{1.8}{GW190924A}{1.7}{GW190915A}{1.7}{GW190910A}{1.7}{GW190909A}{1.7}{GW190828B}{1.7}{GW190828A}{1.7}{GW190814A}{1.6}{GW190803A}{1.7}{GW190731A}{1.8}{GW190728A}{1.7}{GW190727A}{1.7}{GW190720A}{1.7}{GW190719A}{1.7}{GW190708A}{1.7}{GW190707A}{1.7}{GW190706A}{1.7}{GW190701A}{1.7}{GW190630A}{1.7}{GW190620A}{1.7}{GW190602A}{1.7}{GW190527A}{1.8}{GW190521B}{1.7}{GW190521A}{1.7}{GW190519A}{1.7}{GW190517A}{1.7}{GW190514A}{1.7}{GW190513A}{1.7}{GW190512A}{1.7}{GW190503A}{1.7}{GW190426A}{1.8}{GW190425A}{1.7}{GW190424A}{1.7}{GW190421A}{1.7}{GW190413B}{1.7}{GW190413A}{1.8}{GW190412A}{1.7}{GW190408A}{1.7}}}
\newcommand{\phaseIMRminus}[1]{\IfEqCase{#1}{{GW190930A}{2.97}{GW190929A}{2.87}{GW190924A}{2.80}{GW190915A}{3.06}{GW190910A}{2.76}{GW190909A}{2.73}{GW190828B}{2.82}{GW190828A}{2.80}{GW190814A}{3.42}{GW190803A}{2.89}{GW190731A}{2.91}{GW190728A}{2.83}{GW190727A}{2.90}{GW190720A}{2.76}{GW190719A}{2.85}{GW190708A}{2.82}{GW190707A}{2.97}{GW190706A}{2.82}{GW190701A}{2.86}{GW190630A}{2.78}{GW190620A}{2.91}{GW190602A}{2.78}{GW190527A}{2.66}{GW190521B}{2.87}{GW190521A}{3.03}{GW190519A}{2.70}{GW190517A}{2.73}{GW190514A}{2.70}{GW190513A}{2.68}{GW190512A}{2.73}{GW190503A}{2.85}{GW190426A}{2.79}{GW190425A}{2.82}{GW190424A}{2.84}{GW190421A}{2.73}{GW190413B}{2.97}{GW190413A}{2.81}{GW190412A}{1.37}{GW190408A}{2.92}}}
\newcommand{\phaseIMRmed}[1]{\IfEqCase{#1}{{GW190930A}{3.29}{GW190929A}{3.16}{GW190924A}{3.10}{GW190915A}{3.44}{GW190910A}{3.02}{GW190909A}{3.05}{GW190828B}{3.15}{GW190828A}{3.13}{GW190814A}{3.61}{GW190803A}{3.19}{GW190731A}{3.28}{GW190728A}{3.14}{GW190727A}{3.23}{GW190720A}{3.07}{GW190719A}{3.20}{GW190708A}{3.09}{GW190707A}{3.33}{GW190706A}{3.09}{GW190701A}{3.18}{GW190630A}{3.06}{GW190620A}{3.21}{GW190602A}{3.09}{GW190527A}{2.97}{GW190521B}{3.23}{GW190521A}{3.37}{GW190519A}{3.02}{GW190517A}{3.03}{GW190514A}{3.01}{GW190513A}{2.96}{GW190512A}{3.03}{GW190503A}{3.16}{GW190426A}{3.09}{GW190425A}{3.12}{GW190424A}{3.12}{GW190421A}{3.03}{GW190413B}{3.29}{GW190413A}{3.14}{GW190412A}{1.62}{GW190408A}{3.26}}}
\newcommand{\phaseIMRplus}[1]{\IfEqCase{#1}{{GW190930A}{2.71}{GW190929A}{2.82}{GW190924A}{2.85}{GW190915A}{2.58}{GW190910A}{2.94}{GW190909A}{2.91}{GW190828B}{2.82}{GW190828A}{2.83}{GW190814A}{2.48}{GW190803A}{2.78}{GW190731A}{2.71}{GW190728A}{2.82}{GW190727A}{2.74}{GW190720A}{2.91}{GW190719A}{2.74}{GW190708A}{2.89}{GW190707A}{2.65}{GW190706A}{2.84}{GW190701A}{2.80}{GW190630A}{2.91}{GW190620A}{2.77}{GW190602A}{2.83}{GW190527A}{2.98}{GW190521B}{2.78}{GW190521A}{2.59}{GW190519A}{2.91}{GW190517A}{2.90}{GW190514A}{2.93}{GW190513A}{3.00}{GW190512A}{2.95}{GW190503A}{2.81}{GW190426A}{2.85}{GW190425A}{2.87}{GW190424A}{2.85}{GW190421A}{2.86}{GW190413B}{2.69}{GW190413A}{2.81}{GW190412A}{4.32}{GW190408A}{2.75}}}
\newcommand{\phionetwoIMRminus}[1]{\IfEqCase{#1}{{GW190930A}{2.78}{GW190929A}{2.83}{GW190924A}{2.84}{GW190915A}{2.82}{GW190910A}{2.84}{GW190909A}{2.80}{GW190828B}{2.71}{GW190828A}{2.74}{GW190814A}{2.79}{GW190803A}{2.73}{GW190731A}{2.85}{GW190728A}{2.84}{GW190727A}{2.76}{GW190720A}{2.60}{GW190719A}{2.85}{GW190708A}{2.83}{GW190707A}{2.85}{GW190706A}{2.86}{GW190701A}{2.82}{GW190630A}{2.67}{GW190620A}{2.91}{GW190602A}{2.84}{GW190527A}{2.96}{GW190521B}{2.74}{GW190521A}{3.06}{GW190519A}{2.78}{GW190517A}{2.81}{GW190514A}{2.74}{GW190513A}{3.07}{GW190512A}{2.83}{GW190503A}{2.95}{GW190426A}{0.00}{GW190425A}{2.87}{GW190424A}{2.70}{GW190421A}{2.92}{GW190413B}{2.85}{GW190413A}{2.71}{GW190412A}{3.11}{GW190408A}{2.80}}}
\newcommand{\phionetwoIMRmed}[1]{\IfEqCase{#1}{{GW190930A}{3.12}{GW190929A}{3.14}{GW190924A}{3.15}{GW190915A}{3.11}{GW190910A}{3.14}{GW190909A}{3.15}{GW190828B}{3.00}{GW190828A}{3.05}{GW190814A}{3.10}{GW190803A}{3.05}{GW190731A}{3.15}{GW190728A}{3.16}{GW190727A}{3.10}{GW190720A}{2.95}{GW190719A}{3.18}{GW190708A}{3.20}{GW190707A}{3.15}{GW190706A}{3.18}{GW190701A}{3.11}{GW190630A}{2.98}{GW190620A}{3.23}{GW190602A}{3.15}{GW190527A}{3.29}{GW190521B}{3.04}{GW190521A}{3.34}{GW190519A}{3.11}{GW190517A}{3.11}{GW190514A}{3.05}{GW190513A}{3.43}{GW190512A}{3.13}{GW190503A}{3.31}{GW190426A}{0.00}{GW190425A}{3.18}{GW190424A}{3.05}{GW190421A}{3.24}{GW190413B}{3.19}{GW190413A}{3.02}{GW190412A}{3.43}{GW190408A}{3.08}}}
\newcommand{\phionetwoIMRplus}[1]{\IfEqCase{#1}{{GW190930A}{2.85}{GW190929A}{2.83}{GW190924A}{2.82}{GW190915A}{2.88}{GW190910A}{2.82}{GW190909A}{2.82}{GW190828B}{2.93}{GW190828A}{2.86}{GW190814A}{2.87}{GW190803A}{2.91}{GW190731A}{2.81}{GW190728A}{2.82}{GW190727A}{2.85}{GW190720A}{2.99}{GW190719A}{2.81}{GW190708A}{2.75}{GW190707A}{2.83}{GW190706A}{2.84}{GW190701A}{2.85}{GW190630A}{3.00}{GW190620A}{2.74}{GW190602A}{2.82}{GW190527A}{2.69}{GW190521B}{2.94}{GW190521A}{2.65}{GW190519A}{2.83}{GW190517A}{2.85}{GW190514A}{2.90}{GW190513A}{2.56}{GW190512A}{2.86}{GW190503A}{2.65}{GW190426A}{0.00}{GW190425A}{2.76}{GW190424A}{2.93}{GW190421A}{2.74}{GW190413B}{2.78}{GW190413A}{2.93}{GW190412A}{2.56}{GW190408A}{2.88}}}
\newcommand{\phijlIMRminus}[1]{\IfEqCase{#1}{{GW190930A}{2.93}{GW190929A}{3.18}{GW190924A}{2.75}{GW190915A}{2.52}{GW190910A}{3.01}{GW190909A}{2.86}{GW190828B}{2.70}{GW190828A}{2.89}{GW190814A}{1.88}{GW190803A}{2.86}{GW190731A}{2.79}{GW190728A}{2.94}{GW190727A}{2.82}{GW190720A}{2.65}{GW190719A}{2.93}{GW190708A}{2.89}{GW190707A}{2.92}{GW190706A}{2.89}{GW190701A}{4.08}{GW190630A}{2.47}{GW190620A}{3.04}{GW190602A}{2.52}{GW190527A}{2.75}{GW190521B}{3.17}{GW190521A}{3.22}{GW190519A}{2.77}{GW190517A}{1.84}{GW190514A}{2.85}{GW190513A}{2.33}{GW190512A}{2.74}{GW190503A}{2.95}{GW190426A}{1.43}{GW190425A}{2.88}{GW190424A}{2.78}{GW190421A}{2.56}{GW190413B}{2.94}{GW190413A}{2.81}{GW190412A}{1.07}{GW190408A}{3.05}}}
\newcommand{\phijlIMRmed}[1]{\IfEqCase{#1}{{GW190930A}{3.26}{GW190929A}{3.42}{GW190924A}{3.06}{GW190915A}{3.39}{GW190910A}{3.33}{GW190909A}{3.26}{GW190828B}{3.03}{GW190828A}{3.16}{GW190814A}{2.29}{GW190803A}{3.19}{GW190731A}{3.17}{GW190728A}{3.28}{GW190727A}{3.15}{GW190720A}{2.95}{GW190719A}{3.22}{GW190708A}{3.19}{GW190707A}{3.24}{GW190706A}{3.31}{GW190701A}{4.47}{GW190630A}{2.70}{GW190620A}{3.30}{GW190602A}{2.97}{GW190527A}{3.02}{GW190521B}{3.54}{GW190521A}{3.58}{GW190519A}{3.02}{GW190517A}{2.26}{GW190514A}{3.17}{GW190513A}{2.88}{GW190512A}{3.06}{GW190503A}{3.13}{GW190426A}{1.70}{GW190425A}{3.23}{GW190424A}{3.11}{GW190421A}{3.02}{GW190413B}{3.22}{GW190413A}{3.13}{GW190412A}{1.15}{GW190408A}{3.33}}}
\newcommand{\phijlIMRplus}[1]{\IfEqCase{#1}{{GW190930A}{2.75}{GW190929A}{2.58}{GW190924A}{2.90}{GW190915A}{2.27}{GW190910A}{2.64}{GW190909A}{2.64}{GW190828B}{2.93}{GW190828A}{2.80}{GW190814A}{3.42}{GW190803A}{2.80}{GW190731A}{2.74}{GW190728A}{2.69}{GW190727A}{2.82}{GW190720A}{3.01}{GW190719A}{2.77}{GW190708A}{2.79}{GW190707A}{2.75}{GW190706A}{2.63}{GW190701A}{1.54}{GW190630A}{3.30}{GW190620A}{2.81}{GW190602A}{2.91}{GW190527A}{3.00}{GW190521B}{2.26}{GW190521A}{2.34}{GW190519A}{2.97}{GW190517A}{3.41}{GW190514A}{2.75}{GW190513A}{3.00}{GW190512A}{2.82}{GW190503A}{2.99}{GW190426A}{1.19}{GW190425A}{2.76}{GW190424A}{2.82}{GW190421A}{2.82}{GW190413B}{2.72}{GW190413A}{2.83}{GW190412A}{5.01}{GW190408A}{2.68}}}
\newcommand{\massratioIMRminus}[1]{\IfEqCase{#1}{{GW190930A}{0.46}{GW190929A}{0.14}{GW190924A}{0.36}{GW190915A}{0.24}{GW190910A}{0.25}{GW190909A}{0.37}{GW190828B}{0.20}{GW190828A}{0.26}{GW190814A}{0.010}{GW190803A}{0.32}{GW190731A}{0.33}{GW190728A}{0.41}{GW190727A}{0.35}{GW190720A}{0.30}{GW190719A}{0.26}{GW190708A}{0.30}{GW190707A}{0.29}{GW190706A}{0.33}{GW190701A}{0.30}{GW190630A}{0.19}{GW190620A}{0.25}{GW190602A}{0.29}{GW190527A}{0.32}{GW190521B}{0.22}{GW190521A}{0.29}{GW190519A}{0.29}{GW190517A}{0.36}{GW190514A}{0.34}{GW190513A}{0.18}{GW190512A}{0.24}{GW190503A}{0.29}{GW190426A}{0.15}{GW190425A}{0.25}{GW190424A}{0.30}{GW190421A}{0.31}{GW190413B}{0.35}{GW190413A}{0.30}{GW190412A}{0.08}{GW190408A}{0.26}}}
\newcommand{\massratioIMRmed}[1]{\IfEqCase{#1}{{GW190930A}{0.61}{GW190929A}{0.27}{GW190924A}{0.51}{GW190915A}{0.65}{GW190910A}{0.80}{GW190909A}{0.66}{GW190828B}{0.47}{GW190828A}{0.80}{GW190814A}{0.111}{GW190803A}{0.73}{GW190731A}{0.71}{GW190728A}{0.66}{GW190727A}{0.74}{GW190720A}{0.57}{GW190719A}{0.48}{GW190708A}{0.70}{GW190707A}{0.72}{GW190706A}{0.62}{GW190701A}{0.71}{GW190630A}{0.60}{GW190620A}{0.56}{GW190602A}{0.54}{GW190527A}{0.53}{GW190521B}{0.74}{GW190521A}{0.72}{GW190519A}{0.64}{GW190517A}{0.67}{GW190514A}{0.75}{GW190513A}{0.48}{GW190512A}{0.57}{GW190503A}{0.67}{GW190426A}{0.25}{GW190425A}{0.67}{GW190424A}{0.77}{GW190421A}{0.72}{GW190413B}{0.64}{GW190413A}{0.74}{GW190412A}{0.31}{GW190408A}{0.73}}}
\newcommand{\massratioIMRplus}[1]{\IfEqCase{#1}{{GW190930A}{0.34}{GW190929A}{0.48}{GW190924A}{0.42}{GW190915A}{0.30}{GW190910A}{0.18}{GW190909A}{0.30}{GW190828B}{0.41}{GW190828A}{0.18}{GW190814A}{0.009}{GW190803A}{0.24}{GW190731A}{0.26}{GW190728A}{0.31}{GW190727A}{0.23}{GW190720A}{0.39}{GW190719A}{0.44}{GW190708A}{0.26}{GW190707A}{0.25}{GW190706A}{0.34}{GW190701A}{0.25}{GW190630A}{0.34}{GW190620A}{0.37}{GW190602A}{0.40}{GW190527A}{0.40}{GW190521B}{0.23}{GW190521A}{0.25}{GW190519A}{0.32}{GW190517A}{0.29}{GW190514A}{0.22}{GW190513A}{0.43}{GW190512A}{0.37}{GW190503A}{0.29}{GW190426A}{0.41}{GW190425A}{0.29}{GW190424A}{0.20}{GW190421A}{0.25}{GW190413B}{0.32}{GW190413A}{0.23}{GW190412A}{0.12}{GW190408A}{0.24}}}
\newcommand{\geocenttimeIMRminus}[1]{\IfEqCase{#1}{{GW190930A}{0.009}{GW190929A}{0.04}{GW190924A}{0.004}{GW190915A}{0.004}{GW190910A}{0.04}{GW190909A}{0.03}{GW190828B}{0.002}{GW190828A}{0.002}{GW190814A}{0.003}{GW190803A}{0.009}{GW190731A}{0.009}{GW190728A}{0.04}{GW190727A}{0.004}{GW190720A}{0.02}{GW190719A}{0.03}{GW190708A}{0.04}{GW190707A}{0.01}{GW190706A}{0.04}{GW190701A}{0.003}{GW190630A}{0.005}{GW190620A}{0.04}{GW190602A}{0.006}{GW190527A}{0.02}{GW190521B}{0.007}{GW190521A}{0.01}{GW190519A}{0.04}{GW190517A}{0.02}{GW190514A}{0.04}{GW190513A}{0.002}{GW190512A}{0.001}{GW190503A}{0.002}{GW190426A}{0.03}{GW190425A}{0.009}{GW190424A}{0.02}{GW190421A}{0.003}{GW190413B}{0.004}{GW190413A}{0.006}{GW190412A}{0.005}{GW190408A}{0.004}}}
\newcommand{\geocenttimeIMRmed}[1]{\IfEqCase{#1}{{GW190930A}{1253885759.2}{GW190929A}{1253755327.5}{GW190924A}{1253326744.8}{GW190915A}{1252627040.7}{GW190910A}{1252150105.3}{GW190909A}{1252064527.7}{GW190828B}{1251010527.9}{GW190828A}{1251009263.7}{GW190814A}{1249852257.0}{GW190803A}{1248834439.9}{GW190731A}{1248617394.6}{GW190728A}{1248331528.5}{GW190727A}{1248242632.0}{GW190720A}{1247616534.7}{GW190719A}{1247608532.9}{GW190708A}{1246663515.4}{GW190707A}{1246527224.2}{GW190706A}{1246487219.4}{GW190701A}{1246048404.6}{GW190630A}{1245955943.2}{GW190620A}{1245035079.3}{GW190602A}{1243533585.1}{GW190527A}{1242984073.8}{GW190521B}{1242459857.5}{GW190521A}{1242442967.4}{GW190519A}{1242315362.4}{GW190517A}{1242107479.8}{GW190514A}{1241852074.9}{GW190513A}{1241816086.7}{GW190512A}{1241719652.4}{GW190503A}{1240944862.3}{GW190426A}{1240327333.4}{GW190425A}{1240215503.0}{GW190424A}{1240164426.1}{GW190421A}{1239917954.2}{GW190413B}{1239198206.7}{GW190413A}{1239168612.5}{GW190412A}{1239082262.2}{GW190408A}{1238782700.3}}}
\newcommand{\geocenttimeIMRplus}[1]{\IfEqCase{#1}{{GW190930A}{0.01}{GW190929A}{0.02}{GW190924A}{0.004}{GW190915A}{0.002}{GW190910A}{0.007}{GW190909A}{0.010}{GW190828B}{0.04}{GW190828A}{0.04}{GW190814A}{0.003}{GW190803A}{0.02}{GW190731A}{0.03}{GW190728A}{0.001}{GW190727A}{0.04}{GW190720A}{0.008}{GW190719A}{0.01}{GW190708A}{0.004}{GW190707A}{0.03}{GW190706A}{0.006}{GW190701A}{0.003}{GW190630A}{0.03}{GW190620A}{0.006}{GW190602A}{0.009}{GW190527A}{0.02}{GW190521B}{0.002}{GW190521A}{0.04}{GW190519A}{0.007}{GW190517A}{0.01}{GW190514A}{0.005}{GW190513A}{0.04}{GW190512A}{0.005}{GW190503A}{0.003}{GW190426A}{0.02}{GW190425A}{0.03}{GW190424A}{0.02}{GW190421A}{0.009}{GW190413B}{0.04}{GW190413A}{0.03}{GW190412A}{0.005}{GW190408A}{0.001}}}
\newcommand{\raIMRminus}[1]{\IfEqCase{#1}{{GW190930A}{5.30187}{GW190929A}{2.90716}{GW190924A}{0.11450}{GW190915A}{0.09186}{GW190910A}{1.32977}{GW190909A}{1.44176}{GW190828B}{0.24432}{GW190828A}{0.11562}{GW190814A}{0.03020}{GW190803A}{0.30342}{GW190731A}{1.62088}{GW190728A}{3.94631}{GW190727A}{1.00546}{GW190720A}{4.69082}{GW190719A}{2.52616}{GW190708A}{2.56101}{GW190707A}{2.80344}{GW190706A}{0.12785}{GW190701A}{0.02858}{GW190630A}{3.00962}{GW190620A}{3.89597}{GW190602A}{0.16106}{GW190527A}{4.87431}{GW190521B}{0.65316}{GW190521A}{3.53306}{GW190519A}{3.22456}{GW190517A}{0.09560}{GW190514A}{2.74884}{GW190513A}{4.24807}{GW190512A}{0.45841}{GW190503A}{0.06967}{GW190426A}{5.11703}{GW190425A}{1.14713}{GW190424A}{2.92902}{GW190421A}{1.98508}{GW190413B}{0.51111}{GW190413A}{0.86077}{GW190412A}{0.36956}{GW190408A}{5.78710}}}
\newcommand{\raIMRmed}[1]{\IfEqCase{#1}{{GW190930A}{5.56626}{GW190929A}{4.52976}{GW190924A}{2.23783}{GW190915A}{3.41447}{GW190910A}{1.88063}{GW190909A}{1.80555}{GW190828B}{2.45490}{GW190828A}{2.48782}{GW190814A}{0.22046}{GW190803A}{1.62682}{GW190731A}{2.62361}{GW190728A}{5.48446}{GW190727A}{2.50075}{GW190720A}{5.19392}{GW190719A}{2.85941}{GW190708A}{3.04830}{GW190707A}{3.69054}{GW190706A}{2.59840}{GW190701A}{0.66316}{GW190630A}{5.87324}{GW190620A}{4.31970}{GW190602A}{1.29270}{GW190527A}{5.16266}{GW190521B}{5.00715}{GW190521A}{3.56762}{GW190519A}{3.26337}{GW190517A}{4.05235}{GW190514A}{3.51453}{GW190513A}{4.99127}{GW190512A}{4.37724}{GW190503A}{1.65256}{GW190426A}{5.27391}{GW190425A}{1.62833}{GW190424A}{3.15377}{GW190421A}{3.46633}{GW190413B}{2.71115}{GW190413A}{1.17004}{GW190412A}{3.81292}{GW190408A}{5.99953}}}
\newcommand{\raIMRplus}[1]{\IfEqCase{#1}{{GW190930A}{0.49890}{GW190929A}{0.95496}{GW190924A}{0.21465}{GW190915A}{0.07280}{GW190910A}{2.90915}{GW190909A}{4.03198}{GW190828B}{3.50244}{GW190828A}{3.27136}{GW190814A}{0.17455}{GW190803A}{1.69156}{GW190731A}{0.76888}{GW190728A}{0.71196}{GW190727A}{3.68550}{GW190720A}{0.71951}{GW190719A}{3.21725}{GW190708A}{2.86189}{GW190707A}{2.10330}{GW190706A}{3.42235}{GW190701A}{0.03131}{GW190630A}{0.11225}{GW190620A}{0.28935}{GW190602A}{0.29864}{GW190527A}{0.82398}{GW190521B}{0.58797}{GW190521A}{2.68052}{GW190519A}{2.97976}{GW190517A}{1.58008}{GW190514A}{1.87892}{GW190513A}{0.11611}{GW190512A}{0.57197}{GW190503A}{0.07425}{GW190426A}{0.85181}{GW190425A}{3.12911}{GW190424A}{2.78982}{GW190421A}{0.17619}{GW190413B}{2.10904}{GW190413A}{2.27912}{GW190412A}{0.03668}{GW190408A}{0.15567}}}
\newcommand{\decIMRminus}[1]{\IfEqCase{#1}{{GW190930A}{0.69744}{GW190929A}{1.03996}{GW190924A}{0.37638}{GW190915A}{0.46141}{GW190910A}{0.98249}{GW190909A}{1.21610}{GW190828B}{0.48258}{GW190828A}{0.47136}{GW190814A}{0.13140}{GW190803A}{0.61258}{GW190731A}{0.76462}{GW190728A}{1.45373}{GW190727A}{0.21353}{GW190720A}{1.54766}{GW190719A}{1.34010}{GW190708A}{1.27055}{GW190707A}{0.78951}{GW190706A}{1.26217}{GW190701A}{0.08511}{GW190630A}{0.81825}{GW190620A}{1.14171}{GW190602A}{0.24139}{GW190527A}{0.73358}{GW190521B}{0.80120}{GW190521A}{0.31872}{GW190519A}{0.89619}{GW190517A}{0.21140}{GW190514A}{1.34361}{GW190513A}{0.06842}{GW190512A}{0.15610}{GW190503A}{0.08094}{GW190426A}{1.53579}{GW190425A}{0.89977}{GW190424A}{1.12293}{GW190421A}{0.57456}{GW190413B}{0.19980}{GW190413A}{0.08405}{GW190412A}{0.25558}{GW190408A}{0.61019}}}
\newcommand{\decIMRmed}[1]{\IfEqCase{#1}{{GW190930A}{0.65524}{GW190929A}{0.08602}{GW190924A}{0.24165}{GW190915A}{0.63294}{GW190910A}{0.05486}{GW190909A}{0.28574}{GW190828B}{-0.64372}{GW190828A}{-0.39759}{GW190814A}{-0.43687}{GW190803A}{0.40407}{GW190731A}{-0.63750}{GW190728A}{0.16190}{GW190727A}{-1.03131}{GW190720A}{0.62896}{GW190719A}{0.34485}{GW190708A}{0.32622}{GW190707A}{-0.14200}{GW190706A}{0.46456}{GW190701A}{-0.11203}{GW190630A}{-0.12269}{GW190620A}{0.33303}{GW190602A}{-0.58336}{GW190527A}{-0.57177}{GW190521B}{0.40716}{GW190521A}{-0.88872}{GW190519A}{0.24979}{GW190517A}{-0.76968}{GW190514A}{0.62000}{GW190513A}{-0.49061}{GW190512A}{-0.46654}{GW190503A}{-0.87887}{GW190426A}{0.90282}{GW190425A}{-0.13006}{GW190424A}{0.01445}{GW190421A}{-0.82553}{GW190413B}{-0.52708}{GW190413A}{-0.74413}{GW190412A}{0.63171}{GW190408A}{0.84614}}}
\newcommand{\decIMRplus}[1]{\IfEqCase{#1}{{GW190930A}{0.45251}{GW190929A}{0.94880}{GW190924A}{0.20954}{GW190915A}{0.51929}{GW190910A}{0.90077}{GW190909A}{0.95134}{GW190828B}{1.27320}{GW190828A}{1.21142}{GW190814A}{0.03517}{GW190803A}{0.78424}{GW190731A}{1.04803}{GW190728A}{0.44605}{GW190727A}{1.95957}{GW190720A}{0.04883}{GW190719A}{0.83400}{GW190708A}{0.86955}{GW190707A}{1.35253}{GW190706A}{0.48044}{GW190701A}{0.09559}{GW190630A}{0.67598}{GW190620A}{0.82034}{GW190602A}{0.59155}{GW190527A}{0.93448}{GW190521B}{0.19354}{GW190521A}{1.59538}{GW190519A}{0.76745}{GW190517A}{1.17654}{GW190514A}{0.76564}{GW190513A}{1.48376}{GW190512A}{0.48829}{GW190503A}{0.11333}{GW190426A}{0.61722}{GW190425A}{0.96811}{GW190424A}{1.09684}{GW190421A}{0.64278}{GW190413B}{1.84696}{GW190413A}{2.10974}{GW190412A}{0.02860}{GW190408A}{0.35802}}}
\newcommand{\luminositydistanceIMRminus}[1]{\IfEqCase{#1}{{GW190930A}{0.32}{GW190929A}{0.82}{GW190924A}{0.21}{GW190915A}{0.59}{GW190910A}{0.83}{GW190909A}{2.07}{GW190828B}{0.70}{GW190828A}{0.86}{GW190814A}{0.04}{GW190803A}{1.42}{GW190731A}{1.59}{GW190728A}{0.38}{GW190727A}{1.15}{GW190720A}{0.29}{GW190719A}{1.91}{GW190708A}{0.38}{GW190707A}{0.34}{GW190706A}{2.12}{GW190701A}{0.72}{GW190630A}{0.40}{GW190620A}{1.29}{GW190602A}{1.16}{GW190527A}{1.12}{GW190521B}{0.49}{GW190521A}{1.80}{GW190519A}{1.43}{GW190517A}{0.81}{GW190514A}{1.98}{GW190513A}{0.81}{GW190512A}{0.60}{GW190503A}{0.61}{GW190426A}{0.16}{GW190425A}{0.07}{GW190424A}{1.09}{GW190421A}{1.12}{GW190413B}{1.82}{GW190413A}{1.29}{GW190412A}{0.21}{GW190408A}{0.56}}}
\newcommand{\luminositydistanceIMRmed}[1]{\IfEqCase{#1}{{GW190930A}{0.76}{GW190929A}{1.85}{GW190924A}{0.56}{GW190915A}{1.60}{GW190910A}{1.80}{GW190909A}{3.60}{GW190828B}{1.67}{GW190828A}{2.03}{GW190814A}{0.25}{GW190803A}{2.96}{GW190731A}{3.11}{GW190728A}{0.88}{GW190727A}{2.75}{GW190720A}{0.74}{GW190719A}{3.41}{GW190708A}{0.84}{GW190707A}{0.73}{GW190706A}{4.56}{GW190701A}{1.87}{GW190630A}{0.93}{GW190620A}{2.63}{GW190602A}{2.33}{GW190527A}{2.24}{GW190521B}{1.13}{GW190521A}{4.25}{GW190519A}{3.45}{GW190517A}{1.83}{GW190514A}{3.68}{GW190513A}{1.92}{GW190512A}{1.42}{GW190503A}{1.46}{GW190426A}{0.37}{GW190425A}{0.16}{GW190424A}{1.96}{GW190421A}{2.43}{GW190413B}{3.69}{GW190413A}{3.15}{GW190412A}{0.74}{GW190408A}{1.40}}}
\newcommand{\luminositydistanceIMRplus}[1]{\IfEqCase{#1}{{GW190930A}{0.34}{GW190929A}{3.06}{GW190924A}{0.21}{GW190915A}{0.72}{GW190910A}{0.88}{GW190909A}{3.12}{GW190828B}{0.65}{GW190828A}{0.62}{GW190814A}{0.04}{GW190803A}{1.75}{GW190731A}{2.30}{GW190728A}{0.24}{GW190727A}{1.27}{GW190720A}{0.55}{GW190719A}{2.88}{GW190708A}{0.34}{GW190707A}{0.37}{GW190706A}{2.36}{GW190701A}{0.78}{GW190630A}{0.49}{GW190620A}{1.64}{GW190602A}{1.75}{GW190527A}{2.30}{GW190521B}{0.39}{GW190521A}{1.62}{GW190519A}{1.70}{GW190517A}{1.52}{GW190514A}{2.65}{GW190513A}{0.73}{GW190512A}{0.54}{GW190503A}{0.61}{GW190426A}{0.18}{GW190425A}{0.07}{GW190424A}{1.50}{GW190421A}{1.41}{GW190413B}{2.40}{GW190413A}{1.93}{GW190412A}{0.15}{GW190408A}{0.44}}}
\newcommand{\psiIMRminus}[1]{\IfEqCase{#1}{{GW190930A}{1.36}{GW190929A}{1.44}{GW190924A}{1.39}{GW190915A}{1.40}{GW190910A}{1.41}{GW190909A}{1.46}{GW190828B}{1.47}{GW190828A}{1.36}{GW190814A}{0.23}{GW190803A}{1.47}{GW190731A}{1.43}{GW190728A}{1.41}{GW190727A}{1.36}{GW190720A}{1.38}{GW190719A}{1.37}{GW190708A}{1.43}{GW190707A}{1.33}{GW190706A}{1.48}{GW190701A}{1.43}{GW190630A}{1.47}{GW190620A}{1.42}{GW190602A}{1.46}{GW190527A}{1.44}{GW190521B}{1.44}{GW190521A}{1.00}{GW190519A}{1.45}{GW190517A}{1.46}{GW190514A}{1.44}{GW190513A}{1.27}{GW190512A}{1.45}{GW190503A}{1.44}{GW190426A}{1.40}{GW190425A}{1.46}{GW190424A}{1.38}{GW190421A}{1.42}{GW190413B}{1.44}{GW190413A}{1.46}{GW190412A}{2.20}{GW190408A}{1.38}}}
\newcommand{\psiIMRmed}[1]{\IfEqCase{#1}{{GW190930A}{1.51}{GW190929A}{1.62}{GW190924A}{1.55}{GW190915A}{1.55}{GW190910A}{1.55}{GW190909A}{1.61}{GW190828B}{1.63}{GW190828A}{1.51}{GW190814A}{0.27}{GW190803A}{1.63}{GW190731A}{1.60}{GW190728A}{1.57}{GW190727A}{1.49}{GW190720A}{1.52}{GW190719A}{1.56}{GW190708A}{1.55}{GW190707A}{1.50}{GW190706A}{1.67}{GW190701A}{1.58}{GW190630A}{1.59}{GW190620A}{1.55}{GW190602A}{1.62}{GW190527A}{1.61}{GW190521B}{1.61}{GW190521A}{1.11}{GW190519A}{1.60}{GW190517A}{1.67}{GW190514A}{1.61}{GW190513A}{1.41}{GW190512A}{1.54}{GW190503A}{1.61}{GW190426A}{1.58}{GW190425A}{1.62}{GW190424A}{1.51}{GW190421A}{1.56}{GW190413B}{1.60}{GW190413A}{1.63}{GW190412A}{2.30}{GW190408A}{1.59}}}
\newcommand{\psiIMRplus}[1]{\IfEqCase{#1}{{GW190930A}{1.44}{GW190929A}{1.35}{GW190924A}{1.41}{GW190915A}{1.40}{GW190910A}{1.43}{GW190909A}{1.38}{GW190828B}{1.32}{GW190828A}{1.47}{GW190814A}{2.81}{GW190803A}{1.34}{GW190731A}{1.40}{GW190728A}{1.42}{GW190727A}{1.47}{GW190720A}{1.46}{GW190719A}{1.40}{GW190708A}{1.43}{GW190707A}{1.45}{GW190706A}{1.29}{GW190701A}{1.42}{GW190630A}{1.46}{GW190620A}{1.45}{GW190602A}{1.39}{GW190527A}{1.38}{GW190521B}{1.38}{GW190521A}{1.91}{GW190519A}{1.40}{GW190517A}{1.28}{GW190514A}{1.37}{GW190513A}{1.52}{GW190512A}{1.47}{GW190503A}{1.36}{GW190426A}{1.36}{GW190425A}{1.38}{GW190424A}{1.46}{GW190421A}{1.41}{GW190413B}{1.38}{GW190413A}{1.36}{GW190412A}{0.72}{GW190408A}{1.37}}}
\newcommand{\chirpmassdetIMRminus}[1]{\IfEqCase{#1}{{GW190930A}{0.3}{GW190929A}{13.9}{GW190924A}{0.03}{GW190915A}{3.6}{GW190910A}{3.9}{GW190909A}{11.1}{GW190828B}{0.8}{GW190828A}{2.6}{GW190814A}{0.02}{GW190803A}{6.5}{GW190731A}{8.6}{GW190728A}{0.08}{GW190727A}{7.0}{GW190720A}{0.1}{GW190719A}{7.0}{GW190708A}{0.3}{GW190707A}{0.08}{GW190706A}{17.0}{GW190701A}{8.7}{GW190630A}{1.6}{GW190620A}{12.0}{GW190602A}{19.3}{GW190527A}{6.5}{GW190521B}{2.7}{GW190521A}{19.2}{GW190519A}{11.7}{GW190517A}{4.2}{GW190514A}{8.1}{GW190513A}{2.2}{GW190512A}{0.9}{GW190503A}{6.7}{GW190426A}{0.01}{GW190425A}{0.0006}{GW190424A}{5.5}{GW190421A}{7.0}{GW190413B}{13.8}{GW190413A}{6.0}{GW190412A}{0.2}{GW190408A}{1.7}}}
\newcommand{\chirpmassdetIMRmed}[1]{\IfEqCase{#1}{{GW190930A}{9.8}{GW190929A}{49.5}{GW190924A}{6.44}{GW190915A}{33.2}{GW190910A}{43.7}{GW190909A}{47.6}{GW190828B}{17.3}{GW190828A}{33.9}{GW190814A}{6.41}{GW190803A}{42.8}{GW190731A}{47.0}{GW190728A}{10.1}{GW190727A}{44.7}{GW190720A}{10.4}{GW190719A}{38.0}{GW190708A}{15.4}{GW190707A}{9.87}{GW190706A}{77.3}{GW190701A}{55.1}{GW190630A}{28.9}{GW190620A}{56.6}{GW190602A}{69.4}{GW190527A}{34.1}{GW190521B}{39.0}{GW190521A}{112.5}{GW190519A}{67.4}{GW190517A}{35.9}{GW190514A}{48.7}{GW190513A}{28.7}{GW190512A}{18.5}{GW190503A}{38.4}{GW190426A}{2.60}{GW190425A}{1.49}{GW190424A}{43.0}{GW190421A}{45.7}{GW190413B}{57.0}{GW190413A}{39.0}{GW190412A}{15.2}{GW190408A}{23.7}}}
\newcommand{\chirpmassdetIMRplus}[1]{\IfEqCase{#1}{{GW190930A}{0.2}{GW190929A}{17.7}{GW190924A}{0.06}{GW190915A}{3.3}{GW190910A}{4.0}{GW190909A}{14.4}{GW190828B}{0.7}{GW190828A}{2.8}{GW190814A}{0.02}{GW190803A}{6.2}{GW190731A}{7.1}{GW190728A}{0.09}{GW190727A}{5.3}{GW190720A}{0.1}{GW190719A}{30.6}{GW190708A}{0.3}{GW190707A}{0.1}{GW190706A}{9.5}{GW190701A}{7.6}{GW190630A}{2.1}{GW190620A}{8.9}{GW190602A}{12.9}{GW190527A}{26.9}{GW190521B}{2.6}{GW190521A}{15.3}{GW190519A}{6.8}{GW190517A}{3.2}{GW190514A}{7.0}{GW190513A}{4.0}{GW190512A}{0.9}{GW190503A}{5.7}{GW190426A}{0.01}{GW190425A}{0.0008}{GW190424A}{5.6}{GW190421A}{6.0}{GW190413B}{9.5}{GW190413A}{6.7}{GW190412A}{0.3}{GW190408A}{1.3}}}
\newcommand{\spinoneIMRminus}[1]{\IfEqCase{#1}{{GW190930A}{0.36}{GW190929A}{0.53}{GW190924A}{0.27}{GW190915A}{0.49}{GW190910A}{0.34}{GW190909A}{0.44}{GW190828B}{0.21}{GW190828A}{0.37}{GW190814A}{0.03}{GW190803A}{0.39}{GW190731A}{0.37}{GW190728A}{0.31}{GW190727A}{0.46}{GW190720A}{0.38}{GW190719A}{0.51}{GW190708A}{0.21}{GW190707A}{0.22}{GW190706A}{0.54}{GW190701A}{0.37}{GW190630A}{0.27}{GW190620A}{0.50}{GW190602A}{0.36}{GW190527A}{0.42}{GW190521B}{0.31}{GW190521A}{0.57}{GW190519A}{0.51}{GW190517A}{0.34}{GW190514A}{0.46}{GW190513A}{0.28}{GW190512A}{0.20}{GW190503A}{0.31}{GW190426A}{0.14}{GW190425A}{0.25}{GW190424A}{0.52}{GW190421A}{0.42}{GW190413B}{0.49}{GW190413A}{0.37}{GW190412A}{0.25}{GW190408A}{0.29}}}
\newcommand{\spinoneIMRmed}[1]{\IfEqCase{#1}{{GW190930A}{0.41}{GW190929A}{0.66}{GW190924A}{0.30}{GW190915A}{0.55}{GW190910A}{0.38}{GW190909A}{0.50}{GW190828B}{0.23}{GW190828A}{0.41}{GW190814A}{0.03}{GW190803A}{0.44}{GW190731A}{0.41}{GW190728A}{0.35}{GW190727A}{0.50}{GW190720A}{0.45}{GW190719A}{0.58}{GW190708A}{0.23}{GW190707A}{0.25}{GW190706A}{0.69}{GW190701A}{0.42}{GW190630A}{0.30}{GW190620A}{0.61}{GW190602A}{0.40}{GW190527A}{0.47}{GW190521B}{0.34}{GW190521A}{0.64}{GW190519A}{0.75}{GW190517A}{0.86}{GW190514A}{0.51}{GW190513A}{0.31}{GW190512A}{0.22}{GW190503A}{0.35}{GW190426A}{0.14}{GW190425A}{0.27}{GW190424A}{0.59}{GW190421A}{0.47}{GW190413B}{0.54}{GW190413A}{0.41}{GW190412A}{0.40}{GW190408A}{0.32}}}
\newcommand{\spinoneIMRplus}[1]{\IfEqCase{#1}{{GW190930A}{0.44}{GW190929A}{0.29}{GW190924A}{0.45}{GW190915A}{0.40}{GW190910A}{0.53}{GW190909A}{0.44}{GW190828B}{0.46}{GW190828A}{0.49}{GW190814A}{0.06}{GW190803A}{0.48}{GW190731A}{0.51}{GW190728A}{0.40}{GW190727A}{0.44}{GW190720A}{0.42}{GW190719A}{0.38}{GW190708A}{0.47}{GW190707A}{0.48}{GW190706A}{0.27}{GW190701A}{0.49}{GW190630A}{0.46}{GW190620A}{0.34}{GW190602A}{0.50}{GW190527A}{0.46}{GW190521B}{0.52}{GW190521A}{0.32}{GW190519A}{0.22}{GW190517A}{0.12}{GW190514A}{0.43}{GW190513A}{0.52}{GW190512A}{0.48}{GW190503A}{0.51}{GW190426A}{0.40}{GW190425A}{0.51}{GW190424A}{0.37}{GW190421A}{0.47}{GW190413B}{0.40}{GW190413A}{0.51}{GW190412A}{0.22}{GW190408A}{0.52}}}
\newcommand{\spintwoIMRminus}[1]{\IfEqCase{#1}{{GW190930A}{0.41}{GW190929A}{0.45}{GW190924A}{0.36}{GW190915A}{0.46}{GW190910A}{0.37}{GW190909A}{0.46}{GW190828B}{0.36}{GW190828A}{0.37}{GW190814A}{0.45}{GW190803A}{0.41}{GW190731A}{0.41}{GW190728A}{0.37}{GW190727A}{0.42}{GW190720A}{0.47}{GW190719A}{0.47}{GW190708A}{0.30}{GW190707A}{0.31}{GW190706A}{0.52}{GW190701A}{0.41}{GW190630A}{0.36}{GW190620A}{0.48}{GW190602A}{0.42}{GW190527A}{0.45}{GW190521B}{0.36}{GW190521A}{0.48}{GW190519A}{0.49}{GW190517A}{0.61}{GW190514A}{0.46}{GW190513A}{0.39}{GW190512A}{0.37}{GW190503A}{0.39}{GW190426A}{0.009}{GW190425A}{0.25}{GW190424A}{0.44}{GW190421A}{0.42}{GW190413B}{0.47}{GW190413A}{0.41}{GW190412A}{0.45}{GW190408A}{0.38}}}
\newcommand{\spintwoIMRmed}[1]{\IfEqCase{#1}{{GW190930A}{0.46}{GW190929A}{0.50}{GW190924A}{0.40}{GW190915A}{0.52}{GW190910A}{0.41}{GW190909A}{0.51}{GW190828B}{0.40}{GW190828A}{0.41}{GW190814A}{0.50}{GW190803A}{0.46}{GW190731A}{0.45}{GW190728A}{0.41}{GW190727A}{0.47}{GW190720A}{0.55}{GW190719A}{0.52}{GW190708A}{0.33}{GW190707A}{0.34}{GW190706A}{0.58}{GW190701A}{0.46}{GW190630A}{0.40}{GW190620A}{0.53}{GW190602A}{0.47}{GW190527A}{0.50}{GW190521B}{0.40}{GW190521A}{0.53}{GW190519A}{0.55}{GW190517A}{0.70}{GW190514A}{0.52}{GW190513A}{0.43}{GW190512A}{0.41}{GW190503A}{0.43}{GW190426A}{0.009}{GW190425A}{0.28}{GW190424A}{0.49}{GW190421A}{0.46}{GW190413B}{0.52}{GW190413A}{0.45}{GW190412A}{0.50}{GW190408A}{0.42}}}
\newcommand{\spintwoIMRplus}[1]{\IfEqCase{#1}{{GW190930A}{0.46}{GW190929A}{0.44}{GW190924A}{0.51}{GW190915A}{0.43}{GW190910A}{0.52}{GW190909A}{0.43}{GW190828B}{0.52}{GW190828A}{0.50}{GW190814A}{0.43}{GW190803A}{0.47}{GW190731A}{0.47}{GW190728A}{0.50}{GW190727A}{0.47}{GW190720A}{0.40}{GW190719A}{0.43}{GW190708A}{0.58}{GW190707A}{0.53}{GW190706A}{0.37}{GW190701A}{0.48}{GW190630A}{0.50}{GW190620A}{0.41}{GW190602A}{0.46}{GW190527A}{0.44}{GW190521B}{0.51}{GW190521A}{0.42}{GW190519A}{0.40}{GW190517A}{0.26}{GW190514A}{0.42}{GW190513A}{0.49}{GW190512A}{0.49}{GW190503A}{0.49}{GW190426A}{0.03}{GW190425A}{0.51}{GW190424A}{0.45}{GW190421A}{0.47}{GW190413B}{0.42}{GW190413A}{0.47}{GW190412A}{0.45}{GW190408A}{0.49}}}
\newcommand{\tiltoneIMRminus}[1]{\IfEqCase{#1}{{GW190930A}{0.73}{GW190929A}{0.72}{GW190924A}{0.98}{GW190915A}{0.84}{GW190910A}{0.97}{GW190909A}{1.12}{GW190828B}{0.98}{GW190828A}{0.76}{GW190814A}{1.14}{GW190803A}{1.04}{GW190731A}{0.91}{GW190728A}{0.75}{GW190727A}{0.83}{GW190720A}{0.72}{GW190719A}{0.62}{GW190708A}{0.95}{GW190707A}{1.10}{GW190706A}{0.53}{GW190701A}{1.11}{GW190630A}{0.83}{GW190620A}{0.61}{GW190602A}{0.89}{GW190527A}{0.90}{GW190521B}{0.85}{GW190521A}{0.88}{GW190519A}{0.56}{GW190517A}{0.47}{GW190514A}{1.11}{GW190513A}{0.93}{GW190512A}{1.05}{GW190503A}{1.10}{GW190426A}{0.00}{GW190425A}{0.80}{GW190424A}{0.81}{GW190421A}{1.00}{GW190413B}{1.03}{GW190413A}{1.12}{GW190412A}{0.48}{GW190408A}{1.11}}}
\newcommand{\tiltoneIMRmed}[1]{\IfEqCase{#1}{{GW190930A}{0.99}{GW190929A}{1.65}{GW190924A}{1.24}{GW190915A}{1.51}{GW190910A}{1.53}{GW190909A}{1.89}{GW190828B}{1.46}{GW190828A}{1.09}{GW190814A}{1.56}{GW190803A}{1.62}{GW190731A}{1.37}{GW190728A}{1.02}{GW190727A}{1.27}{GW190720A}{0.99}{GW190719A}{0.87}{GW190708A}{1.48}{GW190707A}{1.78}{GW190706A}{0.75}{GW190701A}{1.80}{GW190630A}{1.32}{GW190620A}{0.87}{GW190602A}{1.34}{GW190527A}{1.39}{GW190521B}{1.40}{GW190521A}{1.46}{GW190519A}{0.80}{GW190517A}{0.66}{GW190514A}{1.92}{GW190513A}{1.35}{GW190512A}{1.57}{GW190503A}{1.69}{GW190426A}{0.00}{GW190425A}{1.31}{GW190424A}{1.24}{GW190421A}{1.77}{GW190413B}{1.67}{GW190413A}{1.70}{GW190412A}{0.91}{GW190408A}{1.75}}}
\newcommand{\tiltoneIMRplus}[1]{\IfEqCase{#1}{{GW190930A}{1.22}{GW190929A}{0.89}{GW190924A}{1.17}{GW190915A}{0.86}{GW190910A}{1.02}{GW190909A}{0.89}{GW190828B}{1.05}{GW190828A}{1.22}{GW190814A}{1.15}{GW190803A}{1.02}{GW190731A}{1.13}{GW190728A}{1.19}{GW190727A}{1.08}{GW190720A}{0.87}{GW190719A}{1.12}{GW190708A}{1.01}{GW190707A}{0.80}{GW190706A}{0.93}{GW190701A}{0.93}{GW190630A}{0.98}{GW190620A}{0.92}{GW190602A}{1.04}{GW190527A}{1.06}{GW190521B}{1.00}{GW190521A}{1.13}{GW190519A}{0.69}{GW190517A}{0.52}{GW190514A}{0.92}{GW190513A}{1.07}{GW190512A}{1.01}{GW190503A}{1.02}{GW190426A}{3.14}{GW190425A}{0.66}{GW190424A}{1.00}{GW190421A}{0.93}{GW190413B}{0.97}{GW190413A}{1.02}{GW190412A}{0.58}{GW190408A}{0.97}}}
\newcommand{\tilttwoIMRminus}[1]{\IfEqCase{#1}{{GW190930A}{0.97}{GW190929A}{1.11}{GW190924A}{1.06}{GW190915A}{1.00}{GW190910A}{1.11}{GW190909A}{1.22}{GW190828B}{1.09}{GW190828A}{0.90}{GW190814A}{1.07}{GW190803A}{1.11}{GW190731A}{1.02}{GW190728A}{0.90}{GW190727A}{0.99}{GW190720A}{1.03}{GW190719A}{0.86}{GW190708A}{1.02}{GW190707A}{1.19}{GW190706A}{0.82}{GW190701A}{1.18}{GW190630A}{0.99}{GW190620A}{0.85}{GW190602A}{0.98}{GW190527A}{1.03}{GW190521B}{1.00}{GW190521A}{1.02}{GW190519A}{0.84}{GW190517A}{0.63}{GW190514A}{1.17}{GW190513A}{1.01}{GW190512A}{1.03}{GW190503A}{1.10}{GW190426A}{0.00}{GW190425A}{0.87}{GW190424A}{0.92}{GW190421A}{1.15}{GW190413B}{1.06}{GW190413A}{1.15}{GW190412A}{0.87}{GW190408A}{1.07}}}
\newcommand{\tilttwoIMRmed}[1]{\IfEqCase{#1}{{GW190930A}{1.38}{GW190929A}{1.56}{GW190924A}{1.53}{GW190915A}{1.54}{GW190910A}{1.69}{GW190909A}{1.84}{GW190828B}{1.60}{GW190828A}{1.26}{GW190814A}{1.68}{GW190803A}{1.64}{GW190731A}{1.49}{GW190728A}{1.28}{GW190727A}{1.43}{GW190720A}{1.46}{GW190719A}{1.21}{GW190708A}{1.49}{GW190707A}{1.84}{GW190706A}{1.10}{GW190701A}{1.79}{GW190630A}{1.48}{GW190620A}{1.16}{GW190602A}{1.35}{GW190527A}{1.46}{GW190521B}{1.52}{GW190521A}{1.57}{GW190519A}{1.13}{GW190517A}{0.89}{GW190514A}{1.91}{GW190513A}{1.39}{GW190512A}{1.53}{GW190503A}{1.71}{GW190426A}{0.00}{GW190425A}{1.41}{GW190424A}{1.37}{GW190421A}{1.82}{GW190413B}{1.60}{GW190413A}{1.71}{GW190412A}{1.24}{GW190408A}{1.75}}}
\newcommand{\tilttwoIMRplus}[1]{\IfEqCase{#1}{{GW190930A}{1.16}{GW190929A}{1.12}{GW190924A}{1.08}{GW190915A}{1.08}{GW190910A}{0.98}{GW190909A}{0.94}{GW190828B}{1.06}{GW190828A}{1.24}{GW190814A}{1.00}{GW190803A}{1.04}{GW190731A}{1.14}{GW190728A}{1.25}{GW190727A}{1.16}{GW190720A}{1.13}{GW190719A}{1.27}{GW190708A}{1.07}{GW190707A}{0.88}{GW190706A}{1.28}{GW190701A}{0.97}{GW190630A}{1.11}{GW190620A}{1.23}{GW190602A}{1.20}{GW190527A}{1.20}{GW190521B}{1.03}{GW190521A}{1.09}{GW190519A}{1.26}{GW190517A}{1.15}{GW190514A}{0.89}{GW190513A}{1.21}{GW190512A}{1.07}{GW190503A}{1.01}{GW190426A}{3.14}{GW190425A}{0.94}{GW190424A}{1.15}{GW190421A}{0.92}{GW190413B}{1.03}{GW190413A}{1.04}{GW190412A}{1.31}{GW190408A}{0.93}}}
\newcommand{\thetajnIMRminus}[1]{\IfEqCase{#1}{{GW190930A}{0.76}{GW190929A}{1.11}{GW190924A}{0.47}{GW190915A}{1.27}{GW190910A}{1.41}{GW190909A}{1.15}{GW190828B}{1.46}{GW190828A}{2.00}{GW190814A}{0.23}{GW190803A}{1.04}{GW190731A}{1.15}{GW190728A}{1.02}{GW190727A}{0.82}{GW190720A}{1.91}{GW190719A}{1.28}{GW190708A}{1.17}{GW190707A}{1.94}{GW190706A}{1.13}{GW190701A}{0.50}{GW190630A}{0.82}{GW190620A}{1.77}{GW190602A}{1.53}{GW190527A}{0.89}{GW190521B}{0.74}{GW190521A}{0.66}{GW190519A}{1.36}{GW190517A}{1.50}{GW190514A}{1.24}{GW190513A}{0.54}{GW190512A}{1.33}{GW190503A}{0.64}{GW190426A}{1.43}{GW190425A}{0.85}{GW190424A}{1.23}{GW190421A}{1.35}{GW190413B}{1.29}{GW190413A}{0.62}{GW190412A}{0.29}{GW190408A}{1.31}}}
\newcommand{\thetajnIMRmed}[1]{\IfEqCase{#1}{{GW190930A}{0.95}{GW190929A}{1.73}{GW190924A}{0.61}{GW190915A}{1.79}{GW190910A}{1.67}{GW190909A}{1.48}{GW190828B}{1.70}{GW190828A}{2.32}{GW190814A}{0.83}{GW190803A}{1.34}{GW190731A}{1.43}{GW190728A}{1.21}{GW190727A}{1.08}{GW190720A}{2.59}{GW190719A}{1.59}{GW190708A}{1.39}{GW190707A}{2.21}{GW190706A}{1.44}{GW190701A}{0.69}{GW190630A}{1.04}{GW190620A}{2.14}{GW190602A}{1.84}{GW190527A}{1.19}{GW190521B}{1.01}{GW190521A}{0.91}{GW190519A}{1.70}{GW190517A}{2.24}{GW190514A}{1.56}{GW190513A}{0.72}{GW190512A}{1.58}{GW190503A}{2.50}{GW190426A}{1.70}{GW190425A}{1.08}{GW190424A}{1.53}{GW190421A}{1.68}{GW190413B}{1.70}{GW190413A}{0.83}{GW190412A}{0.74}{GW190408A}{1.57}}}
\newcommand{\thetajnIMRplus}[1]{\IfEqCase{#1}{{GW190930A}{1.91}{GW190929A}{0.90}{GW190924A}{2.14}{GW190915A}{0.92}{GW190910A}{1.24}{GW190909A}{1.31}{GW190828B}{1.21}{GW190828A}{0.64}{GW190814A}{1.59}{GW190803A}{1.48}{GW190731A}{1.42}{GW190728A}{1.70}{GW190727A}{1.73}{GW190720A}{0.41}{GW190719A}{1.25}{GW190708A}{1.52}{GW190707A}{0.73}{GW190706A}{1.37}{GW190701A}{0.54}{GW190630A}{1.83}{GW190620A}{0.75}{GW190602A}{1.00}{GW190527A}{1.58}{GW190521B}{1.79}{GW190521A}{1.92}{GW190519A}{1.12}{GW190517A}{0.61}{GW190514A}{1.26}{GW190513A}{2.00}{GW190512A}{1.32}{GW190503A}{0.47}{GW190426A}{1.19}{GW190425A}{1.77}{GW190424A}{1.33}{GW190421A}{1.14}{GW190413B}{1.08}{GW190413A}{1.94}{GW190412A}{1.37}{GW190408A}{1.31}}}
\newcommand{\massonedetIMRminus}[1]{\IfEqCase{#1}{{GW190930A}{2.9}{GW190929A}{30.5}{GW190924A}{2.8}{GW190915A}{8.1}{GW190910A}{6.7}{GW190909A}{14.3}{GW190828B}{7.9}{GW190828A}{5.1}{GW190814A}{1.2}{GW190803A}{9.6}{GW190731A}{10.7}{GW190728A}{2.6}{GW190727A}{9.0}{GW190720A}{3.8}{GW190719A}{19.3}{GW190708A}{3.2}{GW190707A}{1.9}{GW190706A}{20.7}{GW190701A}{11.6}{GW190630A}{8.2}{GW190620A}{16.6}{GW190602A}{20.8}{GW190527A}{14.4}{GW190521B}{6.5}{GW190521A}{20.9}{GW190519A}{16.8}{GW190517A}{8.4}{GW190514A}{10.4}{GW190513A}{12.7}{GW190512A}{6.3}{GW190503A}{9.4}{GW190426A}{2.5}{GW190425A}{0.4}{GW190424A}{9.0}{GW190421A}{9.9}{GW190413B}{15.2}{GW190413A}{10.1}{GW190412A}{5.3}{GW190408A}{4.6}}}
\newcommand{\massonedetIMRmed}[1]{\IfEqCase{#1}{{GW190930A}{14.4}{GW190929A}{112.7}{GW190924A}{10.5}{GW190915A}{47.5}{GW190910A}{56.9}{GW190909A}{69.2}{GW190828B}{29.4}{GW190828A}{43.9}{GW190814A}{24.4}{GW190803A}{58.6}{GW190731A}{65.3}{GW190728A}{14.4}{GW190727A}{60.7}{GW190720A}{15.9}{GW190719A}{63.9}{GW190708A}{21.2}{GW190707A}{13.4}{GW190706A}{113.9}{GW190701A}{76.1}{GW190630A}{43.3}{GW190620A}{86.9}{GW190602A}{109.6}{GW190527A}{56.3}{GW190521B}{52.6}{GW190521A}{153.2}{GW190519A}{97.8}{GW190517A}{50.9}{GW190514A}{65.6}{GW190513A}{48.8}{GW190512A}{28.6}{GW190503A}{54.7}{GW190426A}{6.2}{GW190425A}{2.1}{GW190424A}{57.1}{GW190421A}{62.8}{GW190413B}{83.2}{GW190413A}{53.0}{GW190412A}{32.4}{GW190408A}{32.1}}}
\newcommand{\massonedetIMRplus}[1]{\IfEqCase{#1}{{GW190930A}{17.1}{GW190929A}{36.8}{GW190924A}{10.7}{GW190915A}{10.9}{GW190910A}{9.6}{GW190909A}{30.8}{GW190828B}{9.3}{GW190828A}{8.6}{GW190814A}{1.3}{GW190803A}{14.5}{GW190731A}{16.0}{GW190728A}{10.1}{GW190727A}{14.7}{GW190720A}{8.3}{GW190719A}{69.1}{GW190708A}{7.4}{GW190707A}{4.1}{GW190706A}{29.7}{GW190701A}{16.0}{GW190630A}{8.7}{GW190620A}{19.4}{GW190602A}{23.6}{GW190527A}{41.5}{GW190521B}{7.9}{GW190521A}{38.0}{GW190519A}{22.3}{GW190517A}{17.1}{GW190514A}{19.1}{GW190513A}{12.1}{GW190512A}{8.9}{GW190503A}{12.2}{GW190426A}{4.2}{GW190425A}{0.6}{GW190424A}{14.4}{GW190421A}{15.5}{GW190413B}{21.4}{GW190413A}{14.7}{GW190412A}{5.5}{GW190408A}{7.2}}}
\newcommand{\masstwodetIMRminus}[1]{\IfEqCase{#1}{{GW190930A}{4.0}{GW190929A}{14.4}{GW190924A}{2.3}{GW190915A}{8.4}{GW190910A}{9.9}{GW190909A}{20.8}{GW190828B}{3.3}{GW190828A}{7.3}{GW190814A}{0.10}{GW190803A}{14.0}{GW190731A}{17.8}{GW190728A}{3.3}{GW190727A}{16.5}{GW190720A}{2.7}{GW190719A}{12.1}{GW190708A}{3.5}{GW190707A}{2.0}{GW190706A}{31.4}{GW190701A}{18.4}{GW190630A}{5.1}{GW190620A}{19.0}{GW190602A}{29.0}{GW190527A}{13.7}{GW190521B}{8.0}{GW190521A}{36.8}{GW190519A}{23.4}{GW190517A}{12.2}{GW190514A}{16.9}{GW190513A}{5.6}{GW190512A}{4.1}{GW190503A}{12.9}{GW190426A}{0.5}{GW190425A}{0.3}{GW190424A}{12.1}{GW190421A}{15.5}{GW190413B}{25.7}{GW190413A}{11.3}{GW190412A}{1.2}{GW190408A}{5.2}}}
\newcommand{\masstwodetIMRmed}[1]{\IfEqCase{#1}{{GW190930A}{8.9}{GW190929A}{30.8}{GW190924A}{5.3}{GW190915A}{31.1}{GW190910A}{45.0}{GW190909A}{45.5}{GW190828B}{13.8}{GW190828A}{34.9}{GW190814A}{2.72}{GW190803A}{42.3}{GW190731A}{46.0}{GW190728A}{9.5}{GW190727A}{44.6}{GW190720A}{9.0}{GW190719A}{32.0}{GW190708A}{14.9}{GW190707A}{9.6}{GW190706A}{71.3}{GW190701A}{54.2}{GW190630A}{26.0}{GW190620A}{49.6}{GW190602A}{59.7}{GW190527A}{29.7}{GW190521B}{38.8}{GW190521A}{110.7}{GW190519A}{62.5}{GW190517A}{33.9}{GW190514A}{48.7}{GW190513A}{23.1}{GW190512A}{16.2}{GW190503A}{36.6}{GW190426A}{1.6}{GW190425A}{1.4}{GW190424A}{43.8}{GW190421A}{45.0}{GW190413B}{53.3}{GW190413A}{38.6}{GW190412A}{10.1}{GW190408A}{23.3}}}
\newcommand{\masstwodetIMRplus}[1]{\IfEqCase{#1}{{GW190930A}{2.1}{GW190929A}{33.0}{GW190924A}{1.8}{GW190915A}{7.5}{GW190910A}{7.2}{GW190909A}{17.9}{GW190828B}{5.1}{GW190828A}{5.3}{GW190814A}{0.09}{GW190803A}{10.1}{GW190731A}{11.8}{GW190728A}{2.0}{GW190727A}{9.2}{GW190720A}{2.6}{GW190719A}{17.8}{GW190708A}{2.5}{GW190707A}{1.5}{GW190706A}{21.8}{GW190701A}{13.3}{GW190630A}{7.2}{GW190620A}{19.2}{GW190602A}{27.7}{GW190527A}{26.4}{GW190521B}{6.7}{GW190521A}{26.1}{GW190519A}{17.0}{GW190517A}{8.3}{GW190514A}{11.1}{GW190513A}{10.1}{GW190512A}{4.6}{GW190503A}{10.3}{GW190426A}{0.9}{GW190425A}{0.3}{GW190424A}{8.3}{GW190421A}{10.5}{GW190413B}{17.4}{GW190413A}{10.0}{GW190412A}{1.6}{GW190408A}{4.0}}}
\newcommand{\totalmassdetIMRminus}[1]{\IfEqCase{#1}{{GW190930A}{1.1}{GW190929A}{24.9}{GW190924A}{1.0}{GW190915A}{6.6}{GW190910A}{8.0}{GW190909A}{19.0}{GW190828B}{3.2}{GW190828A}{5.3}{GW190814A}{1.1}{GW190803A}{12.2}{GW190731A}{14.2}{GW190728A}{0.7}{GW190727A}{11.1}{GW190720A}{1.3}{GW190719A}{14.9}{GW190708A}{1.0}{GW190707A}{0.5}{GW190706A}{22.5}{GW190701A}{14.9}{GW190630A}{3.9}{GW190620A}{18.9}{GW190602A}{24.3}{GW190527A}{10.6}{GW190521B}{4.5}{GW190521A}{34.0}{GW190519A}{15.8}{GW190517A}{6.8}{GW190514A}{14.5}{GW190513A}{6.3}{GW190512A}{2.8}{GW190503A}{11.3}{GW190426A}{1.6}{GW190425A}{0.08}{GW190424A}{11.3}{GW190421A}{12.5}{GW190413B}{18.4}{GW190413A}{13.3}{GW190412A}{3.7}{GW190408A}{3.5}}}
\newcommand{\totalmassdetIMRmed}[1]{\IfEqCase{#1}{{GW190930A}{23.4}{GW190929A}{145.9}{GW190924A}{15.8}{GW190915A}{78.8}{GW190910A}{101.6}{GW190909A}{113.9}{GW190828B}{43.1}{GW190828A}{78.8}{GW190814A}{27.2}{GW190803A}{100.8}{GW190731A}{110.7}{GW190728A}{23.9}{GW190727A}{104.9}{GW190720A}{25.0}{GW190719A}{94.7}{GW190708A}{36.3}{GW190707A}{23.1}{GW190706A}{185.6}{GW190701A}{129.9}{GW190630A}{69.7}{GW190620A}{137.2}{GW190602A}{169.1}{GW190527A}{84.9}{GW190521B}{91.2}{GW190521A}{263.8}{GW190519A}{160.4}{GW190517A}{85.5}{GW190514A}{114.3}{GW190513A}{72.1}{GW190512A}{44.9}{GW190503A}{91.2}{GW190426A}{7.8}{GW190425A}{3.50}{GW190424A}{100.8}{GW190421A}{107.7}{GW190413B}{136.0}{GW190413A}{92.0}{GW190412A}{42.5}{GW190408A}{55.7}}}
\newcommand{\totalmassdetIMRplus}[1]{\IfEqCase{#1}{{GW190930A}{13.0}{GW190929A}{33.3}{GW190924A}{8.5}{GW190915A}{8.1}{GW190910A}{9.2}{GW190909A}{35.4}{GW190828B}{6.2}{GW190828A}{6.6}{GW190814A}{1.3}{GW190803A}{14.0}{GW190731A}{15.8}{GW190728A}{6.7}{GW190727A}{11.9}{GW190720A}{5.7}{GW190719A}{91.0}{GW190708A}{3.8}{GW190707A}{2.0}{GW190706A}{19.3}{GW190701A}{16.3}{GW190630A}{5.1}{GW190620A}{18.5}{GW190602A}{24.8}{GW190527A}{64.4}{GW190521B}{5.5}{GW190521A}{36.6}{GW190519A}{15.4}{GW190517A}{8.5}{GW190514A}{16.8}{GW190513A}{9.6}{GW190512A}{5.0}{GW190503A}{12.1}{GW190426A}{3.7}{GW190425A}{0.3}{GW190424A}{12.9}{GW190421A}{13.7}{GW190413B}{20.7}{GW190413A}{15.3}{GW190412A}{4.4}{GW190408A}{3.4}}}
\newcommand{\symmetricmassratioIMRminus}[1]{\IfEqCase{#1}{{GW190930A}{0.12}{GW190929A}{0.07}{GW190924A}{0.11}{GW190915A}{0.03}{GW190910A}{0.02}{GW190909A}{0.06}{GW190828B}{0.05}{GW190828A}{0.02}{GW190814A}{0.006}{GW190803A}{0.04}{GW190731A}{0.04}{GW190728A}{0.08}{GW190727A}{0.04}{GW190720A}{0.07}{GW190719A}{0.07}{GW190708A}{0.04}{GW190707A}{0.03}{GW190706A}{0.06}{GW190701A}{0.04}{GW190630A}{0.03}{GW190620A}{0.05}{GW190602A}{0.07}{GW190527A}{0.08}{GW190521B}{0.02}{GW190521A}{0.03}{GW190519A}{0.05}{GW190517A}{0.06}{GW190514A}{0.04}{GW190513A}{0.04}{GW190512A}{0.05}{GW190503A}{0.04}{GW190426A}{0.08}{GW190425A}{0.03}{GW190424A}{0.03}{GW190421A}{0.04}{GW190413B}{0.07}{GW190413A}{0.03}{GW190412A}{0.03}{GW190408A}{0.03}}}
\newcommand{\symmetricmassratioIMRmed}[1]{\IfEqCase{#1}{{GW190930A}{0.24}{GW190929A}{0.17}{GW190924A}{0.22}{GW190915A}{0.24}{GW190910A}{0.247}{GW190909A}{0.24}{GW190828B}{0.22}{GW190828A}{0.247}{GW190814A}{0.090}{GW190803A}{0.244}{GW190731A}{0.243}{GW190728A}{0.24}{GW190727A}{0.244}{GW190720A}{0.23}{GW190719A}{0.22}{GW190708A}{0.242}{GW190707A}{0.243}{GW190706A}{0.24}{GW190701A}{0.243}{GW190630A}{0.23}{GW190620A}{0.23}{GW190602A}{0.23}{GW190527A}{0.23}{GW190521B}{0.244}{GW190521A}{0.243}{GW190519A}{0.24}{GW190517A}{0.240}{GW190514A}{0.245}{GW190513A}{0.22}{GW190512A}{0.23}{GW190503A}{0.240}{GW190426A}{0.16}{GW190425A}{0.240}{GW190424A}{0.246}{GW190421A}{0.243}{GW190413B}{0.24}{GW190413A}{0.245}{GW190412A}{0.18}{GW190408A}{0.244}}}
\newcommand{\symmetricmassratioIMRplus}[1]{\IfEqCase{#1}{{GW190930A}{0.01}{GW190929A}{0.08}{GW190924A}{0.03}{GW190915A}{0.01}{GW190910A}{0.003}{GW190909A}{0.01}{GW190828B}{0.03}{GW190828A}{0.003}{GW190814A}{0.006}{GW190803A}{0.006}{GW190731A}{0.007}{GW190728A}{0.01}{GW190727A}{0.006}{GW190720A}{0.02}{GW190719A}{0.03}{GW190708A}{0.007}{GW190707A}{0.007}{GW190706A}{0.01}{GW190701A}{0.007}{GW190630A}{0.02}{GW190620A}{0.02}{GW190602A}{0.02}{GW190527A}{0.02}{GW190521B}{0.006}{GW190521A}{0.007}{GW190519A}{0.01}{GW190517A}{0.010}{GW190514A}{0.005}{GW190513A}{0.03}{GW190512A}{0.02}{GW190503A}{0.010}{GW190426A}{0.08}{GW190425A}{0.010}{GW190424A}{0.004}{GW190421A}{0.007}{GW190413B}{0.01}{GW190413A}{0.005}{GW190412A}{0.03}{GW190408A}{0.006}}}
\newcommand{\iotaIMRminus}[1]{\IfEqCase{#1}{{GW190930A}{0.76}{GW190929A}{1.35}{GW190924A}{0.46}{GW190915A}{1.52}{GW190910A}{1.42}{GW190909A}{1.14}{GW190828B}{1.46}{GW190828A}{2.02}{GW190814A}{0.26}{GW190803A}{1.03}{GW190731A}{1.14}{GW190728A}{1.01}{GW190727A}{0.83}{GW190720A}{1.88}{GW190719A}{1.25}{GW190708A}{1.16}{GW190707A}{1.96}{GW190706A}{1.14}{GW190701A}{0.56}{GW190630A}{0.82}{GW190620A}{1.70}{GW190602A}{1.30}{GW190527A}{0.90}{GW190521B}{0.73}{GW190521A}{0.80}{GW190519A}{1.31}{GW190517A}{1.36}{GW190514A}{1.20}{GW190513A}{0.51}{GW190512A}{1.33}{GW190503A}{0.67}{GW190426A}{1.43}{GW190425A}{0.85}{GW190424A}{1.18}{GW190421A}{1.26}{GW190413B}{1.21}{GW190413A}{0.62}{GW190412A}{0.32}{GW190408A}{1.36}}}
\newcommand{\iotaIMRmed}[1]{\IfEqCase{#1}{{GW190930A}{0.96}{GW190929A}{1.80}{GW190924A}{0.62}{GW190915A}{1.90}{GW190910A}{1.70}{GW190909A}{1.52}{GW190828B}{1.70}{GW190828A}{2.33}{GW190814A}{0.81}{GW190803A}{1.33}{GW190731A}{1.43}{GW190728A}{1.22}{GW190727A}{1.11}{GW190720A}{2.56}{GW190719A}{1.58}{GW190708A}{1.37}{GW190707A}{2.21}{GW190706A}{1.49}{GW190701A}{0.78}{GW190630A}{1.01}{GW190620A}{2.11}{GW190602A}{1.69}{GW190527A}{1.22}{GW190521B}{1.09}{GW190521A}{1.06}{GW190519A}{1.69}{GW190517A}{2.15}{GW190514A}{1.56}{GW190513A}{0.71}{GW190512A}{1.58}{GW190503A}{2.47}{GW190426A}{1.70}{GW190425A}{1.09}{GW190424A}{1.53}{GW190421A}{1.67}{GW190413B}{1.68}{GW190413A}{0.84}{GW190412A}{0.73}{GW190408A}{1.60}}}
\newcommand{\iotaIMRplus}[1]{\IfEqCase{#1}{{GW190930A}{1.89}{GW190929A}{0.95}{GW190924A}{2.12}{GW190915A}{0.95}{GW190910A}{1.19}{GW190909A}{1.22}{GW190828B}{1.20}{GW190828A}{0.63}{GW190814A}{1.63}{GW190803A}{1.48}{GW190731A}{1.40}{GW190728A}{1.68}{GW190727A}{1.68}{GW190720A}{0.42}{GW190719A}{1.25}{GW190708A}{1.56}{GW190707A}{0.73}{GW190706A}{1.27}{GW190701A}{0.66}{GW190630A}{1.87}{GW190620A}{0.77}{GW190602A}{1.09}{GW190527A}{1.51}{GW190521B}{1.63}{GW190521A}{1.74}{GW190519A}{1.08}{GW190517A}{0.64}{GW190514A}{1.24}{GW190513A}{1.96}{GW190512A}{1.30}{GW190503A}{0.49}{GW190426A}{1.19}{GW190425A}{1.77}{GW190424A}{1.29}{GW190421A}{1.10}{GW190413B}{1.06}{GW190413A}{1.95}{GW190412A}{1.40}{GW190408A}{1.31}}}
\newcommand{\spinonexIMRminus}[1]{\IfEqCase{#1}{{GW190930A}{0.44}{GW190929A}{0.71}{GW190924A}{0.39}{GW190915A}{0.68}{GW190910A}{0.55}{GW190909A}{0.59}{GW190828B}{0.38}{GW190828A}{0.50}{GW190814A}{0.04}{GW190803A}{0.59}{GW190731A}{0.57}{GW190728A}{0.41}{GW190727A}{0.62}{GW190720A}{0.46}{GW190719A}{0.56}{GW190708A}{0.41}{GW190707A}{0.36}{GW190706A}{0.55}{GW190701A}{0.54}{GW190630A}{0.44}{GW190620A}{0.55}{GW190602A}{0.56}{GW190527A}{0.59}{GW190521B}{0.51}{GW190521A}{0.71}{GW190519A}{0.61}{GW190517A}{0.59}{GW190514A}{0.60}{GW190513A}{0.47}{GW190512A}{0.37}{GW190503A}{0.48}{GW190426A}{0.00}{GW190425A}{0.50}{GW190424A}{0.65}{GW190421A}{0.64}{GW190413B}{0.64}{GW190413A}{0.53}{GW190412A}{0.31}{GW190408A}{0.48}}}
\newcommand{\spinonexIMRmed}[1]{\IfEqCase{#1}{{GW190930A}{0.002}{GW190929A}{0.00}{GW190924A}{0.0001}{GW190915A}{0.00}{GW190910A}{0.00}{GW190909A}{0.00}{GW190828B}{0.00005}{GW190828A}{0.0006}{GW190814A}{0.00}{GW190803A}{0.00}{GW190731A}{0.002}{GW190728A}{0.00}{GW190727A}{0.00}{GW190720A}{0.002}{GW190719A}{-0.01}{GW190708A}{0.0004}{GW190707A}{0.003}{GW190706A}{0.002}{GW190701A}{0.001}{GW190630A}{-0.01}{GW190620A}{0.009}{GW190602A}{0.0004}{GW190527A}{0.00}{GW190521B}{0.00}{GW190521A}{-0.04}{GW190519A}{0.006}{GW190517A}{-0.03}{GW190514A}{0.00}{GW190513A}{0.0007}{GW190512A}{0.0006}{GW190503A}{0.00}{GW190426A}{0.00}{GW190425A}{0.00}{GW190424A}{0.00}{GW190421A}{0.00}{GW190413B}{0.0002}{GW190413A}{0.0007}{GW190412A}{-0.04}{GW190408A}{0.0008}}}
\newcommand{\spinonexIMRplus}[1]{\IfEqCase{#1}{{GW190930A}{0.50}{GW190929A}{0.71}{GW190924A}{0.38}{GW190915A}{0.68}{GW190910A}{0.56}{GW190909A}{0.58}{GW190828B}{0.36}{GW190828A}{0.56}{GW190814A}{0.04}{GW190803A}{0.57}{GW190731A}{0.56}{GW190728A}{0.41}{GW190727A}{0.63}{GW190720A}{0.48}{GW190719A}{0.52}{GW190708A}{0.37}{GW190707A}{0.43}{GW190706A}{0.57}{GW190701A}{0.56}{GW190630A}{0.43}{GW190620A}{0.56}{GW190602A}{0.55}{GW190527A}{0.61}{GW190521B}{0.49}{GW190521A}{0.68}{GW190519A}{0.61}{GW190517A}{0.63}{GW190514A}{0.60}{GW190513A}{0.43}{GW190512A}{0.37}{GW190503A}{0.46}{GW190426A}{0.00}{GW190425A}{0.47}{GW190424A}{0.68}{GW190421A}{0.63}{GW190413B}{0.67}{GW190413A}{0.54}{GW190412A}{0.44}{GW190408A}{0.50}}}
\newcommand{\spinoneyIMRminus}[1]{\IfEqCase{#1}{{GW190930A}{0.46}{GW190929A}{0.73}{GW190924A}{0.39}{GW190915A}{0.69}{GW190910A}{0.55}{GW190909A}{0.59}{GW190828B}{0.37}{GW190828A}{0.52}{GW190814A}{0.04}{GW190803A}{0.57}{GW190731A}{0.55}{GW190728A}{0.41}{GW190727A}{0.64}{GW190720A}{0.55}{GW190719A}{0.57}{GW190708A}{0.38}{GW190707A}{0.41}{GW190706A}{0.55}{GW190701A}{0.54}{GW190630A}{0.46}{GW190620A}{0.54}{GW190602A}{0.56}{GW190527A}{0.58}{GW190521B}{0.52}{GW190521A}{0.71}{GW190519A}{0.61}{GW190517A}{0.61}{GW190514A}{0.59}{GW190513A}{0.46}{GW190512A}{0.37}{GW190503A}{0.49}{GW190426A}{0.00}{GW190425A}{0.48}{GW190424A}{0.67}{GW190421A}{0.60}{GW190413B}{0.66}{GW190413A}{0.53}{GW190412A}{0.40}{GW190408A}{0.46}}}
\newcommand{\spinoneyIMRmed}[1]{\IfEqCase{#1}{{GW190930A}{0.0009}{GW190929A}{0.01}{GW190924A}{0.0008}{GW190915A}{0.003}{GW190910A}{0.002}{GW190909A}{0.0010}{GW190828B}{0.0009}{GW190828A}{0.00}{GW190814A}{0.00}{GW190803A}{0.002}{GW190731A}{0.003}{GW190728A}{0.0008}{GW190727A}{0.00}{GW190720A}{0.0007}{GW190719A}{0.00}{GW190708A}{0.0010}{GW190707A}{0.0002}{GW190706A}{0.005}{GW190701A}{0.00}{GW190630A}{0.00}{GW190620A}{0.00}{GW190602A}{0.002}{GW190527A}{0.00}{GW190521B}{0.003}{GW190521A}{-0.01}{GW190519A}{0.00}{GW190517A}{0.00}{GW190514A}{0.003}{GW190513A}{0.00}{GW190512A}{0.00}{GW190503A}{0.00}{GW190426A}{0.00}{GW190425A}{0.003}{GW190424A}{-0.01}{GW190421A}{0.003}{GW190413B}{0.004}{GW190413A}{0.00}{GW190412A}{0.09}{GW190408A}{0.00}}}
\newcommand{\spinoneyIMRplus}[1]{\IfEqCase{#1}{{GW190930A}{0.48}{GW190929A}{0.71}{GW190924A}{0.39}{GW190915A}{0.67}{GW190910A}{0.58}{GW190909A}{0.59}{GW190828B}{0.37}{GW190828A}{0.51}{GW190814A}{0.04}{GW190803A}{0.59}{GW190731A}{0.57}{GW190728A}{0.41}{GW190727A}{0.62}{GW190720A}{0.49}{GW190719A}{0.54}{GW190708A}{0.40}{GW190707A}{0.39}{GW190706A}{0.55}{GW190701A}{0.55}{GW190630A}{0.44}{GW190620A}{0.55}{GW190602A}{0.57}{GW190527A}{0.62}{GW190521B}{0.51}{GW190521A}{0.70}{GW190519A}{0.61}{GW190517A}{0.60}{GW190514A}{0.58}{GW190513A}{0.49}{GW190512A}{0.37}{GW190503A}{0.48}{GW190426A}{0.00}{GW190425A}{0.48}{GW190424A}{0.69}{GW190421A}{0.60}{GW190413B}{0.66}{GW190413A}{0.54}{GW190412A}{0.34}{GW190408A}{0.47}}}
\newcommand{\spinonezIMRminus}[1]{\IfEqCase{#1}{{GW190930A}{0.31}{GW190929A}{0.43}{GW190924A}{0.23}{GW190915A}{0.38}{GW190910A}{0.39}{GW190909A}{0.57}{GW190828B}{0.24}{GW190828A}{0.31}{GW190814A}{0.06}{GW190803A}{0.46}{GW190731A}{0.34}{GW190728A}{0.28}{GW190727A}{0.35}{GW190720A}{0.30}{GW190719A}{0.39}{GW190708A}{0.22}{GW190707A}{0.28}{GW190706A}{0.47}{GW190701A}{0.50}{GW190630A}{0.22}{GW190620A}{0.40}{GW190602A}{0.32}{GW190527A}{0.36}{GW190521B}{0.26}{GW190521A}{0.61}{GW190519A}{0.46}{GW190517A}{0.38}{GW190514A}{0.55}{GW190513A}{0.28}{GW190512A}{0.30}{GW190503A}{0.43}{GW190426A}{0.51}{GW190425A}{0.12}{GW190424A}{0.38}{GW190421A}{0.46}{GW190413B}{0.51}{GW190413A}{0.55}{GW190412A}{0.22}{GW190408A}{0.36}}}
\newcommand{\spinonezIMRmed}[1]{\IfEqCase{#1}{{GW190930A}{0.18}{GW190929A}{-0.04}{GW190924A}{0.06}{GW190915A}{0.02}{GW190910A}{0.007}{GW190909A}{-0.11}{GW190828B}{0.01}{GW190828A}{0.16}{GW190814A}{0.00010}{GW190803A}{-0.01}{GW190731A}{0.05}{GW190728A}{0.16}{GW190727A}{0.11}{GW190720A}{0.23}{GW190719A}{0.32}{GW190708A}{0.01}{GW190707A}{-0.03}{GW190706A}{0.45}{GW190701A}{-0.06}{GW190630A}{0.05}{GW190620A}{0.35}{GW190602A}{0.06}{GW190527A}{0.05}{GW190521B}{0.03}{GW190521A}{0.04}{GW190519A}{0.48}{GW190517A}{0.64}{GW190514A}{-0.12}{GW190513A}{0.04}{GW190512A}{0.00005}{GW190503A}{-0.02}{GW190426A}{-0.03}{GW190425A}{0.06}{GW190424A}{0.14}{GW190421A}{-0.06}{GW190413B}{-0.03}{GW190413A}{-0.02}{GW190412A}{0.24}{GW190408A}{-0.03}}}
\newcommand{\spinonezIMRplus}[1]{\IfEqCase{#1}{{GW190930A}{0.42}{GW190929A}{0.46}{GW190924A}{0.43}{GW190915A}{0.40}{GW190910A}{0.40}{GW190909A}{0.38}{GW190828B}{0.26}{GW190828A}{0.42}{GW190814A}{0.05}{GW190803A}{0.44}{GW190731A}{0.48}{GW190728A}{0.32}{GW190727A}{0.50}{GW190720A}{0.31}{GW190719A}{0.46}{GW190708A}{0.26}{GW190707A}{0.24}{GW190706A}{0.38}{GW190701A}{0.35}{GW190630A}{0.34}{GW190620A}{0.42}{GW190602A}{0.44}{GW190527A}{0.51}{GW190521B}{0.34}{GW190521A}{0.50}{GW190519A}{0.34}{GW190517A}{0.26}{GW190514A}{0.41}{GW190513A}{0.39}{GW190512A}{0.24}{GW190503A}{0.33}{GW190426A}{0.36}{GW190425A}{0.18}{GW190424A}{0.48}{GW190421A}{0.37}{GW190413B}{0.43}{GW190413A}{0.41}{GW190412A}{0.17}{GW190408A}{0.31}}}
\newcommand{\spintwoxIMRminus}[1]{\IfEqCase{#1}{{GW190930A}{0.57}{GW190929A}{0.59}{GW190924A}{0.53}{GW190915A}{0.63}{GW190910A}{0.57}{GW190909A}{0.56}{GW190828B}{0.56}{GW190828A}{0.52}{GW190814A}{0.61}{GW190803A}{0.59}{GW190731A}{0.58}{GW190728A}{0.53}{GW190727A}{0.59}{GW190720A}{0.63}{GW190719A}{0.59}{GW190708A}{0.52}{GW190707A}{0.49}{GW190706A}{0.57}{GW190701A}{0.57}{GW190630A}{0.51}{GW190620A}{0.58}{GW190602A}{0.56}{GW190527A}{0.57}{GW190521B}{0.56}{GW190521A}{0.67}{GW190519A}{0.59}{GW190517A}{0.61}{GW190514A}{0.59}{GW190513A}{0.54}{GW190512A}{0.55}{GW190503A}{0.56}{GW190426A}{0.00}{GW190425A}{0.47}{GW190424A}{0.59}{GW190421A}{0.57}{GW190413B}{0.62}{GW190413A}{0.56}{GW190412A}{0.56}{GW190408A}{0.57}}}
\newcommand{\spintwoxIMRmed}[1]{\IfEqCase{#1}{{GW190930A}{0.00}{GW190929A}{0.00}{GW190924A}{0.0004}{GW190915A}{0.00}{GW190910A}{0.00}{GW190909A}{0.0007}{GW190828B}{0.00}{GW190828A}{-0.01}{GW190814A}{0.0009}{GW190803A}{0.00}{GW190731A}{0.0006}{GW190728A}{0.0003}{GW190727A}{0.00004}{GW190720A}{0.00}{GW190719A}{0.00}{GW190708A}{0.001}{GW190707A}{0.0002}{GW190706A}{0.007}{GW190701A}{0.00}{GW190630A}{0.004}{GW190620A}{0.002}{GW190602A}{0.00}{GW190527A}{0.00}{GW190521B}{0.00}{GW190521A}{-0.01}{GW190519A}{0.0001}{GW190517A}{0.00003}{GW190514A}{0.0009}{GW190513A}{0.00}{GW190512A}{0.0002}{GW190503A}{0.00}{GW190426A}{0.00}{GW190425A}{0.0006}{GW190424A}{0.002}{GW190421A}{-0.01}{GW190413B}{0.00}{GW190413A}{0.002}{GW190412A}{0.002}{GW190408A}{0.00}}}
\newcommand{\spintwoxIMRplus}[1]{\IfEqCase{#1}{{GW190930A}{0.57}{GW190929A}{0.60}{GW190924A}{0.54}{GW190915A}{0.63}{GW190910A}{0.58}{GW190909A}{0.60}{GW190828B}{0.53}{GW190828A}{0.49}{GW190814A}{0.61}{GW190803A}{0.58}{GW190731A}{0.55}{GW190728A}{0.53}{GW190727A}{0.59}{GW190720A}{0.59}{GW190719A}{0.58}{GW190708A}{0.52}{GW190707A}{0.51}{GW190706A}{0.58}{GW190701A}{0.58}{GW190630A}{0.53}{GW190620A}{0.58}{GW190602A}{0.59}{GW190527A}{0.59}{GW190521B}{0.55}{GW190521A}{0.61}{GW190519A}{0.56}{GW190517A}{0.60}{GW190514A}{0.57}{GW190513A}{0.55}{GW190512A}{0.55}{GW190503A}{0.57}{GW190426A}{0.00}{GW190425A}{0.47}{GW190424A}{0.58}{GW190421A}{0.56}{GW190413B}{0.63}{GW190413A}{0.57}{GW190412A}{0.58}{GW190408A}{0.54}}}
\newcommand{\spintwoyIMRminus}[1]{\IfEqCase{#1}{{GW190930A}{0.58}{GW190929A}{0.59}{GW190924A}{0.53}{GW190915A}{0.63}{GW190910A}{0.57}{GW190909A}{0.58}{GW190828B}{0.55}{GW190828A}{0.54}{GW190814A}{0.60}{GW190803A}{0.59}{GW190731A}{0.55}{GW190728A}{0.53}{GW190727A}{0.59}{GW190720A}{0.61}{GW190719A}{0.56}{GW190708A}{0.51}{GW190707A}{0.48}{GW190706A}{0.59}{GW190701A}{0.57}{GW190630A}{0.54}{GW190620A}{0.60}{GW190602A}{0.58}{GW190527A}{0.58}{GW190521B}{0.60}{GW190521A}{0.63}{GW190519A}{0.59}{GW190517A}{0.61}{GW190514A}{0.56}{GW190513A}{0.54}{GW190512A}{0.53}{GW190503A}{0.56}{GW190426A}{0.00}{GW190425A}{0.48}{GW190424A}{0.60}{GW190421A}{0.57}{GW190413B}{0.62}{GW190413A}{0.56}{GW190412A}{0.53}{GW190408A}{0.56}}}
\newcommand{\spintwoyIMRmed}[1]{\IfEqCase{#1}{{GW190930A}{0.00}{GW190929A}{0.0003}{GW190924A}{0.0007}{GW190915A}{0.00}{GW190910A}{0.002}{GW190909A}{0.002}{GW190828B}{0.004}{GW190828A}{0.00}{GW190814A}{0.003}{GW190803A}{0.00}{GW190731A}{0.0006}{GW190728A}{0.00}{GW190727A}{0.0008}{GW190720A}{0.004}{GW190719A}{0.0004}{GW190708A}{0.00007}{GW190707A}{0.00}{GW190706A}{0.0006}{GW190701A}{0.0008}{GW190630A}{0.002}{GW190620A}{0.002}{GW190602A}{0.005}{GW190527A}{0.002}{GW190521B}{0.003}{GW190521A}{0.01}{GW190519A}{0.00}{GW190517A}{0.00}{GW190514A}{0.003}{GW190513A}{0.00009}{GW190512A}{0.003}{GW190503A}{0.0007}{GW190426A}{0.00}{GW190425A}{0.00}{GW190424A}{0.00}{GW190421A}{0.0002}{GW190413B}{0.0009}{GW190413A}{0.0001}{GW190412A}{0.01}{GW190408A}{0.0008}}}
\newcommand{\spintwoyIMRplus}[1]{\IfEqCase{#1}{{GW190930A}{0.56}{GW190929A}{0.59}{GW190924A}{0.55}{GW190915A}{0.64}{GW190910A}{0.58}{GW190909A}{0.59}{GW190828B}{0.57}{GW190828A}{0.50}{GW190814A}{0.61}{GW190803A}{0.59}{GW190731A}{0.59}{GW190728A}{0.52}{GW190727A}{0.60}{GW190720A}{0.58}{GW190719A}{0.60}{GW190708A}{0.49}{GW190707A}{0.50}{GW190706A}{0.59}{GW190701A}{0.57}{GW190630A}{0.53}{GW190620A}{0.56}{GW190602A}{0.59}{GW190527A}{0.59}{GW190521B}{0.57}{GW190521A}{0.66}{GW190519A}{0.58}{GW190517A}{0.60}{GW190514A}{0.58}{GW190513A}{0.54}{GW190512A}{0.56}{GW190503A}{0.55}{GW190426A}{0.00}{GW190425A}{0.48}{GW190424A}{0.60}{GW190421A}{0.57}{GW190413B}{0.60}{GW190413A}{0.56}{GW190412A}{0.59}{GW190408A}{0.57}}}
\newcommand{\spintwozIMRminus}[1]{\IfEqCase{#1}{{GW190930A}{0.47}{GW190929A}{0.56}{GW190924A}{0.44}{GW190915A}{0.53}{GW190910A}{0.47}{GW190909A}{0.62}{GW190828B}{0.46}{GW190828A}{0.39}{GW190814A}{0.56}{GW190803A}{0.53}{GW190731A}{0.48}{GW190728A}{0.46}{GW190727A}{0.45}{GW190720A}{0.60}{GW190719A}{0.48}{GW190708A}{0.39}{GW190707A}{0.41}{GW190706A}{0.47}{GW190701A}{0.57}{GW190630A}{0.46}{GW190620A}{0.47}{GW190602A}{0.48}{GW190527A}{0.52}{GW190521B}{0.42}{GW190521A}{0.55}{GW190519A}{0.44}{GW190517A}{0.50}{GW190514A}{0.58}{GW190513A}{0.51}{GW190512A}{0.44}{GW190503A}{0.56}{GW190426A}{0.03}{GW190425A}{0.18}{GW190424A}{0.47}{GW190421A}{0.57}{GW190413B}{0.57}{GW190413A}{0.57}{GW190412A}{0.51}{GW190408A}{0.51}}}
\newcommand{\spintwozIMRmed}[1]{\IfEqCase{#1}{{GW190930A}{0.05}{GW190929A}{0.002}{GW190924A}{0.005}{GW190915A}{0.006}{GW190910A}{-0.02}{GW190909A}{-0.08}{GW190828B}{0.00}{GW190828A}{0.08}{GW190814A}{-0.03}{GW190803A}{-0.01}{GW190731A}{0.02}{GW190728A}{0.08}{GW190727A}{0.03}{GW190720A}{0.03}{GW190719A}{0.12}{GW190708A}{0.01}{GW190707A}{-0.06}{GW190706A}{0.19}{GW190701A}{-0.06}{GW190630A}{0.02}{GW190620A}{0.15}{GW190602A}{0.06}{GW190527A}{0.02}{GW190521B}{0.01}{GW190521A}{0.000002}{GW190519A}{0.18}{GW190517A}{0.39}{GW190514A}{-0.12}{GW190513A}{0.04}{GW190512A}{0.007}{GW190503A}{-0.03}{GW190426A}{0.00}{GW190425A}{0.03}{GW190424A}{0.05}{GW190421A}{-0.07}{GW190413B}{-0.01}{GW190413A}{-0.03}{GW190412A}{0.10}{GW190408A}{-0.04}}}
\newcommand{\spintwozIMRplus}[1]{\IfEqCase{#1}{{GW190930A}{0.53}{GW190929A}{0.57}{GW190924A}{0.51}{GW190915A}{0.51}{GW190910A}{0.40}{GW190909A}{0.46}{GW190828B}{0.48}{GW190828A}{0.54}{GW190814A}{0.47}{GW190803A}{0.46}{GW190731A}{0.52}{GW190728A}{0.51}{GW190727A}{0.53}{GW190720A}{0.59}{GW190719A}{0.61}{GW190708A}{0.46}{GW190707A}{0.35}{GW190706A}{0.59}{GW190701A}{0.44}{GW190630A}{0.44}{GW190620A}{0.60}{GW190602A}{0.57}{GW190527A}{0.58}{GW190521B}{0.39}{GW190521A}{0.50}{GW190519A}{0.59}{GW190517A}{0.43}{GW190514A}{0.44}{GW190513A}{0.51}{GW190512A}{0.46}{GW190503A}{0.42}{GW190426A}{0.03}{GW190425A}{0.30}{GW190424A}{0.51}{GW190421A}{0.42}{GW190413B}{0.49}{GW190413A}{0.47}{GW190412A}{0.61}{GW190408A}{0.41}}}
\newcommand{\phioneIMRminus}[1]{\IfEqCase{#1}{{GW190930A}{2.83}{GW190929A}{2.72}{GW190924A}{2.81}{GW190915A}{2.79}{GW190910A}{2.75}{GW190909A}{2.81}{GW190828B}{2.79}{GW190828A}{2.89}{GW190814A}{3.11}{GW190803A}{2.78}{GW190731A}{2.77}{GW190728A}{2.81}{GW190727A}{2.85}{GW190720A}{2.82}{GW190719A}{2.84}{GW190708A}{2.76}{GW190707A}{2.82}{GW190706A}{2.82}{GW190701A}{2.86}{GW190630A}{2.89}{GW190620A}{2.88}{GW190602A}{2.79}{GW190527A}{2.87}{GW190521B}{2.73}{GW190521A}{2.82}{GW190519A}{2.84}{GW190517A}{2.83}{GW190514A}{2.77}{GW190513A}{2.97}{GW190512A}{2.84}{GW190503A}{2.86}{GW190426A}{0.00}{GW190425A}{2.73}{GW190424A}{2.94}{GW190421A}{2.76}{GW190413B}{2.75}{GW190413A}{2.85}{GW190412A}{2.13}{GW190408A}{2.84}}}
\newcommand{\phioneIMRmed}[1]{\IfEqCase{#1}{{GW190930A}{3.12}{GW190929A}{3.05}{GW190924A}{3.11}{GW190915A}{3.09}{GW190910A}{3.08}{GW190909A}{3.12}{GW190828B}{3.10}{GW190828A}{3.19}{GW190814A}{3.43}{GW190803A}{3.09}{GW190731A}{3.05}{GW190728A}{3.12}{GW190727A}{3.16}{GW190720A}{3.13}{GW190719A}{3.22}{GW190708A}{3.08}{GW190707A}{3.13}{GW190706A}{3.10}{GW190701A}{3.19}{GW190630A}{3.24}{GW190620A}{3.17}{GW190602A}{3.09}{GW190527A}{3.18}{GW190521B}{3.06}{GW190521A}{3.22}{GW190519A}{3.16}{GW190517A}{3.16}{GW190514A}{3.08}{GW190513A}{3.29}{GW190512A}{3.15}{GW190503A}{3.19}{GW190426A}{0.00}{GW190425A}{3.05}{GW190424A}{3.28}{GW190421A}{3.07}{GW190413B}{3.07}{GW190413A}{3.17}{GW190412A}{2.50}{GW190408A}{3.15}}}
\newcommand{\phioneIMRplus}[1]{\IfEqCase{#1}{{GW190930A}{2.84}{GW190929A}{2.91}{GW190924A}{2.85}{GW190915A}{2.88}{GW190910A}{2.88}{GW190909A}{2.81}{GW190828B}{2.86}{GW190828A}{2.80}{GW190814A}{2.54}{GW190803A}{2.86}{GW190731A}{2.92}{GW190728A}{2.85}{GW190727A}{2.81}{GW190720A}{2.83}{GW190719A}{2.75}{GW190708A}{2.87}{GW190707A}{2.85}{GW190706A}{2.88}{GW190701A}{2.80}{GW190630A}{2.69}{GW190620A}{2.81}{GW190602A}{2.86}{GW190527A}{2.78}{GW190521B}{2.90}{GW190521A}{2.69}{GW190519A}{2.82}{GW190517A}{2.77}{GW190514A}{2.87}{GW190513A}{2.70}{GW190512A}{2.83}{GW190503A}{2.76}{GW190426A}{0.00}{GW190425A}{2.90}{GW190424A}{2.68}{GW190421A}{2.89}{GW190413B}{2.88}{GW190413A}{2.80}{GW190412A}{3.29}{GW190408A}{2.84}}}
\newcommand{\phitwoIMRminus}[1]{\IfEqCase{#1}{{GW190930A}{2.88}{GW190929A}{2.82}{GW190924A}{2.81}{GW190915A}{2.86}{GW190910A}{2.78}{GW190909A}{2.79}{GW190828B}{2.71}{GW190828A}{2.82}{GW190814A}{2.78}{GW190803A}{2.84}{GW190731A}{2.82}{GW190728A}{2.84}{GW190727A}{2.79}{GW190720A}{2.76}{GW190719A}{2.83}{GW190708A}{2.81}{GW190707A}{2.82}{GW190706A}{2.81}{GW190701A}{2.81}{GW190630A}{2.79}{GW190620A}{2.80}{GW190602A}{2.74}{GW190527A}{2.78}{GW190521B}{2.73}{GW190521A}{2.60}{GW190519A}{2.84}{GW190517A}{2.90}{GW190514A}{2.76}{GW190513A}{2.82}{GW190512A}{2.76}{GW190503A}{2.81}{GW190426A}{0.00}{GW190425A}{2.84}{GW190424A}{2.86}{GW190421A}{2.78}{GW190413B}{2.81}{GW190413A}{2.83}{GW190412A}{2.62}{GW190408A}{2.80}}}
\newcommand{\phitwoIMRmed}[1]{\IfEqCase{#1}{{GW190930A}{3.22}{GW190929A}{3.13}{GW190924A}{3.11}{GW190915A}{3.18}{GW190910A}{3.10}{GW190909A}{3.10}{GW190828B}{3.04}{GW190828A}{3.17}{GW190814A}{3.08}{GW190803A}{3.16}{GW190731A}{3.11}{GW190728A}{3.15}{GW190727A}{3.11}{GW190720A}{3.08}{GW190719A}{3.13}{GW190708A}{3.14}{GW190707A}{3.15}{GW190706A}{3.12}{GW190701A}{3.11}{GW190630A}{3.09}{GW190620A}{3.11}{GW190602A}{3.05}{GW190527A}{3.11}{GW190521B}{3.06}{GW190521A}{2.93}{GW190519A}{3.17}{GW190517A}{3.20}{GW190514A}{3.10}{GW190513A}{3.14}{GW190512A}{3.07}{GW190503A}{3.12}{GW190426A}{0.00}{GW190425A}{3.15}{GW190424A}{3.17}{GW190421A}{3.13}{GW190413B}{3.12}{GW190413A}{3.14}{GW190412A}{2.91}{GW190408A}{3.12}}}
\newcommand{\phitwoIMRplus}[1]{\IfEqCase{#1}{{GW190930A}{2.73}{GW190929A}{2.84}{GW190924A}{2.84}{GW190915A}{2.78}{GW190910A}{2.87}{GW190909A}{2.85}{GW190828B}{2.92}{GW190828A}{2.77}{GW190814A}{2.88}{GW190803A}{2.82}{GW190731A}{2.85}{GW190728A}{2.82}{GW190727A}{2.84}{GW190720A}{2.88}{GW190719A}{2.87}{GW190708A}{2.83}{GW190707A}{2.83}{GW190706A}{2.86}{GW190701A}{2.86}{GW190630A}{2.88}{GW190620A}{2.86}{GW190602A}{2.91}{GW190527A}{2.87}{GW190521B}{2.88}{GW190521A}{3.01}{GW190519A}{2.79}{GW190517A}{2.78}{GW190514A}{2.89}{GW190513A}{2.84}{GW190512A}{2.90}{GW190503A}{2.81}{GW190426A}{0.00}{GW190425A}{2.83}{GW190424A}{2.77}{GW190421A}{2.83}{GW190413B}{2.83}{GW190413A}{2.83}{GW190412A}{3.05}{GW190408A}{2.83}}}
\newcommand{\chieffIMRminus}[1]{\IfEqCase{#1}{{GW190930A}{0.16}{GW190929A}{0.33}{GW190924A}{0.11}{GW190915A}{0.23}{GW190910A}{0.20}{GW190909A}{0.35}{GW190828B}{0.15}{GW190828A}{0.16}{GW190814A}{0.06}{GW190803A}{0.28}{GW190731A}{0.24}{GW190728A}{0.07}{GW190727A}{0.25}{GW190720A}{0.11}{GW190719A}{0.31}{GW190708A}{0.08}{GW190707A}{0.07}{GW190706A}{0.29}{GW190701A}{0.29}{GW190630A}{0.14}{GW190620A}{0.27}{GW190602A}{0.24}{GW190527A}{0.28}{GW190521B}{0.12}{GW190521A}{0.43}{GW190519A}{0.21}{GW190517A}{0.22}{GW190514A}{0.32}{GW190513A}{0.17}{GW190512A}{0.16}{GW190503A}{0.24}{GW190426A}{0.30}{GW190425A}{0.05}{GW190424A}{0.25}{GW190421A}{0.25}{GW190413B}{0.35}{GW190413A}{0.33}{GW190412A}{0.10}{GW190408A}{0.18}}}
\newcommand{\chieffIMRmed}[1]{\IfEqCase{#1}{{GW190930A}{0.15}{GW190929A}{-0.03}{GW190924A}{0.05}{GW190915A}{0.02}{GW190910A}{-0.01}{GW190909A}{-0.13}{GW190828B}{0.01}{GW190828A}{0.16}{GW190814A}{-0.01}{GW190803A}{-0.02}{GW190731A}{0.06}{GW190728A}{0.12}{GW190727A}{0.11}{GW190720A}{0.16}{GW190719A}{0.28}{GW190708A}{0.01}{GW190707A}{-0.07}{GW190706A}{0.38}{GW190701A}{-0.08}{GW190630A}{0.06}{GW190620A}{0.30}{GW190602A}{0.09}{GW190527A}{0.07}{GW190521B}{0.04}{GW190521A}{0.03}{GW190519A}{0.37}{GW190517A}{0.54}{GW190514A}{-0.15}{GW190513A}{0.05}{GW190512A}{0.002}{GW190503A}{-0.04}{GW190426A}{-0.03}{GW190425A}{0.06}{GW190424A}{0.13}{GW190421A}{-0.09}{GW190413B}{-0.03}{GW190413A}{-0.05}{GW190412A}{0.22}{GW190408A}{-0.05}}}
\newcommand{\chieffIMRplus}[1]{\IfEqCase{#1}{{GW190930A}{0.35}{GW190929A}{0.37}{GW190924A}{0.37}{GW190915A}{0.20}{GW190910A}{0.18}{GW190909A}{0.29}{GW190828B}{0.16}{GW190828A}{0.15}{GW190814A}{0.06}{GW190803A}{0.24}{GW190731A}{0.25}{GW190728A}{0.24}{GW190727A}{0.26}{GW190720A}{0.16}{GW190719A}{0.31}{GW190708A}{0.13}{GW190707A}{0.12}{GW190706A}{0.21}{GW190701A}{0.23}{GW190630A}{0.15}{GW190620A}{0.22}{GW190602A}{0.24}{GW190527A}{0.29}{GW190521B}{0.13}{GW190521A}{0.31}{GW190519A}{0.17}{GW190517A}{0.15}{GW190514A}{0.30}{GW190513A}{0.23}{GW190512A}{0.14}{GW190503A}{0.21}{GW190426A}{0.32}{GW190425A}{0.11}{GW190424A}{0.23}{GW190421A}{0.23}{GW190413B}{0.27}{GW190413A}{0.27}{GW190412A}{0.08}{GW190408A}{0.14}}}
\newcommand{\chipIMRminus}[1]{\IfEqCase{#1}{{GW190930A}{0.25}{GW190929A}{0.46}{GW190924A}{0.19}{GW190915A}{0.38}{GW190910A}{0.34}{GW190909A}{0.32}{GW190828B}{0.19}{GW190828A}{0.30}{GW190814A}{0.03}{GW190803A}{0.34}{GW190731A}{0.31}{GW190728A}{0.21}{GW190727A}{0.38}{GW190720A}{0.24}{GW190719A}{0.30}{GW190708A}{0.22}{GW190707A}{0.23}{GW190706A}{0.29}{GW190701A}{0.31}{GW190630A}{0.24}{GW190620A}{0.29}{GW190602A}{0.30}{GW190527A}{0.34}{GW190521B}{0.30}{GW190521A}{0.43}{GW190519A}{0.33}{GW190517A}{0.29}{GW190514A}{0.33}{GW190513A}{0.23}{GW190512A}{0.21}{GW190503A}{0.26}{GW190426A}{0.00}{GW190425A}{0.27}{GW190424A}{0.41}{GW190421A}{0.34}{GW190413B}{0.37}{GW190413A}{0.31}{GW190412A}{0.17}{GW190408A}{0.28}}}
\newcommand{\chipIMRmed}[1]{\IfEqCase{#1}{{GW190930A}{0.34}{GW190929A}{0.60}{GW190924A}{0.25}{GW190915A}{0.55}{GW190910A}{0.43}{GW190909A}{0.45}{GW190828B}{0.24}{GW190828A}{0.41}{GW190814A}{0.04}{GW190803A}{0.45}{GW190731A}{0.42}{GW190728A}{0.30}{GW190727A}{0.50}{GW190720A}{0.35}{GW190719A}{0.40}{GW190708A}{0.28}{GW190707A}{0.29}{GW190706A}{0.45}{GW190701A}{0.42}{GW190630A}{0.31}{GW190620A}{0.44}{GW190602A}{0.39}{GW190527A}{0.43}{GW190521B}{0.39}{GW190521A}{0.59}{GW190519A}{0.52}{GW190517A}{0.54}{GW190514A}{0.47}{GW190513A}{0.30}{GW190512A}{0.26}{GW190503A}{0.35}{GW190426A}{0.00}{GW190425A}{0.34}{GW190424A}{0.55}{GW190421A}{0.46}{GW190413B}{0.51}{GW190413A}{0.41}{GW190412A}{0.30}{GW190408A}{0.37}}}
\newcommand{\chipIMRplus}[1]{\IfEqCase{#1}{{GW190930A}{0.42}{GW190929A}{0.30}{GW190924A}{0.46}{GW190915A}{0.36}{GW190910A}{0.43}{GW190909A}{0.42}{GW190828B}{0.43}{GW190828A}{0.41}{GW190814A}{0.04}{GW190803A}{0.41}{GW190731A}{0.44}{GW190728A}{0.42}{GW190727A}{0.39}{GW190720A}{0.46}{GW190719A}{0.40}{GW190708A}{0.41}{GW190707A}{0.42}{GW190706A}{0.36}{GW190701A}{0.43}{GW190630A}{0.40}{GW190620A}{0.36}{GW190602A}{0.45}{GW190527A}{0.43}{GW190521B}{0.43}{GW190521A}{0.33}{GW190519A}{0.32}{GW190517A}{0.26}{GW190514A}{0.41}{GW190513A}{0.46}{GW190512A}{0.42}{GW190503A}{0.44}{GW190426A}{0.00}{GW190425A}{0.43}{GW190424A}{0.35}{GW190421A}{0.43}{GW190413B}{0.39}{GW190413A}{0.42}{GW190412A}{0.25}{GW190408A}{0.45}}}
\newcommand{\costiltoneIMRminus}[1]{\IfEqCase{#1}{{GW190930A}{1.15}{GW190929A}{0.75}{GW190924A}{1.07}{GW190915A}{0.78}{GW190910A}{0.87}{GW190909A}{0.62}{GW190828B}{0.92}{GW190828A}{1.13}{GW190814A}{0.92}{GW190803A}{0.83}{GW190731A}{1.00}{GW190728A}{1.12}{GW190727A}{1.00}{GW190720A}{0.84}{GW190719A}{1.05}{GW190708A}{0.89}{GW190707A}{0.64}{GW190706A}{0.84}{GW190701A}{0.69}{GW190630A}{0.91}{GW190620A}{0.86}{GW190602A}{0.96}{GW190527A}{0.95}{GW190521B}{0.91}{GW190521A}{0.96}{GW190519A}{0.61}{GW190517A}{0.41}{GW190514A}{0.61}{GW190513A}{0.97}{GW190512A}{0.85}{GW190503A}{0.79}{GW190426A}{0.00}{GW190425A}{0.65}{GW190424A}{0.95}{GW190421A}{0.71}{GW190413B}{0.78}{GW190413A}{0.78}{GW190412A}{0.53}{GW190408A}{0.74}}}
\newcommand{\costiltoneIMRmed}[1]{\IfEqCase{#1}{{GW190930A}{0.55}{GW190929A}{-0.08}{GW190924A}{0.32}{GW190915A}{0.06}{GW190910A}{0.04}{GW190909A}{-0.32}{GW190828B}{0.11}{GW190828A}{0.46}{GW190814A}{0.01}{GW190803A}{-0.05}{GW190731A}{0.20}{GW190728A}{0.52}{GW190727A}{0.30}{GW190720A}{0.55}{GW190719A}{0.65}{GW190708A}{0.09}{GW190707A}{-0.21}{GW190706A}{0.73}{GW190701A}{-0.23}{GW190630A}{0.24}{GW190620A}{0.65}{GW190602A}{0.23}{GW190527A}{0.18}{GW190521B}{0.17}{GW190521A}{0.11}{GW190519A}{0.70}{GW190517A}{0.79}{GW190514A}{-0.34}{GW190513A}{0.22}{GW190512A}{0.001}{GW190503A}{-0.12}{GW190426A}{-1.00}{GW190425A}{0.26}{GW190424A}{0.32}{GW190421A}{-0.20}{GW190413B}{-0.10}{GW190413A}{-0.13}{GW190412A}{0.61}{GW190408A}{-0.18}}}
\newcommand{\costiltoneIMRplus}[1]{\IfEqCase{#1}{{GW190930A}{0.42}{GW190929A}{0.68}{GW190924A}{0.64}{GW190915A}{0.72}{GW190910A}{0.81}{GW190909A}{1.04}{GW190828B}{0.78}{GW190828A}{0.48}{GW190814A}{0.90}{GW190803A}{0.89}{GW190731A}{0.69}{GW190728A}{0.44}{GW190727A}{0.61}{GW190720A}{0.42}{GW190719A}{0.32}{GW190708A}{0.77}{GW190707A}{0.99}{GW190706A}{0.24}{GW190701A}{1.00}{GW190630A}{0.64}{GW190620A}{0.32}{GW190602A}{0.67}{GW190527A}{0.70}{GW190521B}{0.68}{GW190521A}{0.72}{GW190519A}{0.27}{GW190517A}{0.19}{GW190514A}{1.03}{GW190513A}{0.69}{GW190512A}{0.87}{GW190503A}{0.95}{GW190426A}{2.00}{GW190425A}{0.61}{GW190424A}{0.59}{GW190421A}{0.92}{GW190413B}{0.90}{GW190413A}{0.97}{GW190412A}{0.30}{GW190408A}{0.98}}}
\newcommand{\costilttwoIMRminus}[1]{\IfEqCase{#1}{{GW190930A}{1.02}{GW190929A}{0.91}{GW190924A}{0.90}{GW190915A}{0.90}{GW190910A}{0.77}{GW190909A}{0.67}{GW190828B}{0.85}{GW190828A}{1.11}{GW190814A}{0.79}{GW190803A}{0.83}{GW190731A}{0.95}{GW190728A}{1.11}{GW190727A}{0.99}{GW190720A}{0.96}{GW190719A}{1.14}{GW190708A}{0.92}{GW190707A}{0.65}{GW190706A}{1.18}{GW190701A}{0.71}{GW190630A}{0.94}{GW190620A}{1.13}{GW190602A}{1.04}{GW190527A}{0.99}{GW190521B}{0.88}{GW190521A}{0.89}{GW190519A}{1.15}{GW190517A}{1.08}{GW190514A}{0.61}{GW190513A}{1.03}{GW190512A}{0.90}{GW190503A}{0.78}{GW190426A}{0.00}{GW190425A}{0.87}{GW190424A}{1.01}{GW190421A}{0.67}{GW190413B}{0.84}{GW190413A}{0.79}{GW190412A}{1.15}{GW190408A}{0.72}}}
\newcommand{\costilttwoIMRmed}[1]{\IfEqCase{#1}{{GW190930A}{0.19}{GW190929A}{0.02}{GW190924A}{0.04}{GW190915A}{0.03}{GW190910A}{-0.12}{GW190909A}{-0.26}{GW190828B}{-0.03}{GW190828A}{0.31}{GW190814A}{-0.11}{GW190803A}{-0.07}{GW190731A}{0.08}{GW190728A}{0.29}{GW190727A}{0.14}{GW190720A}{0.11}{GW190719A}{0.36}{GW190708A}{0.08}{GW190707A}{-0.26}{GW190706A}{0.46}{GW190701A}{-0.21}{GW190630A}{0.09}{GW190620A}{0.40}{GW190602A}{0.21}{GW190527A}{0.11}{GW190521B}{0.05}{GW190521A}{0.0006}{GW190519A}{0.42}{GW190517A}{0.63}{GW190514A}{-0.33}{GW190513A}{0.18}{GW190512A}{0.04}{GW190503A}{-0.14}{GW190426A}{-1.00}{GW190425A}{0.16}{GW190424A}{0.20}{GW190421A}{-0.25}{GW190413B}{-0.03}{GW190413A}{-0.14}{GW190412A}{0.33}{GW190408A}{-0.18}}}
\newcommand{\costilttwoIMRplus}[1]{\IfEqCase{#1}{{GW190930A}{0.73}{GW190929A}{0.89}{GW190924A}{0.85}{GW190915A}{0.83}{GW190910A}{0.95}{GW190909A}{1.08}{GW190828B}{0.91}{GW190828A}{0.63}{GW190814A}{0.93}{GW190803A}{0.93}{GW190731A}{0.81}{GW190728A}{0.64}{GW190727A}{0.77}{GW190720A}{0.80}{GW190719A}{0.58}{GW190708A}{0.81}{GW190707A}{1.06}{GW190706A}{0.51}{GW190701A}{1.03}{GW190630A}{0.79}{GW190620A}{0.55}{GW190602A}{0.72}{GW190527A}{0.80}{GW190521B}{0.82}{GW190521A}{0.85}{GW190519A}{0.53}{GW190517A}{0.34}{GW190514A}{1.07}{GW190513A}{0.75}{GW190512A}{0.84}{GW190503A}{0.96}{GW190426A}{2.00}{GW190425A}{0.70}{GW190424A}{0.70}{GW190421A}{1.03}{GW190413B}{0.89}{GW190413A}{0.99}{GW190412A}{0.61}{GW190408A}{0.96}}}
\newcommand{\comovingdistIMRminus}[1]{\IfEqCase{#1}{{GW190930A}{256}{GW190929A}{521}{GW190924A}{174}{GW190915A}{390}{GW190910A}{537}{GW190909A}{1072}{GW190828B}{458}{GW190828A}{533}{GW190814A}{39}{GW190803A}{773}{GW190731A}{856}{GW190728A}{293}{GW190727A}{634}{GW190720A}{234}{GW190719A}{1002}{GW190708A}{301}{GW190707A}{274}{GW190706A}{940}{GW190701A}{455}{GW190630A}{310}{GW190620A}{734}{GW190602A}{697}{GW190527A}{679}{GW190521B}{359}{GW190521A}{815}{GW190519A}{714}{GW190517A}{523}{GW190514A}{997}{GW190513A}{511}{GW190512A}{414}{GW190503A}{420}{GW190426A}{143}{GW190425A}{67}{GW190424A}{706}{GW190421A}{658}{GW190413B}{905}{GW190413A}{668}{GW190412A}{164}{GW190408A}{387}}}
\newcommand{\comovingdistIMRmed}[1]{\IfEqCase{#1}{{GW190930A}{659}{GW190929A}{1382}{GW190924A}{500}{GW190915A}{1233}{GW190910A}{1351}{GW190909A}{2254}{GW190828B}{1274}{GW190828A}{1485}{GW190814A}{235}{GW190803A}{1967}{GW190731A}{2038}{GW190728A}{749}{GW190727A}{1861}{GW190720A}{643}{GW190719A}{2173}{GW190708A}{722}{GW190707A}{634}{GW190706A}{2646}{GW190701A}{1394}{GW190630A}{782}{GW190620A}{1805}{GW190602A}{1646}{GW190527A}{1598}{GW190521B}{928}{GW190521A}{2526}{GW190519A}{2189}{GW190517A}{1368}{GW190514A}{2289}{GW190513A}{1421}{GW190512A}{1120}{GW190503A}{1145}{GW190426A}{344}{GW190425A}{151}{GW190424A}{1442}{GW190421A}{1698}{GW190413B}{2293}{GW190413A}{2056}{GW190412A}{643}{GW190408A}{1104}}}
\newcommand{\comovingdistIMRplus}[1]{\IfEqCase{#1}{{GW190930A}{249}{GW190929A}{1394}{GW190924A}{165}{GW190915A}{412}{GW190910A}{475}{GW190909A}{1119}{GW190828B}{368}{GW190828A}{329}{GW190814A}{35}{GW190803A}{737}{GW190731A}{916}{GW190728A}{174}{GW190727A}{572}{GW190720A}{391}{GW190719A}{1070}{GW190708A}{240}{GW190707A}{269}{GW190706A}{787}{GW190701A}{422}{GW190630A}{331}{GW190620A}{730}{GW190602A}{811}{GW190527A}{1040}{GW190521B}{258}{GW190521A}{584}{GW190519A}{673}{GW190517A}{777}{GW190514A}{964}{GW190513A}{391}{GW190512A}{327}{GW190503A}{362}{GW190426A}{154}{GW190425A}{64}{GW190424A}{750}{GW190421A}{658}{GW190413B}{884}{GW190413A}{782}{GW190412A}{113}{GW190408A}{271}}}
\newcommand{\cosiotaIMRminus}[1]{\IfEqCase{#1}{{GW190930A}{1.53}{GW190929A}{0.69}{GW190924A}{1.73}{GW190915A}{0.63}{GW190910A}{0.84}{GW190909A}{0.97}{GW190828B}{0.84}{GW190828A}{0.29}{GW190814A}{1.45}{GW190803A}{1.18}{GW190731A}{1.09}{GW190728A}{1.31}{GW190727A}{1.39}{GW190720A}{0.15}{GW190719A}{0.95}{GW190708A}{1.18}{GW190707A}{0.39}{GW190706A}{1.01}{GW190701A}{0.58}{GW190630A}{1.50}{GW190620A}{0.45}{GW190602A}{0.81}{GW190527A}{1.26}{GW190521B}{1.37}{GW190521A}{1.43}{GW190519A}{0.81}{GW190517A}{0.39}{GW190514A}{0.96}{GW190513A}{1.65}{GW190512A}{0.96}{GW190503A}{0.20}{GW190426A}{0.84}{GW190425A}{1.42}{GW190424A}{0.99}{GW190421A}{0.83}{GW190413B}{0.81}{GW190413A}{1.61}{GW190412A}{1.27}{GW190408A}{0.94}}}
\newcommand{\cosiotaIMRmed}[1]{\IfEqCase{#1}{{GW190930A}{0.57}{GW190929A}{-0.23}{GW190924A}{0.81}{GW190915A}{-0.32}{GW190910A}{-0.13}{GW190909A}{0.05}{GW190828B}{-0.13}{GW190828A}{-0.69}{GW190814A}{0.69}{GW190803A}{0.24}{GW190731A}{0.14}{GW190728A}{0.34}{GW190727A}{0.45}{GW190720A}{-0.84}{GW190719A}{0.00}{GW190708A}{0.20}{GW190707A}{-0.59}{GW190706A}{0.08}{GW190701A}{0.71}{GW190630A}{0.53}{GW190620A}{-0.51}{GW190602A}{-0.12}{GW190527A}{0.35}{GW190521B}{0.46}{GW190521A}{0.48}{GW190519A}{-0.12}{GW190517A}{-0.55}{GW190514A}{0.01}{GW190513A}{0.76}{GW190512A}{-0.01}{GW190503A}{-0.78}{GW190426A}{-0.13}{GW190425A}{0.46}{GW190424A}{0.04}{GW190421A}{-0.10}{GW190413B}{-0.11}{GW190413A}{0.67}{GW190412A}{0.74}{GW190408A}{-0.03}}}
\newcommand{\cosiotaIMRplus}[1]{\IfEqCase{#1}{{GW190930A}{0.41}{GW190929A}{1.13}{GW190924A}{0.17}{GW190915A}{1.25}{GW190910A}{1.09}{GW190909A}{0.87}{GW190828B}{1.10}{GW190828A}{1.64}{GW190814A}{0.16}{GW190803A}{0.72}{GW190731A}{0.82}{GW190728A}{0.63}{GW190727A}{0.51}{GW190720A}{1.61}{GW190719A}{0.95}{GW190708A}{0.78}{GW190707A}{1.56}{GW190706A}{0.86}{GW190701A}{0.26}{GW190630A}{0.45}{GW190620A}{1.43}{GW190602A}{1.05}{GW190527A}{0.61}{GW190521B}{0.47}{GW190521A}{0.48}{GW190519A}{1.04}{GW190517A}{1.25}{GW190514A}{0.92}{GW190513A}{0.22}{GW190512A}{0.98}{GW190503A}{0.55}{GW190426A}{1.09}{GW190425A}{0.51}{GW190424A}{0.90}{GW190421A}{1.02}{GW190413B}{1.00}{GW190413A}{0.31}{GW190412A}{0.17}{GW190408A}{1.00}}}
\newcommand{\finalspinIMRminus}[1]{\IfEqCase{#1}{{GW190930A}{0.07}{GW190929A}{0.30}{GW190924A}{0.05}{GW190915A}{0.12}{GW190910A}{0.08}{GW190909A}{0.19}{GW190828A}{0.07}{GW190814A}{0.03}{GW190803A}{0.13}{GW190731A}{0.14}{GW190728A}{0.04}{GW190727A}{0.12}{GW190720A}{0.05}{GW190719A}{0.20}{GW190708A}{0.05}{GW190707A}{0.04}{GW190706A}{0.13}{GW190701A}{0.14}{GW190630A}{0.08}{GW190620A}{0.17}{GW190602A}{0.20}{GW190527A}{0.22}{GW190521B}{0.07}{GW190521A}{0.16}{GW190519A}{0.10}{GW190517A}{0.09}{GW190514A}{0.14}{GW190513A}{0.12}{GW190512A}{0.10}{GW190424A}{0.10}{GW190421A}{0.13}{GW190413B}{0.22}{GW190413A}{0.13}{GW190412A}{0.07}}}
\newcommand{\finalspinIMRmed}[1]{\IfEqCase{#1}{{GW190930A}{0.72}{GW190929A}{0.63}{GW190924A}{0.67}{GW190915A}{0.70}{GW190910A}{0.69}{GW190909A}{0.63}{GW190828A}{0.74}{GW190814A}{0.28}{GW190803A}{0.68}{GW190731A}{0.71}{GW190728A}{0.71}{GW190727A}{0.73}{GW190720A}{0.72}{GW190719A}{0.77}{GW190708A}{0.68}{GW190707A}{0.66}{GW190706A}{0.81}{GW190701A}{0.66}{GW190630A}{0.69}{GW190620A}{0.78}{GW190602A}{0.69}{GW190527A}{0.69}{GW190521B}{0.70}{GW190521A}{0.71}{GW190519A}{0.82}{GW190517A}{0.87}{GW190514A}{0.64}{GW190513A}{0.66}{GW190512A}{0.65}{GW190424A}{0.74}{GW190421A}{0.66}{GW190413B}{0.67}{GW190413A}{0.67}{GW190412A}{0.67}}}
\newcommand{\finalspinIMRplus}[1]{\IfEqCase{#1}{{GW190930A}{0.08}{GW190929A}{0.21}{GW190924A}{0.06}{GW190915A}{0.10}{GW190910A}{0.08}{GW190909A}{0.13}{GW190828A}{0.07}{GW190814A}{0.02}{GW190803A}{0.11}{GW190731A}{0.11}{GW190728A}{0.05}{GW190727A}{0.10}{GW190720A}{0.07}{GW190719A}{0.13}{GW190708A}{0.05}{GW190707A}{0.04}{GW190706A}{0.08}{GW190701A}{0.10}{GW190630A}{0.08}{GW190620A}{0.09}{GW190602A}{0.12}{GW190527A}{0.14}{GW190521B}{0.07}{GW190521A}{0.12}{GW190519A}{0.06}{GW190517A}{0.04}{GW190514A}{0.12}{GW190513A}{0.13}{GW190512A}{0.07}{GW190424A}{0.09}{GW190421A}{0.11}{GW190413B}{0.12}{GW190413A}{0.11}{GW190412A}{0.07}}}
\newcommand{\finalmassdetIMRminus}[1]{\IfEqCase{#1}{{GW190930A}{1.0}{GW190929A}{23.7}{GW190924A}{1.1}{GW190915A}{6.0}{GW190910A}{7.1}{GW190909A}{17.8}{GW190828A}{4.7}{GW190814A}{1.1}{GW190803A}{11.0}{GW190731A}{12.6}{GW190728A}{0.7}{GW190727A}{9.8}{GW190720A}{1.3}{GW190719A}{14.0}{GW190708A}{1.0}{GW190707A}{0.5}{GW190706A}{19.0}{GW190701A}{13.3}{GW190630A}{3.8}{GW190620A}{16.2}{GW190602A}{21.0}{GW190527A}{10.0}{GW190521B}{4.1}{GW190521A}{29.7}{GW190519A}{13.3}{GW190517A}{5.8}{GW190514A}{13.3}{GW190513A}{6.4}{GW190512A}{2.8}{GW190424A}{10.3}{GW190421A}{11.4}{GW190413B}{16.6}{GW190413A}{12.3}{GW190412A}{3.9}}}
\newcommand{\finalmassdetIMRmed}[1]{\IfEqCase{#1}{{GW190930A}{22.2}{GW190929A}{142.1}{GW190924A}{15.2}{GW190915A}{75.3}{GW190910A}{96.9}{GW190909A}{109.4}{GW190828A}{74.8}{GW190814A}{26.9}{GW190803A}{96.3}{GW190731A}{105.6}{GW190728A}{22.7}{GW190727A}{99.8}{GW190720A}{23.8}{GW190719A}{90.3}{GW190708A}{34.6}{GW190707A}{22.0}{GW190706A}{175.2}{GW190701A}{124.3}{GW190630A}{66.6}{GW190620A}{130.1}{GW190602A}{161.9}{GW190527A}{81.4}{GW190521B}{87.0}{GW190521A}{251.6}{GW190519A}{151.3}{GW190517A}{79.8}{GW190514A}{109.6}{GW190513A}{69.3}{GW190512A}{43.0}{GW190424A}{95.8}{GW190421A}{103.1}{GW190413B}{130.3}{GW190413A}{88.0}{GW190412A}{41.2}}}
\newcommand{\finalmassdetIMRplus}[1]{\IfEqCase{#1}{{GW190930A}{13.4}{GW190929A}{32.3}{GW190924A}{8.7}{GW190915A}{7.5}{GW190910A}{8.2}{GW190909A}{34.0}{GW190828A}{5.8}{GW190814A}{1.3}{GW190803A}{12.8}{GW190731A}{14.5}{GW190728A}{6.9}{GW190727A}{10.9}{GW190720A}{5.9}{GW190719A}{86.5}{GW190708A}{4.0}{GW190707A}{2.1}{GW190706A}{17.6}{GW190701A}{14.9}{GW190630A}{5.1}{GW190620A}{16.4}{GW190602A}{22.2}{GW190527A}{61.3}{GW190521B}{4.8}{GW190521A}{32.9}{GW190519A}{14.3}{GW190517A}{8.6}{GW190514A}{15.6}{GW190513A}{9.2}{GW190512A}{5.3}{GW190424A}{11.6}{GW190421A}{12.8}{GW190413B}{19.4}{GW190413A}{13.9}{GW190412A}{4.5}}}
\newcommand{\redshiftIMRminus}[1]{\IfEqCase{#1}{{GW190930A}{0.06}{GW190929A}{0.14}{GW190924A}{0.04}{GW190915A}{0.10}{GW190910A}{0.14}{GW190909A}{0.31}{GW190828B}{0.12}{GW190828A}{0.14}{GW190814A}{0.009}{GW190803A}{0.22}{GW190731A}{0.24}{GW190728A}{0.07}{GW190727A}{0.18}{GW190720A}{0.06}{GW190719A}{0.29}{GW190708A}{0.07}{GW190707A}{0.07}{GW190706A}{0.29}{GW190701A}{0.12}{GW190630A}{0.08}{GW190620A}{0.20}{GW190602A}{0.19}{GW190527A}{0.18}{GW190521B}{0.09}{GW190521A}{0.25}{GW190519A}{0.21}{GW190517A}{0.14}{GW190514A}{0.29}{GW190513A}{0.13}{GW190512A}{0.10}{GW190503A}{0.11}{GW190426A}{0.03}{GW190425A}{0.02}{GW190424A}{0.18}{GW190421A}{0.18}{GW190413B}{0.27}{GW190413A}{0.19}{GW190412A}{0.04}{GW190408A}{0.10}}}
\newcommand{\redshiftIMRmed}[1]{\IfEqCase{#1}{{GW190930A}{0.16}{GW190929A}{0.34}{GW190924A}{0.12}{GW190915A}{0.30}{GW190910A}{0.33}{GW190909A}{0.60}{GW190828B}{0.31}{GW190828A}{0.37}{GW190814A}{0.05}{GW190803A}{0.51}{GW190731A}{0.53}{GW190728A}{0.18}{GW190727A}{0.48}{GW190720A}{0.15}{GW190719A}{0.57}{GW190708A}{0.17}{GW190707A}{0.15}{GW190706A}{0.72}{GW190701A}{0.34}{GW190630A}{0.19}{GW190620A}{0.46}{GW190602A}{0.41}{GW190527A}{0.40}{GW190521B}{0.22}{GW190521A}{0.68}{GW190519A}{0.58}{GW190517A}{0.34}{GW190514A}{0.61}{GW190513A}{0.35}{GW190512A}{0.27}{GW190503A}{0.28}{GW190426A}{0.08}{GW190425A}{0.03}{GW190424A}{0.36}{GW190421A}{0.43}{GW190413B}{0.61}{GW190413A}{0.53}{GW190412A}{0.15}{GW190408A}{0.27}}}
\newcommand{\redshiftIMRplus}[1]{\IfEqCase{#1}{{GW190930A}{0.06}{GW190929A}{0.43}{GW190924A}{0.04}{GW190915A}{0.11}{GW190910A}{0.13}{GW190909A}{0.40}{GW190828B}{0.10}{GW190828A}{0.09}{GW190814A}{0.008}{GW190803A}{0.24}{GW190731A}{0.30}{GW190728A}{0.04}{GW190727A}{0.18}{GW190720A}{0.10}{GW190719A}{0.37}{GW190708A}{0.06}{GW190707A}{0.07}{GW190706A}{0.29}{GW190701A}{0.12}{GW190630A}{0.08}{GW190620A}{0.23}{GW190602A}{0.25}{GW190527A}{0.32}{GW190521B}{0.07}{GW190521A}{0.21}{GW190519A}{0.22}{GW190517A}{0.22}{GW190514A}{0.34}{GW190513A}{0.11}{GW190512A}{0.09}{GW190503A}{0.10}{GW190426A}{0.04}{GW190425A}{0.01}{GW190424A}{0.22}{GW190421A}{0.20}{GW190413B}{0.31}{GW190413A}{0.26}{GW190412A}{0.03}{GW190408A}{0.07}}}
\newcommand{\massonesourceIMRminus}[1]{\IfEqCase{#1}{{GW190930A}{2.6}{GW190929A}{34.7}{GW190924A}{2.5}{GW190915A}{7.0}{GW190910A}{6.3}{GW190909A}{11.3}{GW190828B}{6.4}{GW190828A}{4.4}{GW190814A}{1.1}{GW190803A}{7.9}{GW190731A}{9.5}{GW190728A}{2.3}{GW190727A}{7.2}{GW190720A}{3.3}{GW190719A}{13.8}{GW190708A}{3.0}{GW190707A}{1.8}{GW190706A}{15.9}{GW190701A}{9.6}{GW190630A}{7.5}{GW190620A}{14.2}{GW190602A}{18.6}{GW190527A}{11.8}{GW190521B}{5.9}{GW190521A}{16.8}{GW190519A}{13.6}{GW190517A}{7.6}{GW190514A}{8.9}{GW190513A}{9.9}{GW190512A}{5.3}{GW190503A}{7.8}{GW190426A}{2.3}{GW190425A}{0.3}{GW190424A}{7.7}{GW190421A}{8.3}{GW190413B}{12.6}{GW190413A}{6.6}{GW190412A}{4.2}{GW190408A}{4.0}}}
\newcommand{\massonesourceIMRmed}[1]{\IfEqCase{#1}{{GW190930A}{12.5}{GW190929A}{84.9}{GW190924A}{9.4}{GW190915A}{36.6}{GW190910A}{42.8}{GW190909A}{43.4}{GW190828B}{22.5}{GW190828A}{32.5}{GW190814A}{23.2}{GW190803A}{38.9}{GW190731A}{42.7}{GW190728A}{12.3}{GW190727A}{41.1}{GW190720A}{13.8}{GW190719A}{41.1}{GW190708A}{18.2}{GW190707A}{11.7}{GW190706A}{66.2}{GW190701A}{56.6}{GW190630A}{36.6}{GW190620A}{59.7}{GW190602A}{77.1}{GW190527A}{40.4}{GW190521B}{43.2}{GW190521A}{92.7}{GW190519A}{62.1}{GW190517A}{37.8}{GW190514A}{40.7}{GW190513A}{36.1}{GW190512A}{22.6}{GW190503A}{42.8}{GW190426A}{5.7}{GW190425A}{2.0}{GW190424A}{42.0}{GW190421A}{43.9}{GW190413B}{51.5}{GW190413A}{34.3}{GW190412A}{28.2}{GW190408A}{25.4}}}
\newcommand{\massonesourceIMRplus}[1]{\IfEqCase{#1}{{GW190930A}{14.9}{GW190929A}{29.2}{GW190924A}{9.6}{GW190915A}{9.3}{GW190910A}{8.9}{GW190909A}{23.3}{GW190828B}{7.9}{GW190828A}{7.0}{GW190814A}{1.3}{GW190803A}{12.9}{GW190731A}{14.0}{GW190728A}{8.7}{GW190727A}{12.9}{GW190720A}{7.4}{GW190719A}{36.8}{GW190708A}{6.5}{GW190707A}{3.7}{GW190706A}{24.2}{GW190701A}{14.0}{GW190630A}{7.6}{GW190620A}{17.4}{GW190602A}{23.0}{GW190527A}{26.2}{GW190521B}{7.5}{GW190521A}{22.5}{GW190519A}{19.2}{GW190517A}{14.9}{GW190514A}{17.2}{GW190513A}{10.6}{GW190512A}{7.5}{GW190503A}{11.3}{GW190426A}{3.9}{GW190425A}{0.6}{GW190424A}{12.5}{GW190421A}{13.8}{GW190413B}{20.2}{GW190413A}{10.8}{GW190412A}{4.6}{GW190408A}{6.3}}}
\newcommand{\masstwosourceIMRminus}[1]{\IfEqCase{#1}{{GW190930A}{3.5}{GW190929A}{9.8}{GW190924A}{2.0}{GW190915A}{6.0}{GW190910A}{7.2}{GW190909A}{11.4}{GW190828B}{2.5}{GW190828A}{5.2}{GW190814A}{0.10}{GW190803A}{8.0}{GW190731A}{10.2}{GW190728A}{2.8}{GW190727A}{10.0}{GW190720A}{2.3}{GW190719A}{6.8}{GW190708A}{3.1}{GW190707A}{1.8}{GW190706A}{15.8}{GW190701A}{12.4}{GW190630A}{4.4}{GW190620A}{11.4}{GW190602A}{17.9}{GW190527A}{8.8}{GW190521B}{6.5}{GW190521A}{19.9}{GW190519A}{13.3}{GW190517A}{9.0}{GW190514A}{9.4}{GW190513A}{3.9}{GW190512A}{3.3}{GW190503A}{9.2}{GW190426A}{0.5}{GW190425A}{0.3}{GW190424A}{8.3}{GW190421A}{9.6}{GW190413B}{13.0}{GW190413A}{6.7}{GW190412A}{1.1}{GW190408A}{4.1}}}
\newcommand{\masstwosourceIMRmed}[1]{\IfEqCase{#1}{{GW190930A}{7.7}{GW190929A}{22.6}{GW190924A}{4.8}{GW190915A}{23.8}{GW190910A}{33.6}{GW190909A}{27.8}{GW190828B}{10.5}{GW190828A}{25.7}{GW190814A}{2.58}{GW190803A}{27.6}{GW190731A}{29.4}{GW190728A}{8.1}{GW190727A}{29.9}{GW190720A}{7.8}{GW190719A}{20.1}{GW190708A}{12.8}{GW190707A}{8.4}{GW190706A}{40.6}{GW190701A}{40.0}{GW190630A}{21.9}{GW190620A}{33.5}{GW190602A}{41.2}{GW190527A}{21.0}{GW190521B}{31.9}{GW190521A}{65.3}{GW190519A}{39.3}{GW190517A}{25.1}{GW190514A}{29.6}{GW190513A}{17.2}{GW190512A}{12.8}{GW190503A}{28.5}{GW190426A}{1.5}{GW190425A}{1.4}{GW190424A}{31.8}{GW190421A}{31.0}{GW190413B}{32.3}{GW190413A}{24.9}{GW190412A}{8.8}{GW190408A}{18.4}}}
\newcommand{\masstwosourceIMRplus}[1]{\IfEqCase{#1}{{GW190930A}{1.9}{GW190929A}{16.8}{GW190924A}{1.6}{GW190915A}{6.0}{GW190910A}{6.8}{GW190909A}{12.2}{GW190828B}{3.8}{GW190828A}{4.5}{GW190814A}{0.09}{GW190803A}{8.0}{GW190731A}{9.6}{GW190728A}{1.8}{GW190727A}{7.1}{GW190720A}{2.4}{GW190719A}{11.0}{GW190708A}{2.3}{GW190707A}{1.5}{GW190706A}{15.4}{GW190701A}{10.2}{GW190630A}{6.1}{GW190620A}{13.2}{GW190602A}{19.0}{GW190527A}{13.1}{GW190521B}{5.9}{GW190521A}{18.4}{GW190519A}{12.0}{GW190517A}{6.9}{GW190514A}{9.7}{GW190513A}{7.4}{GW190512A}{3.7}{GW190503A}{7.8}{GW190426A}{0.8}{GW190425A}{0.3}{GW190424A}{7.8}{GW190421A}{8.6}{GW190413B}{12.2}{GW190413A}{6.4}{GW190412A}{1.6}{GW190408A}{3.5}}}
\newcommand{\totalmasssourceIMRminus}[1]{\IfEqCase{#1}{{GW190930A}{1.6}{GW190929A}{26.1}{GW190924A}{1.2}{GW190915A}{6.1}{GW190910A}{7.6}{GW190909A}{14.4}{GW190828B}{3.9}{GW190828A}{4.5}{GW190814A}{1.0}{GW190803A}{9.5}{GW190731A}{12.1}{GW190728A}{1.4}{GW190727A}{8.0}{GW190720A}{1.9}{GW190719A}{12.7}{GW190708A}{2.1}{GW190707A}{1.4}{GW190706A}{15.4}{GW190701A}{10.0}{GW190630A}{4.7}{GW190620A}{13.0}{GW190602A}{16.0}{GW190527A}{10.7}{GW190521B}{5.0}{GW190521A}{16.8}{GW190519A}{12.4}{GW190517A}{8.8}{GW190514A}{12.0}{GW190513A}{6.3}{GW190512A}{3.4}{GW190503A}{7.5}{GW190426A}{1.5}{GW190425A}{0.1}{GW190424A}{11.1}{GW190421A}{9.9}{GW190413B}{13.7}{GW190413A}{8.2}{GW190412A}{2.8}{GW190408A}{3.3}}}
\newcommand{\totalmasssourceIMRmed}[1]{\IfEqCase{#1}{{GW190930A}{20.4}{GW190929A}{106.6}{GW190924A}{14.2}{GW190915A}{60.5}{GW190910A}{76.2}{GW190909A}{71.6}{GW190828B}{33.2}{GW190828A}{57.9}{GW190814A}{25.8}{GW190803A}{66.7}{GW190731A}{72.1}{GW190728A}{20.6}{GW190727A}{70.8}{GW190720A}{21.7}{GW190719A}{61.2}{GW190708A}{31.3}{GW190707A}{20.3}{GW190706A}{106.8}{GW190701A}{96.4}{GW190630A}{58.8}{GW190620A}{93.5}{GW190602A}{118.6}{GW190527A}{61.6}{GW190521B}{74.9}{GW190521A}{156.3}{GW190519A}{101.3}{GW190517A}{63.7}{GW190514A}{70.6}{GW190513A}{53.7}{GW190512A}{35.6}{GW190503A}{71.3}{GW190426A}{7.2}{GW190425A}{3.4}{GW190424A}{74.2}{GW190421A}{75.0}{GW190413B}{84.2}{GW190413A}{59.5}{GW190412A}{37.0}{GW190408A}{43.9}}}
\newcommand{\totalmasssourceIMRplus}[1]{\IfEqCase{#1}{{GW190930A}{11.2}{GW190929A}{30.0}{GW190924A}{7.5}{GW190915A}{7.5}{GW190910A}{10.5}{GW190909A}{24.7}{GW190828B}{5.8}{GW190828A}{7.0}{GW190814A}{1.2}{GW190803A}{13.1}{GW190731A}{15.9}{GW190728A}{5.7}{GW190727A}{11.7}{GW190720A}{4.9}{GW190719A}{40.5}{GW190708A}{3.6}{GW190707A}{2.0}{GW190706A}{22.6}{GW190701A}{13.2}{GW190630A}{5.5}{GW190620A}{16.3}{GW190602A}{20.0}{GW190527A}{29.3}{GW190521B}{6.8}{GW190521A}{30.7}{GW190519A}{17.1}{GW190517A}{9.2}{GW190514A}{18.4}{GW190513A}{8.8}{GW190512A}{4.9}{GW190503A}{9.4}{GW190426A}{3.5}{GW190425A}{0.3}{GW190424A}{13.1}{GW190421A}{13.9}{GW190413B}{18.2}{GW190413A}{10.9}{GW190412A}{3.6}{GW190408A}{4.6}}}
\newcommand{\chirpmasssourceIMRminus}[1]{\IfEqCase{#1}{{GW190930A}{0.5}{GW190929A}{8.1}{GW190924A}{0.2}{GW190915A}{2.6}{GW190910A}{3.4}{GW190909A}{6.3}{GW190828B}{0.9}{GW190828A}{1.9}{GW190814A}{0.05}{GW190803A}{4.0}{GW190731A}{5.3}{GW190728A}{0.3}{GW190727A}{4.0}{GW190720A}{0.7}{GW190719A}{4.4}{GW190708A}{0.7}{GW190707A}{0.5}{GW190706A}{6.9}{GW190701A}{5.2}{GW190630A}{1.8}{GW190620A}{6.0}{GW190602A}{9.9}{GW190527A}{4.0}{GW190521B}{2.4}{GW190521A}{8.4}{GW190519A}{5.8}{GW190517A}{3.9}{GW190514A}{5.2}{GW190513A}{1.7}{GW190512A}{1.0}{GW190503A}{4.1}{GW190426A}{0.08}{GW190425A}{0.02}{GW190424A}{4.7}{GW190421A}{4.3}{GW190413B}{6.1}{GW190413A}{3.4}{GW190412A}{0.4}{GW190408A}{1.4}}}
\newcommand{\chirpmasssourceIMRmed}[1]{\IfEqCase{#1}{{GW190930A}{8.5}{GW190929A}{35.5}{GW190924A}{5.8}{GW190915A}{25.3}{GW190910A}{32.7}{GW190909A}{29.6}{GW190828B}{13.1}{GW190828A}{24.8}{GW190814A}{6.08}{GW190803A}{28.1}{GW190731A}{30.2}{GW190728A}{8.6}{GW190727A}{30.0}{GW190720A}{9.0}{GW190719A}{24.2}{GW190708A}{13.2}{GW190707A}{8.6}{GW190706A}{43.6}{GW190701A}{40.8}{GW190630A}{24.4}{GW190620A}{38.1}{GW190602A}{47.7}{GW190527A}{24.3}{GW190521B}{32.0}{GW190521A}{66.1}{GW190519A}{42.0}{GW190517A}{26.6}{GW190514A}{29.9}{GW190513A}{21.2}{GW190512A}{14.6}{GW190503A}{29.9}{GW190426A}{2.41}{GW190425A}{1.44}{GW190424A}{31.6}{GW190421A}{31.6}{GW190413B}{34.6}{GW190413A}{25.2}{GW190412A}{13.2}{GW190408A}{18.6}}}
\newcommand{\chirpmasssourceIMRplus}[1]{\IfEqCase{#1}{{GW190930A}{0.5}{GW190929A}{12.1}{GW190924A}{0.2}{GW190915A}{3.2}{GW190910A}{4.6}{GW190909A}{9.9}{GW190828B}{1.4}{GW190828A}{3.1}{GW190814A}{0.06}{GW190803A}{5.6}{GW190731A}{7.1}{GW190728A}{0.6}{GW190727A}{5.0}{GW190720A}{0.5}{GW190719A}{11.9}{GW190708A}{0.9}{GW190707A}{0.5}{GW190706A}{10.4}{GW190701A}{5.9}{GW190630A}{2.2}{GW190620A}{7.6}{GW190602A}{10.2}{GW190527A}{12.2}{GW190521B}{2.9}{GW190521A}{13.7}{GW190519A}{7.4}{GW190517A}{3.6}{GW190514A}{7.6}{GW190513A}{3.2}{GW190512A}{1.4}{GW190503A}{4.2}{GW190426A}{0.08}{GW190425A}{0.02}{GW190424A}{5.6}{GW190421A}{5.9}{GW190413B}{8.2}{GW190413A}{4.6}{GW190412A}{0.5}{GW190408A}{1.9}}}
\newcommand{\finalmasssourceIMRminus}[1]{\IfEqCase{#1}{{GW190930A}{1.5}{GW190929A}{26.6}{GW190924A}{1.2}{GW190915A}{5.8}{GW190910A}{7.2}{GW190909A}{13.9}{GW190828A}{4.2}{GW190814A}{1.0}{GW190803A}{9.0}{GW190731A}{11.6}{GW190728A}{1.3}{GW190727A}{7.6}{GW190720A}{1.8}{GW190719A}{12.2}{GW190708A}{2.0}{GW190707A}{1.3}{GW190706A}{14.8}{GW190701A}{9.3}{GW190630A}{4.6}{GW190620A}{12.4}{GW190602A}{15.5}{GW190527A}{10.5}{GW190521B}{4.6}{GW190521A}{15.5}{GW190519A}{11.9}{GW190517A}{8.4}{GW190514A}{11.4}{GW190513A}{6.3}{GW190512A}{3.4}{GW190424A}{10.5}{GW190421A}{9.4}{GW190413B}{13.5}{GW190413A}{7.7}{GW190412A}{2.9}}}
\newcommand{\finalmasssourceIMRmed}[1]{\IfEqCase{#1}{{GW190930A}{19.5}{GW190929A}{104.2}{GW190924A}{13.6}{GW190915A}{57.8}{GW190910A}{72.7}{GW190909A}{68.8}{GW190828A}{54.9}{GW190814A}{25.5}{GW190803A}{63.7}{GW190731A}{68.8}{GW190728A}{19.6}{GW190727A}{67.4}{GW190720A}{20.7}{GW190719A}{58.4}{GW190708A}{29.8}{GW190707A}{19.4}{GW190706A}{101.1}{GW190701A}{92.3}{GW190630A}{56.3}{GW190620A}{88.9}{GW190602A}{113.9}{GW190527A}{59.1}{GW190521B}{71.3}{GW190521A}{149.1}{GW190519A}{95.7}{GW190517A}{59.6}{GW190514A}{67.6}{GW190513A}{51.6}{GW190512A}{34.1}{GW190424A}{70.5}{GW190421A}{71.8}{GW190413B}{80.8}{GW190413A}{56.9}{GW190412A}{35.8}}}
\newcommand{\finalmasssourceIMRplus}[1]{\IfEqCase{#1}{{GW190930A}{11.5}{GW190929A}{29.0}{GW190924A}{7.7}{GW190915A}{7.1}{GW190910A}{9.9}{GW190909A}{23.7}{GW190828A}{6.6}{GW190814A}{1.2}{GW190803A}{12.4}{GW190731A}{15.0}{GW190728A}{5.9}{GW190727A}{11.0}{GW190720A}{5.1}{GW190719A}{38.9}{GW190708A}{3.7}{GW190707A}{2.1}{GW190706A}{21.4}{GW190701A}{12.4}{GW190630A}{5.3}{GW190620A}{15.3}{GW190602A}{18.7}{GW190527A}{28.2}{GW190521B}{6.4}{GW190521A}{28.4}{GW190519A}{16.3}{GW190517A}{9.3}{GW190514A}{17.7}{GW190513A}{8.7}{GW190512A}{5.1}{GW190424A}{12.4}{GW190421A}{13.4}{GW190413B}{17.7}{GW190413A}{10.3}{GW190412A}{3.7}}}
\newcommand{\radiatedenergyIMRminus}[1]{\IfEqCase{#1}{{GW190930A}{0.3}{GW190929A}{1.2}{GW190924A}{0.2}{GW190915A}{0.8}{GW190910A}{0.7}{GW190909A}{1.2}{GW190828A}{0.6}{GW190814A}{0.007}{GW190803A}{0.9}{GW190731A}{1.2}{GW190728A}{0.2}{GW190727A}{1.0}{GW190720A}{0.2}{GW190719A}{1.2}{GW190708A}{0.2}{GW190707A}{0.10}{GW190706A}{2.3}{GW190701A}{1.3}{GW190630A}{0.5}{GW190620A}{1.9}{GW190602A}{2.4}{GW190527A}{1.2}{GW190521B}{0.6}{GW190521A}{2.4}{GW190519A}{2.1}{GW190517A}{1.7}{GW190514A}{0.9}{GW190513A}{0.5}{GW190512A}{0.3}{GW190424A}{1.0}{GW190421A}{1.0}{GW190413B}{1.5}{GW190413A}{0.7}{GW190412A}{0.1}}}
\newcommand{\radiatedenergyIMRmed}[1]{\IfEqCase{#1}{{GW190930A}{0.9}{GW190929A}{2.5}{GW190924A}{0.6}{GW190915A}{2.7}{GW190910A}{3.5}{GW190909A}{2.8}{GW190828A}{3.0}{GW190814A}{0.2}{GW190803A}{3.0}{GW190731A}{3.3}{GW190728A}{1.0}{GW190727A}{3.4}{GW190720A}{1.0}{GW190719A}{2.8}{GW190708A}{1.4}{GW190707A}{0.9}{GW190706A}{5.8}{GW190701A}{4.1}{GW190630A}{2.6}{GW190620A}{4.7}{GW190602A}{5.0}{GW190527A}{2.5}{GW190521B}{3.5}{GW190521A}{7.2}{GW190519A}{5.7}{GW190517A}{4.1}{GW190514A}{2.9}{GW190513A}{2.1}{GW190512A}{1.5}{GW190424A}{3.7}{GW190421A}{3.2}{GW190413B}{3.5}{GW190413A}{2.6}{GW190412A}{1.2}}}
\newcommand{\radiatedenergyIMRplus}[1]{\IfEqCase{#1}{{GW190930A}{0.1}{GW190929A}{2.4}{GW190924A}{0.07}{GW190915A}{0.8}{GW190910A}{0.8}{GW190909A}{1.4}{GW190828A}{0.6}{GW190814A}{0.006}{GW190803A}{1.0}{GW190731A}{1.3}{GW190728A}{0.09}{GW190727A}{1.1}{GW190720A}{0.1}{GW190719A}{2.1}{GW190708A}{0.1}{GW190707A}{0.08}{GW190706A}{2.2}{GW190701A}{1.2}{GW190630A}{0.6}{GW190620A}{2.0}{GW190602A}{2.2}{GW190527A}{1.8}{GW190521B}{0.6}{GW190521A}{3.0}{GW190519A}{1.6}{GW190517A}{1.1}{GW190514A}{1.2}{GW190513A}{0.9}{GW190512A}{0.3}{GW190424A}{1.2}{GW190421A}{1.0}{GW190413B}{1.4}{GW190413A}{0.9}{GW190412A}{0.2}}}
\newcommand{\PEpercentBNSIMR}[1]{\IfEqCase{#1}{{GW190930A}{0}{GW190929A}{0}{GW190924A}{0}{GW190915A}{0}{GW190910A}{0}{GW190909A}{0}{GW190828B}{0}{GW190828A}{0}{GW190814A}{0}{GW190803A}{0}{GW190731A}{0}{GW190728A}{0}{GW190727A}{0}{GW190720A}{0}{GW190719A}{0}{GW190708A}{0}{GW190707A}{0}{GW190706A}{0}{GW190701A}{0}{GW190630A}{0}{GW190620A}{0}{GW190602A}{0}{GW190527A}{0}{GW190521B}{0}{GW190521A}{0}{GW190519A}{0}{GW190517A}{0}{GW190514A}{0}{GW190513A}{0}{GW190512A}{0}{GW190503A}{0}{GW190426A}{1}{GW190425A}{100}{GW190424A}{0}{GW190421A}{0}{GW190413B}{0}{GW190413A}{0}{GW190412A}{0}{GW190408A}{0}}}
\newcommand{\PEpercentNSBHIMR}[1]{\IfEqCase{#1}{{GW190930A}{0}{GW190929A}{0}{GW190924A}{8}{GW190915A}{0}{GW190910A}{0}{GW190909A}{0}{GW190828B}{0}{GW190828A}{0}{GW190814A}{100}{GW190803A}{0}{GW190731A}{0}{GW190728A}{0}{GW190727A}{0}{GW190720A}{0}{GW190719A}{0}{GW190708A}{0}{GW190707A}{0}{GW190706A}{0}{GW190701A}{0}{GW190630A}{0}{GW190620A}{0}{GW190602A}{0}{GW190527A}{0}{GW190521B}{0}{GW190521A}{0}{GW190519A}{0}{GW190517A}{0}{GW190514A}{0}{GW190513A}{0}{GW190512A}{0}{GW190503A}{0}{GW190426A}{99}{GW190425A}{0}{GW190424A}{0}{GW190421A}{0}{GW190413B}{0}{GW190413A}{0}{GW190412A}{0}{GW190408A}{0}}}
\newcommand{\PEpercentBBHIMR}[1]{\IfEqCase{#1}{{GW190930A}{100}{GW190929A}{100}{GW190924A}{92}{GW190915A}{100}{GW190910A}{100}{GW190909A}{100}{GW190828B}{100}{GW190828A}{100}{GW190814A}{0}{GW190803A}{100}{GW190731A}{100}{GW190728A}{100}{GW190727A}{100}{GW190720A}{100}{GW190719A}{100}{GW190708A}{100}{GW190707A}{100}{GW190706A}{100}{GW190701A}{100}{GW190630A}{100}{GW190620A}{100}{GW190602A}{100}{GW190527A}{100}{GW190521B}{100}{GW190521A}{100}{GW190519A}{100}{GW190517A}{100}{GW190514A}{100}{GW190513A}{100}{GW190512A}{100}{GW190503A}{100}{GW190426A}{0}{GW190425A}{0}{GW190424A}{100}{GW190421A}{100}{GW190413B}{100}{GW190413A}{100}{GW190412A}{100}{GW190408A}{100}}}
\newcommand{\PEpercentMassGapIMR}[1]{\IfEqCase{#1}{{GW190930A}{0}{GW190929A}{0}{GW190924A}{0}{GW190915A}{0}{GW190910A}{0}{GW190909A}{0}{GW190828B}{0}{GW190828A}{0}{GW190814A}{0}{GW190803A}{0}{GW190731A}{0}{GW190728A}{0}{GW190727A}{0}{GW190720A}{0}{GW190719A}{0}{GW190708A}{0}{GW190707A}{0}{GW190706A}{0}{GW190701A}{0}{GW190630A}{0}{GW190620A}{0}{GW190602A}{0}{GW190527A}{0}{GW190521B}{0}{GW190521A}{0}{GW190519A}{0}{GW190517A}{0}{GW190514A}{0}{GW190513A}{0}{GW190512A}{0}{GW190503A}{0}{GW190426A}{0}{GW190425A}{0}{GW190424A}{0}{GW190421A}{0}{GW190413B}{0}{GW190413A}{0}{GW190412A}{0}{GW190408A}{0}}}

%% file: GW190425A_macros.tex
\newcommand{\loglikelihoodfourtwofiveminus}[1]{\IfEqCase{#1}{{AlignedSpinInspiralTidalHS}{5.5}{AlignedSpinInspiralTidalLS}{5.4}{AlignedSpinTidalHS}{6.5}{AlignedSpinTidalLS}{6.3}{IMRPhenomDNRTidal-HS}{5.4}{IMRPhenomDNRTidal-LS}{5.5}{IMRPhenomPv2NRTidal-HS}{5.7}{IMRPhenomPv2NRTidal-LS}{5.5}{SEOBNRv4TsurrogateHS}{5.2}{SEOBNRv4TsurrogateLS}{5.4}{SEOBNRv4TsurrogatehighspinRIFT}{18.6}{SEOBNRv4TsurrogatelowspinRIFT}{5.0}{TEOBResumS-HS}{15.8}{TEOBResumS-LS}{5.1}{TaylorF2-HS}{5.5}{TaylorF2-LS}{5.4}{PrecessingSpinIMRTidalHS}{5.7}{PrecessingSpinIMRTidalLS}{5.5}{PublicationSamples}{5.6}}}
\newcommand{\loglikelihoodfourtwofivemed}[1]{\IfEqCase{#1}{{AlignedSpinInspiralTidalHS}{-240965.7}{AlignedSpinInspiralTidalLS}{-240964.4}{AlignedSpinTidalHS}{-1045826.3}{AlignedSpinTidalLS}{-1045825.4}{IMRPhenomDNRTidal-HS}{-1045828.5}{IMRPhenomDNRTidal-LS}{-1045827.1}{IMRPhenomPv2NRTidal-HS}{-500483.9}{IMRPhenomPv2NRTidal-LS}{-500482.8}{SEOBNRv4TsurrogateHS}{-1045828.0}{SEOBNRv4TsurrogateLS}{-1045827.1}{SEOBNRv4TsurrogatehighspinRIFT}{66.9}{SEOBNRv4TsurrogatelowspinRIFT}{68.5}{TEOBResumS-HS}{67.1}{TEOBResumS-LS}{68.8}{TaylorF2-HS}{-240965.7}{TaylorF2-LS}{-240964.4}{PrecessingSpinIMRTidalHS}{-500483.9}{PrecessingSpinIMRTidalLS}{-500482.8}{PublicationSamples}{-500483.9}}}
\newcommand{\loglikelihoodfourtwofiveplus}[1]{\IfEqCase{#1}{{AlignedSpinInspiralTidalHS}{3.8}{AlignedSpinInspiralTidalLS}{3.2}{AlignedSpinTidalHS}{1045896.4}{AlignedSpinTidalLS}{1045896.5}{IMRPhenomDNRTidal-HS}{4.5}{IMRPhenomDNRTidal-LS}{3.7}{IMRPhenomPv2NRTidal-HS}{4.5}{IMRPhenomPv2NRTidal-LS}{3.8}{SEOBNRv4TsurrogateHS}{3.9}{SEOBNRv4TsurrogateLS}{3.5}{SEOBNRv4TsurrogatehighspinRIFT}{4.3}{SEOBNRv4TsurrogatelowspinRIFT}{3.4}{TEOBResumS-HS}{4.4}{TEOBResumS-LS}{3.3}{TaylorF2-HS}{3.8}{TaylorF2-LS}{3.2}{PrecessingSpinIMRTidalHS}{4.5}{PrecessingSpinIMRTidalLS}{3.8}{PublicationSamples}{4.5}}}
\newcommand{\chiefffourtwofiveminus}[1]{\IfEqCase{#1}{{AlignedSpinInspiralTidalHS}{0.04}{AlignedSpinInspiralTidalLS}{0.01}{AlignedSpinTidalHS}{0.03}{AlignedSpinTidalLS}{0.01}{IMRPhenomDNRTidal-HS}{0.03}{IMRPhenomDNRTidal-LS}{0.01}{IMRPhenomPv2NRTidal-HS}{0.05}{IMRPhenomPv2NRTidal-LS}{0.01}{SEOBNRv4TsurrogateHS}{0.03}{SEOBNRv4TsurrogateLS}{0.01}{SEOBNRv4TsurrogatehighspinRIFT}{0.03}{SEOBNRv4TsurrogatelowspinRIFT}{0.01}{TEOBResumS-HS}{0.03}{TEOBResumS-LS}{0.01}{TaylorF2-HS}{0.04}{TaylorF2-LS}{0.01}{PrecessingSpinIMRTidalHS}{0.05}{PrecessingSpinIMRTidalLS}{0.01}{PublicationSamples}{0.05}}}
\newcommand{\chiefffourtwofivemed}[1]{\IfEqCase{#1}{{AlignedSpinInspiralTidalHS}{0.05}{AlignedSpinInspiralTidalLS}{0.01}{AlignedSpinTidalHS}{0.04}{AlignedSpinTidalLS}{0.01}{IMRPhenomDNRTidal-HS}{0.04}{IMRPhenomDNRTidal-LS}{0.01}{IMRPhenomPv2NRTidal-HS}{0.06}{IMRPhenomPv2NRTidal-LS}{0.01}{SEOBNRv4TsurrogateHS}{0.04}{SEOBNRv4TsurrogateLS}{0.01}{SEOBNRv4TsurrogatehighspinRIFT}{0.03}{SEOBNRv4TsurrogatelowspinRIFT}{0.01}{TEOBResumS-HS}{0.04}{TEOBResumS-LS}{0.01}{TaylorF2-HS}{0.05}{TaylorF2-LS}{0.01}{PrecessingSpinIMRTidalHS}{0.06}{PrecessingSpinIMRTidalLS}{0.01}{PublicationSamples}{0.06}}}
\newcommand{\chiefffourtwofiveplus}[1]{\IfEqCase{#1}{{AlignedSpinInspiralTidalHS}{0.09}{AlignedSpinInspiralTidalLS}{0.01}{AlignedSpinTidalHS}{0.08}{AlignedSpinTidalLS}{0.01}{IMRPhenomDNRTidal-HS}{0.10}{IMRPhenomDNRTidal-LS}{0.01}{IMRPhenomPv2NRTidal-HS}{0.11}{IMRPhenomPv2NRTidal-LS}{0.01}{SEOBNRv4TsurrogateHS}{0.07}{SEOBNRv4TsurrogateLS}{0.01}{SEOBNRv4TsurrogatehighspinRIFT}{0.07}{SEOBNRv4TsurrogatelowspinRIFT}{0.02}{TEOBResumS-HS}{0.06}{TEOBResumS-LS}{0.02}{TaylorF2-HS}{0.09}{TaylorF2-LS}{0.01}{PrecessingSpinIMRTidalHS}{0.11}{PrecessingSpinIMRTidalLS}{0.01}{PublicationSamples}{0.11}}}
\newcommand{\totalmasssourcefourtwofiveminus}[1]{\IfEqCase{#1}{{AlignedSpinInspiralTidalHS}{0.09}{AlignedSpinInspiralTidalLS}{0.05}{AlignedSpinTidalHS}{0.08}{AlignedSpinTidalLS}{0.05}{IMRPhenomDNRTidal-HS}{0.09}{IMRPhenomDNRTidal-LS}{0.05}{IMRPhenomPv2NRTidal-HS}{0.1}{IMRPhenomPv2NRTidal-LS}{0.05}{SEOBNRv4TsurrogateHS}{0.07}{SEOBNRv4TsurrogateLS}{0.05}{SEOBNRv4TsurrogatehighspinRIFT}{0.07}{SEOBNRv4TsurrogatelowspinRIFT}{0.05}{TEOBResumS-HS}{0.08}{TEOBResumS-LS}{0.05}{TaylorF2-HS}{0.09}{TaylorF2-LS}{0.05}{PrecessingSpinIMRTidalHS}{0.1}{PrecessingSpinIMRTidalLS}{0.05}{PublicationSamples}{0.1}}}
\newcommand{\totalmasssourcefourtwofivemed}[1]{\IfEqCase{#1}{{AlignedSpinInspiralTidalHS}{3.37}{AlignedSpinInspiralTidalLS}{3.31}{AlignedSpinTidalHS}{3.35}{AlignedSpinTidalLS}{3.31}{IMRPhenomDNRTidal-HS}{3.36}{IMRPhenomDNRTidal-LS}{3.31}{IMRPhenomPv2NRTidal-HS}{3.4}{IMRPhenomPv2NRTidal-LS}{3.31}{SEOBNRv4TsurrogateHS}{3.35}{SEOBNRv4TsurrogateLS}{3.31}{SEOBNRv4TsurrogatehighspinRIFT}{3.34}{SEOBNRv4TsurrogatelowspinRIFT}{3.31}{TEOBResumS-HS}{3.35}{TEOBResumS-LS}{3.31}{TaylorF2-HS}{3.37}{TaylorF2-LS}{3.31}{PrecessingSpinIMRTidalHS}{3.4}{PrecessingSpinIMRTidalLS}{3.31}{PublicationSamples}{3.4}}}
\newcommand{\totalmasssourcefourtwofiveplus}[1]{\IfEqCase{#1}{{AlignedSpinInspiralTidalHS}{0.2}{AlignedSpinInspiralTidalLS}{0.06}{AlignedSpinTidalHS}{0.3}{AlignedSpinTidalLS}{0.06}{IMRPhenomDNRTidal-HS}{0.4}{IMRPhenomDNRTidal-LS}{0.06}{IMRPhenomPv2NRTidal-HS}{0.3}{IMRPhenomPv2NRTidal-LS}{0.06}{SEOBNRv4TsurrogateHS}{0.2}{SEOBNRv4TsurrogateLS}{0.06}{SEOBNRv4TsurrogatehighspinRIFT}{0.2}{SEOBNRv4TsurrogatelowspinRIFT}{0.06}{TEOBResumS-HS}{0.2}{TEOBResumS-LS}{0.06}{TaylorF2-HS}{0.2}{TaylorF2-LS}{0.06}{PrecessingSpinIMRTidalHS}{0.3}{PrecessingSpinIMRTidalLS}{0.06}{PublicationSamples}{0.3}}}
\newcommand{\chipfourtwofiveminus}[1]{\IfEqCase{#1}{{AlignedSpinInspiralTidalHS}{0.00}{AlignedSpinInspiralTidalLS}{0.00}{AlignedSpinTidalHS}{0.00}{AlignedSpinTidalLS}{0.00}{IMRPhenomDNRTidal-HS}{0.00}{IMRPhenomDNRTidal-LS}{0.00}{IMRPhenomPv2NRTidal-HS}{0.27}{IMRPhenomPv2NRTidal-LS}{0.02}{SEOBNRv4TsurrogateHS}{0.00}{SEOBNRv4TsurrogateLS}{0.00}{SEOBNRv4TsurrogatehighspinRIFT}{0.00}{SEOBNRv4TsurrogatelowspinRIFT}{0.00}{TEOBResumS-HS}{0.00}{TEOBResumS-LS}{0.00}{TaylorF2-HS}{0.00}{TaylorF2-LS}{0.00}{PrecessingSpinIMRTidalHS}{0.27}{PrecessingSpinIMRTidalLS}{0.02}{PublicationSamples}{0.27}}}
\newcommand{\chipfourtwofivemed}[1]{\IfEqCase{#1}{{AlignedSpinInspiralTidalHS}{0.00}{AlignedSpinInspiralTidalLS}{0.00}{AlignedSpinTidalHS}{0.00}{AlignedSpinTidalLS}{0.00}{IMRPhenomDNRTidal-HS}{0.00}{IMRPhenomDNRTidal-LS}{0.00}{IMRPhenomPv2NRTidal-HS}{0.34}{IMRPhenomPv2NRTidal-LS}{0.03}{SEOBNRv4TsurrogateHS}{0.00}{SEOBNRv4TsurrogateLS}{0.00}{SEOBNRv4TsurrogatehighspinRIFT}{0.00}{SEOBNRv4TsurrogatelowspinRIFT}{0.00}{TEOBResumS-HS}{0.00}{TEOBResumS-LS}{0.00}{TaylorF2-HS}{0.00}{TaylorF2-LS}{0.00}{PrecessingSpinIMRTidalHS}{0.34}{PrecessingSpinIMRTidalLS}{0.03}{PublicationSamples}{0.34}}}
\newcommand{\chipfourtwofiveplus}[1]{\IfEqCase{#1}{{AlignedSpinInspiralTidalHS}{0.00}{AlignedSpinInspiralTidalLS}{0.00}{AlignedSpinTidalHS}{0.00}{AlignedSpinTidalLS}{0.00}{IMRPhenomDNRTidal-HS}{0.00}{IMRPhenomDNRTidal-LS}{0.00}{IMRPhenomPv2NRTidal-HS}{0.43}{IMRPhenomPv2NRTidal-LS}{0.02}{SEOBNRv4TsurrogateHS}{0.00}{SEOBNRv4TsurrogateLS}{0.00}{SEOBNRv4TsurrogatehighspinRIFT}{0.00}{SEOBNRv4TsurrogatelowspinRIFT}{0.00}{TEOBResumS-HS}{0.00}{TEOBResumS-LS}{0.00}{TaylorF2-HS}{0.00}{TaylorF2-LS}{0.00}{PrecessingSpinIMRTidalHS}{0.43}{PrecessingSpinIMRTidalLS}{0.02}{PublicationSamples}{0.43}}}
\newcommand{\spinoneyfourtwofiveminus}[1]{\IfEqCase{#1}{{AlignedSpinInspiralTidalHS}{0.00}{AlignedSpinInspiralTidalLS}{0.00}{AlignedSpinTidalHS}{0.00}{AlignedSpinTidalLS}{0.00}{IMRPhenomDNRTidal-HS}{0.00}{IMRPhenomDNRTidal-LS}{0.00}{IMRPhenomPv2NRTidal-HS}{0.49}{IMRPhenomPv2NRTidal-LS}{0.03}{SEOBNRv4TsurrogateHS}{0.00}{SEOBNRv4TsurrogateLS}{0.00}{SEOBNRv4TsurrogatehighspinRIFT}{0.00}{SEOBNRv4TsurrogatelowspinRIFT}{0.00}{TEOBResumS-HS}{0.00}{TEOBResumS-LS}{0.00}{TaylorF2-HS}{0.00}{TaylorF2-LS}{0.00}{PrecessingSpinIMRTidalHS}{0.48}{PrecessingSpinIMRTidalLS}{0.03}{PublicationSamples}{0.48}}}
\newcommand{\spinoneyfourtwofivemed}[1]{\IfEqCase{#1}{{AlignedSpinInspiralTidalHS}{0.00}{AlignedSpinInspiralTidalLS}{0.00}{AlignedSpinTidalHS}{0.00}{AlignedSpinTidalLS}{0.00}{IMRPhenomDNRTidal-HS}{0.00}{IMRPhenomDNRTidal-LS}{0.00}{IMRPhenomPv2NRTidal-HS}{0.003}{IMRPhenomPv2NRTidal-LS}{0.00}{SEOBNRv4TsurrogateHS}{0.00}{SEOBNRv4TsurrogateLS}{0.00}{SEOBNRv4TsurrogatehighspinRIFT}{0.00}{SEOBNRv4TsurrogatelowspinRIFT}{0.00}{TEOBResumS-HS}{0.00}{TEOBResumS-LS}{0.00}{TaylorF2-HS}{0.00}{TaylorF2-LS}{0.00}{PrecessingSpinIMRTidalHS}{0.003}{PrecessingSpinIMRTidalLS}{0.00}{PublicationSamples}{0.003}}}
\newcommand{\spinoneyfourtwofiveplus}[1]{\IfEqCase{#1}{{AlignedSpinInspiralTidalHS}{0.00}{AlignedSpinInspiralTidalLS}{0.00}{AlignedSpinTidalHS}{0.00}{AlignedSpinTidalLS}{0.00}{IMRPhenomDNRTidal-HS}{0.00}{IMRPhenomDNRTidal-LS}{0.00}{IMRPhenomPv2NRTidal-HS}{0.48}{IMRPhenomPv2NRTidal-LS}{0.03}{SEOBNRv4TsurrogateHS}{0.00}{SEOBNRv4TsurrogateLS}{0.00}{SEOBNRv4TsurrogatehighspinRIFT}{0.00}{SEOBNRv4TsurrogatelowspinRIFT}{0.00}{TEOBResumS-HS}{0.00}{TEOBResumS-LS}{0.00}{TaylorF2-HS}{0.00}{TaylorF2-LS}{0.00}{PrecessingSpinIMRTidalHS}{0.48}{PrecessingSpinIMRTidalLS}{0.03}{PublicationSamples}{0.48}}}
\newcommand{\phitwofourtwofiveminus}[1]{\IfEqCase{#1}{{AlignedSpinInspiralTidalHS}{0.00}{AlignedSpinInspiralTidalLS}{0.00}{AlignedSpinTidalHS}{0.00}{AlignedSpinTidalLS}{0.00}{IMRPhenomDNRTidal-HS}{0.00}{IMRPhenomDNRTidal-LS}{0.00}{IMRPhenomPv2NRTidal-HS}{2.83}{IMRPhenomPv2NRTidal-LS}{2.82}{SEOBNRv4TsurrogateHS}{0.00}{SEOBNRv4TsurrogateLS}{0.00}{SEOBNRv4TsurrogatehighspinRIFT}{0.00}{SEOBNRv4TsurrogatelowspinRIFT}{0.00}{TEOBResumS-HS}{0.00}{TEOBResumS-LS}{0.00}{TaylorF2-HS}{0.00}{TaylorF2-LS}{0.00}{PrecessingSpinIMRTidalHS}{2.84}{PrecessingSpinIMRTidalLS}{2.82}{PublicationSamples}{2.84}}}
\newcommand{\phitwofourtwofivemed}[1]{\IfEqCase{#1}{{AlignedSpinInspiralTidalHS}{0.00}{AlignedSpinInspiralTidalLS}{0.00}{AlignedSpinTidalHS}{0.00}{AlignedSpinTidalLS}{0.00}{IMRPhenomDNRTidal-HS}{0.00}{IMRPhenomDNRTidal-LS}{0.00}{IMRPhenomPv2NRTidal-HS}{3.14}{IMRPhenomPv2NRTidal-LS}{3.13}{SEOBNRv4TsurrogateHS}{0.00}{SEOBNRv4TsurrogateLS}{0.00}{SEOBNRv4TsurrogatehighspinRIFT}{0.00}{SEOBNRv4TsurrogatelowspinRIFT}{0.00}{TEOBResumS-HS}{0.00}{TEOBResumS-LS}{0.00}{TaylorF2-HS}{0.00}{TaylorF2-LS}{0.00}{PrecessingSpinIMRTidalHS}{3.15}{PrecessingSpinIMRTidalLS}{3.13}{PublicationSamples}{3.15}}}
\newcommand{\phitwofourtwofiveplus}[1]{\IfEqCase{#1}{{AlignedSpinInspiralTidalHS}{0.00}{AlignedSpinInspiralTidalLS}{0.00}{AlignedSpinTidalHS}{0.00}{AlignedSpinTidalLS}{0.00}{IMRPhenomDNRTidal-HS}{0.00}{IMRPhenomDNRTidal-LS}{0.00}{IMRPhenomPv2NRTidal-HS}{2.84}{IMRPhenomPv2NRTidal-LS}{2.83}{SEOBNRv4TsurrogateHS}{0.00}{SEOBNRv4TsurrogateLS}{0.00}{SEOBNRv4TsurrogatehighspinRIFT}{0.00}{SEOBNRv4TsurrogatelowspinRIFT}{0.00}{TEOBResumS-HS}{0.00}{TEOBResumS-LS}{0.00}{TaylorF2-HS}{0.00}{TaylorF2-LS}{0.00}{PrecessingSpinIMRTidalHS}{2.83}{PrecessingSpinIMRTidalLS}{2.83}{PublicationSamples}{2.83}}}
\newcommand{\phionetwofourtwofiveminus}[1]{\IfEqCase{#1}{{AlignedSpinInspiralTidalHS}{0.00}{AlignedSpinInspiralTidalLS}{0.00}{AlignedSpinTidalHS}{0.00}{AlignedSpinTidalLS}{0.00}{IMRPhenomDNRTidal-HS}{0.00}{IMRPhenomDNRTidal-LS}{0.00}{IMRPhenomPv2NRTidal-HS}{2.88}{IMRPhenomPv2NRTidal-LS}{2.74}{SEOBNRv4TsurrogateHS}{0.00}{SEOBNRv4TsurrogateLS}{0.00}{SEOBNRv4TsurrogatehighspinRIFT}{0.00}{SEOBNRv4TsurrogatelowspinRIFT}{0.00}{TEOBResumS-HS}{0.00}{TEOBResumS-LS}{0.00}{TaylorF2-HS}{0.00}{TaylorF2-LS}{0.00}{PrecessingSpinIMRTidalHS}{2.87}{PrecessingSpinIMRTidalLS}{2.75}{PublicationSamples}{2.88}}}
\newcommand{\phionetwofourtwofivemed}[1]{\IfEqCase{#1}{{AlignedSpinInspiralTidalHS}{0.00}{AlignedSpinInspiralTidalLS}{0.00}{AlignedSpinTidalHS}{0.00}{AlignedSpinTidalLS}{0.00}{IMRPhenomDNRTidal-HS}{0.00}{IMRPhenomDNRTidal-LS}{0.00}{IMRPhenomPv2NRTidal-HS}{3.19}{IMRPhenomPv2NRTidal-LS}{3.05}{SEOBNRv4TsurrogateHS}{0.00}{SEOBNRv4TsurrogateLS}{0.00}{SEOBNRv4TsurrogatehighspinRIFT}{0.00}{SEOBNRv4TsurrogatelowspinRIFT}{0.00}{TEOBResumS-HS}{0.00}{TEOBResumS-LS}{0.00}{TaylorF2-HS}{0.00}{TaylorF2-LS}{0.00}{PrecessingSpinIMRTidalHS}{3.18}{PrecessingSpinIMRTidalLS}{3.05}{PublicationSamples}{3.18}}}
\newcommand{\phionetwofourtwofiveplus}[1]{\IfEqCase{#1}{{AlignedSpinInspiralTidalHS}{0.00}{AlignedSpinInspiralTidalLS}{0.00}{AlignedSpinTidalHS}{0.00}{AlignedSpinTidalLS}{0.00}{IMRPhenomDNRTidal-HS}{0.00}{IMRPhenomDNRTidal-LS}{0.00}{IMRPhenomPv2NRTidal-HS}{2.75}{IMRPhenomPv2NRTidal-LS}{2.88}{SEOBNRv4TsurrogateHS}{0.00}{SEOBNRv4TsurrogateLS}{0.00}{SEOBNRv4TsurrogatehighspinRIFT}{0.00}{SEOBNRv4TsurrogatelowspinRIFT}{0.00}{TEOBResumS-HS}{0.00}{TEOBResumS-LS}{0.00}{TaylorF2-HS}{0.00}{TaylorF2-LS}{0.00}{PrecessingSpinIMRTidalHS}{2.76}{PrecessingSpinIMRTidalLS}{2.88}{PublicationSamples}{2.76}}}
\newcommand{\rafourtwofiveminus}[1]{\IfEqCase{#1}{{AlignedSpinInspiralTidalHS}{1.04004}{AlignedSpinInspiralTidalLS}{1.10804}{AlignedSpinTidalHS}{1.04466}{AlignedSpinTidalLS}{1.05437}{IMRPhenomDNRTidal-HS}{1.04415}{IMRPhenomDNRTidal-LS}{1.34502}{IMRPhenomPv2NRTidal-HS}{1.14841}{IMRPhenomPv2NRTidal-LS}{1.49247}{SEOBNRv4TsurrogateHS}{1.04813}{SEOBNRv4TsurrogateLS}{0.99881}{SEOBNRv4TsurrogatehighspinRIFT}{1.03845}{SEOBNRv4TsurrogatelowspinRIFT}{1.07158}{TEOBResumS-HS}{1.04161}{TEOBResumS-LS}{1.06268}{TaylorF2-HS}{1.03744}{TaylorF2-LS}{1.11687}{PrecessingSpinIMRTidalHS}{1.14713}{PrecessingSpinIMRTidalLS}{1.36500}{PublicationSamples}{1.14187}}}
\newcommand{\rafourtwofivemed}[1]{\IfEqCase{#1}{{AlignedSpinInspiralTidalHS}{1.47807}{AlignedSpinInspiralTidalLS}{1.57610}{AlignedSpinTidalHS}{1.52327}{AlignedSpinTidalLS}{1.53218}{IMRPhenomDNRTidal-HS}{1.52282}{IMRPhenomDNRTidal-LS}{1.86790}{IMRPhenomPv2NRTidal-HS}{1.62902}{IMRPhenomPv2NRTidal-LS}{1.99502}{SEOBNRv4TsurrogateHS}{1.52665}{SEOBNRv4TsurrogateLS}{1.46873}{SEOBNRv4TsurrogatehighspinRIFT}{1.50830}{SEOBNRv4TsurrogatelowspinRIFT}{1.53881}{TEOBResumS-HS}{1.51000}{TEOBResumS-LS}{1.50223}{TaylorF2-HS}{1.47697}{TaylorF2-LS}{1.57735}{PrecessingSpinIMRTidalHS}{1.62833}{PrecessingSpinIMRTidalLS}{1.86958}{PublicationSamples}{1.62336}}}
\newcommand{\rafourtwofiveplus}[1]{\IfEqCase{#1}{{AlignedSpinInspiralTidalHS}{3.24632}{AlignedSpinInspiralTidalLS}{3.13801}{AlignedSpinTidalHS}{3.25020}{AlignedSpinTidalLS}{3.18422}{IMRPhenomDNRTidal-HS}{3.29588}{IMRPhenomDNRTidal-LS}{2.86054}{IMRPhenomPv2NRTidal-HS}{3.13029}{IMRPhenomPv2NRTidal-LS}{2.74917}{SEOBNRv4TsurrogateHS}{3.18219}{SEOBNRv4TsurrogateLS}{3.22628}{SEOBNRv4TsurrogatehighspinRIFT}{3.24313}{SEOBNRv4TsurrogatelowspinRIFT}{3.18050}{TEOBResumS-HS}{3.24215}{TEOBResumS-LS}{3.23570}{TaylorF2-HS}{3.25577}{TaylorF2-LS}{3.13908}{PrecessingSpinIMRTidalHS}{3.12911}{PrecessingSpinIMRTidalLS}{2.87696}{PublicationSamples}{3.13407}}}
\newcommand{\phijlfourtwofiveminus}[1]{\IfEqCase{#1}{{AlignedSpinInspiralTidalHS}{0.97}{AlignedSpinInspiralTidalLS}{0.88}{AlignedSpinTidalHS}{0.78}{AlignedSpinTidalLS}{0.77}{IMRPhenomDNRTidal-HS}{0.74}{IMRPhenomDNRTidal-LS}{0.78}{IMRPhenomPv2NRTidal-HS}{2.89}{IMRPhenomPv2NRTidal-LS}{2.58}{SEOBNRv4TsurrogateHS}{0.89}{SEOBNRv4TsurrogateLS}{0.85}{SEOBNRv4TsurrogatehighspinRIFT}{0.00}{SEOBNRv4TsurrogatelowspinRIFT}{0.00}{TEOBResumS-HS}{0.00}{TEOBResumS-LS}{0.00}{TaylorF2-HS}{0.97}{TaylorF2-LS}{0.88}{PrecessingSpinIMRTidalHS}{2.88}{PrecessingSpinIMRTidalLS}{2.57}{PublicationSamples}{2.89}}}
\newcommand{\phijlfourtwofivemed}[1]{\IfEqCase{#1}{{AlignedSpinInspiralTidalHS}{1.22}{AlignedSpinInspiralTidalLS}{1.11}{AlignedSpinTidalHS}{0.78}{AlignedSpinTidalLS}{0.77}{IMRPhenomDNRTidal-HS}{0.96}{IMRPhenomDNRTidal-LS}{1.02}{IMRPhenomPv2NRTidal-HS}{3.23}{IMRPhenomPv2NRTidal-LS}{2.87}{SEOBNRv4TsurrogateHS}{1.13}{SEOBNRv4TsurrogateLS}{1.09}{SEOBNRv4TsurrogatehighspinRIFT}{0.00}{SEOBNRv4TsurrogatelowspinRIFT}{0.00}{TEOBResumS-HS}{0.00}{TEOBResumS-LS}{0.00}{TaylorF2-HS}{1.22}{TaylorF2-LS}{1.11}{PrecessingSpinIMRTidalHS}{3.23}{PrecessingSpinIMRTidalLS}{2.86}{PublicationSamples}{3.23}}}
\newcommand{\phijlfourtwofiveplus}[1]{\IfEqCase{#1}{{AlignedSpinInspiralTidalHS}{1.64}{AlignedSpinInspiralTidalLS}{1.75}{AlignedSpinTidalHS}{2.37}{AlignedSpinTidalLS}{2.37}{IMRPhenomDNRTidal-HS}{1.88}{IMRPhenomDNRTidal-LS}{1.83}{IMRPhenomPv2NRTidal-HS}{2.75}{IMRPhenomPv2NRTidal-LS}{3.05}{SEOBNRv4TsurrogateHS}{1.72}{SEOBNRv4TsurrogateLS}{1.76}{SEOBNRv4TsurrogatehighspinRIFT}{3.14}{SEOBNRv4TsurrogatelowspinRIFT}{3.14}{TEOBResumS-HS}{3.14}{TEOBResumS-LS}{3.14}{TaylorF2-HS}{1.63}{TaylorF2-LS}{1.75}{PrecessingSpinIMRTidalHS}{2.76}{PrecessingSpinIMRTidalLS}{3.07}{PublicationSamples}{2.75}}}
\newcommand{\tilttwofourtwofiveminus}[1]{\IfEqCase{#1}{{AlignedSpinInspiralTidalHS}{0.00}{AlignedSpinInspiralTidalLS}{0.00}{AlignedSpinTidalHS}{0.00}{AlignedSpinTidalLS}{0.00}{IMRPhenomDNRTidal-HS}{0.00}{IMRPhenomDNRTidal-LS}{0.00}{IMRPhenomPv2NRTidal-HS}{0.87}{IMRPhenomPv2NRTidal-LS}{0.78}{SEOBNRv4TsurrogateHS}{0.00}{SEOBNRv4TsurrogateLS}{0.00}{SEOBNRv4TsurrogatehighspinRIFT}{0.00}{SEOBNRv4TsurrogatelowspinRIFT}{0.00}{TEOBResumS-HS}{0.00}{TEOBResumS-LS}{0.00}{TaylorF2-HS}{0.00}{TaylorF2-LS}{0.00}{PrecessingSpinIMRTidalHS}{0.87}{PrecessingSpinIMRTidalLS}{0.79}{PublicationSamples}{0.87}}}
\newcommand{\tilttwofourtwofivemed}[1]{\IfEqCase{#1}{{AlignedSpinInspiralTidalHS}{0.00}{AlignedSpinInspiralTidalLS}{0.00}{AlignedSpinTidalHS}{0.00}{AlignedSpinTidalLS}{0.00}{IMRPhenomDNRTidal-HS}{0.00}{IMRPhenomDNRTidal-LS}{0.00}{IMRPhenomPv2NRTidal-HS}{1.41}{IMRPhenomPv2NRTidal-LS}{1.09}{SEOBNRv4TsurrogateHS}{0.00}{SEOBNRv4TsurrogateLS}{0.00}{SEOBNRv4TsurrogatehighspinRIFT}{0.00}{SEOBNRv4TsurrogatelowspinRIFT}{0.00}{TEOBResumS-HS}{0.00}{TEOBResumS-LS}{0.00}{TaylorF2-HS}{0.00}{TaylorF2-LS}{0.00}{PrecessingSpinIMRTidalHS}{1.41}{PrecessingSpinIMRTidalLS}{1.09}{PublicationSamples}{1.41}}}
\newcommand{\tilttwofourtwofiveplus}[1]{\IfEqCase{#1}{{AlignedSpinInspiralTidalHS}{3.14}{AlignedSpinInspiralTidalLS}{3.14}{AlignedSpinTidalHS}{3.14}{AlignedSpinTidalLS}{3.14}{IMRPhenomDNRTidal-HS}{3.14}{IMRPhenomDNRTidal-LS}{3.14}{IMRPhenomPv2NRTidal-HS}{0.94}{IMRPhenomPv2NRTidal-LS}{1.21}{SEOBNRv4TsurrogateHS}{3.14}{SEOBNRv4TsurrogateLS}{3.14}{SEOBNRv4TsurrogatehighspinRIFT}{3.14}{SEOBNRv4TsurrogatelowspinRIFT}{3.14}{TEOBResumS-HS}{3.14}{TEOBResumS-LS}{3.14}{TaylorF2-HS}{3.14}{TaylorF2-LS}{3.14}{PrecessingSpinIMRTidalHS}{0.94}{PrecessingSpinIMRTidalLS}{1.20}{PublicationSamples}{0.94}}}
\newcommand{\costhetajnfourtwofiveminus}[1]{\IfEqCase{#1}{{AlignedSpinInspiralTidalHS}{1.30}{AlignedSpinInspiralTidalLS}{1.40}{AlignedSpinTidalHS}{1.44}{AlignedSpinTidalLS}{1.44}{IMRPhenomDNRTidal-HS}{1.53}{IMRPhenomDNRTidal-LS}{1.48}{IMRPhenomPv2NRTidal-HS}{1.43}{IMRPhenomPv2NRTidal-LS}{1.44}{SEOBNRv4TsurrogateHS}{1.38}{SEOBNRv4TsurrogateLS}{1.42}{SEOBNRv4TsurrogatehighspinRIFT}{1.41}{SEOBNRv4TsurrogatelowspinRIFT}{1.42}{TEOBResumS-HS}{1.42}{TEOBResumS-LS}{1.40}{TaylorF2-HS}{1.30}{TaylorF2-LS}{1.40}{PrecessingSpinIMRTidalHS}{1.43}{PrecessingSpinIMRTidalLS}{1.44}{PublicationSamples}{1.43}}}
\newcommand{\costhetajnfourtwofivemed}[1]{\IfEqCase{#1}{{AlignedSpinInspiralTidalHS}{0.34}{AlignedSpinInspiralTidalLS}{0.44}{AlignedSpinTidalHS}{0.49}{AlignedSpinTidalLS}{0.49}{IMRPhenomDNRTidal-HS}{0.58}{IMRPhenomDNRTidal-LS}{0.53}{IMRPhenomPv2NRTidal-HS}{0.47}{IMRPhenomPv2NRTidal-LS}{0.48}{SEOBNRv4TsurrogateHS}{0.43}{SEOBNRv4TsurrogateLS}{0.46}{SEOBNRv4TsurrogatehighspinRIFT}{0.45}{SEOBNRv4TsurrogatelowspinRIFT}{0.46}{TEOBResumS-HS}{0.46}{TEOBResumS-LS}{0.44}{TaylorF2-HS}{0.34}{TaylorF2-LS}{0.44}{PrecessingSpinIMRTidalHS}{0.47}{PrecessingSpinIMRTidalLS}{0.48}{PublicationSamples}{0.47}}}
\newcommand{\costhetajnfourtwofiveplus}[1]{\IfEqCase{#1}{{AlignedSpinInspiralTidalHS}{0.62}{AlignedSpinInspiralTidalLS}{0.53}{AlignedSpinTidalHS}{0.49}{AlignedSpinTidalLS}{0.49}{IMRPhenomDNRTidal-HS}{0.40}{IMRPhenomDNRTidal-LS}{0.45}{IMRPhenomPv2NRTidal-HS}{0.50}{IMRPhenomPv2NRTidal-LS}{0.49}{SEOBNRv4TsurrogateHS}{0.54}{SEOBNRv4TsurrogateLS}{0.51}{SEOBNRv4TsurrogatehighspinRIFT}{0.52}{SEOBNRv4TsurrogatelowspinRIFT}{0.51}{TEOBResumS-HS}{0.51}{TEOBResumS-LS}{0.54}{TaylorF2-HS}{0.63}{TaylorF2-LS}{0.53}{PrecessingSpinIMRTidalHS}{0.50}{PrecessingSpinIMRTidalLS}{0.49}{PublicationSamples}{0.50}}}
\newcommand{\spintwofourtwofiveminus}[1]{\IfEqCase{#1}{{AlignedSpinInspiralTidalHS}{0.07}{AlignedSpinInspiralTidalLS}{0.01}{AlignedSpinTidalHS}{0.07}{AlignedSpinTidalLS}{0.01}{IMRPhenomDNRTidal-HS}{0.10}{IMRPhenomDNRTidal-LS}{0.01}{IMRPhenomPv2NRTidal-HS}{0.25}{IMRPhenomPv2NRTidal-LS}{0.02}{SEOBNRv4TsurrogateHS}{0.06}{SEOBNRv4TsurrogateLS}{0.01}{SEOBNRv4TsurrogatehighspinRIFT}{0.06}{SEOBNRv4TsurrogatelowspinRIFT}{0.01}{TEOBResumS-HS}{0.06}{TEOBResumS-LS}{0.01}{TaylorF2-HS}{0.07}{TaylorF2-LS}{0.01}{PrecessingSpinIMRTidalHS}{0.25}{PrecessingSpinIMRTidalLS}{0.02}{PublicationSamples}{0.25}}}
\newcommand{\spintwofourtwofivemed}[1]{\IfEqCase{#1}{{AlignedSpinInspiralTidalHS}{0.08}{AlignedSpinInspiralTidalLS}{0.01}{AlignedSpinTidalHS}{0.07}{AlignedSpinTidalLS}{0.01}{IMRPhenomDNRTidal-HS}{0.11}{IMRPhenomDNRTidal-LS}{0.01}{IMRPhenomPv2NRTidal-HS}{0.28}{IMRPhenomPv2NRTidal-LS}{0.03}{SEOBNRv4TsurrogateHS}{0.06}{SEOBNRv4TsurrogateLS}{0.01}{SEOBNRv4TsurrogatehighspinRIFT}{0.06}{SEOBNRv4TsurrogatelowspinRIFT}{0.01}{TEOBResumS-HS}{0.06}{TEOBResumS-LS}{0.01}{TaylorF2-HS}{0.08}{TaylorF2-LS}{0.01}{PrecessingSpinIMRTidalHS}{0.28}{PrecessingSpinIMRTidalLS}{0.03}{PublicationSamples}{0.28}}}
\newcommand{\spintwofourtwofiveplus}[1]{\IfEqCase{#1}{{AlignedSpinInspiralTidalHS}{0.26}{AlignedSpinInspiralTidalLS}{0.03}{AlignedSpinTidalHS}{0.27}{AlignedSpinTidalLS}{0.03}{IMRPhenomDNRTidal-HS}{0.38}{IMRPhenomDNRTidal-LS}{0.03}{IMRPhenomPv2NRTidal-HS}{0.51}{IMRPhenomPv2NRTidal-LS}{0.02}{SEOBNRv4TsurrogateHS}{0.19}{SEOBNRv4TsurrogateLS}{0.03}{SEOBNRv4TsurrogatehighspinRIFT}{0.19}{SEOBNRv4TsurrogatelowspinRIFT}{0.03}{TEOBResumS-HS}{0.19}{TEOBResumS-LS}{0.03}{TaylorF2-HS}{0.26}{TaylorF2-LS}{0.03}{PrecessingSpinIMRTidalHS}{0.51}{PrecessingSpinIMRTidalLS}{0.02}{PublicationSamples}{0.51}}}
\newcommand{\massonedetfourtwofiveminus}[1]{\IfEqCase{#1}{{AlignedSpinInspiralTidalHS}{0.3}{AlignedSpinInspiralTidalLS}{0.09}{AlignedSpinTidalHS}{0.2}{AlignedSpinTidalLS}{0.09}{IMRPhenomDNRTidal-HS}{0.3}{IMRPhenomDNRTidal-LS}{0.09}{IMRPhenomPv2NRTidal-HS}{0.4}{IMRPhenomPv2NRTidal-LS}{0.09}{SEOBNRv4TsurrogateHS}{0.2}{SEOBNRv4TsurrogateLS}{0.09}{SEOBNRv4TsurrogatehighspinRIFT}{0.2}{SEOBNRv4TsurrogatelowspinRIFT}{0.09}{TEOBResumS-HS}{0.2}{TEOBResumS-LS}{0.09}{TaylorF2-HS}{0.3}{TaylorF2-LS}{0.09}{PrecessingSpinIMRTidalHS}{0.4}{PrecessingSpinIMRTidalLS}{0.09}{PublicationSamples}{0.4}}}
\newcommand{\massonedetfourtwofivemed}[1]{\IfEqCase{#1}{{AlignedSpinInspiralTidalHS}{2.0}{AlignedSpinInspiralTidalLS}{1.81}{AlignedSpinTidalHS}{2.0}{AlignedSpinTidalLS}{1.81}{IMRPhenomDNRTidal-HS}{2.0}{IMRPhenomDNRTidal-LS}{1.81}{IMRPhenomPv2NRTidal-HS}{2.1}{IMRPhenomPv2NRTidal-LS}{1.80}{SEOBNRv4TsurrogateHS}{2.0}{SEOBNRv4TsurrogateLS}{1.80}{SEOBNRv4TsurrogatehighspinRIFT}{1.9}{SEOBNRv4TsurrogatelowspinRIFT}{1.81}{TEOBResumS-HS}{2.0}{TEOBResumS-LS}{1.81}{TaylorF2-HS}{2.0}{TaylorF2-LS}{1.81}{PrecessingSpinIMRTidalHS}{2.1}{PrecessingSpinIMRTidalLS}{1.80}{PublicationSamples}{2.1}}}
\newcommand{\massonedetfourtwofiveplus}[1]{\IfEqCase{#1}{{AlignedSpinInspiralTidalHS}{0.5}{AlignedSpinInspiralTidalLS}{0.2}{AlignedSpinTidalHS}{0.6}{AlignedSpinTidalLS}{0.2}{IMRPhenomDNRTidal-HS}{0.7}{IMRPhenomDNRTidal-LS}{0.2}{IMRPhenomPv2NRTidal-HS}{0.6}{IMRPhenomPv2NRTidal-LS}{0.2}{SEOBNRv4TsurrogateHS}{0.5}{SEOBNRv4TsurrogateLS}{0.2}{SEOBNRv4TsurrogatehighspinRIFT}{0.5}{SEOBNRv4TsurrogatelowspinRIFT}{0.2}{TEOBResumS-HS}{0.5}{TEOBResumS-LS}{0.2}{TaylorF2-HS}{0.5}{TaylorF2-LS}{0.2}{PrecessingSpinIMRTidalHS}{0.6}{PrecessingSpinIMRTidalLS}{0.2}{PublicationSamples}{0.6}}}
\newcommand{\spintwoxfourtwofiveminus}[1]{\IfEqCase{#1}{{AlignedSpinInspiralTidalHS}{0.00}{AlignedSpinInspiralTidalLS}{0.00}{AlignedSpinTidalHS}{0.00}{AlignedSpinTidalLS}{0.00}{IMRPhenomDNRTidal-HS}{0.00}{IMRPhenomDNRTidal-LS}{0.00}{IMRPhenomPv2NRTidal-HS}{0.47}{IMRPhenomPv2NRTidal-LS}{0.03}{SEOBNRv4TsurrogateHS}{0.00}{SEOBNRv4TsurrogateLS}{0.00}{SEOBNRv4TsurrogatehighspinRIFT}{0.00}{SEOBNRv4TsurrogatelowspinRIFT}{0.00}{TEOBResumS-HS}{0.00}{TEOBResumS-LS}{0.00}{TaylorF2-HS}{0.00}{TaylorF2-LS}{0.00}{PrecessingSpinIMRTidalHS}{0.47}{PrecessingSpinIMRTidalLS}{0.03}{PublicationSamples}{0.47}}}
\newcommand{\spintwoxfourtwofivemed}[1]{\IfEqCase{#1}{{AlignedSpinInspiralTidalHS}{0.00}{AlignedSpinInspiralTidalLS}{0.00}{AlignedSpinTidalHS}{0.00}{AlignedSpinTidalLS}{0.00}{IMRPhenomDNRTidal-HS}{0.00}{IMRPhenomDNRTidal-LS}{0.00}{IMRPhenomPv2NRTidal-HS}{0.0007}{IMRPhenomPv2NRTidal-LS}{0.00}{SEOBNRv4TsurrogateHS}{0.00}{SEOBNRv4TsurrogateLS}{0.00}{SEOBNRv4TsurrogatehighspinRIFT}{0.00}{SEOBNRv4TsurrogatelowspinRIFT}{0.00}{TEOBResumS-HS}{0.00}{TEOBResumS-LS}{0.00}{TaylorF2-HS}{0.00}{TaylorF2-LS}{0.00}{PrecessingSpinIMRTidalHS}{0.0006}{PrecessingSpinIMRTidalLS}{0.00}{PublicationSamples}{0.0007}}}
\newcommand{\spintwoxfourtwofiveplus}[1]{\IfEqCase{#1}{{AlignedSpinInspiralTidalHS}{0.00}{AlignedSpinInspiralTidalLS}{0.00}{AlignedSpinTidalHS}{0.00}{AlignedSpinTidalLS}{0.00}{IMRPhenomDNRTidal-HS}{0.00}{IMRPhenomDNRTidal-LS}{0.00}{IMRPhenomPv2NRTidal-HS}{0.48}{IMRPhenomPv2NRTidal-LS}{0.03}{SEOBNRv4TsurrogateHS}{0.00}{SEOBNRv4TsurrogateLS}{0.00}{SEOBNRv4TsurrogatehighspinRIFT}{0.00}{SEOBNRv4TsurrogatelowspinRIFT}{0.00}{TEOBResumS-HS}{0.00}{TEOBResumS-LS}{0.00}{TaylorF2-HS}{0.00}{TaylorF2-LS}{0.00}{PrecessingSpinIMRTidalHS}{0.47}{PrecessingSpinIMRTidalLS}{0.03}{PublicationSamples}{0.47}}}
\newcommand{\massratiofourtwofiveminus}[1]{\IfEqCase{#1}{{AlignedSpinInspiralTidalHS}{0.24}{AlignedSpinInspiralTidalLS}{0.15}{AlignedSpinTidalHS}{0.30}{AlignedSpinTidalLS}{0.15}{IMRPhenomDNRTidal-HS}{0.31}{IMRPhenomDNRTidal-LS}{0.15}{IMRPhenomPv2NRTidal-HS}{0.25}{IMRPhenomPv2NRTidal-LS}{0.15}{SEOBNRv4TsurrogateHS}{0.29}{SEOBNRv4TsurrogateLS}{0.15}{SEOBNRv4TsurrogatehighspinRIFT}{0.27}{SEOBNRv4TsurrogatelowspinRIFT}{0.15}{TEOBResumS-HS}{0.27}{TEOBResumS-LS}{0.15}{TaylorF2-HS}{0.24}{TaylorF2-LS}{0.15}{PrecessingSpinIMRTidalHS}{0.25}{PrecessingSpinIMRTidalLS}{0.15}{PublicationSamples}{0.25}}}
\newcommand{\massratiofourtwofivemed}[1]{\IfEqCase{#1}{{AlignedSpinInspiralTidalHS}{0.70}{AlignedSpinInspiralTidalLS}{0.89}{AlignedSpinTidalHS}{0.74}{AlignedSpinTidalLS}{0.89}{IMRPhenomDNRTidal-HS}{0.72}{IMRPhenomDNRTidal-LS}{0.89}{IMRPhenomPv2NRTidal-HS}{0.67}{IMRPhenomPv2NRTidal-LS}{0.90}{SEOBNRv4TsurrogateHS}{0.77}{SEOBNRv4TsurrogateLS}{0.90}{SEOBNRv4TsurrogatehighspinRIFT}{0.78}{SEOBNRv4TsurrogatelowspinRIFT}{0.89}{TEOBResumS-HS}{0.75}{TEOBResumS-LS}{0.89}{TaylorF2-HS}{0.70}{TaylorF2-LS}{0.89}{PrecessingSpinIMRTidalHS}{0.67}{PrecessingSpinIMRTidalLS}{0.90}{PublicationSamples}{0.67}}}
\newcommand{\massratiofourtwofiveplus}[1]{\IfEqCase{#1}{{AlignedSpinInspiralTidalHS}{0.26}{AlignedSpinInspiralTidalLS}{0.10}{AlignedSpinTidalHS}{0.22}{AlignedSpinTidalLS}{0.09}{IMRPhenomDNRTidal-HS}{0.25}{IMRPhenomDNRTidal-LS}{0.09}{IMRPhenomPv2NRTidal-HS}{0.29}{IMRPhenomPv2NRTidal-LS}{0.09}{SEOBNRv4TsurrogateHS}{0.20}{SEOBNRv4TsurrogateLS}{0.09}{SEOBNRv4TsurrogatehighspinRIFT}{0.19}{SEOBNRv4TsurrogatelowspinRIFT}{0.10}{TEOBResumS-HS}{0.22}{TEOBResumS-LS}{0.10}{TaylorF2-HS}{0.26}{TaylorF2-LS}{0.10}{PrecessingSpinIMRTidalHS}{0.29}{PrecessingSpinIMRTidalLS}{0.09}{PublicationSamples}{0.29}}}
\newcommand{\spinonefourtwofiveminus}[1]{\IfEqCase{#1}{{AlignedSpinInspiralTidalHS}{0.06}{AlignedSpinInspiralTidalLS}{0.01}{AlignedSpinTidalHS}{0.06}{AlignedSpinTidalLS}{0.01}{IMRPhenomDNRTidal-HS}{0.08}{IMRPhenomDNRTidal-LS}{0.01}{IMRPhenomPv2NRTidal-HS}{0.25}{IMRPhenomPv2NRTidal-LS}{0.03}{SEOBNRv4TsurrogateHS}{0.05}{SEOBNRv4TsurrogateLS}{0.01}{SEOBNRv4TsurrogatehighspinRIFT}{0.05}{SEOBNRv4TsurrogatelowspinRIFT}{0.01}{TEOBResumS-HS}{0.05}{TEOBResumS-LS}{0.01}{TaylorF2-HS}{0.06}{TaylorF2-LS}{0.01}{PrecessingSpinIMRTidalHS}{0.25}{PrecessingSpinIMRTidalLS}{0.03}{PublicationSamples}{0.25}}}
\newcommand{\spinonefourtwofivemed}[1]{\IfEqCase{#1}{{AlignedSpinInspiralTidalHS}{0.06}{AlignedSpinInspiralTidalLS}{0.01}{AlignedSpinTidalHS}{0.06}{AlignedSpinTidalLS}{0.01}{IMRPhenomDNRTidal-HS}{0.09}{IMRPhenomDNRTidal-LS}{0.01}{IMRPhenomPv2NRTidal-HS}{0.27}{IMRPhenomPv2NRTidal-LS}{0.03}{SEOBNRv4TsurrogateHS}{0.06}{SEOBNRv4TsurrogateLS}{0.01}{SEOBNRv4TsurrogatehighspinRIFT}{0.06}{SEOBNRv4TsurrogatelowspinRIFT}{0.01}{TEOBResumS-HS}{0.06}{TEOBResumS-LS}{0.01}{TaylorF2-HS}{0.06}{TaylorF2-LS}{0.01}{PrecessingSpinIMRTidalHS}{0.27}{PrecessingSpinIMRTidalLS}{0.03}{PublicationSamples}{0.27}}}
\newcommand{\spinonefourtwofiveplus}[1]{\IfEqCase{#1}{{AlignedSpinInspiralTidalHS}{0.17}{AlignedSpinInspiralTidalLS}{0.03}{AlignedSpinTidalHS}{0.19}{AlignedSpinTidalLS}{0.03}{IMRPhenomDNRTidal-HS}{0.25}{IMRPhenomDNRTidal-LS}{0.03}{IMRPhenomPv2NRTidal-HS}{0.51}{IMRPhenomPv2NRTidal-LS}{0.02}{SEOBNRv4TsurrogateHS}{0.15}{SEOBNRv4TsurrogateLS}{0.03}{SEOBNRv4TsurrogatehighspinRIFT}{0.15}{SEOBNRv4TsurrogatelowspinRIFT}{0.03}{TEOBResumS-HS}{0.14}{TEOBResumS-LS}{0.03}{TaylorF2-HS}{0.17}{TaylorF2-LS}{0.03}{PrecessingSpinIMRTidalHS}{0.51}{PrecessingSpinIMRTidalLS}{0.02}{PublicationSamples}{0.51}}}
\newcommand{\costiltonefourtwofiveminus}[1]{\IfEqCase{#1}{{AlignedSpinInspiralTidalHS}{2.00}{AlignedSpinInspiralTidalLS}{2.00}{AlignedSpinTidalHS}{2.00}{AlignedSpinTidalLS}{2.00}{IMRPhenomDNRTidal-HS}{2.00}{IMRPhenomDNRTidal-LS}{2.00}{IMRPhenomPv2NRTidal-HS}{0.65}{IMRPhenomPv2NRTidal-LS}{1.10}{SEOBNRv4TsurrogateHS}{2.00}{SEOBNRv4TsurrogateLS}{2.00}{SEOBNRv4TsurrogatehighspinRIFT}{2.00}{SEOBNRv4TsurrogatelowspinRIFT}{2.00}{TEOBResumS-HS}{2.00}{TEOBResumS-LS}{2.00}{TaylorF2-HS}{2.00}{TaylorF2-LS}{2.00}{PrecessingSpinIMRTidalHS}{0.65}{PrecessingSpinIMRTidalLS}{1.10}{PublicationSamples}{0.65}}}
\newcommand{\costiltonefourtwofivemed}[1]{\IfEqCase{#1}{{AlignedSpinInspiralTidalHS}{1.00}{AlignedSpinInspiralTidalLS}{1.00}{AlignedSpinTidalHS}{1.00}{AlignedSpinTidalLS}{1.00}{IMRPhenomDNRTidal-HS}{1.00}{IMRPhenomDNRTidal-LS}{1.00}{IMRPhenomPv2NRTidal-HS}{0.26}{IMRPhenomPv2NRTidal-LS}{0.51}{SEOBNRv4TsurrogateHS}{1.00}{SEOBNRv4TsurrogateLS}{1.00}{SEOBNRv4TsurrogatehighspinRIFT}{1.00}{SEOBNRv4TsurrogatelowspinRIFT}{1.00}{TEOBResumS-HS}{1.00}{TEOBResumS-LS}{1.00}{TaylorF2-HS}{1.00}{TaylorF2-LS}{1.00}{PrecessingSpinIMRTidalHS}{0.26}{PrecessingSpinIMRTidalLS}{0.51}{PublicationSamples}{0.26}}}
\newcommand{\costiltonefourtwofiveplus}[1]{\IfEqCase{#1}{{AlignedSpinInspiralTidalHS}{0.00}{AlignedSpinInspiralTidalLS}{0.00}{AlignedSpinTidalHS}{0.00}{AlignedSpinTidalLS}{0.00}{IMRPhenomDNRTidal-HS}{0.00}{IMRPhenomDNRTidal-LS}{0.00}{IMRPhenomPv2NRTidal-HS}{0.61}{IMRPhenomPv2NRTidal-LS}{0.45}{SEOBNRv4TsurrogateHS}{0.00}{SEOBNRv4TsurrogateLS}{0.00}{SEOBNRv4TsurrogatehighspinRIFT}{0.00}{SEOBNRv4TsurrogatelowspinRIFT}{0.00}{TEOBResumS-HS}{0.00}{TEOBResumS-LS}{0.00}{TaylorF2-HS}{0.00}{TaylorF2-LS}{0.00}{PrecessingSpinIMRTidalHS}{0.61}{PrecessingSpinIMRTidalLS}{0.44}{PublicationSamples}{0.61}}}
\newcommand{\phasefourtwofiveminus}[1]{\IfEqCase{#1}{{AlignedSpinInspiralTidalHS}{3.10}{AlignedSpinInspiralTidalLS}{2.74}{AlignedSpinTidalHS}{2.81}{AlignedSpinTidalLS}{2.76}{IMRPhenomDNRTidal-HS}{2.83}{IMRPhenomDNRTidal-LS}{2.62}{IMRPhenomPv2NRTidal-HS}{2.82}{IMRPhenomPv2NRTidal-LS}{2.85}{SEOBNRv4TsurrogateHS}{2.74}{SEOBNRv4TsurrogateLS}{2.81}{SEOBNRv4TsurrogatehighspinRIFT}{2.78}{SEOBNRv4TsurrogatelowspinRIFT}{2.79}{TEOBResumS-HS}{2.78}{TEOBResumS-LS}{2.83}{TaylorF2-HS}{3.13}{TaylorF2-LS}{2.74}{PrecessingSpinIMRTidalHS}{2.82}{PrecessingSpinIMRTidalLS}{2.86}{PublicationSamples}{2.82}}}
\newcommand{\phasefourtwofivemed}[1]{\IfEqCase{#1}{{AlignedSpinInspiralTidalHS}{3.44}{AlignedSpinInspiralTidalLS}{3.01}{AlignedSpinTidalHS}{3.13}{AlignedSpinTidalLS}{3.07}{IMRPhenomDNRTidal-HS}{3.13}{IMRPhenomDNRTidal-LS}{2.88}{IMRPhenomPv2NRTidal-HS}{3.13}{IMRPhenomPv2NRTidal-LS}{3.20}{SEOBNRv4TsurrogateHS}{3.11}{SEOBNRv4TsurrogateLS}{3.16}{SEOBNRv4TsurrogatehighspinRIFT}{3.12}{SEOBNRv4TsurrogatelowspinRIFT}{3.12}{TEOBResumS-HS}{3.09}{TEOBResumS-LS}{3.14}{TaylorF2-HS}{3.46}{TaylorF2-LS}{3.00}{PrecessingSpinIMRTidalHS}{3.12}{PrecessingSpinIMRTidalLS}{3.20}{PublicationSamples}{3.13}}}
\newcommand{\phasefourtwofiveplus}[1]{\IfEqCase{#1}{{AlignedSpinInspiralTidalHS}{2.49}{AlignedSpinInspiralTidalLS}{2.93}{AlignedSpinTidalHS}{2.82}{AlignedSpinTidalLS}{2.88}{IMRPhenomDNRTidal-HS}{2.82}{IMRPhenomDNRTidal-LS}{3.04}{IMRPhenomPv2NRTidal-HS}{2.86}{IMRPhenomPv2NRTidal-LS}{2.74}{SEOBNRv4TsurrogateHS}{2.81}{SEOBNRv4TsurrogateLS}{2.74}{SEOBNRv4TsurrogatehighspinRIFT}{2.81}{SEOBNRv4TsurrogatelowspinRIFT}{2.79}{TEOBResumS-HS}{2.87}{TEOBResumS-LS}{2.86}{TaylorF2-HS}{2.47}{TaylorF2-LS}{2.94}{PrecessingSpinIMRTidalHS}{2.87}{PrecessingSpinIMRTidalLS}{2.73}{PublicationSamples}{2.86}}}
\newcommand{\masstwodetfourtwofiveminus}[1]{\IfEqCase{#1}{{AlignedSpinInspiralTidalHS}{0.3}{AlignedSpinInspiralTidalLS}{0.1}{AlignedSpinTidalHS}{0.3}{AlignedSpinTidalLS}{0.1}{IMRPhenomDNRTidal-HS}{0.3}{IMRPhenomDNRTidal-LS}{0.1}{IMRPhenomPv2NRTidal-HS}{0.3}{IMRPhenomPv2NRTidal-LS}{0.1}{SEOBNRv4TsurrogateHS}{0.3}{SEOBNRv4TsurrogateLS}{0.1}{SEOBNRv4TsurrogatehighspinRIFT}{0.3}{SEOBNRv4TsurrogatelowspinRIFT}{0.1}{TEOBResumS-HS}{0.3}{TEOBResumS-LS}{0.1}{TaylorF2-HS}{0.3}{TaylorF2-LS}{0.1}{PrecessingSpinIMRTidalHS}{0.3}{PrecessingSpinIMRTidalLS}{0.1}{PublicationSamples}{0.3}}}
\newcommand{\masstwodetfourtwofivemed}[1]{\IfEqCase{#1}{{AlignedSpinInspiralTidalHS}{1.4}{AlignedSpinInspiralTidalLS}{1.61}{AlignedSpinTidalHS}{1.5}{AlignedSpinTidalLS}{1.61}{IMRPhenomDNRTidal-HS}{1.5}{IMRPhenomDNRTidal-LS}{1.62}{IMRPhenomPv2NRTidal-HS}{1.4}{IMRPhenomPv2NRTidal-LS}{1.62}{SEOBNRv4TsurrogateHS}{1.5}{SEOBNRv4TsurrogateLS}{1.62}{SEOBNRv4TsurrogatehighspinRIFT}{1.5}{SEOBNRv4TsurrogatelowspinRIFT}{1.61}{TEOBResumS-HS}{1.5}{TEOBResumS-LS}{1.61}{TaylorF2-HS}{1.4}{TaylorF2-LS}{1.61}{PrecessingSpinIMRTidalHS}{1.4}{PrecessingSpinIMRTidalLS}{1.62}{PublicationSamples}{1.4}}}
\newcommand{\masstwodetfourtwofiveplus}[1]{\IfEqCase{#1}{{AlignedSpinInspiralTidalHS}{0.2}{AlignedSpinInspiralTidalLS}{0.09}{AlignedSpinTidalHS}{0.2}{AlignedSpinTidalLS}{0.08}{IMRPhenomDNRTidal-HS}{0.2}{IMRPhenomDNRTidal-LS}{0.08}{IMRPhenomPv2NRTidal-HS}{0.3}{IMRPhenomPv2NRTidal-LS}{0.08}{SEOBNRv4TsurrogateHS}{0.2}{SEOBNRv4TsurrogateLS}{0.08}{SEOBNRv4TsurrogatehighspinRIFT}{0.2}{SEOBNRv4TsurrogatelowspinRIFT}{0.08}{TEOBResumS-HS}{0.2}{TEOBResumS-LS}{0.09}{TaylorF2-HS}{0.2}{TaylorF2-LS}{0.09}{PrecessingSpinIMRTidalHS}{0.3}{PrecessingSpinIMRTidalLS}{0.08}{PublicationSamples}{0.3}}}
\newcommand{\masstwosourcefourtwofiveminus}[1]{\IfEqCase{#1}{{AlignedSpinInspiralTidalHS}{0.3}{AlignedSpinInspiralTidalLS}{0.1}{AlignedSpinTidalHS}{0.3}{AlignedSpinTidalLS}{0.1}{IMRPhenomDNRTidal-HS}{0.3}{IMRPhenomDNRTidal-LS}{0.1}{IMRPhenomPv2NRTidal-HS}{0.3}{IMRPhenomPv2NRTidal-LS}{0.1}{SEOBNRv4TsurrogateHS}{0.3}{SEOBNRv4TsurrogateLS}{0.1}{SEOBNRv4TsurrogatehighspinRIFT}{0.3}{SEOBNRv4TsurrogatelowspinRIFT}{0.1}{TEOBResumS-HS}{0.3}{TEOBResumS-LS}{0.1}{TaylorF2-HS}{0.3}{TaylorF2-LS}{0.1}{PrecessingSpinIMRTidalHS}{0.3}{PrecessingSpinIMRTidalLS}{0.1}{PublicationSamples}{0.3}}}
\newcommand{\masstwosourcefourtwofivemed}[1]{\IfEqCase{#1}{{AlignedSpinInspiralTidalHS}{1.4}{AlignedSpinInspiralTidalLS}{1.56}{AlignedSpinTidalHS}{1.4}{AlignedSpinTidalLS}{1.56}{IMRPhenomDNRTidal-HS}{1.4}{IMRPhenomDNRTidal-LS}{1.56}{IMRPhenomPv2NRTidal-HS}{1.4}{IMRPhenomPv2NRTidal-LS}{1.57}{SEOBNRv4TsurrogateHS}{1.4}{SEOBNRv4TsurrogateLS}{1.56}{SEOBNRv4TsurrogatehighspinRIFT}{1.5}{SEOBNRv4TsurrogatelowspinRIFT}{1.56}{TEOBResumS-HS}{1.4}{TEOBResumS-LS}{1.56}{TaylorF2-HS}{1.4}{TaylorF2-LS}{1.56}{PrecessingSpinIMRTidalHS}{1.4}{PrecessingSpinIMRTidalLS}{1.57}{PublicationSamples}{1.4}}}
\newcommand{\masstwosourcefourtwofiveplus}[1]{\IfEqCase{#1}{{AlignedSpinInspiralTidalHS}{0.2}{AlignedSpinInspiralTidalLS}{0.09}{AlignedSpinTidalHS}{0.2}{AlignedSpinTidalLS}{0.08}{IMRPhenomDNRTidal-HS}{0.2}{IMRPhenomDNRTidal-LS}{0.08}{IMRPhenomPv2NRTidal-HS}{0.3}{IMRPhenomPv2NRTidal-LS}{0.08}{SEOBNRv4TsurrogateHS}{0.2}{SEOBNRv4TsurrogateLS}{0.08}{SEOBNRv4TsurrogatehighspinRIFT}{0.2}{SEOBNRv4TsurrogatelowspinRIFT}{0.08}{TEOBResumS-HS}{0.2}{TEOBResumS-LS}{0.09}{TaylorF2-HS}{0.2}{TaylorF2-LS}{0.09}{PrecessingSpinIMRTidalHS}{0.3}{PrecessingSpinIMRTidalLS}{0.08}{PublicationSamples}{0.3}}}
\newcommand{\decfourtwofiveminus}[1]{\IfEqCase{#1}{{AlignedSpinInspiralTidalHS}{0.92098}{AlignedSpinInspiralTidalLS}{0.94910}{AlignedSpinTidalHS}{0.88455}{AlignedSpinTidalLS}{0.88762}{IMRPhenomDNRTidal-HS}{0.92222}{IMRPhenomDNRTidal-LS}{0.90469}{IMRPhenomPv2NRTidal-HS}{0.90104}{IMRPhenomPv2NRTidal-LS}{0.97042}{SEOBNRv4TsurrogateHS}{0.87704}{SEOBNRv4TsurrogateLS}{0.89335}{SEOBNRv4TsurrogatehighspinRIFT}{0.89796}{SEOBNRv4TsurrogatelowspinRIFT}{0.90269}{TEOBResumS-HS}{0.88789}{TEOBResumS-LS}{0.89751}{TaylorF2-HS}{0.92183}{TaylorF2-LS}{0.95570}{PrecessingSpinIMRTidalHS}{0.89977}{PrecessingSpinIMRTidalLS}{0.96778}{PublicationSamples}{0.89807}}}
\newcommand{\decfourtwofivemed}[1]{\IfEqCase{#1}{{AlignedSpinInspiralTidalHS}{-0.14949}{AlignedSpinInspiralTidalLS}{-0.10865}{AlignedSpinTidalHS}{-0.15883}{AlignedSpinTidalLS}{-0.13405}{IMRPhenomDNRTidal-HS}{-0.18418}{IMRPhenomDNRTidal-LS}{-0.06597}{IMRPhenomPv2NRTidal-HS}{-0.12984}{IMRPhenomPv2NRTidal-LS}{-0.05685}{SEOBNRv4TsurrogateHS}{-0.14893}{SEOBNRv4TsurrogateLS}{-0.17562}{SEOBNRv4TsurrogatehighspinRIFT}{-0.15053}{SEOBNRv4TsurrogatelowspinRIFT}{-0.12883}{TEOBResumS-HS}{-0.15241}{TEOBResumS-LS}{-0.16207}{TaylorF2-HS}{-0.14824}{TaylorF2-LS}{-0.10463}{PrecessingSpinIMRTidalHS}{-0.13006}{PrecessingSpinIMRTidalLS}{-0.06120}{PublicationSamples}{-0.13133}}}
\newcommand{\decfourtwofiveplus}[1]{\IfEqCase{#1}{{AlignedSpinInspiralTidalHS}{0.99909}{AlignedSpinInspiralTidalLS}{0.93177}{AlignedSpinTidalHS}{0.97810}{AlignedSpinTidalLS}{0.98373}{IMRPhenomDNRTidal-HS}{0.95052}{IMRPhenomDNRTidal-LS}{0.93229}{IMRPhenomPv2NRTidal-HS}{0.96946}{IMRPhenomPv2NRTidal-LS}{0.91574}{SEOBNRv4TsurrogateHS}{1.00436}{SEOBNRv4TsurrogateLS}{1.00055}{SEOBNRv4TsurrogatehighspinRIFT}{0.99086}{SEOBNRv4TsurrogatelowspinRIFT}{0.97101}{TEOBResumS-HS}{0.98792}{TEOBResumS-LS}{1.00135}{TaylorF2-HS}{0.99948}{TaylorF2-LS}{0.92354}{PrecessingSpinIMRTidalHS}{0.96811}{PrecessingSpinIMRTidalLS}{0.91709}{PublicationSamples}{0.96897}}}
\newcommand{\psifourtwofiveminus}[1]{\IfEqCase{#1}{{AlignedSpinInspiralTidalHS}{1.40}{AlignedSpinInspiralTidalLS}{1.41}{AlignedSpinTidalHS}{1.63}{AlignedSpinTidalLS}{1.69}{IMRPhenomDNRTidal-HS}{1.36}{IMRPhenomDNRTidal-LS}{1.41}{IMRPhenomPv2NRTidal-HS}{1.46}{IMRPhenomPv2NRTidal-LS}{1.40}{SEOBNRv4TsurrogateHS}{1.36}{SEOBNRv4TsurrogateLS}{1.43}{SEOBNRv4TsurrogatehighspinRIFT}{2.83}{SEOBNRv4TsurrogatelowspinRIFT}{2.86}{TEOBResumS-HS}{2.86}{TEOBResumS-LS}{2.86}{TaylorF2-HS}{1.41}{TaylorF2-LS}{1.41}{PrecessingSpinIMRTidalHS}{1.46}{PrecessingSpinIMRTidalLS}{1.39}{PublicationSamples}{1.46}}}
\newcommand{\psifourtwofivemed}[1]{\IfEqCase{#1}{{AlignedSpinInspiralTidalHS}{1.55}{AlignedSpinInspiralTidalLS}{1.56}{AlignedSpinTidalHS}{1.80}{AlignedSpinTidalLS}{1.86}{IMRPhenomDNRTidal-HS}{1.50}{IMRPhenomDNRTidal-LS}{1.56}{IMRPhenomPv2NRTidal-HS}{1.61}{IMRPhenomPv2NRTidal-LS}{1.54}{SEOBNRv4TsurrogateHS}{1.50}{SEOBNRv4TsurrogateLS}{1.56}{SEOBNRv4TsurrogatehighspinRIFT}{3.13}{SEOBNRv4TsurrogatelowspinRIFT}{3.14}{TEOBResumS-HS}{3.16}{TEOBResumS-LS}{3.16}{TaylorF2-HS}{1.56}{TaylorF2-LS}{1.55}{PrecessingSpinIMRTidalHS}{1.62}{PrecessingSpinIMRTidalLS}{1.54}{PublicationSamples}{1.61}}}
\newcommand{\psifourtwofiveplus}[1]{\IfEqCase{#1}{{AlignedSpinInspiralTidalHS}{1.42}{AlignedSpinInspiralTidalLS}{1.44}{AlignedSpinTidalHS}{3.54}{AlignedSpinTidalLS}{3.48}{IMRPhenomDNRTidal-HS}{1.50}{IMRPhenomDNRTidal-LS}{1.43}{IMRPhenomPv2NRTidal-HS}{1.38}{IMRPhenomPv2NRTidal-LS}{1.43}{SEOBNRv4TsurrogateHS}{1.47}{SEOBNRv4TsurrogateLS}{1.42}{SEOBNRv4TsurrogatehighspinRIFT}{2.85}{SEOBNRv4TsurrogatelowspinRIFT}{2.83}{TEOBResumS-HS}{2.82}{TEOBResumS-LS}{2.84}{TaylorF2-HS}{1.41}{TaylorF2-LS}{1.44}{PrecessingSpinIMRTidalHS}{1.38}{PrecessingSpinIMRTidalLS}{1.43}{PublicationSamples}{1.38}}}
\newcommand{\networkoptimalsnrfourtwofiveminus}[1]{\IfEqCase{#1}{{AlignedSpinInspiralTidalHS}{1.7}{AlignedSpinInspiralTidalLS}{1.7}{IMRPhenomDNRTidal-HS}{1.7}{IMRPhenomDNRTidal-LS}{1.7}{IMRPhenomPv2NRTidal-HS}{1.7}{IMRPhenomPv2NRTidal-LS}{1.7}{SEOBNRv4TsurrogateHS}{1.7}{SEOBNRv4TsurrogateLS}{1.7}{TaylorF2-HS}{1.7}{TaylorF2-LS}{1.7}{PrecessingSpinIMRTidalHS}{1.7}{PrecessingSpinIMRTidalLS}{1.7}{PublicationSamples}{1.7}}}
\newcommand{\networkoptimalsnrfourtwofivemed}[1]{\IfEqCase{#1}{{AlignedSpinInspiralTidalHS}{12.1}{AlignedSpinInspiralTidalLS}{12.2}{IMRPhenomDNRTidal-HS}{12.0}{IMRPhenomDNRTidal-LS}{12.1}{IMRPhenomPv2NRTidal-HS}{12.0}{IMRPhenomPv2NRTidal-LS}{12.1}{SEOBNRv4TsurrogateHS}{12.0}{SEOBNRv4TsurrogateLS}{12.1}{TaylorF2-HS}{12.1}{TaylorF2-LS}{12.2}{PrecessingSpinIMRTidalHS}{12.0}{PrecessingSpinIMRTidalLS}{12.1}{PublicationSamples}{12.0}}}
\newcommand{\networkoptimalsnrfourtwofiveplus}[1]{\IfEqCase{#1}{{AlignedSpinInspiralTidalHS}{1.7}{AlignedSpinInspiralTidalLS}{1.7}{IMRPhenomDNRTidal-HS}{1.7}{IMRPhenomDNRTidal-LS}{1.7}{IMRPhenomPv2NRTidal-HS}{1.7}{IMRPhenomPv2NRTidal-LS}{1.7}{SEOBNRv4TsurrogateHS}{1.7}{SEOBNRv4TsurrogateLS}{1.7}{TaylorF2-HS}{1.7}{TaylorF2-LS}{1.7}{PrecessingSpinIMRTidalHS}{1.7}{PrecessingSpinIMRTidalLS}{1.7}{PublicationSamples}{1.7}}}
\newcommand{\thetajnfourtwofiveminus}[1]{\IfEqCase{#1}{{AlignedSpinInspiralTidalHS}{0.97}{AlignedSpinInspiralTidalLS}{0.88}{AlignedSpinTidalHS}{0.83}{AlignedSpinTidalLS}{0.83}{IMRPhenomDNRTidal-HS}{0.74}{IMRPhenomDNRTidal-LS}{0.78}{IMRPhenomPv2NRTidal-HS}{0.85}{IMRPhenomPv2NRTidal-LS}{0.84}{SEOBNRv4TsurrogateHS}{0.89}{SEOBNRv4TsurrogateLS}{0.85}{SEOBNRv4TsurrogatehighspinRIFT}{0.86}{SEOBNRv4TsurrogatelowspinRIFT}{0.86}{TEOBResumS-HS}{0.86}{TEOBResumS-LS}{0.88}{TaylorF2-HS}{0.97}{TaylorF2-LS}{0.88}{PrecessingSpinIMRTidalHS}{0.85}{PrecessingSpinIMRTidalLS}{0.84}{PublicationSamples}{0.85}}}
\newcommand{\thetajnfourtwofivemed}[1]{\IfEqCase{#1}{{AlignedSpinInspiralTidalHS}{1.22}{AlignedSpinInspiralTidalLS}{1.11}{AlignedSpinTidalHS}{1.06}{AlignedSpinTidalLS}{1.06}{IMRPhenomDNRTidal-HS}{0.96}{IMRPhenomDNRTidal-LS}{1.02}{IMRPhenomPv2NRTidal-HS}{1.08}{IMRPhenomPv2NRTidal-LS}{1.07}{SEOBNRv4TsurrogateHS}{1.13}{SEOBNRv4TsurrogateLS}{1.09}{SEOBNRv4TsurrogatehighspinRIFT}{1.10}{SEOBNRv4TsurrogatelowspinRIFT}{1.09}{TEOBResumS-HS}{1.09}{TEOBResumS-LS}{1.12}{TaylorF2-HS}{1.22}{TaylorF2-LS}{1.11}{PrecessingSpinIMRTidalHS}{1.08}{PrecessingSpinIMRTidalLS}{1.07}{PublicationSamples}{1.08}}}
\newcommand{\thetajnfourtwofiveplus}[1]{\IfEqCase{#1}{{AlignedSpinInspiralTidalHS}{1.64}{AlignedSpinInspiralTidalLS}{1.75}{AlignedSpinTidalHS}{1.79}{AlignedSpinTidalLS}{1.78}{IMRPhenomDNRTidal-HS}{1.88}{IMRPhenomDNRTidal-LS}{1.83}{IMRPhenomPv2NRTidal-HS}{1.77}{IMRPhenomPv2NRTidal-LS}{1.78}{SEOBNRv4TsurrogateHS}{1.72}{SEOBNRv4TsurrogateLS}{1.76}{SEOBNRv4TsurrogatehighspinRIFT}{1.76}{SEOBNRv4TsurrogatelowspinRIFT}{1.75}{TEOBResumS-HS}{1.77}{TEOBResumS-LS}{1.74}{TaylorF2-HS}{1.63}{TaylorF2-LS}{1.75}{PrecessingSpinIMRTidalHS}{1.77}{PrecessingSpinIMRTidalLS}{1.78}{PublicationSamples}{1.77}}}
\newcommand{\totalmassdetfourtwofiveminus}[1]{\IfEqCase{#1}{{AlignedSpinInspiralTidalHS}{0.06}{AlignedSpinInspiralTidalLS}{0.007}{AlignedSpinTidalHS}{0.04}{AlignedSpinTidalLS}{0.007}{IMRPhenomDNRTidal-HS}{0.06}{IMRPhenomDNRTidal-LS}{0.006}{IMRPhenomPv2NRTidal-HS}{0.08}{IMRPhenomPv2NRTidal-LS}{0.006}{SEOBNRv4TsurrogateHS}{0.03}{SEOBNRv4TsurrogateLS}{0.006}{SEOBNRv4TsurrogatehighspinRIFT}{0.03}{SEOBNRv4TsurrogatelowspinRIFT}{0.007}{TEOBResumS-HS}{0.04}{TEOBResumS-LS}{0.007}{TaylorF2-HS}{0.06}{TaylorF2-LS}{0.007}{PrecessingSpinIMRTidalHS}{0.08}{PrecessingSpinIMRTidalLS}{0.006}{PublicationSamples}{0.08}}}
\newcommand{\totalmassdetfourtwofivemed}[1]{\IfEqCase{#1}{{AlignedSpinInspiralTidalHS}{3.48}{AlignedSpinInspiralTidalLS}{3.42}{AlignedSpinTidalHS}{3.46}{AlignedSpinTidalLS}{3.42}{IMRPhenomDNRTidal-HS}{3.47}{IMRPhenomDNRTidal-LS}{3.42}{IMRPhenomPv2NRTidal-HS}{3.50}{IMRPhenomPv2NRTidal-LS}{3.42}{SEOBNRv4TsurrogateHS}{3.45}{SEOBNRv4TsurrogateLS}{3.42}{SEOBNRv4TsurrogatehighspinRIFT}{3.45}{SEOBNRv4TsurrogatelowspinRIFT}{3.42}{TEOBResumS-HS}{3.46}{TEOBResumS-LS}{3.42}{TaylorF2-HS}{3.48}{TaylorF2-LS}{3.42}{PrecessingSpinIMRTidalHS}{3.50}{PrecessingSpinIMRTidalLS}{3.42}{PublicationSamples}{3.50}}}
\newcommand{\totalmassdetfourtwofiveplus}[1]{\IfEqCase{#1}{{AlignedSpinInspiralTidalHS}{0.3}{AlignedSpinInspiralTidalLS}{0.04}{AlignedSpinTidalHS}{0.3}{AlignedSpinTidalLS}{0.04}{IMRPhenomDNRTidal-HS}{0.4}{IMRPhenomDNRTidal-LS}{0.04}{IMRPhenomPv2NRTidal-HS}{0.3}{IMRPhenomPv2NRTidal-LS}{0.04}{SEOBNRv4TsurrogateHS}{0.2}{SEOBNRv4TsurrogateLS}{0.04}{SEOBNRv4TsurrogatehighspinRIFT}{0.2}{SEOBNRv4TsurrogatelowspinRIFT}{0.04}{TEOBResumS-HS}{0.2}{TEOBResumS-LS}{0.04}{TaylorF2-HS}{0.3}{TaylorF2-LS}{0.04}{PrecessingSpinIMRTidalHS}{0.3}{PrecessingSpinIMRTidalLS}{0.04}{PublicationSamples}{0.3}}}
\newcommand{\redshiftfourtwofiveminus}[1]{\IfEqCase{#1}{{AlignedSpinInspiralTidalHS}{0.02}{AlignedSpinInspiralTidalLS}{0.02}{AlignedSpinTidalHS}{0.02}{AlignedSpinTidalLS}{0.02}{IMRPhenomDNRTidal-HS}{0.02}{IMRPhenomDNRTidal-LS}{0.02}{IMRPhenomPv2NRTidal-HS}{0.02}{IMRPhenomPv2NRTidal-LS}{0.02}{SEOBNRv4TsurrogateHS}{0.02}{SEOBNRv4TsurrogateLS}{0.02}{SEOBNRv4TsurrogatehighspinRIFT}{0.02}{SEOBNRv4TsurrogatelowspinRIFT}{0.02}{TEOBResumS-HS}{0.02}{TEOBResumS-LS}{0.02}{TaylorF2-HS}{0.02}{TaylorF2-LS}{0.02}{PrecessingSpinIMRTidalHS}{0.02}{PrecessingSpinIMRTidalLS}{0.02}{PublicationSamples}{0.02}}}
\newcommand{\redshiftfourtwofivemed}[1]{\IfEqCase{#1}{{AlignedSpinInspiralTidalHS}{0.04}{AlignedSpinInspiralTidalLS}{0.04}{AlignedSpinTidalHS}{0.04}{AlignedSpinTidalLS}{0.03}{IMRPhenomDNRTidal-HS}{0.04}{IMRPhenomDNRTidal-LS}{0.03}{IMRPhenomPv2NRTidal-HS}{0.03}{IMRPhenomPv2NRTidal-LS}{0.03}{SEOBNRv4TsurrogateHS}{0.03}{SEOBNRv4TsurrogateLS}{0.03}{SEOBNRv4TsurrogatehighspinRIFT}{0.04}{SEOBNRv4TsurrogatelowspinRIFT}{0.03}{TEOBResumS-HS}{0.04}{TEOBResumS-LS}{0.03}{TaylorF2-HS}{0.04}{TaylorF2-LS}{0.04}{PrecessingSpinIMRTidalHS}{0.03}{PrecessingSpinIMRTidalLS}{0.03}{PublicationSamples}{0.03}}}
\newcommand{\redshiftfourtwofiveplus}[1]{\IfEqCase{#1}{{AlignedSpinInspiralTidalHS}{0.02}{AlignedSpinInspiralTidalLS}{0.01}{AlignedSpinTidalHS}{0.02}{AlignedSpinTidalLS}{0.01}{IMRPhenomDNRTidal-HS}{0.01}{IMRPhenomDNRTidal-LS}{0.01}{IMRPhenomPv2NRTidal-HS}{0.01}{IMRPhenomPv2NRTidal-LS}{0.01}{SEOBNRv4TsurrogateHS}{0.01}{SEOBNRv4TsurrogateLS}{0.01}{SEOBNRv4TsurrogatehighspinRIFT}{0.02}{SEOBNRv4TsurrogatelowspinRIFT}{0.02}{TEOBResumS-HS}{0.02}{TEOBResumS-LS}{0.01}{TaylorF2-HS}{0.02}{TaylorF2-LS}{0.02}{PrecessingSpinIMRTidalHS}{0.01}{PrecessingSpinIMRTidalLS}{0.01}{PublicationSamples}{0.01}}}
\newcommand{\iotafourtwofiveminus}[1]{\IfEqCase{#1}{{AlignedSpinInspiralTidalHS}{0.97}{AlignedSpinInspiralTidalLS}{0.88}{AlignedSpinTidalHS}{0.83}{AlignedSpinTidalLS}{0.83}{IMRPhenomDNRTidal-HS}{0.74}{IMRPhenomDNRTidal-LS}{0.78}{IMRPhenomPv2NRTidal-HS}{0.85}{IMRPhenomPv2NRTidal-LS}{0.84}{SEOBNRv4TsurrogateHS}{0.89}{SEOBNRv4TsurrogateLS}{0.85}{SEOBNRv4TsurrogatehighspinRIFT}{0.86}{SEOBNRv4TsurrogatelowspinRIFT}{0.86}{TEOBResumS-HS}{0.86}{TEOBResumS-LS}{0.88}{TaylorF2-HS}{0.97}{TaylorF2-LS}{0.88}{PrecessingSpinIMRTidalHS}{0.85}{PrecessingSpinIMRTidalLS}{0.84}{PublicationSamples}{0.85}}}
\newcommand{\iotafourtwofivemed}[1]{\IfEqCase{#1}{{AlignedSpinInspiralTidalHS}{1.22}{AlignedSpinInspiralTidalLS}{1.11}{AlignedSpinTidalHS}{1.06}{AlignedSpinTidalLS}{1.06}{IMRPhenomDNRTidal-HS}{0.96}{IMRPhenomDNRTidal-LS}{1.02}{IMRPhenomPv2NRTidal-HS}{1.09}{IMRPhenomPv2NRTidal-LS}{1.07}{SEOBNRv4TsurrogateHS}{1.13}{SEOBNRv4TsurrogateLS}{1.09}{SEOBNRv4TsurrogatehighspinRIFT}{1.10}{SEOBNRv4TsurrogatelowspinRIFT}{1.09}{TEOBResumS-HS}{1.09}{TEOBResumS-LS}{1.12}{TaylorF2-HS}{1.22}{TaylorF2-LS}{1.11}{PrecessingSpinIMRTidalHS}{1.09}{PrecessingSpinIMRTidalLS}{1.07}{PublicationSamples}{1.09}}}
\newcommand{\iotafourtwofiveplus}[1]{\IfEqCase{#1}{{AlignedSpinInspiralTidalHS}{1.64}{AlignedSpinInspiralTidalLS}{1.75}{AlignedSpinTidalHS}{1.79}{AlignedSpinTidalLS}{1.78}{IMRPhenomDNRTidal-HS}{1.88}{IMRPhenomDNRTidal-LS}{1.83}{IMRPhenomPv2NRTidal-HS}{1.77}{IMRPhenomPv2NRTidal-LS}{1.78}{SEOBNRv4TsurrogateHS}{1.72}{SEOBNRv4TsurrogateLS}{1.76}{SEOBNRv4TsurrogatehighspinRIFT}{1.76}{SEOBNRv4TsurrogatelowspinRIFT}{1.75}{TEOBResumS-HS}{1.77}{TEOBResumS-LS}{1.74}{TaylorF2-HS}{1.63}{TaylorF2-LS}{1.75}{PrecessingSpinIMRTidalHS}{1.77}{PrecessingSpinIMRTidalLS}{1.78}{PublicationSamples}{1.77}}}
\newcommand{\spinonexfourtwofiveminus}[1]{\IfEqCase{#1}{{AlignedSpinInspiralTidalHS}{0.00}{AlignedSpinInspiralTidalLS}{0.00}{AlignedSpinTidalHS}{0.00}{AlignedSpinTidalLS}{0.00}{IMRPhenomDNRTidal-HS}{0.00}{IMRPhenomDNRTidal-LS}{0.00}{IMRPhenomPv2NRTidal-HS}{0.50}{IMRPhenomPv2NRTidal-LS}{0.03}{SEOBNRv4TsurrogateHS}{0.00}{SEOBNRv4TsurrogateLS}{0.00}{SEOBNRv4TsurrogatehighspinRIFT}{0.00}{SEOBNRv4TsurrogatelowspinRIFT}{0.00}{TEOBResumS-HS}{0.00}{TEOBResumS-LS}{0.00}{TaylorF2-HS}{0.00}{TaylorF2-LS}{0.00}{PrecessingSpinIMRTidalHS}{0.50}{PrecessingSpinIMRTidalLS}{0.03}{PublicationSamples}{0.50}}}
\newcommand{\spinonexfourtwofivemed}[1]{\IfEqCase{#1}{{AlignedSpinInspiralTidalHS}{0.00}{AlignedSpinInspiralTidalLS}{0.00}{AlignedSpinTidalHS}{0.00}{AlignedSpinTidalLS}{0.00}{IMRPhenomDNRTidal-HS}{0.00}{IMRPhenomDNRTidal-LS}{0.00}{IMRPhenomPv2NRTidal-HS}{0.00}{IMRPhenomPv2NRTidal-LS}{0.00009}{SEOBNRv4TsurrogateHS}{0.00}{SEOBNRv4TsurrogateLS}{0.00}{SEOBNRv4TsurrogatehighspinRIFT}{0.00}{SEOBNRv4TsurrogatelowspinRIFT}{0.00}{TEOBResumS-HS}{0.00}{TEOBResumS-LS}{0.00}{TaylorF2-HS}{0.00}{TaylorF2-LS}{0.00}{PrecessingSpinIMRTidalHS}{0.00}{PrecessingSpinIMRTidalLS}{0.0001}{PublicationSamples}{0.00}}}
\newcommand{\spinonexfourtwofiveplus}[1]{\IfEqCase{#1}{{AlignedSpinInspiralTidalHS}{0.00}{AlignedSpinInspiralTidalLS}{0.00}{AlignedSpinTidalHS}{0.00}{AlignedSpinTidalLS}{0.00}{IMRPhenomDNRTidal-HS}{0.00}{IMRPhenomDNRTidal-LS}{0.00}{IMRPhenomPv2NRTidal-HS}{0.47}{IMRPhenomPv2NRTidal-LS}{0.03}{SEOBNRv4TsurrogateHS}{0.00}{SEOBNRv4TsurrogateLS}{0.00}{SEOBNRv4TsurrogatehighspinRIFT}{0.00}{SEOBNRv4TsurrogatelowspinRIFT}{0.00}{TEOBResumS-HS}{0.00}{TEOBResumS-LS}{0.00}{TaylorF2-HS}{0.00}{TaylorF2-LS}{0.00}{PrecessingSpinIMRTidalHS}{0.47}{PrecessingSpinIMRTidalLS}{0.03}{PublicationSamples}{0.47}}}
\newcommand{\chirpmassdetfourtwofiveminus}[1]{\IfEqCase{#1}{{AlignedSpinInspiralTidalHS}{0.0005}{AlignedSpinInspiralTidalLS}{0.0003}{AlignedSpinTidalHS}{0.0005}{AlignedSpinTidalLS}{0.0003}{IMRPhenomDNRTidal-HS}{0.0005}{IMRPhenomDNRTidal-LS}{0.0003}{IMRPhenomPv2NRTidal-HS}{0.0006}{IMRPhenomPv2NRTidal-LS}{0.0003}{SEOBNRv4TsurrogateHS}{0.0005}{SEOBNRv4TsurrogateLS}{0.0003}{SEOBNRv4TsurrogatehighspinRIFT}{0.0005}{SEOBNRv4TsurrogatelowspinRIFT}{0.0004}{TEOBResumS-HS}{0.0005}{TEOBResumS-LS}{0.0003}{TaylorF2-HS}{0.0005}{TaylorF2-LS}{0.0003}{PrecessingSpinIMRTidalHS}{0.0006}{PrecessingSpinIMRTidalLS}{0.0003}{PublicationSamples}{0.0006}}}
\newcommand{\chirpmassdetfourtwofivemed}[1]{\IfEqCase{#1}{{AlignedSpinInspiralTidalHS}{1.49}{AlignedSpinInspiralTidalLS}{1.49}{AlignedSpinTidalHS}{1.49}{AlignedSpinTidalLS}{1.49}{IMRPhenomDNRTidal-HS}{1.49}{IMRPhenomDNRTidal-LS}{1.49}{IMRPhenomPv2NRTidal-HS}{1.49}{IMRPhenomPv2NRTidal-LS}{1.49}{SEOBNRv4TsurrogateHS}{1.49}{SEOBNRv4TsurrogateLS}{1.49}{SEOBNRv4TsurrogatehighspinRIFT}{1.49}{SEOBNRv4TsurrogatelowspinRIFT}{1.49}{TEOBResumS-HS}{1.49}{TEOBResumS-LS}{1.49}{TaylorF2-HS}{1.49}{TaylorF2-LS}{1.49}{PrecessingSpinIMRTidalHS}{1.49}{PrecessingSpinIMRTidalLS}{1.49}{PublicationSamples}{1.49}}}
\newcommand{\chirpmassdetfourtwofiveplus}[1]{\IfEqCase{#1}{{AlignedSpinInspiralTidalHS}{0.0007}{AlignedSpinInspiralTidalLS}{0.0003}{AlignedSpinTidalHS}{0.0007}{AlignedSpinTidalLS}{0.0003}{IMRPhenomDNRTidal-HS}{0.0008}{IMRPhenomDNRTidal-LS}{0.0004}{IMRPhenomPv2NRTidal-HS}{0.0008}{IMRPhenomPv2NRTidal-LS}{0.0003}{SEOBNRv4TsurrogateHS}{0.0006}{SEOBNRv4TsurrogateLS}{0.0003}{SEOBNRv4TsurrogatehighspinRIFT}{0.0006}{SEOBNRv4TsurrogatelowspinRIFT}{0.0003}{TEOBResumS-HS}{0.0005}{TEOBResumS-LS}{0.0003}{TaylorF2-HS}{0.0007}{TaylorF2-LS}{0.0003}{PrecessingSpinIMRTidalHS}{0.0008}{PrecessingSpinIMRTidalLS}{0.0003}{PublicationSamples}{0.0008}}}
\newcommand{\cosiotafourtwofiveminus}[1]{\IfEqCase{#1}{{AlignedSpinInspiralTidalHS}{1.30}{AlignedSpinInspiralTidalLS}{1.40}{AlignedSpinTidalHS}{1.44}{AlignedSpinTidalLS}{1.44}{IMRPhenomDNRTidal-HS}{1.53}{IMRPhenomDNRTidal-LS}{1.48}{IMRPhenomPv2NRTidal-HS}{1.42}{IMRPhenomPv2NRTidal-LS}{1.44}{SEOBNRv4TsurrogateHS}{1.38}{SEOBNRv4TsurrogateLS}{1.42}{SEOBNRv4TsurrogatehighspinRIFT}{1.41}{SEOBNRv4TsurrogatelowspinRIFT}{1.42}{TEOBResumS-HS}{1.42}{TEOBResumS-LS}{1.40}{TaylorF2-HS}{1.30}{TaylorF2-LS}{1.40}{PrecessingSpinIMRTidalHS}{1.42}{PrecessingSpinIMRTidalLS}{1.44}{PublicationSamples}{1.42}}}
\newcommand{\cosiotafourtwofivemed}[1]{\IfEqCase{#1}{{AlignedSpinInspiralTidalHS}{0.34}{AlignedSpinInspiralTidalLS}{0.44}{AlignedSpinTidalHS}{0.49}{AlignedSpinTidalLS}{0.49}{IMRPhenomDNRTidal-HS}{0.58}{IMRPhenomDNRTidal-LS}{0.53}{IMRPhenomPv2NRTidal-HS}{0.46}{IMRPhenomPv2NRTidal-LS}{0.48}{SEOBNRv4TsurrogateHS}{0.43}{SEOBNRv4TsurrogateLS}{0.46}{SEOBNRv4TsurrogatehighspinRIFT}{0.45}{SEOBNRv4TsurrogatelowspinRIFT}{0.46}{TEOBResumS-HS}{0.46}{TEOBResumS-LS}{0.44}{TaylorF2-HS}{0.34}{TaylorF2-LS}{0.44}{PrecessingSpinIMRTidalHS}{0.46}{PrecessingSpinIMRTidalLS}{0.48}{PublicationSamples}{0.46}}}
\newcommand{\cosiotafourtwofiveplus}[1]{\IfEqCase{#1}{{AlignedSpinInspiralTidalHS}{0.62}{AlignedSpinInspiralTidalLS}{0.53}{AlignedSpinTidalHS}{0.49}{AlignedSpinTidalLS}{0.49}{IMRPhenomDNRTidal-HS}{0.40}{IMRPhenomDNRTidal-LS}{0.45}{IMRPhenomPv2NRTidal-HS}{0.51}{IMRPhenomPv2NRTidal-LS}{0.49}{SEOBNRv4TsurrogateHS}{0.54}{SEOBNRv4TsurrogateLS}{0.51}{SEOBNRv4TsurrogatehighspinRIFT}{0.52}{SEOBNRv4TsurrogatelowspinRIFT}{0.51}{TEOBResumS-HS}{0.51}{TEOBResumS-LS}{0.54}{TaylorF2-HS}{0.63}{TaylorF2-LS}{0.53}{PrecessingSpinIMRTidalHS}{0.51}{PrecessingSpinIMRTidalLS}{0.49}{PublicationSamples}{0.51}}}
\newcommand{\comovingdistfourtwofiveminus}[1]{\IfEqCase{#1}{{AlignedSpinInspiralTidalHS}{72}{AlignedSpinInspiralTidalLS}{71}{AlignedSpinTidalHS}{69}{AlignedSpinTidalLS}{70}{IMRPhenomDNRTidal-HS}{69}{IMRPhenomDNRTidal-LS}{69}{IMRPhenomPv2NRTidal-HS}{67}{IMRPhenomPv2NRTidal-LS}{68}{SEOBNRv4TsurrogateHS}{68}{SEOBNRv4TsurrogateLS}{69}{SEOBNRv4TsurrogatehighspinRIFT}{70}{SEOBNRv4TsurrogatelowspinRIFT}{68}{TEOBResumS-HS}{70}{TEOBResumS-LS}{69}{TaylorF2-HS}{72}{TaylorF2-LS}{71}{PrecessingSpinIMRTidalHS}{67}{PrecessingSpinIMRTidalLS}{68}{PublicationSamples}{67}}}
\newcommand{\comovingdistfourtwofivemed}[1]{\IfEqCase{#1}{{AlignedSpinInspiralTidalHS}{156}{AlignedSpinInspiralTidalLS}{157}{AlignedSpinTidalHS}{155}{AlignedSpinTidalLS}{153}{IMRPhenomDNRTidal-HS}{155}{IMRPhenomDNRTidal-LS}{153}{IMRPhenomPv2NRTidal-HS}{151}{IMRPhenomPv2NRTidal-LS}{151}{SEOBNRv4TsurrogateHS}{153}{SEOBNRv4TsurrogateLS}{153}{SEOBNRv4TsurrogatehighspinRIFT}{157}{SEOBNRv4TsurrogatelowspinRIFT}{152}{TEOBResumS-HS}{157}{TEOBResumS-LS}{153}{TaylorF2-HS}{156}{TaylorF2-LS}{157}{PrecessingSpinIMRTidalHS}{151}{PrecessingSpinIMRTidalLS}{151}{PublicationSamples}{151}}}
\newcommand{\comovingdistfourtwofiveplus}[1]{\IfEqCase{#1}{{AlignedSpinInspiralTidalHS}{67}{AlignedSpinInspiralTidalLS}{65}{AlignedSpinTidalHS}{65}{AlignedSpinTidalLS}{63}{IMRPhenomDNRTidal-HS}{63}{IMRPhenomDNRTidal-LS}{64}{IMRPhenomPv2NRTidal-HS}{64}{IMRPhenomPv2NRTidal-LS}{64}{SEOBNRv4TsurrogateHS}{64}{SEOBNRv4TsurrogateLS}{64}{SEOBNRv4TsurrogatehighspinRIFT}{70}{SEOBNRv4TsurrogatelowspinRIFT}{65}{TEOBResumS-HS}{68}{TEOBResumS-LS}{64}{TaylorF2-HS}{67}{TaylorF2-LS}{65}{PrecessingSpinIMRTidalHS}{64}{PrecessingSpinIMRTidalLS}{64}{PublicationSamples}{64}}}
\newcommand{\logpriorfourtwofiveminus}[1]{\IfEqCase{#1}{{AlignedSpinInspiralTidalHS}{8.6}{AlignedSpinInspiralTidalLS}{8.5}{IMRPhenomDNRTidal-HS}{8.6}{IMRPhenomDNRTidal-LS}{8.6}{IMRPhenomPv2NRTidal-HS}{8.6}{IMRPhenomPv2NRTidal-LS}{8.4}{SEOBNRv4TsurrogateHS}{8.4}{SEOBNRv4TsurrogateLS}{8.8}{TaylorF2-HS}{8.6}{TaylorF2-LS}{8.5}{PrecessingSpinIMRTidalHS}{8.6}{PrecessingSpinIMRTidalLS}{8.4}{PublicationSamples}{8.6}}}
\newcommand{\logpriorfourtwofivemed}[1]{\IfEqCase{#1}{{AlignedSpinInspiralTidalHS}{102.5}{AlignedSpinInspiralTidalLS}{106.7}{IMRPhenomDNRTidal-HS}{94.5}{IMRPhenomDNRTidal-LS}{99.2}{IMRPhenomPv2NRTidal-HS}{98.4}{IMRPhenomPv2NRTidal-LS}{97.8}{SEOBNRv4TsurrogateHS}{95.6}{SEOBNRv4TsurrogateLS}{99.0}{TaylorF2-HS}{102.5}{TaylorF2-LS}{106.7}{PrecessingSpinIMRTidalHS}{98.4}{PrecessingSpinIMRTidalLS}{97.8}{PublicationSamples}{98.4}}}
\newcommand{\logpriorfourtwofiveplus}[1]{\IfEqCase{#1}{{AlignedSpinInspiralTidalHS}{6.8}{AlignedSpinInspiralTidalLS}{6.9}{IMRPhenomDNRTidal-HS}{6.8}{IMRPhenomDNRTidal-LS}{6.9}{IMRPhenomPv2NRTidal-HS}{6.7}{IMRPhenomPv2NRTidal-LS}{6.7}{SEOBNRv4TsurrogateHS}{6.9}{SEOBNRv4TsurrogateLS}{6.9}{TaylorF2-HS}{6.8}{TaylorF2-LS}{6.9}{PrecessingSpinIMRTidalHS}{6.7}{PrecessingSpinIMRTidalLS}{6.7}{PublicationSamples}{6.7}}}
\newcommand{\tiltonefourtwofiveminus}[1]{\IfEqCase{#1}{{AlignedSpinInspiralTidalHS}{0.00}{AlignedSpinInspiralTidalLS}{0.00}{AlignedSpinTidalHS}{0.00}{AlignedSpinTidalLS}{0.00}{IMRPhenomDNRTidal-HS}{0.00}{IMRPhenomDNRTidal-LS}{0.00}{IMRPhenomPv2NRTidal-HS}{0.80}{IMRPhenomPv2NRTidal-LS}{0.74}{SEOBNRv4TsurrogateHS}{0.00}{SEOBNRv4TsurrogateLS}{0.00}{SEOBNRv4TsurrogatehighspinRIFT}{0.00}{SEOBNRv4TsurrogatelowspinRIFT}{0.00}{TEOBResumS-HS}{0.00}{TEOBResumS-LS}{0.00}{TaylorF2-HS}{0.00}{TaylorF2-LS}{0.00}{PrecessingSpinIMRTidalHS}{0.80}{PrecessingSpinIMRTidalLS}{0.74}{PublicationSamples}{0.79}}}
\newcommand{\tiltonefourtwofivemed}[1]{\IfEqCase{#1}{{AlignedSpinInspiralTidalHS}{0.00}{AlignedSpinInspiralTidalLS}{0.00}{AlignedSpinTidalHS}{0.00}{AlignedSpinTidalLS}{0.00}{IMRPhenomDNRTidal-HS}{0.00}{IMRPhenomDNRTidal-LS}{0.00}{IMRPhenomPv2NRTidal-HS}{1.31}{IMRPhenomPv2NRTidal-LS}{1.03}{SEOBNRv4TsurrogateHS}{0.00}{SEOBNRv4TsurrogateLS}{0.00}{SEOBNRv4TsurrogatehighspinRIFT}{0.00}{SEOBNRv4TsurrogatelowspinRIFT}{0.00}{TEOBResumS-HS}{0.00}{TEOBResumS-LS}{0.00}{TaylorF2-HS}{0.00}{TaylorF2-LS}{0.00}{PrecessingSpinIMRTidalHS}{1.31}{PrecessingSpinIMRTidalLS}{1.03}{PublicationSamples}{1.31}}}
\newcommand{\tiltonefourtwofiveplus}[1]{\IfEqCase{#1}{{AlignedSpinInspiralTidalHS}{3.14}{AlignedSpinInspiralTidalLS}{3.14}{AlignedSpinTidalHS}{3.14}{AlignedSpinTidalLS}{3.14}{IMRPhenomDNRTidal-HS}{3.14}{IMRPhenomDNRTidal-LS}{3.14}{IMRPhenomPv2NRTidal-HS}{0.66}{IMRPhenomPv2NRTidal-LS}{1.16}{SEOBNRv4TsurrogateHS}{3.14}{SEOBNRv4TsurrogateLS}{3.14}{SEOBNRv4TsurrogatehighspinRIFT}{3.14}{SEOBNRv4TsurrogatelowspinRIFT}{3.14}{TEOBResumS-HS}{3.14}{TEOBResumS-LS}{3.14}{TaylorF2-HS}{3.14}{TaylorF2-LS}{3.14}{PrecessingSpinIMRTidalHS}{0.66}{PrecessingSpinIMRTidalLS}{1.17}{PublicationSamples}{0.66}}}
\newcommand{\spintwoyfourtwofiveminus}[1]{\IfEqCase{#1}{{AlignedSpinInspiralTidalHS}{0.00}{AlignedSpinInspiralTidalLS}{0.00}{AlignedSpinTidalHS}{0.00}{AlignedSpinTidalLS}{0.00}{IMRPhenomDNRTidal-HS}{0.00}{IMRPhenomDNRTidal-LS}{0.00}{IMRPhenomPv2NRTidal-HS}{0.48}{IMRPhenomPv2NRTidal-LS}{0.03}{SEOBNRv4TsurrogateHS}{0.00}{SEOBNRv4TsurrogateLS}{0.00}{SEOBNRv4TsurrogatehighspinRIFT}{0.00}{SEOBNRv4TsurrogatelowspinRIFT}{0.00}{TEOBResumS-HS}{0.00}{TEOBResumS-LS}{0.00}{TaylorF2-HS}{0.00}{TaylorF2-LS}{0.00}{PrecessingSpinIMRTidalHS}{0.48}{PrecessingSpinIMRTidalLS}{0.03}{PublicationSamples}{0.48}}}
\newcommand{\spintwoyfourtwofivemed}[1]{\IfEqCase{#1}{{AlignedSpinInspiralTidalHS}{0.00}{AlignedSpinInspiralTidalLS}{0.00}{AlignedSpinTidalHS}{0.00}{AlignedSpinTidalLS}{0.00}{IMRPhenomDNRTidal-HS}{0.00}{IMRPhenomDNRTidal-LS}{0.00}{IMRPhenomPv2NRTidal-HS}{0.00002}{IMRPhenomPv2NRTidal-LS}{0.00003}{SEOBNRv4TsurrogateHS}{0.00}{SEOBNRv4TsurrogateLS}{0.00}{SEOBNRv4TsurrogatehighspinRIFT}{0.00}{SEOBNRv4TsurrogatelowspinRIFT}{0.00}{TEOBResumS-HS}{0.00}{TEOBResumS-LS}{0.00}{TaylorF2-HS}{0.00}{TaylorF2-LS}{0.00}{PrecessingSpinIMRTidalHS}{0.00}{PrecessingSpinIMRTidalLS}{0.00002}{PublicationSamples}{0.00}}}
\newcommand{\spintwoyfourtwofiveplus}[1]{\IfEqCase{#1}{{AlignedSpinInspiralTidalHS}{0.00}{AlignedSpinInspiralTidalLS}{0.00}{AlignedSpinTidalHS}{0.00}{AlignedSpinTidalLS}{0.00}{IMRPhenomDNRTidal-HS}{0.00}{IMRPhenomDNRTidal-LS}{0.00}{IMRPhenomPv2NRTidal-HS}{0.48}{IMRPhenomPv2NRTidal-LS}{0.03}{SEOBNRv4TsurrogateHS}{0.00}{SEOBNRv4TsurrogateLS}{0.00}{SEOBNRv4TsurrogatehighspinRIFT}{0.00}{SEOBNRv4TsurrogatelowspinRIFT}{0.00}{TEOBResumS-HS}{0.00}{TEOBResumS-LS}{0.00}{TaylorF2-HS}{0.00}{TaylorF2-LS}{0.00}{PrecessingSpinIMRTidalHS}{0.48}{PrecessingSpinIMRTidalLS}{0.03}{PublicationSamples}{0.48}}}
\newcommand{\spintwozfourtwofiveminus}[1]{\IfEqCase{#1}{{AlignedSpinInspiralTidalHS}{0.18}{AlignedSpinInspiralTidalLS}{0.02}{AlignedSpinTidalHS}{0.24}{AlignedSpinTidalLS}{0.02}{IMRPhenomDNRTidal-HS}{0.39}{IMRPhenomDNRTidal-LS}{0.02}{IMRPhenomPv2NRTidal-HS}{0.18}{IMRPhenomPv2NRTidal-LS}{0.02}{SEOBNRv4TsurrogateHS}{0.16}{SEOBNRv4TsurrogateLS}{0.02}{SEOBNRv4TsurrogatehighspinRIFT}{0.18}{SEOBNRv4TsurrogatelowspinRIFT}{0.02}{TEOBResumS-HS}{0.18}{TEOBResumS-LS}{0.02}{TaylorF2-HS}{0.18}{TaylorF2-LS}{0.02}{PrecessingSpinIMRTidalHS}{0.18}{PrecessingSpinIMRTidalLS}{0.02}{PublicationSamples}{0.18}}}
\newcommand{\spintwozfourtwofivemed}[1]{\IfEqCase{#1}{{AlignedSpinInspiralTidalHS}{0.04}{AlignedSpinInspiralTidalLS}{0.008}{AlignedSpinTidalHS}{0.03}{AlignedSpinTidalLS}{0.009}{IMRPhenomDNRTidal-HS}{0.03}{IMRPhenomDNRTidal-LS}{0.009}{IMRPhenomPv2NRTidal-HS}{0.03}{IMRPhenomPv2NRTidal-LS}{0.009}{SEOBNRv4TsurrogateHS}{0.02}{SEOBNRv4TsurrogateLS}{0.009}{SEOBNRv4TsurrogatehighspinRIFT}{0.03}{SEOBNRv4TsurrogatelowspinRIFT}{0.01}{TEOBResumS-HS}{0.03}{TEOBResumS-LS}{0.009}{TaylorF2-HS}{0.04}{TaylorF2-LS}{0.008}{PrecessingSpinIMRTidalHS}{0.03}{PrecessingSpinIMRTidalLS}{0.009}{PublicationSamples}{0.03}}}
\newcommand{\spintwozfourtwofiveplus}[1]{\IfEqCase{#1}{{AlignedSpinInspiralTidalHS}{0.30}{AlignedSpinInspiralTidalLS}{0.03}{AlignedSpinTidalHS}{0.26}{AlignedSpinTidalLS}{0.03}{IMRPhenomDNRTidal-HS}{0.37}{IMRPhenomDNRTidal-LS}{0.03}{IMRPhenomPv2NRTidal-HS}{0.30}{IMRPhenomPv2NRTidal-LS}{0.03}{SEOBNRv4TsurrogateHS}{0.20}{SEOBNRv4TsurrogateLS}{0.03}{SEOBNRv4TsurrogatehighspinRIFT}{0.21}{SEOBNRv4TsurrogatelowspinRIFT}{0.03}{TEOBResumS-HS}{0.21}{TEOBResumS-LS}{0.03}{TaylorF2-HS}{0.30}{TaylorF2-LS}{0.03}{PrecessingSpinIMRTidalHS}{0.30}{PrecessingSpinIMRTidalLS}{0.03}{PublicationSamples}{0.30}}}
\newcommand{\massonesourcefourtwofiveminus}[1]{\IfEqCase{#1}{{AlignedSpinInspiralTidalHS}{0.3}{AlignedSpinInspiralTidalLS}{0.10}{AlignedSpinTidalHS}{0.2}{AlignedSpinTidalLS}{0.09}{IMRPhenomDNRTidal-HS}{0.3}{IMRPhenomDNRTidal-LS}{0.09}{IMRPhenomPv2NRTidal-HS}{0.3}{IMRPhenomPv2NRTidal-LS}{0.09}{SEOBNRv4TsurrogateHS}{0.2}{SEOBNRv4TsurrogateLS}{0.09}{SEOBNRv4TsurrogatehighspinRIFT}{0.2}{SEOBNRv4TsurrogatelowspinRIFT}{0.09}{TEOBResumS-HS}{0.2}{TEOBResumS-LS}{0.09}{TaylorF2-HS}{0.3}{TaylorF2-LS}{0.10}{PrecessingSpinIMRTidalHS}{0.3}{PrecessingSpinIMRTidalLS}{0.09}{PublicationSamples}{0.3}}}
\newcommand{\massonesourcefourtwofivemed}[1]{\IfEqCase{#1}{{AlignedSpinInspiralTidalHS}{2.0}{AlignedSpinInspiralTidalLS}{1.75}{AlignedSpinTidalHS}{1.9}{AlignedSpinTidalLS}{1.75}{IMRPhenomDNRTidal-HS}{2.0}{IMRPhenomDNRTidal-LS}{1.75}{IMRPhenomPv2NRTidal-HS}{2.0}{IMRPhenomPv2NRTidal-LS}{1.74}{SEOBNRv4TsurrogateHS}{1.9}{SEOBNRv4TsurrogateLS}{1.74}{SEOBNRv4TsurrogatehighspinRIFT}{1.9}{SEOBNRv4TsurrogatelowspinRIFT}{1.75}{TEOBResumS-HS}{1.9}{TEOBResumS-LS}{1.75}{TaylorF2-HS}{2.0}{TaylorF2-LS}{1.75}{PrecessingSpinIMRTidalHS}{2.0}{PrecessingSpinIMRTidalLS}{1.74}{PublicationSamples}{2.0}}}
\newcommand{\massonesourcefourtwofiveplus}[1]{\IfEqCase{#1}{{AlignedSpinInspiralTidalHS}{0.5}{AlignedSpinInspiralTidalLS}{0.2}{AlignedSpinTidalHS}{0.6}{AlignedSpinTidalLS}{0.2}{IMRPhenomDNRTidal-HS}{0.7}{IMRPhenomDNRTidal-LS}{0.2}{IMRPhenomPv2NRTidal-HS}{0.6}{IMRPhenomPv2NRTidal-LS}{0.2}{SEOBNRv4TsurrogateHS}{0.5}{SEOBNRv4TsurrogateLS}{0.2}{SEOBNRv4TsurrogatehighspinRIFT}{0.5}{SEOBNRv4TsurrogatelowspinRIFT}{0.2}{TEOBResumS-HS}{0.5}{TEOBResumS-LS}{0.2}{TaylorF2-HS}{0.5}{TaylorF2-LS}{0.2}{PrecessingSpinIMRTidalHS}{0.6}{PrecessingSpinIMRTidalLS}{0.2}{PublicationSamples}{0.6}}}
\newcommand{\geocenttimefourtwofiveminus}[1]{\IfEqCase{#1}{{AlignedSpinInspiralTidalHS}{0.007}{AlignedSpinInspiralTidalLS}{0.008}{AlignedSpinTidalHS}{0.03}{AlignedSpinTidalLS}{0.02}{IMRPhenomDNRTidal-HS}{0.008}{IMRPhenomDNRTidal-LS}{0.01}{IMRPhenomPv2NRTidal-HS}{0.009}{IMRPhenomPv2NRTidal-LS}{0.01}{SEOBNRv4TsurrogateHS}{0.01}{SEOBNRv4TsurrogateLS}{0.008}{SEOBNRv4TsurrogatehighspinRIFT}{0.0}{SEOBNRv4TsurrogatelowspinRIFT}{0.0}{TEOBResumS-HS}{0.0}{TEOBResumS-LS}{0.0}{TaylorF2-HS}{0.007}{TaylorF2-LS}{0.008}{PrecessingSpinIMRTidalHS}{0.009}{PrecessingSpinIMRTidalLS}{0.01}{PublicationSamples}{0.009}}}
\newcommand{\geocenttimefourtwofivemed}[1]{\IfEqCase{#1}{{AlignedSpinInspiralTidalHS}{1240215503.0}{AlignedSpinInspiralTidalLS}{1240215503.0}{AlignedSpinTidalHS}{1240215503.0}{AlignedSpinTidalLS}{1240215503.0}{IMRPhenomDNRTidal-HS}{1240215503.0}{IMRPhenomDNRTidal-LS}{1240215503.0}{IMRPhenomPv2NRTidal-HS}{1240215503.0}{IMRPhenomPv2NRTidal-LS}{1240215503.0}{SEOBNRv4TsurrogateHS}{1240215503.0}{SEOBNRv4TsurrogateLS}{1240215503.0}{SEOBNRv4TsurrogatehighspinRIFT}{1240215503.0}{SEOBNRv4TsurrogatelowspinRIFT}{1240215503.0}{TEOBResumS-HS}{1240215503.0}{TEOBResumS-LS}{1240215503.0}{TaylorF2-HS}{1240215503.0}{TaylorF2-LS}{1240215503.0}{PrecessingSpinIMRTidalHS}{1240215503.0}{PrecessingSpinIMRTidalLS}{1240215503.0}{PublicationSamples}{1240215503.0}}}
\newcommand{\geocenttimefourtwofiveplus}[1]{\IfEqCase{#1}{{AlignedSpinInspiralTidalHS}{0.04}{AlignedSpinInspiralTidalLS}{0.03}{AlignedSpinTidalHS}{0.02}{AlignedSpinTidalLS}{0.02}{IMRPhenomDNRTidal-HS}{0.04}{IMRPhenomDNRTidal-LS}{0.03}{IMRPhenomPv2NRTidal-HS}{0.03}{IMRPhenomPv2NRTidal-LS}{0.03}{SEOBNRv4TsurrogateHS}{0.04}{SEOBNRv4TsurrogateLS}{0.04}{SEOBNRv4TsurrogatehighspinRIFT}{0.0}{SEOBNRv4TsurrogatelowspinRIFT}{0.0}{TEOBResumS-HS}{0.0}{TEOBResumS-LS}{0.0}{TaylorF2-HS}{0.04}{TaylorF2-LS}{0.03}{PrecessingSpinIMRTidalHS}{0.03}{PrecessingSpinIMRTidalLS}{0.03}{PublicationSamples}{0.03}}}
\newcommand{\costilttwofourtwofiveminus}[1]{\IfEqCase{#1}{{AlignedSpinInspiralTidalHS}{2.00}{AlignedSpinInspiralTidalLS}{2.00}{AlignedSpinTidalHS}{2.00}{AlignedSpinTidalLS}{2.00}{IMRPhenomDNRTidal-HS}{2.00}{IMRPhenomDNRTidal-LS}{2.00}{IMRPhenomPv2NRTidal-HS}{0.87}{IMRPhenomPv2NRTidal-LS}{1.13}{SEOBNRv4TsurrogateHS}{2.00}{SEOBNRv4TsurrogateLS}{2.00}{SEOBNRv4TsurrogatehighspinRIFT}{2.00}{SEOBNRv4TsurrogatelowspinRIFT}{2.00}{TEOBResumS-HS}{2.00}{TEOBResumS-LS}{2.00}{TaylorF2-HS}{2.00}{TaylorF2-LS}{2.00}{PrecessingSpinIMRTidalHS}{0.87}{PrecessingSpinIMRTidalLS}{1.12}{PublicationSamples}{0.86}}}
\newcommand{\costilttwofourtwofivemed}[1]{\IfEqCase{#1}{{AlignedSpinInspiralTidalHS}{1.00}{AlignedSpinInspiralTidalLS}{1.00}{AlignedSpinTidalHS}{1.00}{AlignedSpinTidalLS}{1.00}{IMRPhenomDNRTidal-HS}{1.00}{IMRPhenomDNRTidal-LS}{1.00}{IMRPhenomPv2NRTidal-HS}{0.16}{IMRPhenomPv2NRTidal-LS}{0.46}{SEOBNRv4TsurrogateHS}{1.00}{SEOBNRv4TsurrogateLS}{1.00}{SEOBNRv4TsurrogatehighspinRIFT}{1.00}{SEOBNRv4TsurrogatelowspinRIFT}{1.00}{TEOBResumS-HS}{1.00}{TEOBResumS-LS}{1.00}{TaylorF2-HS}{1.00}{TaylorF2-LS}{1.00}{PrecessingSpinIMRTidalHS}{0.16}{PrecessingSpinIMRTidalLS}{0.46}{PublicationSamples}{0.16}}}
\newcommand{\costilttwofourtwofiveplus}[1]{\IfEqCase{#1}{{AlignedSpinInspiralTidalHS}{0.00}{AlignedSpinInspiralTidalLS}{0.00}{AlignedSpinTidalHS}{0.00}{AlignedSpinTidalLS}{0.00}{IMRPhenomDNRTidal-HS}{0.00}{IMRPhenomDNRTidal-LS}{0.00}{IMRPhenomPv2NRTidal-HS}{0.70}{IMRPhenomPv2NRTidal-LS}{0.49}{SEOBNRv4TsurrogateHS}{0.00}{SEOBNRv4TsurrogateLS}{0.00}{SEOBNRv4TsurrogatehighspinRIFT}{0.00}{SEOBNRv4TsurrogatelowspinRIFT}{0.00}{TEOBResumS-HS}{0.00}{TEOBResumS-LS}{0.00}{TaylorF2-HS}{0.00}{TaylorF2-LS}{0.00}{PrecessingSpinIMRTidalHS}{0.70}{PrecessingSpinIMRTidalLS}{0.49}{PublicationSamples}{0.70}}}
\newcommand{\luminositydistancefourtwofiveminus}[1]{\IfEqCase{#1}{{AlignedSpinInspiralTidalHS}{0.08}{AlignedSpinInspiralTidalLS}{0.07}{AlignedSpinTidalHS}{0.07}{AlignedSpinTidalLS}{0.07}{IMRPhenomDNRTidal-HS}{0.07}{IMRPhenomDNRTidal-LS}{0.07}{IMRPhenomPv2NRTidal-HS}{0.07}{IMRPhenomPv2NRTidal-LS}{0.07}{SEOBNRv4TsurrogateHS}{0.07}{SEOBNRv4TsurrogateLS}{0.07}{SEOBNRv4TsurrogatehighspinRIFT}{0.07}{SEOBNRv4TsurrogatelowspinRIFT}{0.07}{TEOBResumS-HS}{0.07}{TEOBResumS-LS}{0.07}{TaylorF2-HS}{0.08}{TaylorF2-LS}{0.08}{PrecessingSpinIMRTidalHS}{0.07}{PrecessingSpinIMRTidalLS}{0.07}{PublicationSamples}{0.07}}}
\newcommand{\luminositydistancefourtwofivemed}[1]{\IfEqCase{#1}{{AlignedSpinInspiralTidalHS}{0.16}{AlignedSpinInspiralTidalLS}{0.16}{AlignedSpinTidalHS}{0.16}{AlignedSpinTidalLS}{0.16}{IMRPhenomDNRTidal-HS}{0.16}{IMRPhenomDNRTidal-LS}{0.16}{IMRPhenomPv2NRTidal-HS}{0.16}{IMRPhenomPv2NRTidal-LS}{0.16}{SEOBNRv4TsurrogateHS}{0.16}{SEOBNRv4TsurrogateLS}{0.16}{SEOBNRv4TsurrogatehighspinRIFT}{0.16}{SEOBNRv4TsurrogatelowspinRIFT}{0.16}{TEOBResumS-HS}{0.16}{TEOBResumS-LS}{0.16}{TaylorF2-HS}{0.16}{TaylorF2-LS}{0.16}{PrecessingSpinIMRTidalHS}{0.16}{PrecessingSpinIMRTidalLS}{0.16}{PublicationSamples}{0.16}}}
\newcommand{\luminositydistancefourtwofiveplus}[1]{\IfEqCase{#1}{{AlignedSpinInspiralTidalHS}{0.07}{AlignedSpinInspiralTidalLS}{0.07}{AlignedSpinTidalHS}{0.07}{AlignedSpinTidalLS}{0.07}{IMRPhenomDNRTidal-HS}{0.07}{IMRPhenomDNRTidal-LS}{0.07}{IMRPhenomPv2NRTidal-HS}{0.07}{IMRPhenomPv2NRTidal-LS}{0.07}{SEOBNRv4TsurrogateHS}{0.07}{SEOBNRv4TsurrogateLS}{0.07}{SEOBNRv4TsurrogatehighspinRIFT}{0.08}{SEOBNRv4TsurrogatelowspinRIFT}{0.07}{TEOBResumS-HS}{0.07}{TEOBResumS-LS}{0.07}{TaylorF2-HS}{0.07}{TaylorF2-LS}{0.07}{PrecessingSpinIMRTidalHS}{0.07}{PrecessingSpinIMRTidalLS}{0.07}{PublicationSamples}{0.07}}}
\newcommand{\spinonezfourtwofiveminus}[1]{\IfEqCase{#1}{{AlignedSpinInspiralTidalHS}{0.14}{AlignedSpinInspiralTidalLS}{0.02}{AlignedSpinTidalHS}{0.15}{AlignedSpinTidalLS}{0.02}{IMRPhenomDNRTidal-HS}{0.22}{IMRPhenomDNRTidal-LS}{0.02}{IMRPhenomPv2NRTidal-HS}{0.12}{IMRPhenomPv2NRTidal-LS}{0.02}{SEOBNRv4TsurrogateHS}{0.11}{SEOBNRv4TsurrogateLS}{0.02}{SEOBNRv4TsurrogatehighspinRIFT}{0.14}{SEOBNRv4TsurrogatelowspinRIFT}{0.02}{TEOBResumS-HS}{0.13}{TEOBResumS-LS}{0.02}{TaylorF2-HS}{0.14}{TaylorF2-LS}{0.02}{PrecessingSpinIMRTidalHS}{0.12}{PrecessingSpinIMRTidalLS}{0.02}{PublicationSamples}{0.12}}}
\newcommand{\spinonezfourtwofivemed}[1]{\IfEqCase{#1}{{AlignedSpinInspiralTidalHS}{0.04}{AlignedSpinInspiralTidalLS}{0.01}{AlignedSpinTidalHS}{0.04}{AlignedSpinTidalLS}{0.01}{IMRPhenomDNRTidal-HS}{0.06}{IMRPhenomDNRTidal-LS}{0.01}{IMRPhenomPv2NRTidal-HS}{0.06}{IMRPhenomPv2NRTidal-LS}{0.01}{SEOBNRv4TsurrogateHS}{0.04}{SEOBNRv4TsurrogateLS}{0.01}{SEOBNRv4TsurrogatehighspinRIFT}{0.04}{SEOBNRv4TsurrogatelowspinRIFT}{0.01}{TEOBResumS-HS}{0.04}{TEOBResumS-LS}{0.01}{TaylorF2-HS}{0.04}{TaylorF2-LS}{0.01}{PrecessingSpinIMRTidalHS}{0.06}{PrecessingSpinIMRTidalLS}{0.01}{PublicationSamples}{0.06}}}
\newcommand{\spinonezfourtwofiveplus}[1]{\IfEqCase{#1}{{AlignedSpinInspiralTidalHS}{0.19}{AlignedSpinInspiralTidalLS}{0.03}{AlignedSpinTidalHS}{0.20}{AlignedSpinTidalLS}{0.03}{IMRPhenomDNRTidal-HS}{0.26}{IMRPhenomDNRTidal-LS}{0.03}{IMRPhenomPv2NRTidal-HS}{0.18}{IMRPhenomPv2NRTidal-LS}{0.03}{SEOBNRv4TsurrogateHS}{0.16}{SEOBNRv4TsurrogateLS}{0.03}{SEOBNRv4TsurrogatehighspinRIFT}{0.16}{SEOBNRv4TsurrogatelowspinRIFT}{0.03}{TEOBResumS-HS}{0.16}{TEOBResumS-LS}{0.03}{TaylorF2-HS}{0.19}{TaylorF2-LS}{0.03}{PrecessingSpinIMRTidalHS}{0.18}{PrecessingSpinIMRTidalLS}{0.03}{PublicationSamples}{0.18}}}
\newcommand{\networkmatchedfiltersnrfourtwofiveminus}[1]{\IfEqCase{#1}{{AlignedSpinInspiralTidalHS}{0.4}{AlignedSpinInspiralTidalLS}{0.4}{IMRPhenomDNRTidal-HS}{0.4}{IMRPhenomDNRTidal-LS}{0.4}{IMRPhenomPv2NRTidal-HS}{0.4}{IMRPhenomPv2NRTidal-LS}{0.4}{SEOBNRv4TsurrogateHS}{0.4}{SEOBNRv4TsurrogateLS}{0.4}{TaylorF2-HS}{0.4}{TaylorF2-LS}{0.4}{PrecessingSpinIMRTidalHS}{0.4}{PrecessingSpinIMRTidalLS}{0.4}{PublicationSamples}{0.4}}}
\newcommand{\networkmatchedfiltersnrfourtwofivemed}[1]{\IfEqCase{#1}{{AlignedSpinInspiralTidalHS}{12.4}{AlignedSpinInspiralTidalLS}{12.5}{IMRPhenomDNRTidal-HS}{12.3}{IMRPhenomDNRTidal-LS}{12.4}{IMRPhenomPv2NRTidal-HS}{12.4}{IMRPhenomPv2NRTidal-LS}{12.5}{SEOBNRv4TsurrogateHS}{12.4}{SEOBNRv4TsurrogateLS}{12.4}{TaylorF2-HS}{12.4}{TaylorF2-LS}{12.5}{PrecessingSpinIMRTidalHS}{12.4}{PrecessingSpinIMRTidalLS}{12.5}{PublicationSamples}{12.4}}}
\newcommand{\networkmatchedfiltersnrfourtwofiveplus}[1]{\IfEqCase{#1}{{AlignedSpinInspiralTidalHS}{0.3}{AlignedSpinInspiralTidalLS}{0.2}{IMRPhenomDNRTidal-HS}{0.3}{IMRPhenomDNRTidal-LS}{0.3}{IMRPhenomPv2NRTidal-HS}{0.3}{IMRPhenomPv2NRTidal-LS}{0.3}{SEOBNRv4TsurrogateHS}{0.3}{SEOBNRv4TsurrogateLS}{0.3}{TaylorF2-HS}{0.3}{TaylorF2-LS}{0.2}{PrecessingSpinIMRTidalHS}{0.3}{PrecessingSpinIMRTidalLS}{0.3}{PublicationSamples}{0.3}}}
\newcommand{\chirpmasssourcefourtwofiveminus}[1]{\IfEqCase{#1}{{AlignedSpinInspiralTidalHS}{0.02}{AlignedSpinInspiralTidalLS}{0.02}{AlignedSpinTidalHS}{0.02}{AlignedSpinTidalLS}{0.02}{IMRPhenomDNRTidal-HS}{0.02}{IMRPhenomDNRTidal-LS}{0.02}{IMRPhenomPv2NRTidal-HS}{0.02}{IMRPhenomPv2NRTidal-LS}{0.02}{SEOBNRv4TsurrogateHS}{0.02}{SEOBNRv4TsurrogateLS}{0.02}{SEOBNRv4TsurrogatehighspinRIFT}{0.02}{SEOBNRv4TsurrogatelowspinRIFT}{0.02}{TEOBResumS-HS}{0.02}{TEOBResumS-LS}{0.02}{TaylorF2-HS}{0.02}{TaylorF2-LS}{0.02}{PrecessingSpinIMRTidalHS}{0.02}{PrecessingSpinIMRTidalLS}{0.02}{PublicationSamples}{0.02}}}
\newcommand{\chirpmasssourcefourtwofivemed}[1]{\IfEqCase{#1}{{AlignedSpinInspiralTidalHS}{1.44}{AlignedSpinInspiralTidalLS}{1.44}{AlignedSpinTidalHS}{1.44}{AlignedSpinTidalLS}{1.44}{IMRPhenomDNRTidal-HS}{1.44}{IMRPhenomDNRTidal-LS}{1.44}{IMRPhenomPv2NRTidal-HS}{1.44}{IMRPhenomPv2NRTidal-LS}{1.44}{SEOBNRv4TsurrogateHS}{1.44}{SEOBNRv4TsurrogateLS}{1.44}{SEOBNRv4TsurrogatehighspinRIFT}{1.44}{SEOBNRv4TsurrogatelowspinRIFT}{1.44}{TEOBResumS-HS}{1.44}{TEOBResumS-LS}{1.44}{TaylorF2-HS}{1.44}{TaylorF2-LS}{1.44}{PrecessingSpinIMRTidalHS}{1.44}{PrecessingSpinIMRTidalLS}{1.44}{PublicationSamples}{1.44}}}
\newcommand{\chirpmasssourcefourtwofiveplus}[1]{\IfEqCase{#1}{{AlignedSpinInspiralTidalHS}{0.02}{AlignedSpinInspiralTidalLS}{0.02}{AlignedSpinTidalHS}{0.02}{AlignedSpinTidalLS}{0.02}{IMRPhenomDNRTidal-HS}{0.02}{IMRPhenomDNRTidal-LS}{0.02}{IMRPhenomPv2NRTidal-HS}{0.02}{IMRPhenomPv2NRTidal-LS}{0.02}{SEOBNRv4TsurrogateHS}{0.02}{SEOBNRv4TsurrogateLS}{0.02}{SEOBNRv4TsurrogatehighspinRIFT}{0.02}{SEOBNRv4TsurrogatelowspinRIFT}{0.02}{TEOBResumS-HS}{0.02}{TEOBResumS-LS}{0.02}{TaylorF2-HS}{0.02}{TaylorF2-LS}{0.02}{PrecessingSpinIMRTidalHS}{0.02}{PrecessingSpinIMRTidalLS}{0.02}{PublicationSamples}{0.02}}}
\newcommand{\phionefourtwofiveminus}[1]{\IfEqCase{#1}{{AlignedSpinInspiralTidalHS}{0.00}{AlignedSpinInspiralTidalLS}{0.00}{AlignedSpinTidalHS}{0.00}{AlignedSpinTidalLS}{0.00}{IMRPhenomDNRTidal-HS}{0.00}{IMRPhenomDNRTidal-LS}{0.00}{IMRPhenomPv2NRTidal-HS}{2.73}{IMRPhenomPv2NRTidal-LS}{2.85}{SEOBNRv4TsurrogateHS}{0.00}{SEOBNRv4TsurrogateLS}{0.00}{SEOBNRv4TsurrogatehighspinRIFT}{0.00}{SEOBNRv4TsurrogatelowspinRIFT}{0.00}{TEOBResumS-HS}{0.00}{TEOBResumS-LS}{0.00}{TaylorF2-HS}{0.00}{TaylorF2-LS}{0.00}{PrecessingSpinIMRTidalHS}{2.73}{PrecessingSpinIMRTidalLS}{2.85}{PublicationSamples}{2.73}}}
\newcommand{\phionefourtwofivemed}[1]{\IfEqCase{#1}{{AlignedSpinInspiralTidalHS}{0.00}{AlignedSpinInspiralTidalLS}{0.00}{AlignedSpinTidalHS}{0.00}{AlignedSpinTidalLS}{0.00}{IMRPhenomDNRTidal-HS}{0.00}{IMRPhenomDNRTidal-LS}{0.00}{IMRPhenomPv2NRTidal-HS}{3.05}{IMRPhenomPv2NRTidal-LS}{3.15}{SEOBNRv4TsurrogateHS}{0.00}{SEOBNRv4TsurrogateLS}{0.00}{SEOBNRv4TsurrogatehighspinRIFT}{0.00}{SEOBNRv4TsurrogatelowspinRIFT}{0.00}{TEOBResumS-HS}{0.00}{TEOBResumS-LS}{0.00}{TaylorF2-HS}{0.00}{TaylorF2-LS}{0.00}{PrecessingSpinIMRTidalHS}{3.05}{PrecessingSpinIMRTidalLS}{3.15}{PublicationSamples}{3.06}}}
\newcommand{\phionefourtwofiveplus}[1]{\IfEqCase{#1}{{AlignedSpinInspiralTidalHS}{0.00}{AlignedSpinInspiralTidalLS}{0.00}{AlignedSpinTidalHS}{0.00}{AlignedSpinTidalLS}{0.00}{IMRPhenomDNRTidal-HS}{0.00}{IMRPhenomDNRTidal-LS}{0.00}{IMRPhenomPv2NRTidal-HS}{2.90}{IMRPhenomPv2NRTidal-LS}{2.83}{SEOBNRv4TsurrogateHS}{0.00}{SEOBNRv4TsurrogateLS}{0.00}{SEOBNRv4TsurrogatehighspinRIFT}{0.00}{SEOBNRv4TsurrogatelowspinRIFT}{0.00}{TEOBResumS-HS}{0.00}{TEOBResumS-LS}{0.00}{TaylorF2-HS}{0.00}{TaylorF2-LS}{0.00}{PrecessingSpinIMRTidalHS}{2.90}{PrecessingSpinIMRTidalLS}{2.83}{PublicationSamples}{2.90}}}
\newcommand{\symmetricmassratiofourtwofiveminus}[1]{\IfEqCase{#1}{{AlignedSpinInspiralTidalHS}{0.03}{AlignedSpinInspiralTidalLS}{0.005}{AlignedSpinTidalHS}{0.03}{AlignedSpinTidalLS}{0.005}{IMRPhenomDNRTidal-HS}{0.04}{IMRPhenomDNRTidal-LS}{0.005}{IMRPhenomPv2NRTidal-HS}{0.03}{IMRPhenomPv2NRTidal-LS}{0.005}{SEOBNRv4TsurrogateHS}{0.03}{SEOBNRv4TsurrogateLS}{0.004}{SEOBNRv4TsurrogatehighspinRIFT}{0.02}{SEOBNRv4TsurrogatelowspinRIFT}{0.005}{TEOBResumS-HS}{0.03}{TEOBResumS-LS}{0.005}{TaylorF2-HS}{0.03}{TaylorF2-LS}{0.005}{PrecessingSpinIMRTidalHS}{0.03}{PrecessingSpinIMRTidalLS}{0.005}{PublicationSamples}{0.03}}}
\newcommand{\symmetricmassratiofourtwofivemed}[1]{\IfEqCase{#1}{{AlignedSpinInspiralTidalHS}{0.242}{AlignedSpinInspiralTidalLS}{0.249}{AlignedSpinTidalHS}{0.245}{AlignedSpinTidalLS}{0.249}{IMRPhenomDNRTidal-HS}{0.243}{IMRPhenomDNRTidal-LS}{0.249}{IMRPhenomPv2NRTidal-HS}{0.240}{IMRPhenomPv2NRTidal-LS}{0.249}{SEOBNRv4TsurrogateHS}{0.246}{SEOBNRv4TsurrogateLS}{0.249}{SEOBNRv4TsurrogatehighspinRIFT}{0.246}{SEOBNRv4TsurrogatelowspinRIFT}{0.249}{TEOBResumS-HS}{0.245}{TEOBResumS-LS}{0.249}{TaylorF2-HS}{0.242}{TaylorF2-LS}{0.249}{PrecessingSpinIMRTidalHS}{0.240}{PrecessingSpinIMRTidalLS}{0.249}{PublicationSamples}{0.240}}}
\newcommand{\symmetricmassratiofourtwofiveplus}[1]{\IfEqCase{#1}{{AlignedSpinInspiralTidalHS}{0.007}{AlignedSpinInspiralTidalLS}{0.0008}{AlignedSpinTidalHS}{0.005}{AlignedSpinTidalLS}{0.0008}{IMRPhenomDNRTidal-HS}{0.007}{IMRPhenomDNRTidal-LS}{0.0008}{IMRPhenomPv2NRTidal-HS}{0.010}{IMRPhenomPv2NRTidal-LS}{0.0007}{SEOBNRv4TsurrogateHS}{0.004}{SEOBNRv4TsurrogateLS}{0.0007}{SEOBNRv4TsurrogatehighspinRIFT}{0.004}{SEOBNRv4TsurrogatelowspinRIFT}{0.0008}{TEOBResumS-HS}{0.005}{TEOBResumS-LS}{0.0009}{TaylorF2-HS}{0.007}{TaylorF2-LS}{0.0008}{PrecessingSpinIMRTidalHS}{0.010}{PrecessingSpinIMRTidalLS}{0.0007}{PublicationSamples}{0.010}}}

%% file: GW190425_lambda_macros_RW.tex
\newcommand{\lambdatildeboundonefourtwofiveLS}[1]{\IfEqCase{#1}{{PhenomP}{580}{PhenomD}{560}{TF2}{630}{SEOB}{630}{TEOB}{600}}}
\newcommand{\lambdatildeboundonefourtwofiveHS}[1]{\IfEqCase{#1}{{PhenomP}{1080}{PhenomD}{1090}{TF2}{940}{SEOB}{1040}{TEOB}{870}}}
\newcommand{\lowerboundmassoneonefourtwofiveLS}[1]{\IfEqCase{#1}{{SEOB}{1.6}{PhenomD}{1.6}{PhenomP}{1.6}{TEOB}{1.6}{TF2}{1.6}}}
\newcommand{\upperboundmassoneonefourtwofiveLS}[1]{\IfEqCase{#1}{{SEOB}{1.9}{PhenomD}{1.9}{PhenomP}{1.9}{TEOB}{1.9}{TF2}{1.9}}}
\newcommand{\lowerboundmassoneonefourtwofiveHS}[1]{\IfEqCase{#1}{{SEOB}{1.6}{PhenomD}{1.6}{PhenomP}{1.6}{TEOB}{1.6}{TF2}{1.6}}}
\newcommand{\upperboundmassoneonefourtwofiveHS}[1]{\IfEqCase{#1}{{SEOB}{2.3}{PhenomD}{2.5}{PhenomP}{2.5}{TEOB}{2.3}{TF2}{2.4}}}
\newcommand{\lowerboundmasstwoonefourtwofiveLS}[1]{\IfEqCase{#1}{{SEOB}{1.5}{PhenomD}{1.5}{PhenomP}{1.5}{TEOB}{1.5}{TF2}{1.4}}}
\newcommand{\upperboundmasstwoonefourtwofiveLS}[1]{\IfEqCase{#1}{{SEOB}{1.7}{PhenomD}{1.7}{PhenomP}{1.7}{TEOB}{1.7}{TF2}{1.7}}}
\newcommand{\lowerboundmasstwoonefourtwofiveHS}[1]{\IfEqCase{#1}{{SEOB}{1.2}{PhenomD}{1.1}{PhenomP}{1.1}{TEOB}{1.2}{TF2}{1.2}}}
\newcommand{\upperboundmasstwoonefourtwofiveHS}[1]{\IfEqCase{#1}{{SEOB}{1.7}{PhenomD}{1.7}{PhenomP}{1.7}{TEOB}{1.7}{TF2}{1.7}}}
\newcommand{\lowerboundmassratioonefourtwofiveLS}[1]{\IfEqCase{#1}{{SEOB}{0.78}{PhenomD}{0.77}{PhenomP}{0.78}{TEOB}{0.77}{TF2}{0.77}}}
\newcommand{\upperboundmassratioonefourtwofiveLS}[1]{\IfEqCase{#1}{{SEOB}{1.0}{PhenomD}{1.0}{PhenomP}{1.0}{TEOB}{1.0}{TF2}{1.0}}}
\newcommand{\lowerboundmassratioonefourtwofiveHS}[1]{\IfEqCase{#1}{{SEOB}{0.54}{PhenomD}{0.46}{PhenomP}{0.45}{TEOB}{0.54}{TF2}{0.5}}}
\newcommand{\upperboundmassratioonefourtwofiveHS}[1]{\IfEqCase{#1}{{SEOB}{1.0}{PhenomD}{1.0}{PhenomP}{1.0}{TEOB}{1.0}{TF2}{1.0}}}

%% file: chip_js.tex
\newcommand{\chippripostJS}[1]{\IfEqCase{#1}{{GW190408A}{0.03}{GW190412A}{0.18}{GW190413A}{0.01}{GW190413B}{0.04}{GW190421A}{0.02}{GW190424A}{0.03}{GW190503A}{0.02}{GW190512A}{0.08}{GW190513A}{0.05}{GW190514A}{0.01}{GW190517A}{0.01}{GW190519A}{0.01}{GW190521A}{0.15}{GW190521B}{0.05}{GW190527A}{0.01}{GW190602A}{0.01}{GW190620A}{0.01}{GW190630A}{0.07}{GW190701A}{0.01}{GW190706A}{0.02}{GW190707A}{0.03}{GW190708A}{0.02}{GW190719A}{0.0}{GW190720A}{0.03}{GW190727A}{0.01}{GW190728A}{0.06}{GW190731A}{0.01}{GW190803A}{0.01}{GW190814A}{0.72}{GW190828A}{0.02}{GW190828B}{0.05}{GW190909A}{0.04}{GW190910A}{0.01}{GW190915A}{0.06}{GW190924A}{0.04}{GW190929A}{0.04}{GW190930A}{0.03}}}

%% file: extremal_spin_macros.tex
\newcommand{\percenthighspin}[1]{\IfEqCase{#1}{{GW190517A}{77}{GW190521A}{58}{GW190719A}{43}{GW190620A}{42}{GW190929A}{40}{GW190909A}{40}{GW190519A}{39}{GW190413B}{39}{GW190514A}{38}{GW190706A}{36}{GW190915A}{35}{GW190424A}{34}{GW190527A}{33}{GW190421A}{29}{GW190727A}{27}{GW190803A}{27}{GW190602A}{26}{GW190413A}{26}{GW190731A}{25}{GW190701A}{24}{GW190503A}{21}{GW190720A}{21}{GW190828A}{20}{GW190814A}{18}{GW190513A}{18}{GW190412A}{16}{GW190930A}{15}{GW190828B}{15}{GW190910A}{15}{GW190408A}{14}{GW190728A}{11}{GW190924A}{10}{GW190512A}{9}{GW190425A}{8}{GW190707A}{8}{GW190521B}{7}{GW190630A}{7}{GW190708A}{7}{GW190426A}{0}}}
\newcommand{\highspinfirst}{GW190517A}
\newcommand{\highspinsecond}{GW190521A}
\newcommand{\highspinthird}{GW190719A}

%% file: o3a_catalog_authors.tex
\author{R.~Abbott}
\affiliation{LIGO, California Institute of Technology, Pasadena, CA 91125, USA}
\author{T.~D.~Abbott}
\affiliation{Louisiana State University, Baton Rouge, LA 70803, USA}
\author{S.~Abraham}
\affiliation{Inter-University Centre for Astronomy and Astrophysics, Pune 411007, India}
\author{F.~Acernese}
\affiliation{Dipartimento di Farmacia, Universit\`a di Salerno, I-84084 Fisciano, Salerno, Italy  }
\affiliation{INFN, Sezione di Napoli, Complesso Universitario di Monte S.Angelo, I-80126 Napoli, Italy  }
\author{K.~Ackley}
\affiliation{OzGrav, School of Physics \& Astronomy, Monash University, Clayton 3800, Victoria, Australia}
\author{A.~Adams}
\affiliation{Christopher Newport University, Newport News, VA 23606, USA}
\author{C.~Adams}
\affiliation{LIGO Livingston Observatory, Livingston, LA 70754, USA}
\author{R.~X.~Adhikari}
\affiliation{LIGO, California Institute of Technology, Pasadena, CA 91125, USA}
\author{V.~B.~Adya}
\affiliation{OzGrav, Australian National University, Canberra, Australian Capital Territory 0200, Australia}
\author{C.~Affeldt}
\affiliation{Max Planck Institute for Gravitational Physics (Albert Einstein Institute), D-30167 Hannover, Germany}
\affiliation{Leibniz Universit\"at Hannover, D-30167 Hannover, Germany}
% ...

\collaboration{The LIGO Scientific Collaboration and the Virgo Collaboration}

%% file: introduction.tex
\section{Introduction}
\resetlinenumber

Since the discovery of gravitational waves from a \ac{BBH} coalescence in
2015~\cite{Abbott:2016blz}, the Advanced LIGO~\cite{TheLIGOScientific:2014jea}
and Advanced Virgo~\cite{TheVirgo:2014hva} gravitational wave detectors have
opened a new window on our Universe~\cite{Abbott:2016nmj, Abbott:2017vtc,
Abbott:2017gyy, Abbott:2017oio, LIGOScientific:2018mvr}.  Binary black hole
observations have allowed us to probe gravity in the strong-field
regime~\cite{TheLIGOScientific:2016src, LIGOScientific:2019fpa} and to
establish the rate and population properties of \ac{BBH}
coalescences~\cite{LIGOScientific:2018jsj}. In addition to \acp{BBH}, Advanced
LIGO and Advanced Virgo detected the first gravitational wave signal from a
\ac{BNS} coalescence, GW170817~\cite{TheLIGOScientific:2017qsa}, which was also
the first joint detection of gravitational waves and electromagnetic
emission~\cite{Monitor:2017mdv, GBM:2017lvd}.  Gravitational wave discoveries
have had a profound impact on physics, astronomy and
astrophysics~\cite{Abbott:2018exr, LIGOScientific:2019eut, Monitor:2017mdv,
Abbott:2017xzu, Abbott:2019yzh, Abbott:2018lct}, and the public release of LIGO
and Virgo data~\cite{Abbott:2019ebz, Trovato:2019liz} has enabled groups other
than the \LVC to perform analyses searching for gravitational wave
signals~\cite{Zackay:2019btq, Venumadhav:2019lyq, Venumadhav:2019tad,
Zackay:2019tzo, Nitz:2018imz, Nitz:2019hdf, Magee:2019vmb} and to report
additional candidate events in some cases.

\begin{figure}[tb] 
\begin{center}
\includegraphics[width=\columnwidth]{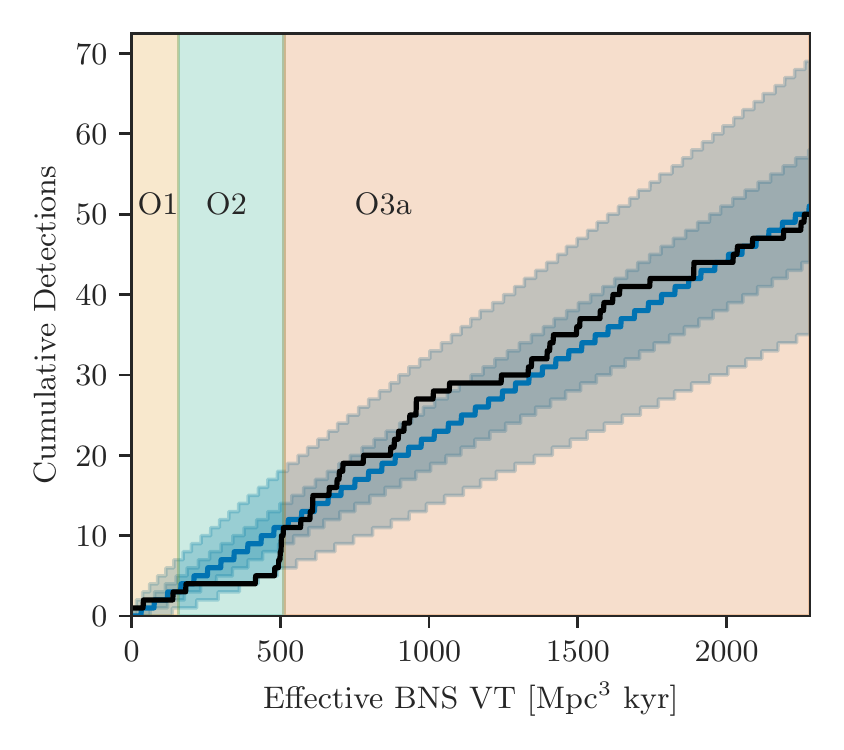}
\end{center}
\caption{\label{fig:vt_det} The number of compact binary coalescence 
detections versus the effective
volume--time (VT) to which the gravitational wave network is sensitive to
\ac{BNS} coalescences.
The effective VT is defined as the Euclidean sensitive volume \cite{Chen:2017wpg}
of the second-most sensitive detector in the network at a given time, multiplied
by the live time of that network configuration.
The Euclidean sensitive volume of the network is the volume of a sphere with a radius
given by the 
\ac{BNS} inspiral range~\cite{Allen:2005fk, Chen:2017wpg}
(shown in Fig.~\ref{fig:range}) of the second most sensitive detector in the network. 
To account for the addition of single detector candidates in O3a, 
a single-detector Euclidean sensitive volume was also included, defined using the 
inspiral range of the most sensitive detector divided by 1.5.
The effective BNS VT does not account for differences in sensitivity across
the entire population of signals detected, necessary cosmological corrections, or changes to analysis pipeline efficiency between observing runs, 
but, as shown in this figure, is consistent with the currently observed rate of detections. 
The colored bands indicate the three runs, O1, O2 and O3a.
The black line is the cumulative number of confident detections of
all compact binary coalescences (including black holes and neutron stars)
for GWTC-1~\cite{LIGOScientific:2018mvr} and this catalog. The blue line,
dark blue band, and light blue band are the median, 50\% confidence interval,
and 90\% confidence interval of draws
from a Poisson fit to the number of detections at the end of O3a.  The increase in detection rate is dominated by improvements to the sensitivity of the LIGO and Virgo detectors with changes in analysis methods between observing runs being sub-dominant.
}
\end{figure}

We present the results of searches for compact binaries in Advanced LIGO and
Advanced Virgo data taken between \RUNSTART{} and \RUNEND{}. This period,
referred to as O3a, is the first six months of Advanced LIGO and Advanced
Virgo's eleven month long third observing run.  
The first Gravitational-Wave Transient Catalog (GWTC-1) of compact binary coalescences
(CBC)  includes candidate events observed by Advanced LIGO and Advanced Virgo during 
the first (O1) and second (O2) observing runs~\cite{LIGOScientific:2018mvr}.
The increased sensitivity of
Advanced LIGO and Advanced Virgo during O3a has enabled us to increase the number of
confident gravitational wave detections more than three-fold over
GWTC-1. Together, GWTC-1 and
the new candidate events presented here comprise GWTC-2. Fig.~\ref{fig:vt_det} shows
this consistent increase in both the effective binary neutron star volume--time
(BNS VT) of the gravitational wave network, 
and the number of detections across
these observing runs.  The BNS VT is Euclidean sensitive volume of the detector
network \cite{Allen:2005fk, Chen:2017wpg} multiplied by the live time of the network 
and should be approximately proportional to the total number of detections.  
Our analysis of the O3a dataset has resulted in
\NUMEVENTS{} gravitational wave candidate events passing our \ac{FAR} threshold of
\FARTHRESHYR{} per year.  Given our use of multiple search pipelines to
identify candidate events, we expect $\sim 3$ false alarms, i.e.\ candidate events caused
by instrumental noise, to be present in this catalog.  It is not possible to
determine with certainty which specific candidate events are due to noise; instead we
provide statistical measures of false alarm rate and probability of
astrophysical origin.  Among these candidate events, \fixme{\PREVIOUSLYREPORTED{}}
have been reported previously in real-time processing via GCN Notices and
Circulars~\cite{LIGOPublicGCNs}.  Furthermore, four gravitational wave candidate events
from O3a have already been published separately due to their interesting
properties: \NAME{GW190425A}~\cite{Abbott:2020uma} is the second gravitational
wave event consistent with a \ac{BNS} coalescence;
\NAME{GW190412A}~\cite{Abbott:2020uma} is the first \ac{BBH} observation with
definitively asymmetric component masses, which also produced detectable
gravitational radiation beyond the leading quadrupolar order;
\NAME{GW190814A}~\cite{GW190814A} is an even more asymmetric system having a
$\sim$23\,\Msun\ object merging with a $\sim$2.6\,\Msun\ object, making the
latter either the lightest black hole or heaviest neutron star known to be in a
double compact object system; \NAME{GW190521A}~\cite{GW190521Adiscovery,
GW190521Aastro} is a \ac{BBH} with total mass $\sim150\,\Msun$ having a primary
mass above $65\,\Msun$ at 99\% credibility.

Here we present \fixme{\NEWEVENTS{}} candidate events for the first time along
with the \fixme{\PREVIOUSLYREPORTED{}} previously reported candidates. Among
the \NUMEVENTS{} candidates, we find gravitational wave emission consistent
with the coalescence of \acp{BBH}, \acp{BNS},
and \acp{NSBH}.

We report on the status of the Advanced LIGO and Advanced Virgo gravitational
wave detectors (Sec.~\ref{sec:instruments}) and the properties and quality
of the data taken during the analyzed period (Sec.~\ref{sec:data}).  We then
describe the analysis methods that led to the identification of the
\NUMEVENTS{} gravitational wave candidates (Sec.~\ref{sec:searches}), as well
as the inference of their parameters (Sec.~\ref{sec:PEmethods}).  Next, we
report the significance of the identified candidates, as well as a comparison
to the public gravitational wave alerts (Sec.~\ref{s:candidates}).  Finally,
we discuss the properties (Sec.~\ref{sec:peresults}) and the reconstructed waveforms (Sec.~\ref{sec:wfreconstructions}) of each event .  Further
interpretation of the binary population is conducted in companion
papers~\cite{o3tgr, o3apop}.  We will analyze the second half of Advanced LIGO
and Advanced Virgo's third observing run (O3b) in future publications.

We provide a public data release associated with the results contained in this paper
at \cite{datarelease}. This includes the data behind the figures, the simulation
data used in estimating search sensitivity, and the posterior samples used in
estimating the source properties.

%% file: instruments.tex
\section{Instruments}
\label{sec:instruments}
\resetlinenumber

The Advanced LIGO~\cite{TheLIGOScientific:2014jea} and Advanced Virgo~\cite{TheVirgo:2014hva}
detectors are kilometer-scale laser interferometers~\cite{Vajente:2019a}. The current
generation of detectors started operations in 2015, and since then
have been alternating periods of observation with periods of tuning and
improvement. 
Since O1~\cite{TheLIGOScientific:2016pea} and O2~\cite{LIGOScientific:2018mvr}, the sensitivity and robustness of
the detectors improved significantly. 

The LIGO detectors underwent several 
upgrades between the end of O2 and the start of O3a~\cite{Buikema:2020dlj}. 
The main laser sources were replaced to allow for higher operating powers.  
The LIGO Hanford detector operated with \HANFORDPOWER{}~W of input power and the Livingston detector with \LIVINGSTONPOWER{}~W. Those levels can be compared to \HANFORDPOWEROTWO{}~W and \LIVINGSTONPOWEROTWO{}~W during O2 for Hanford and Livingston respectively. The laser sources replacement also reduced fluctuations in the input beam pointing and size that were previously detrimental for the detector
sensitivity~\cite{Driggers:2018gii}. At both LIGO detectors the two end test masses
were replaced with mirrors with lower scattering losses~\cite{Billingsley:2017a}, 
allowing for higher circulating power. Additionally, annular test masses were installed to reduce noise
induced by residual gas damping \cite{Dolesi:2011}. In the Hanford
interferometer, one of the two input test masses was also replaced, because the
one which was previously installed had a large point absorber~\cite{point-absorber} 
that limited the amount of power that could be handled
in the arm cavities. 

The build up of electric charge on the test masses was also an issue during previous runs, therefore several actions were 
undertaken to mitigate the contribution of this noise source: electric field meters were installed in end stations to monitor the local eletric field; baffles were installed in front of the vacuum system ion pumps to mitigate charging, and a test mass discharging system was put in operation and successfully deployed on all LIGO Hanford test masses.

Parametric instabilities~\cite{Evans:2015raa}, 
i.e., radiation-pressure-induced excitation of the test
masses' mechanical modes, also limited the maximum power allowed into the
interferometer. This problem was mitigated with the installation
of acoustic mode dampers~\cite{Biscans:2019akh} that reduce the mechanical
quality factor of the test mass resonant modes and thus suppress parametric
instabilities. 

The high frequency ($\gtrsim 1$~kHz) sensitivity of both
detectors was significantly improved compared to the 
O2 observing run (a factor \HANFORDIMPROVEMENTHF{} for the 
Hanford detector, and \LIVINGSTONIMPROVEMENTHF{} for the Livingston detector), 
partially due to the increased circulating
power made possible by the improvements already discussed, and by the
installation of squeezed light sources~\cite{Snabel:2010sq, Barsotti:2019squeez} 
to reduce the quantum vacuum noise
entering the interferometers~\cite{Tse:2019wcy}, making O3 the first observing 
run of the LIGO detectors that routinely implement quantum noise reduction techniques. 
GEO600 has been using the same approach since 2011~\cite{Ab:2011squeez, Grote:2013squeez} 
but has not detected gravitational waves so far due to an overall too-low sensitivity.

Additionally, many beam dumps
and baffles were installed at both LIGO sites, to mitigate the effect of scattered
light~\cite{Ottaway:2012a} that can be the source of non-stationary disturbances. 
Finally, the feedback control systems for the seismic
isolation and for the angular and longitudinal control of the instruments were
improved, increasing the detectors' duty cycle and robustness against external
disturbances. With respect to O2, the LIGO Hanford median \ac{BNS} inspiral range, as defined in \cite{Allen:2005fk},
increased by a factor \HANFORDRANGEINCREASE{} (from \HANFORDRANGEOTWO{} to 
\HANFORDRANGE{}) and the LIGO Livingston median range by \LIVINGSTONRANGEINCREASE{}
(from \LIVINGSTONRANGEOTWO{} to \LIVINGSTONRANGE{}).

Also in Virgo between the O2 and the O3a observing runs, many upgrades have been implemented to boost the sensitivity..
The most important upgrade was the replacement of the steel wires suspending the four
test masses with fused-silica fibers~\cite{Aisa:2016rsg} to improve the sensitivity below 100~Hz.
This was achieved by changing the design of the final stage of the mirror suspension to improve the screening of the fused-silica fibers~\cite{Naticchioni:2018enb} from residual particles injected by the vacuum system.
In parallel, the vacuum system was modified to avoid dust pollution of the environment.

Another upgrade to improve the low-frequency sensitivity was the suspension of the external injection bench (used to manipulate and steer the input laser beam into the interferometer). In this way, the seismic motion of the optics was reduced, and, consequently, the beam jitter noise contribution~\cite{Blom:2015fna, Blom:2015a}.

The major upgrade to boost the Virgo high-frequency sensitivity was the installation of a more powerful laser that can deliver more than \VIRGOPOWERMAX{}\,W output power. After some commissioning activities at different power values, the laser power injected into the interferometer was set to \VIRGOPOWER{}~W, improving both the sensitivity and the stability of the interferometer at the same time. The laser power was almost doubled compared to the \VIRGOPOWEROTWO{}~W injected during O2.\footnote{In \cite{Andrea:2018strain} it was
reported that the injected power during O2 was 14~W; however, measurements taken during end-of-run commissioning activities 
determined that the correct value was approximately 30\% lower than expected.}
In this configuration, due to the marginally stable power recycling cavity,
the aberration induced by thermal effects prevented a reliable and robust interferometer longitudinal
control, and worsened the alignment performances. Therefore, the thermal
compensation system actuators were used to stabilize the power
recycling cavity. The end test masses' radii of curvature were tuned with the
ring heaters, maximizing the power circulating in the arm cavities~\cite{Rocchi:2012a, Nardecchia:2020a}.  

The squeezing technique adopted by LIGO to improve the high frequency sensitivity of the detectors, was also implemented in Virgo interferometer~\cite{Acernese:2019sbr}. This system was fully operational also in Virgo for the first time during O3.
To reduce the optical losses optimizing the squeezing performance, the photodiodes installed at the interferometer dark port to measure the gravitational wave signals were replaced by high quantum efficiency ones~\cite{Vahlbruch:2016qwp}. 

Moreover, complementary activities have been carried out in parallel to the main upgrades, reducing the overall contribution of the various technical noises.  In particular, the improvement of the control strategy for the suspended benches allowed the reduction of the noise contribution below 30 Hz, the installation of baffles and diaphragms on the optical benches and inside the vacuum tanks reduced the impact of the scattered light on the sensitivity.
Finally, sources of environmental noises have been identified and removed.
All these upgrades increased the Virgo median \ac{BNS} range by \VIRGORANGEINCREASE{} (from \VIRGORANGEOTWO{} to 
\VIRGORANGE{}) with respect to the O2 run.

\begin{figure}[tb] 
\begin{center}
\includegraphics[width=\columnwidth]{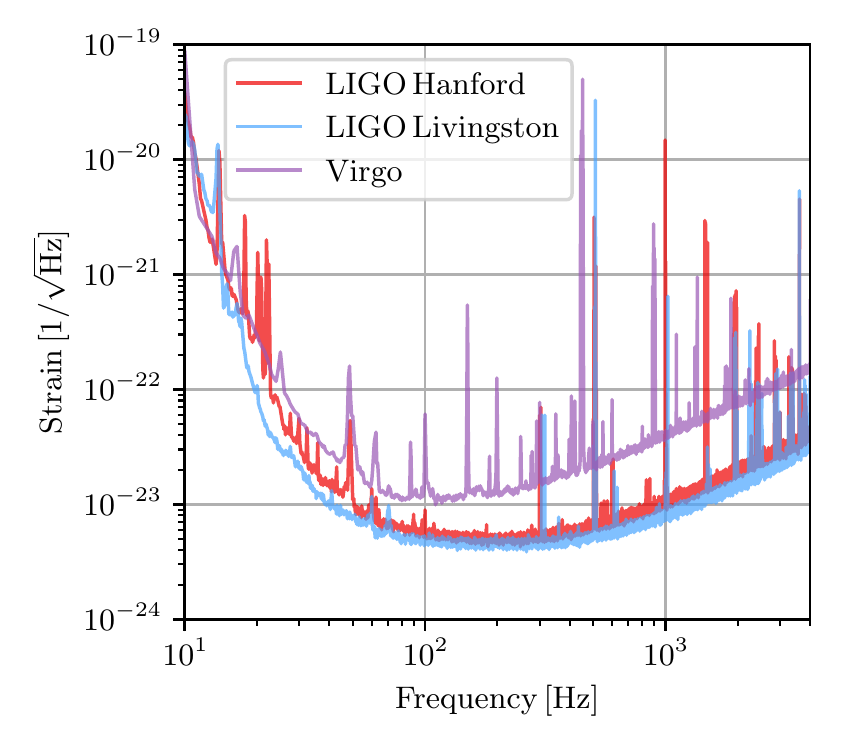}
\end{center}
\caption{\label{fig:strain} Representative amplitude spectral density of the three detectors' strain sensitivity (LIGO Livingston 5 September 2019 20:53 UTC, LIGO Hanford 29 April 2019 11:47 UTC, Virgo 10 April 2019 00:34 UTC). From these spectra we compute \ac{BNS} inspiral ranges of \HANFORDRANGEPLOT{}, \LIVINGSTONRANGEPLOT{}, and \VIRGORANGEPLOT{} for LIGO Hanford, LIGO Livingston, and Virgo, respectively.}
\end{figure}

\begin{figure*}[tb] 
\begin{center}
\includegraphics[width=\columnwidth]{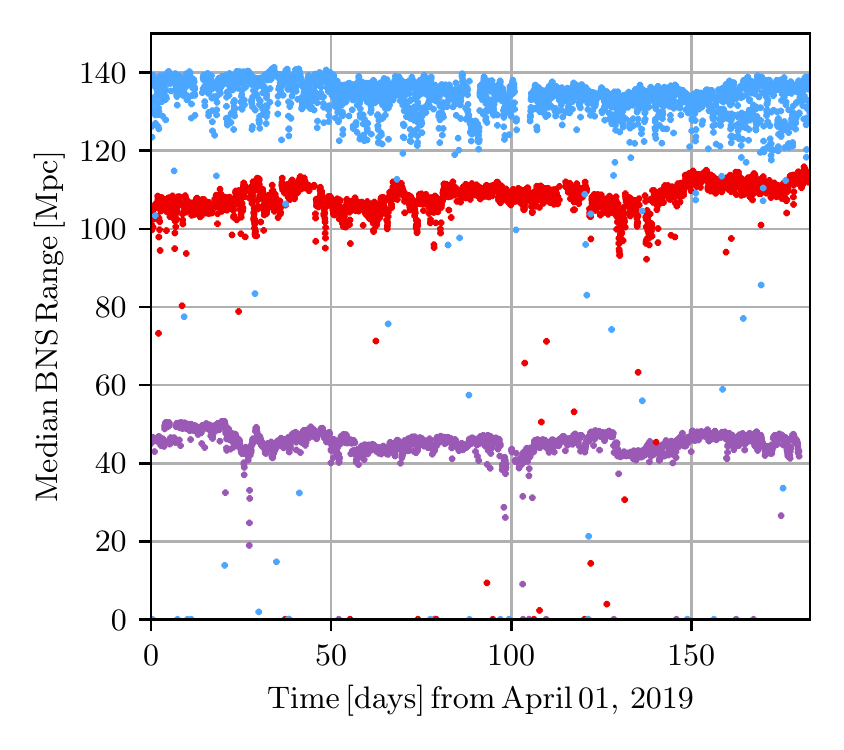}
\includegraphics[width=\columnwidth]{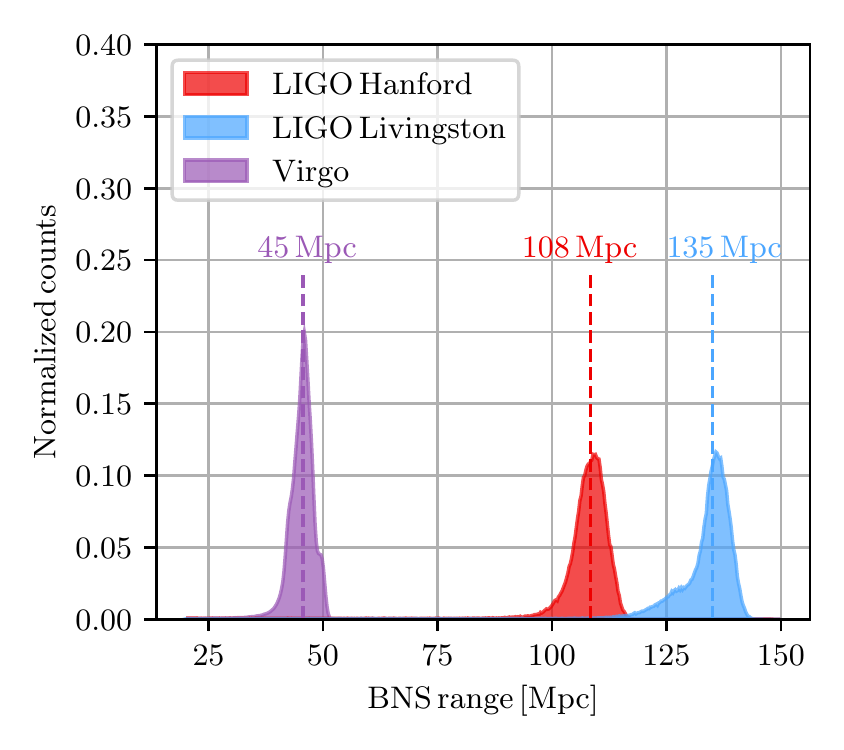} 
\end{center}
\caption{\label{fig:range} The \ac{BNS} range of the LIGO and Virgo detectors. (Left) The evolution in time of the range over the entire duration of O3a. Each data point corresponds to the median value of the range over one-hour-long time segments. (Right) Distribution of the range and the median values for the entire duration of O3a. }
\end{figure*}

Fig.~\ref{fig:strain}  shows representative sensitivities of the three detectors
during O3a, as measured by the amplitude spectral density of the calibrated strain output.
Fig.~\ref{fig:range} shows the evolution of the detectors' sensitivity over time, as measured by the binary neutron star
range.  The 
up-time of the detectors was kept as high as possible, but was
nevertheless limited by many factors, such as earthquakes, instrumental
failures and planned maintenance periods. The duty cycle for the three
detectors was \VIRGODUTYCYCLE{}\% (\VIRGODAYS{} days) for Virgo, \HANFORDDUTYCYCLE{}\% (\HANFORDDAYS{} days)
for LIGO Hanford, and \LIVINGSTONDUTYCYCLE{}\% (\LIVINGSTONDAYS{} days) for LIGO
Livingston. With these duty cycles, the full 3-detector network was in observing mode
for \THREEDETECTORSDUTYCYCLE{}\% of the time (\THREEDETECTORSDAYS{} days). Moreover, for \ONEDETECTORDUTYCYCLE{}\% of the
time (\ONEDETECTORDAYS{} days) at least one detector was observing and for \TWODETECTORSDUTYCYCLE{}\% 
(\TWODETECTORSDAYS{} days) at least two detectors were observing. For comparison, during the O2 run the duty cycles were 
\HANFORDDUTYCYCLEOTWO{}\% for LIGO Hanford and \LIVINGSTONDUTYCYCLEOTWO{}\% for LIGO Livingston, so that two detectors were 
in observing mode \TWODETECTORSDUTYCYCLEOTWO{}\% of the time and at least one detector was in observing mode \ONEDETECTORDUTYCYCLEOTWO{}\% of the time. The goal was of course to obtain the highest possible duty cycle. The current performance is limited by planned maintenance periods and the time needed to recover the control of the interferometers after large transients triggered, for example, by earthquakes. Work is on-going at all detectors to improve the duty cycle.

%% file: data.tex
\section{data}
\label{sec:data}
\resetlinenumber

Before analyzing LIGO and Virgo time-domain data for gravitational waves,
we apply multiple data conditioning steps to accurately calibrate 
the data into strain and mitigate periods 
of poor data quality~\cite{LIGOScientific:2019hgc}.  
Segments of data where each
interferometer was operating in a nominal state, free from external
intervention, are recorded~\cite{Fisher:2020}.  
Data from outside these time periods 
is not used in analyses unless additional investigations are 
completed to understand the state of the interferometer~\cite{Abbott:2017gyy,GW190814A}.
The data conditioning process involves calibration
of the data, both in near-real time and in higher latency; 
subtraction of noise from known instrumental sources; 
and identification of short-duration noise transients, 
which we refer to as glitches~\cite{Christensen:2004kh}, 
that should be excluded from any searches for astrophysical candidates either 
by not considering the data containing the glitches or 
mitigating the glitches with methods such as gating. 

For time periods containing gravitational wave candidate events, additional
investigation of the data quality is completed as a part of event validation to
evaluate if instrumental artifacts could impact the detection and analysis of the
candidate events~\cite{TheLIGOScientific:2016zmo}.  
These investigations sometimes lead to additional data processing
steps such as modeling and subtraction of glitches beyond what is
completed to mitigate glitches before searching the data. 

This section further outlines the procedures used to calibrate the 
data collected by LIGO and Virgo, characterize the data quality, 
validate any identified gravitational-wave candidates, and
subtract glitches that may impact the analysis of candidates.  

\subsection{Calibration and noise subtraction}\label{ss:calibration}
The optical power variations at the gravitational wave readout ports of the 
LIGO and Virgo detectors are calibrated into a time series of dimensionless strain measured by the detectors before 
use by astrophysical analyses~\cite{Viets2018,Acernese:2018bfl}.
The calibration process requires data conditioning filters whose response is complex-valued, 
frequency-dependent, and informed by detailed modeling of the feedback control system 
along with the interferometric, opto-mechanical response of 
the detectors~\cite{Abbott:2016jsd}.  
Some control system model parameters vary slowly with time throughout operation of the 
interferometer.
These parameters must be monitored and, when possible, the filters are corrected in near real-time (low-latency)~\cite{Tuyenbayev2016}. 
Other parameters may change at discrete times and cause systematic error in the data 
stream that cannot be accounted for in low latency. 
Examples of such error can arise from poorly compensated changes in electronics configurations, 
accidental application of incorrect control parameter values,
model errors not-yet-known at the start of the observing period, 
and hardware problems such as failures of analog  
electronics within the control system. 
Most of these sources of error are subtle, and can only be assessed once they are 
measured and quantified a posteriori. 

All three detectors use photon calibrator (Pcal) 
systems~\cite{Karki2016,Bhattacharjee:2020yxe,Estevez:2020pvj} for absolute
reference.
These reference systems are used to develop each static detector model, 
measure parametric time dependence, 
and establish residual levels of systematic error in each strain data stream once constructed. 
Each of these measurements required to evaluate the systematic error 
is done by using the Pcal systems to drive forces on the end test masses via 
radiation pressure, creating displacement above other detector noise.  
Validating the strain data stream in low latency for all time, by establishing 
carefully quantified estimates of the systematic error with these excitations, 
competes with the desire for unhampered astrophysical sensitivity. 
As a compromise, systematic error is measured continuously only at a select 
few frequencies at the edges of the sensitive frequency band of the detector with 
monochromatic excitations during observation. 
The data stream is only validated at high frequency resolution, 
and across the entire detection band, at roughly weekly cadence: 
the detectors are fully functional, but are declared out of observation mode, 
and swept-sinusoid and colored-random-noise Pcal excitations are driven above the noise.
These measurements can only provide approximate, point-estimate bounds on the data 
stream's error in low-latency; they cannot reflect the error distribution for all time.

Once an observing period with a stable detector configuration has completed,
estimates of the probability distribution of systematic error for all observation 
time are created~\cite{Sun:2020wke}. 
These estimates leverage the power of hind-sight, the collection of measurements 
mentioned above, and other measurements of individual components gathered while 
the detector is offline. 
During this error characterization process, if any identified systematic error 
is egregious and well-quantified, where possible, the control system model 
and data conditioning filters are modified to remove the error. 
The data stream is then regenerated offline from the optical power variations 
and control signals, and the systematic error estimate is updated accordingly~\cite{Viets2018}.

Results in this paper are derived from either low-latency (C00) or offline, 
recalibrated (C01) strain data, depending on whether offline data were available 
and whether the results are sensitive to calibration error.
Detection algorithms for gravitational wave candidates, 
described in Sec.~\ref{sec:searches}, are insensitive to typical levels of 
systematic error in calibration~\cite{TheLIGOScientific:2016qqj}, 
so low-latency data may be used for additional offline analyses without concern. 
However, if available, offline data are preferred for its improved accuracy and completeness.
At the time the data were searched for candidate events, 
offline data were only available for a portion of O3a. 
The candidate events presented in this paper detected prior to 5 June 2019 
were identified using LIGO offline data, 
whereas those from 5 June 2019 until 1 October 2019 were identified using LIGO low-latency data.

Once candidate events are found by detection algorithms using either LIGO data stream, 
all estimations of the candidates' astrophysical parameters use the C01 LIGO version 
of strain data using methods described in Sec.~\ref{sec:PEmethods}.
The C01 LIGO version of strain data was available for the entirety of O3a at the
time these additional analyses were completed. 
As such analyses are more sensitive to calibration error~\cite{Vitale2012}, 
it is advantageous to use the definitive characterization of error at the time of 
each event available with LIGO C01 data.
The probability distribution of error for LIGO C01 strain data in 
O3a are characterized in \cite{Sun:2020wke}.
Analysis of Virgo's collection of validation measurements during the run did 
not lead to a significant improvement to the low-latency strain data stream offline. 
As such, only low-latency strain and its bounds on systematic error from point-estimate 
measurements are used for all detection and astrophysical parameter estimation results presented in this paper. 
The bounds of systematic error of Virgo strain in O3a are reported in \cite{VIR-0652B-19}.

Numerous noise sources that limit detector sensitivity
are measured and linearly subtracted
from the data using witness auxiliary sensors that measure
the source of the noise~\cite{Davis:2018yrz, Driggers:2018gii, T2100058}. 
In O3a, the sinusoidal excitations used for calibration,
and noise from the harmonics of the power mains were subtracted
from LIGO data as a part of the calibration procedure~\cite{T2100058}.
This subtraction was completed for both 
online and offline versions of the calibration. 
For time periods around a subset of identified candidate events, 
additional noise contributions due to non-stationary couplings of the 
power mains were subtracted~\cite{Vajente:2019ycy}.

The Virgo online strain data production 
also performed broadband noise subtraction during O3a~\cite{VIR-0652B-19}. 
The subtracted noises included frequency noise of the input laser, 
noise introduced controlling the displacement of the beam splitter,
and amplitude noise of the 56 MHz modulation frequency.
An additional offline strain data set was produced for 
14 September 2019 through 1 October 2019 using the same calibration 
as the online data but with improved noise subtraction resulting 
in a \ac{BNS} range increase of up to 
\VIRGONOISESUBRANGEINCREASE{}~\cite{VIR-1201A-19}
and is used in source parameter estimation of candidate events that occurred during this time period.

\subsection{Data quality}\label{ss:DQdiscussion}

During O3a, the data quality was closely monitored using 
summarized information from the detectors
and their subsystems~\cite{Davis:2021ecd,gwsumm-software}. 
Deeper studies were conducted to identify the
causes of data quality issues,
which enabled instrumental mitigation of the sources
during the run.
For example, at Livingston, glitches from a mechanical camera shutter
and beats of varying radio-frequency
signals were identified and eliminated. 
At Hanford and Livingston, strong frequency peaks that wandered in time 
were tracked down to the amplitude stabilizer
for the 
laser used to provide squeezed light. 
At Hanford, broad features in the spectrum at 48\,Hz 
and multiples were tracked to scattered light from vacuum chamber doors 
and mitigated with absorptive black glass.
These studies were a part of ongoing efforts to improve
the data quality and the up-time of
the detectors~\cite{Buikema:2020dlj,Davis:2021ecd}.

\begin{figure}[tb] 
\begin{center}
\includegraphics[width=\columnwidth]{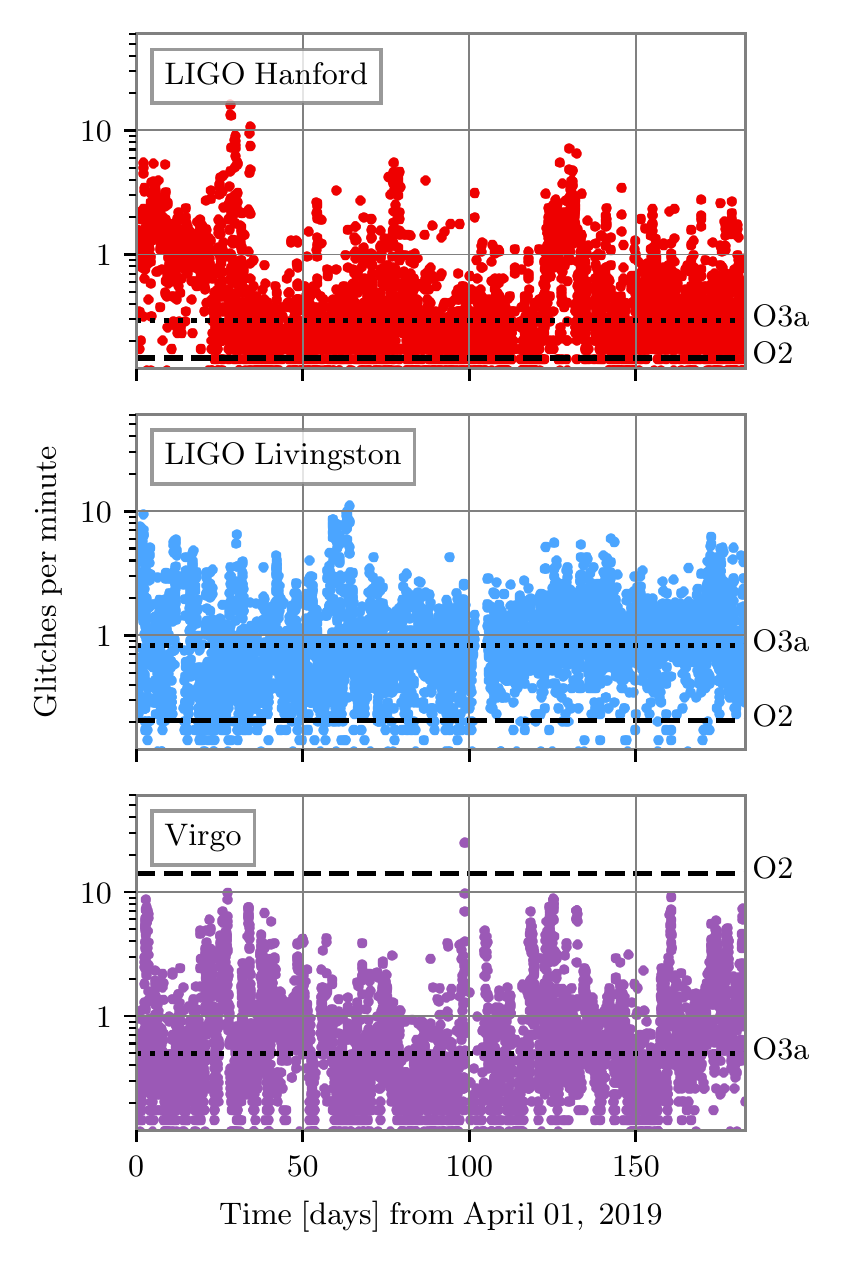}
\end{center}
\caption{\label{fig:glitch_rate}  The rate of single interferometer 
glitches with SNR~$>$~6.5
and frequency between 10~Hz and 2048~Hz identified by Omicron~\cite{omicron,Omicron:2019}
in each detector during O3a. 
Each point represents the average rate over a 2048~s interval. 
Dotted black lines show the median glitch rate for each detector
in O2 and O3a.
}
\end{figure}

For analyses of gravitational wave transients, 
the most common data quality issue is the presence of glitches.
The rate of glitches with
\ac{SNR}~$>$~6.5
in the LIGO and Virgo detectors in O3a is shown in 
Fig.~\ref{fig:glitch_rate}. 
In all three detectors, the glitch rate is dominated by
glitches with peak frequencies below 100~Hz.
This rate was higher than in previous 
observing runs~\cite{TheLIGOScientific:2016zmo} for both LIGO detectors and was
especially problematic at LIGO Livingston, where the rate of glitches
was \LLOOTHREEGLITCHRATE{} in O3a,
compared to \LLOOTWOGLITCHRATE{} in O2.
The Virgo glitch rate decreased significantly between O2 and O3a, 
thanks to the work done during the O2--O3a shutdown to improve the accuracy of Virgo's 
operating point control and to identify, fix, or mitigate several sources of noise.
The increased rate of glitches in the LIGO detectors
limited the overall sensitivity of searches for
gravitational waves in O3a and created challenges for analysis of candidate
events.

\begin{figure}[tb] 
\begin{center}
\includegraphics[width=\columnwidth]{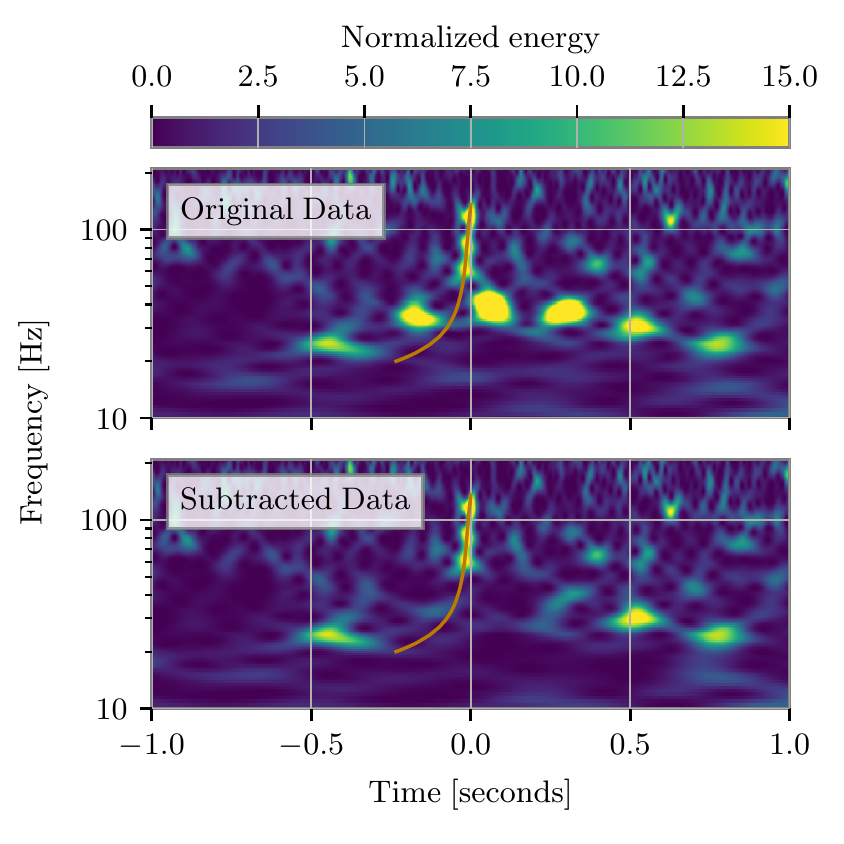}
\end{center}
\caption{\label{fig:scatter} Top: time--frequency representation of the data
surrounding event \protect\NAME{GW190701A}\ at LIGO Livingston, showing the presence of 
scattered light glitches (modulated arches).  
Bottom: The same data after glitch identification and
subtraction as described in Sec.~\ref{ss:denoise}. 
In both plots, the time--frequency track 
of the matched filter template used to identify 
\protect\NAME{GW190701A}\ is overlaid in orange.
Investigations identified \SCATTERINGEVENTS{} 
candidate events in coincidence with 
similar scattering glitches, and required 
mitigation before further analysis. 
Despite the clear overlap of the signal with the glitch, 
the excess power from the glitch is successfully modeled
and subtracted.
}
\end{figure}

The most problematic source of glitches in O3a was caused 
by laser light scattered out of the main laser beam, which is reflected off
walls of the vacuum systems and other equipment back into the main beam~\cite{Ottaway:2012a,
Accadia:2010zzb, flanagan_thorne_bs, Soni:2020rbu}. 
Scattered light noise is
correlated with periods of high seismic activity.  For this reason, daily
cycles of scattered light glitches were present throughout O3a,
especially at LIGO Livingston, tied to ground motion driven by human activity.
This noise is often visible as
arch-shaped features in time--frequency~\cite{Accadia:2010zzb}, as shown in
Fig.~\ref{fig:scatter}, and was present at or near the time of many of the
candidate events in this catalog. 
A significant portion of the increase in glitch rate between O2 and O3a
at the LIGO detectors can be accounted for by the increased rate of scattered light
glitches~\cite{Soni:2020rbu}.
A potential major source of this noise for O3a was
light scattered from the gold-coated electrostatic drives mounted to the
fused silica reaction masses that are suspended directly behind the LIGO test
masses to provide a stable platform for low-noise actuation~\cite{Soni:2020rbu}.  
Changes were implemented during O3b that reduced scattered light noise 
entering through this path~\cite{Soni:2020rbu}.

Broadband short-duration glitches also occurred often in all detectors during O3a. 
A sub-class of those, blip glitches~\cite{Cabero:2019orq}, 
was one of the most problematic sources of transient noise
in previous observing runs, and is still present in O3a 
at a rate of \BLIPRATE.  
These glitches are one of the limiting sources
of noise for searches for gravitational waves from 
high mass compact binaries~\cite{Davis:2020nyf}
and no sources or witnesses for the majority of these glitches 
have been identified. 
In O3a, there was also
an additional population of short duration glitches with 
\ac{SNR} $> 100$.  
These loud glitches occurred
in both LIGO detectors, with unknown origin. 
Additional description of these glitches, along
with details of potential sources that have been ruled out, 
can be found in~\cite{Davis:2021ecd}.

Many glitches
in LIGO and Virgo data have well-understood sources and couplings, making it
possible to identify short time periods where excess power from
environmental or technical sources will be present in the strain data.
Flagging these time periods as containing poor data quality, either by removing
the data from the search or decreasing the significance of any candidate events
identified, has been shown to improve the overall sensitivity of searches for
gravitational waves from compact binaries~\cite{TheLIGOScientific:2017lwt,idq-in-gstlal,Davis:2021ecd}.
While a select number of LIGO and Virgo data quality issues are 
flagged in low latency, such as hardware injections and
digital signal overflows, the majority of data quality flags are only
available for offline searches.

Before performing searches for gravitational wave signals, periods of poor data
quality are flagged at various levels, called
categories~\cite{TheLIGOScientific:2016zmo,LIGOScientific:2019hgc,Davis:2021ecd}.
In O3a, data from an observatory not operating in a nominal state are 
flagged (category 1) and not used in any search.  
Additional periods
likely to contain excess power in each LIGO detector 
were flagged based on detailed follow up of
identified sources of noise (category 2), statistical correlation between
witness sensors (category 3), and machine-learning based predictions
(iDQ)~\cite{Essick:2020qpo,idq-in-gstlal}.
No additional data quality products for Virgo were used beyond category 1.
The specific set of data quality products used in
O3a is search-specific, as described in Sec.~\ref{sec:searches}. 
Category 2 flags are tuned separately for searches for both
gravitational waves from CBC and minimally-modeled (Burst) sources.  
Category 3 flags are only tuned for Burst searches.
The amount of time removed by each category of veto in O3a is shown in Table~\ref{tab:veto_sum}.

\begin{event_table}
\begin{table*}[!tbh]
	\begin{tabularx}{\textwidth}{l@{\extracolsep{\fill}}rrrr}
        \textbf{Detector} & \textbf{Category 1} & \textbf{CBC category 2} & \textbf{Burst category 2} & \textbf{Burst category 3} \\
        \hline
        \makebox[0pt][l]{\fboxsep0pt\colorbox{lightgray}{\mystrut\hspace*{1.0\linewidth}}}LIGO Hanford & \LHOCATONE \% & \LHOCATTWOCBC \% & \LHOCATTWOBURST \% & \LHOCATTHREEBURST \% \\
        LIGO Livingston & \LLOCATONE \% & \LLOCATTWOCBC \% & \LLOCATTWOBURST \% & \LLOCATTHREEBURST \% \\
        \makebox[0pt][l]{\fboxsep0pt\colorbox{lightgray}{\mystrut\hspace*{1.0\linewidth}}}Virgo & \VIRGOCATONE \% & \VIRGOCATTWOCBC & \VIRGOCATTWOBURST & \VIRGOCATTHREEBURST \\
        \hline
	\end{tabularx}
	\caption{Percent of single-detector time removed by each category of veto for each detector. 
                 Category 1 vetoes were applied in all analyses described in Sec.~\ref{sec:searches}. 
                 CBC category 2 vetoes were applied only by the PyCBC search. 
                 Burst categories 2 and 3 were applied only by the cWB search.}
	\label{tab:veto_sum}
\end{table*}
\end{event_table}

\subsection{Event Validation}\label{ss:validation}

Event validation procedures similar to those used
for previous gravitational wave candidate events~\cite{TheLIGOScientific:2016zmo,LIGOScientific:2018mvr,Davis:2021ecd}
were used for all candidate events in this catalog
to evaluate if instrumental artifacts could impact analysis. 
Within tens of minutes of low-latency candidate event identification, 
time--frequency visualizations and monitors
of the gravitational wave strain data~\cite{Chatterji:2004qg,Zevin:2016qwy,Mozzon:2020gwa,
Areeda:2016mee,VIR-0237A-15,Verkindt:2019}
were used to identify any data quality issues present.
On the same timescale, 
data from hundreds of auxiliary sensors monitoring the detectors and their
environments
were used to identify potential auxiliary witnesses to instrumental 
artifacts~\cite{Effler:2014zpa,Nguyen:2021ybi,Essick:2020qpo}.
Tools that relied upon deeper information 
about glitches and data non-stationarity gathered offline, 
such as long-term monitors of the instruments and their
subsystems~\cite{gwsumm-software,VIR-0546A-16,VIR-0554A-16}
and identification of likely sources of glitches by correlation with
auxiliary sensors~\cite{Smith:2011an,Isogai:2010zz,Omicron:2019},
were also used to vet candidate events in this catalog.
These procedures did not identify evidence of instrumental origin
for any of the candidate events in this catalog, 
but did identify a number of data quality issues 
that could potentially impact analyses of these candidate events.

Candidate events with data quality issues identified by these event validation procedures
required further mitigation before analysis. 
In cases when
glitches occurred in time coincidence with a candidate event
(but could not account for the candidate event itself), additional data processing
steps were completed to mitigate the effect of those glitches on
estimation of the candidate event parameters.
If possible, the identified glitches were
subtracted using the methods described in Sec.~\ref{ss:denoise}.  
In cases when
sufficient subtraction was not possible, customized configurations of parameter
estimation analyses were used to exclude the time period or frequency
bandwidth impacted by glitches.
An example of a glitch coincident with a signal that required
glitch subtraction is shown in 
Fig.~\ref{fig:scatter}.
While the presence of excess power from transient noise 
did not prevent confident identification
of this event, glitch subtraction was required before the source
properties of the event could be evaluated. 
Although only data recorded from detectors in observing mode were used 
to identify candidate events in this catalog, 
some candidate events occurred at times when one detector in the 
network was operating, but not in observing mode. 
If it was concluded that data from the additional detector
would substantially impact the scientific conclusions reached from 
analyzing the candidate, the additional data were investigated. 
For those cases, the data quality and calibration for the non-observing 
detector was evaluated to determine whether the data could be used 
in the estimation of candidate event source properties. 
Such data were used for one candidate event, \protect\NAME{GW190814A}~\cite{GW190814A}.
The full list of candidate events requiring specific mitigation steps, due to either
the presence of glitches or the state of a detector, is found in Sec.~\ref{sec:peresults}.
The total number of candidate events requiring mitigation is 
consistant with the number expected based on the glitch rate in each detector.

\subsection{Glitch subtraction}\label{ss:denoise}

Data containing gravitational wave candidate events and glitches in the same
time--frequency volume are pre-processed through a glitch-subtraction procedure
prior to being analyzed by the parameter estimation pipelines.  
The glitch-subtraction procedure evolved from the 
BayesWave (BW) algorithm~\cite{Cornish:2014kda,Littenberg:2015kpb,Cornish:2020dwh} 
used for glitch subtraction in the Livingston detector at the time of the GW170817
binary neutron star merger~\cite{Pankow:2018qpo,TheLIGOScientific:2017qsa}, where the
non-Gaussian, incoherent, noise was modeled as a linear combination of wavelets
which was subtracted from the data. The number of wavelets used in the fit was
determined using a trans-dimensional Markov chain Monte Carlo 
(MCMC) algorithm that balances using fewer wavelets 
against the quality of the fit~\cite{Cornish:2014kda}.

To prevent the glitch-subtraction
procedure from corrupting the signal candidate event, the time segment and bandwidth
of the wavelet-based analysis are chosen, when possible, 
to exclude from subtraction the strongest part of the signal. 
For cases where the signal and glitch overlapped
in time--frequency space, a more robust application of the glitch-subtraction
algorithm was used which simultaneously fits for the signal and the glitch.
In the signal-plus-glitch application, signal wavelets are included in the
model if they are coherent over the detector network (marginalizing over sky
location, etc.), while the glitch wavelets are independent 
in each detector~\cite{Littenberg:2015kpb}.
Only the glitch model wavelets are then used in the subtraction.  
The signal-plus-glitch procedure was tested by injecting 
simulated coherent \ac{BBH} signals onto known single detector glitches
from O2 and verifying that the signals were unaffected in the
process.

The glitch-subtraction procedure is only used as a pre-processing step for the
parameter estimation analysis 
(described in Sec.~\ref{sec:PEmethods}), 
and is not part of the analyses that determine
the presence, or significance, of gravitational wave candidate events. 
As shown in Fig.~\ref{fig:scatter}, the glitch subtraction methods described here 
are able to successfully remove excess power caused by glitches
present near the time of candidate events.

%% file: methods.tex
\section{Candidate identification \note{(Chad)}}
\label{sec:searches}
\resetlinenumber

Candidate identification happens on two timescales. First, five low-latency
gravitational wave pipelines~\cite{Messick:2016aqy, Klimenko:2015ypf,
DalCanton:2020vpm, Adams:2015ulm, chu2017low} process the data immediately
after acquisition with the goal of generating public detection alerts to the
broader astronomical community within minutes~\cite{LIGOScientific:2019gag}.
Second, an offline reanalysis of gravitational wave data is conducted to
produce the curated candidate event list here.  The offline analysis may
benefit from updated data calibration, data quality vetoes, the ability to
estimate event significance from the full data, and further algorithmic
development that takes place over the course of an observing run.  Although the
candidate events presented here are derived from offline analysis, we provide a
comparison with the public alerts in Sec.~\ref{s:candidates}. Candidates are
identified using two methods.  The first method searches for minimally-modeled
sources.  The second method searches for signals from a bank of template
waveforms~\cite{Owen:1998dk} modeled after the expected gravitational wave
emission from coalescing compact binaries in general relativity.  In
Sec.~\ref{s:candidates}, we present results from one search for
minimally-modeled transient sources, Coherent WaveBurst
(\CWB{})~\cite{Klimenko:2004qh,Klimenko:2005xv,Klimenko:2006rh,
Klimenko:2011hz,Klimenko:2015ypf}, and two searches for modeled sources,
\GSTLAL{}~\cite{Sachdev:2019vvd, Hanna:2019ezx, Messick:2016aqy} and
\PYCBC{}~\cite{Allen:2005fk, Allen:2004gu, Canton:2014ena, Usman:2015kfa,
Nitz:2017svb}.  \CWB{}, \GSTLAL{}, and \PYCBC{} were also three of the five
low-latency pipelines in O3. The two remaining low latency pipelines,
MBTAOnline~\cite{Adams:2015ulm,Aubin:2020goo} and SPIIR~\cite{chu2017low} were
not configured for offline reanalysis at the time of this publication and are
therefore not included in GWTC-2.  Below, we summarize the methods used by each
of \CWB{}, \GSTLAL{}, and \PYCBC{} to identify candidate events.

\subsection{Coherent WaveBurst search for minimally-modeled transient sources \note{(Edoardo)}}
\label{ss:cwbmethods}

\CWB{} is a search pipeline for detection and reconstruction of transient GW
signals that operates without a specific waveform model~\cite{Abbott:2019prv} and   
was used in
previous searches by the 
\LVC~\cite{Abbott:2016blz, TheLIGOScientific:2016uux, LIGOScientific:2018mvr}. 
\CWB{} identifies
coincident signal power in  multiple detectors, searching for transient signals
with durations up to a few seconds in the detector bandwidth. The analysis is
performed on the time--frequency data obtained with the Wilson--Daubechies--Meyer wavelet
transform~\cite{Klimenko:2004qh, Necula:2012zz} and normalized by the amplitude spectral
density of the detector noise. \CWB{} selects the time--frequency data samples
above fluctuations of the detector noise and groups them into clusters.  For
clusters correlated in multiple detectors, \CWB{} reconstructs the source sky
location and signal waveforms with the constrained maximum likelihood method ~\cite{Klimenko:2015ypf}.
The signal SNR is estimated from the signal waveforms reconstructed by cWB, and the 
network SNR is calculated combining the SNRs of individual detectors. 

The \CWB{} detection statistic is based on the coherent energy $E_{\rm c}$
obtained by cross-correlating the normalized signal waveforms reconstructed in
different detectors. It is normalized by a chi-squared statistic 
$\chi^2 = E_{\rm n}/N_{\rm df}$, where $E_{\rm n}$ is the residual noise energy estimated
after the reconstructed waveforms are subtracted from the data, and $N_{\rm
df}$ is the number of independent wavelet amplitudes describing the event.  The
\CWB{} detection statistic is $\eta_{\rm c}\propto [{E_{\rm
c}/\max(\chi^2,1)}]^{1/2}$, where the $\chi^2$ correction is applied to reduce
the contribution of non-Gaussian noise.  To improve the robustness of the
algorithm against glitches, \CWB{} uses signal-independent vetoes, which reduce
the \ac{FAR} of the pipeline; this includes category 2 Burst data 
quality flags in the processing step and hierarchical vetoes in the 
post-production phase \cite{Smith:2011an,Aasi:2012wd}.  Other vetos applied to candidate events
are on the network correlation coefficient $c_{\rm c} = E_{\rm c}/(E_{\rm
c}+E_{\rm n})$ and the $\chi^2$. To further reduce the background, the \CWB{}
analysis employs additional signal-dependent vetos based on the properties of
the time--frequency evolution of compact binary signals: a) the frequency of
the signal is increasing in time ~\cite{Tiwari:2015bda} and b) the central frequency of the signal
$f_\mathrm{c}$ is inversely proportional to the total mass of the system ~\cite{Szczepanczyk:2020osv}.
\CWB{} searches are performed with two pipeline configurations targeting
detection of high-mass ($f_\mathrm{c}<80$~Hz) and low-mass
($f_\mathrm{c}>80$~Hz) \ac{BBH} systems.  They use different signal-dependent vetos
defined a priori to alleviate the large variability of non-stationary noise in
the detectors' bandwidth. 

We estimate the significance of candidate events by systematically time-shifting 
the data of one detector with respect to the other in each detector pair, 
with a time lag so large that actual astrophysical events are excluded, and repeating this for a large 
number of different time lags over a total time $T_\mathrm{bkg}$. We count the number
of events $N$ due to instrumental noise that have a ranking statistic value such as the SNR 
that is at least as large as that of the candidate event and we compute the \ac{FAR} as the number 
of background events divided by the total background time~\cite{Was:2009vh}. 
The detection significance of a candidate event identified by either configuration in a single frequency range is
determined by its \ac{FAR} measured by the corresponding \CWB{} configuration. 
Whenever the low-mass and high-mass configurations overlap, the trials factor of two is included to determine 
the final \ac{FAR}~\cite{GW190521Adiscovery}. In the end, each configuration reports the selected 
candidate events and their \ac{FAR}.

The sensitivity of the \CWB{} search pipeline approaches that of matched-filter methods for coalescing stellar mass \acp{BBH} with high chirp masses, where most of the signal energy is concentrated in just a few wavelets of the \CWB{} representation~\cite{CalderonBustillo:2017skv}. It is less competitive for low chirp mass events, where the signal power is spread over large time–frequency areas. \CWB{} can also detect sources that are not well represented in current template banks (e.g., eccentric systems or high-mass ratio precessing systems)~\cite{Salemi:2019owp}.
Tests with \CWB{} showed that the detection efficiency for a given \ac{FAR} threshold is slightly smaller with the inclusion of Virgo. Therefore, also to reduce computing time, all \CWB{} detection candidates and waveform consistency tests reported in this catalog use the Hanford--Livingston network only.


\subsection{\GSTLAL{} and \PYCBC{} searches for modeled sources}\label{sec:mf_searches}

Both the \GSTLAL{} pipeline~\cite{Sachdev:2019vvd, Hanna:2019ezx,
Messick:2016aqy} and the \PYCBC{}~\cite{alex_nitz_2019_3561157}
pipeline~\cite{Allen:2005fk, Allen:2004gu, Canton:2014ena, Usman:2015kfa,
Nitz:2017svb} implement independently designed matched-filter analyses. Both
were used in previous \LVC searches for gravitational
waves~\cite{Abbott:2016blz, Abbott:2016nmj, TheLIGOScientific:2016pea,
Abbott:2017vtc, Abbott:2017gyy, Abbott:2017oio, LIGOScientific:2018mvr}. 

The matched filter method relies on a model of the signal, dependent on the source
physical parameters. Most important for the phase evolution of the source (and
therefore the matched filter) are the intrinsic parameters: two individual component
masses $m_1,\, m_2$, and two dimensionless spin vectors
$\vec{\chi}_{\{1,2\}}$, where the dimensionless spin is related to each
component's spin angular momentum $\vec{S}$ by $\vec{\chi}_i = c\vec{S}_i/(Gm_i^2)$.

We also make use of combinations of these intrinsic parameters that are typically
well-constrained by gravitational wave measurements; the binary chirp mass~\cite{Blanchet:1995ez},
\begin{equation}
    \Mc = \frac{(m_1 m_2)^{3/5}}{(m_1 + m_2)^{1/5}},
\end{equation}
determines to lowest order the phase evolution during the inspiral, and is
typically better constrained than the component masses.  At higher
orders, the mass ratio $q = {m_2}/{m_1}$ (where $m_2 \leq m_1$) and
effective inspiral spin $\chieff$ affect the binary phase evolution.
The effective inspiral spin is defined
as~\cite{Ajith:2009bn}:
\begin{equation}
    \chieff = \frac{(m_1 \vec{\chi}_1 + m_2 \vec{\chi}_2)\cdot\hat{L}_\mathrm{N}}{M},
\end{equation}
where $M = m_1+m_2$ is the total mass and $\hat{L}_\mathrm{N}$ is the unit vector along the Newtonian
orbital angular momentum. The spin tilt angle for each component object
$\theta_{\rm LS_i}=\cos^{-1} \left(\vec{\chi}_i\cdot\hat{L}_\mathrm{N}/|\vec{\chi}_i|\right)$ quantifies the
angle between the orbital angular momentum vector and its spin vector. Since the
spin and angular momentum vectors vary if the system precesses, by convention we use
the spin parameters at a reference frequency of 20\,Hz, with the exception
of GW190521 where we use 11\,Hz for consistency with previous publications~\cite{GW190521Adiscovery,GW190521Aastro}.
Additional intrinsic parameters are needed to
describe eccentricity, which we assume to be zero in our modeled analyses.
The timescale for circularization of isolated binaries with non-zero
eccentricity at birth is sufficiently short that sources
are expected to have negligible eccentricity when they enter the sensitive
bands of the LIGO and Virgo detectors~\cite{Peters:1964zz}. However,
dynamically formed binaries may have residual eccentricity as the signal enters
the sensitive band of the detector. These systems have been the target of unmodelled
searches of previous observing runs, but with no candidate events reported~\cite{Salemi:2019owp}.

Seven extrinsic parameters provide the orientation and position of the
source in relation to the Earth: the luminosity distance $\DL$, two-dimensional sky
position (right ascension $\alpha$ and declination $\delta$), 
inclination between total angular momentum and line-of-sight
$\theta_{JN}$, time of merger $t_c$, a reference phase $\phi$, and
polarization angle $\psi$. 

In general, as the signal travels from the source to the
detector its frequency is redshifted by a factor $(1+z)$.
For a system involving only black holes, the observed signal is
identical to that from a source in the rest frame of the detector with total
mass $M^\mathrm{det}=(1+z)M$~\cite{Krolak:1987ofj, Cutler:1994ys}.  For convenience, the
templates used by the modeled searches are defined in the rest frame of the
detectors which subsumes the factor $(1+z)$ into the definition of masses.

For this work, the \GSTLAL{} analysis used a template bank with component
masses between $1\,\Msun$ and $400\,\Msun$ with total masses, $M^\mathrm{det}$, between
$2\,\Msun$ and $758\,\Msun$ and spins that are aligned or anti-aligned with the
binary's orbital angular momentum, such that only the spin components
$\chi_{i,z} = \vec{\chi}_i\cdot \hat{L}_\mathrm{N}$ are non-zero.  The bank was constructed in five regions via a
stochastic placement algorithm~\cite{Privitera:2013xza,Harry:2009ea} satisfying
different minimal match (mm)~\cite{Owen:1998dk} criteria with waveforms
starting at $f_{\text{min}}$ as described in Table~\ref{t:GstLAL-tb}.
Template placement was augmented
to improve the collection of background statistics in the last region shown in Table~\ref{t:GstLAL-tb} by a grid
of templates distributed uniformly in the logarithm of component mass to
improve detection efficiency for systems with primary mass $m_1^\mathrm{det}$ above $50\,\Msun$~\cite{LIGOScientific:2018mvr, Mukherjee:2018yra}.  The TaylorF2 waveform
approximant~\cite{Sathyaprakash:1991mt, Blanchet:1995ez, Poisson:1997ha, Damour:2001bu, Mikoczi:2005dn, Blanchet:2005tk,
Arun:2008kb,Buonanno:2009zt,Bohe:2013cla,Bohe:2015ana,Mishra:2016whh}.
was used for templates with $\Mc^\mathrm{det} < 1.73\,\Msun$ and the
SEOBNRv4\_ROM waveform approximant~\cite{Bohe:2016gbl} for templates with $\Mc
\geq 1.73\,\Msun$.

\begin{event_table}
\begin{table}
{\small
\noindent\begin{tabularx}{\columnwidth}{l@{\extracolsep{\fill}}rrrrrrrr}
\textbf{mm} & $\boldsymbol{m_1}$ & $\boldsymbol{m_2}$ & $\boldsymbol{M} $& $\boldsymbol{q}$ & $\boldsymbol{\chi_{1,z}}$ &  $\boldsymbol{\chi_{2,z}}$ & $\boldsymbol{f_{\text{min}}}(\mathrm{Hz})$ \\
\hline
\makebox[0pt][l]{\fboxsep0pt\colorbox{lightgray}{\mystrut\hspace*{1.0\linewidth}}}0.99 & $1,3$ & $1,3$ & $<6$ & $0.33,1$ & low & low & 15\\
0.97 & $3,150$ & $1,3$ & $<153$ & $0.02,1$ & high & low & 15 \\
\makebox[0pt][l]{\fboxsep0pt\colorbox{lightgray}{\mystrut\hspace*{1.0\linewidth}}}0.99 & $3,91$ & $3,50$ & $<100$ & $0.1,1$ & high & high & 15 \\
0.97 & $30,392$ & $3,36$ & $<400$ & $0.02,0.1$ & high & high & 15 \\
\makebox[0pt][l]{\fboxsep0pt\colorbox{lightgray}{\mystrut\hspace*{1.0\linewidth}}}0.99 & $50,400$ & $9,400$ & $>100$ & $0.1,1$ & high & high & 10 \\
\hline
\end{tabularx}
}
\caption{\label{t:GstLAL-tb} \GSTLAL{} template bank parameters. Low spin denotes the range $-0.05$ to $0.05$ and
high spin denotes the range $-0.999$ to $0.999$. }
\end{table}
\end{event_table}

The \PYCBC{} analysis used a template bank covering the same parameter space as
for GWTC-1~\cite{LIGOScientific:2018mvr} shown in Figs.~3 and 7 of
\cite{DalCanton:2017ala}.  Unlike the previous work, the template bank here was
created using a hybrid geometric-random method described in \cite{Roy:2017qgg,
Roy:2017oul}.  This new method provides a more efficient template bank---in
terms of covering the full parameter space with fewer template waveforms---than
the stochastic method~\cite{Privitera:2013xza,Harry:2009ea}. This bank is
broadly similar to the parameter space covered in the \GSTLAL{} search
described above with a key difference being that only templates longer than
0.15~s were kept.  Details about this cut and its effect on the explored mass
and spin range can be found in~\cite{DalCanton:2017ala}.

Both \GSTLAL{} and \PYCBC{} scan data from each gravitational wave detector
against the above-described banks of template waveforms to produce SNR
time-series~\cite{Allen:2005fk}.  The \ac{SNR} time series are maximized over short time windows to
produce a set of triggers for each template and each detector.  Triggers that
pass an \ac{SNR} threshold of 4 in one detector form the basis of candidate events
according to the procedures for each pipeline described below.  \PYCBC{}
removes the time period during category 2 veto flags from the final results,
while \GSTLAL{} uses only iDQ for single-detector triggers and no data quality
products for coincident triggers.

\GSTLAL{} defines a candidate event as consisting of triggers from one or more
gravitational wave detectors ranked by the SNRs of the triggers, signal-consistency
tests, time delays between each detector, phase differences between detectors,
the (possibly zero) time-averaged volumetric sensitivity of each detector, and
the signal population model. These parameters are used as variables in the
likelihood-ratio ranking statistic $\mathcal{L}$~\cite{Cannon:2015gha, Messick:2016aqy, Sachdev:2019vvd, Hanna:2019ezx} which
is a monotonic function of the inverse false alarm
probability~\cite{Cannon:2012zt}. 

There are two differences between the ranking
statistic used here and in O2~\cite{LIGOScientific:2018mvr}. First, we
implemented a template likelihood
$p(\text{T}| \mathrm{signal, SNR})$~\cite{Fong:2018elx}, which is the probability that a trigger
is recovered by a template $\text{T}$, given the trigger \ac{SNR} and that the signal
belongs to some population.  Previous versions of \GSTLAL{} approximated
$p(\text{T} | \mathrm{signal, SNR})$ by a constant in
$\mathcal{L}$~\cite{Sachdev:2019vvd, Messick:2016aqy, Cannon:2015gha}, implying
all templates were equally likely to recover a signal.  Now, the template
likelihood is informed by the template bank (to account for the fact that
templates are not uniformly distributed in parameter space~\cite{Dent:2013cva})
and a signal-population model, which for this search considers $\vec{\theta} = \{m_1^\mathrm{det},
m_2^\mathrm{det}, \chi_{1,z}, \chi_{2,z} \}$ and is given by $p(\vec{\theta}| \mathrm{signal})
\, \mathrm{d}\vec{\theta} \propto 1/(4\,m_1^\mathrm{det}\,m_2^\mathrm{det}) \, \mathrm{d}\vec{\theta}$.
The distribution $p(\vec{\theta} | \mathrm{signal})$ is deliberately broad to minimize the number of
missed signals.  Second, single-detector candidate events are ranked using both an
empirically determined penalty and information from iDQ~\cite{Essick:2020qpo} as
described previously.  The penalty from iDQ is added to the denominator of
$\mathcal{L}$.  Candidates from the \GSTLAL{} search had their likelihood
ratios and significance estimated using the entire $\sim6$ month data set.

\PYCBC{} identifies candidate events by requiring triggers in both LIGO Hanford and
LIGO Livingston with a time delay smaller than the light travel time between
observatories. These candidate events are then ranked using a set of signal-based
vetoes, data quality information, and by comparing the properties of the event
against those expected from astrophysical signals~\cite{Nitz:2017svb}.  A
\ac{FAR} is then computed for each of these candidate events by estimating the
background noise distribution using time-shifted analyses similarly to
\CWB{}~\cite{Allen:2005fk, Babak:2012zx, Usman:2015kfa}, triggers from LIGO
Livingston being timeshifted relative to LIGO Hanford by multiples of $0.1$~s.
Virgo data were not searched with the \PYCBC{} pipeline due to three-detector
searches not being completely integrated in the version of code used.

A focused search for \ac{BBH} coalescences~\cite{Nitz:2019hdf} is also used here,
denoted later as \PYCBC{} BBH.  This was motivated by the fact that all signals
observed in O1 and O2, with the single exception of the binary neutron star
merger GW170817, were consistent with \ac{BBH} coalescences with mass ratio close to
1 and effective inspiral spins close to 0.  The full parameter space search
used for GWTC-1~\cite{LIGOScientific:2018mvr}, in contrast, was tuned to
observe signals anywhere in the possible space of signal parameters, which
might include signals that do not have very high matches with search templates. 
The \PYCBC{} BBH search uses a recently developed detection statistic~\cite{Nitz:2019hdf,
Davies:2020tsx} which includes a number of tuning choices to reject triggers that
do not match the filter waveforms well, and also includes a template weighting 
implementing a prior that signals detectable in any given range of \ac{SNR} 
are uniformly distributed in chirp mass. 
This search enabled \PYCBC{} to identify more BBHs in the O1 and O2 datasets
than reported in the GWTC-1 paper~\cite{LIGOScientific:2018mvr,Nitz:2019hdf}.
This included some of the BBHs first reported 
in \cite{Venumadhav:2019lyq,Zackay:2019tzo} by independently-developed searches~\cite{Venumadhav:2019tad,Zackay:2019kkv}.
We use the focused \ac{BBH} search in this work to better extract \ac{BBH} 
coalescences from the data: this search
considers only a reduced set of filter templates defined prior to the analysis of
O3 data, namely systems with mass ratio $q>1/3$, and with both component masses 
(in detector frame) larger than $5\,\Msun$.

\subsection{Estimation of modeled searches sensitivity}
\label{sec:simulation}

In order to estimate the sensitivity of the \GSTLAL{} and \PYCBC{} searches, we
conducted a campaign of 
simulated signals injected
into the O3a gravitational wave data and analyzed by both matched filter
pipelines.  The simulated population, intended to cover (or over-cover) the detected 
population of stellar-mass \acp{BBH} \cite{LIGOScientific:2018jsj,o3apop}, contains
component masses $m_1$, $m_2$ between 2\,\Msun{} and 100\,\Msun{} and extends out
to a maximum redshift of 2.3.  In order to reduce statistical uncertainties, the 
mass, spin and redshift distributions should be sufficiently similar to population
models for which we intend to estimate merger rates: see \cite{o3apop} (Appendix A)
for further discussion of selection functions in population inference. 
For the simulations, we chose $p(m_1)
\propto m_1^{-2.35}$, $p(m_2\,|\,m_1) \propto m_2^{2}$ (for $m_2<m_1$), and $\chi_{i,z}$
values distributed uniformly between $-0.998$ and $0.998$.  The
cosmological distribution of sources simulates a merger rate in the comoving
frame that evolves as $R(z) = R(0)(1+z)^2$, thus the source redshift
distribution follows $p(z) \propto (1+z) \mathrm{d} V_\mathrm{c} / \mathrm{d}z$, where $V_\mathrm{c}$ is the
comoving volume (see, e.g., \cite{Abbott:2016nhf} for further discussion of 
cosmological effects).  The simulation set was generated in two stages: first, points
were chosen according to this distribution out to $z=2.3$; then, these \NUMINJORIGAPPX{}
samples were reduced to a set of potentially detectable signals by imposing that the
expected LIGO Hanford--LIGO Livingston network \ac{SNR}, calculated using representative
noise \acp{PSD}, be above a threshold of $6$.  The \NUMINJAPPX{} signals remaining 
after this cut were assigned merger times ranging uniformly over the duration of O3 
for analysis by the searches.  The SEOBNRv4\_opt aligned-spin waveform 
model~\cite{Bohe:2016gbl} was used for the simulated signals.

The expected number of signals from such a population detected by a given
analysis may be written as
\begin{equation}
  \hat{N} = \mathcal{V} R(0),
\end{equation}
where $R(0)$ is the rate of signals per unit volume and unit observing time at
present, and $\mathcal{V}$ is the effective surveyed hypervolume, a measure of
analysis sensitivity for the injected population.\footnote{The measure
$\mathcal{V}$ is not equivalent to a geometric volume--time (VT), because both
the injected population density and the comoving volume element vary over
redshift.} 
Since each analysis recovered over $10^4$ injections, statistical counting
uncertainties in the hypervolumes $\mathcal{V}$ are at the sub-percent level.
Our estimates of sensitivity are, though, affected by possible systematic
uncertainties in calibration of the strain data.  Strain calibration affects
the detectability of simulated signals via the magnitude of the response
function, which is affected by uncertainties of at most a few percent over the
frequency range where the SNR of binary merger signals is accumulated; see
\cite{Sun:2020wke} for details. 
The hypervolume $\mathcal{V}$ surveyed by each analysis, at a detection \ac{FAR}
threshold of \FARTHRESHYR{} per year, is \GSTLALVT{} for \GSTLAL{},
\PYCBCVT{} for \PYCBC{}, and \PYCBCBBHVT{} for \PYCBC{} \ac{BBH}.  For a combination of all matched
filter analyses presented in this catalog, with any injection found in \emph{one
or more} analysis below the \ac{FAR} threshold considered as detected, we find a
surveyed hypervolume \ANYCBCVT{}. \fixme{Full results from the injection campaign 
are available via a data release at \cite{software-inj-release}.}

\subsection{Estimation of signal probability \note{(Shasvath)}}
\label{ss:pastro}

For each candidate event, the probability of origin from an astrophysical source 
$p_{\rm astro}$ and corresponding probability of terrestrial noise origin 
$p_{\rm terr} = 1 - p_{\rm astro}$ may be estimated using the outputs of search
pipelines. 
We obtain these probabilities for the candidate events in this catalog consistent with a BBH,
via the Poisson mixture model formalism~\cite{Farr:2013yna} used in O1 ~\cite{Abbott:2016nhf, TheLIGOScientific:2016pea, Abbott:2016drs}. Only BBH candidate events 
are considered because, other than GW190425~\cite{GCN24168} whose component masses are consistent with those of NSs, 
all significant detections can be classified as \acp{BBH}.\footnote{There is some ambiguity regarding the BBH nature of GW190814~\cite{GW190814A}. 
Nevertheless, its chirp mass and total mass are not inconsistent with those of BBHs.} A low significance candidate NSBH event, \NAME{GW190426A},
was reported in low-latency~\cite{GCN24237}; however, its astrophysical probability is strongly dependent 
on prior assumptions of the rate of such signals. We therefore do not estimate its $p_{\rm astro}$ here.

We start by collecting the ranking statistics $\vec{x} = \lbrace{x_1, x_2, \ldots, x_N \rbrace}$ 
of all candidate events more significant than a predefined threshold: for the \GSTLAL{} pipeline events 
are thresholded on FAR, while for \PYCBC{}, a ranking statistic threshold is applied.  
The threshold for \GSTLAL{} is chosen to ensure that the total number of background events considered
exceeds the number of signals by a large ($\sim 100$) factor, which enables an accurate estimate of 
the total rate of background events above threshold. On the other hand, \PYCBC{} estimates the
background rate from time-shifted analyses, as outlined above in Section~\ref{sec:mf_searches}, thus
the requirement to include a large number of background events is relaxed. The statistic 
threshold is then set low enough to include (at least) all events with $p_{\rm astro} \gtrsim 0.1$. 

Additionally, for both \GSTLAL{} and \PYCBC{} searches, a threshold of $4.35\,M_{\odot}$ is applied on the chirp mass of the templates, 
corresponding to a $5\,M_{\odot} + 5\,M_{\odot}$ binary, 
ensuring that the selected candidate events have template masses consistent with those of putative BBHs.
Using the distribution of ranking statistics $x$ under the foreground model, 
$f(x) = p(x | \mathrm{signal})$, and the distribution under the background model, 
$b(x) = p(x | \mathrm{noise})$, estimated by each matched filter pipeline, 
we assign a Bayes factor $k(x) = f(x)/b(x)$ to each event.
Assuming that foreground and background triggers are drawn from independent Poisson 
processes, one can then calculate the posterior over the Poisson expected counts 
for each process, $\Lambda_1$ and $\Lambda_0$. 

For the \PYCBC{} searches presented here we proceed as for O1 and O2
\cite{LIGOScientific:2018mvr} and estimate foreground and background event
densities empirically for all putative BBH candidate events with 
ranking statistic above a given threshold.
The \PYCBC\ full parameter space and BBH
focused searches differ in how their ranking statistics are calculated:
for the full search~\cite{Nitz:2017svb}, a threshold of $7.9$ is 
applied to the ranking statistic; while for the BBH search~\cite{Nitz:2019hdf, Davies:2020tsx}, a threshold value of $9$ is applied.
We empirically measure the rate of
noise events satisfying these cuts via time-shifted analyses, and infer the 
posterior over the rate of
signals; finally we marginalize over the signal rate to obtain probabilities of 
astrophysical and terrestrial origin for each event~\cite{T1700029}.
Since the \PYCBC{} BBH search is more sensitive to realistic BBH signal
populations, implying a more accurate estimate of the relative densities of
signal and noise events within its targeted mass region, we consider the 
$p_{\rm astro}$ values from the BBH analysis to be more accurate for events 
recovered by both searches. 

For the \GSTLAL\ analysis, we estimate the astrophysical and terrestrial probabilities,
$p_{\mathrm{astro}}(x | \vec{x})$, from the joint posterior on the Poisson expected counts,
$p(\Lambda_0, \Lambda_1 | \vec{x})$, where the set of triggers $\vec{x}$ have a chirp mass $\Mc > 4.35\,\Msun$,
and a FAR $< 8766$~yr$^{-1}$. The prior used to construct the joint counts posterior
is taken to be the corresponding posterior from O1 and O2~\cite{Kapadia:2019uut}.

\section{Estimation of source parameters \note{(John, Neil, Zoheyr, Salvo)}}\label{sec:PEmethods}

Once triggers of interest have been identified, the physical parameters
of the candidate event gravitational wave signals are inferred by computing
their posterior probability density functions. 
The uncertainty in the source parameters is quantified by the posterior probability 
distribution $p(\vec{\vartheta}|\vec{d})$, which is calculated using Bayes' theorem as
\begin{equation}
p(\vec{\vartheta}|\vec{d}) \propto p(\vec{d}|\vec{\vartheta}) \pi(\vec{\vartheta})\,,
\end{equation}
where $p( \vec{d}|\vec{\vartheta})$ is the likelihood of the 
data given the model parameters $\vec{\vartheta}$,
and $\pi(\vec{\vartheta})$ is the prior probability distribution for the parameters.
The likelihood is calculated from a coherent analysis of data from each of the 
detectors. 
As in our previous analyses, e.g., \cite{TheLIGOScientific:2016wfe},
we assume that the noise can be treated as Gaussian, stationary, and uncorrelated 
between detectors~\cite{LIGOScientific:2019hgc,Berry:2014jja} in the stretch of data used to calculate the likelihood and to 
measure the noise \PSD. This yields a Gaussian 
likelihood~\cite{Cutler:1994ys,Veitch:2014wba} for the data from a single detector,
\begin{equation}
    \label{eqn:pe_likelihood}
    p(d^i | \vec{\vartheta}) \propto \exp\left[- \frac{1}{2}  \left\langle d^i - h_M^i(\vec\vartheta) \middle| d^i - 
h_M^i(\vec\vartheta) \right\rangle \right],
\end{equation}
where $d^i$ is the data of the $i$-th instrument,  
$h_M^i(\vec\vartheta)$ is the waveform model calculated at $\vec\vartheta$ projected
on the $i$-th detector and adjusted to account for the uncertainty in offline
calibration described in Sec.~\ref{ss:calibration}. 
The noise-weighted inner product $\langle a | b \rangle$~\cite{Finn:1992wt,Cutler:1994ys}
requires specifying the frequency range in which the analysis is performed
as well as the noise \PSD. 

Upon detection of a binary merger,
exploratory analyses are first conducted to identify which models and settings
are most suitable for use in our production analyses.
In general, we use a low frequency cutoff of $f_\mathrm{low}=20$~Hz, unless data 
quality requirements at the time of specific candidate events are different, in which
case a specific range is noted in Sec.~\ref{sec:peresults}. 
The high frequency cutoff is always equal to the Nyquist frequency of the 
analysis for each event, which is determined by the sampling rate.
The sampling rate is tailored for specific candidate events, since the signals (especially the higher-mass \ac{BBH} signals)
do not require the full bandwidth available at the native sampling rate of 16\,kHz.
The \PSD characterizing the noise at the time of each event is measured by 
BW using the same data that is used for likelihood
computation~\cite{Littenberg:2014oda,Chatziioannou:2019zvs}.
We then obtain the final joint likelihood over all detectors 
by multiplying together the likelihood from each detector in Eq.~\eqref{eqn:pe_likelihood}.

The gravitational wave signal emitted by a circularized compact binary composed 
of two black holes depends on fifteen unknown parameters, defined in Sec.~\ref{sec:mf_searches}.
The initial masses and spins of the inspiraling black holes
determine the peak gravitational wave luminosity and mass and spin of 
the post-merger remnant black hole, which we 
calculate from fits to \ac{NR}~\cite{Abbott:2017vtc,Hofmann:2016yih,Jimenez-Forteza:2016oae,Healy:2016lce,JohnsonMcDaniel:2016,Keitel:2016krm}. 
When one or both objects are neutron stars, matter effects modify the binary
inspiral and are included via the dimensionless quadrupole tidal deformability $\Lambda_i$,
adding one extra parameter for each neutron star in the binary.
Other matter effects, such as octupolar and higher tidal deformabilities,
non-black hole spin-induced multipole moments, and \textit{f}-mode resonances, are
parameterized by the $\Lambda_i$ using quasiuniversal relations~\cite{Yagi:2016bkt}.
The dominant tidal contribution to the waveform is given by the dimensionless tidal deformability parameter~\cite{Favata:2013rwa,Wade:2014vqa}
\begin{align}
    \tilde \Lambda &= \frac{16}{13} \frac{\left[(m_1 + 12m_2)m_1^4\Lambda_1 + (m_2 + 12 m_1)m_2^4\Lambda_2\right]}{(m_1 + m_2)^5}.
\end{align}
Non-spinning black holes have $\Lambda_i = 0$~\cite{Binnington:2009bb,Damour:2009vw},
and the waveform models we use adopt the convention that this is true for all black holes~\cite{Landry:2015zfa,Pani:2015hfa}.
We do not calculate final masses and spins when matter effects are included
in the analyses, as this requires an accurate prediction of the ejected mass
following possible tidal disruption, taking into account the equation
of state and uncertain details of the dynamics of the merger.

\begin{PE_table}
\begin{table*}
  \centering

\input{combined_waveform_table_new.tex}

        \caption[]{Waveform models used in this paper. We indicate
          which multipoles are included for each model. For precessing
          models, the multipoles correspond to those in the
          co-precessing frame. The Combined key column specifies which
          results generated with these waveforms are combined in our data
          release under a common key. The models below the horizontal
          line include matter effects.  \\ $^{*}$ For these datasets we
          enforce $|\vec{\chi}_{i}|=0$. \\ $^{\dagger}$ The data
          release contains versions of these keys with
          HS and LS in the name, which correspond to the high spin
          ($|\vec{\chi}_{i}|\leq0.89$) and low spin
          ($|\vec{\chi}_{i}|\leq0.05$) priors respectively. }

    \label{tab:combined_waveforms}
\end{table*}
\end{PE_table}
\subsection{Waveform models}\label{ss:waveforms}
We characterize the detected binaries using multiple waveform models, each
of which uses a different set of modeling techniques 
and includes different physical effects.  For every event
with inferred component masses above $3\,\Msun$ in preliminary analyses,
we perform production parameter estimation runs using a subset of BBH waveforms. 
IMRPhenomPv2~\cite{Hannam:2013oca,Khan:2015jqa,Husa:2015iqa} is a phenomenological model for gravitational waves 
from precessing \ac{BBH} systems, calibrated to \ac{NR} and using an 
effective single-spin description to model effects from spin-precession\cite{Schmidt:2014iyl}.
SEOBNRv4P~\cite{Ossokine:2020kjp,Bohe:2016gbl} is
based on the effective-one-body (EOB) formalism~\cite{Buonanno:1998gg,Buonanno:2000ef} and calibrated to NR, 
with a generic two-spin treatment of the precession dynamics.
These models rely on twisting up procedures,
where aligned-spin, NR-calibrated waveform models
defined in the co-precessing frame are mapped (through a suitable frame rotation)
to approximate the multipoles of a precessing system in the inertial frame~\cite{Schmidt:2010it,Boyle:2011gg,OShaughnessy:2011pmr,Ochsner:2012dj,Schmidt:2012rh}.
These models do not include contributions to the strain
from spherical harmonic modes beyond $\ell=2$, so we also
analyze each event with at least one of the following
models that incorporate higher-order multipole (HM) moments and precession effects:
IMRPhenomPv3HM~\cite{Khan:2019kot,Khan:2018fmp},
SEOBNRv4PHM~\cite{Ossokine:2020kjp,Babak:2016tgq}
and NRSur7dq4~\cite{Varma:2019csw}.
IMRPhenomPv3HM (based on IMRPhenomHM~\cite{London:2017bcn}) and SEOBNRv4PHM 
(based on SEOBNRv4HM~\cite{Cotesta:2018fcv}) both rely on the twisting-up
approach described above.
NRSur7dq4 is a surrogate waveform model for \ac{BBH} systems that directly 
interpolates a large set of precessing \ac{NR} simulations. Unlike the other two HM models, 
NRSur7dq4 waveforms are restricted by the length of the \ac{NR} simulations in the training set, 
covering only $\sim 20$ orbits before merger.

Any sources with evidence for at least one binary component below $3\,\Msun$
are characterized using several waveforms
capable of modeling matter effects. For the \ac{BNS} system \NAME{GW190425A}\
\cite{Abbott:2020uma} we use the following: IMRPhenomD\_NRTidal and
IMRPhenomPv2\_NRTidal~\cite{Dietrich:2018uni,Dietrich:2017aum}, which are  based
on the \ac{BBH} models IMRPhenomD  and IMRPhenomPv2 respectively,  and
incorporate \ac{NR} and tidal EOB-tuned contributions from tidal interaction as well
as equation-of-state dependent self-spin effects;
TaylorF2~\cite{Sathyaprakash:1991mt, Blanchet:1995ez, Poisson:1997ha, Damour:2001bu, Mikoczi:2005dn, Blanchet:2005tk,
Arun:2008kb,Buonanno:2009zt,Bohe:2013cla,Bohe:2015ana,Mishra:2016whh},
which describes waveforms from the inspiral of non-precessing compact binaries,
with matter effects derived in the post-Newtonian formalism, including
quadrupole-monopole coupling parameterized in terms of the tidal
deformabilities~\cite{Damour:2012yf,Maselli:2013mva,Yagi:2016bkt};
TEOBResumS~\cite{Bernuzzi:2014owa}, an aligned-spin EOB model
that incorporates post-Newtonian and self-force contributions to the tidal
potential;  and finally a frequency-domain surrogate model of aligned-spin
SEOBNRv4T
waveforms~\cite{Lackey:2018zvw,Hinderer:2016eia,Steinhoff:2016rfi,Bohe:2016gbl},
which were derived in the EOB approach and include dynamical
tides.

For potential NSBH sources with $m_1>3\,\Msun>m_2$, we use both BBH waveforms and
the NSBH-specific aligned-spin
waveform models SEOBNRv4\_ROM\_NRTidalv2\_NSBH~\cite{Matas:2020wab}
and IMRPhenomNSBH~\cite{Thompson:2020nei} which feature the dominant
quadrupole modes. SEOBNRv4\_ROM\_NRTidalv2\_NSBH uses SEOBNRv4\_ROM as
the \ac{BBH} baseline, while IMRPhenomNSBH employs phase evolution from
IMRPhenomD~\cite{Khan:2015jqa,Husa:2015iqa} and amplitude from IMRPhenomC~\cite{Santamaria:2010yb}.
Both models contain a phenomenological description of the tidal effects tuned
to \ac{NR} simulations~\cite{Dietrich:2019kaq} and include corrections to the
amplitude through inspiral, merger, and ringdown to account for a
possibility of tidal disruption.

To account for the systematic uncertainties in the waveform models, 
we combine equal numbers of posterior samples (described in Sec.~\ref{ss:sampling}) 
from all parameter estimation
runs for an event that use waveforms
with comparable physics. This treats
the constituent waveform models in the combined
results as having equal weight
rather than weighting them by marginal likelihood
as suggested in \cite{Ashton:2019leq}.
Table~\ref{tab:combined_waveforms} 
shows the waveforms employed in this work, the keys under which we 
group results using these waveforms, and descriptions of the 
physical effects incorporated in the models. 

\subsection{Sampling methods}\label{ss:sampling}
We use several methods to draw samples from the posterior distributions
on source parameters using the models described above.
The \LALINFERENCE~\cite{Veitch:2014wba} package was used
for most analyses presented in this paper. This package provides two independent stochastic 
sampling algorithms: a MCMC algorithm and a nested 
sampling~\cite{skilling2006} algorithm.  We employ \LALINFERENCE's nested sampling algorithm
for most of the \ac{BBH} analyses performed with the IMRPhenomD and IMRPhenomPv2
waveforms, and the MCMC algorithm for those performed with SEOBNRv4P. 
However, the serial nature of these methods makes them unsuitable for use with some of
the more computationally costly waveform models, such as waveforms with HMs and 
precession effects,  especially for long-duration
signals. For these, we also use RIFT, which performs a hybrid exploration of the parameter space 
split into intrinsic and extrinsic parameters~\cite{Pankow:2015cra,Lange:2017wki,Wysocki:2019grj}, and Parallel Bilby,
based on a distributed implementation of nested sampling~\cite{Smith:2019ucc,Ashton:2018jfp,Romero-Shaw:2020owr,Speagle:2020},
which was also used in previously published analyses of \NAME{GW190412A}~\cite{GW190412},
\NAME{GW190425A}~\cite{Abbott:2020uma}, and \NAME{GW190814A}~\cite{GW190814A}. The raw 
posterior samples from the analyses described above
are then collated to a common format using the PESummary
package~\cite{Hoy:2020vys}.

\subsection{Priors} \label{ss:priors}
Each event is analysed independently using a prior distribution on the source parameters
that is chosen to ensure adequate sampling of the parameter space and simplicity in using the 
posterior samples for further analyses. We choose a prior that is uniform in
spin magnitudes and redshifted component masses, and isotropic in spin orientations,
sky location and binary orientation.
The prior on luminosity distance corresponds to a uniform merger
rate in the co-moving frame of the source, using a flat $\mathrm{\Lambda{}CDM}$
cosmology with Hubble constant $H_0=67.9~\mathrm{km\,s^{-1}\,Mpc^{-1}}$ and
matter density $\Omega_{\rm m}=0.3065$~\cite{Ade:2015xua}; this physically
motivated prior differs from that used in previous published results which used
a prior $\propto \DL^2$. However, our data release includes parameter estimation
samples for both the flat-in-comoving-volume and $\propto \DL^2$ priors. 
For details on the conversion see Appendix~\ref{appendix:sourceparams}.
Intrinsic source masses are computed by dividing the redshifted masses measured in the detector frame by $(1+z)$,
where $z$ is calculated using the same cosmological model.

For the LALInference and Parallel Bilby analyses, we 
marginalize over uncertainty in the 
strain calibration.  The calibration errors in amplitude
and phase are described by frequency-dependent splines, whose coefficients are
allowed to vary alongside signal parameters in the inference. The prior
distribution on the calibration error at each spline node
is set by the measured uncertainty at each node~\cite{TheLIGOScientific:2016wfe}.

%% file: combined_waveform_table_new.tex
 \begin{tabularx}{\textwidth}{@{\extracolsep{\fill}}llccK}
 \textbf{Combined key} & \textbf{Waveform name}
& \textbf{Precession} & \textbf{Multipoles $(\ell,\,|m|)$} & \textbf{Ref.}\\
\hline
\rowcolor{lightgray}
\multirow{1}{*}{ZeroSpinIMR}$^{*}$ & IMRPhenomD & $\times$ & (2, 2) &
\cite{Husa:2015iqa, Khan:2015jqa}
\\[4pt]
\multirow{1}{*}{AlignedSpinIMR} &
SEOBNRv4\_ROM  &  $\times$ &  (2, 2) &
\cite{Bohe:2016gbl}
\\[4pt]

\rowcolor{lightgray}
& IMRPhenomHM  &  $\times$ & (2, 2), (2, 1), (3, 3), (3, 2), (4,
4), (4, 3) & \cite{London:2017bcn} \\
\rowcolor{lightgray}
\multirow{-2}{*}{AlignedSpinIMRHM} &
 SEOBNRv4HM\_ROM  & $\times$ &  (2, 2), (2, 1), (3, 3), (4, 4), (5,
5) &
\cite{Cotesta:2018fcv, Cotesta:2020qhw} 
\\[4pt]

\multirow{2}{*}{PrecessingSpinIMR} & 
SEOBNRv4P   & \checkmark &  (2, 2), (2, 1) & \cite{Ossokine:2020kjp,Babak:2016tgq,Pan:2013rra}\\
 &  IMRPhenomPv2   &
    \checkmark & (2, 2) & \cite{Hannam:2013oca, Khan:2018fmp} 
  \\[4pt]
 
\rowcolor{lightgray}
 & IMRPhenomPv3HM &    \checkmark & (2, 2), (2, 1), (3, 3), (3, 2),
(4, 4), (4, 3) & \cite{Khan:2019kot}\\
\rowcolor{lightgray}
& NRSur7dq4 & \checkmark & $\ell \leq 4$&
\cite{Varma:2019csw}\\ 
\rowcolor{lightgray}
 
\multirow{-3}{*}{PrecessingSpinIMRHM} & 
 SEOBNRv4PHM    &  \checkmark & (2, 2), (2, 1), (3, 3), (4, 4), (5,
5)& \cite{Ossokine:2020kjp,Babak:2016tgq,Pan:2013rra}\\
\hline
\multirow{4}{*}{AlignedSpinTidal$^{\dagger}$}&
 IMRPhenomD\_NRTidal  & $\times$ & (2, 2) & \cite{Dietrich:2018uni,Dietrich:2017aum} \\
& TEOBResumS  &$\times$ & (2, 2) & \cite{Bernuzzi:2014owa}\\
& SEOBNRv4T\_surrogate  & $\times$ & (2, 2) & \cite{Lackey:2018zvw,Hinderer:2016eia,Steinhoff:2016rfi,Bohe:2016gbl} \\
\rowcolor{lightgray}
\multirow{1}{*} {PrecessingSpinIMRTidal$^{\dagger}$} &
IMRPhenomP\_NRTidal   & \checkmark & (2, 2) & \cite{Dietrich:2018uni,Dietrich:2017aum} \\

\multirow{1}{*} {AlignedSpinInspiralTidal$^{\dagger}$} 
& TaylorF2  & $\times$ & (2, 2) & \cite{Sathyaprakash:1991mt, Blanchet:1995ez, Poisson:1997ha, Damour:2001bu, Mikoczi:2005dn, Blanchet:2005tk,
Arun:2008kb,Buonanno:2009zt,Bohe:2013cla,Bohe:2015ana,Mishra:2016whh}
\\[4pt]
\rowcolor{lightgray}
& SEOBNRv4\_ROM\_NRTidalv2\_NSBH  & $\times$ & (2, 2) & \cite{Matas:2020wab} \\
\rowcolor{lightgray}
\multirow{-2}{*}{{\footnotesize AlignedSpinIMRTidal\_NSBH}} &
IMRPhenomNSBH &   $\times$ & (2, 2) & \cite{Thompson:2020nei} \\
\hline
 \end{tabularx}

%% file: candidate_list.tex
\section{Candidate event list \note{(Becca, Liting, Ryan, ???)}}
\label{s:candidates}
\resetlinenumber

Table~\ref{tab:events} presents the results from each of \CWB{}, \GSTLAL{}, and
\PYCBC{} passing a FAR threshold of \FARTHRESHYR~yr$^{-1}$; the
full gravitational wave name encodes the UTC date with the time of
the event given after the underscore.  \NAME{GW190521A}~\cite{GW190521Adiscovery, GW190521Aastro},
\NAME{GW190425A}~\cite{Abbott:2020uma}, \NAME{GW190412A}~\cite{GW190412}, and
\NAME{GW190814A}~\cite{GW190814A} were published previously and these names are used
here verbatim.
 The
\FARTHRESHYR~yr$^{-1}$ threshold was chosen to be higher (more permissive) than
the threshold used for public alerts,
1.2~yr$^{-1}$,\footnote{\protect\url{https://emfollow.docs.ligo.org/userguide/analysis/index.html\#alert-threshold}}
but sufficiently low to provide an expected contamination fraction below 10\%.  An
extended candidate event list, which may contain marginally significant triggers, will
be provided later once final data calibration and quality checks are available.
Unlike GWTC-1~\cite{LIGOScientific:2018mvr}, a separate $p_{\text{astro}}$
threshold was not applied, however, $p_{\text{astro}}$ is greater
than $50\%$ for all candidate events for which $p_{\text{astro}}$ was calculated in
this work, satisfying the same criteria as GWTC-1.\footnote{Had we only applied a \FARTHRESHYR~yr$^{-1}$ \ac{FAR} threshold in GWTC-1 and not a $p_{\mathrm{astro}}$ threshold, one additional marginal candidate event from GWTC-1 would have passed the criteria we use here: 170616.} Among the \NUMEVENTS{}
reported candidate events passing the FAR threshold of \FARTHRESHYR~yr$^{-1}$,
\NUMCWB{} were detected by \CWB{}, \NUMGSTLAL{} candidate events were detected
by \GSTLAL{}, and \NUMPYCBC{} candidate events were detected by \PYCBC{};
\TWOPIPES{} candidate events were recovered by at least two pipelines.  Given
the FAR threshold and number of candidate events detected, the expectation value for noise events is $<4$
in this list.  Based on the FAR or the
probability of being a signal described in Sec.~\ref{ss:pastro},
\NAME{GW190426A}, \NAME{GW190719A}, and \NAME{GW190909A}\ are the most likely
to be noise among the candidate event list.  

\CWB{} recovered fewer candidate events than either \GSTLAL{} or \PYCBC{}. This is
expected because \CWB{} has highest sensitivity for short
duration, high mass signals, and its sensitivity decreases for lower mass
systems with longer duration.  \CWB{} also required candidate events to be found in
coincidence between at least two detectors.

The difference in candidate event recovery between \GSTLAL{} and \PYCBC{} is
primarily due to \PYCBC{} analyzing only times when both LIGO
Hanford and LIGO Livingston were operating and requiring signals to be observed
in both.  \PYCBC{} did not analyze Virgo data due to the fact that the code
version used for this catalog had not fully integrated 3-detector analysis
including Virgo.  \GSTLAL{} analyzed LIGO and Virgo data and allowed for the
detection of candidate events from one, two, or three gravitational wave detectors.
These algorithmic choices account for the \GSTLAL{}-only detection of
\NAME{GW190424A}, \NAME{GW190425A}, \NAME{GW190620A}, \NAME{GW190708A}\, and
\NAME{GW190910A}, which were detected above the required \ac{SNR} threshold only in
the LIGO Livingston detector, and for \NAME{GW190630A}, \NAME{GW190701A}, and
\NAME{GW190814A}, where the inclusion of Virgo was essential for determining
event significance.  The difference in candidate event recovery between \PYCBC{} and
\GSTLAL{} is also consistent with the results of the simulation presented in
Sec.~\ref{sec:simulation}.  After accounting for differences in analyzed data,
the \GSTLAL\ and \PYCBC\ methods detect a comparable number of candidate events. 

The remaining differences between the candidate event lists from \GSTLAL{}
(\NAME{GW190426A}, \NAME{GW190527A}, \NAME{GW190909A}, and \NAME{GW190929A}) and
\PYCBC{} (\NAME{GW190413A}, \NAME{GW190514A}, and \NAME{GW190719A}) arise from
the low \ac{SNR} of each event (\ac{SNR} $\lesssim 10$). Small
fluctuations in \ac{SNR} caused by different \PSD estimation and data
segmentation between pipelines leads to differences in significance estimation.
This list of low \ac{SNR} candidate events, which were only identified by one
pipeline, also contains the three candidate events with the highest minimum
\ac{FAR} among the pipelines (\NAME{GW190426A}, \NAME{GW190719A},
and \NAME{GW190909A}) which have the highest likelihood among the full candidate event
list of being caused by noise.

\begin{event_table}
\begin{table*}
\input{event_table}
\caption{
\label{tab:events} Gravitational wave candidate event list.  We find \NUMEVENTS{}
candidate events passing the \ac{FAR} threshold of \FARTHRESHYR~yr$^{-1}$ in at least
one of the four searches. Except for previously published events, the gravitational wave name encodes the UTC date with
the time of the event given after the underscore. Bold-faced names indicate the events that were not previously reported. 
The second column denotes the observing instruments.
For each of the four pipelines, \CWB, \GSTLAL, \PYCBC, and \PYCBC~BBH, we
provide the \ac{FAR} and network \ac{SNR}. Of the \NUMEVENTS{} candidate events, the
\NUMSNGLS{} that were found above the required \ac{SNR} threshold in only one
of the gravitational wave detectors are denoted by a dagger ($^\dagger$).  For
candidate events found above threshold in only one detector (single-detector
candidate events), the \ac{FAR} estimate involves extrapolation.  All single-detector
candidate events in this list by definition are rarer than the background data
collected in this analysis.  Therefore, a conservative bound on the \ac{FAR}
for triggers denoted by $^\dagger$ is $\sim 2~$yr$^{-1}$.
\protect\NAME{GW190521A}, \protect\NAME{GW190602A}, \protect\NAME{GW190701A},
and \protect\NAME{GW190706A}\ were identified by the \CWB{} high-mass search as
described in Sec.~\ref{ss:cwbmethods}.  \GSTLAL{} \acp{FAR} have been capped at
$1\times10^{-5}$~yr$^{-1}$ to be consistent with the limiting \acp{FAR} from other
pipelines. Dashes indicate that a pipeline did not find the event below the
specified \FARTHRESHYR~yr$^{-1}$ threshold.  Blank entries indicate that the
data were not searched by a pipeline.  The probability that an event is
astrophysical in origin as described in Sec.~\ref{ss:pastro} is indicated in
the column $p_{\mathrm{astro}}$.  $^*$~\PYCBC{} and \CWB{} \acp{SNR} do not include
Virgo. S190510g, S190718y, S190901ap, S190910d, S190910h, S190923y, and
S190930t~\cite{GCN24442,GCN25087,GCN25606,GCN25695,GCN25707,GCN25814,GCN25876}
were candidate events disseminated via GCNs which are not recovered here. The
SNR for \protect\NAME{GW190814A}\ (22.2) differs from the previously published
value~\cite{GW190814A} because the non-observing-mode LIGO Hanford data were not
analyzed in this work. The \ac{FAR} of \protect\NAME{GW190425A}\ differs from~\cite{Abbott:2020uma} due
to different background data and pipeline configuration.  The \ac{SNR} of \protect\NAME{GW190521A}\ differs from~\cite{GW190521Adiscovery} due to the inclusion of sub-threshold Virgo \ac{SNR}.
}
\end{table*}
\end{event_table}

Since 2 April 2019 20:00 UTC, the \LVC produced automated, public preliminary
GCN Notices for gravitational wave candidate events appearing in two or more
interferometers with \acp{FAR} less than 6 per year before a multiple analysis
trials factor was applied resulting in an effective threshold of 1.2~yr$^{-1}$~\cite{GCN24045}. On 11 June 2019, this was extended to
include gravitational wave candidate events appearing in only one interferometer and
satisfying the same \ac{FAR} threshold.  During O3a, \NUMOPA{} candidate events were
disseminated as plausible astrophysical signals;  \NUMOPANOTFOUND{} were not
recovered above the threshold considered in this work.\footnote{\url{https://gracedb.ligo.org/superevents/public/O3/}}

S190510g, S190718y, S190901ap, S190910d, S190910h, S190923y, and
S190930t~\cite{GCN24442,GCN25087,GCN25606,GCN25695,GCN25707,GCN25814,GCN25876}
are the \NUMOPANOTFOUND{} candidate events disseminated via GCNs which are not
recovered here.  S190718y, S190901ap, S190910h, and S190930t were initially
identified as single-detector candidate events (with an \ac{SNR} above threshold in
only one detector) with \acp{FAR} of 1.14~yr$^{-1}$, 0.22~yr$^{-1}$,
1.14~yr$^{-1}$, and 0.47~yr$^{-1}$, respectively.  Relaxing the demand for
coincident observation across interferometers allows \LVC analyses to report on
additional astrophysically interesting candidate events~\cite{GCN24168,G298048}, but
also removes a powerful check on the search background and leads to larger
uncertainties in the FAR. All public alerts were subsequently followed-up in
low-latency to assess whether the analysis pipelines and detectors were
operating as expected.  The low-latency followup of S190718y, S190901ap,
S190910h, and S190930t did not uncover any reason to retract these candidate
events based on data quality. However, after offline re-analysis with
additional background statistics, these four single detector candidate events are no
longer significant enough to merit inclusion in Table~\ref{tab:events}.

The remaining three public alert candidate events not recovered here---S190510g,
S190910d, and S190923y---were found in coincidence in low-latency, albeit at
modest significance.  S190510g was found in low-latency by \GSTLAL{} and assigned a
\ac{FAR} of 0.28~yr$^{-1}$. There were initially data quality concerns with
S190510g~\cite{GCN24462} and offline follow-up using an additional 24 hours of
background collection revealed the candidate event to be less significant than
originally estimated~\cite{GCN24489}.  Comparison to the full O3a background
corroborates that the candidate event no longer passes the \ac{FAR} threshold of
\FARTHRESHYR{}~yr$^{-1}$.  S190910d was identified in low-latency by the SPIIR compact
binary search pipeline~\cite{chu2017low} with a \ac{FAR} of 0.12~yr$^{-1}$.
The compact binary search pipeline MBTAOnline also
recorded a low significance candidate event at this time.  However, the candidate event was
not observed in low-latency or offline by \GSTLAL{}, \PYCBC{}\, or \CWB{} as significant.
Presently, MBTAOnline and SPIIR are configured to run in low-latency only.
S190923y was reported in low-latency by \PYCBC{} and assigned a \ac{FAR} of
1.51~yr$^{-1}$.  \GSTLAL{} and MBTAOnline also recorded low significance
candidate events at this time.  No pipeline retains this candidate event in the offline
analysis below the threshold of \FARTHRESHYR{}~yr$^{-1}$ and it is therefore
excluded from Table~\ref{tab:events}. These and other sub-threshold events will
be explored further in a future publication.

The remaining \PREVIOUSLYREPORTED{} public alerts are recovered in our offline
analysis and included in the candidate event list presented in
Table~\ref{tab:events}.  The table also includes \NEWEVENTS{} gravitational
wave detections not previously reported.  Four of these detections,
\NAME{GW190424A}, \NAME{GW190620A}, \NAME{GW190708A}\, and \NAME{GW190910A},
had detection-level \ac{SNR} ($\geq 4$) in only one of the two LIGO detectors,
making them single-detector observations from the perspective of candidate event
significance.  The incorporation of iDQ data quality into event ranking,
combined with tuning of the signal consistency tests to further reject the O3a
glitch background, improved the sensitivity of the offline \GSTLAL{} analysis to
single-detector candidate events compared to the low-latency configuration, which accounts for
these new discoveries.  Nine new detections were observed in two or more
interferometers and appear for the first time in Table \ref{tab:events}:
\NAME{GW190413A}, \NAME{GW190413B}, \NAME{GW190514A}, \NAME{GW190527A},
\NAME{GW190719A}, \NAME{GW190731A}, \NAME{GW190803A}, \NAME{GW190909A}, and
\NAME{GW190929A}.  Several of these detections were observed in low-latency,
but did not meet the criteria for public release. They all exhibit moderate
network SNRs ($\lesssim 10$).  The offline analyses here differ from their
low-latency counterparts through having improved template banks, improved use of
data quality information, improved data calibration, data cleaning, and
improved tuning to reject the non-stationary noise background observed in O3a.
These differences account for the new moderate \ac{SNR} candidate events. 

During O3a, the \LVC issued \RETRACTIONS{} retractions for public alerts that
were promptly determined to be unlikely to have originated from astrophysical
systems. The LIGO Livingston detector was more problematic for low-latency
analyses than in previous observing runs with the rate of noise glitches being
significantly higher than in O2 (see Sec.~\ref{ss:DQdiscussion}).  Low-latency
detection, especially for candidate events originating in only one interferometer, was
especially challenging during O3a.  

S190405ar was the first retraction~\cite{GCN24109}.  This Notice was
distributed in error and was never considered to be of astrophysical origin
because the event's \ac{FAR} was significantly above threshold at
6800~yr$^{-1}$.  The remaining retractions were caused by severely
non-stationary noise.  A glitch in LIGO Hanford data led to the identification
and subsequent retraction of S190518bb~\cite{GCN24591}.  S190524q, S190808ae,
S190816i, S190822c, S190829u, and S190928c all exhibited extreme non-stationary
noise in the LIGO Livingston detector~\cite{GCN24656, GCN25301, GCN25367,
GCN25442, GCN25554, GCN25883}. S190822c was assigned a significant FAR at
identification but was only observed in a single interferometer. As previously
mentioned, single detector triggers are subject to higher uncertainties in FAR.
Out of these triggers, an automatic gating method deployed later in the
observing run used by searches in order to mitigate non-stationary noise in the
detectors would have vetoed S190822c. However, at the time, this automatic
gating was not being employed. FARs calculated by the pipelines are correct if
the data collected so far is representative of the future data. Online
pipelines assign FARs using previously collected data, while offline analyses
have access to asynchronous background. If a new non-stationary noise source
emerges in the data during an observing run, it is possible for it to be
misidentified as a candidate event. Although this can also affect candidates
observed in multiple interferometers, single detector candidates are especially
susceptible to novel noise sources seen for the first time in low-latency.  In
some cases where the impact of glitches was unknown, follow-up analyses were
performed to remove instrumental artifacts and reassess the candidate event
significance. After follow-up, none of these \RETRACTIONS{} retracted candidate events
remained significant. Subsequent offline analyses did not identify them as
significant either.

%% file: event_table.tex
\begin{tabularx}{\textwidth}{l @{\extracolsep{\fill}} r l l l l l l l l l l l}
\textbf{Name} & \textbf{Inst.} & \multicolumn{2}{c}{\textbf{cWB}} & \multicolumn{3}{c}{\textbf{GstLAL}} & \multicolumn{3}{c}{\textbf{PyCBC}} & \multicolumn{3}{c}{\textbf{PyCBC BBH}} \\
               & & FAR ($\mathrm{yr}^{-1}$) & SNR$^*$ & FAR ($\mathrm{yr}^{-1}$) & SNR & $p_{\text{astro}}$ & FAR ($\mathrm{yr}^{-1}$) & SNR$^*$ & $p_{\text{astro}}$ & FAR ($\mathrm{yr}^{-1}$) & SNR$^*$ & $p_{\text{astro}}$ \\
\hline
\makebox[0pt][l]{\fboxsep0pt\colorbox{lightgray}{\mystrut\hspace*{1.0\linewidth}}}{\PUBLIC{GW190408A} \NAME{GW190408A}} & \OBSERVINGINSTRUMENTS{GW190408A} & \CWBALLSKYFAR{GW190408A} &\CWBALLSKYSNR{GW190408A} & \GSTLALALLSKYFAR{GW190408A} & \GSTLALALLSKYSNR{GW190408A} & \GSTLALALLSKYPASTRO{GW190408A} & \PYCBCALLSKYFAR{GW190408A} & \PYCBCALLSKYSNR{GW190408A} & \PYCBCALLSKYPASTRO{GW190408A} & \PYCBCHIGHMASSFAR{GW190408A} & \PYCBCHIGHMASSSNR{GW190408A} & \PYCBCHIGHMASSPASTRO{GW190408A} \\
{\PUBLIC{GW190412A} \NAME{GW190412A}} & \OBSERVINGINSTRUMENTS{GW190412A} & \CWBALLSKYFAR{GW190412A} &\CWBALLSKYSNR{GW190412A} & \GSTLALALLSKYFAR{GW190412A} & \GSTLALALLSKYSNR{GW190412A} & \GSTLALALLSKYPASTRO{GW190412A} & \PYCBCALLSKYFAR{GW190412A} & \PYCBCALLSKYSNR{GW190412A} & \PYCBCALLSKYPASTRO{GW190412A} & \PYCBCHIGHMASSFAR{GW190412A} & \PYCBCHIGHMASSSNR{GW190412A} & \PYCBCHIGHMASSPASTRO{GW190412A} \\
\makebox[0pt][l]{\fboxsep0pt\colorbox{lightgray}{\mystrut\hspace*{1.0\linewidth}}}{\PUBLIC{GW190413A} \NAME{GW190413A}} & \OBSERVINGINSTRUMENTS{GW190413A} & \CWBALLSKYFAR{GW190413A} &\CWBALLSKYSNR{GW190413A} & \GSTLALALLSKYFAR{GW190413A} & \GSTLALALLSKYSNR{GW190413A} & \GSTLALALLSKYPASTRO{GW190413A} & \PYCBCALLSKYFAR{GW190413A} & \PYCBCALLSKYSNR{GW190413A} & \PYCBCALLSKYPASTRO{GW190413A} & \PYCBCHIGHMASSFAR{GW190413A} & \PYCBCHIGHMASSSNR{GW190413A} & \PYCBCHIGHMASSPASTRO{GW190413A} \\
{\PUBLIC{GW190413B} \NAME{GW190413B}} & \OBSERVINGINSTRUMENTS{GW190413B} & \CWBALLSKYFAR{GW190413B} &\CWBALLSKYSNR{GW190413B} & \GSTLALALLSKYFAR{GW190413B} & \GSTLALALLSKYSNR{GW190413B} & \GSTLALALLSKYPASTRO{GW190413B} & \PYCBCALLSKYFAR{GW190413B} & \PYCBCALLSKYSNR{GW190413B} & \PYCBCALLSKYPASTRO{GW190413B} & \PYCBCHIGHMASSFAR{GW190413B} & \PYCBCHIGHMASSSNR{GW190413B} & \PYCBCHIGHMASSPASTRO{GW190413B} \\
\makebox[0pt][l]{\fboxsep0pt\colorbox{lightgray}{\mystrut\hspace*{1.0\linewidth}}}{\PUBLIC{GW190421A} \NAME{GW190421A}} & \OBSERVINGINSTRUMENTS{GW190421A} & \CWBALLSKYFAR{GW190421A} &\CWBALLSKYSNR{GW190421A} & \GSTLALALLSKYFAR{GW190421A} & \GSTLALALLSKYSNR{GW190421A} & \GSTLALALLSKYPASTRO{GW190421A} & \PYCBCALLSKYFAR{GW190421A} & \PYCBCALLSKYSNR{GW190421A} & \PYCBCALLSKYPASTRO{GW190421A} & \PYCBCHIGHMASSFAR{GW190421A} & \PYCBCHIGHMASSSNR{GW190421A} & \PYCBCHIGHMASSPASTRO{GW190421A} \\
{\PUBLIC{GW190424A} \NAME{GW190424A}} & \OBSERVINGINSTRUMENTS{GW190424A} & \CWBALLSKYFAR{GW190424A} &\CWBALLSKYSNR{GW190424A} & \GSTLALALLSKYFAR{GW190424A} & \GSTLALALLSKYSNR{GW190424A} & \GSTLALALLSKYPASTRO{GW190424A} & \PYCBCALLSKYFAR{GW190424A} & \PYCBCALLSKYSNR{GW190424A} & \PYCBCALLSKYPASTRO{GW190424A} & \PYCBCHIGHMASSFAR{GW190424A} & \PYCBCHIGHMASSSNR{GW190424A} & \PYCBCHIGHMASSPASTRO{GW190424A} \\
\makebox[0pt][l]{\fboxsep0pt\colorbox{lightgray}{\mystrut\hspace*{1.0\linewidth}}}{\PUBLIC{GW190425A} \NAME{GW190425A}} & \OBSERVINGINSTRUMENTS{GW190425A} & \CWBALLSKYFAR{GW190425A} &\CWBALLSKYSNR{GW190425A} & \GSTLALALLSKYFAR{GW190425A} & \GSTLALALLSKYSNR{GW190425A} & \GSTLALALLSKYPASTRO{GW190425A} & \PYCBCALLSKYFAR{GW190425A} & \PYCBCALLSKYSNR{GW190425A} & \PYCBCALLSKYPASTRO{GW190425A} & \PYCBCHIGHMASSFAR{GW190425A} & \PYCBCHIGHMASSSNR{GW190425A} & \PYCBCHIGHMASSPASTRO{GW190425A} \\
{\PUBLIC{GW190426A} \NAME{GW190426A}} & \OBSERVINGINSTRUMENTS{GW190426A} & \CWBALLSKYFAR{GW190426A} &\CWBALLSKYSNR{GW190426A} & \GSTLALALLSKYFAR{GW190426A} & \GSTLALALLSKYSNR{GW190426A} & \GSTLALALLSKYPASTRO{GW190426A} & \PYCBCALLSKYFAR{GW190426A} & \PYCBCALLSKYSNR{GW190426A} & \PYCBCALLSKYPASTRO{GW190426A} & \PYCBCHIGHMASSFAR{GW190426A} & \PYCBCHIGHMASSSNR{GW190426A} & \PYCBCHIGHMASSPASTRO{GW190426A} \\
\makebox[0pt][l]{\fboxsep0pt\colorbox{lightgray}{\mystrut\hspace*{1.0\linewidth}}}{\PUBLIC{GW190503A} \NAME{GW190503A}} & \OBSERVINGINSTRUMENTS{GW190503A} & \CWBALLSKYFAR{GW190503A} &\CWBALLSKYSNR{GW190503A} & \GSTLALALLSKYFAR{GW190503A} & \GSTLALALLSKYSNR{GW190503A} & \GSTLALALLSKYPASTRO{GW190503A} & \PYCBCALLSKYFAR{GW190503A} & \PYCBCALLSKYSNR{GW190503A} & \PYCBCALLSKYPASTRO{GW190503A} & \PYCBCHIGHMASSFAR{GW190503A} & \PYCBCHIGHMASSSNR{GW190503A} & \PYCBCHIGHMASSPASTRO{GW190503A} \\
{\PUBLIC{GW190512A} \NAME{GW190512A}} & \OBSERVINGINSTRUMENTS{GW190512A} & \CWBALLSKYFAR{GW190512A} &\CWBALLSKYSNR{GW190512A} & \GSTLALALLSKYFAR{GW190512A} & \GSTLALALLSKYSNR{GW190512A} & \GSTLALALLSKYPASTRO{GW190512A} & \PYCBCALLSKYFAR{GW190512A} & \PYCBCALLSKYSNR{GW190512A} & \PYCBCALLSKYPASTRO{GW190512A} & \PYCBCHIGHMASSFAR{GW190512A} & \PYCBCHIGHMASSSNR{GW190512A} & \PYCBCHIGHMASSPASTRO{GW190512A} \\
\makebox[0pt][l]{\fboxsep0pt\colorbox{lightgray}{\mystrut\hspace*{1.0\linewidth}}}{\PUBLIC{GW190513A} \NAME{GW190513A}} & \OBSERVINGINSTRUMENTS{GW190513A} & \CWBALLSKYFAR{GW190513A} &\CWBALLSKYSNR{GW190513A} & \GSTLALALLSKYFAR{GW190513A} & \GSTLALALLSKYSNR{GW190513A} & \GSTLALALLSKYPASTRO{GW190513A} & \PYCBCALLSKYFAR{GW190513A} & \PYCBCALLSKYSNR{GW190513A} & \PYCBCALLSKYPASTRO{GW190513A} & \PYCBCHIGHMASSFAR{GW190513A} & \PYCBCHIGHMASSSNR{GW190513A} & \PYCBCHIGHMASSPASTRO{GW190513A} \\
{\PUBLIC{GW190514A} \NAME{GW190514A}} & \OBSERVINGINSTRUMENTS{GW190514A} & \CWBALLSKYFAR{GW190514A} &\CWBALLSKYSNR{GW190514A} & \GSTLALALLSKYFAR{GW190514A} & \GSTLALALLSKYSNR{GW190514A} & \GSTLALALLSKYPASTRO{GW190514A} & \PYCBCALLSKYFAR{GW190514A} & \PYCBCALLSKYSNR{GW190514A} & \PYCBCALLSKYPASTRO{GW190514A} & \PYCBCHIGHMASSFAR{GW190514A} & \PYCBCHIGHMASSSNR{GW190514A} & \PYCBCHIGHMASSPASTRO{GW190514A} \\
\makebox[0pt][l]{\fboxsep0pt\colorbox{lightgray}{\mystrut\hspace*{1.0\linewidth}}}{\PUBLIC{GW190517A} \NAME{GW190517A}} & \OBSERVINGINSTRUMENTS{GW190517A} & \CWBALLSKYFAR{GW190517A} &\CWBALLSKYSNR{GW190517A} & \GSTLALALLSKYFAR{GW190517A} & \GSTLALALLSKYSNR{GW190517A} & \GSTLALALLSKYPASTRO{GW190517A} & \PYCBCALLSKYFAR{GW190517A} & \PYCBCALLSKYSNR{GW190517A} & \PYCBCALLSKYPASTRO{GW190517A} & \PYCBCHIGHMASSFAR{GW190517A} & \PYCBCHIGHMASSSNR{GW190517A} & \PYCBCHIGHMASSPASTRO{GW190517A} \\
{\PUBLIC{GW190519A} \NAME{GW190519A}} & \OBSERVINGINSTRUMENTS{GW190519A} & \CWBALLSKYFAR{GW190519A} &\CWBALLSKYSNR{GW190519A} & \GSTLALALLSKYFAR{GW190519A} & \GSTLALALLSKYSNR{GW190519A} & \GSTLALALLSKYPASTRO{GW190519A} & \PYCBCALLSKYFAR{GW190519A} & \PYCBCALLSKYSNR{GW190519A} & \PYCBCALLSKYPASTRO{GW190519A} & \PYCBCHIGHMASSFAR{GW190519A} & \PYCBCHIGHMASSSNR{GW190519A} & \PYCBCHIGHMASSPASTRO{GW190519A} \\
\makebox[0pt][l]{\fboxsep0pt\colorbox{lightgray}{\mystrut\hspace*{1.0\linewidth}}}{\PUBLIC{GW190521A} \NAME{GW190521A}} & \OBSERVINGINSTRUMENTS{GW190521A} & \CWBALLSKYFAR{GW190521A} &\CWBALLSKYSNR{GW190521A} & \GSTLALALLSKYFAR{GW190521A} & \GSTLALALLSKYSNR{GW190521A} & \GSTLALALLSKYPASTRO{GW190521A} & \PYCBCALLSKYFAR{GW190521A} & \PYCBCALLSKYSNR{GW190521A} & \PYCBCALLSKYPASTRO{GW190521A} & \PYCBCHIGHMASSFAR{GW190521A} & \PYCBCHIGHMASSSNR{GW190521A} & \PYCBCHIGHMASSPASTRO{GW190521A} \\
{\PUBLIC{GW190521B} \NAME{GW190521B}} & \OBSERVINGINSTRUMENTS{GW190521B} & \CWBALLSKYFAR{GW190521B} &\CWBALLSKYSNR{GW190521B} & \GSTLALALLSKYFAR{GW190521B} & \GSTLALALLSKYSNR{GW190521B} & \GSTLALALLSKYPASTRO{GW190521B} & \PYCBCALLSKYFAR{GW190521B} & \PYCBCALLSKYSNR{GW190521B} & \PYCBCALLSKYPASTRO{GW190521B} & \PYCBCHIGHMASSFAR{GW190521B} & \PYCBCHIGHMASSSNR{GW190521B} & \PYCBCHIGHMASSPASTRO{GW190521B} \\
\makebox[0pt][l]{\fboxsep0pt\colorbox{lightgray}{\mystrut\hspace*{1.0\linewidth}}}{\PUBLIC{GW190527A} \NAME{GW190527A}} & \OBSERVINGINSTRUMENTS{GW190527A} & \CWBALLSKYFAR{GW190527A} &\CWBALLSKYSNR{GW190527A} & \GSTLALALLSKYFAR{GW190527A} & \GSTLALALLSKYSNR{GW190527A} & \GSTLALALLSKYPASTRO{GW190527A} & \PYCBCALLSKYFAR{GW190527A} & \PYCBCALLSKYSNR{GW190527A} & \PYCBCALLSKYPASTRO{GW190527A} & \PYCBCHIGHMASSFAR{GW190527A} & \PYCBCHIGHMASSSNR{GW190527A} & \PYCBCHIGHMASSPASTRO{GW190527A} \\
{\PUBLIC{GW190602A} \NAME{GW190602A}} & \OBSERVINGINSTRUMENTS{GW190602A} & \CWBALLSKYFAR{GW190602A} &\CWBALLSKYSNR{GW190602A} & \GSTLALALLSKYFAR{GW190602A} & \GSTLALALLSKYSNR{GW190602A} & \GSTLALALLSKYPASTRO{GW190602A} & \PYCBCALLSKYFAR{GW190602A} & \PYCBCALLSKYSNR{GW190602A} & \PYCBCALLSKYPASTRO{GW190602A} & \PYCBCHIGHMASSFAR{GW190602A} & \PYCBCHIGHMASSSNR{GW190602A} & \PYCBCHIGHMASSPASTRO{GW190602A} \\
\makebox[0pt][l]{\fboxsep0pt\colorbox{lightgray}{\mystrut\hspace*{1.0\linewidth}}}{\PUBLIC{GW190620A} \NAME{GW190620A}} & \OBSERVINGINSTRUMENTS{GW190620A} & \CWBALLSKYFAR{GW190620A} &\CWBALLSKYSNR{GW190620A} & \GSTLALALLSKYFAR{GW190620A} & \GSTLALALLSKYSNR{GW190620A} & \GSTLALALLSKYPASTRO{GW190620A} & \PYCBCALLSKYFAR{GW190620A} & \PYCBCALLSKYSNR{GW190620A} & \PYCBCALLSKYPASTRO{GW190620A} & \PYCBCHIGHMASSFAR{GW190620A} & \PYCBCHIGHMASSSNR{GW190620A} & \PYCBCHIGHMASSPASTRO{GW190620A} \\
{\PUBLIC{GW190630A} \NAME{GW190630A}} & \OBSERVINGINSTRUMENTS{GW190630A} & \CWBALLSKYFAR{GW190630A} &\CWBALLSKYSNR{GW190630A} & \GSTLALALLSKYFAR{GW190630A} & \GSTLALALLSKYSNR{GW190630A} & \GSTLALALLSKYPASTRO{GW190630A} & \PYCBCALLSKYFAR{GW190630A} & \PYCBCALLSKYSNR{GW190630A} & \PYCBCALLSKYPASTRO{GW190630A} & \PYCBCHIGHMASSFAR{GW190630A} & \PYCBCHIGHMASSSNR{GW190630A} & \PYCBCHIGHMASSPASTRO{GW190630A} \\
\makebox[0pt][l]{\fboxsep0pt\colorbox{lightgray}{\mystrut\hspace*{1.0\linewidth}}}{\PUBLIC{GW190701A} \NAME{GW190701A}} & \OBSERVINGINSTRUMENTS{GW190701A} & \CWBALLSKYFAR{GW190701A} &\CWBALLSKYSNR{GW190701A} & \GSTLALALLSKYFAR{GW190701A} & \GSTLALALLSKYSNR{GW190701A} & \GSTLALALLSKYPASTRO{GW190701A} & \PYCBCALLSKYFAR{GW190701A} & \PYCBCALLSKYSNR{GW190701A} & \PYCBCALLSKYPASTRO{GW190701A} & \PYCBCHIGHMASSFAR{GW190701A} & \PYCBCHIGHMASSSNR{GW190701A} & \PYCBCHIGHMASSPASTRO{GW190701A} \\
{\PUBLIC{GW190706A} \NAME{GW190706A}} & \OBSERVINGINSTRUMENTS{GW190706A} & \CWBALLSKYFAR{GW190706A} &\CWBALLSKYSNR{GW190706A} & \GSTLALALLSKYFAR{GW190706A} & \GSTLALALLSKYSNR{GW190706A} & \GSTLALALLSKYPASTRO{GW190706A} & \PYCBCALLSKYFAR{GW190706A} & \PYCBCALLSKYSNR{GW190706A} & \PYCBCALLSKYPASTRO{GW190706A} & \PYCBCHIGHMASSFAR{GW190706A} & \PYCBCHIGHMASSSNR{GW190706A} & \PYCBCHIGHMASSPASTRO{GW190706A} \\
\makebox[0pt][l]{\fboxsep0pt\colorbox{lightgray}{\mystrut\hspace*{1.0\linewidth}}}{\PUBLIC{GW190707A} \NAME{GW190707A}} & \OBSERVINGINSTRUMENTS{GW190707A} & \CWBALLSKYFAR{GW190707A} &\CWBALLSKYSNR{GW190707A} & \GSTLALALLSKYFAR{GW190707A} & \GSTLALALLSKYSNR{GW190707A} & \GSTLALALLSKYPASTRO{GW190707A} & \PYCBCALLSKYFAR{GW190707A} & \PYCBCALLSKYSNR{GW190707A} & \PYCBCALLSKYPASTRO{GW190707A} & \PYCBCHIGHMASSFAR{GW190707A} & \PYCBCHIGHMASSSNR{GW190707A} & \PYCBCHIGHMASSPASTRO{GW190707A} \\
{\PUBLIC{GW190708A} \NAME{GW190708A}} & \OBSERVINGINSTRUMENTS{GW190708A} & \CWBALLSKYFAR{GW190708A} &\CWBALLSKYSNR{GW190708A} & \GSTLALALLSKYFAR{GW190708A} & \GSTLALALLSKYSNR{GW190708A} & \GSTLALALLSKYPASTRO{GW190708A} & \PYCBCALLSKYFAR{GW190708A} & \PYCBCALLSKYSNR{GW190708A} & \PYCBCALLSKYPASTRO{GW190708A} & \PYCBCHIGHMASSFAR{GW190708A} & \PYCBCHIGHMASSSNR{GW190708A} & \PYCBCHIGHMASSPASTRO{GW190708A} \\
\makebox[0pt][l]{\fboxsep0pt\colorbox{lightgray}{\mystrut\hspace*{1.0\linewidth}}}{\PUBLIC{GW190719A} \NAME{GW190719A}} & \OBSERVINGINSTRUMENTS{GW190719A} & \CWBALLSKYFAR{GW190719A} &\CWBALLSKYSNR{GW190719A} & \GSTLALALLSKYFAR{GW190719A} & \GSTLALALLSKYSNR{GW190719A} & \GSTLALALLSKYPASTRO{GW190719A} & \PYCBCALLSKYFAR{GW190719A} & \PYCBCALLSKYSNR{GW190719A} & \PYCBCALLSKYPASTRO{GW190719A} & \PYCBCHIGHMASSFAR{GW190719A} & \PYCBCHIGHMASSSNR{GW190719A} & \PYCBCHIGHMASSPASTRO{GW190719A} \\
{\PUBLIC{GW190720A} \NAME{GW190720A}} & \OBSERVINGINSTRUMENTS{GW190720A} & \CWBALLSKYFAR{GW190720A} &\CWBALLSKYSNR{GW190720A} & \GSTLALALLSKYFAR{GW190720A} & \GSTLALALLSKYSNR{GW190720A} & \GSTLALALLSKYPASTRO{GW190720A} & \PYCBCALLSKYFAR{GW190720A} & \PYCBCALLSKYSNR{GW190720A} & \PYCBCALLSKYPASTRO{GW190720A} & \PYCBCHIGHMASSFAR{GW190720A} & \PYCBCHIGHMASSSNR{GW190720A} & \PYCBCHIGHMASSPASTRO{GW190720A} \\
\makebox[0pt][l]{\fboxsep0pt\colorbox{lightgray}{\mystrut\hspace*{1.0\linewidth}}}{\PUBLIC{GW190727A} \NAME{GW190727A}} & \OBSERVINGINSTRUMENTS{GW190727A} & \CWBALLSKYFAR{GW190727A} &\CWBALLSKYSNR{GW190727A} & \GSTLALALLSKYFAR{GW190727A} & \GSTLALALLSKYSNR{GW190727A} & \GSTLALALLSKYPASTRO{GW190727A} & \PYCBCALLSKYFAR{GW190727A} & \PYCBCALLSKYSNR{GW190727A} & \PYCBCALLSKYPASTRO{GW190727A} & \PYCBCHIGHMASSFAR{GW190727A} & \PYCBCHIGHMASSSNR{GW190727A} & \PYCBCHIGHMASSPASTRO{GW190727A} \\
{\PUBLIC{GW190728A} \NAME{GW190728A}} & \OBSERVINGINSTRUMENTS{GW190728A} & \CWBALLSKYFAR{GW190728A} &\CWBALLSKYSNR{GW190728A} & \GSTLALALLSKYFAR{GW190728A} & \GSTLALALLSKYSNR{GW190728A} & \GSTLALALLSKYPASTRO{GW190728A} & \PYCBCALLSKYFAR{GW190728A} & \PYCBCALLSKYSNR{GW190728A} & \PYCBCALLSKYPASTRO{GW190728A} & \PYCBCHIGHMASSFAR{GW190728A} & \PYCBCHIGHMASSSNR{GW190728A} & \PYCBCHIGHMASSPASTRO{GW190728A} \\
\makebox[0pt][l]{\fboxsep0pt\colorbox{lightgray}{\mystrut\hspace*{1.0\linewidth}}}{\PUBLIC{GW190731A} \NAME{GW190731A}} & \OBSERVINGINSTRUMENTS{GW190731A} & \CWBALLSKYFAR{GW190731A} &\CWBALLSKYSNR{GW190731A} & \GSTLALALLSKYFAR{GW190731A} & \GSTLALALLSKYSNR{GW190731A} & \GSTLALALLSKYPASTRO{GW190731A} & \PYCBCALLSKYFAR{GW190731A} & \PYCBCALLSKYSNR{GW190731A} & \PYCBCALLSKYPASTRO{GW190731A} & \PYCBCHIGHMASSFAR{GW190731A} & \PYCBCHIGHMASSSNR{GW190731A} & \PYCBCHIGHMASSPASTRO{GW190731A} \\
{\PUBLIC{GW190803A} \NAME{GW190803A}} & \OBSERVINGINSTRUMENTS{GW190803A} & \CWBALLSKYFAR{GW190803A} &\CWBALLSKYSNR{GW190803A} & \GSTLALALLSKYFAR{GW190803A} & \GSTLALALLSKYSNR{GW190803A} & \GSTLALALLSKYPASTRO{GW190803A} & \PYCBCALLSKYFAR{GW190803A} & \PYCBCALLSKYSNR{GW190803A} & \PYCBCALLSKYPASTRO{GW190803A} & \PYCBCHIGHMASSFAR{GW190803A} & \PYCBCHIGHMASSSNR{GW190803A} & \PYCBCHIGHMASSPASTRO{GW190803A} \\
\makebox[0pt][l]{\fboxsep0pt\colorbox{lightgray}{\mystrut\hspace*{1.0\linewidth}}}{\PUBLIC{GW190814A} \NAME{GW190814A}} & \OBSERVINGINSTRUMENTS{GW190814A} & \CWBALLSKYFAR{GW190814A} &\CWBALLSKYSNR{GW190814A} & \GSTLALALLSKYFAR{GW190814A} & \GSTLALALLSKYSNR{GW190814A} & \GSTLALALLSKYPASTRO{GW190814A} & \PYCBCALLSKYFAR{GW190814A} & \PYCBCALLSKYSNR{GW190814A} & \PYCBCALLSKYPASTRO{GW190814A} & \PYCBCHIGHMASSFAR{GW190814A} & \PYCBCHIGHMASSSNR{GW190814A} & \PYCBCHIGHMASSPASTRO{GW190814A} \\
{\PUBLIC{GW190828A} \NAME{GW190828A}} & \OBSERVINGINSTRUMENTS{GW190828A} & \CWBALLSKYFAR{GW190828A} &\CWBALLSKYSNR{GW190828A} & \GSTLALALLSKYFAR{GW190828A} & \GSTLALALLSKYSNR{GW190828A} & \GSTLALALLSKYPASTRO{GW190828A} & \PYCBCALLSKYFAR{GW190828A} & \PYCBCALLSKYSNR{GW190828A} & \PYCBCALLSKYPASTRO{GW190828A} & \PYCBCHIGHMASSFAR{GW190828A} & \PYCBCHIGHMASSSNR{GW190828A} & \PYCBCHIGHMASSPASTRO{GW190828A} \\
\makebox[0pt][l]{\fboxsep0pt\colorbox{lightgray}{\mystrut\hspace*{1.0\linewidth}}}{\PUBLIC{GW190828B} \NAME{GW190828B}} & \OBSERVINGINSTRUMENTS{GW190828B} & \CWBALLSKYFAR{GW190828B} &\CWBALLSKYSNR{GW190828B} & \GSTLALALLSKYFAR{GW190828B} & \GSTLALALLSKYSNR{GW190828B} & \GSTLALALLSKYPASTRO{GW190828B} & \PYCBCALLSKYFAR{GW190828B} & \PYCBCALLSKYSNR{GW190828B} & \PYCBCALLSKYPASTRO{GW190828B} & \PYCBCHIGHMASSFAR{GW190828B} & \PYCBCHIGHMASSSNR{GW190828B} & \PYCBCHIGHMASSPASTRO{GW190828B} \\
{\PUBLIC{GW190909A} \NAME{GW190909A}} & \OBSERVINGINSTRUMENTS{GW190909A} & \CWBALLSKYFAR{GW190909A} &\CWBALLSKYSNR{GW190909A} & \GSTLALALLSKYFAR{GW190909A} & \GSTLALALLSKYSNR{GW190909A} & \GSTLALALLSKYPASTRO{GW190909A} & \PYCBCALLSKYFAR{GW190909A} & \PYCBCALLSKYSNR{GW190909A} & \PYCBCALLSKYPASTRO{GW190909A} & \PYCBCHIGHMASSFAR{GW190909A} & \PYCBCHIGHMASSSNR{GW190909A} & \PYCBCHIGHMASSPASTRO{GW190909A} \\
\makebox[0pt][l]{\fboxsep0pt\colorbox{lightgray}{\mystrut\hspace*{1.0\linewidth}}}{\PUBLIC{GW190910A} \NAME{GW190910A}} & \OBSERVINGINSTRUMENTS{GW190910A} & \CWBALLSKYFAR{GW190910A} &\CWBALLSKYSNR{GW190910A} & \GSTLALALLSKYFAR{GW190910A} & \GSTLALALLSKYSNR{GW190910A} & \GSTLALALLSKYPASTRO{GW190910A} & \PYCBCALLSKYFAR{GW190910A} & \PYCBCALLSKYSNR{GW190910A} & \PYCBCALLSKYPASTRO{GW190910A} & \PYCBCHIGHMASSFAR{GW190910A} & \PYCBCHIGHMASSSNR{GW190910A} & \PYCBCHIGHMASSPASTRO{GW190910A} \\
{\PUBLIC{GW190915A} \NAME{GW190915A}} & \OBSERVINGINSTRUMENTS{GW190915A} & \CWBALLSKYFAR{GW190915A} &\CWBALLSKYSNR{GW190915A} & \GSTLALALLSKYFAR{GW190915A} & \GSTLALALLSKYSNR{GW190915A} & \GSTLALALLSKYPASTRO{GW190915A} & \PYCBCALLSKYFAR{GW190915A} & \PYCBCALLSKYSNR{GW190915A} & \PYCBCALLSKYPASTRO{GW190915A} & \PYCBCHIGHMASSFAR{GW190915A} & \PYCBCHIGHMASSSNR{GW190915A} & \PYCBCHIGHMASSPASTRO{GW190915A} \\
\makebox[0pt][l]{\fboxsep0pt\colorbox{lightgray}{\mystrut\hspace*{1.0\linewidth}}}{\PUBLIC{GW190924A} \NAME{GW190924A}} & \OBSERVINGINSTRUMENTS{GW190924A} & \CWBALLSKYFAR{GW190924A} &\CWBALLSKYSNR{GW190924A} & \GSTLALALLSKYFAR{GW190924A} & \GSTLALALLSKYSNR{GW190924A} & \GSTLALALLSKYPASTRO{GW190924A} & \PYCBCALLSKYFAR{GW190924A} & \PYCBCALLSKYSNR{GW190924A} & \PYCBCALLSKYPASTRO{GW190924A} & \PYCBCHIGHMASSFAR{GW190924A} & \PYCBCHIGHMASSSNR{GW190924A} & \PYCBCHIGHMASSPASTRO{GW190924A} \\
{\PUBLIC{GW190929A} \NAME{GW190929A}} & \OBSERVINGINSTRUMENTS{GW190929A} & \CWBALLSKYFAR{GW190929A} &\CWBALLSKYSNR{GW190929A} & \GSTLALALLSKYFAR{GW190929A} & \GSTLALALLSKYSNR{GW190929A} & \GSTLALALLSKYPASTRO{GW190929A} & \PYCBCALLSKYFAR{GW190929A} & \PYCBCALLSKYSNR{GW190929A} & \PYCBCALLSKYPASTRO{GW190929A} & \PYCBCHIGHMASSFAR{GW190929A} & \PYCBCHIGHMASSSNR{GW190929A} & \PYCBCHIGHMASSPASTRO{GW190929A} \\
\makebox[0pt][l]{\fboxsep0pt\colorbox{lightgray}{\mystrut\hspace*{1.0\linewidth}}}{\PUBLIC{GW190930A} \NAME{GW190930A}} & \OBSERVINGINSTRUMENTS{GW190930A} & \CWBALLSKYFAR{GW190930A} &\CWBALLSKYSNR{GW190930A} & \GSTLALALLSKYFAR{GW190930A} & \GSTLALALLSKYSNR{GW190930A} & \GSTLALALLSKYPASTRO{GW190930A} & \PYCBCALLSKYFAR{GW190930A} & \PYCBCALLSKYSNR{GW190930A} & \PYCBCALLSKYPASTRO{GW190930A} & \PYCBCHIGHMASSFAR{GW190930A} & \PYCBCHIGHMASSSNR{GW190930A} & \PYCBCHIGHMASSPASTRO{GW190930A} \\
\hline
\end{tabularx}

%% file: source_properties.tex
\section{Source properties \note{(Zoheyr, John, ???)}}
\label{sec:peresults}
\resetlinenumber

We analyze the \NUMEVENTS{} candidate events shown in Table~\ref{tab:events}
with the parameter estimation techniques
described in Sec.~\ref{sec:PEmethods}. For a subset of candidate events, event validation procedures
outlined in Sec.~\ref{ss:validation} identified transient noise
that may have impacted the results of parameter inference.
In order to minimize the effect of this transient noise,
candidate-specific procedures were explored for all
impacted candidate events.
In most cases, transient noise was mitigated through the glitch subtraction
methods outlined in Sec.~\ref{ss:denoise}.
After application of these methods,
the identified transient noise was considered mitigated
if the data surrounding the event was consistent with Gaussian noise,
as measured by the variance of the measured
power spectral density
during the time period containing the identified transient~\cite{Mozzon:2020gwa}.
If data were not Gaussian after glitch subtraction, we
evaluated the SNR lost by restricting the frequency range
of data considered in parameter inference to fully excise
the identified transient noise.
In cases where the single detector SNR loss was below 10\%,
this reduced frequency range was used in analyses.
Otherwise, the nominal frequency range was used.
The full list of candidate events using candidate-specific mitigation,
along with the mitigation configuration,
is found in Table~\ref{tab:detchar_events}.

\begin{PE_table}
\begin{table}
{\rowcolors{1}{}{lightgray}
\input{detchar_table}
}
\caption{
\label{tab:detchar_events}
List of candidate-specific data usage and mitigation methods
for parameter estimates.
Only candidate events for which mitigation of instrumental artifacts was performed are listed.
The glitch subtraction methods
used for these candidate events are detailed in Sec.~\ref{ss:denoise}.
The minimum frequency is the lower limit of data
used in analyses of gravitational wave source
properties for the listed interferometer.
}
\end{table}
\end{PE_table}

Based on the investigations described in Appendix~\ref{app:systematics}, we
find that most gravitational wave candidate events in this catalog exhibit
small changes in source parameter
estimates when spherical harmonic modes
above $\ell=2$ are included. However, these
differences in aggregate 
could still affect population-level studies, so we present as fiducial results
the combined posterior samples of HM runs for all \acp{BBH} candidate events
except \NAME{GW190707A}, \NAME{GW190720A}, \NAME{GW190728A}, \NAME{GW190915A},
\NAME{GW190924A}, and \NAME{GW190930A}. For these six
exceptions, we present combined IMRPhenomPv2--SEOBNRv4P
samples, because the effect of higher modes is 
either negligible or subdominant to the systematics
between IMRPhenomPv2 and SEOBNRv4P results, as detailed in
Appendix~\ref{app:systematics}.
\NAME{GW190412A}~\cite{GW190412}, \NAME{GW190521A}~\cite{GW190521Adiscovery,GW190521Aastro},  and
\NAME{GW190814A}~\cite{GW190814A} were analyzed extensively in separate publications with
the HM waveform families SEOBNRv4PHM, IMRPhenomPv3HM,
or NRSur7dq4, so we present HM runs
for these candidate events as fiducial results here and
defer readers to those publications for details on those candidate events.\footnote{
Where posterior samples from existing publications are used,
we still impose the uniform-in-comoving volume prior described in Sec.~\ref{ss:priors} for
consistency with the new results in this publication.}

\NAME{GW190425A}, \NAME{GW190426A}, and \NAME{GW190814A}\
showed indications of including at least one neutron
star, and so were also analyzed using tidal waveforms in addition to IMRPhenomPv2 and SEOBNRv4P,
and discussed in Sec.~\ref{ss:pe_non_bbh}.
Further details on waveform systematics and 
the waveforms employed in this work can be found 
in Appendix~\ref{app:systematics}, and the full suite 
of posterior samples is publicly available
at \cite{lvc:datadoi}.

In the following subsections, we summarize the results of
our parameter estimation analyses and highlight candidate events of
particular interest.
To identify candidate events with the most extreme parameter
values, we repeatedly select one posterior sample at random from each event and record
which candidate events have the lowest and highest values of each parameter. From these
repeated trials, we determine each event's probability of having the lowest
or highest value for a given parameter.
Table~\ref{tab:PE} shows 90\% credible
intervals on the source parameters of all \NUMEVENTS{} candidate events using
the priors described in Sec.~\ref{ss:priors} and waveforms
specified above.

\begin{figure*}[tb]
    \centering
    \includegraphics[width=0.8\textwidth]{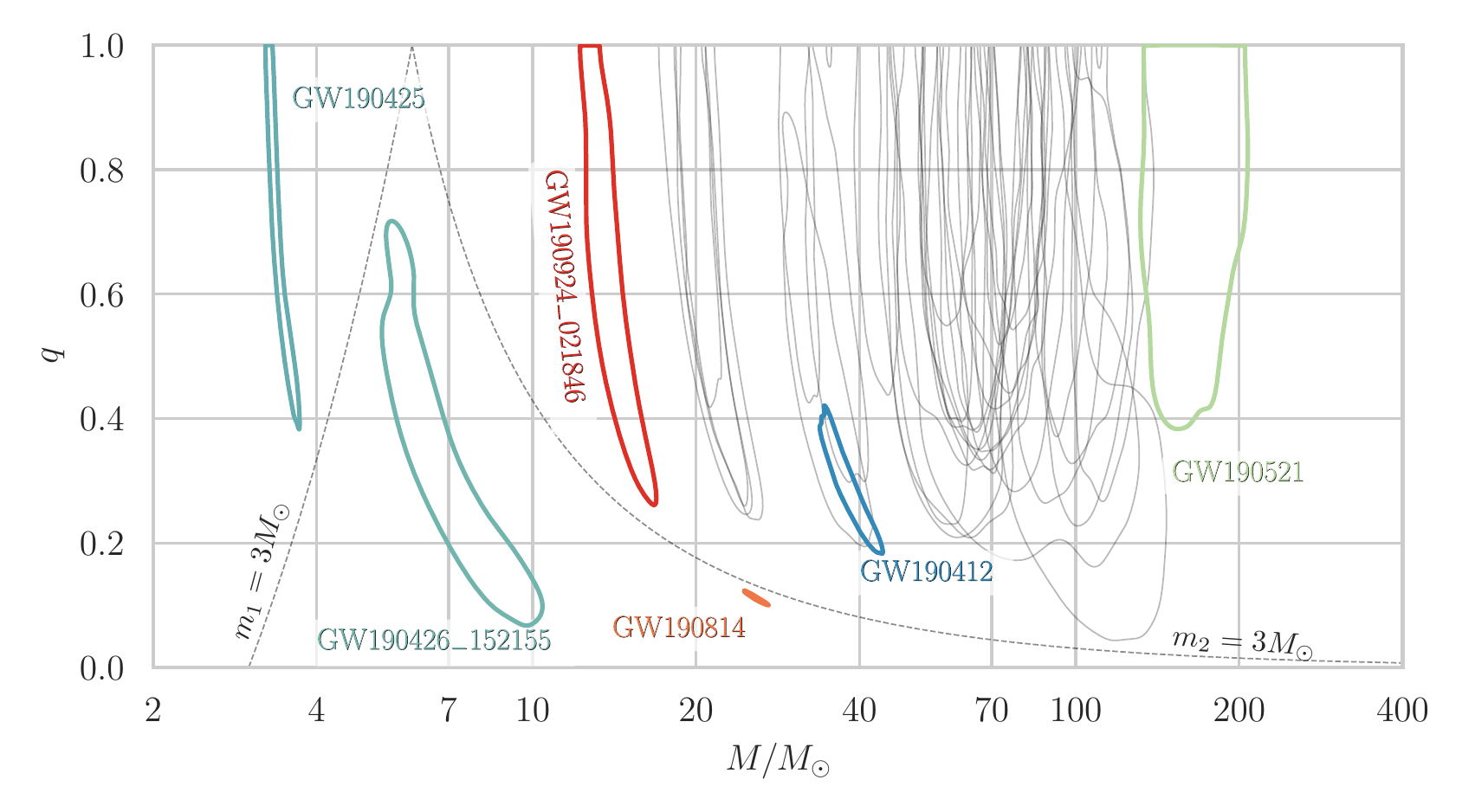}
    \caption{\label{fig:mtotqpost} Credible region contours for all
      candidate events in the plane of total mass $M$ and mass ratio $q$. Each
      contour represents the 90\% credible region for a different
      event. We highlight the previously published candidate events:
      \protect\NAME{GW190412A}, \protect\NAME{GW190425A},
      \protect\NAME{GW190521A}\ and \protect\NAME{GW190814A}, the potential NSBH \protect\NAME{GW190426A}, and finally
      \protect\NAME{GW190924A}, which is most probably the least
      massive system with both masses $>3\,\Msun$.  The dashed lines
      delineate regions where the primary/secondary can have a mass
      below $3\,\Msun$. For the region above the $m_2=3
      \,\Msun$ line, both objects in the binary have masses above
      $3\,\Msun$. }
\end{figure*}

\begin{figure*}[tb]
    \centering
    \includegraphics[width=0.8\textwidth]{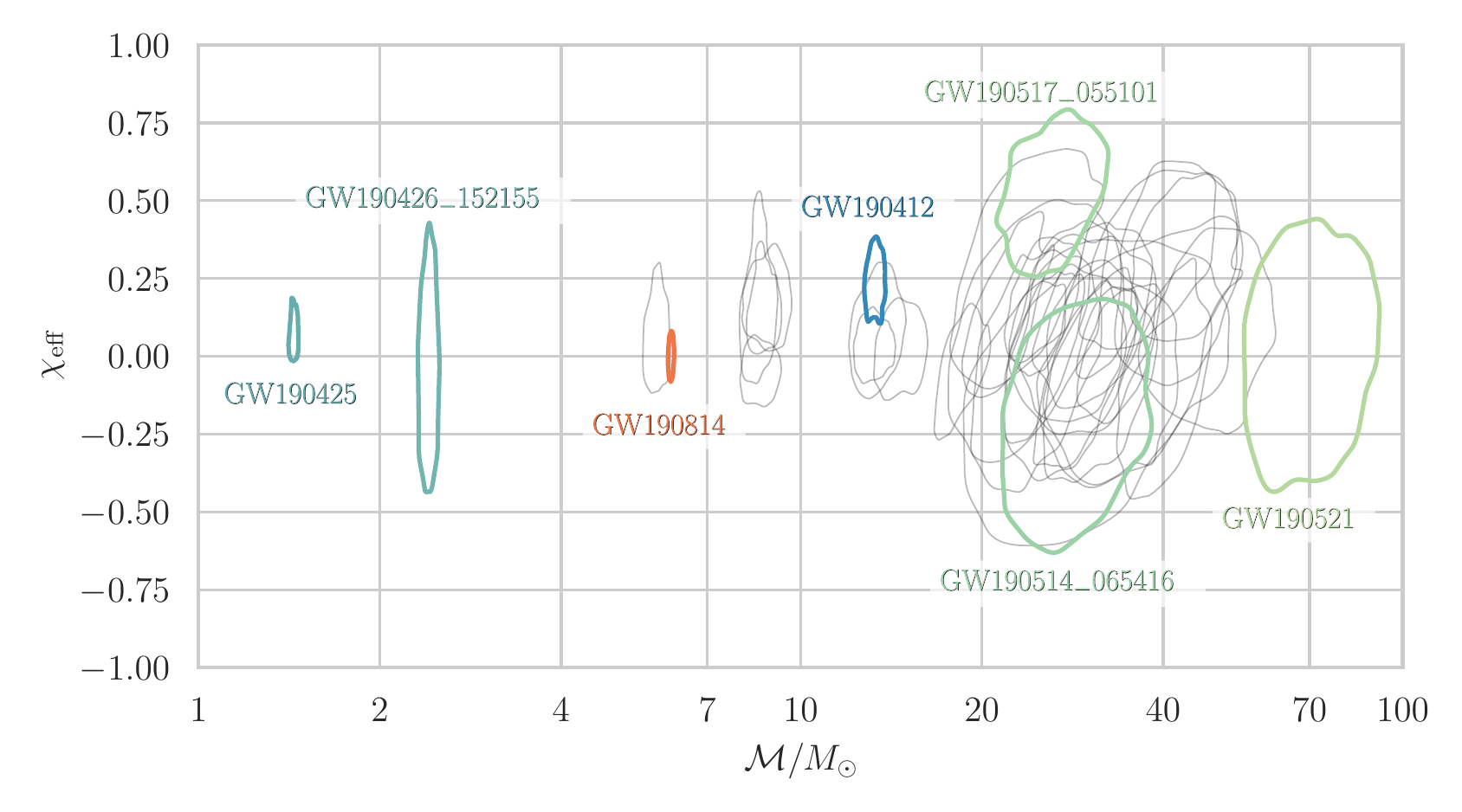}
    \caption{\label{fig:mcchieffpost} Credible region contours for all
      candidate events in the plane of chirp mass $\mathcal{M}$ and effective
      inspiral spin $\chieff$. Each contour represents the 90\%
      credible region for a different event. We highlighted the
      previously published candidate events (cf.\ Fig.~\ref{fig:mtotqpost}), as
      well as \protect\NAME{GW190517A}\ and \protect\NAME{GW190514A},
      which have the highest probabilities of having the largest and
      smallest $\chieff$ respectively.}
\end{figure*}

\begin{figure*}[t]
    \includegraphics[width=\textwidth]{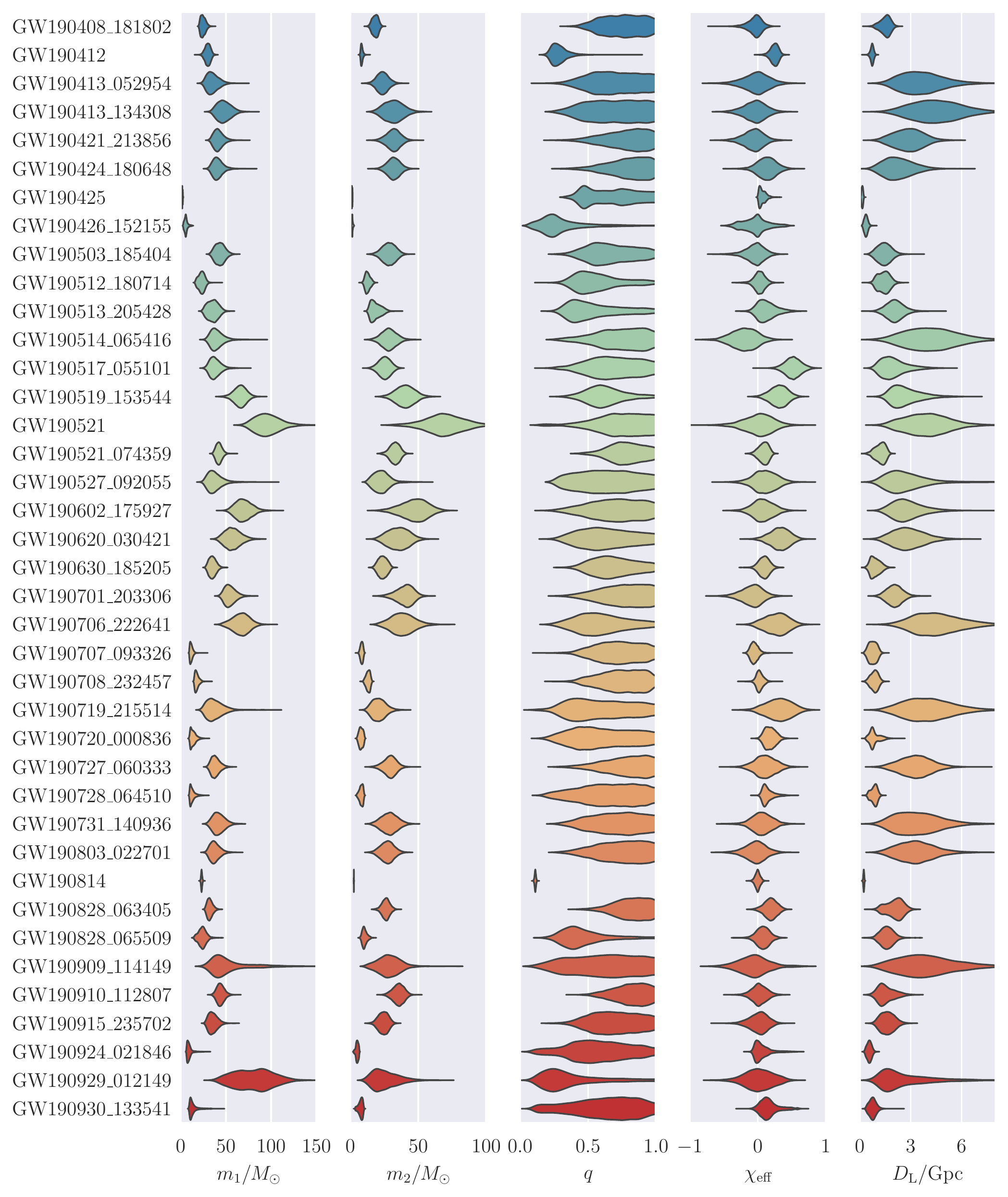}
    \caption{Marginal posterior distributions on primary mass $m_1$, secondary
        mass $m_{2}$, mass ratio $q$, effective inspiral spin $\chieff$, and the
        luminosity distance $d_{\rm L}$ for all candidate events in O3a. The
        vertical extent of each colored region is proportional to
        one-dimensional marginal posterior distribution at a given parameter
        value for the corresponding event.} \label{fig:multi_pdfs}%
\end{figure*}

To provide an overview of the posterior distributions of the source
parameters for all GWTC-2 candidate events,
we show 90\% credible regions for all candidate events in the $M$--$q$
and $\mathcal{M}$--$\chieff$ planes
in Figs.~\ref{fig:mtotqpost} and \ref{fig:mcchieffpost}, respectively,
and the corresponding
one-dimensional marginal
posterior distributions on $m_1$, $q$, and $\chieff$ in Fig.~\ref{fig:multi_pdfs}.

\begin{PE_table}
\begin{table*}
    \centering
    {\rowcolors{1}{}{lightgray}
    \input{PE_table.tex}
    }
    \caption{Median and 90\% symmetric credible intervals on selected source
      parameters. The columns show source total mass $M$, chirp
      mass $\mathcal{M}$ and component masses $m_{i}$, dimensionless
      effective inspiral spin $\chieff$, luminosity distance
      $\DL$, redshift $z$, final mass $M_{\rm f}$,
      final spin $\chi_\mathrm{f}$, and sky localization $\Delta \Omega$.
      The sky localization is the
      area of the 90\% credible region. For \protect\NAME{GW190425A}\
      we show the results using the high-spin prior
      ($|\vec{\chi}_{i}|\leq0.89$). We also report the
      network matched filter SNR for all events. These SNRs are from 
      LALInference IMRPhenomPv2 runs since RIFT
      does not produce the SNRs automatically, except for 
      \protect\NAME{GW190425A}\ and \protect\NAME{GW190426A}\, which
      use the SNRs from fiducial runs, and \protect\NAME{GW190412A}, \protect\NAME{GW190521A}, and 
      \protect\NAME{GW190814A}, which use IMRPhenomPv3HM SNRs.
      For GW190521 we report results averaged over three waveform
      families, in contrast to the results highlighting one waveform
      family in~\cite{GW190521Adiscovery}.
      }
    \label{tab:PE}
\end{table*}
\end{PE_table}

\subsection{Masses of sources with $m_2>3\,\Msun$}
Our candidate event list includes compact binary mergers
that have higher total masses than those
in GWTC-1~\cite{LIGOScientific:2018mvr} as well as mergers with component masses
in the purported lower mass gap of
$\sim 2.5$--$5\,\Msun$~\cite{Bailyn:1997xt,Ozel:2010su,Farr:2010tu,Ozel:2012ax}.
Here we describe the masses for candidate events with $m_2>3\,\Msun$, which we can confidently
expect to be \acp{BBH}. The remaining candidate events are described separately in Sec.~\ref{ss:pe_non_bbh}.

A majority of the masses of black holes reported herein are
larger than those reported via electromagnetic observations~\cite{Miller:2014aaa,Heida:2017shu,Thompson:2018ycv}.
By repeatedly selecting one posterior sample at random from each event and recording the most
massive among the ensemble, we find that the most massive binary system is probably
the one associated with \NAME{\totalmasssourcemost}~\cite{GW190521Adiscovery,GW190521Aastro}.
This system has a \totalmasssourcemostpercent\%
probability of being the most massive, with a total mass of
\fixme{$\totalmasssourcemed{\totalmasssourcemost}_{-\totalmasssourceminus{\totalmasssourcemost}}^{+\totalmasssourceplus{\totalmasssourcemost}}\,\Msun$}
and remnant mass $\finalmasssourcemed{\totalmasssourcemost}_{-\finalmasssourceminus{\totalmasssourcemost}}^{+\finalmasssourceplus{\totalmasssourcemost}}\,\Msun$,
where we have averaged over SEOBNRv4PHM, NRSur7dq4 and IMRPhenomPv3HM
waveform families. This averaging is done for consistency with other
sources contained in this catalog,
in contrast to the individual results reported in \cite{GW190521Aastro}, where
the NRSur7dq4 results are highlighted.
The more massive component in the source of \NAME{\massonesourcemost}\
has an \massonesourcemostpercent \% probability
of being the most massive BH detected in gravitational waves to date
(\fixme{$m_1 = \massonesourcemed{\massonesourcemost}_
    {-\massonesourceminus{\massonesourcemost}}^
{+\massonesourceplus{\massonesourcemost}}\,\Msun$}).
\NAME{GW190519A}, \NAME{GW190602A}, and \NAME{GW190706A}\ also have notably
high total masses with over 50\% posterior support for total mass $M>100\,\Msun$.

The least massive O3a system with $m_2 > 3\,\Msun$
is probably (\fixme{\totalmasssourceleastpercent}\%) the one
associated with \NAME{\totalmasssourceleast}\
($M=\totalmasssourcemed{\totalmasssourceleast}_
{-\totalmasssourceminus{\totalmasssourceleast}}^
{+\totalmasssourceplus{\totalmasssourceleast}}\,\Msun$),
and likely also has the least massive
object over $3\,\Msun$ (\masstwosourceleastpercent \%
probability and
$m_2 = \masstwosourcemed{\masstwosourceleast}_
{-\masstwosourceminus{\masstwosourceleast}}^
{+\masstwosourceplus{\masstwosourceleast}}\,\Msun$).

For most sources detected in O3a, the mass ratio
posteriors have support at unity and 
therefore are consistent with equal mass mergers.
An exception is the source of \NAME{GW190412A}\
which was the first event detected that had a 
confidently unequal mass ratio 
($q=\massratiomed{GW190412A}_
{- \massratiominus{GW190412A}}^
{+ \massratioplus{GW190412A}}$)
and exhibited strong
signs of HM contributions to the waveform~\cite{GW190412}.
Although its mass ratio
is confidently bounded away from unity, \NAME{GW190412A}\ only has a 
\massratioleastpercent\% chance of having the smallest
mass ratio among O3a sources with $m_2>3\,\Msun$.  
As seen in Figs.~\ref{fig:mtotqpost} and \ref{fig:multi_pdfs},
the mass ratios are not well constrained for many systems, so 
one or more
could have a smaller mass ratio than \NAME{GW190412A}.

\subsection{Sources with $m_2<3\,\Msun$}\label{ss:pe_non_bbh}
\subsubsection{GW190425}\label{sss:GW190425A}

The least massive O3a system is associated with \NAME{GW190425A}\ and is likely a binary
neutron star system given the inferred masses
($m_1=\massonesourcemed{GW190425A}_
{-\massonesourceminus{GW190425A}}^
{+\massonesourceplus{GW190425A}}\,\Msun$
and $m_2=\masstwosourcemed{GW190425A}_
{-\masstwosourceminus{GW190425A}}^
{+\masstwosourceplus{GW190425A}}\,\Msun$),
but constraints on the tidal parameters do not rule out a NSBH or \ac{BBH} origin.
These estimates were obtained with the PhenomPv2NRT
model with a high-spin prior that restricts dimensionless spin
magnitudes of the compact objects to be less than 0.89, and previously reported
in \cite{Abbott:2020uma}.

Although the inferred component masses of \NAME{GW190425A}'s source are consistent with
masses of known neutron stars~\cite{Lattimer:2012nd,Ozel:2016oaf,Alsing:2017bbc,Chatziioannou:2020msi},
the total mass $\totalmasssourcemed{GW190425A}_{-\totalmasssourceminus{GW190425A}}^{+\totalmasssourceplus{GW190425A}}\,\Msun$
is greater than that of observed Galactic \acp{BNS}~\cite{Tauris:2017omb,Farrow:2019xnc}. This raises the question of whether
\NAME{GW190425A}'s source was formed in a different environment from the double
neutron star systems observed to date
\cite{Abbott:2020uma,Safarzadeh:2020efa,Gupta:2019nwj,Romero-Shaw:2020aaj,Mandel:2020cig}.

\begin{figure}[htb!]
\centering
\includegraphics[width=\columnwidth]{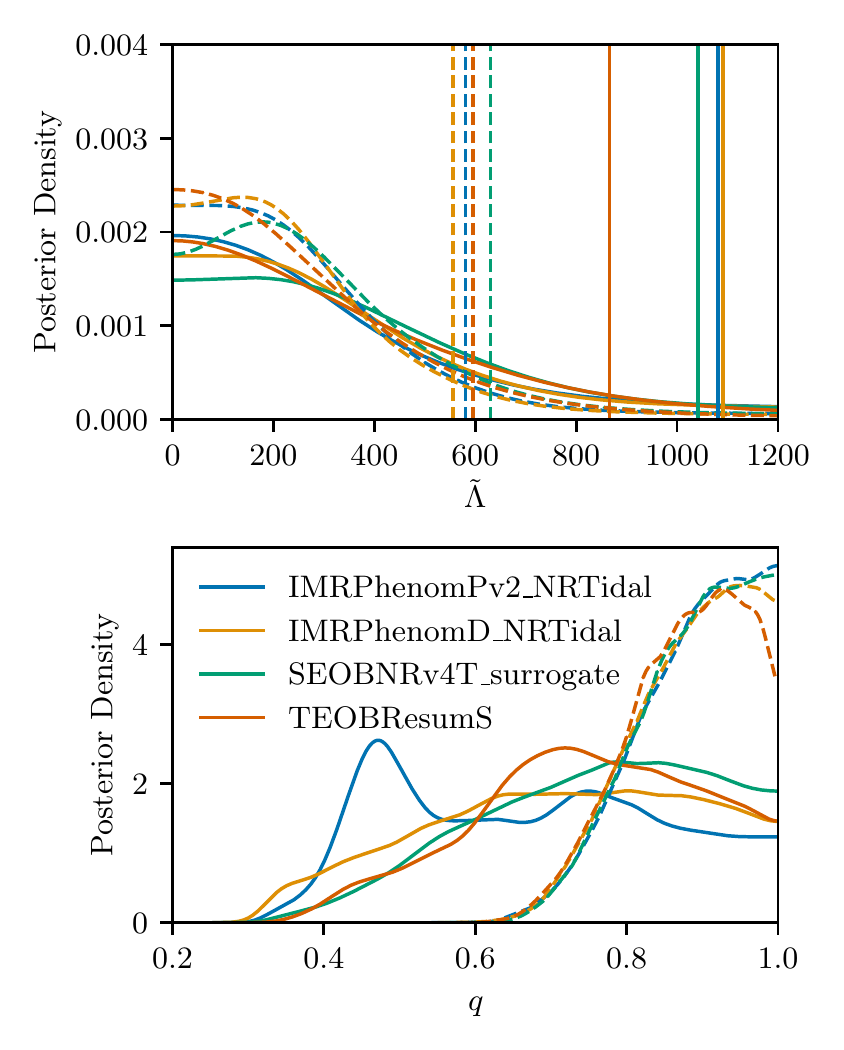}

\caption{\label{Fig:GW190425A_q_lambda} Marginalised distributions for
the combined dimensionless tidal deformability parameter $\tilde{\Lambda}$ (top)  and for mass ratio (bottom)
for BNS event \protect\NAME{GW190425A}. The solid lines correspond to the high spin prior, while the dashed to the low spin prior.
In the top panel, the vertical lines mark the corresponding 90\% upper bound on $\tilde{\Lambda}$.
SEOBNRv4T\_surrogate constrains mass ratio better
towards equal masses for high larger spins. The plot and quoted 90\% upper limits for $\tilde{\Lambda}$
are obtained by reweighing posterior distribution as described in Appendix F of~\cite{Abbott:2020uma}.}
\end{figure}

\begin{PE_table}
\begin{table}[htb!]
  \centering
  {\rowcolors{1}{}{lightgray}
       \input{GW190425_source_properties_table}

  } \caption{Source properties of \protect\NAME{GW190425A}\ with different waveform
    families. For the primary mass we give the 0\%--90\% interval, while for the
    secondary mass and mass ratio we give the 10\%--100\% confidence intervals.
    The quoted 90\% upper limits for $\tilde{\Lambda}$ are obtained by
    reweighing its posterior distribution as detailed in Appendix F of
    \cite{Abbott:2020uma}.  The top half of the table describes values from low
    spin prior (LS), while bottom half for high spin prior (HS). For LS, all the
    results are consistent with each other, while for HS there are slight
    differences among different waveforms.}
    \label{table:GW190425A}
\end{table}
\end{PE_table}
Extending the discussion of waveform systematics reported with the discovery of
\NAME{GW190425A}~\cite{Abbott:2020uma}, we perform four supplementary analyses
using the non-precessing EOB models SEOBNRv4T\_surrogate and
TEOBResumS for low spin and high spin priors. The results are summarised in
the Table~\ref{table:GW190425A}. All the analyses produce quantitatively similar
results to the corresponding non-precessing analysis reported in the discovery
paper, and the two EOB models produce consistent results between
them.

For the low-spin
prior, the results agree with the non-precessing IMRPhenomDNRTidal analysis reported in the
discovery paper, with SEOBNRv4T\_surrogate and TEOBResumS recovering chirp mass as
$\mathcal{M}=
\chirpmasssourcefourtwofivemed{SEOBNRv4TsurrogateLS}
_{-\chirpmasssourcefourtwofiveminus{SEOBNRv4TsurrogateLS}}
^{+\chirpmasssourcefourtwofiveplus{SEOBNRv4TsurrogateLS}}\,\Msun$
and effective inspiral spin of
$\chieff=
\chiefffourtwofivemed{SEOBNRv4TsurrogateLS}
_{-\chiefffourtwofiveminus{SEOBNRv4TsurrogateLS}}
^{+\chiefffourtwofiveplus{SEOBNRv4TsurrogateLS}}$. When allowing larger compact
object spins, the results with IMRPhenomDNRTidal and EOB models exhibit some
differences, but overall give consistent posteriors.

Our inferences about tidal parameters are likewise consistent with the previous
non-precessing analyses. We follow the procedure of \cite{Abbott:2020uma}
and re-weight the posteriors to a flat in $\tilde{\Lambda}$ prior.  For the
low-spin prior, the two EOB models give very similar bounds, with 
$\tilde\Lambda$ constrained below $\lambdatildeboundonefourtwofiveLS{SEOB}$. For
the high spin prior, the TEOBResumS waveform model constrains the dimensionless
tidal deformability parameter better ($\tilde\Lambda \le
\lambdatildeboundonefourtwofiveHS{TEOB}$) as compared to the other waveform
models, as seen in the top panel of Fig.~\ref{Fig:GW190425A_q_lambda}.

The two EOB models also constrain the mass
ratio better than other waveforms
(\lowerboundmassratioonefourtwofiveHS{SEOB}--\upperboundmassratioonefourtwofiveHS{SEOB})
as can be seen in the bottom panel of the
Fig.~\ref{Fig:GW190425A_q_lambda}.  The previously-reported analyses
that allow for significant precessing spins have much greater
flexibility and thus, for the high-spin prior, produce a more
asymmetric mass ratio posterior distribution than the two
non-precessing updates reported here.

Finally, both the EOB models find the luminosity distance of
$D_\mathrm{L} = \luminositydistancefourtwofivemed{SEOBNRv4TsurrogateHS}_
{-\luminositydistancefourtwofiveminus{SEOBNRv4TsurrogateHS}}^
{+\luminositydistancefourtwofiveplus{SEOBNRv4TsurrogateHS}}$
Gpc independent of the spin prior used.

\subsubsection{GW190814}\label{sss:GW190814}

Amongst the O3a events, \NAME{GW190814A}'s source~\cite{GW190814A} has the least massive secondary
component after the sources of \NAME{GW190425A}\ and \NAME{GW190426A}.
\NAME{GW190814A}'s less massive component has mass
$m_2=\masstwosourcemed{GW190814A}_
{-\masstwosourceminus{GW190814A}}^
{+\masstwosourceplus{GW190814A}}\,\Msun$,
making its interpretation as
as a black hole or a neutron star
unclear~\cite{GW190814A}.
\NAME{GW190814A}\ also has the most extreme mass ratio of all the candidate events,
$q=\massratiomed{GW190814A}_{-\massratiominus{GW190814A}}^{+\massratioplus{GW190814A}}$~\cite{GW190814A}.

\subsubsection{GW190426\_152155}\label{sss:GW190426}
\NAME{GW190426A}\ is the candidate event with the highest \ac{FAR}: 1.4~yr$^{-1}$.
Assuming it is a real signal of astrophysical origin, we estimate its component masses to be
$m_1=\massonesourcemed{GW190426A}_
{-\massonesourceminus{GW190426A}}^
{+\massonesourceplus{GW190426A}}\,\Msun$
and $m_2=\masstwosourcemed{GW190426A}_
{-\masstwosourceminus{GW190426A}}^
{+\masstwosourceplus{GW190426A}}\,\Msun$,
raising the possibility that it could have originated from either a \ac{BBH} or an \ac{NSBH}
source.
The mass of the secondary component is consistent with masses
of (previously) reported neutron stars~\cite{Ozel:2016oaf,Tauris:2017omb,Alsing:2017bbc,LIGOScientific:2019fpa}, but
the data are uninformative about potential tidal effects, showing
essentially no difference between the prior and posterior on $\Lambda_2$
obtained from NSBH waveforms SEOBNRv4\_ROM\_NRTidalv2\_NSBH or IMRPhenomNSBH which
we use for our fiducial results.
A more definitive assessment of this event likely requires further observations
to establish the rate of astrophysical signals with comparable properties.

\subsection{Spins}

\begin{figure*}[t]
\centering
\begin{subfloat}
  \centering
  \includegraphics[width=0.3\textwidth]{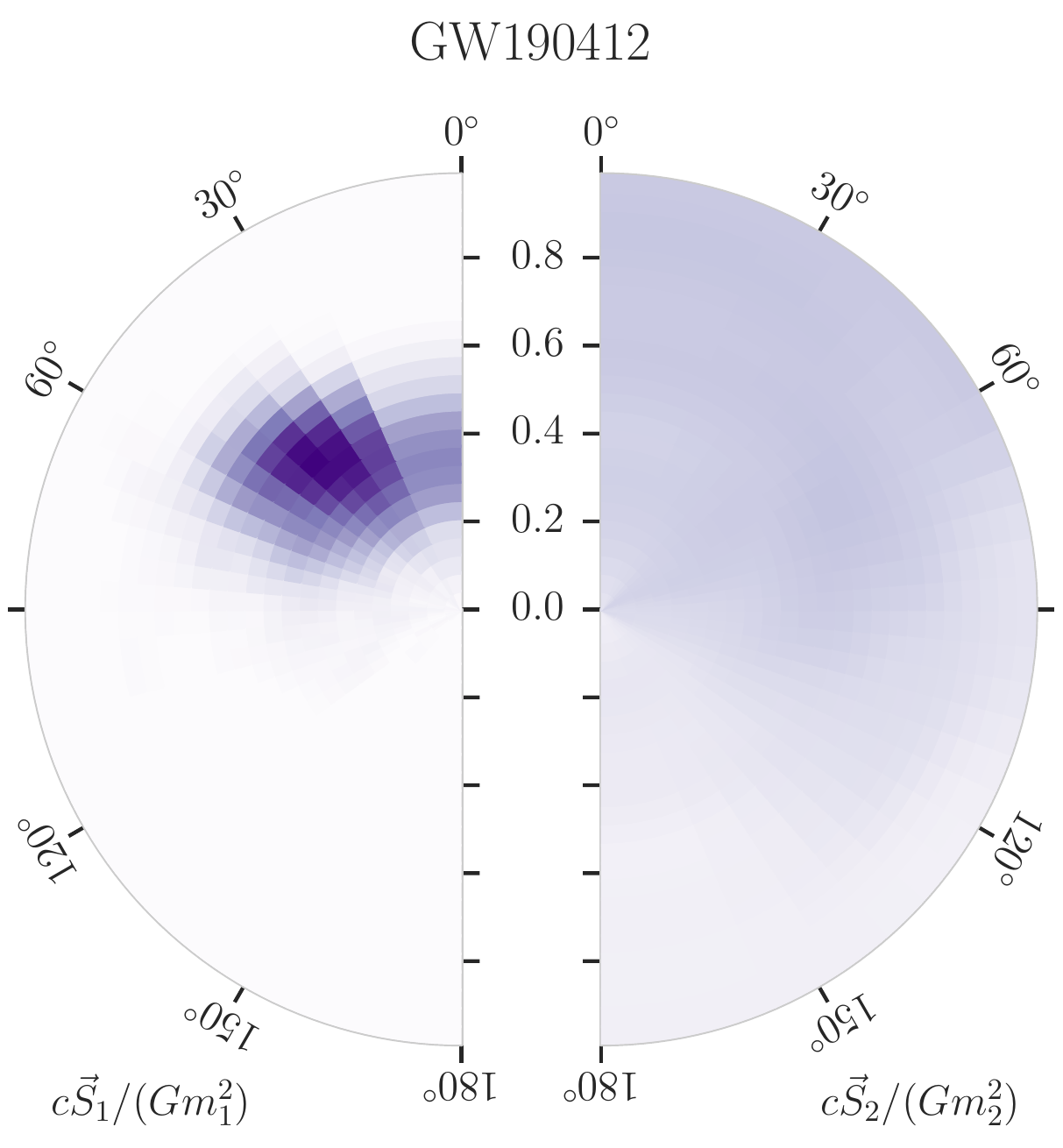}
\end{subfloat}
\hfill
\begin{subfloat}
  \centering
  \includegraphics[width=0.3\textwidth]{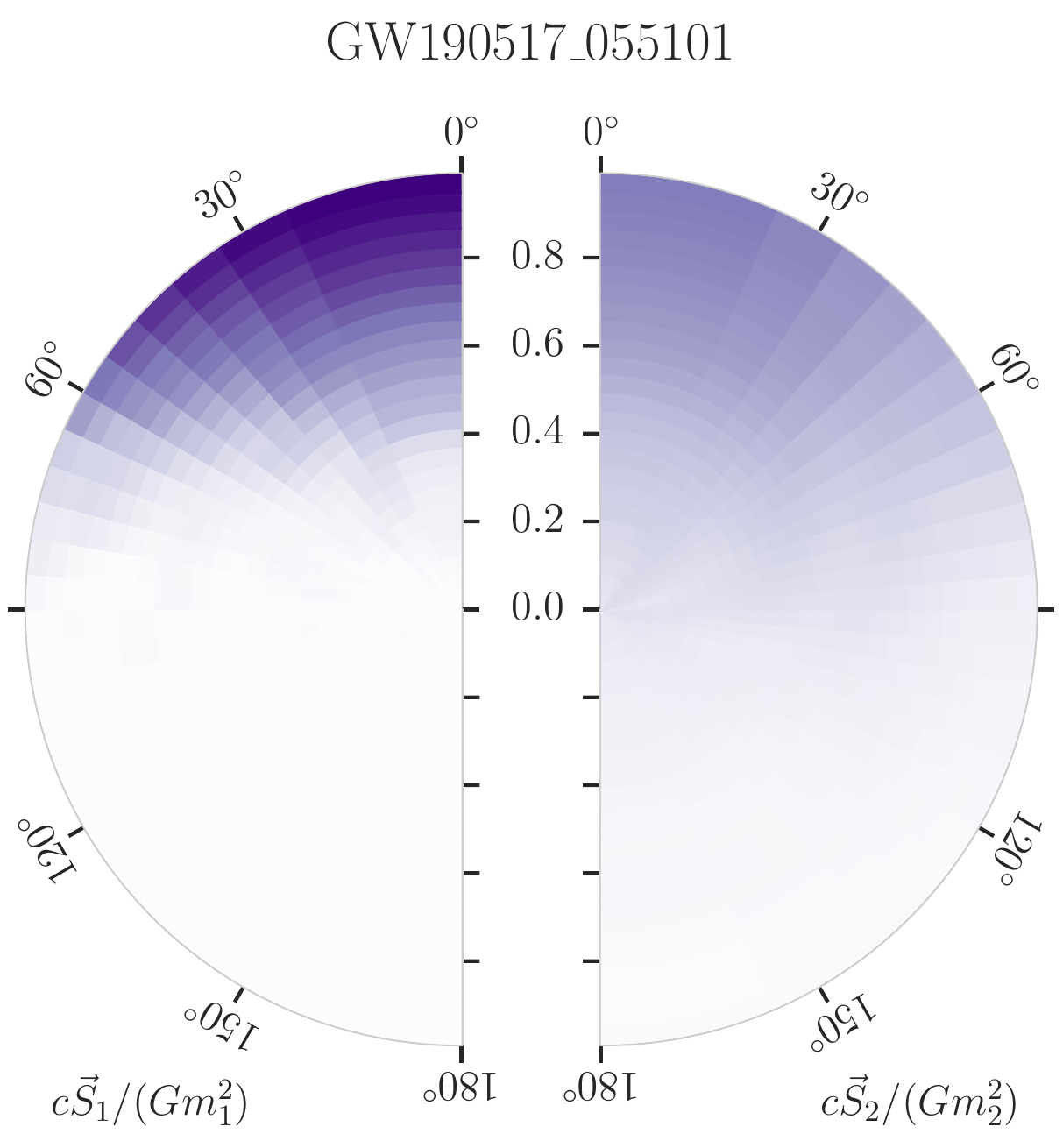}
\end{subfloat}
\hfill
\begin{subfloat}
  \centering
  \includegraphics[width=0.3\textwidth]{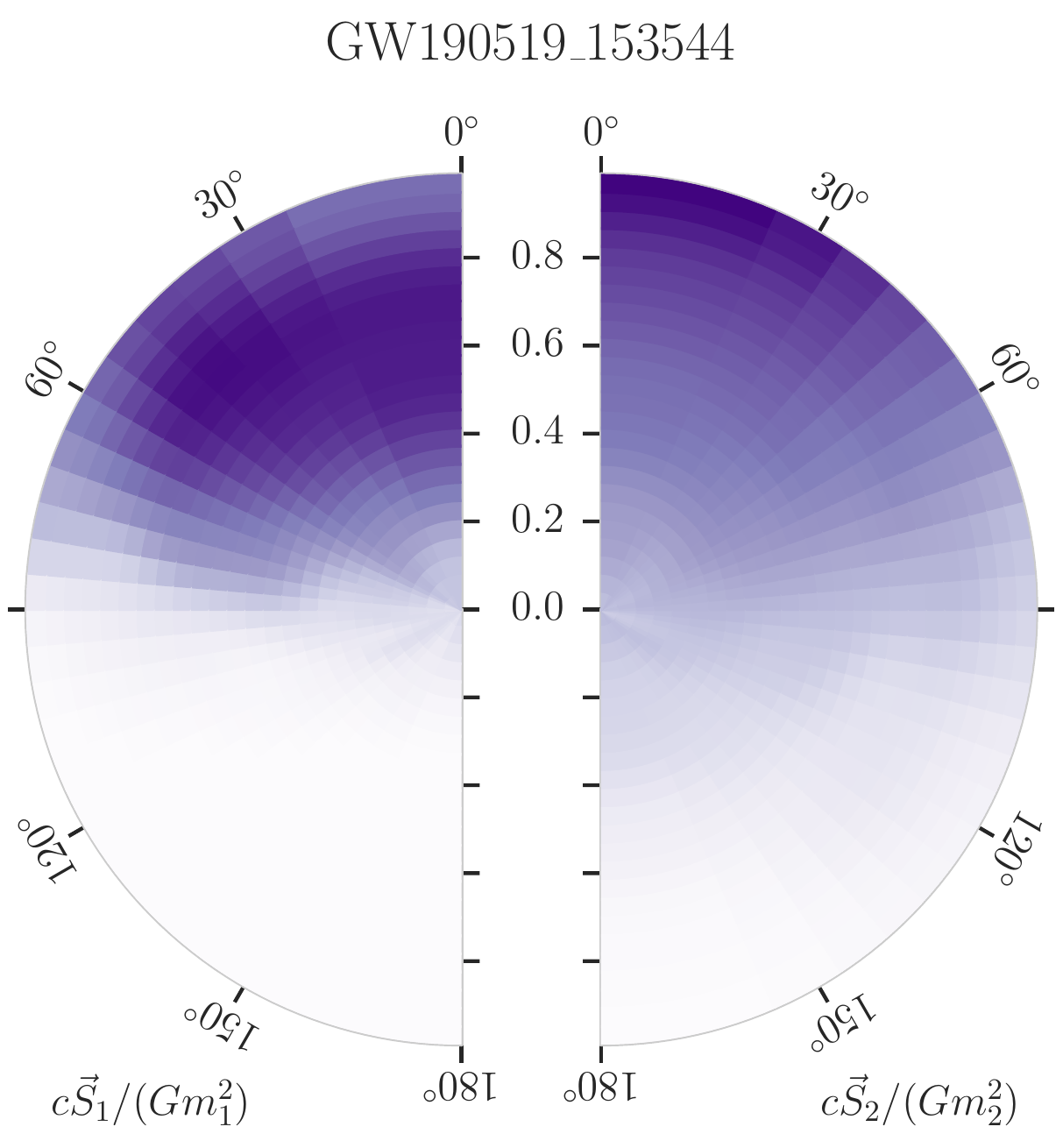}
\end{subfloat}
\newline
\begin{subfloat}
  \centering
  \includegraphics[width=0.3\textwidth]{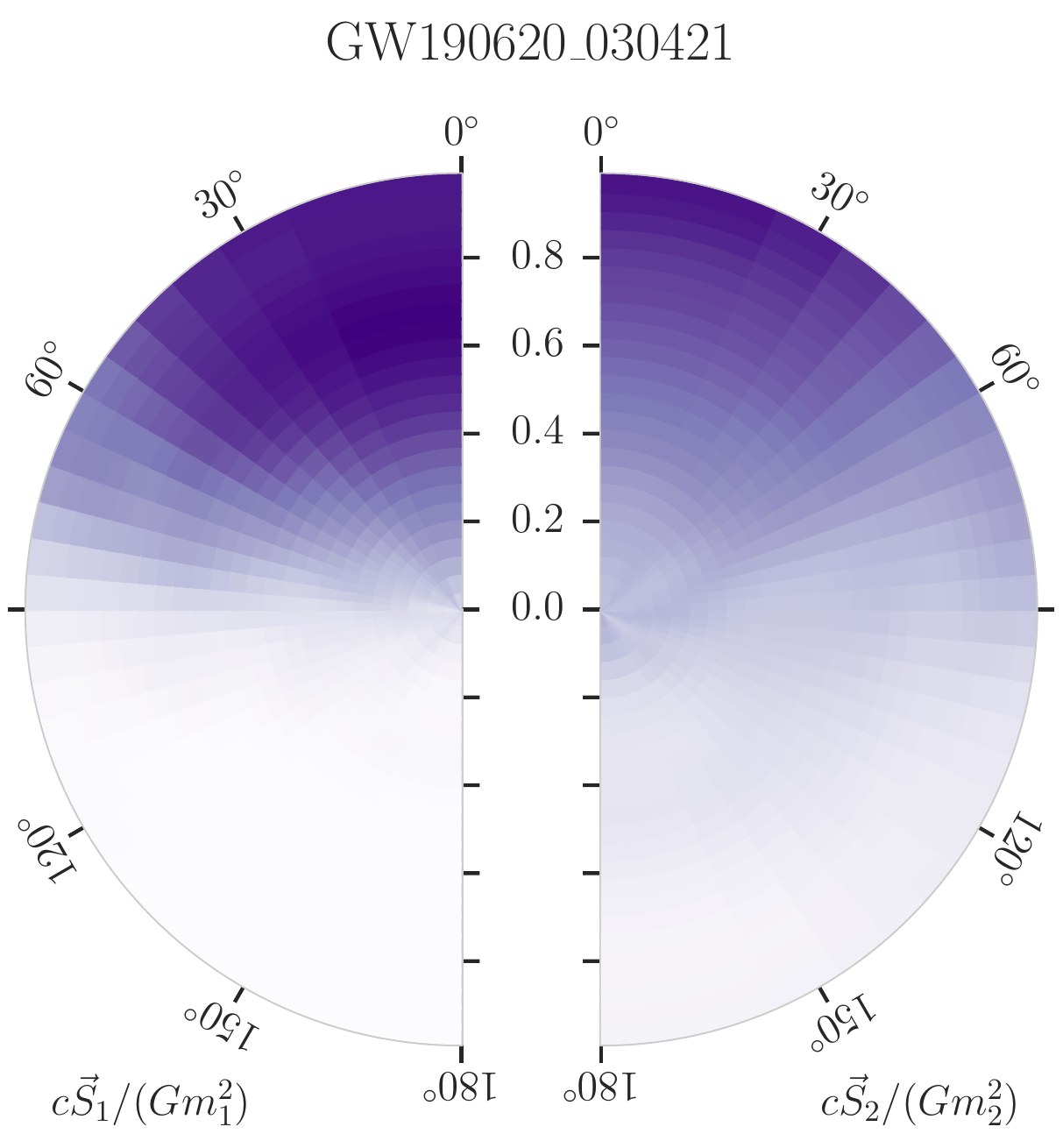}
\end{subfloat}
\hfill
\begin{subfloat}
  \centering
  \includegraphics[width=0.3\textwidth]{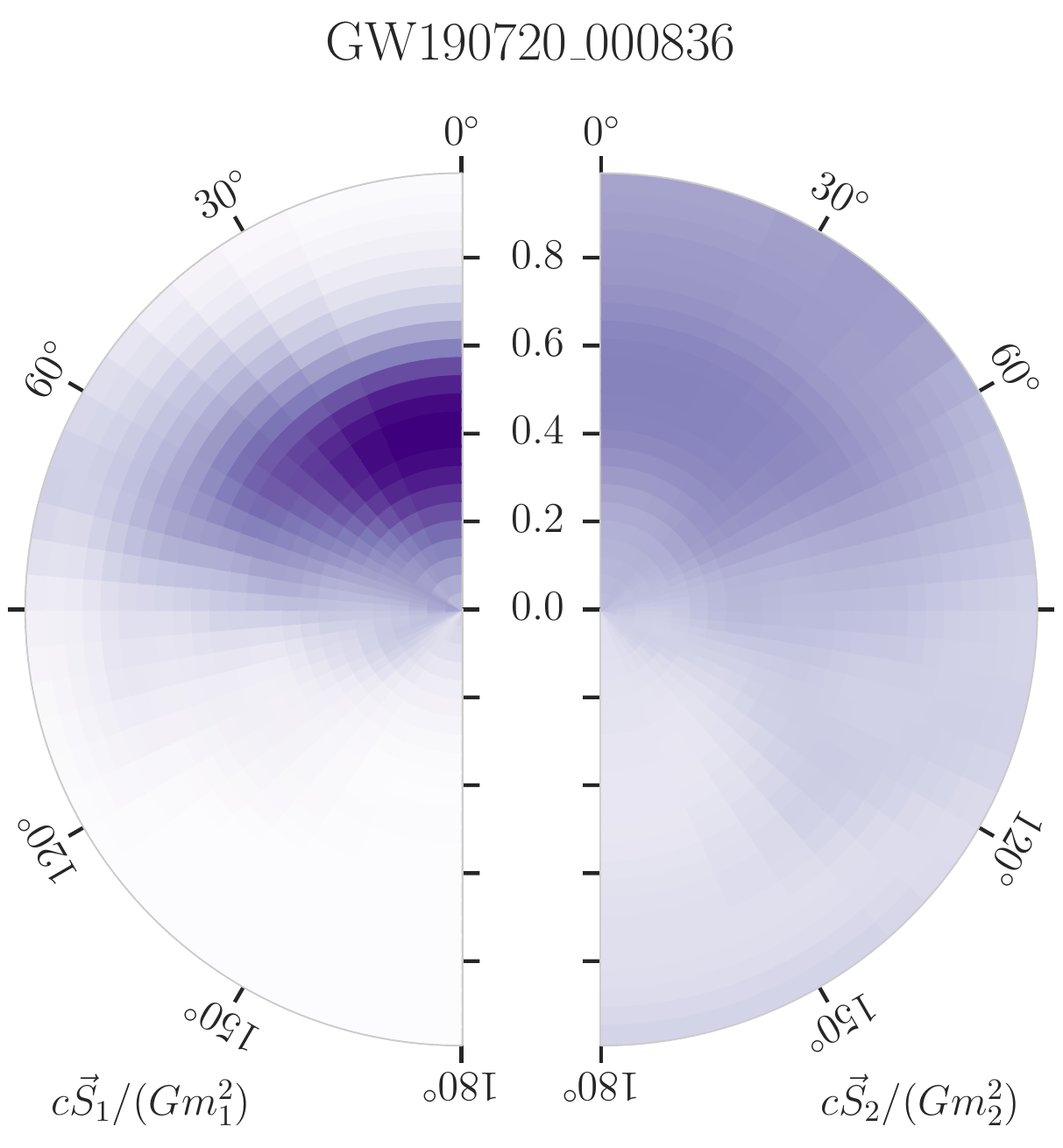}
\end{subfloat}
\hfill
\begin{subfloat}
  \centering
  \includegraphics[width=0.3\textwidth]{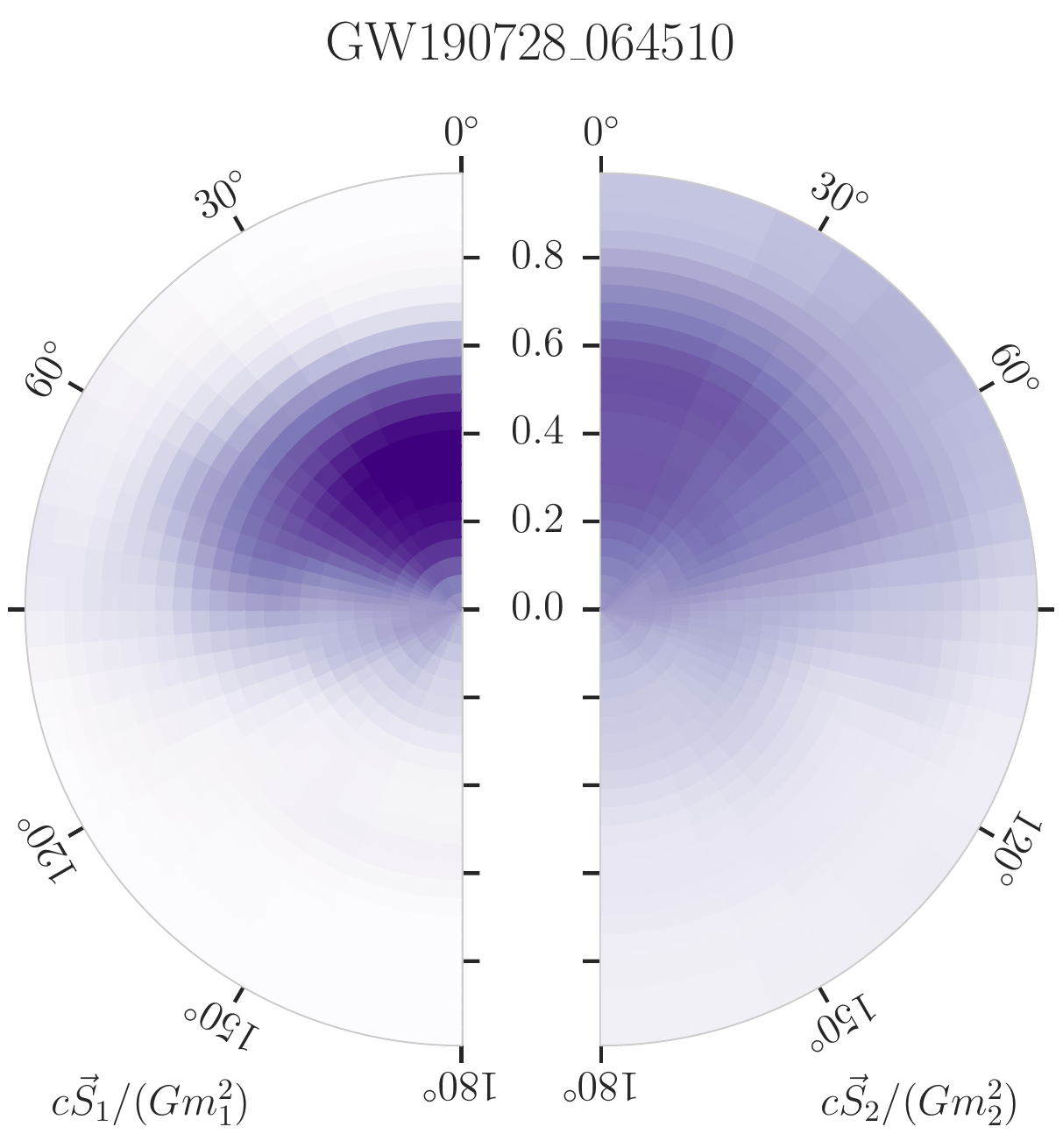}
\end{subfloat}

\caption{Estimates of the dimensionless spin parameters
  $\vec{\chi}_{i}=c\vec{S}_i/(Gm_{i}^2)$ for merger components of
  selected sources.  Each pixel's radial distance from the circle's
  center on the left (right) side of each disk corresponds to the spin
  magnitude $|\vec{\chi}|$ of the more (less) massive component, and
  the pixel's angle from the vertical axis indicates the tilt angle
  $\theta_{\rm LS}$ between  each spin and the Newtonian orbital
  angular momentum.  Pixels have equal prior probability, and shading
  denotes the posterior probability of each pixel, after marginalizing
  over azimuthal angles.  }
\label{fig:spin_disks}
\end{figure*}

Most of the compact objects detected in O3a have
spin magnitudes consistent with zero, within uncertainties, but in
some cases the spins can be constrained away from zero.
Two systems have a $> 50\%$ chance of having at least one
black hole with dimensionless spin magnitude $\chi_{i=\{1,2\}}> 0.8$:
\NAME{\highspinfirst}\ has $\chi_{i=\{1,2\}}> 0.8$ with 
\percenthighspin{\highspinfirst}\% credibility, and \NAME{\highspinsecond}\
with \percenthighspin{\highspinsecond}\% credibility. In addition to 
spin magnitudes, we also consider the effective inspiral spins.
As described in Sec.~\ref{sec:mf_searches}, the effective inspiral
spin $\chieff$ is the mass-weighted combination of aligned
spins and is approximately conserved under precession.
Effective inspiral spin posterior distributions for all candidate events
are shown in Figs.~\ref{fig:mcchieffpost} and ~\ref{fig:multi_pdfs}.
We find 11 systems that show signs of non-zero $\chieff$.
At 90\% credibility, \NAME{GW190412A}, \NAME{GW190425A},
\NAME{GW190517A}, \NAME{GW190519A},
\NAME{GW190620A}, \NAME{GW190706A}, \NAME{GW190719A}, \NAME{GW190720A},
\NAME{GW190728A}, \NAME{GW190828A}, and \NAME{GW190930A}\ have sources with $\chieff>0$.
No individual systems
were found to have $\chieff<0$ with $\geq 95\%$ probability,
but the event with the lowest $\chieff$ in O3a is
probably \NAME{\chieffleast}\ with $\chieff = \chieffmed{\chieffleast}_
{-\chieffminus{\chieffleast}}^{+\chieffplus{\chieffleast}}$.
The surplus of events with $\chieff>0$ compared to none 
with $\chieff<0$ suggests that the spin orientations of 
black holes in binaries are not isotropically distributed with respect
to their orbital angular momenta. This possibility is 
explored further in \cite{o3apop}.

The \ac{BBH} which most likely has the largest measured $\chieff$
is \NAME{\chieffmost}\
($\chieff = \chieffmed{\chieffmost}_
{-\chieffminus{\chieffmost}}^{+\chieffplus{\chieffmost}}$).
It has a \chieffmostpercent \% posterior probability of having the highest $\chieff$,
followed by \NAME{\chieffmostsecond}\
($\chieff = \chieffmed{\chieffmostsecond}_
{-\chieffminus{\chieffmostsecond}}^{+\chieffplus{\chieffmostsecond}}$)
with \chieffmostpercentsecond \%. 

In some cases,
the joint $\chieff$ and mass-ratio measurement for an event
enables a tighter measurement of the spin magnitude
of the primary mass than for candidate events with mass ratios closer to  unity.
For example, we find primary spin magnitudes of
$\chi_1 = \spinonemed{GW190720A}_{-\spinoneminus{GW190720A}}^{+\spinoneplus{GW190720A}}$
for the source of \NAME{GW190720A}, and
$\chi_1 = \spinonemed{GW190728A}_{-\spinoneminus{GW190728A}}^{+\spinoneplus{GW190728A}}$
for the source of \NAME{GW190728A}.
Posterior distributions on spin magnitudes and
tilt angles are shown in Fig.~\ref{fig:spin_disks}
for these two candidate events and other select systems that exhibit
non-zero spins.

\begin{figure}[tb]
\begin{center}
\includegraphics[width=\columnwidth]{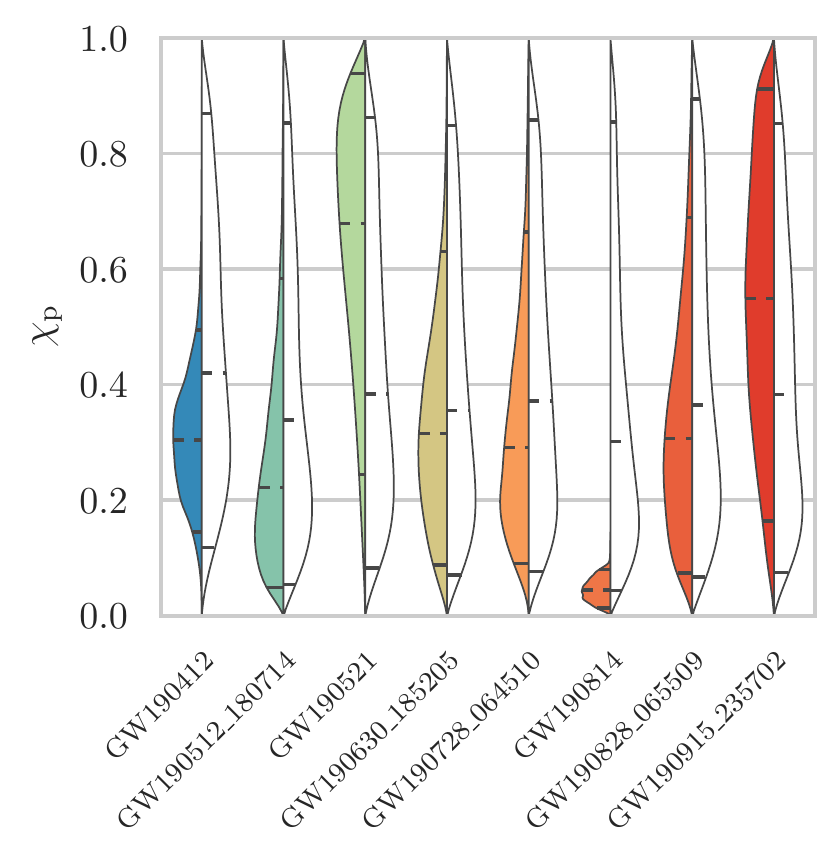}
\end{center}
\caption{\label{fig:chip} Posterior and prior distributions on the
    effective precession spin parameter $\chip$ for select events.
    The $\chip$ prior shown on the right half of each leaf is
    conditioned on the posterior of $\chieff$ since the default
    prior in sampling contains correlations between $\chip$ and $\chieff$.
}
\end{figure}

The magnitude of spin-precession in the waveform can be
partially captured by the effective precession spin
parameter $\chip$, which includes the
projection of component spin vectors onto
the orbital plane~\cite{Hannam:2013oca,Schmidt:2014iyl}. To
identify events for which the data constrain $\chip$, we
compare the $\chip$
priors---conditioned on the posteriors of $\chieff$---to the $\chip$
posteriors,
as done in \cite{LIGOScientific:2018mvr}. Conditioning the $\chip$
prior on the $\chieff$ posterior accounts for the correlated
prior between $\chieff$ and $\chip$ in the default spin prior
choices presented in Sec.~\ref{sec:PEmethods}.
Fig.~\ref{fig:chip} shows the one-dimensional posterior and ($\chieff$-conditioned) prior
distributions on $\chip$ for events with Jensen--Shannon (JS)
divergence~\cite{Lin:1991} $D_{\rm JS}^{\chip}>0.05$~bit, where
$D_{\rm JS}^{\chip}$ is calculated between the $\chip$ posterior and conditioned prior.
For most of the candidate events, the posterior on $\chip$
is similar to the prior, indicating that the data are largely uninformative
about precession, but there are a few notable exceptions.
The $\chip$ inference on \NAME{GW190814A}\ is most striking
with $D_{\rm JS}^{\chi_p}=\chippripostJS{GW190814A}~\mathrm{bit}$:
the spin magnitude of the primary mass of the system is constrained to be
near zero, resulting in a correspondingly small $\chip$ value.
After \NAME{GW190814A}, $D_{\rm JS}^{\chip}$ is
largest for \NAME{GW190412A}\ \cite{GW190412,Zevin:2020gxf}
and \NAME{GW190521A}\ \cite{GW190521Adiscovery,GW190521Aastro},
with $D_{\rm JS}^{\chi_p}=\chippripostJS{GW190412A}~\mathrm{bit}$
and $D_{\rm JS}^{\chi_p}=\chippripostJS{GW190521A}~\mathrm{bit}$,
respectively. 
Unlike \NAME{GW190814A},
the $\chip$ posterior distributions for these events are
constrained away from zero, showing preference for
precession in these systems. 
The tilt angle of \NAME{GW190412A}'s
more massive component's spin with respect to the Newtonian
orbital angular momentum is particularly well constrained
to $\theta_{{LS_1}} = \tiltonemed{GW190412A}_{-\tiltoneminus{GW190412A}}^{+\tiltoneplus{GW190412A}}$,
as seen in Fig.~\ref{fig:spin_disks}. 
To further investigate possible precession in these signals, we 
compute the precession \ac{SNR} $\rho_p$ \cite{Fairhurst:2019vut, Fairhurst:2019srr}, which characterizes the 
observability of precession in gravitational wave data. 
\NAME{GW190412A}\ has the largest precession SNR in O3a with a median
$\rho_p = 3.0$. Despite parameter estimation preferring high $\chi_p$
for \NAME{GW190521A}, we calculate $\rho_p = 1.6$ showing that the measurable
precession in the signal is small. Using empirical relations between $\rho_p$ 
and Bayes' factors for precession \cite{Green:2020ptm,Pratten:2020}, we 
see that the corresponding Bayes' factors for precession also only
mildly favor precession.
A population-level analysis of the
GWTC-2 spins is presented in \cite{o3apop} and finds
\fixme{evidence for the presence of spin precession in the population}.

\subsection{Three-Dimensional Localization}
The most distant event, after accounting for measurement uncertainties in distance, is most probably
\NAME{\luminositydistancemost}, with an estimated
luminosity distance and redshift of
$D_\mathrm{L} = \luminositydistancemed{\luminositydistancemost}_
{-\luminositydistanceminus{\luminositydistancemost}}^
{+\luminositydistanceplus{\luminositydistancemost}}$ Gpc and
$z = \redshiftmed{\luminositydistancemost}_
{-\redshiftminus{\luminositydistancemost}}^
{+\redshiftplus{\luminositydistancemost}}$,
respectively, approximately twice the luminosity distance of the most distant
source from GWTC-1, GW170729~\cite{LIGOScientific:2018mvr},
and comparable to the distance for gravitational wave candidate GW170817A \cite{Zackay:2019btq} (not to be confused 
with the \ac{BNS} signal GW170817). 
However, \NAME{GW190909A}, \NAME{GW190514A}, \NAME{GW190521A},
\NAME{GW190706A}, and \NAME{GW190719A}\
have similarly large distances to \NAME{\luminositydistancemost}.
With candidate events at these cosmological distances,
we can more readily measure the Hubble constant, and the evolution of the \ac{BBH}
merger rate over cosmic time.  Such
analyses are performed in \cite{o3apop}. The closest
source detected in O3a is \NAME{GW190425A},
with an inferred luminosity distance of
$\DL = \luminositydistancemed{GW190425A}_
{-\luminositydistanceminus{GW190425A}}^
{+\luminositydistanceplus{GW190425A}}$~Gpc,
about four times the distance for GW170817.

Overall, \NAME{\minareaevent}\ is the best localized event detected in O3a.
The contour encompassing 90\% of this event's two-dimensional sky position posterior is $\Delta \Omega = \skyarea{\minareaevent}$~deg$^2$,
and 90\% of this event's 3D sky position posterior is contained in $\Delta V_{90} = \skyvol{\minareaevent}$~Gpc$^3$.
Although \NAME{\minareaevent}\ was initially detected in only LIGO Livingston and Virgo, it was reanalyzed
with LIGO Hanford data, enabling the strong constraint on 3D source position.
\NAME{GW190412A}\ and \NAME{GW190701A}\ were also relatively well localized with
$\Delta \Omega = \skyarea{GW190412A}$~deg$^2$, $\Delta V_{90} = \skyvol{GW190701A}$~Gpc$^3$ and
$\Delta \Omega = \skyarea{GW190701A}$~deg$^2$, $\Delta V_{90} = \skyvol{GW190701A}$~Gpc$^3$, respectively,
and were both detected in all three detectors.  \NAME{\maxareaevent}\ was detected and analyzed only in
Livingston data, and therefore had the largest localization area and volume with
$\Delta\Omega = \skyarea{\maxareaevent}$~deg$^2$ and $\Delta V_{90} = \skyvol{\maxareaevent}$~Gpc$^3$.
Credible intervals on each source's distance and sky area are shown in
Table~\ref{tab:PE}.
Probability density sky-maps for all events are available as part of the data release~\cite{lvc:datadoi}.

\section{Waveform reconstructions}\label{sec:wfreconstructions}

\resetlinenumber

Template-based~\cite{PhysRevD.58.082001,Veitch:2014wba}
and minimally-modeled methods~\cite{Klimenko:2015ypf,Cornish:2014kda,Salemi:2019uea}
are complementary techniques for producing waveform reconstructions.
Waveform templates provide a mapping between
the shape of the waveforms and the parameters of the source, such as the
masses and spins of a binary system, but are limited to those sources for
which we have models. Minimally-modeled reconstructions make it possible to
discover unexpected phenomena, but they do not provide a direct mapping to
the physical properties of the source. Currently available waveform templates
for binary mergers are based on various approximations and numerical solutions
to Einstein's equations that cover a subset of the full parameter space.
These waveform templates may thus fail to capture some features of the signal.
A more exotic possibility is that gravity behaves differently than predicted
by general relativity. One way to test these scenarios is to compare the
template-based and minimally-modeled waveform reconstructions.

A standard measure of the agreement between two waveforms $h_1, h_2$ is the
\textit{match}, or \textit{overlap}
\begin{equation}
{\cal O}\langle h_1,h_2\rangle = \frac{\langle h_1 | h_2\rangle}{\sqrt{\langle h_1 | h_1\rangle \langle h_2 | h_2\rangle }}\, ,
\end{equation}
where $\langle a|b\rangle$ denotes the noise-weighted inner product~\cite{Finn:1992wt,Cutler:1994ys}.  
The match is constrained to be $\le 1$.

For each event, the matches were computed between the maximum likelihood
template-based waveforms and two minimally-modeled waveform reconstruction
methods, \CWB~\cite{Klimenko:2015ypf} and BW~\cite{Cornish:2014kda}, using
data that contain the event (on-source data).
To ascertain whether these match values are in line with expectations, waveforms
from the template based analysis were added to data near, but not including,
each event (off-source data). The minimally-modeled waveform reconstructions were repeated
multiple times on these off-source data to estimate the distribution of match
values we would expect for each event. For each event, these distributions were
used to compute a p-value, given by the fraction of off-source match values that
are below the  on-source match.
For some of the lower \ac{SNR} candidate events the minimally-modeled methods were unable
to reconstruct the signals,  and these candidate events were excluded from the analysis.
Details of the analysis procedure, and additional results, can be found in
Appendix~\ref{appendix:WCT}.

\begin{figure}[t]
\centering
\includegraphics[width=\columnwidth]{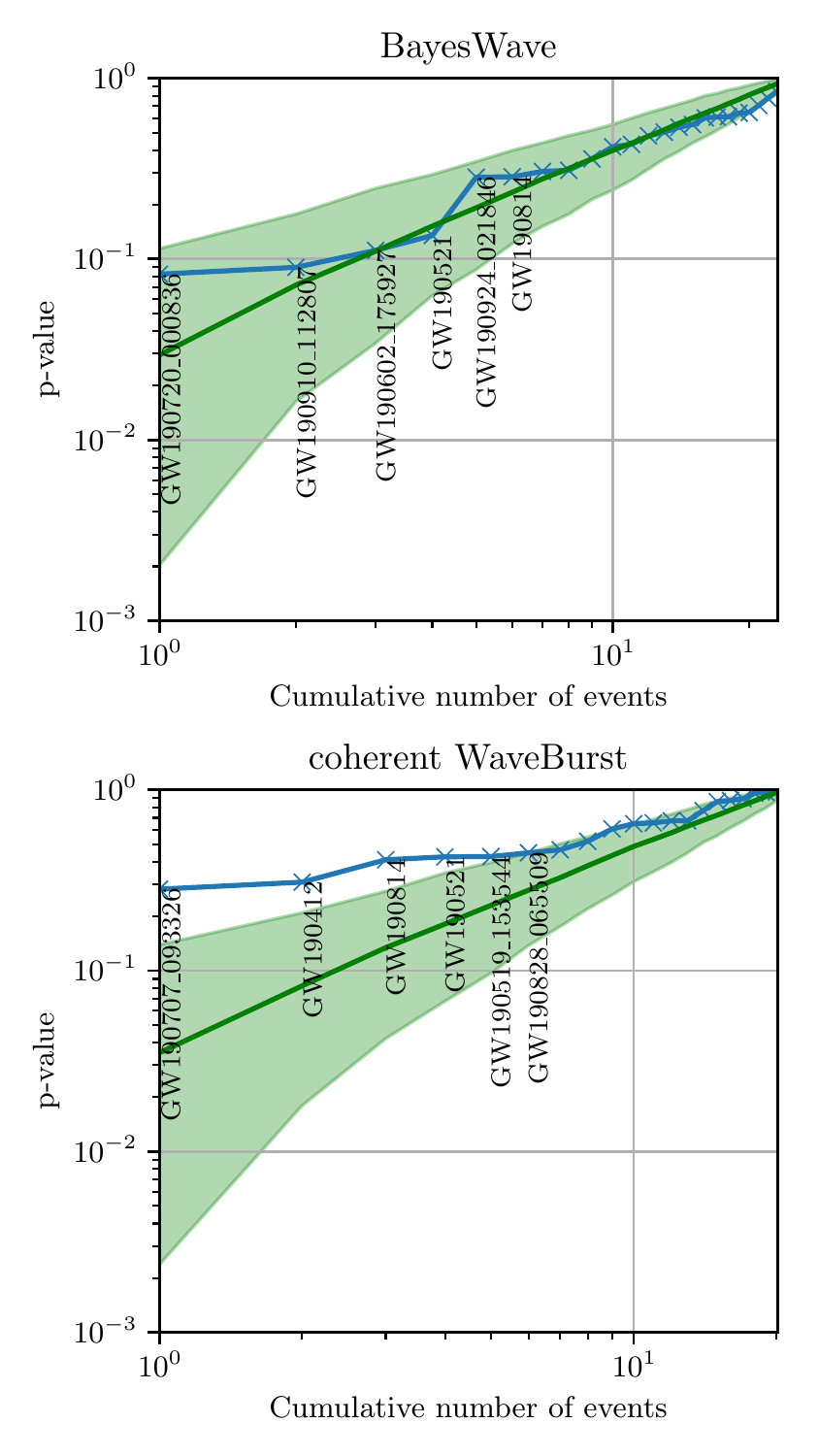}
\caption{P-value plot for the candidate events
reconstructed by the minimally modeled pipelines in O3a.
The upper panel is from the BW analysis and the lower panel
is from the \CWB\ analysis. The \CWB\ analysis includes the candidate events that have
been both detected and reconstructed (listed in Table \ref{tab:events}),
and those that have not been detected but have been reconstructed offline.
(\protect\NAME{GW190513A}, \protect\NAME{GW190707A}, \protect\NAME{GW190728A},
\protect\NAME{GW190814A}, \protect\NAME{GW190828B}).
The BW analysis uses the same selection of
\ac{BBH} candidate events as were used for testing general relativity~\cite{o3tgr}.
The p-values are sorted in increasing order
and plotted vs. the order number (which is also the cumulative number
of candidate events). Each p-value is obtained from the observed on-source match
value and the corresponding off-source distribution of the match values from
off-source injections. The green band indicates the theoretical 90\% symmetric
distribution about the null hypothesis (dark green line).  Only deviations below
the green 90\% confidence band indicate disagreement, and we see that there are
none. 
A few of the  \CWB\ p-values are above the band, indicating either a statistical fluctuation, 
or an overfitting,
which we attribute to a small asymmetry between the way the on-source and off-
source matches are computed.}
\label{Fig:pvalue}
\end{figure}

Fig.~\ref{Fig:pvalue} shows the p-values for the analyzed candidate events sorted in
increasing order~\cite{Salemi:2019uea,Ghonge:2020suv}. Events with p-values above the diagonal
have on-source matches that are higher than expected, while those below the line
have matches that are lower than expected. 
The \CWB\ analysis shows some events
outside the theoretical 90\% band for the single-order statistic, 
which can be due to statistical fluctuations. Because of correlations between p-values 
in the ordered plot, for 20 p-values the probability of finding exactly four p-values outside 
the 90\% band as in figure \ref{Fig:pvalue} is 5.2\%, although it may also point to
a systematic effect in the analysis. Since the p-values are higher than expected,
there is no evidence of a discrepancy with the template-based analysis. Our tests 
also indicates that if there is indeed a systematic effect, it may originate from an 
overestimate of the off-source matches no larger than 2\%. 
Further details are given in Appendix~\ref{appendix:WCT}. 
Overall, the p-value distributions support the null hypothesis that the minimally-modeled
waveform reconstructions are consistent with the GR derived waveform templates.

%% file: detchar_table.tex
\begin{tabularx}{\columnwidth}{@{\extracolsep{\fill}}p{3cm} Z}
\textbf{Name} & \textbf{Mitigation}\\
\hline
\NAME{GW190413B} & \MITIGATIONMETHOD{GW190413B}\\
\NAME{GW190424A} & \MITIGATIONMETHOD{GW190424A}\\
\NAME{GW190425A} & \MITIGATIONMETHOD{GW190425A}\\
\NAME{GW190503A} & \MITIGATIONMETHOD{GW190503A}\\
\NAME{GW190513A} & \MITIGATIONMETHOD{GW190513A}\\
\NAME{GW190514A} & \MITIGATIONMETHOD{GW190514A}\\
\NAME{GW190701A} & \MITIGATIONMETHOD{GW190701A}\\
\NAME{GW190727A} & \MITIGATIONMETHOD{GW190727A}\\
\NAME{GW190814A} & \MITIGATIONMETHOD{GW190814A}\\
\NAME{GW190924A} & \MITIGATIONMETHOD{GW190924A}\\
\hline
\end{tabularx}

%% file: PE_table.tex
\begin{tabularx}{\textwidth}{@{\extracolsep{\fill}}p{2.85cm} U U U U V U U U U W U}
Event & $\underset{\displaystyle (M_\odot)}{M}$ & $\underset{\displaystyle (M_\odot)}{\mathcal{M}}$ & $\underset{\displaystyle (M_\odot)}{m_1}$ & $\underset{\displaystyle (M_\odot)}{m_2}$ & $\chi_{{\rm eff}}$ & $\underset{\displaystyle ({\rm Gpc})}{D_\mathrm{L}}$ & $z$ & $\underset{\displaystyle (M_\odot)}{M_\mathrm{f}}$ & $\chi_\mathrm{f}$ & $\underset{\displaystyle (\mathrm{deg}^2)}{\Delta\Omega}$ & $\mathrm{SNR}$\\ \hline
\NNAME{GW190408A} & $\totalmasssourcemed{GW190408A}_{ -\totalmasssourceminus{GW190408A} }^{ +\totalmasssourceplus{GW190408A} }$ & $\chirpmasssourcemed{GW190408A}_{ -\chirpmasssourceminus{GW190408A} }^{ +\chirpmasssourceplus{GW190408A} }$ & $\massonesourcemed{GW190408A}_{ -\massonesourceminus{GW190408A} }^{ +\massonesourceplus{GW190408A} }$ & $\masstwosourcemed{GW190408A}_{ -\masstwosourceminus{GW190408A} }^{ +\masstwosourceplus{GW190408A} }$ & $\chieffmed{GW190408A}_{ -\chieffminus{GW190408A} }^{ +\chieffplus{GW190408A} }$ & $\luminositydistancemed{GW190408A}_{ -\luminositydistanceminus{GW190408A} }^{ +\luminositydistanceplus{GW190408A} }$ & $\redshiftmed{GW190408A}_{ -\redshiftminus{GW190408A} }^{ +\redshiftplus{GW190408A} }$ & $\finalmasssourcemed{GW190408A}_{ -\finalmasssourceminus{GW190408A} }^{ +\finalmasssourceplus{GW190408A} }$ & $\finalspinmed{GW190408A}_{ -\finalspinminus{GW190408A} }^{ +\finalspinplus{GW190408A} }$ & $\skyarea{GW190408A}$ & $\networkmatchedfiltersnrIMRmed{GW190408A}_{ -\networkmatchedfiltersnrIMRminus{GW190408A} }^{ +\networkmatchedfiltersnrIMRplus{GW190408A} }$\\
\NNAME{GW190412A} & $\totalmasssourcemed{GW190412A}_{ -\totalmasssourceminus{GW190412A} }^{ +\totalmasssourceplus{GW190412A} }$ & $\chirpmasssourcemed{GW190412A}_{ -\chirpmasssourceminus{GW190412A} }^{ +\chirpmasssourceplus{GW190412A} }$ & $\massonesourcemed{GW190412A}_{ -\massonesourceminus{GW190412A} }^{ +\massonesourceplus{GW190412A} }$ & $\masstwosourcemed{GW190412A}_{ -\masstwosourceminus{GW190412A} }^{ +\masstwosourceplus{GW190412A} }$ & $\chieffmed{GW190412A}_{ -\chieffminus{GW190412A} }^{ +\chieffplus{GW190412A} }$ & $\luminositydistancemed{GW190412A}_{ -\luminositydistanceminus{GW190412A} }^{ +\luminositydistanceplus{GW190412A} }$ & $\redshiftmed{GW190412A}_{ -\redshiftminus{GW190412A} }^{ +\redshiftplus{GW190412A} }$ & $\finalmasssourcemed{GW190412A}_{ -\finalmasssourceminus{GW190412A} }^{ +\finalmasssourceplus{GW190412A} }$ & $\finalspinmed{GW190412A}_{ -\finalspinminus{GW190412A} }^{ +\finalspinplus{GW190412A} }$ & $\skyarea{GW190412A}$ & $\networkmatchedfiltersnrIMRmed{GW190412A}_{ -\networkmatchedfiltersnrIMRminus{GW190412A} }^{ +\networkmatchedfiltersnrIMRplus{GW190412A} }$\\
\NNAME{GW190413A} & $\totalmasssourcemed{GW190413A}_{ -\totalmasssourceminus{GW190413A} }^{ +\totalmasssourceplus{GW190413A} }$ & $\chirpmasssourcemed{GW190413A}_{ -\chirpmasssourceminus{GW190413A} }^{ +\chirpmasssourceplus{GW190413A} }$ & $\massonesourcemed{GW190413A}_{ -\massonesourceminus{GW190413A} }^{ +\massonesourceplus{GW190413A} }$ & $\masstwosourcemed{GW190413A}_{ -\masstwosourceminus{GW190413A} }^{ +\masstwosourceplus{GW190413A} }$ & $\chieffmed{GW190413A}_{ -\chieffminus{GW190413A} }^{ +\chieffplus{GW190413A} }$ & $\luminositydistancemed{GW190413A}_{ -\luminositydistanceminus{GW190413A} }^{ +\luminositydistanceplus{GW190413A} }$ & $\redshiftmed{GW190413A}_{ -\redshiftminus{GW190413A} }^{ +\redshiftplus{GW190413A} }$ & $\finalmasssourcemed{GW190413A}_{ -\finalmasssourceminus{GW190413A} }^{ +\finalmasssourceplus{GW190413A} }$ & $\finalspinmed{GW190413A}_{ -\finalspinminus{GW190413A} }^{ +\finalspinplus{GW190413A} }$ & $\skyarea{GW190413A}$ & $\networkmatchedfiltersnrIMRmed{GW190413A}_{ -\networkmatchedfiltersnrIMRminus{GW190413A} }^{ +\networkmatchedfiltersnrIMRplus{GW190413A} }$\\
\NNAME{GW190413B} & $\totalmasssourcemed{GW190413B}_{ -\totalmasssourceminus{GW190413B} }^{ +\totalmasssourceplus{GW190413B} }$ & $\chirpmasssourcemed{GW190413B}_{ -\chirpmasssourceminus{GW190413B} }^{ +\chirpmasssourceplus{GW190413B} }$ & $\massonesourcemed{GW190413B}_{ -\massonesourceminus{GW190413B} }^{ +\massonesourceplus{GW190413B} }$ & $\masstwosourcemed{GW190413B}_{ -\masstwosourceminus{GW190413B} }^{ +\masstwosourceplus{GW190413B} }$ & $\chieffmed{GW190413B}_{ -\chieffminus{GW190413B} }^{ +\chieffplus{GW190413B} }$ & $\luminositydistancemed{GW190413B}_{ -\luminositydistanceminus{GW190413B} }^{ +\luminositydistanceplus{GW190413B} }$ & $\redshiftmed{GW190413B}_{ -\redshiftminus{GW190413B} }^{ +\redshiftplus{GW190413B} }$ & $\finalmasssourcemed{GW190413B}_{ -\finalmasssourceminus{GW190413B} }^{ +\finalmasssourceplus{GW190413B} }$ & $\finalspinmed{GW190413B}_{ -\finalspinminus{GW190413B} }^{ +\finalspinplus{GW190413B} }$ & $\skyarea{GW190413B}$ & $\networkmatchedfiltersnrIMRmed{GW190413B}_{ -\networkmatchedfiltersnrIMRminus{GW190413B} }^{ +\networkmatchedfiltersnrIMRplus{GW190413B} }$\\
\NNAME{GW190421A} & $\totalmasssourcemed{GW190421A}_{ -\totalmasssourceminus{GW190421A} }^{ +\totalmasssourceplus{GW190421A} }$ & $\chirpmasssourcemed{GW190421A}_{ -\chirpmasssourceminus{GW190421A} }^{ +\chirpmasssourceplus{GW190421A} }$ & $\massonesourcemed{GW190421A}_{ -\massonesourceminus{GW190421A} }^{ +\massonesourceplus{GW190421A} }$ & $\masstwosourcemed{GW190421A}_{ -\masstwosourceminus{GW190421A} }^{ +\masstwosourceplus{GW190421A} }$ & $\chieffmed{GW190421A}_{ -\chieffminus{GW190421A} }^{ +\chieffplus{GW190421A} }$ & $\luminositydistancemed{GW190421A}_{ -\luminositydistanceminus{GW190421A} }^{ +\luminositydistanceplus{GW190421A} }$ & $\redshiftmed{GW190421A}_{ -\redshiftminus{GW190421A} }^{ +\redshiftplus{GW190421A} }$ & $\finalmasssourcemed{GW190421A}_{ -\finalmasssourceminus{GW190421A} }^{ +\finalmasssourceplus{GW190421A} }$ & $\finalspinmed{GW190421A}_{ -\finalspinminus{GW190421A} }^{ +\finalspinplus{GW190421A} }$ & $\skyarea{GW190421A}$ & $\networkmatchedfiltersnrIMRmed{GW190421A}_{ -\networkmatchedfiltersnrIMRminus{GW190421A} }^{ +\networkmatchedfiltersnrIMRplus{GW190421A} }$\\
\NNAME{GW190424A} & $\totalmasssourcemed{GW190424A}_{ -\totalmasssourceminus{GW190424A} }^{ +\totalmasssourceplus{GW190424A} }$ & $\chirpmasssourcemed{GW190424A}_{ -\chirpmasssourceminus{GW190424A} }^{ +\chirpmasssourceplus{GW190424A} }$ & $\massonesourcemed{GW190424A}_{ -\massonesourceminus{GW190424A} }^{ +\massonesourceplus{GW190424A} }$ & $\masstwosourcemed{GW190424A}_{ -\masstwosourceminus{GW190424A} }^{ +\masstwosourceplus{GW190424A} }$ & $\chieffmed{GW190424A}_{ -\chieffminus{GW190424A} }^{ +\chieffplus{GW190424A} }$ & $\luminositydistancemed{GW190424A}_{ -\luminositydistanceminus{GW190424A} }^{ +\luminositydistanceplus{GW190424A} }$ & $\redshiftmed{GW190424A}_{ -\redshiftminus{GW190424A} }^{ +\redshiftplus{GW190424A} }$ & $\finalmasssourcemed{GW190424A}_{ -\finalmasssourceminus{GW190424A} }^{ +\finalmasssourceplus{GW190424A} }$ & $\finalspinmed{GW190424A}_{ -\finalspinminus{GW190424A} }^{ +\finalspinplus{GW190424A} }$ & $\skyarea{GW190424A}$ & $\networkmatchedfiltersnrIMRmed{GW190424A}_{ -\networkmatchedfiltersnrIMRminus{GW190424A} }^{ +\networkmatchedfiltersnrIMRplus{GW190424A} }$\\
\NNAME{GW190425A} & $\totalmasssourcemed{GW190425A}_{ -\totalmasssourceminus{GW190425A} }^{ +\totalmasssourceplus{GW190425A} }$ & $\chirpmasssourcemed{GW190425A}_{ -\chirpmasssourceminus{GW190425A} }^{ +\chirpmasssourceplus{GW190425A} }$ & $\massonesourcemed{GW190425A}_{ -\massonesourceminus{GW190425A} }^{ +\massonesourceplus{GW190425A} }$ & $\masstwosourcemed{GW190425A}_{ -\masstwosourceminus{GW190425A} }^{ +\masstwosourceplus{GW190425A} }$ & $\chieffmed{GW190425A}_{ -\chieffminus{GW190425A} }^{ +\chieffplus{GW190425A} }$ & $\luminositydistancemed{GW190425A}_{ -\luminositydistanceminus{GW190425A} }^{ +\luminositydistanceplus{GW190425A} }$ & $\redshiftmed{GW190425A}_{ -\redshiftminus{GW190425A} }^{ +\redshiftplus{GW190425A} }$ & -- & -- & $\skyarea{GW190425A}$ & $\networkmatchedfiltersnrIMRmed{GW190425A}_{ -\networkmatchedfiltersnrIMRminus{GW190425A} }^{ +\networkmatchedfiltersnrIMRplus{GW190425A} }$\\
\NNAME{GW190426A} & $\totalmasssourcemed{GW190426A}_{ -\totalmasssourceminus{GW190426A} }^{ +\totalmasssourceplus{GW190426A} }$ & $\chirpmasssourcemed{GW190426A}_{ -\chirpmasssourceminus{GW190426A} }^{ +\chirpmasssourceplus{GW190426A} }$ & $\massonesourcemed{GW190426A}_{ -\massonesourceminus{GW190426A} }^{ +\massonesourceplus{GW190426A} }$ & $\masstwosourcemed{GW190426A}_{ -\masstwosourceminus{GW190426A} }^{ +\masstwosourceplus{GW190426A} }$ & $\chieffmed{GW190426A}_{ -\chieffminus{GW190426A} }^{ +\chieffplus{GW190426A} }$ & $\luminositydistancemed{GW190426A}_{ -\luminositydistanceminus{GW190426A} }^{ +\luminositydistanceplus{GW190426A} }$ & $\redshiftmed{GW190426A}_{ -\redshiftminus{GW190426A} }^{ +\redshiftplus{GW190426A} }$ & -- & -- & $\skyarea{GW190426A}$ & $\networkmatchedfiltersnrIMRmed{GW190426A}_{ -\networkmatchedfiltersnrIMRminus{GW190426A} }^{ +\networkmatchedfiltersnrIMRplus{GW190426A} }$\\
\NNAME{GW190503A} & $\totalmasssourcemed{GW190503A}_{ -\totalmasssourceminus{GW190503A} }^{ +\totalmasssourceplus{GW190503A} }$ & $\chirpmasssourcemed{GW190503A}_{ -\chirpmasssourceminus{GW190503A} }^{ +\chirpmasssourceplus{GW190503A} }$ & $\massonesourcemed{GW190503A}_{ -\massonesourceminus{GW190503A} }^{ +\massonesourceplus{GW190503A} }$ & $\masstwosourcemed{GW190503A}_{ -\masstwosourceminus{GW190503A} }^{ +\masstwosourceplus{GW190503A} }$ & $\chieffmed{GW190503A}_{ -\chieffminus{GW190503A} }^{ +\chieffplus{GW190503A} }$ & $\luminositydistancemed{GW190503A}_{ -\luminositydistanceminus{GW190503A} }^{ +\luminositydistanceplus{GW190503A} }$ & $\redshiftmed{GW190503A}_{ -\redshiftminus{GW190503A} }^{ +\redshiftplus{GW190503A} }$ & $\finalmasssourcemed{GW190503A}_{ -\finalmasssourceminus{GW190503A} }^{ +\finalmasssourceplus{GW190503A} }$ & $\finalspinmed{GW190503A}_{ -\finalspinminus{GW190503A} }^{ +\finalspinplus{GW190503A} }$ & $\skyarea{GW190503A}$ & $\networkmatchedfiltersnrIMRmed{GW190503A}_{ -\networkmatchedfiltersnrIMRminus{GW190503A} }^{ +\networkmatchedfiltersnrIMRplus{GW190503A} }$\\
\NNAME{GW190512A} & $\totalmasssourcemed{GW190512A}_{ -\totalmasssourceminus{GW190512A} }^{ +\totalmasssourceplus{GW190512A} }$ & $\chirpmasssourcemed{GW190512A}_{ -\chirpmasssourceminus{GW190512A} }^{ +\chirpmasssourceplus{GW190512A} }$ & $\massonesourcemed{GW190512A}_{ -\massonesourceminus{GW190512A} }^{ +\massonesourceplus{GW190512A} }$ & $\masstwosourcemed{GW190512A}_{ -\masstwosourceminus{GW190512A} }^{ +\masstwosourceplus{GW190512A} }$ & $\chieffmed{GW190512A}_{ -\chieffminus{GW190512A} }^{ +\chieffplus{GW190512A} }$ & $\luminositydistancemed{GW190512A}_{ -\luminositydistanceminus{GW190512A} }^{ +\luminositydistanceplus{GW190512A} }$ & $\redshiftmed{GW190512A}_{ -\redshiftminus{GW190512A} }^{ +\redshiftplus{GW190512A} }$ & $\finalmasssourcemed{GW190512A}_{ -\finalmasssourceminus{GW190512A} }^{ +\finalmasssourceplus{GW190512A} }$ & $\finalspinmed{GW190512A}_{ -\finalspinminus{GW190512A} }^{ +\finalspinplus{GW190512A} }$ & $\skyarea{GW190512A}$ & $\networkmatchedfiltersnrIMRmed{GW190512A}_{ -\networkmatchedfiltersnrIMRminus{GW190512A} }^{ +\networkmatchedfiltersnrIMRplus{GW190512A} }$\\
\NNAME{GW190513A} & $\totalmasssourcemed{GW190513A}_{ -\totalmasssourceminus{GW190513A} }^{ +\totalmasssourceplus{GW190513A} }$ & $\chirpmasssourcemed{GW190513A}_{ -\chirpmasssourceminus{GW190513A} }^{ +\chirpmasssourceplus{GW190513A} }$ & $\massonesourcemed{GW190513A}_{ -\massonesourceminus{GW190513A} }^{ +\massonesourceplus{GW190513A} }$ & $\masstwosourcemed{GW190513A}_{ -\masstwosourceminus{GW190513A} }^{ +\masstwosourceplus{GW190513A} }$ & $\chieffmed{GW190513A}_{ -\chieffminus{GW190513A} }^{ +\chieffplus{GW190513A} }$ & $\luminositydistancemed{GW190513A}_{ -\luminositydistanceminus{GW190513A} }^{ +\luminositydistanceplus{GW190513A} }$ & $\redshiftmed{GW190513A}_{ -\redshiftminus{GW190513A} }^{ +\redshiftplus{GW190513A} }$ & $\finalmasssourcemed{GW190513A}_{ -\finalmasssourceminus{GW190513A} }^{ +\finalmasssourceplus{GW190513A} }$ & $\finalspinmed{GW190513A}_{ -\finalspinminus{GW190513A} }^{ +\finalspinplus{GW190513A} }$ & $\skyarea{GW190513A}$ & $\networkmatchedfiltersnrIMRmed{GW190513A}_{ -\networkmatchedfiltersnrIMRminus{GW190513A} }^{ +\networkmatchedfiltersnrIMRplus{GW190513A} }$\\
\NNAME{GW190514A} & $\totalmasssourcemed{GW190514A}_{ -\totalmasssourceminus{GW190514A} }^{ +\totalmasssourceplus{GW190514A} }$ & $\chirpmasssourcemed{GW190514A}_{ -\chirpmasssourceminus{GW190514A} }^{ +\chirpmasssourceplus{GW190514A} }$ & $\massonesourcemed{GW190514A}_{ -\massonesourceminus{GW190514A} }^{ +\massonesourceplus{GW190514A} }$ & $\masstwosourcemed{GW190514A}_{ -\masstwosourceminus{GW190514A} }^{ +\masstwosourceplus{GW190514A} }$ & $\chieffmed{GW190514A}_{ -\chieffminus{GW190514A} }^{ +\chieffplus{GW190514A} }$ & $\luminositydistancemed{GW190514A}_{ -\luminositydistanceminus{GW190514A} }^{ +\luminositydistanceplus{GW190514A} }$ & $\redshiftmed{GW190514A}_{ -\redshiftminus{GW190514A} }^{ +\redshiftplus{GW190514A} }$ & $\finalmasssourcemed{GW190514A}_{ -\finalmasssourceminus{GW190514A} }^{ +\finalmasssourceplus{GW190514A} }$ & $\finalspinmed{GW190514A}_{ -\finalspinminus{GW190514A} }^{ +\finalspinplus{GW190514A} }$ & $\skyarea{GW190514A}$ & $\networkmatchedfiltersnrIMRmed{GW190514A}_{ -\networkmatchedfiltersnrIMRminus{GW190514A} }^{ +\networkmatchedfiltersnrIMRplus{GW190514A} }$\\
\NNAME{GW190517A} & $\totalmasssourcemed{GW190517A}_{ -\totalmasssourceminus{GW190517A} }^{ +\totalmasssourceplus{GW190517A} }$ & $\chirpmasssourcemed{GW190517A}_{ -\chirpmasssourceminus{GW190517A} }^{ +\chirpmasssourceplus{GW190517A} }$ & $\massonesourcemed{GW190517A}_{ -\massonesourceminus{GW190517A} }^{ +\massonesourceplus{GW190517A} }$ & $\masstwosourcemed{GW190517A}_{ -\masstwosourceminus{GW190517A} }^{ +\masstwosourceplus{GW190517A} }$ & $\chieffmed{GW190517A}_{ -\chieffminus{GW190517A} }^{ +\chieffplus{GW190517A} }$ & $\luminositydistancemed{GW190517A}_{ -\luminositydistanceminus{GW190517A} }^{ +\luminositydistanceplus{GW190517A} }$ & $\redshiftmed{GW190517A}_{ -\redshiftminus{GW190517A} }^{ +\redshiftplus{GW190517A} }$ & $\finalmasssourcemed{GW190517A}_{ -\finalmasssourceminus{GW190517A} }^{ +\finalmasssourceplus{GW190517A} }$ & $\finalspinmed{GW190517A}_{ -\finalspinminus{GW190517A} }^{ +\finalspinplus{GW190517A} }$ & $\skyarea{GW190517A}$ & $\networkmatchedfiltersnrIMRmed{GW190517A}_{ -\networkmatchedfiltersnrIMRminus{GW190517A} }^{ +\networkmatchedfiltersnrIMRplus{GW190517A} }$\\
\NNAME{GW190519A} & $\totalmasssourcemed{GW190519A}_{ -\totalmasssourceminus{GW190519A} }^{ +\totalmasssourceplus{GW190519A} }$ & $\chirpmasssourcemed{GW190519A}_{ -\chirpmasssourceminus{GW190519A} }^{ +\chirpmasssourceplus{GW190519A} }$ & $\massonesourcemed{GW190519A}_{ -\massonesourceminus{GW190519A} }^{ +\massonesourceplus{GW190519A} }$ & $\masstwosourcemed{GW190519A}_{ -\masstwosourceminus{GW190519A} }^{ +\masstwosourceplus{GW190519A} }$ & $\chieffmed{GW190519A}_{ -\chieffminus{GW190519A} }^{ +\chieffplus{GW190519A} }$ & $\luminositydistancemed{GW190519A}_{ -\luminositydistanceminus{GW190519A} }^{ +\luminositydistanceplus{GW190519A} }$ & $\redshiftmed{GW190519A}_{ -\redshiftminus{GW190519A} }^{ +\redshiftplus{GW190519A} }$ & $\finalmasssourcemed{GW190519A}_{ -\finalmasssourceminus{GW190519A} }^{ +\finalmasssourceplus{GW190519A} }$ & $\finalspinmed{GW190519A}_{ -\finalspinminus{GW190519A} }^{ +\finalspinplus{GW190519A} }$ & $\skyarea{GW190519A}$ & $\networkmatchedfiltersnrIMRmed{GW190519A}_{ -\networkmatchedfiltersnrIMRminus{GW190519A} }^{ +\networkmatchedfiltersnrIMRplus{GW190519A} }$\\
\NNAME{GW190521A} & $\totalmasssourcemed{GW190521A}_{ -\totalmasssourceminus{GW190521A} }^{ +\totalmasssourceplus{GW190521A} }$ & $\chirpmasssourcemed{GW190521A}_{ -\chirpmasssourceminus{GW190521A} }^{ +\chirpmasssourceplus{GW190521A} }$ & $\massonesourcemed{GW190521A}_{ -\massonesourceminus{GW190521A} }^{ +\massonesourceplus{GW190521A} }$ & $\masstwosourcemed{GW190521A}_{ -\masstwosourceminus{GW190521A} }^{ +\masstwosourceplus{GW190521A} }$ & $\chieffmed{GW190521A}_{ -\chieffminus{GW190521A} }^{ +\chieffplus{GW190521A} }$ & $\luminositydistancemed{GW190521A}_{ -\luminositydistanceminus{GW190521A} }^{ +\luminositydistanceplus{GW190521A} }$ & $\redshiftmed{GW190521A}_{ -\redshiftminus{GW190521A} }^{ +\redshiftplus{GW190521A} }$ & $\finalmasssourcemed{GW190521A}_{ -\finalmasssourceminus{GW190521A} }^{ +\finalmasssourceplus{GW190521A} }$ & $\finalspinmed{GW190521A}_{ -\finalspinminus{GW190521A} }^{ +\finalspinplus{GW190521A} }$ & $\skyarea{GW190521A}$ & $\networkmatchedfiltersnrIMRmed{GW190521A}_{ -\networkmatchedfiltersnrIMRminus{GW190521A} }^{ +\networkmatchedfiltersnrIMRplus{GW190521A} }$\\
\NNAME{GW190521B} & $\totalmasssourcemed{GW190521B}_{ -\totalmasssourceminus{GW190521B} }^{ +\totalmasssourceplus{GW190521B} }$ & $\chirpmasssourcemed{GW190521B}_{ -\chirpmasssourceminus{GW190521B} }^{ +\chirpmasssourceplus{GW190521B} }$ & $\massonesourcemed{GW190521B}_{ -\massonesourceminus{GW190521B} }^{ +\massonesourceplus{GW190521B} }$ & $\masstwosourcemed{GW190521B}_{ -\masstwosourceminus{GW190521B} }^{ +\masstwosourceplus{GW190521B} }$ & $\chieffmed{GW190521B}_{ -\chieffminus{GW190521B} }^{ +\chieffplus{GW190521B} }$ & $\luminositydistancemed{GW190521B}_{ -\luminositydistanceminus{GW190521B} }^{ +\luminositydistanceplus{GW190521B} }$ & $\redshiftmed{GW190521B}_{ -\redshiftminus{GW190521B} }^{ +\redshiftplus{GW190521B} }$ & $\finalmasssourcemed{GW190521B}_{ -\finalmasssourceminus{GW190521B} }^{ +\finalmasssourceplus{GW190521B} }$ & $\finalspinmed{GW190521B}_{ -\finalspinminus{GW190521B} }^{ +\finalspinplus{GW190521B} }$ & $\skyarea{GW190521B}$ & $\networkmatchedfiltersnrIMRmed{GW190521B}_{ -\networkmatchedfiltersnrIMRminus{GW190521B} }^{ +\networkmatchedfiltersnrIMRplus{GW190521B} }$\\
\NNAME{GW190527A} & $\totalmasssourcemed{GW190527A}_{ -\totalmasssourceminus{GW190527A} }^{ +\totalmasssourceplus{GW190527A} }$ & $\chirpmasssourcemed{GW190527A}_{ -\chirpmasssourceminus{GW190527A} }^{ +\chirpmasssourceplus{GW190527A} }$ & $\massonesourcemed{GW190527A}_{ -\massonesourceminus{GW190527A} }^{ +\massonesourceplus{GW190527A} }$ & $\masstwosourcemed{GW190527A}_{ -\masstwosourceminus{GW190527A} }^{ +\masstwosourceplus{GW190527A} }$ & $\chieffmed{GW190527A}_{ -\chieffminus{GW190527A} }^{ +\chieffplus{GW190527A} }$ & $\luminositydistancemed{GW190527A}_{ -\luminositydistanceminus{GW190527A} }^{ +\luminositydistanceplus{GW190527A} }$ & $\redshiftmed{GW190527A}_{ -\redshiftminus{GW190527A} }^{ +\redshiftplus{GW190527A} }$ & $\finalmasssourcemed{GW190527A}_{ -\finalmasssourceminus{GW190527A} }^{ +\finalmasssourceplus{GW190527A} }$ & $\finalspinmed{GW190527A}_{ -\finalspinminus{GW190527A} }^{ +\finalspinplus{GW190527A} }$ & $\skyarea{GW190527A}$ & $\networkmatchedfiltersnrIMRmed{GW190527A}_{ -\networkmatchedfiltersnrIMRminus{GW190527A} }^{ +\networkmatchedfiltersnrIMRplus{GW190527A} }$\\
\NNAME{GW190602A} & $\totalmasssourcemed{GW190602A}_{ -\totalmasssourceminus{GW190602A} }^{ +\totalmasssourceplus{GW190602A} }$ & $\chirpmasssourcemed{GW190602A}_{ -\chirpmasssourceminus{GW190602A} }^{ +\chirpmasssourceplus{GW190602A} }$ & $\massonesourcemed{GW190602A}_{ -\massonesourceminus{GW190602A} }^{ +\massonesourceplus{GW190602A} }$ & $\masstwosourcemed{GW190602A}_{ -\masstwosourceminus{GW190602A} }^{ +\masstwosourceplus{GW190602A} }$ & $\chieffmed{GW190602A}_{ -\chieffminus{GW190602A} }^{ +\chieffplus{GW190602A} }$ & $\luminositydistancemed{GW190602A}_{ -\luminositydistanceminus{GW190602A} }^{ +\luminositydistanceplus{GW190602A} }$ & $\redshiftmed{GW190602A}_{ -\redshiftminus{GW190602A} }^{ +\redshiftplus{GW190602A} }$ & $\finalmasssourcemed{GW190602A}_{ -\finalmasssourceminus{GW190602A} }^{ +\finalmasssourceplus{GW190602A} }$ & $\finalspinmed{GW190602A}_{ -\finalspinminus{GW190602A} }^{ +\finalspinplus{GW190602A} }$ & $\skyarea{GW190602A}$ & $\networkmatchedfiltersnrIMRmed{GW190602A}_{ -\networkmatchedfiltersnrIMRminus{GW190602A} }^{ +\networkmatchedfiltersnrIMRplus{GW190602A} }$\\
\NNAME{GW190620A} & $\totalmasssourcemed{GW190620A}_{ -\totalmasssourceminus{GW190620A} }^{ +\totalmasssourceplus{GW190620A} }$ & $\chirpmasssourcemed{GW190620A}_{ -\chirpmasssourceminus{GW190620A} }^{ +\chirpmasssourceplus{GW190620A} }$ & $\massonesourcemed{GW190620A}_{ -\massonesourceminus{GW190620A} }^{ +\massonesourceplus{GW190620A} }$ & $\masstwosourcemed{GW190620A}_{ -\masstwosourceminus{GW190620A} }^{ +\masstwosourceplus{GW190620A} }$ & $\chieffmed{GW190620A}_{ -\chieffminus{GW190620A} }^{ +\chieffplus{GW190620A} }$ & $\luminositydistancemed{GW190620A}_{ -\luminositydistanceminus{GW190620A} }^{ +\luminositydistanceplus{GW190620A} }$ & $\redshiftmed{GW190620A}_{ -\redshiftminus{GW190620A} }^{ +\redshiftplus{GW190620A} }$ & $\finalmasssourcemed{GW190620A}_{ -\finalmasssourceminus{GW190620A} }^{ +\finalmasssourceplus{GW190620A} }$ & $\finalspinmed{GW190620A}_{ -\finalspinminus{GW190620A} }^{ +\finalspinplus{GW190620A} }$ & $\skyarea{GW190620A}$ & $\networkmatchedfiltersnrIMRmed{GW190620A}_{ -\networkmatchedfiltersnrIMRminus{GW190620A} }^{ +\networkmatchedfiltersnrIMRplus{GW190620A} }$\\
\NNAME{GW190630A} & $\totalmasssourcemed{GW190630A}_{ -\totalmasssourceminus{GW190630A} }^{ +\totalmasssourceplus{GW190630A} }$ & $\chirpmasssourcemed{GW190630A}_{ -\chirpmasssourceminus{GW190630A} }^{ +\chirpmasssourceplus{GW190630A} }$ & $\massonesourcemed{GW190630A}_{ -\massonesourceminus{GW190630A} }^{ +\massonesourceplus{GW190630A} }$ & $\masstwosourcemed{GW190630A}_{ -\masstwosourceminus{GW190630A} }^{ +\masstwosourceplus{GW190630A} }$ & $\chieffmed{GW190630A}_{ -\chieffminus{GW190630A} }^{ +\chieffplus{GW190630A} }$ & $\luminositydistancemed{GW190630A}_{ -\luminositydistanceminus{GW190630A} }^{ +\luminositydistanceplus{GW190630A} }$ & $\redshiftmed{GW190630A}_{ -\redshiftminus{GW190630A} }^{ +\redshiftplus{GW190630A} }$ & $\finalmasssourcemed{GW190630A}_{ -\finalmasssourceminus{GW190630A} }^{ +\finalmasssourceplus{GW190630A} }$ & $\finalspinmed{GW190630A}_{ -\finalspinminus{GW190630A} }^{ +\finalspinplus{GW190630A} }$ & $\skyarea{GW190630A}$ & $\networkmatchedfiltersnrIMRmed{GW190630A}_{ -\networkmatchedfiltersnrIMRminus{GW190630A} }^{ +\networkmatchedfiltersnrIMRplus{GW190630A} }$\\
\NNAME{GW190701A} & $\totalmasssourcemed{GW190701A}_{ -\totalmasssourceminus{GW190701A} }^{ +\totalmasssourceplus{GW190701A} }$ & $\chirpmasssourcemed{GW190701A}_{ -\chirpmasssourceminus{GW190701A} }^{ +\chirpmasssourceplus{GW190701A} }$ & $\massonesourcemed{GW190701A}_{ -\massonesourceminus{GW190701A} }^{ +\massonesourceplus{GW190701A} }$ & $\masstwosourcemed{GW190701A}_{ -\masstwosourceminus{GW190701A} }^{ +\masstwosourceplus{GW190701A} }$ & $\chieffmed{GW190701A}_{ -\chieffminus{GW190701A} }^{ +\chieffplus{GW190701A} }$ & $\luminositydistancemed{GW190701A}_{ -\luminositydistanceminus{GW190701A} }^{ +\luminositydistanceplus{GW190701A} }$ & $\redshiftmed{GW190701A}_{ -\redshiftminus{GW190701A} }^{ +\redshiftplus{GW190701A} }$ & $\finalmasssourcemed{GW190701A}_{ -\finalmasssourceminus{GW190701A} }^{ +\finalmasssourceplus{GW190701A} }$ & $\finalspinmed{GW190701A}_{ -\finalspinminus{GW190701A} }^{ +\finalspinplus{GW190701A} }$ & $\skyarea{GW190701A}$ & $\networkmatchedfiltersnrIMRmed{GW190701A}_{ -\networkmatchedfiltersnrIMRminus{GW190701A} }^{ +\networkmatchedfiltersnrIMRplus{GW190701A} }$\\
\NNAME{GW190706A} & $\totalmasssourcemed{GW190706A}_{ -\totalmasssourceminus{GW190706A} }^{ +\totalmasssourceplus{GW190706A} }$ & $\chirpmasssourcemed{GW190706A}_{ -\chirpmasssourceminus{GW190706A} }^{ +\chirpmasssourceplus{GW190706A} }$ & $\massonesourcemed{GW190706A}_{ -\massonesourceminus{GW190706A} }^{ +\massonesourceplus{GW190706A} }$ & $\masstwosourcemed{GW190706A}_{ -\masstwosourceminus{GW190706A} }^{ +\masstwosourceplus{GW190706A} }$ & $\chieffmed{GW190706A}_{ -\chieffminus{GW190706A} }^{ +\chieffplus{GW190706A} }$ & $\luminositydistancemed{GW190706A}_{ -\luminositydistanceminus{GW190706A} }^{ +\luminositydistanceplus{GW190706A} }$ & $\redshiftmed{GW190706A}_{ -\redshiftminus{GW190706A} }^{ +\redshiftplus{GW190706A} }$ & $\finalmasssourcemed{GW190706A}_{ -\finalmasssourceminus{GW190706A} }^{ +\finalmasssourceplus{GW190706A} }$ & $\finalspinmed{GW190706A}_{ -\finalspinminus{GW190706A} }^{ +\finalspinplus{GW190706A} }$ & $\skyarea{GW190706A}$ & $\networkmatchedfiltersnrIMRmed{GW190706A}_{ -\networkmatchedfiltersnrIMRminus{GW190706A} }^{ +\networkmatchedfiltersnrIMRplus{GW190706A} }$\\
\NNAME{GW190707A} & $\totalmasssourcemed{GW190707A}_{ -\totalmasssourceminus{GW190707A} }^{ +\totalmasssourceplus{GW190707A} }$ & $\chirpmasssourcemed{GW190707A}_{ -\chirpmasssourceminus{GW190707A} }^{ +\chirpmasssourceplus{GW190707A} }$ & $\massonesourcemed{GW190707A}_{ -\massonesourceminus{GW190707A} }^{ +\massonesourceplus{GW190707A} }$ & $\masstwosourcemed{GW190707A}_{ -\masstwosourceminus{GW190707A} }^{ +\masstwosourceplus{GW190707A} }$ & $\chieffmed{GW190707A}_{ -\chieffminus{GW190707A} }^{ +\chieffplus{GW190707A} }$ & $\luminositydistancemed{GW190707A}_{ -\luminositydistanceminus{GW190707A} }^{ +\luminositydistanceplus{GW190707A} }$ & $\redshiftmed{GW190707A}_{ -\redshiftminus{GW190707A} }^{ +\redshiftplus{GW190707A} }$ & $\finalmasssourcemed{GW190707A}_{ -\finalmasssourceminus{GW190707A} }^{ +\finalmasssourceplus{GW190707A} }$ & $\finalspinmed{GW190707A}_{ -\finalspinminus{GW190707A} }^{ +\finalspinplus{GW190707A} }$ & $\skyarea{GW190707A}$ & $\networkmatchedfiltersnrIMRmed{GW190707A}_{ -\networkmatchedfiltersnrIMRminus{GW190707A} }^{ +\networkmatchedfiltersnrIMRplus{GW190707A} }$\\
\NNAME{GW190708A} & $\totalmasssourcemed{GW190708A}_{ -\totalmasssourceminus{GW190708A} }^{ +\totalmasssourceplus{GW190708A} }$ & $\chirpmasssourcemed{GW190708A}_{ -\chirpmasssourceminus{GW190708A} }^{ +\chirpmasssourceplus{GW190708A} }$ & $\massonesourcemed{GW190708A}_{ -\massonesourceminus{GW190708A} }^{ +\massonesourceplus{GW190708A} }$ & $\masstwosourcemed{GW190708A}_{ -\masstwosourceminus{GW190708A} }^{ +\masstwosourceplus{GW190708A} }$ & $\chieffmed{GW190708A}_{ -\chieffminus{GW190708A} }^{ +\chieffplus{GW190708A} }$ & $\luminositydistancemed{GW190708A}_{ -\luminositydistanceminus{GW190708A} }^{ +\luminositydistanceplus{GW190708A} }$ & $\redshiftmed{GW190708A}_{ -\redshiftminus{GW190708A} }^{ +\redshiftplus{GW190708A} }$ & $\finalmasssourcemed{GW190708A}_{ -\finalmasssourceminus{GW190708A} }^{ +\finalmasssourceplus{GW190708A} }$ & $\finalspinmed{GW190708A}_{ -\finalspinminus{GW190708A} }^{ +\finalspinplus{GW190708A} }$ & $\skyarea{GW190708A}$ & $\networkmatchedfiltersnrIMRmed{GW190708A}_{ -\networkmatchedfiltersnrIMRminus{GW190708A} }^{ +\networkmatchedfiltersnrIMRplus{GW190708A} }$\\
\NNAME{GW190719A} & $\totalmasssourcemed{GW190719A}_{ -\totalmasssourceminus{GW190719A} }^{ +\totalmasssourceplus{GW190719A} }$ & $\chirpmasssourcemed{GW190719A}_{ -\chirpmasssourceminus{GW190719A} }^{ +\chirpmasssourceplus{GW190719A} }$ & $\massonesourcemed{GW190719A}_{ -\massonesourceminus{GW190719A} }^{ +\massonesourceplus{GW190719A} }$ & $\masstwosourcemed{GW190719A}_{ -\masstwosourceminus{GW190719A} }^{ +\masstwosourceplus{GW190719A} }$ & $\chieffmed{GW190719A}_{ -\chieffminus{GW190719A} }^{ +\chieffplus{GW190719A} }$ & $\luminositydistancemed{GW190719A}_{ -\luminositydistanceminus{GW190719A} }^{ +\luminositydistanceplus{GW190719A} }$ & $\redshiftmed{GW190719A}_{ -\redshiftminus{GW190719A} }^{ +\redshiftplus{GW190719A} }$ & $\finalmasssourcemed{GW190719A}_{ -\finalmasssourceminus{GW190719A} }^{ +\finalmasssourceplus{GW190719A} }$ & $\finalspinmed{GW190719A}_{ -\finalspinminus{GW190719A} }^{ +\finalspinplus{GW190719A} }$ & $\skyarea{GW190719A}$ & $\networkmatchedfiltersnrIMRmed{GW190719A}_{ -\networkmatchedfiltersnrIMRminus{GW190719A} }^{ +\networkmatchedfiltersnrIMRplus{GW190719A} }$\\
\NNAME{GW190720A} & $\totalmasssourcemed{GW190720A}_{ -\totalmasssourceminus{GW190720A} }^{ +\totalmasssourceplus{GW190720A} }$ & $\chirpmasssourcemed{GW190720A}_{ -\chirpmasssourceminus{GW190720A} }^{ +\chirpmasssourceplus{GW190720A} }$ & $\massonesourcemed{GW190720A}_{ -\massonesourceminus{GW190720A} }^{ +\massonesourceplus{GW190720A} }$ & $\masstwosourcemed{GW190720A}_{ -\masstwosourceminus{GW190720A} }^{ +\masstwosourceplus{GW190720A} }$ & $\chieffmed{GW190720A}_{ -\chieffminus{GW190720A} }^{ +\chieffplus{GW190720A} }$ & $\luminositydistancemed{GW190720A}_{ -\luminositydistanceminus{GW190720A} }^{ +\luminositydistanceplus{GW190720A} }$ & $\redshiftmed{GW190720A}_{ -\redshiftminus{GW190720A} }^{ +\redshiftplus{GW190720A} }$ & $\finalmasssourcemed{GW190720A}_{ -\finalmasssourceminus{GW190720A} }^{ +\finalmasssourceplus{GW190720A} }$ & $\finalspinmed{GW190720A}_{ -\finalspinminus{GW190720A} }^{ +\finalspinplus{GW190720A} }$ & $\skyarea{GW190720A}$ & $\networkmatchedfiltersnrIMRmed{GW190720A}_{ -\networkmatchedfiltersnrIMRminus{GW190720A} }^{ +\networkmatchedfiltersnrIMRplus{GW190720A} }$\\
\NNAME{GW190727A} & $\totalmasssourcemed{GW190727A}_{ -\totalmasssourceminus{GW190727A} }^{ +\totalmasssourceplus{GW190727A} }$ & $\chirpmasssourcemed{GW190727A}_{ -\chirpmasssourceminus{GW190727A} }^{ +\chirpmasssourceplus{GW190727A} }$ & $\massonesourcemed{GW190727A}_{ -\massonesourceminus{GW190727A} }^{ +\massonesourceplus{GW190727A} }$ & $\masstwosourcemed{GW190727A}_{ -\masstwosourceminus{GW190727A} }^{ +\masstwosourceplus{GW190727A} }$ & $\chieffmed{GW190727A}_{ -\chieffminus{GW190727A} }^{ +\chieffplus{GW190727A} }$ & $\luminositydistancemed{GW190727A}_{ -\luminositydistanceminus{GW190727A} }^{ +\luminositydistanceplus{GW190727A} }$ & $\redshiftmed{GW190727A}_{ -\redshiftminus{GW190727A} }^{ +\redshiftplus{GW190727A} }$ & $\finalmasssourcemed{GW190727A}_{ -\finalmasssourceminus{GW190727A} }^{ +\finalmasssourceplus{GW190727A} }$ & $\finalspinmed{GW190727A}_{ -\finalspinminus{GW190727A} }^{ +\finalspinplus{GW190727A} }$ & $\skyarea{GW190727A}$ & $\networkmatchedfiltersnrIMRmed{GW190727A}_{ -\networkmatchedfiltersnrIMRminus{GW190727A} }^{ +\networkmatchedfiltersnrIMRplus{GW190727A} }$\\
\NNAME{GW190728A} & $\totalmasssourcemed{GW190728A}_{ -\totalmasssourceminus{GW190728A} }^{ +\totalmasssourceplus{GW190728A} }$ & $\chirpmasssourcemed{GW190728A}_{ -\chirpmasssourceminus{GW190728A} }^{ +\chirpmasssourceplus{GW190728A} }$ & $\massonesourcemed{GW190728A}_{ -\massonesourceminus{GW190728A} }^{ +\massonesourceplus{GW190728A} }$ & $\masstwosourcemed{GW190728A}_{ -\masstwosourceminus{GW190728A} }^{ +\masstwosourceplus{GW190728A} }$ & $\chieffmed{GW190728A}_{ -\chieffminus{GW190728A} }^{ +\chieffplus{GW190728A} }$ & $\luminositydistancemed{GW190728A}_{ -\luminositydistanceminus{GW190728A} }^{ +\luminositydistanceplus{GW190728A} }$ & $\redshiftmed{GW190728A}_{ -\redshiftminus{GW190728A} }^{ +\redshiftplus{GW190728A} }$ & $\finalmasssourcemed{GW190728A}_{ -\finalmasssourceminus{GW190728A} }^{ +\finalmasssourceplus{GW190728A} }$ & $\finalspinmed{GW190728A}_{ -\finalspinminus{GW190728A} }^{ +\finalspinplus{GW190728A} }$ & $\skyarea{GW190728A}$ & $\networkmatchedfiltersnrIMRmed{GW190728A}_{ -\networkmatchedfiltersnrIMRminus{GW190728A} }^{ +\networkmatchedfiltersnrIMRplus{GW190728A} }$\\
\NNAME{GW190731A} & $\totalmasssourcemed{GW190731A}_{ -\totalmasssourceminus{GW190731A} }^{ +\totalmasssourceplus{GW190731A} }$ & $\chirpmasssourcemed{GW190731A}_{ -\chirpmasssourceminus{GW190731A} }^{ +\chirpmasssourceplus{GW190731A} }$ & $\massonesourcemed{GW190731A}_{ -\massonesourceminus{GW190731A} }^{ +\massonesourceplus{GW190731A} }$ & $\masstwosourcemed{GW190731A}_{ -\masstwosourceminus{GW190731A} }^{ +\masstwosourceplus{GW190731A} }$ & $\chieffmed{GW190731A}_{ -\chieffminus{GW190731A} }^{ +\chieffplus{GW190731A} }$ & $\luminositydistancemed{GW190731A}_{ -\luminositydistanceminus{GW190731A} }^{ +\luminositydistanceplus{GW190731A} }$ & $\redshiftmed{GW190731A}_{ -\redshiftminus{GW190731A} }^{ +\redshiftplus{GW190731A} }$ & $\finalmasssourcemed{GW190731A}_{ -\finalmasssourceminus{GW190731A} }^{ +\finalmasssourceplus{GW190731A} }$ & $\finalspinmed{GW190731A}_{ -\finalspinminus{GW190731A} }^{ +\finalspinplus{GW190731A} }$ & $\skyarea{GW190731A}$ & $\networkmatchedfiltersnrIMRmed{GW190731A}_{ -\networkmatchedfiltersnrIMRminus{GW190731A} }^{ +\networkmatchedfiltersnrIMRplus{GW190731A} }$\\
\NNAME{GW190803A} & $\totalmasssourcemed{GW190803A}_{ -\totalmasssourceminus{GW190803A} }^{ +\totalmasssourceplus{GW190803A} }$ & $\chirpmasssourcemed{GW190803A}_{ -\chirpmasssourceminus{GW190803A} }^{ +\chirpmasssourceplus{GW190803A} }$ & $\massonesourcemed{GW190803A}_{ -\massonesourceminus{GW190803A} }^{ +\massonesourceplus{GW190803A} }$ & $\masstwosourcemed{GW190803A}_{ -\masstwosourceminus{GW190803A} }^{ +\masstwosourceplus{GW190803A} }$ & $\chieffmed{GW190803A}_{ -\chieffminus{GW190803A} }^{ +\chieffplus{GW190803A} }$ & $\luminositydistancemed{GW190803A}_{ -\luminositydistanceminus{GW190803A} }^{ +\luminositydistanceplus{GW190803A} }$ & $\redshiftmed{GW190803A}_{ -\redshiftminus{GW190803A} }^{ +\redshiftplus{GW190803A} }$ & $\finalmasssourcemed{GW190803A}_{ -\finalmasssourceminus{GW190803A} }^{ +\finalmasssourceplus{GW190803A} }$ & $\finalspinmed{GW190803A}_{ -\finalspinminus{GW190803A} }^{ +\finalspinplus{GW190803A} }$ & $\skyarea{GW190803A}$ & $\networkmatchedfiltersnrIMRmed{GW190803A}_{ -\networkmatchedfiltersnrIMRminus{GW190803A} }^{ +\networkmatchedfiltersnrIMRplus{GW190803A} }$\\
\NNAME{GW190814A} & $\totalmasssourcemed{GW190814A}_{ -\totalmasssourceminus{GW190814A} }^{ +\totalmasssourceplus{GW190814A} }$ & $\chirpmasssourcemed{GW190814A}_{ -\chirpmasssourceminus{GW190814A} }^{ +\chirpmasssourceplus{GW190814A} }$ & $\massonesourcemed{GW190814A}_{ -\massonesourceminus{GW190814A} }^{ +\massonesourceplus{GW190814A} }$ & $\masstwosourcemed{GW190814A}_{ -\masstwosourceminus{GW190814A} }^{ +\masstwosourceplus{GW190814A} }$ & $\chieffmed{GW190814A}_{ -\chieffminus{GW190814A} }^{ +\chieffplus{GW190814A} }$ & $\luminositydistancemed{GW190814A}_{ -\luminositydistanceminus{GW190814A} }^{ +\luminositydistanceplus{GW190814A} }$ & $\redshiftmed{GW190814A}_{ -\redshiftminus{GW190814A} }^{ +\redshiftplus{GW190814A} }$ & $\finalmasssourcemed{GW190814A}_{ -\finalmasssourceminus{GW190814A} }^{ +\finalmasssourceplus{GW190814A} }$ & $\finalspinmed{GW190814A}_{ -\finalspinminus{GW190814A} }^{ +\finalspinplus{GW190814A} }$ & $\skyarea{GW190814A}$ & $\networkmatchedfiltersnrIMRmed{GW190814A}_{ -\networkmatchedfiltersnrIMRminus{GW190814A} }^{ +\networkmatchedfiltersnrIMRplus{GW190814A} }$\\
\NNAME{GW190828A} & $\totalmasssourcemed{GW190828A}_{ -\totalmasssourceminus{GW190828A} }^{ +\totalmasssourceplus{GW190828A} }$ & $\chirpmasssourcemed{GW190828A}_{ -\chirpmasssourceminus{GW190828A} }^{ +\chirpmasssourceplus{GW190828A} }$ & $\massonesourcemed{GW190828A}_{ -\massonesourceminus{GW190828A} }^{ +\massonesourceplus{GW190828A} }$ & $\masstwosourcemed{GW190828A}_{ -\masstwosourceminus{GW190828A} }^{ +\masstwosourceplus{GW190828A} }$ & $\chieffmed{GW190828A}_{ -\chieffminus{GW190828A} }^{ +\chieffplus{GW190828A} }$ & $\luminositydistancemed{GW190828A}_{ -\luminositydistanceminus{GW190828A} }^{ +\luminositydistanceplus{GW190828A} }$ & $\redshiftmed{GW190828A}_{ -\redshiftminus{GW190828A} }^{ +\redshiftplus{GW190828A} }$ & $\finalmasssourcemed{GW190828A}_{ -\finalmasssourceminus{GW190828A} }^{ +\finalmasssourceplus{GW190828A} }$ & $\finalspinmed{GW190828A}_{ -\finalspinminus{GW190828A} }^{ +\finalspinplus{GW190828A} }$ & $\skyarea{GW190828A}$ & $\networkmatchedfiltersnrIMRmed{GW190828A}_{ -\networkmatchedfiltersnrIMRminus{GW190828A} }^{ +\networkmatchedfiltersnrIMRplus{GW190828A} }$\\
\NNAME{GW190828B} & $\totalmasssourcemed{GW190828B}_{ -\totalmasssourceminus{GW190828B} }^{ +\totalmasssourceplus{GW190828B} }$ & $\chirpmasssourcemed{GW190828B}_{ -\chirpmasssourceminus{GW190828B} }^{ +\chirpmasssourceplus{GW190828B} }$ & $\massonesourcemed{GW190828B}_{ -\massonesourceminus{GW190828B} }^{ +\massonesourceplus{GW190828B} }$ & $\masstwosourcemed{GW190828B}_{ -\masstwosourceminus{GW190828B} }^{ +\masstwosourceplus{GW190828B} }$ & $\chieffmed{GW190828B}_{ -\chieffminus{GW190828B} }^{ +\chieffplus{GW190828B} }$ & $\luminositydistancemed{GW190828B}_{ -\luminositydistanceminus{GW190828B} }^{ +\luminositydistanceplus{GW190828B} }$ & $\redshiftmed{GW190828B}_{ -\redshiftminus{GW190828B} }^{ +\redshiftplus{GW190828B} }$ & $\finalmasssourcemed{GW190828B}_{ -\finalmasssourceminus{GW190828B} }^{ +\finalmasssourceplus{GW190828B} }$ & $\finalspinmed{GW190828B}_{ -\finalspinminus{GW190828B} }^{ +\finalspinplus{GW190828B} }$ & $\skyarea{GW190828B}$ & $\networkmatchedfiltersnrIMRmed{GW190828B}_{ -\networkmatchedfiltersnrIMRminus{GW190828B} }^{ +\networkmatchedfiltersnrIMRplus{GW190828B} }$\\
\NNAME{GW190909A} & $\totalmasssourcemed{GW190909A}_{ -\totalmasssourceminus{GW190909A} }^{ +\totalmasssourceplus{GW190909A} }$ & $\chirpmasssourcemed{GW190909A}_{ -\chirpmasssourceminus{GW190909A} }^{ +\chirpmasssourceplus{GW190909A} }$ & $\massonesourcemed{GW190909A}_{ -\massonesourceminus{GW190909A} }^{ +\massonesourceplus{GW190909A} }$ & $\masstwosourcemed{GW190909A}_{ -\masstwosourceminus{GW190909A} }^{ +\masstwosourceplus{GW190909A} }$ & $\chieffmed{GW190909A}_{ -\chieffminus{GW190909A} }^{ +\chieffplus{GW190909A} }$ & $\luminositydistancemed{GW190909A}_{ -\luminositydistanceminus{GW190909A} }^{ +\luminositydistanceplus{GW190909A} }$ & $\redshiftmed{GW190909A}_{ -\redshiftminus{GW190909A} }^{ +\redshiftplus{GW190909A} }$ & $\finalmasssourcemed{GW190909A}_{ -\finalmasssourceminus{GW190909A} }^{ +\finalmasssourceplus{GW190909A} }$ & $\finalspinmed{GW190909A}_{ -\finalspinminus{GW190909A} }^{ +\finalspinplus{GW190909A} }$ & $\skyarea{GW190909A}$ & $\networkmatchedfiltersnrIMRmed{GW190909A}_{ -\networkmatchedfiltersnrIMRminus{GW190909A} }^{ +\networkmatchedfiltersnrIMRplus{GW190909A} }$\\
\NNAME{GW190910A} & $\totalmasssourcemed{GW190910A}_{ -\totalmasssourceminus{GW190910A} }^{ +\totalmasssourceplus{GW190910A} }$ & $\chirpmasssourcemed{GW190910A}_{ -\chirpmasssourceminus{GW190910A} }^{ +\chirpmasssourceplus{GW190910A} }$ & $\massonesourcemed{GW190910A}_{ -\massonesourceminus{GW190910A} }^{ +\massonesourceplus{GW190910A} }$ & $\masstwosourcemed{GW190910A}_{ -\masstwosourceminus{GW190910A} }^{ +\masstwosourceplus{GW190910A} }$ & $\chieffmed{GW190910A}_{ -\chieffminus{GW190910A} }^{ +\chieffplus{GW190910A} }$ & $\luminositydistancemed{GW190910A}_{ -\luminositydistanceminus{GW190910A} }^{ +\luminositydistanceplus{GW190910A} }$ & $\redshiftmed{GW190910A}_{ -\redshiftminus{GW190910A} }^{ +\redshiftplus{GW190910A} }$ & $\finalmasssourcemed{GW190910A}_{ -\finalmasssourceminus{GW190910A} }^{ +\finalmasssourceplus{GW190910A} }$ & $\finalspinmed{GW190910A}_{ -\finalspinminus{GW190910A} }^{ +\finalspinplus{GW190910A} }$ & $\skyarea{GW190910A}$ & $\networkmatchedfiltersnrIMRmed{GW190910A}_{ -\networkmatchedfiltersnrIMRminus{GW190910A} }^{ +\networkmatchedfiltersnrIMRplus{GW190910A} }$\\
\NNAME{GW190915A} & $\totalmasssourcemed{GW190915A}_{ -\totalmasssourceminus{GW190915A} }^{ +\totalmasssourceplus{GW190915A} }$ & $\chirpmasssourcemed{GW190915A}_{ -\chirpmasssourceminus{GW190915A} }^{ +\chirpmasssourceplus{GW190915A} }$ & $\massonesourcemed{GW190915A}_{ -\massonesourceminus{GW190915A} }^{ +\massonesourceplus{GW190915A} }$ & $\masstwosourcemed{GW190915A}_{ -\masstwosourceminus{GW190915A} }^{ +\masstwosourceplus{GW190915A} }$ & $\chieffmed{GW190915A}_{ -\chieffminus{GW190915A} }^{ +\chieffplus{GW190915A} }$ & $\luminositydistancemed{GW190915A}_{ -\luminositydistanceminus{GW190915A} }^{ +\luminositydistanceplus{GW190915A} }$ & $\redshiftmed{GW190915A}_{ -\redshiftminus{GW190915A} }^{ +\redshiftplus{GW190915A} }$ & $\finalmasssourcemed{GW190915A}_{ -\finalmasssourceminus{GW190915A} }^{ +\finalmasssourceplus{GW190915A} }$ & $\finalspinmed{GW190915A}_{ -\finalspinminus{GW190915A} }^{ +\finalspinplus{GW190915A} }$ & $\skyarea{GW190915A}$ & $\networkmatchedfiltersnrIMRmed{GW190915A}_{ -\networkmatchedfiltersnrIMRminus{GW190915A} }^{ +\networkmatchedfiltersnrIMRplus{GW190915A} }$\\
\NNAME{GW190924A} & $\totalmasssourcemed{GW190924A}_{ -\totalmasssourceminus{GW190924A} }^{ +\totalmasssourceplus{GW190924A} }$ & $\chirpmasssourcemed{GW190924A}_{ -\chirpmasssourceminus{GW190924A} }^{ +\chirpmasssourceplus{GW190924A} }$ & $\massonesourcemed{GW190924A}_{ -\massonesourceminus{GW190924A} }^{ +\massonesourceplus{GW190924A} }$ & $\masstwosourcemed{GW190924A}_{ -\masstwosourceminus{GW190924A} }^{ +\masstwosourceplus{GW190924A} }$ & $\chieffmed{GW190924A}_{ -\chieffminus{GW190924A} }^{ +\chieffplus{GW190924A} }$ & $\luminositydistancemed{GW190924A}_{ -\luminositydistanceminus{GW190924A} }^{ +\luminositydistanceplus{GW190924A} }$ & $\redshiftmed{GW190924A}_{ -\redshiftminus{GW190924A} }^{ +\redshiftplus{GW190924A} }$ & $\finalmasssourcemed{GW190924A}_{ -\finalmasssourceminus{GW190924A} }^{ +\finalmasssourceplus{GW190924A} }$ & $\finalspinmed{GW190924A}_{ -\finalspinminus{GW190924A} }^{ +\finalspinplus{GW190924A} }$ & $\skyarea{GW190924A}$ & $\networkmatchedfiltersnrIMRmed{GW190924A}_{ -\networkmatchedfiltersnrIMRminus{GW190924A} }^{ +\networkmatchedfiltersnrIMRplus{GW190924A} }$\\
\NNAME{GW190929A} & $\totalmasssourcemed{GW190929A}_{ -\totalmasssourceminus{GW190929A} }^{ +\totalmasssourceplus{GW190929A} }$ & $\chirpmasssourcemed{GW190929A}_{ -\chirpmasssourceminus{GW190929A} }^{ +\chirpmasssourceplus{GW190929A} }$ & $\massonesourcemed{GW190929A}_{ -\massonesourceminus{GW190929A} }^{ +\massonesourceplus{GW190929A} }$ & $\masstwosourcemed{GW190929A}_{ -\masstwosourceminus{GW190929A} }^{ +\masstwosourceplus{GW190929A} }$ & $\chieffmed{GW190929A}_{ -\chieffminus{GW190929A} }^{ +\chieffplus{GW190929A} }$ & $\luminositydistancemed{GW190929A}_{ -\luminositydistanceminus{GW190929A} }^{ +\luminositydistanceplus{GW190929A} }$ & $\redshiftmed{GW190929A}_{ -\redshiftminus{GW190929A} }^{ +\redshiftplus{GW190929A} }$ & $\finalmasssourcemed{GW190929A}_{ -\finalmasssourceminus{GW190929A} }^{ +\finalmasssourceplus{GW190929A} }$ & $\finalspinmed{GW190929A}_{ -\finalspinminus{GW190929A} }^{ +\finalspinplus{GW190929A} }$ & $\skyarea{GW190929A}$ & $\networkmatchedfiltersnrIMRmed{GW190929A}_{ -\networkmatchedfiltersnrIMRminus{GW190929A} }^{ +\networkmatchedfiltersnrIMRplus{GW190929A} }$\\
\NNAME{GW190930A} & $\totalmasssourcemed{GW190930A}_{ -\totalmasssourceminus{GW190930A} }^{ +\totalmasssourceplus{GW190930A} }$ & $\chirpmasssourcemed{GW190930A}_{ -\chirpmasssourceminus{GW190930A} }^{ +\chirpmasssourceplus{GW190930A} }$ & $\massonesourcemed{GW190930A}_{ -\massonesourceminus{GW190930A} }^{ +\massonesourceplus{GW190930A} }$ & $\masstwosourcemed{GW190930A}_{ -\masstwosourceminus{GW190930A} }^{ +\masstwosourceplus{GW190930A} }$ & $\chieffmed{GW190930A}_{ -\chieffminus{GW190930A} }^{ +\chieffplus{GW190930A} }$ & $\luminositydistancemed{GW190930A}_{ -\luminositydistanceminus{GW190930A} }^{ +\luminositydistanceplus{GW190930A} }$ & $\redshiftmed{GW190930A}_{ -\redshiftminus{GW190930A} }^{ +\redshiftplus{GW190930A} }$ & $\finalmasssourcemed{GW190930A}_{ -\finalmasssourceminus{GW190930A} }^{ +\finalmasssourceplus{GW190930A} }$ & $\finalspinmed{GW190930A}_{ -\finalspinminus{GW190930A} }^{ +\finalspinplus{GW190930A} }$ & $\skyarea{GW190930A}$ & $\networkmatchedfiltersnrIMRmed{GW190930A}_{ -\networkmatchedfiltersnrIMRminus{GW190930A} }^{ +\networkmatchedfiltersnrIMRplus{GW190930A} }$\\
\hline
\end{tabularx}

%% file: GW190425_source_properties_table.tex
\begin{tabularx}{\columnwidth}{@{\extracolsep{\fill}}lYYYYY}
          Waveform & $m_1/\Msun$ & $m_2/\Msun$ & $q$ &$\tilde{\Lambda}$ \\
         \hline
         IMRPhenomD\_NRTidal &
         $\lowerboundmassoneonefourtwofiveLS{PhenomD}$--$\upperboundmassoneonefourtwofiveLS{PhenomD}$ &
         $\lowerboundmasstwoonefourtwofiveLS{PhenomD}$--$\upperboundmasstwoonefourtwofiveLS{PhenomD}$ &
         $\lowerboundmassratioonefourtwofiveLS{PhenomD}$--$\upperboundmassratioonefourtwofiveLS{PhenomD}$ &
         $\leqslant$ $\lambdatildeboundonefourtwofiveLS{PhenomD}$
         \\
         IMRPhenomPv2\_NRTidal &
         $\lowerboundmassoneonefourtwofiveLS{PhenomP}$--$\upperboundmassoneonefourtwofiveLS{PhenomP}$ &
         $\lowerboundmasstwoonefourtwofiveLS{PhenomP}$--$\upperboundmasstwoonefourtwofiveLS{PhenomP}$ &
         $\lowerboundmassratioonefourtwofiveLS{PhenomP}$--$\upperboundmassratioonefourtwofiveLS{PhenomP}$ &
         $\leqslant$ $\lambdatildeboundonefourtwofiveLS{PhenomP}$
         \\
         SEOBNRv4T\_surrogate &
         $\lowerboundmassoneonefourtwofiveLS{SEOB}$--$\upperboundmassoneonefourtwofiveLS{SEOB}$ &
         $\lowerboundmasstwoonefourtwofiveLS{SEOB}$--$\upperboundmasstwoonefourtwofiveLS{SEOB}$ &
         $\lowerboundmassratioonefourtwofiveLS{SEOB}$--$\upperboundmassratioonefourtwofiveLS{SEOB}$ &
         $\leqslant$ $\lambdatildeboundonefourtwofiveLS{SEOB}$
         \\
         TEOBResumS  &
         $\lowerboundmassoneonefourtwofiveLS{TEOB}$--$\upperboundmassoneonefourtwofiveLS{TEOB}$ &
         $\lowerboundmasstwoonefourtwofiveLS{TEOB}$--$\upperboundmasstwoonefourtwofiveLS{TEOB}$ &
         $\lowerboundmassratioonefourtwofiveLS{TEOB}$--$\upperboundmassratioonefourtwofiveLS{TEOB}$ &
         $\leqslant$ $\lambdatildeboundonefourtwofiveLS{TEOB}$\\
         \rowcolor{white}
         \\[-9pt]
         \hline
         \\[-6pt]
         IMRPhenomD\_NRTidal &
         $\lowerboundmassoneonefourtwofiveHS{PhenomD}$--$\upperboundmassoneonefourtwofiveHS{PhenomD}$ &
         $\lowerboundmasstwoonefourtwofiveHS{PhenomD}$--$\upperboundmasstwoonefourtwofiveHS{PhenomD}$ &
         $\lowerboundmassratioonefourtwofiveHS{PhenomD}$--$\upperboundmassratioonefourtwofiveHS{PhenomD}$ &
         $\leqslant$ $\lambdatildeboundonefourtwofiveHS{PhenomD}$
         \\
         IMRPhenomPv2\_NRTidal &
         $\lowerboundmassoneonefourtwofiveHS{PhenomP}$--$\upperboundmassoneonefourtwofiveHS{PhenomP}$ &
         $\lowerboundmasstwoonefourtwofiveHS{PhenomP}$--$\upperboundmasstwoonefourtwofiveHS{PhenomP}$ &
         $\lowerboundmassratioonefourtwofiveHS{PhenomP}$--$\upperboundmassratioonefourtwofiveHS{PhenomP}$ &
         $\leqslant$ $\lambdatildeboundonefourtwofiveHS{PhenomP}$
         \\
         SEOBNRv4T\_surrogate &
         $\lowerboundmassoneonefourtwofiveHS{SEOB}$--$\upperboundmassoneonefourtwofiveHS{SEOB}$ &
         $\lowerboundmasstwoonefourtwofiveHS{SEOB}$--$\upperboundmasstwoonefourtwofiveHS{SEOB}$ &
         $\lowerboundmassratioonefourtwofiveHS{SEOB}$--$\upperboundmassratioonefourtwofiveHS{SEOB}$ &
         $\leqslant$ $\lambdatildeboundonefourtwofiveHS{SEOB}
         $
         \\
         TEOBResumS  &
         $\lowerboundmassoneonefourtwofiveHS{TEOB}$--$\upperboundmassoneonefourtwofiveHS{TEOB}$ &
         $\lowerboundmasstwoonefourtwofiveHS{TEOB}$--$\upperboundmasstwoonefourtwofiveHS{TEOB}$ &
         $\lowerboundmassratioonefourtwofiveHS{TEOB}$--$\upperboundmassratioonefourtwofiveHS{TEOB}$ &
         $\leqslant$ $\lambdatildeboundonefourtwofiveHS{TEOB}$ 
         \\
         \hline
    \end{tabularx}

%% file: conclusion.tex
\section{Conclusion}
\resetlinenumber
We have presented the results from a search for compact binary coalescence signals
in the first part of the third observing run of Advanced LIGO and Advanced Virgo.
During the period of observations, spanning 1 April to 1 October 2019,
the three detectors had sensitivity that significantly exceeded previous observing runs,
with median \ac{BNS} inspiral ranges of \HANFORDRANGE{} (Hanford), \LIVINGSTONRANGE{} (Livingston)
and \VIRGORANGE{} (Virgo).
This improved sensitivity allowed us to greatly expand the number of known
compact binary mergers, adding 39 new gravitational wave events to the 11 we have previously
reported in GWTC-1~\cite{LIGOScientific:2018mvr}.

We performed parameter estimation on these 39 new GWTC-2 signals using a range of waveform
models, allowing us to incorporate the effect of HMs in the inference of
source parameters for \ac{BBH} systems, and to compare to the systematic differences
between waveform families.
We find that the sources of these signals include \acp{BBH} that are more massive, farther away, and more asymmetric in mass ratio
than any sources in GWTC-1, as well as three binaries with at least
one component of mass $<3\,\Msun$. These latter systems may include the first detected NSBH mergers;
however, there is insufficient \ac{SNR} to perform an informative measurement of the tidal deformability that would definitively indicate whether they had an NSBH or \ac{BBH} origin.
We expanded our analysis of the potential \ac{BNS} signal \NAME{GW190425z} to encompass
further waveforms which include tidal effects, finding agreement with
previously-reported results~\cite{Abbott:2020uma}.

Our analyses recovered 11 sources with positive effective inspiral spin
but zero with negative (at 90\% credibility). Although no individual binary in O3a has $\chi_{\rm eff}<0$ 
with high credibility, a hierarchical population analysis \cite{o3apop} 
of LVC-reported events to date indicates a non-trivial fraction of binaries 
have negative effective inspiral spin.
This would be consistent with the independent observation of 
gravitational-wave candidate GW170121, which was found to have significant 
probability of having $\chi_\mathrm{eff} < 0$ \cite{Venumadhav:2019lyq,Nitz:2019hdf}.
We also examined the evidence for misalignment of orbital and component angular momenta,
which would be an indication of a dynamical formation channel, and produce precession
of the orbital plane. We find only mild evidence in favor of precession in the
most significant case of \NAME{GW190412A}~\cite{GW190412,Zevin:2020gxf}.
We also performed consistency checks between the waveform models and the observed
data, finding no statistically significant differences.

A pair of companion papers make use of the events in GWTC-2 to study
source populations and fundamental physics.
The inferred population distribution of compact object mergers is
described in \cite{o3apop}, which reports an updated \ac{BBH} merger
rate density of \RBBH, and rate density for \ac{BNS} of \RBNS.  This paper
also investigates mass and spin distributions, and finds evidence
for a population of \ac{BBH} systems with spins misaligned from
the orbital plane. Eight tests of general relativity are reported in 
\cite{o3tgr}, showing no evidence for violations of Einstein's theory of gravity,
leading to some of the best constraints on alternative theories to date.

Data products associated with this catalog are available through the
Gravitational Wave Open Science Center (GWOSC) at \url{https://gw-openscience.org}~\cite{lvc:datadoi}.  Data associated with all events described in this paper
are available through the GWOSC Event Portal, including calibrated
strain time-series, parameter estimation posterior samples, tables of 90\%
credible intervals for physical parameters, and search pipeline results.
These data products may be accessed through a web browser or open source client
package~\cite{macleod:gwoscclient}.

The online Gravitational Wave Transient Catalog represents a cumulative
set of events found in LIGO--Virgo data, including detections presented in GWTC-1, and now
also events from the first six months of the O3 observing run.
O3 continued from November 2019 until
March 2020.  Analysis of the second portion (O3b) is currently in
progress, and events found during this period will be added in the
next update.

The Advanced LIGO and Advanced Virgo detectors are currently undergoing
commissioning to further improve their sensitivity, and will be joined in their fourth
observing run by the KAGRA detector~\cite{Akutsu:2018axf}. This should lead to improvements in the detection
rate and source localization, improving the prospects for multimessenger observation
of future sources~\cite{Aasi:2013wyav10,Pankow:2019oxl}.

%% file: LVCack.tex
\resetlinenumber

The authors gratefully acknowledge the support of the United States
National Science Foundation (NSF) for the construction and operation of the
LIGO Laboratory and Advanced LIGO as well as the Science and Technology Facilities Council (STFC) of the
United Kingdom, the Max-Planck-Society (MPS), and the State of
Niedersachsen/Germany for support of the construction of Advanced LIGO 
and construction and operation of the GEO600 detector. 
Additional support for Advanced LIGO was provided by the Australian Research Council.
The authors gratefully acknowledge the Italian Istituto Nazionale di Fisica Nucleare (INFN),  
the French Centre National de la Recherche Scientifique (CNRS) and
the Netherlands Organization for Scientific Research, 
for the construction and operation of the Virgo detector
and the creation and support  of the EGO consortium. 
The authors also gratefully acknowledge research support from these agencies as well as by 
the Council of Scientific and Industrial Research of India, 
the Department of Science and Technology, India,
the Science \& Engineering Research Board (SERB), India,
the Ministry of Human Resource Development, India,
the Spanish Agencia Estatal de Investigaci\'on,
the Vicepresid\`encia i Conselleria d'Innovaci\'o, Recerca i Turisme and the Conselleria d'Educaci\'o i Universitat del Govern de les Illes Balears,
the Conselleria d'Innovaci\'o, Universitats, Ci\`encia i Societat Digital de la Generalitat Valenciana and
the CERCA Programme Generalitat de Catalunya, Spain,
the National Science Centre of Poland and the Foundation for Polish Science (FNP),
the Swiss National Science Foundation (SNSF),
the Russian Foundation for Basic Research, 
the Russian Science Foundation,
the European Commission,
the European Regional Development Funds (ERDF),
the Royal Society, 
the Scottish Funding Council, 
the Scottish Universities Physics Alliance, 
the Hungarian Scientific Research Fund (OTKA),
the French Lyon Institute of Origins (LIO),
the Belgian Fonds de la Recherche Scientifique (FRS-FNRS), 
Actions de Recherche Concertées (ARC) and
Fonds Wetenschappelijk Onderzoek – Vlaanderen (FWO), Belgium,
the Paris \^{I}le-de-France Region, 
the National Research, Development and Innovation Office Hungary (NKFIH), 
the National Research Foundation of Korea,
the Natural Science and Engineering Research Council Canada,
Canadian Foundation for Innovation (CFI),
the Brazilian Ministry of Science, Technology, Innovations, and Communications,
the International Center for Theoretical Physics South American Institute for Fundamental Research (ICTP-SAIFR), 
the Research Grants Council of Hong Kong,
the National Natural Science Foundation of China (NSFC),
the Leverhulme Trust, 
the Research Corporation, 
the Ministry of Science and Technology (MOST), Taiwan
and
the Kavli Foundation.
The authors gratefully acknowledge the support of the NSF, STFC, INFN and CNRS for provision of computational resources.

{\it We would like to thank all of the essential workers who put their health at risk during the COVID-19 pandemic, without whom we would not have been able to complete this work.}

%% file: systematics_appendix.tex
\section{Waveform systematics}\label{app:systematics}
The choice of waveform influences our inferences of source properties
\cite{Abbott:2016wiq,LIGOScientific:2018mvr,Abbott:2016izl}.
As such, we employ multiple waveform families in the inference
of each event's source parameters.  The full list of
models represented in our publicly available results for
each event is shown in Table~\ref{tab:waveforms}.

\begin{PE_table}
\setlength\extrarowheight{-1pt}
\begin{table*}
    \centering
    {\rowcolors{1}{}{lightgray}
    \input{waveform_table.tex}
    }
    \caption{Summary of the waveform models used for the analyses, available in the
    data release. Fiducial results used for the main presentation in Sec.~\ref{sec:peresults}
    are shown in bold. Where multiple bold waveforms are listed, equal numbers of
    posterior samples from runs with those waveforms are combined. For GW190425, HS and LS
    suffixes correspond to high spin ($|\vec{\chi_i}|\leq0.89$) and low spin ($|\vec{\chi_i}|\leq0.05$)
    priors respectively.}
    \label{tab:waveforms}
\end{table*}
\end{PE_table}

There have been many tools designed to quantify difference between probability distributions,
one of which is the Kullback--Leibler (KL) divergence~\cite{Kullback:1951}.
It is defined by
\begin{equation}
    D_{\rm KL}(p|q) = \int p(\vartheta)\log_2 \left(\frac{p(\vartheta)}{q(\vartheta)}\right) \mathrm{d}\vartheta
\end{equation}
where $p(\vartheta),q(\vartheta)$ are two probability
distributions. However, the KL divergence is not symmetric,
$D_{\rm KL}(p|q) \neq D_{\rm KL}(q|p)$ and further $D_{\rm KL}$ can
diverge if the probability distributions are disjoint, which makes it
impractical to use when comparing posterior distributions of parameter estimation runs.

Both of these issues are addressed by the JS divergence~\cite{Lin:1991} which is a smoothed and symmetrized version of $D_{\rm KL}$.
It is defined as 
\begin{equation}
    D_{\rm JS}(p,q)\equiv \frac{1}{2}\left[D_{\rm KL}(p|s)+D_{\rm KL}(q|s)\right]
\end{equation} 
where $s=(p+q)/2$ is the average distribution.
By construction, the JS divergence is symmetric and it satisfies $0\leq D_{\rm JS}\leq 1$\,bit, which follows from
\begin{widetext}
\begin{equation}
D_{\rm KL}(p|s)= \int p(\vartheta)\log_2 \left(\frac{2p(\vartheta)}{p(\vartheta)+q(\vartheta)}\right) \mathrm{d}\vartheta \leq  \int p(\vartheta)\log_2\left(\frac{2p(\vartheta)}{p(\vartheta)}\right)\mathrm{d}\vartheta=1~\mathrm{bit}.
\end{equation}
\end{widetext}
The convenient properties of the JS divergence make it a useful measure of the difference between posterior probability distributions, and we adopt it to quantify the systematics in our results resulting from different analysis choices.

We use the JS divergence to investigate
two sources of systematics that can impact the results of
inference: the differences in the models with equivalent physics due
to modeling choices, and differences due to the inclusion of different
physical effects.  To assess the importance of the former, we compare
the results using our two fiducial precessing waveform models
(IMRPhenomPv2 and SEOBNRv4P). For the latter, we consider inclusion of
HMs beyond the dominant quadrupole by comparing SEOBNRv4P to
models that include HMs: SEOBNRv4PHM and NRSur7dq4. We
compute the JS divergence between one-dimensional marginal distributions on two
key parameters: mass ratio and effective inspiral spin.
We use a threshold of $0.007$~bit to deem the differences significant,
which for a Gaussian corresponds to a 20\% shift in the mean, measured
in units of one standard deviation. This threshold is larger than
$0.002$~bit, which is the variation that has been shown to arise due to
stochastic sampling~\cite{Romero-Shaw:2020owr}. 
In the following subsection, we describe how we use our 
computed JS divergences and JS threshold to choose the
default parameter estimation results shown in this work. 

\subsection{Choice of waveform models for each event}

\begin{figure*}[t]
    \includegraphics[width=\textwidth]{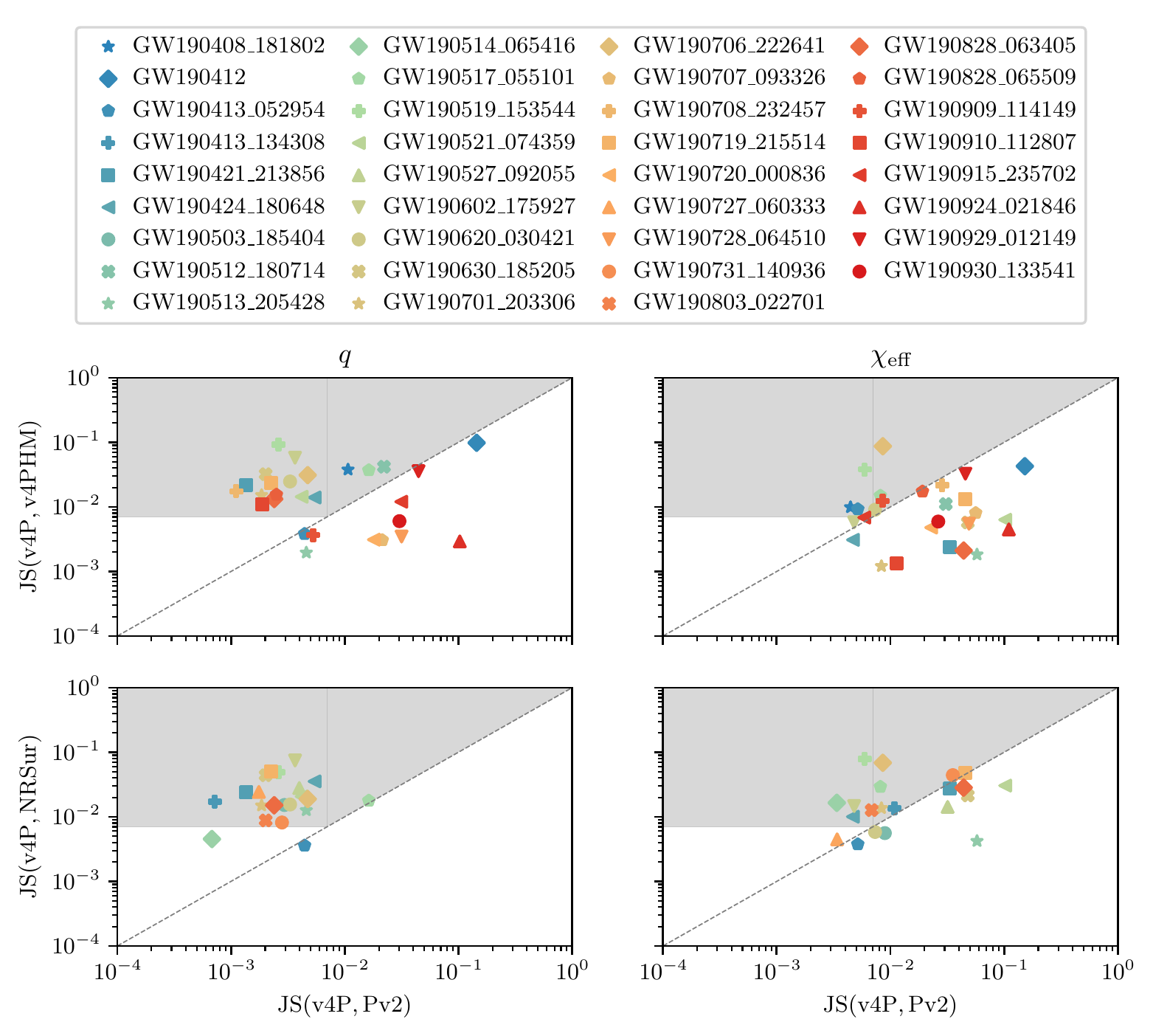}
    \caption{Jensen--Shannon divergence $D_{\rm JS}$ of one-dimensional marginal
      distributions using different waveform models.  The vertical
      axes show $D_{\rm JS}$ between one-dimensional posteriors using SEOBNRv4P and
      either SEOBNRv4PHM or NRSur7dq4.  The horizontal axes show
      $D_{\rm JS}$ between one-dimensional posteriors using SEOBNRv4P and
      IMRPhenomPv2.  The gray regions show the selection criteria from
      Eq.~\eqref{eqn:djs}: any events in these regions are presented
      using higher-order multipole moment results in Sec.~\ref{sec:peresults}.}
    \label{fig:JS}
\end{figure*}

We use the above considerations to select which waveform model(s) are used as the fiducial results presented 
in Sec.~\ref{sec:peresults}.\footnote{Our data release includes results from models with and without HMs.} 
In particular, we present results from models with HMs 
if the following conditions are satisfied for any of the three key parameters and any of the HM 
models: 
\begin{widetext}
\begin{subequations}
\begin{align}
    \label{eqn:djs}
    D_{\rm JS}({\rm SEOBNRv4P,\ HM})&>D_{\rm JS}({\rm SEOBNRv4P,\ IMRPhenomPv2}), \\
    D_{\rm JS}({\rm SEOBNRv4P,\ HM})&>0.007~\mathrm{bit}.
\end{align}
\end{subequations}
\end{widetext}
 When the conditions do not hold, we combine equal number of samples from the
results of IMRPhenomPv2 and SEOBNRv4P and use the joint samples. 
The only exceptions are \NAME{GW190425A}\ and \NAME{GW190426A},
for which we use the tidal waveforms described in Sec.~\ref{sec:PEmethods}, and \NAME{GW190412A}, \NAME{GW190521A}\ and
\NAME{GW190814A}, for which we know HMs are significant~\cite{GW190412,GW190521Adiscovery,GW190814A}.
Fig.~\ref{fig:JS}
shows the JS divergences for every event (except \NAME{GW190425A}, \NAME{GW190426A},
\NAME{GW190521A}, and \NAME{GW190814A}) 
to compare the effects of model systematics 
versus the effects of HMs. The gray regions 
correspond to the criteria in Eq.~\eqref{eqn:djs}: any 
events with $D_{\rm JS}$ values in those regions
are presented using HM results in Sec.~\ref{sec:peresults}.

\subsection{Waveform comparison---Model systematics}

The in-depth studies of \NAME{GW190412A}, \NAME{GW190521A}, and \NAME{GW190814A}\ 
have quantitatively demonstrated how much the choice of
gravitational waveform impacts our interpretation of real gravitational wave events.
We have identified several additional instances where
analyses with our two fiducial precessing waveform models 
(IMRPhenomPv2 and SEOBNRv4P) have notable differences, as
measured by the JS divergence.

\begin{figure*}[t]
    \includegraphics[width=0.7\textwidth]{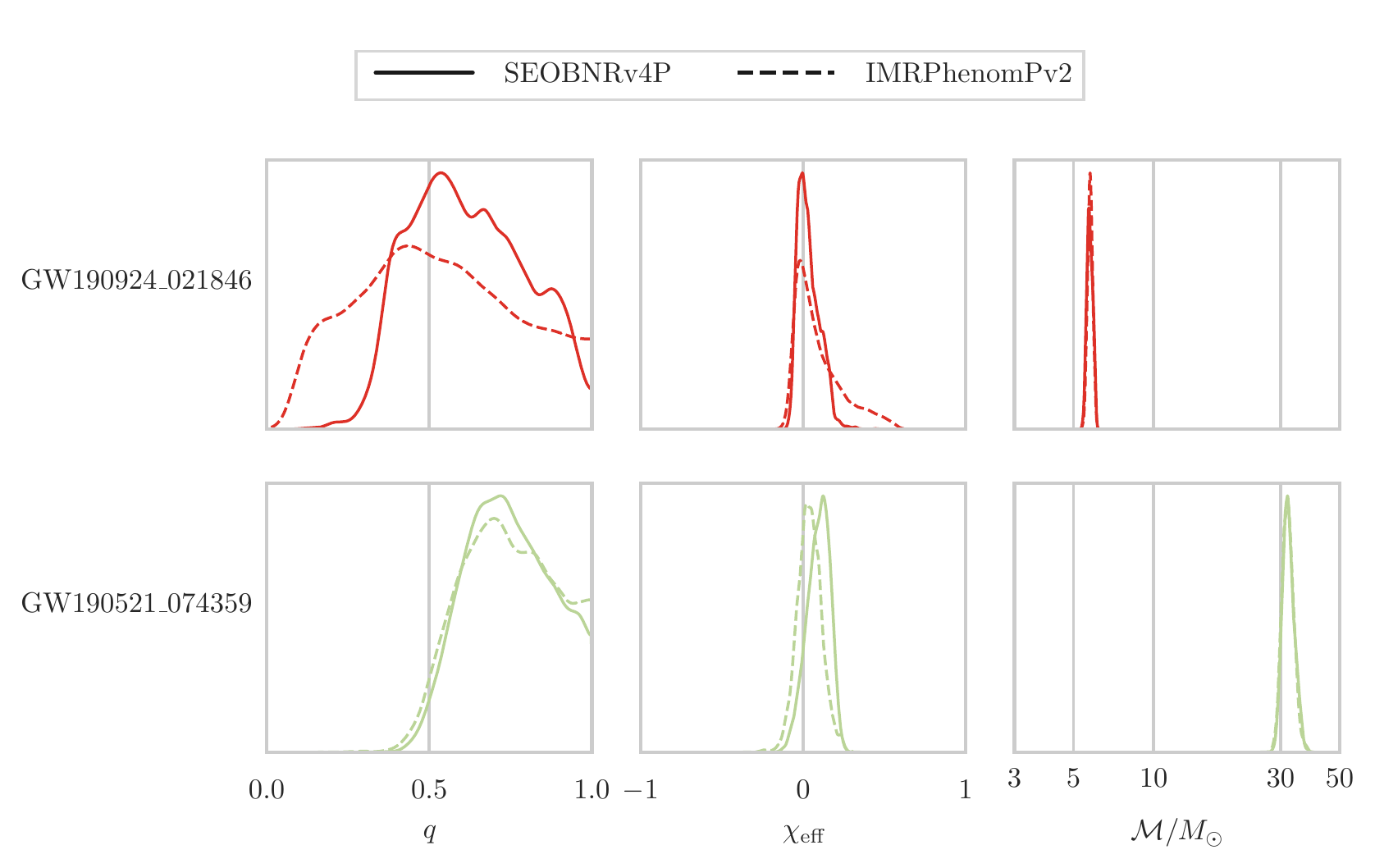}
    \caption{Marginal posterior distributions on the
    mass ratio $q$, effective inspiral spin $\chi_{\rm eff}$, and source-frame chirp mass $\mathcal{M}$ for
    \protect\NAME{GW190924A}\ and \protect\NAME{GW190521B}, with the IMRPhenomPv2 and SEOBNRv4P waveform families.}
    \label{fig:WFcomp}%
\end{figure*}

To illustrate the impact of systematics, we show in Fig.~\ref{fig:WFcomp}  the
posterior distributions for the events with extremal JS divergence in the key parameters $q$ and
$\chieff$: \NAME{GW190924A}\ and \NAME{GW190521B}. 
For \NAME{GW190924A}, SEOBNRv4P prefers more
moderate mass ratios around 1/2, and shows only minor differences in the other
parameters. Meanwhile, \NAME{GW190521B}\ shows the largest differences in the effective inspiral spin, with
SEOBNRv4P more clearly preferring non-zero values.

While the cases above demonstrate the moderate shifts that result from model systematics,
for all the events presented in this paper, there is always substantial 
overlap between the posteriors for all the key quantities we consider. 
Therefore, systematic uncertainty remains subdominant to statistical uncertainty. 

\begin{figure*}[t]
    \includegraphics[width=0.7\textwidth]{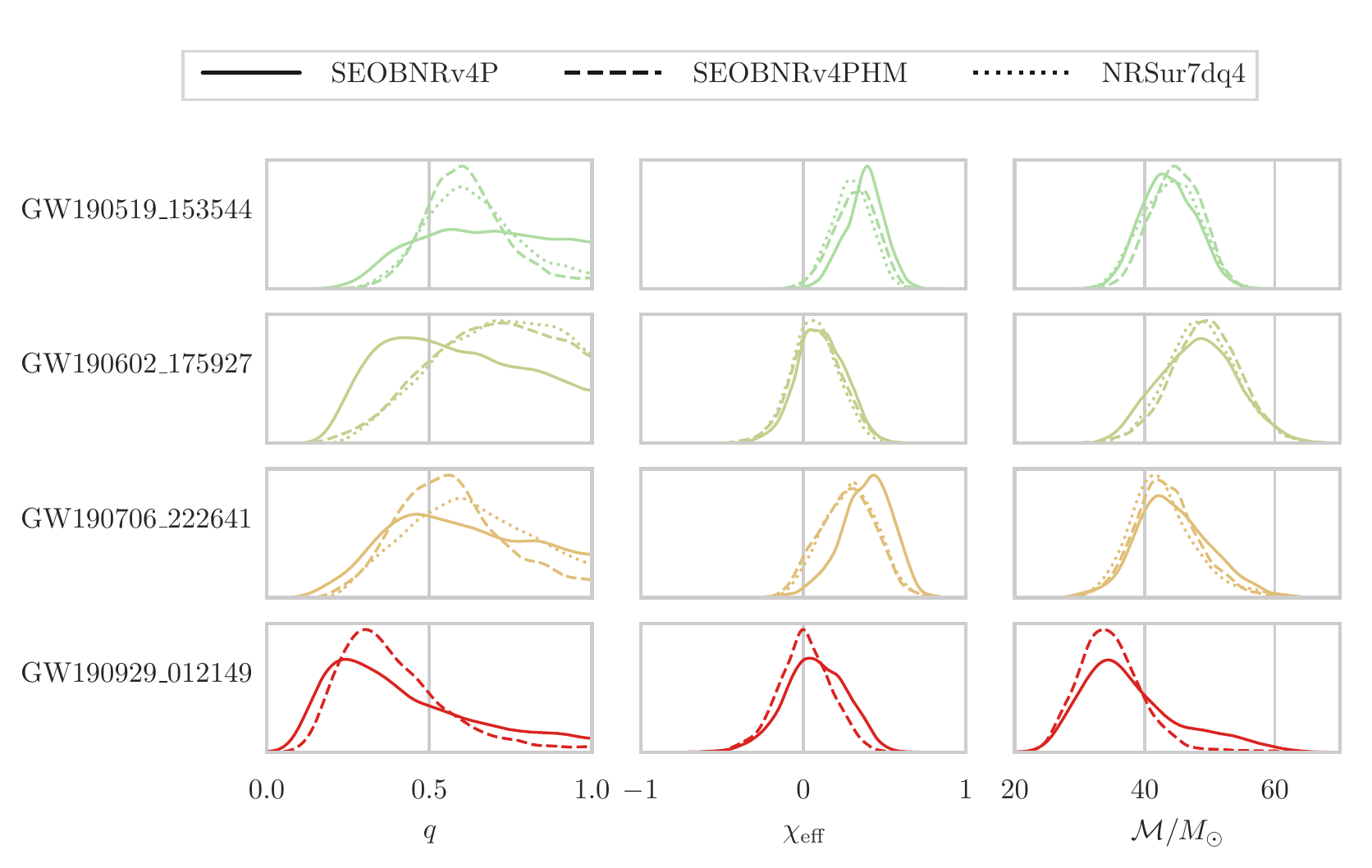}
    \caption{Marginal posterior distributions on  mass ratio $q$,
        effective inspiral spin $\chi_{\rm eff}$, and source-frame chirp mass $\mathcal{M}$, for \protect\NAME{GW190519A}, \protect\NAME{GW190602A},
        \protect\NAME{GW190706A}, and \protect\NAME{GW190929A}\ with
    the SEOBNRv4P, SEOBNRv4PHM, and NRSur7dq4 waveform families.}
    \label{fig:HMcomp}%
\end{figure*}

\subsection{Waveform comparison---Effect of higher-order multipole moments modes}
\note{Roberto Costesta, Richard O'Shaughnessy}

Previous investigations of \NAME{GW190412A}, \NAME{GW190521A}, and \NAME{GW190814A}\ 
showed that gravitational wave radiation beyond the
quadrupole can significantly impact our inferences on individual events~\cite{GW190412,GW190521Adiscovery,GW190814A}. 
Incorporating non-quadrupole modes enables
tighter inferences on the source's distance, component masses, and spins
\cite{Abbott:2016wiq,Shaik:2019dym,2019arXiv191210055P,Varma:2016dnf,Bustillo:2015qty,Littenberg:2012uj,Bustillo:2016gid,Brown:2012nn,Varma:2014jxa, Graff:2015bba,London:2017bcn, OShaughnessy:2014shr,Usman:2018imj,Kumar:2018hml,Mills:2020thr}.
HMs impact inferences about even short-duration and
low-amplitude sources,  because of the weaker constraints our observations provide.  
For example, including non-quadrupole modes can tighten constraints on the source mass ratio for low-amplitude sources
\cite{Shaik:2019dym,2019arXiv191210055P}, because high-mass ratio sources more
efficiently produce non-quadrupole modes.

We systematically applied one or two models with HM to all
sources in our catalog. We find that a majority of the sources investigated have modest shifts in mass ratio or $\chi_{\rm eff}$ due to the impact of HMs.

To illustrate the impact of non-quadrupole modes, Fig.~\ref{fig:HMcomp} shows posterior inferences 
on \NAME{GW190519A}, \NAME{GW190602A}, \NAME{GW190706A}, and \NAME{GW190929A}\
using one or two independent models which include non-quadrupole modes (SEOBNRv4PHM and NRSur7dq4). 
Using  more complete waveform models, the source of \NAME{GW190519A}\ 
is inferred to be edge on, at a smaller distance, a
more asymmetric mass ratio, and thus a higher source-frame mass $m_1$.
Due to this source's favorable orientation, non-quadrupole modes have a significant impact, with a Bayes factor for HMs of
$\sim 15$.  
Similarly, non-quadrupole modes allow us to more strongly exclude both extreme and comparable 
mass ratios for the  source responsible for
\NAME{GW190929A}, and to disfavor comparable masses for \fixme{\NAME{GW190706A}}.
Conversely,  using the same two waveform models, the high-mass source responsible for \fixme{\NAME{GW190602A}}\ is more
confidently inferred to have mass ratio $q$ closer to unity and therefore $m_2$ is skewed to larger values.
Non-quadrupole modes also have a noticeable impact on parameters of \fixme{\NAME{GW190630A}}\ and  \fixme{\NAME{GW190828B}}, in particular the mass ratio and luminosity distance.

%% file: waveform_table.tex
\begin{tabularx}{\textwidth}{@{\extracolsep{\fill}}p{3cm} Z}
Event & Available Runs \\ \hline 
{\small GW190408\_181802} & {\small IMRPhenomD}, {\small IMRPhenomPv2}, {\small SEOBNRv4P}, \textbf{{\small SEOBNRv4PHM}} \\ 
{\small GW190412} & {\small IMRPhenomD}, {\small IMRPhenomHM}, {\small IMRPhenomPv2}, \textbf{{\small IMRPhenomPv3HM}}, {\small SEOBNRv4HM\_ROM}, {\small SEOBNRv4P}, \textbf{{\small SEOBNRv4PHM}}, {\small SEOBNRv4\_ROM} \\ 
{\small GW190413\_052954} & {\small IMRPhenomD}, {\small IMRPhenomPv2}, \textbf{{\small NRSur7dq4}}, {\small SEOBNRv4P}, \textbf{{\small SEOBNRv4PHM}} \\ 
{\small GW190413\_134308} & {\small IMRPhenomD}, {\small IMRPhenomPv2}, \textbf{{\small NRSur7dq4}}, {\small SEOBNRv4P} \\ 
{\small GW190421\_213856} & {\small IMRPhenomD}, {\small IMRPhenomPv2}, \textbf{{\small NRSur7dq4}}, {\small SEOBNRv4P}, \textbf{{\small SEOBNRv4PHM}} \\ 
{\small GW190424\_180648} & {\small IMRPhenomD}, {\small IMRPhenomPv2}, \textbf{{\small NRSur7dq4}}, {\small SEOBNRv4P}, \textbf{{\small SEOBNRv4PHM}} \\ 
{\small GW190425} & {\small IMRPhenomD\_NRTidal-HS}, {\small IMRPhenomD\_NRTidal-LS}, \textbf{{\small IMRPhenomPv2\_NRTidal-HS}}, {\small IMRPhenomPv2\_NRTidal-LS}, {\small SEOBNRv4T\_surrogate\_HS}, {\small SEOBNRv4T\_surrogate\_LS}, {\small TEOBResumS-HS}, {\small TEOBResumS-LS}, {\small TaylorF2-HS}, {\small TaylorF2-LS} \\ 
{\small GW190426\_152155} & \textbf{{\small IMRPhenomNSBH}}, {\small IMRPhenomPv2}, {\small SEOBNRv4PHM}, \textbf{{\small SEOBNRv4\_ROM\_NRTidalv2\_NSBH}}, {\small TaylorF2} \\ 
{\small GW190503\_185404} & {\small IMRPhenomD}, {\small IMRPhenomPv2}, \textbf{{\small NRSur7dq4}}, {\small SEOBNRv4P} \\ 
{\small GW190512\_180714} & {\small IMRPhenomD}, {\small IMRPhenomPv2}, {\small SEOBNRv4P}, \textbf{{\small SEOBNRv4PHM}} \\ 
{\small GW190513\_205428} & {\small IMRPhenomD}, {\small IMRPhenomPv2}, \textbf{{\small NRSur7dq4}}, {\small SEOBNRv4P}, \textbf{{\small SEOBNRv4PHM}} \\ 
{\small GW190514\_065416} & {\small IMRPhenomD}, {\small IMRPhenomPv2}, \textbf{{\small NRSur7dq4}}, {\small SEOBNRv4P} \\ 
{\small GW190517\_055101} & {\small IMRPhenomD}, {\small IMRPhenomPv2}, \textbf{{\small NRSur7dq4}}, {\small SEOBNRv4P}, \textbf{{\small SEOBNRv4PHM}} \\ 
{\small GW190519\_153544} & {\small IMRPhenomD}, {\small IMRPhenomPv2}, \textbf{{\small NRSur7dq4}}, {\small SEOBNRv4P}, \textbf{{\small SEOBNRv4PHM}} \\ 
{\small GW190521} & \textbf{{\small IMRPhenomPv3HM}}, \textbf{{\small NRSur7dq4}}, \textbf{{\small SEOBNRv4PHM}} \\ 
{\small GW190521\_074359} & {\small IMRPhenomD}, {\small IMRPhenomPv2}, \textbf{{\small NRSur7dq4}}, {\small SEOBNRv4P}, \textbf{{\small SEOBNRv4PHM}} \\ 
{\small GW190527\_092055} & {\small IMRPhenomD}, {\small IMRPhenomPv2}, \textbf{{\small NRSur7dq4}}, {\small SEOBNRv4P} \\ 
{\small GW190602\_175927} & {\small IMRPhenomD}, {\small IMRPhenomPv2}, \textbf{{\small NRSur7dq4}}, {\small SEOBNRv4P}, \textbf{{\small SEOBNRv4PHM}} \\ 
{\small GW190620\_030421} & {\small IMRPhenomD}, {\small IMRPhenomPv2}, \textbf{{\small NRSur7dq4}}, {\small SEOBNRv4P}, \textbf{{\small SEOBNRv4PHM}} \\ 
{\small GW190630\_185205} & {\small IMRPhenomD}, {\small IMRPhenomPv2}, \textbf{{\small NRSur7dq4}}, {\small SEOBNRv4P}, \textbf{{\small SEOBNRv4PHM}} \\ 
{\small GW190701\_203306} & {\small IMRPhenomD}, {\small IMRPhenomPv2}, \textbf{{\small NRSur7dq4}}, {\small SEOBNRv4P}, \textbf{{\small SEOBNRv4PHM}} \\ 
{\small GW190706\_222641} & {\small IMRPhenomD}, {\small IMRPhenomPv2}, \textbf{{\small NRSur7dq4}}, {\small SEOBNRv4P}, \textbf{{\small SEOBNRv4PHM}} \\ 
{\small GW190707\_093326} & {\small IMRPhenomD}, \textbf{{\small IMRPhenomPv2}}, \textbf{{\small SEOBNRv4P}}, {\small SEOBNRv4PHM} \\ 
{\small GW190708\_232457} & {\small IMRPhenomD}, {\small IMRPhenomPv2}, {\small SEOBNRv4P}, \textbf{{\small SEOBNRv4PHM}} \\ 
{\small GW190719\_215514} & {\small IMRPhenomD}, {\small IMRPhenomPv2}, \textbf{{\small NRSur7dq4}}, {\small SEOBNRv4P}, \textbf{{\small SEOBNRv4PHM}} \\ 
{\small GW190720\_000836} & {\small IMRPhenomD}, \textbf{{\small IMRPhenomPv2}}, \textbf{{\small SEOBNRv4P}}, {\small SEOBNRv4PHM} \\ 
{\small GW190727\_060333} & {\small IMRPhenomD}, {\small IMRPhenomPv2}, \textbf{{\small NRSur7dq4}}, {\small SEOBNRv4P} \\ 
{\small GW190728\_064510} & {\small IMRPhenomD}, \textbf{{\small IMRPhenomPv2}}, \textbf{{\small SEOBNRv4P}}, {\small SEOBNRv4PHM} \\ 
{\small GW190731\_140936} & {\small IMRPhenomD}, {\small IMRPhenomPv2}, \textbf{{\small NRSur7dq4}}, {\small SEOBNRv4P} \\ 
{\small GW190803\_022701} & {\small IMRPhenomD}, {\small IMRPhenomPv2}, \textbf{{\small NRSur7dq4}}, {\small SEOBNRv4P} \\ 
{\small GW190814} & {\small IMRPhenomD}, {\small IMRPhenomHM}, {\small IMRPhenomNSBH}, \textbf{{\small IMRPhenomPv3HM}}, {\small SEOBNRv4HM\_ROM}, \textbf{{\small SEOBNRv4PHM}}, {\small SEOBNRv4\_ROM}, {\small SEOBNRv4\_ROM\_NRTidalv2\_NSBH} \\ 
{\small GW190828\_063405} & {\small IMRPhenomD}, {\small IMRPhenomPv2}, \textbf{{\small NRSur7dq4}}, {\small SEOBNRv4P}, \textbf{{\small SEOBNRv4PHM}} \\ 
{\small GW190828\_065509} & {\small IMRPhenomD}, {\small IMRPhenomPv2}, {\small SEOBNRv4P}, \textbf{{\small SEOBNRv4PHM}} \\ 
{\small GW190909\_114149} & {\small IMRPhenomD}, {\small IMRPhenomPv2}, {\small SEOBNRv4P}, \textbf{{\small SEOBNRv4PHM}} \\ 
{\small GW190910\_112807} & {\small IMRPhenomD}, {\small IMRPhenomPv2}, {\small SEOBNRv4P}, \textbf{{\small SEOBNRv4PHM}} \\ 
{\small GW190915\_235702} & {\small IMRPhenomD}, \textbf{{\small IMRPhenomPv2}}, \textbf{{\small SEOBNRv4P}}, {\small SEOBNRv4PHM} \\ 
{\small GW190924\_021846} & {\small IMRPhenomD}, \textbf{{\small IMRPhenomPv2}}, \textbf{{\small SEOBNRv4P}}, {\small SEOBNRv4PHM} \\ 
{\small GW190929\_012149} & {\small IMRPhenomD}, \textbf{{\small IMRPhenomPv2}}, \textbf{{\small SEOBNRv4P}}, {\small SEOBNRv4PHM} \\ 
{\small GW190930\_133541} & {\small IMRPhenomD}, \textbf{{\small IMRPhenomPv2}}, \textbf{{\small SEOBNRv4P}}, {\small SEOBNRv4PHM} \\ 
\end{tabularx}

%% file: waveform_recon_appendix.tex
\section{Waveform Consistency Tests}
\label{appendix:WCT}

There are several different quantitative measures that can be used to measure waveform consistency. These include
the residual \ac{SNR}, which is found by subtracting the best-fit waveform template $h_\mathrm{T}$ from the data $d$, then applying
minimally-modeled methods to search for any coherent excess in the noise residual $r=d-h_\mathrm{T}$. Additional measures of waveform
consistency include the distance between waveforms, $\Delta^2\langle h_1,h_2\rangle = \langle h_1 - h_2| h_1 - h_2\rangle$ and the match, or
overlap, ${\cal O}\langle h_1,h_2\rangle$. 

The residuals test has been applied as test of general relativity~\cite{TheLIGOScientific:2016src, LIGOScientific:2019fpa}, 
at least within the precision with which the waveform models approximate general relativity. 
Distance and match provide more sensitive measures of the waveform
consistency than the residuals test since the extrinsic parameters of the source, such as
the arrival time, sky location and polarization, are constrained by the full
signal, while for the residuals test the extrinsic parameters have to be
constrained from the (usually small) difference between the signal and the
template. To compare the signal reconstructions for the current catalog of sources we 
adopt the waveform match as our measure of waveform consistency
since it does not depend on the overall amplitude of the signals, making it a 
convenient choice
when comparing events with a range of amplitudes. 

To make a quantitative assessment of the waveform match values we need to
know how the match depends on quantities such as the \ac{SNR}
and time--frequency volume of the signals. Instrument noise will lead to
non-zero mis-matches, ${\rm MM} = 1 - {\cal O}$, even when using perfect
templates. For example, the maximum likelihood solution for a perfect
template will have a mis-match with mean and variance given by~\cite{Chatziioannou:2017tdw}
\begin{equation}
{\rm E}[{\rm MM}] \simeq  \frac{D-1}{2 \, {\rm SNR}^2}, \quad \quad
{\rm Var}[{\rm MM}] \simeq  \frac{D-1}{2 \, {\rm SNR}^4},
\end{equation}
where $D$ is the number of parameters that define the signal model, and 
where the reduction from $D$ to $D-1$ is because the match is independent from the
overall amplitude of the signal.
For the minimally-modeled waveform
reconstructions the distributions of match values are more difficult
to predict. Using simulations it has been found that the mis-match
decreases with \ac{SNR}, but more slowly than for templates since the
effective dimension of the model increases with SNR~\cite{Littenberg:2015kpb}.
The mis-match also scales with the time--frequency volume. For binary
systems of a given \ac{SNR}, the mis-match will generally be smaller for
high mass systems~\cite{Littenberg:2015kpb,Becsy:2016ofp,Pannarale:2018cct}.
Given these complexities, we chose to empirically estimate the match
distribution from simulations for each event. As a proxy for the signal we use
fair draws from the on-source template-based analysis and inject these into
data surrounding the event; the right ascensions for the simulated signals are
adjusted such that simulated source is at the same sky location in the
frame of the detectors. For the majority of events the waveform model used for
the injections is IMRPhenomPv2. The exceptions are \NAME{GW190412A},
\NAME{GW190521A}, \NAME{GW190814A}\, where IMRPhenomPv3HM
was used.

\begin{figure*}[htb]
\centering
\includegraphics[width=\textwidth]{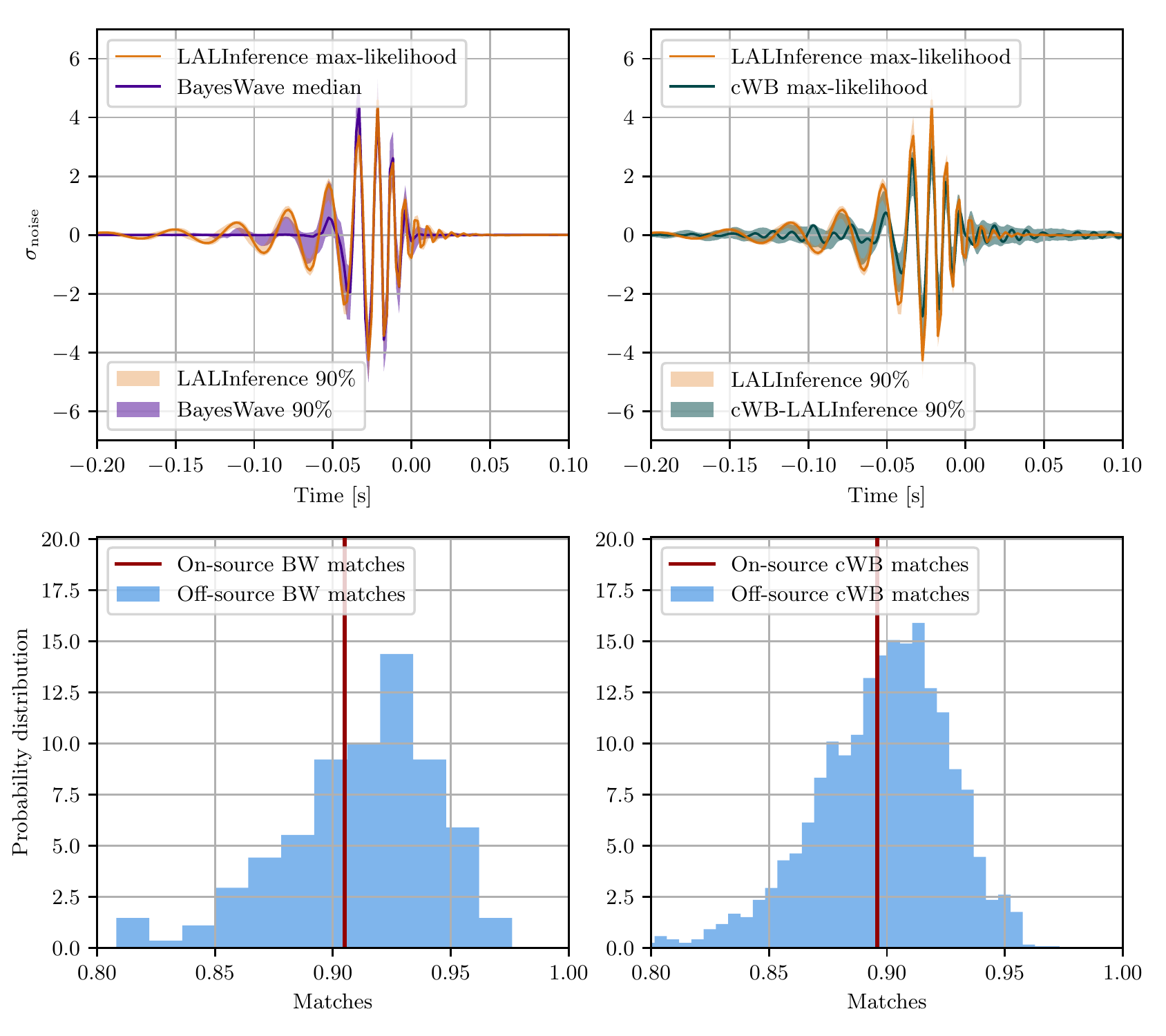} 
\caption{The upper panels show waveform reconstructions for \protect\NAME{GW190519A}\ in the LIGO
Livingston detector. The waveform posterior from the template-based analysis
(shown in orange) is compared to the BW analysis in the upper left panel, and to
the cWB analysis in the upper right panel. 
The panels also show the 90\% credible bands for the Bayesian LALInference and BW algorithms and the 90\% confidence band for \CWB\ derived from off-source injections, i.e., by injecting samples from the template-based analysis into data surrounding the event and repeating the analysis multiple times (these bands are computed on an individual time sample basis).  
The lower two panels show
the distribution of overlap values when running BW and cWB on waveforms drawn
from the template based analysis that are injected into data surrounding the
event. The fraction of runs with matches below that of the on-source analysis
give the p-value for the event.}
\label{Fig:bvf}
\end{figure*}

The \CWB\ and BW
algorithms are used to produce point estimates for the waveform
reconstructions for each of the simulated events. For \CWB\ the point
estimate is a constrained maximum likelihood reconstruction, while for BW
the point estimate is the median of the waveform posterior distribution.
Fig.~\ref{Fig:bvf} illustrates the results obtained for \NAME{GW190519A}. The upper
panels compare the template-based \LALINFERENCE\ waveform reconstruction in the LIGO Livingston detector to the
minimally-modeled BW and  \CWB\  reconstructions. The solid lines are point estimates for the waveforms: for \LALINFERENCE\
the maximum likelihood; for BW the median of posterior draws; and for \CWB\ the constrained maximum likelihood.
The panels also show the 90\% credible bands for the Bayesian \LALINFERENCE\ and BW algorithms and the 90\% confidence band
for  \CWB\  derived by injecting samples
from the template based analysis into data surrounding the event and repeating
the analysis multiple times (these bands are computed on an individual time sample basis). The lower panels of
Fig.~\ref{Fig:bvf} show the distribution of overlaps found when running BW and \CWB\  on simulated data with similar properties
to the event. 
Waveforms drawn from the on-source \LALINFERENCE\ analysis were injected into data surrounding the event.
The overlap between the injected waveform and the point estimates from the BW and \CWB\  analyses of these injections were then used to
produce the histograms seen in Fig.~\ref{Fig:bvf}. The distribution of the match values 
defines a null distribution for each detected event, which takes into account the variability of the \LALINFERENCE\ posterior distribution, the fluctuations of the detector
noise, and the waveform reconstruction errors. The fraction of off-source analyses with overlaps below the on-source
match values, which are shown as vertical lines in the lower panels of Fig.~\ref{Fig:bvf}, define the p-value for this event.

\begin{figure}[t]
\centering
\includegraphics[width=\columnwidth]{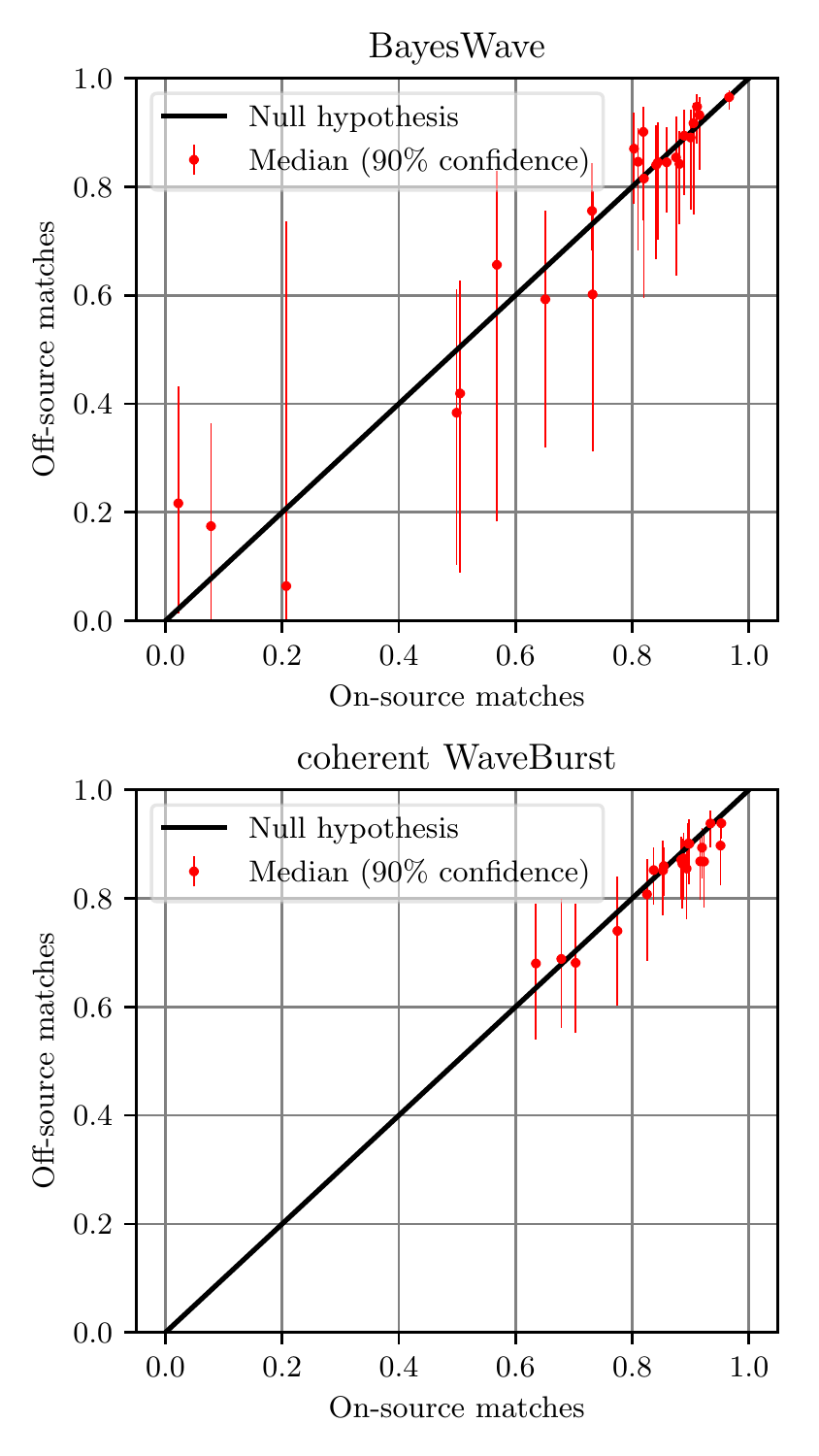} 
\caption{Off-source vs.\ on-source
match values for the candidate events in O3a. The upper panel displays
the results of the BW analysis, and the lower panel shows the results
of the \CWB\ analysis. The on-source match on the horizontal axis is the value
obtained comparing the maximum likelihood waveform from parameter estimation with point estimates
from the minimally-modeled waveform reconstructions. The off-source match on
the vertical axis is the median value of the match distribution obtained from
off-source injection of sample waveforms from the template-based posterior
distribution.  In both panels, the errors bars denote the 90\% equal-tailed 
confidence interval.
}
\label{Fig:match}
\end{figure}

The same analysis procedure was repeated for a subset of additional events. \CWB\  uses 
only events that are above the \CWB\ search thresholds (resulting in a morphology-dependent 
\ac{SNR} threshold which is about 7--10 for the events reported in this catalog), while for BW 
the analysis was restricted to events where the on-source BW analysis yielded  
${\rm SNR} > 7$. 
Fig.~\ref{Fig:match} shows the on-source match values
vs. the off-source median match values with 90\% intervals. The upper panel shows the 
results of the BW analysis, while the lower panel shows the results of the \CWB\ analysis. 
In both cases the p-values point to a  good agreement between 
the minimally-modeled and template-based reconstructions. 

The discrepancies between the two plots may be ascribed to different choices made in the 
two reconstruction algorithms. \CWB\ is both a detection and a reconstruction pipeline. 
For this reason, reconstructions are performed with the same production settings used in 
searches. The production settings are optimized for noise rejection and to enforce strong 
network coherency constraints. To construct the match 
distributions, \CWB\ uses about 2000--3000 waveform injections per event; however, those 
injections that are below the \CWB{} thresholds are not reconstructed: in the majority of 
cases reported here (13) the 
reconstruction efficiency is greater than 80\%, while in a few other cases (7) the efficiency 
ranges from 15\% to 50\%. 
For each event, this efficiency depends on the variability of the noise background. 
In cases of lower efficiency, we have also checked that the waveforms that 
are successfully reconstructed by \CWB{} have parameter distributions that are statistically
indistinguishable from those of the injected waveforms.

BW employs Bayesian inference to characterize detections made by the CBC and \CWB\ search 
pipelines. As such, no cuts are made on the waveform reconstructions, which means that for 
quiet signals, some of the samples will be drawn from the prior distribution, resulting 
in a wide spread in the match distributions.

The distribution of match values for each event are used to compute p-values. 
The overall consistency of the template based and minimally-modeled waveform 
reconstructions can then be summarized by plotting the p-values, ranked in 
ascending order, against the theoretical distribution. In such plots, any 
significant deviations \textit{below} the plot diagonal point to events that should 
undergo further analysis. The p-values for the events reconstructed by \CWB\ 
and BW are shown in Fig.~\ref{Fig:pvalue} in the main text. Applying the
Fisher test~\cite{FisherTest} to the ensemble of p-values yields combined p-values of 0.57 for the
BW analysis and 0.99 for the \CWB\ analysis, indicating that there is no reason to
reject the null hypothesis that the template based and minimally-modeled analysis
are in agreement. The Fisher test is one-sided in that it only penalizes p-values
that are lower than expected. The \CWB\ analysis includes instances where
the p-values are higher than predicted, indicating that the on-source matches are
higher than expected based on the off-source distributions. The cause of this 
bias is likely due to an asymmetry between the on-source and off-source
analyses. The on-source analysis computes the match between the maximum likelihood
template-based waveform and its reconstruction, while the off-source analysis
computes the match between the injected template and its reconstruction. 
Ideally the analysis would be symmetric, with the
maximum likelihood template used both on-source and off-source, but the computational
cost of running the full template-based analysis on the thousands of off-source
injections is prohibitively expensive.

%% file: source_parameters_appendix.tex
\section{Cosmological distance resampling}
\label{appendix:sourceparams}

For the luminosity distance, a prior that goes as $\DL^2$ is enforced in
the sampling, following the same procedure as described in previous publications.
As this assumption becomes increasingly unrealistic as events are detected at
greater distances, the posterior distributions shown in this paper are derived from
a physically motivated prior that incorporates cosmological effects.
We perform rejection sampling on the initial posterior samples to instead use a
prior corresponding to a uniform merger rate per comoving volume in the rest frame of the source.
Using a standard flat $\mathrm{\Lambda{}CDM}$
cosmology, samples are accepted according to the weight
\begin{align}
    w(z)&\propto \frac{1}{(1+z)} \frac{\mathrm{d}V_\mathrm{c}}{\mathrm{d}V_\mathrm{E}}\\ \nonumber
    &\propto (1+z)^{-2} \left(\frac{D_\mathrm{L}E(z)}{D_\mathrm{H}}+(1+z)^2\right)^{-1}.
\end{align}
The initial $1/(1+z)$ factor accounts for time dilation of the observed merger rate,
$\mathrm{d}V_\mathrm{c}$ is the comoving volume element, and $\mathrm{d}V_\mathrm{E}=\DL^2\mathrm{d}\Omega$
is the Euclidean volume element. $D_\mathrm{H}=c/H_0$ is the Hubble distance and
$E(z)\simeq\sqrt{\Omega_{\rm m}(1+z)^3+\Omega_\Lambda}$ for $\mathrm{\Lambda{}CDM}$,
and we use Hubble constant $H_0=67.9~\mathrm{km\,s^{-1}\,Mpc^{-1}}$ and
matter density $\Omega_{\rm m}=0.3065=1-\Omega_\Lambda$~\cite{Ade:2015xua}.